\documentclass[10pt,a4paper]{article}

\usepackage[T1]{fontenc}
\usepackage[utf8]{inputenc}
\usepackage{amssymb,amsmath,amsthm,amsfonts}    %
\usepackage{bm}                                 %
\usepackage{mathtools}                          %
\usepackage{stmaryrd}                           %
\usepackage{breqn}                              %
\usepackage[textfont=footnotesize,labelfont={footnotesize,bf}]{caption}
\usepackage{subcaption}
\usepackage{booktabs}
\usepackage[inline]{enumitem}
\usepackage{graphicx}
\usepackage[dvipsnames]{xcolor}
\usepackage[top=22mm,right=22mm,left=22mm,bottom=22mm]{geometry}
\usepackage[colorlinks=true,linkcolor=blue,urlcolor=blue]{hyperref}
\usepackage[affil-it,auth-sc]{authblk}          %
\usepackage[colorinlistoftodos,textwidth=25mm,textsize=footnotesize]{todonotes}
\usepackage{lipsum}                             %
\usepackage[
    backend=biber,
    style=numeric,
    citestyle=numeric-comp,
    maxbibnames=99,
    minbibnames=99,
    maxcitenames=2,
    mincitenames=1,
    giveninits=true,
    sorting=none,
    sortlocale=auto,
    natbib=true,
    url=false,
    isbn=false,
    doi=true,
    eprint=true
]{biblatex}
\usepackage{fancyhdr}
\DeclareMathOperator{\tr}{tr}

\DeclareMathOperator{\diver}{div}
\DeclareMathOperator{\Div}{Div}
\DeclareMathOperator{\grad}{grad}

\DeclarePairedDelimiter{\abs}{\lvert}{\rvert}
\DeclarePairedDelimiter{\norm}{\lVert}{\rVert}

\newcommand{\scalp}{\bm \cdot}
\newcommand{\trans}[1]{#1^{\intercal}}
\newcommand{\conj}[1]{\overline{#1}}
\newcommand{\conjtrans}[1]{#1^{\mathsf{H}}}
\newcommand{\deriv}[2]{\frac{\partial #1}{\partial #2}}
\newcommand{\nderiv}[3]{\frac{\partial^{\,#3} #1}{\partial #2^{\,#3}}}

\newcommand{\Reals}{\mathbb{R}}

\newcommand{\Integers}{\mathbb{Z}}

\newcommand{\bzero}{\bm 0}
\newcommand{\ba}{\bm a}
\newcommand{\bb}{\bm b}

\newcommand{\be}{\bm e}
\newcommand{\bef}{\bm f}
\newcommand{\bg}{\bm g}

\newcommand{\bk}{\bm k}

\newcommand{\bn}{\bm n}

\newcommand{\bp}{\bm p}
\newcommand{\bq}{\bm q}

\newcommand{\bt}{\bm t}
\newcommand{\bu}{\bm u}

\newcommand{\bx}{\bm x}
\newcommand{\by}{\bm y}

\newcommand{\bA}{\bm A}
\newcommand{\bB}{\bm B}

\newcommand{\bE}{\bm E}

\newcommand{\bI}{\bm I}
\newcommand{\bJ}{\bm J}
\newcommand{\bK}{\bm K}
\newcommand{\bL}{\bm L}
\newcommand{\bM}{\bm M}
\newcommand{\bN}{\bm N}

\newcommand{\bP}{\bm P}

\newcommand{\bS}{\bm S}
\newcommand{\bT}{\bm T}

\newcommand{\bZ}{\bm Z}
\newcommand{\balpha}{\bm \alpha}

\newcommand{\bvarphi}{\bm \varphi}

\newcommand{\bXi}{\bm \Xi}

\newcommand{\bGamma}{\bm \Gamma}

\newcommand{\fC}{\mathbb C}

\newcommand{\fE}{\mathbb E}

\newcommand{\mB}{\mathcal B}
\newcommand{\mC}{\mathcal C}

\newcommand{\mE}{\mathcal E}

\newcommand{\mG}{\mathcal G}

\newcommand{\mL}{\mathcal L}

\newcommand{\mO}{\mathcal O}

\newcommand{\mS}{\mathcal S}
\newcommand{\mT}{\mathcal T}

\newcommand{\mV}{\mathcal V}

\graphicspath{{./}{./figures/}}                 %

\title{Dynamics of prestressed elastic lattices: homogenization, instabilities, and strain localization}

\author[1]{G. Bordiga}
\author[2]{L. Cabras}
\author[1]{A. Piccolroaz}
\author[1]{D. Bigoni\footnote{Corresponding author: e-mail: \href{mailto:bigoni@ing.unitn.it}{bigoni@ing.unitn.it}; phone: +39\,0461\,282507.}}

\affil[1]{DICAM, University of Trento, Trento, Italy}
\affil[2]{DICATAM, University of Brescia, Brescia, Italy}

\date{}

\begin{document}

\maketitle
\thispagestyle{fancy} %

\begin{abstract}
    \noindent
    A lattice (or `grillage') of elastic Rayleigh rods (possessing a distributed mass density, together with rotational inertia) organized in a parallelepiped geometry can be axially loaded up to an arbitrary amount without distortion and then be subject to incremental time-harmonic dynamic motion.
    At certain threshold levels of axial load, the grillage manifests instabilities and displays non-trivial axial and flexural incremental vibrations.
    Including every possible structural geometry and for an arbitrary amount of axial stretching, Floquet-Bloch wave asymptotics is used to homogenize the in-plane mechanical response, so to obtain an equivalent prestressed elastic solid subject to incremental time-harmonic vibration, which includes, as a particular case, the incremental quasi-static response.
    The equivalent elastic solid is obtained from its acoustic tensor, directly derived from homogenization and shown to be independent of the rods’ rotational inertia.
    Loss of strong ellipticity in the equivalent continuum coincides with macro-bifurcation in the lattice, while micro-bifurcation remains undetected in the continuum and corresponds to a vibration of vanishing frequency of the lowest dispersion branch of the lattice, occurring at finite wavelength.
    Dynamic homogenization reveals the structure of the acoustic branches close to ellipticity loss and the analysis of forced vibrations (both in physical space and Fourier space) shows low-frequency wave localizations.
    A perturbative approach based on dynamic Green's function is applied to both the lattice and its equivalent continuum.
    This shows that only macro-instability corresponds to localization of incremental strain, while micro-instabilities occur in modes which spread throughout the whole lattice with an `explosive' character.
    In particular, extremely localized mechanical responses are found both in the lattice and in the solid, with the advantage that the former can be easily realized, for instance via 3D printing.
    In this way, features such as shear band inclination, or the emergence of a single shear band, or competition between micro and macro instabilities become all designable features.
    The comparison between the mechanics of the lattice and its equivalent solid shows that the homogenization technique allows an almost perfect representation, except when micro-bifurcation is the first manifestation of instability.
    Therefore, the presented results pave the way for the design of architected cellular materials to be used in applications where extreme deformations are involved.
\end{abstract}

\paragraph{Keywords}
Dynamic homogenization \textperiodcentered\
Ellipticity loss \textperiodcentered\
Shear bands \textperiodcentered\
Lattice buckling \textperiodcentered\
Bloch waves

\section{Introduction}
\label{sec:introduction}
Shear banding and strain localizations, usually found to emerge before failure of materials, are typically accompanied by large plastic deformation, sudden formation of elastic unloading zones, damage, and possibly fracture.
Mechanical features depend on the tested material, for instance, shear bands in rocks and in metals have different inclinations, and tests cannot usually be repeated on the same specimen, because the sample has to be brought to failure.
Shear bands represent an ultimate instability mode, so that their modeling is complicated by sophisticated, and often phenomenological, elastoplastic constitutive laws, used beyond several bifurcation thresholds.

Imagine now a material in which shear banding and other instabilities may occur well inside the elastic range and far from failure.
A material that can be designed to produce shear bands with the desired inclination, or in which shear bands are the first instability occurring at increasing stress, or in which the anisotropy (not imperfections) allows the formation of only one shear band.
Imagine that this material would be characterized by rigorously determined elastic constitutive laws (thus avoiding complications such as the double branch of the incremental constitutive laws of plasticity) and would be, at least in principle, a material realizable (for instance via 3D printing technology) and testable in laboratory conditions.
This material would be ideal not only to \textit{theoretically} analyze instabilities, but also to \textit{practically} realize the `architected materials' which are preconized to yield extreme mechanical properties such as foldability, channeled response, and surface effects~\cite{overvelde_2017,kochmann_2017,rafsanjani_2019}.
The crucial step towards the definition of a class of these materials was made by Triantafyllidis~\cite{triantafyllidis_1985,geymonat_1993,triantafyllidis_1993,triantafyllidis_1998,nestorovic_2004,santisidavila_2016} and Ponte~Casta\~neda~\cite{pontecastaneda_1989,pontecastaneda_1991,pontecastaneda_1996,pontecastaneda_1997,pontecastaneda_2002,pontecastaneda_2002a,lopez-pamies_2006,lopez-pamies_2006a,avazmohammadi_2016}, who laid down a general framework for the homogenization of elastic composites and for the analysis of bifurcation and strain localization in these materials.
In particular, they showed how to realize an elastic material displaying a prestress-sensitive incremental response, exactly how it is \textit{postulated} for nonlinear elastic solids subject to incremental deformation.
Moreover, they provided a new understanding of strain localization phenomena, showing that a global (called in the following `macro') bifurcation of a lattice structure corresponds to a loss of ellipticity of the equivalent continuum, while the latter is unaffected by a local (called in the following `micro') bifurcation occurring in the composite.

\begin{figure}[htb!]
    \centering
    \includegraphics[width=0.98\linewidth]{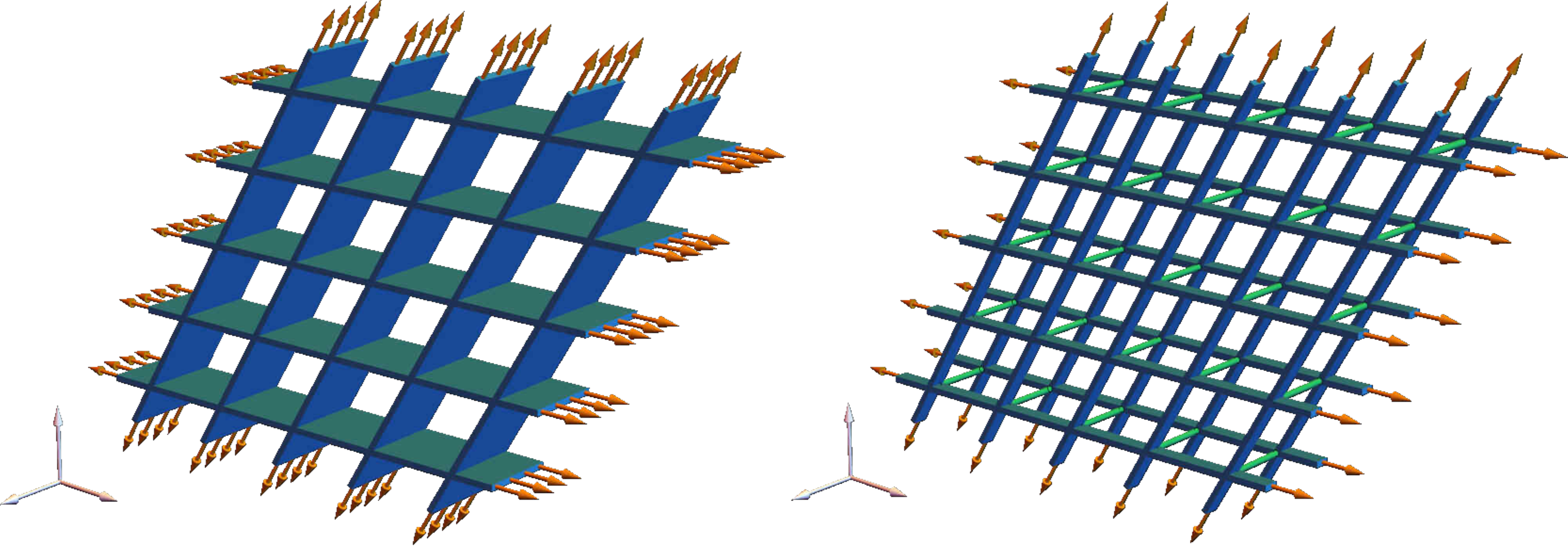}
    \caption{\label{fig:sketch_3D_60}
        Two ways to realize a 3D sample from a 2D lattice.
        Left: the rods are strips with an out-of-plane thickness sufficient to eliminate out-of-plane buckling.
        Right: Two (or more) grillages of rods are transversely connected at the nodes with revolute joints which do not transmit torques.
    }
\end{figure}

\begin{figure}[htb!]
    \centering
    \begin{subfigure}{0.32\textwidth}
        \centering
        \phantomcaption{\label{fig:straws_0}}
        \includegraphics[width=0.98\linewidth]{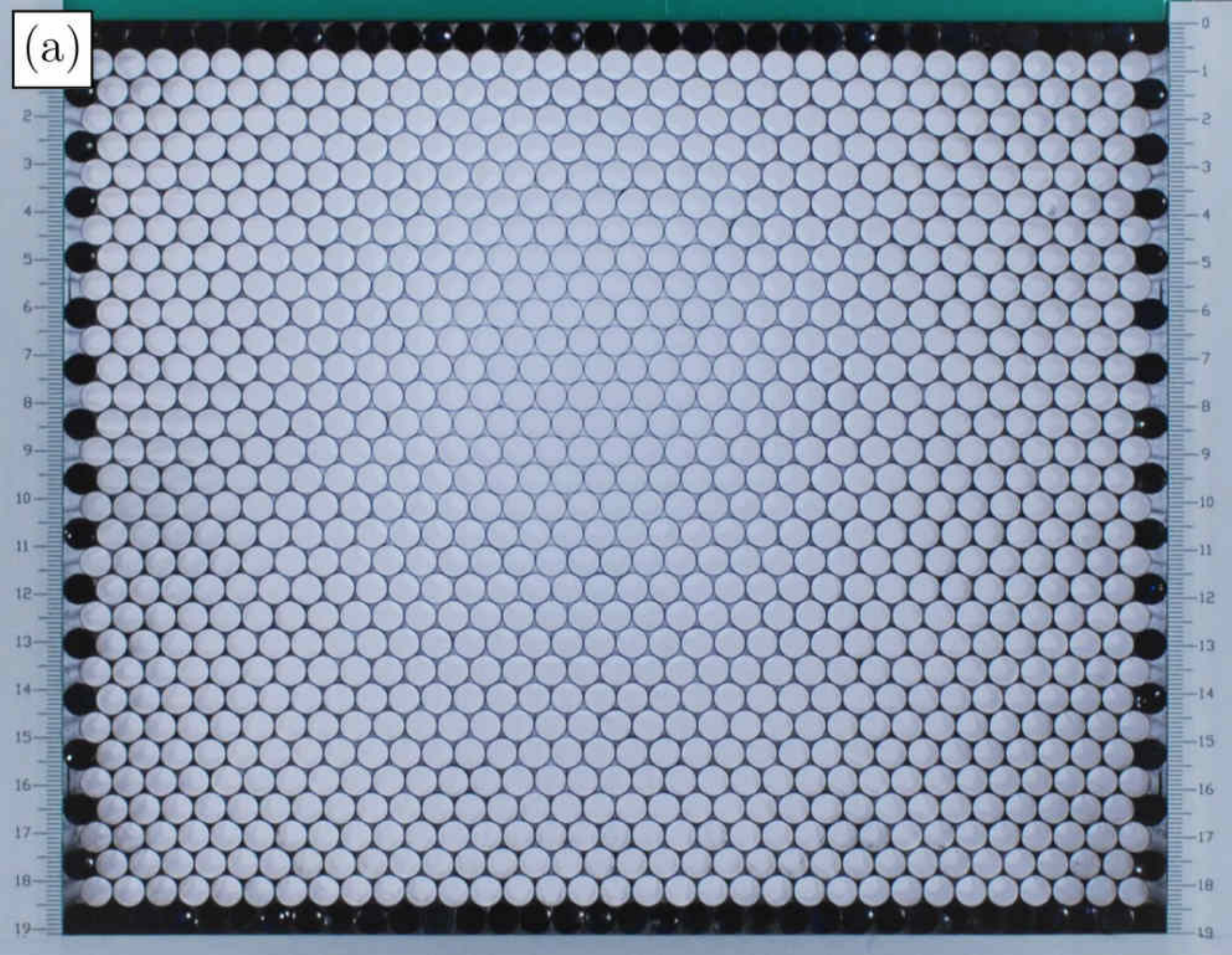}
    \end{subfigure}
    \begin{subfigure}{0.32\textwidth}
        \centering
        \phantomcaption{\label{fig:straws_1}}
        \includegraphics[width=0.98\linewidth]{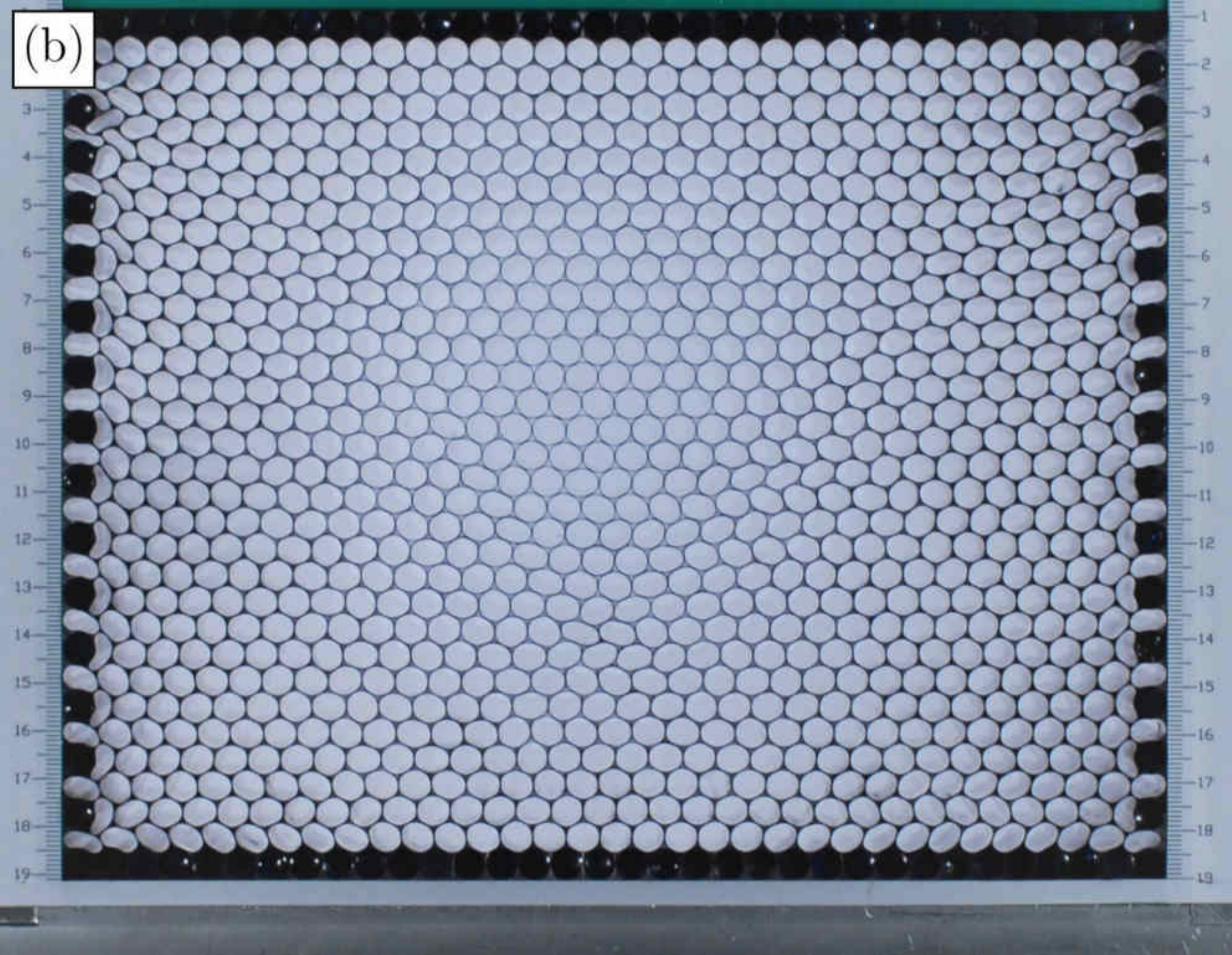}
    \end{subfigure}
    \begin{subfigure}{0.32\textwidth}
        \centering
        \phantomcaption{\label{fig:straws_micro_buckling}}
        \includegraphics[width=0.98\linewidth]{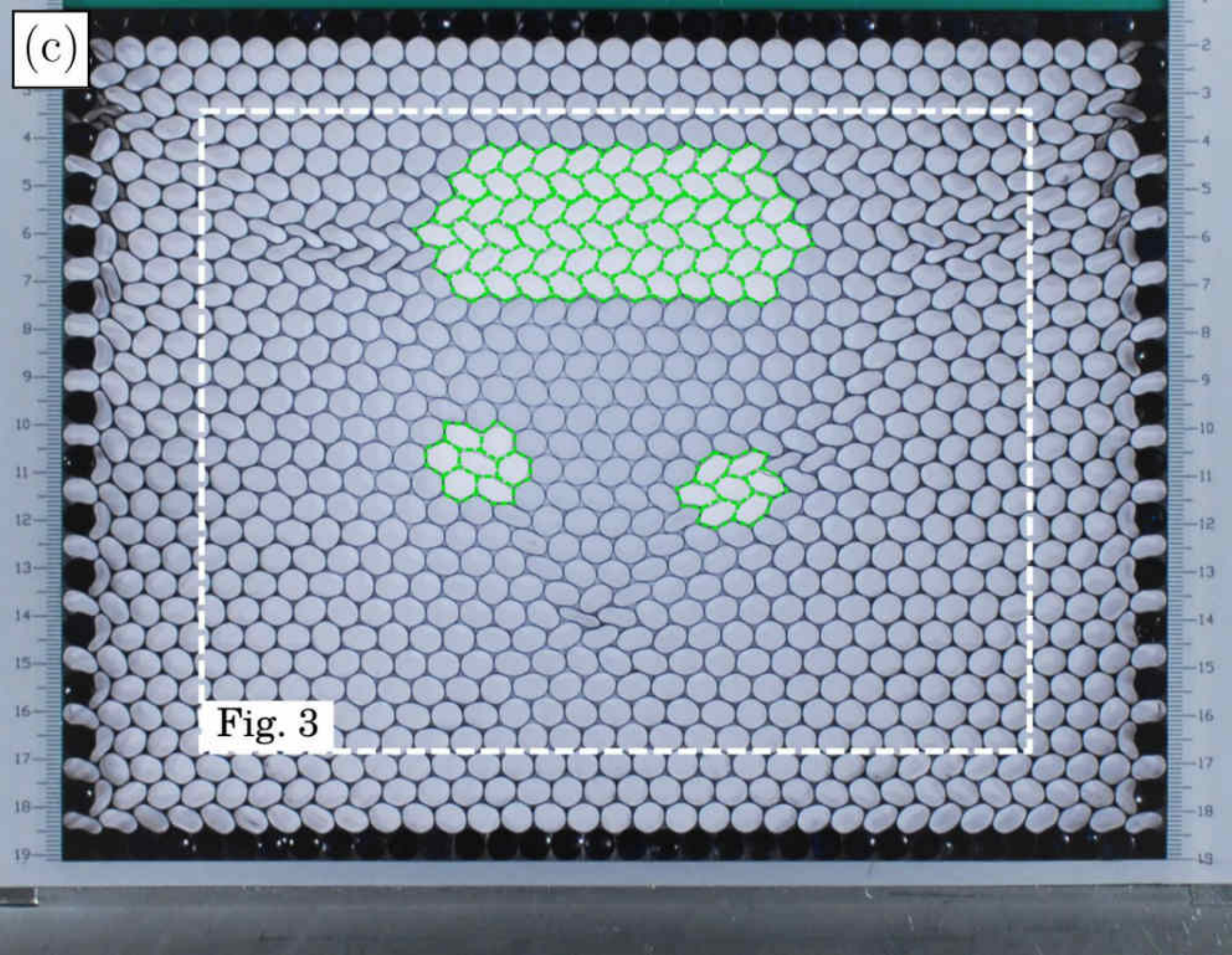}
    \end{subfigure}\\ \vspace{2mm}
    \begin{subfigure}{0.32\textwidth}
        \centering
        \phantomcaption{\label{fig:straws_band_formation}}
        \includegraphics[width=0.98\linewidth]{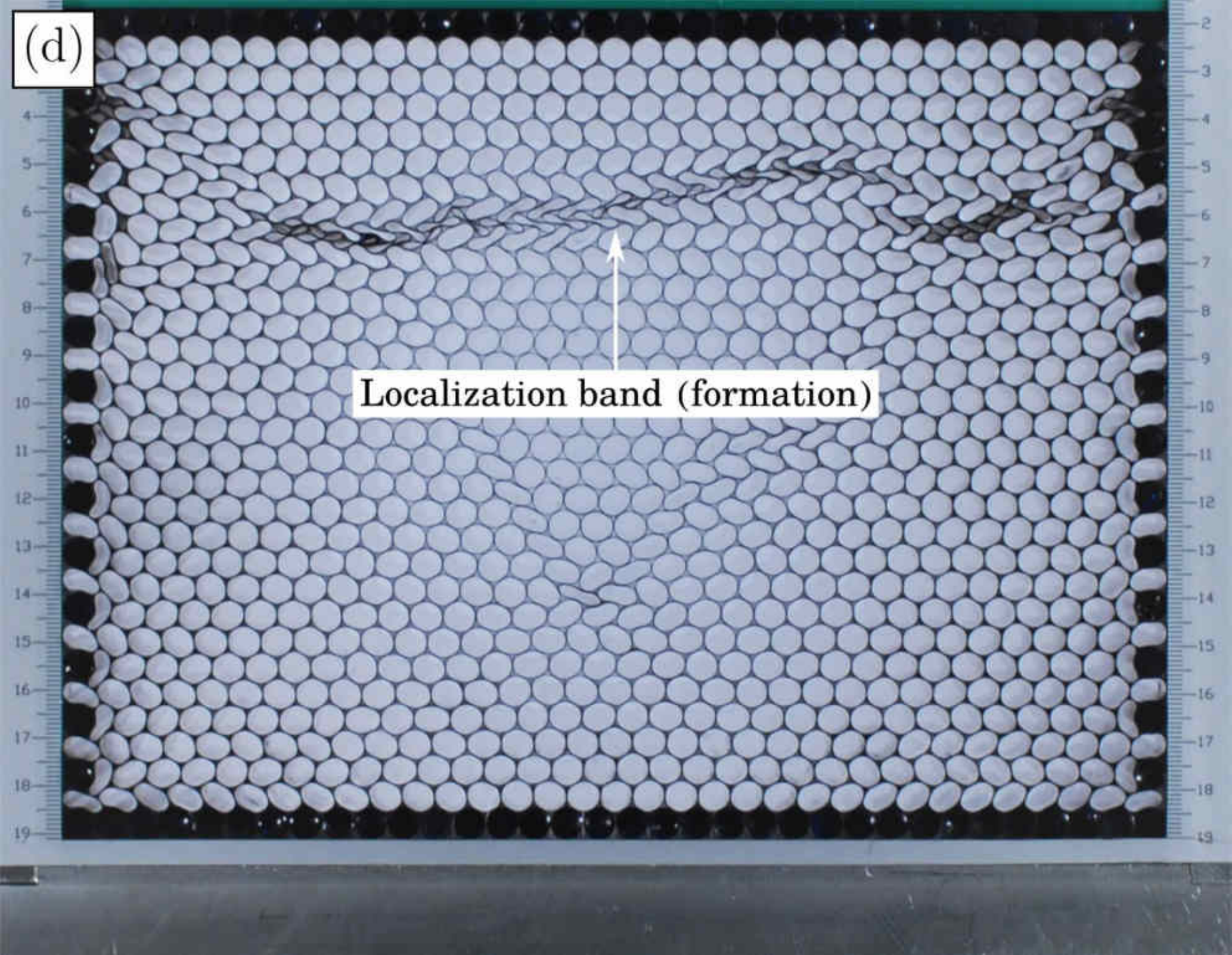}
    \end{subfigure}
    \begin{subfigure}{0.32\textwidth}
        \centering
        \phantomcaption{\label{fig:straws_4}}
        \includegraphics[width=0.98\linewidth]{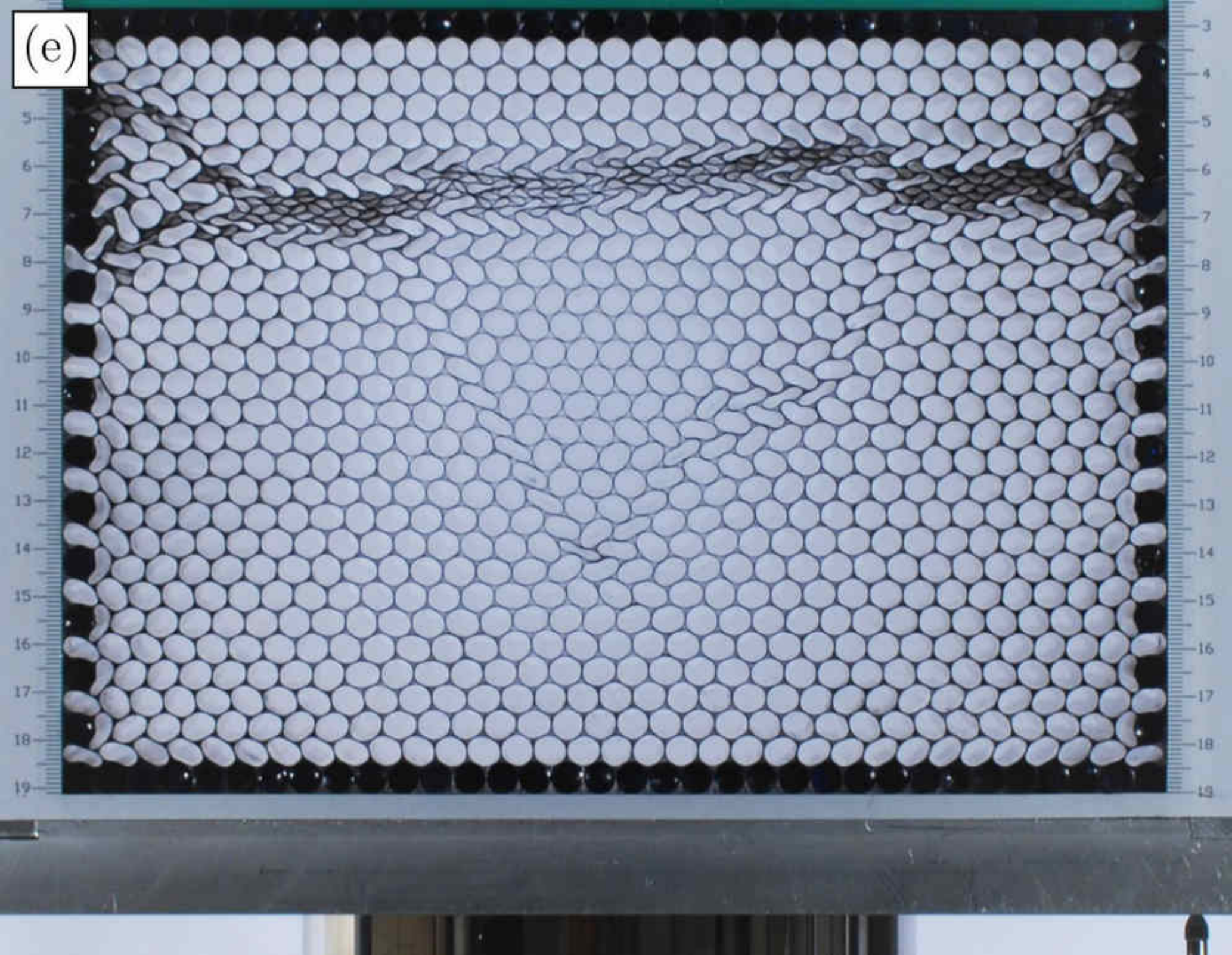}
    \end{subfigure}
    \begin{subfigure}{0.32\textwidth}
        \centering
        \phantomcaption{\label{fig:straws_band_accumulation}}
        \includegraphics[width=0.98\linewidth]{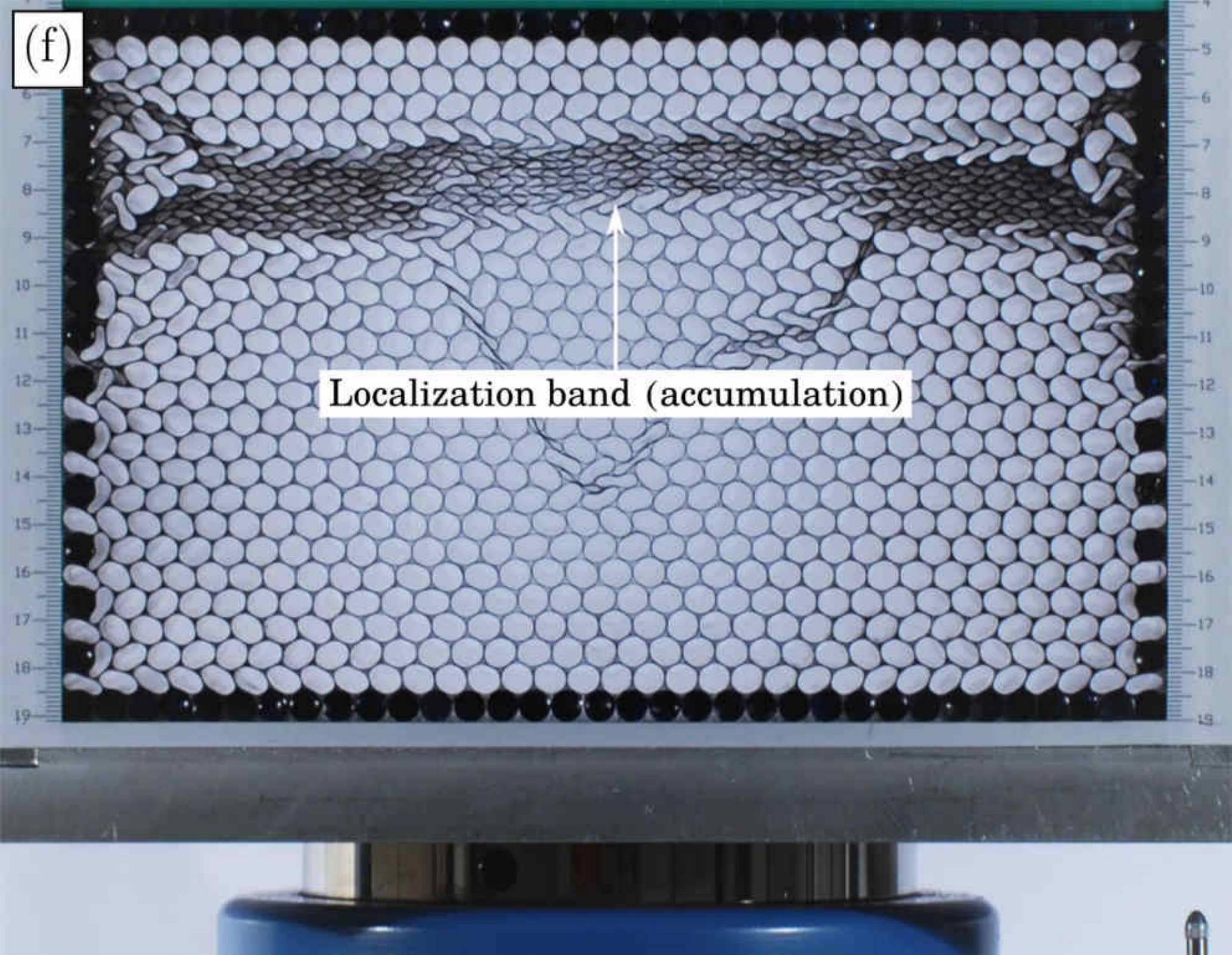}
    \end{subfigure}
    \caption{\label{fig:straws_localization}
        Emergence of a periodic micro-bifurcation (ovalization of the straws' cross-sections, \subref{fig:straws_micro_buckling}), subsequent strain localization (collapse of the straws' cross-sections, \subref{fig:straws_band_formation}), and final strain accumulation (\subref{fig:straws_4} and~\subref{fig:straws_band_accumulation}) during uniaxial deformation of an initially (\subref{fig:straws_0}) hexagonal packing of drinking straws.
        The early deformation (\subref{fig:straws_1}) is almost homogeneous.
    }
\end{figure}

The mentioned findings are extended in the present article to lattices of elastic rods of arbitrary geometry and subject to a nonlinear axial strain of arbitrary amount but leaving the rods in an undistorted state.
The material composing the rods can be an arbitrary nonlinear elastic material, for instance Mooney-Rivlin or characterized by two elastic shear moduli (see Appendix~\ref{sec:linearized_elastica}).
The rods are connected to each other through nodes able to transmit bending moment, so that when the grillage is subject to time-harmonic incremental vibrations, the rods are subject to incremental axial and shear forces and bending moments.
The lattice is idealized as two-dimensional and infinite.
The former assumption does neither mean that the lattice can buckle out-of-plane, nor that a three-dimensional sample cannot be designed.
Rather, the out-of-plane thickness may be large enough to avoid out-of-plane buckling, Fig.~\ref{fig:sketch_3D_60} (left, see also the practical realization reported in Fig. \ref{fig:straws_localization}), or different grids can be connected at nodes with transverse revolute joints unable to transmit torques, Fig.~\ref{fig:sketch_3D_60} (right).
The assumption that the grid is of infinite extent has a twofold origin, namely, it is crucial for the homogenization of periodic media and is the way to the analysis of material instabilities in the equivalent continuum without encountering earlier instabilities.
Technically, this occurs in the so-called van Hove conditions, so that either an infinite medium or a specimen with prescribed displacements over the whole boundary has to be considered~\cite{bigoni_2012}.
Under the mentioned assumptions, a systematic analysis of shear band formation and localization is developed, by applying a perturbative approach \cite{bigoni_2002a}, both to the lattice and to its equivalent continuum.

An asymptotic homogenization scheme, based on the Floquet-Bloch wave technique, is developed for a generic lattice\footnote{
    For the geometries investigated in \cite{triantafyllidis_1993} our homogenization approach provides exactly the same results.
}.
Homogenization techniques based on the asymptotic analysis of wave solutions dates back to Brillouin \cite{brillouin_1946} and Born \cite{born_1955}, and has received significant contributions in recent years when the case of random and periodic media has been considered \cite{parnell_2007,willis_2009,craster_2010,willis_2011,willis_2012,nemat-nasser_2011,nassar_2015,nassar_2017,kutsenko_2017} and extended to the high-frequency regime \cite{norris_2012,meng_2018,guzina_2019}.
With the exception of \cite{kutsenko_2017}, these developments have been so far produced for the analysis of wave propagation in continuous materials, not in structures, so that their practical implementation required the systematic use of numerical techniques (typically finite elements).
It is shown in the present article that low-frequency effective properties can be derived analytically for lattices composed of rods (incrementally loaded in-plane and subject to axial, shear and bending forces) through a direct computation of the wave asymptotics\footnote{
    Our mathematical setting is two-dimensional for simplicity, but the three-dimensional extension is straightforward once the linearized dynamics of the rods is specified.
}.
Recent results on beam grillages \cite{kutsenko_2017} are here extended to the case of elastic lattices, axially stressed up to an arbitrary amount, whose incremental dynamics is derived without restrictions on the rods' constitutive law and without neglecting the rotational inertia of the rods' cross-section.
The low-frequency asymptotics of waves propagating in the lattice is shown to be governed by the spectral properties of the acoustic tensor associated to an effective continuum, equivalent to the discrete structure.
This continuum is prestressed and is obtained from the acoustic tensor, to which the Floquet-Bloch asymptotics directly leads.
In this way, the correspondence between loss of ellipticity in the effective continuum and degeneracy of the acoustic properties in the lattice is \textit{directly} demonstrated.

The elastic rods are characterized by an axial mass density, equipped with rotational inertia (the Rayleigh model~\cite{piccolroaz_2017a,bordiga_2019a,nieves_2019}), so that it is possible to prove that the latter does not influence the vibrational properties at low frequency, expressed by the acoustic tensor of the equivalent continuum.
The homogenization, obtained analytically, is exploited to investigate the lattice response near macro-instability (coincident with the failure of strong ellipticity and thus of ellipticity in the effective continuum), therefore unveiling features of lattice dynamics loaded up to the verge of shear band formation.

The response to the application of a pulsating concentrated force (the infinite-body time-harmonic Green's function for the homogenized solid) is finally analyzed in the spirit of~\cite{bigoni_2005, piccolroaz_2006}.
The force is applied both to the lattice (in the physical and Fourier spaces) and to the equivalent solid at different levels of prestress, with special detail on low-frequency wave localizations.

The comparison between the behavior of the grillage and its equivalent solid reveals features of shear banding, so that this instability is on the one hand given a clear interpretation in terms of global instability of the lattice and on the other sharply discriminated from local instabilities in the composite, which remain undetected in the continuum model.
Therefore, when local instabilities do not occur, the homogenization approach is shown to provide a superb approximation (so that the incremental displacement fields found in the lattice and in the homogenized material are practically coincident).
Furthermore, examples of instabilities `invisible' in the equivalent material are provided.
These are shown to represent a limit for the homogenization approach and are important, as they exhibit an `explosive' character, so that they extend from a localized perturbation to the whole lattice.
\begin{figure}[htb!]
    \centering
    \begin{minipage}[c]{0.274\textwidth}
        \begin{subfigure}{\textwidth}
            \centering
            \caption{\label{fig:geometry_cluster_hexagonal}}
            \includegraphics[width=0.98\linewidth]{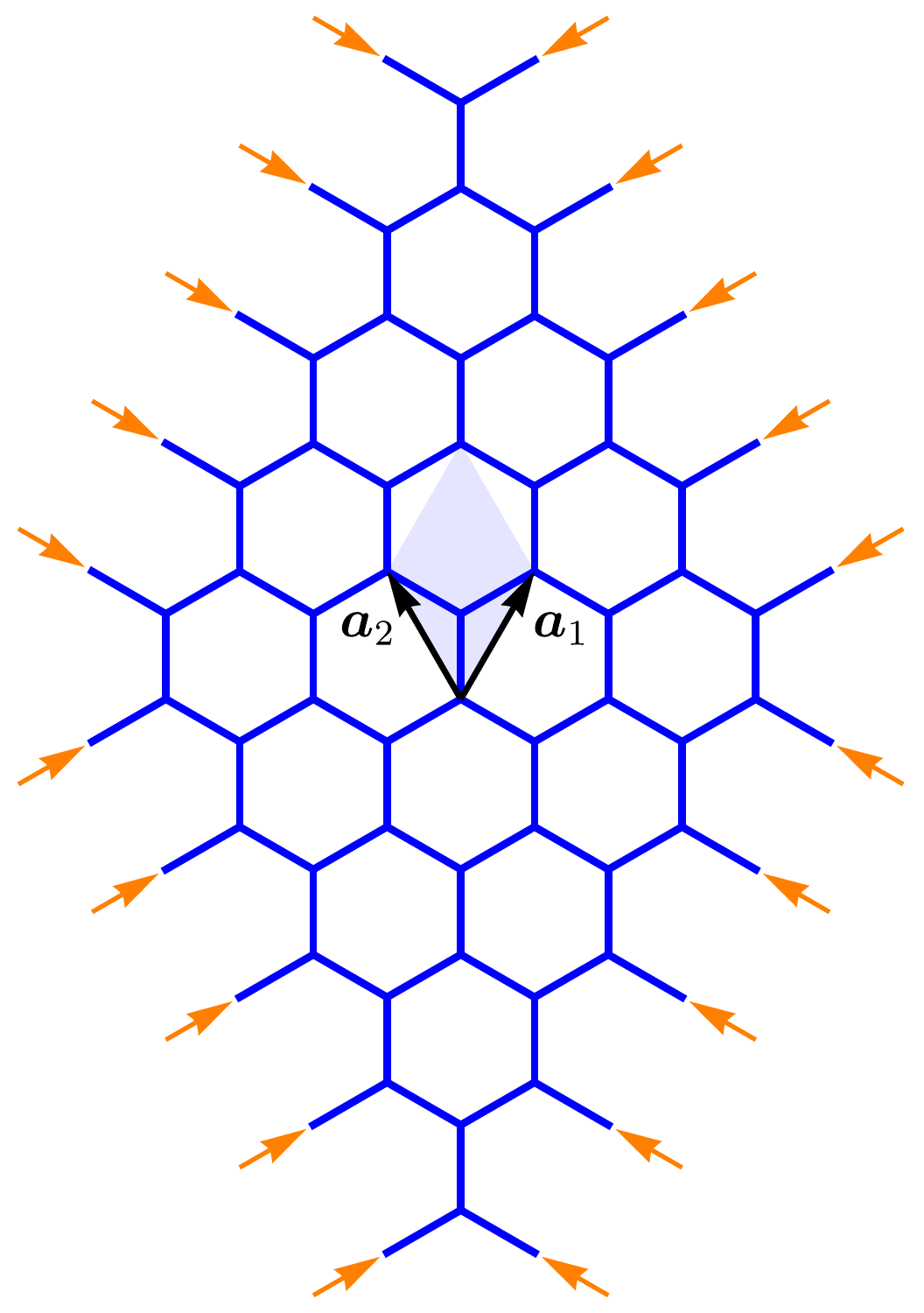}
        \end{subfigure}
    \end{minipage}%
    \begin{minipage}[c]{0.139\textwidth}
        \begin{subfigure}{\textwidth}
            \centering
            \caption{\label{fig:buckling_mode_1_cell_hexagonal}}
            \includegraphics[width=0.98\linewidth]{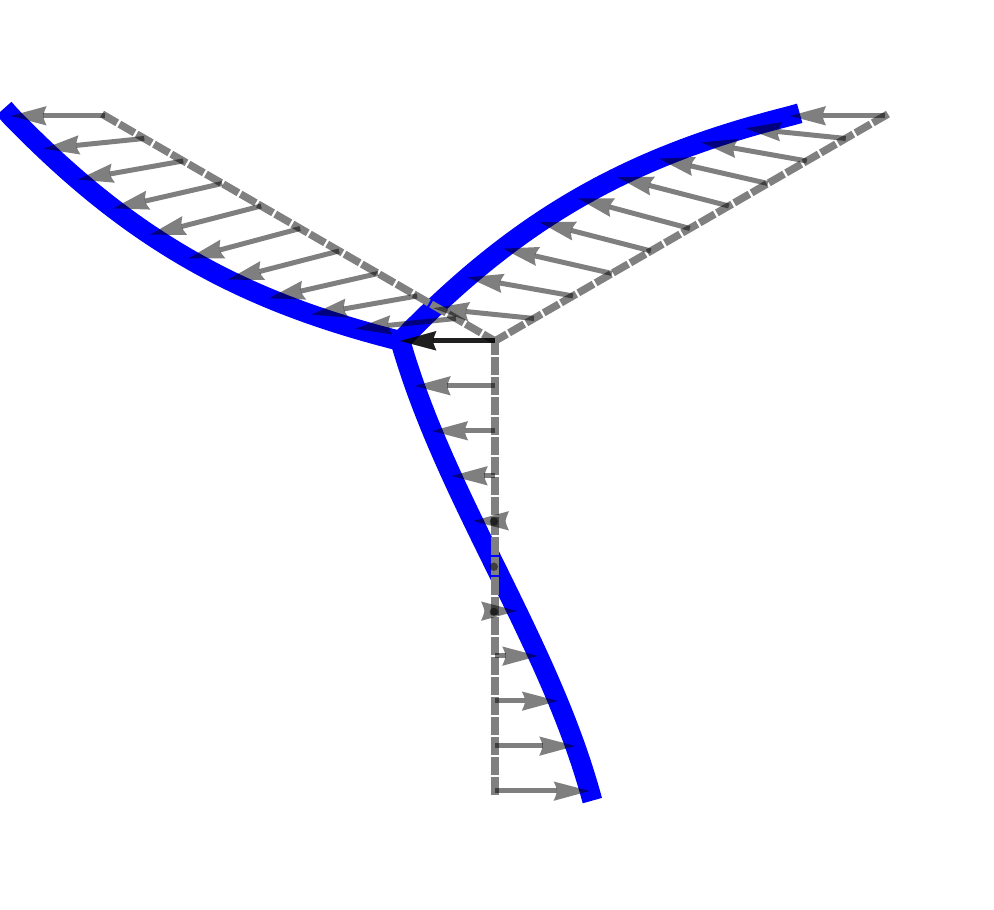}
        \end{subfigure}
        \begin{subfigure}{\textwidth}
            \centering
            \includegraphics[width=0.98\linewidth]{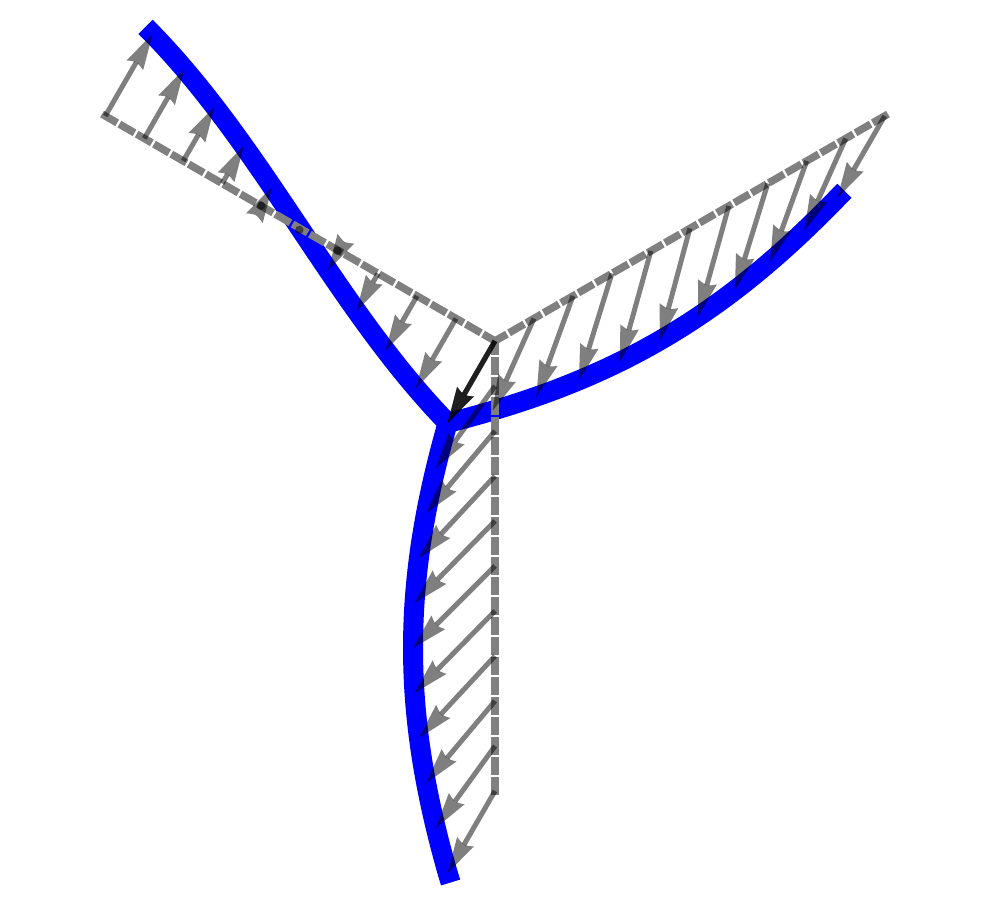}
        \end{subfigure}
        \begin{subfigure}{\textwidth}
            \centering
            \includegraphics[width=0.98\linewidth]{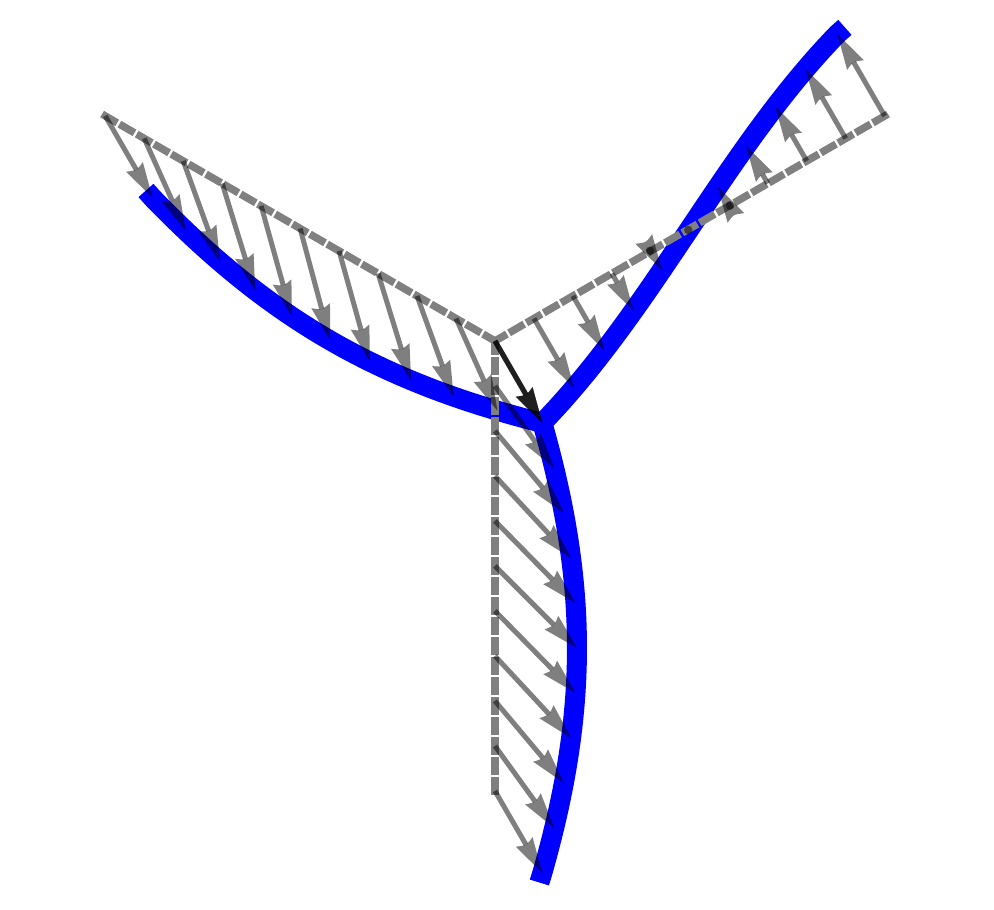}
        \end{subfigure}
    \end{minipage}%
    \begin{minipage}[c]{0.586\textwidth}
        \begin{subfigure}{\textwidth}
            \centering
            \caption{\label{fig:straws_micro_buckling_detailed}}
            \includegraphics[width=0.98\linewidth]{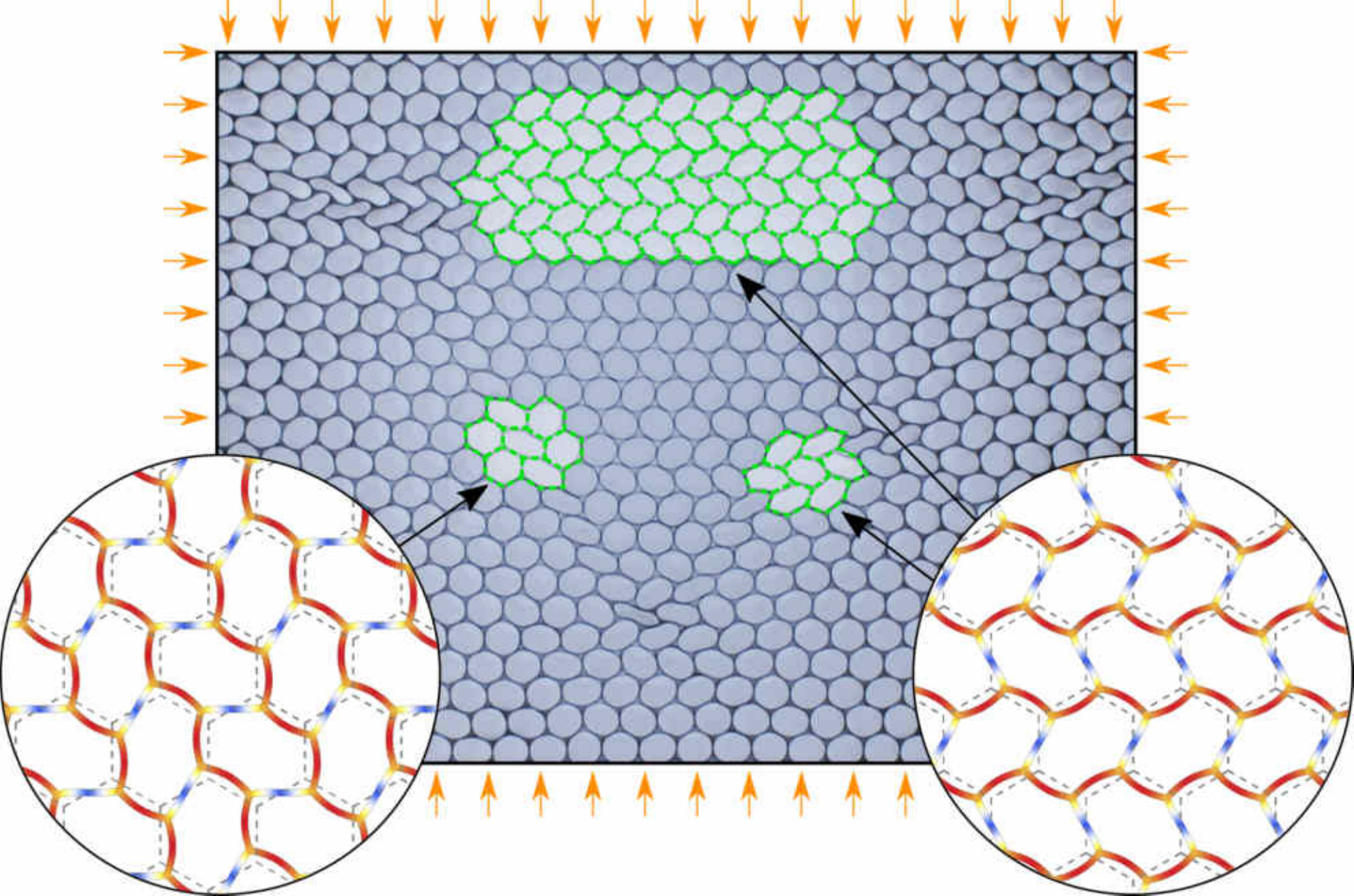}
        \end{subfigure}
    \end{minipage}%
    \caption{\label{fig:straws_buckling detailed}
        The micro-bifurcation mode emerging during the uniaxial deformation of the package of drinking straws shown in Fig.~\ref{fig:straws_localization} is modelled (with the tools provided in this article) as the micro-buckling of a honeycomb lattice of elastic rods, isotropically loaded with compressive forces.
        The equilibrium path of the honeycomb~(\subref{fig:geometry_cluster_hexagonal}) bifurcates displaying three critical modes~(\subref{fig:buckling_mode_1_cell_hexagonal}), which induces a periodic ovalization pattern.
        The latter explains the regular and diffuse buckled zones in the array of drinking straws~(\subref{fig:straws_micro_buckling_detailed}).
    }
\end{figure}

An example of local instability, undetected in the homogenized material, but revealed through the analysis of the microstructure, is provided in Fig.~\ref{fig:straws_localization}, where photos of experiments (performed at the Instabilities Lab of the University of Trento) are shown.
In the experiments, packages of drinking straws, initially in a regular hexagonal disposition, have been subjected to an overall uniaxial strain, so that lateral displacements are prevented and the overall stress is not far from being isotropic.
The drinking straws are 30 cm long, so that out-of-plane phenomena are prevented, as in the realization of Fig.~\ref{fig:sketch_3D_60} (left).
Until the overall strain remains sufficiently low, Fig.~\ref{fig:straws_1}, the deformation is hardly visible (compare with the unloaded configuration, Fig.~\ref{fig:straws_0}), but at higher strain, a micro-bifurcation emerges and displays a periodic ovalization of the straws' cross-sections (Fig.~\ref{fig:straws_micro_buckling}).
Later, the ovalization degenerates into a strain localization (in terms of the collapse of the cross-sections, Fig.~\ref{fig:straws_band_formation}).
Eventually strain band accumulation occurs, Figs.~\ref{fig:straws_4} and \ref{fig:straws_band_accumulation}.
The periodic ovalization is not found in the equivalent continuum, but is very well captured by a bifurcation analysis of an infinite hexagonal grid of rods (Fig.~\ref{fig:straws_buckling detailed}).
The grid is subject to isotropic compression, the stress state closer to that developing during the test.
A periodic bifurcation mode is predicted, which compares well with a detail of the photo shown in Fig.~\ref{fig:straws_micro_buckling}\footnote{
The bifurcation occurs at an axial load in the grid (that was analytically calculated to be $-\arccos^2{(-1/3)} EJ/l^2\approx -3.6EJ/l^2$) smaller than the load corresponding to loss of ellipticity in the equivalent material (which was calculated through the homogenization scheme developed in this article to be $\approx -7.014 EJ/l^2$).
}.

\begin{figure}[htb!]
    \centering
    \includegraphics[width=0.55\linewidth]{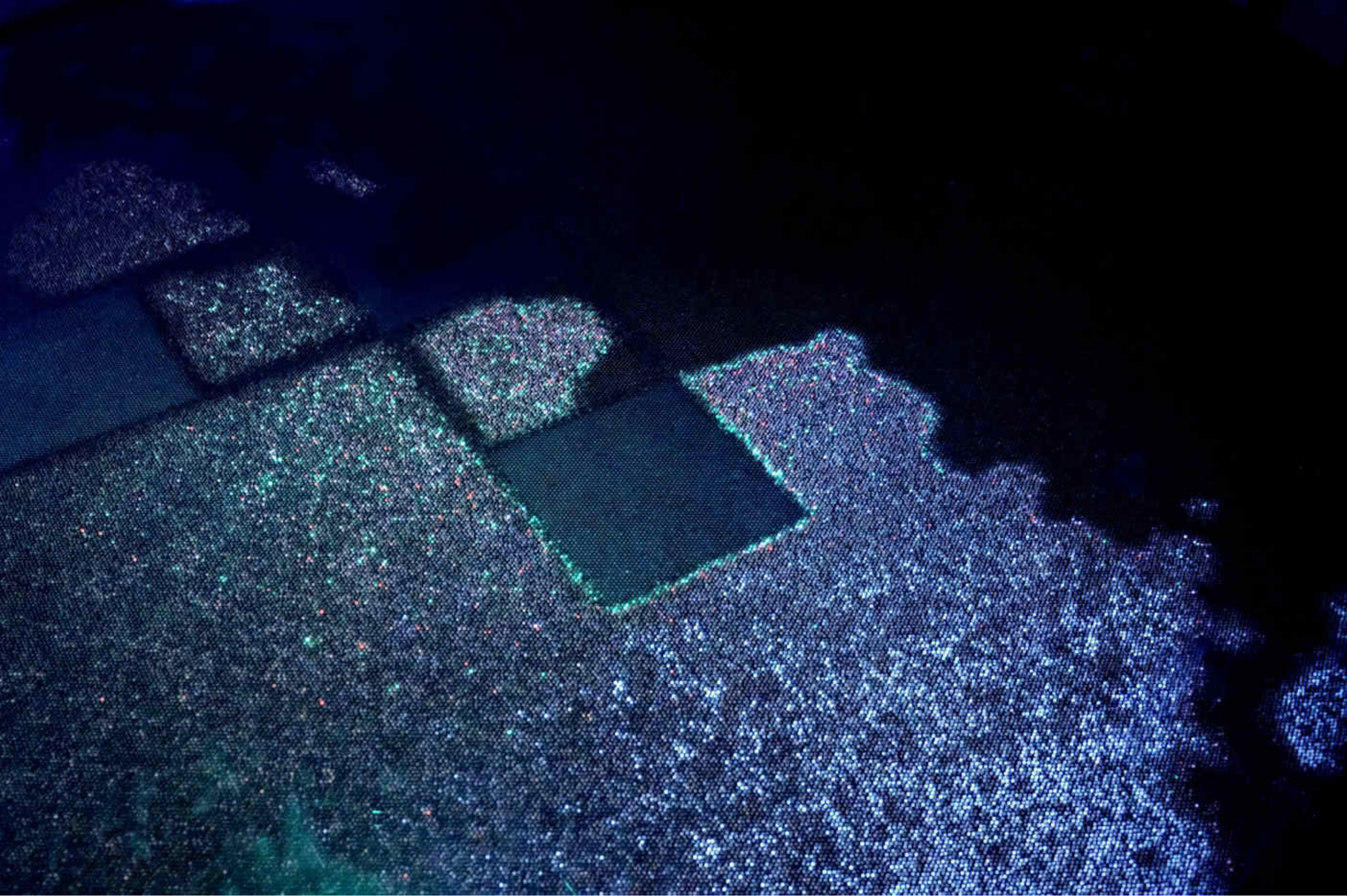}
    \caption{\label{fig:pinscreen}
        Discontinuity wavefronts (some rectilinear, other curvilinear) forming during the (out-of-plane) dynamics
        of a periodic material (used as a toy, the so-called `pinscreen', invented by W. Fleming), which works on the boundary of ellipticity loss (photo taken at the Exploratorium, San Francisco).
    }
\end{figure}

The vibrational properties of a cellular material are deeply affected by the emergence and development of localized signals, edge waves, and topologically protected modes~\cite{mishuris_2009a,wang_2015a,tallarico_2017,carta_2017,garau_2018,pal_2018,mazzotti_2019,bordiga_2019}, an example being that reported in Fig.~\ref{fig:pinscreen}.
In the figure, the dynamic emergence and propagation of discontinuity wavefronts (rectilinear and curvilinear) is shown in the so-called `pinscreen', a material (made up of a perforated plate having each hole filled with a movable pin) on the verge of ellipticity loss.
These localization and discontinuities can be analyzed through dynamic homogenization, which is shown to provide a tool to select the geometry and loading of a lattice in a way to produce an equivalent solid with arbitrary incremental anisotropy, so that the shear band inclination, or the emergence of a single shear band can be designed.
The results that will be presented also demonstrate how lattice models of heterogeneous materials can be highly effective to obtain analytical expressions for homogenized properties, thus allowing an efficient analysis of the influence of the microstructural parameters.
This is a clear advantage over continuum formulations for composites, where analytical results can only be obtained for simple geometries and loading configurations (as for instance in the case of laminated solids~\cite{nestorovic_2004,santisidavila_2016,bacigalupo_2013}).

Several new features are found, including a `super-sensitivity' of the localization direction to the preload state and the conditions in which a perfect correspondence between the lattice and the continuum occurs (so that the discrete system and the equivalent solid share all the same bifurcation modes).
The microscopic features found for the strain localization are shown to share remarkable similarities with the localized failure patterns observed in honeycombs (as Fig.~\ref{fig:straws_localization} demonstrates), foams and wood~\cite{papka_1994,papka_1998,papka_1999,jang_2010}, while the highly localized deformation bands emerging at macroscopic loss of ellipticity are reminiscent of the failure modes observed in balsa wood~\cite{dasilva_2007}.

This article is organized as follows.
The mathematical setting for incremental wave propagation is developed in Section~\ref{sec:system_governing_equations} for a lattice of elastic rods organized in an arbitrary periodic geometry.
The asymptotic analysis of lattice waves is derived in Section~\ref{sec:asymptotic_lattice_waves}, leading to the homogenization result that provides the acoustic tensor associated to the incremental effective elastic continuum, subject to a homogeneous state of prestress.
The stability of the lattice structure and its relation with the strong ellipticity of the equivalent solid is given in Section~\ref{sec:ellipticity_stability}.
The above general treatment is specialized in Section~\ref{sec:grid} to a grid of elastic rods arbitrarily inclined and equipped with diagonal springs, so that specific results on homogenization, stability domains, macroscopic and microscopic bifurcations are presented.
Examples and comparisons of the incremental response are showcased in Section~\ref{sec:dynamic_forced_response} and~\ref{sec:static_forced_response}, where the extreme mechanical behavior of the lattice is unveiled through a perturbative approach, both in time-harmonic and static regimes.

\section{Incremental dynamics of preloaded lattices: governing equations}
\label{sec:system_governing_equations}
The governing equations for incremental wave propagation in an axially-preloaded lattice of elastic rods (connected to each other with joints capable of transmitting bending moment, shear, and axial forces) are presented.
These are obtained (i) by solving for time-harmonic vibrations the incremental dynamics of a single rod (derived in Appendix~\ref{sec:linearized_elastica}),
(ii) by using this solution to formulate the equations of motion for a unit cell, and finally (iii) by applying the Bloch theorem to obtain the equations governing the incremental dynamics of the periodic lattice.
\begin{figure}[htb!]
    \centering
    \includegraphics[width=0.98\linewidth]{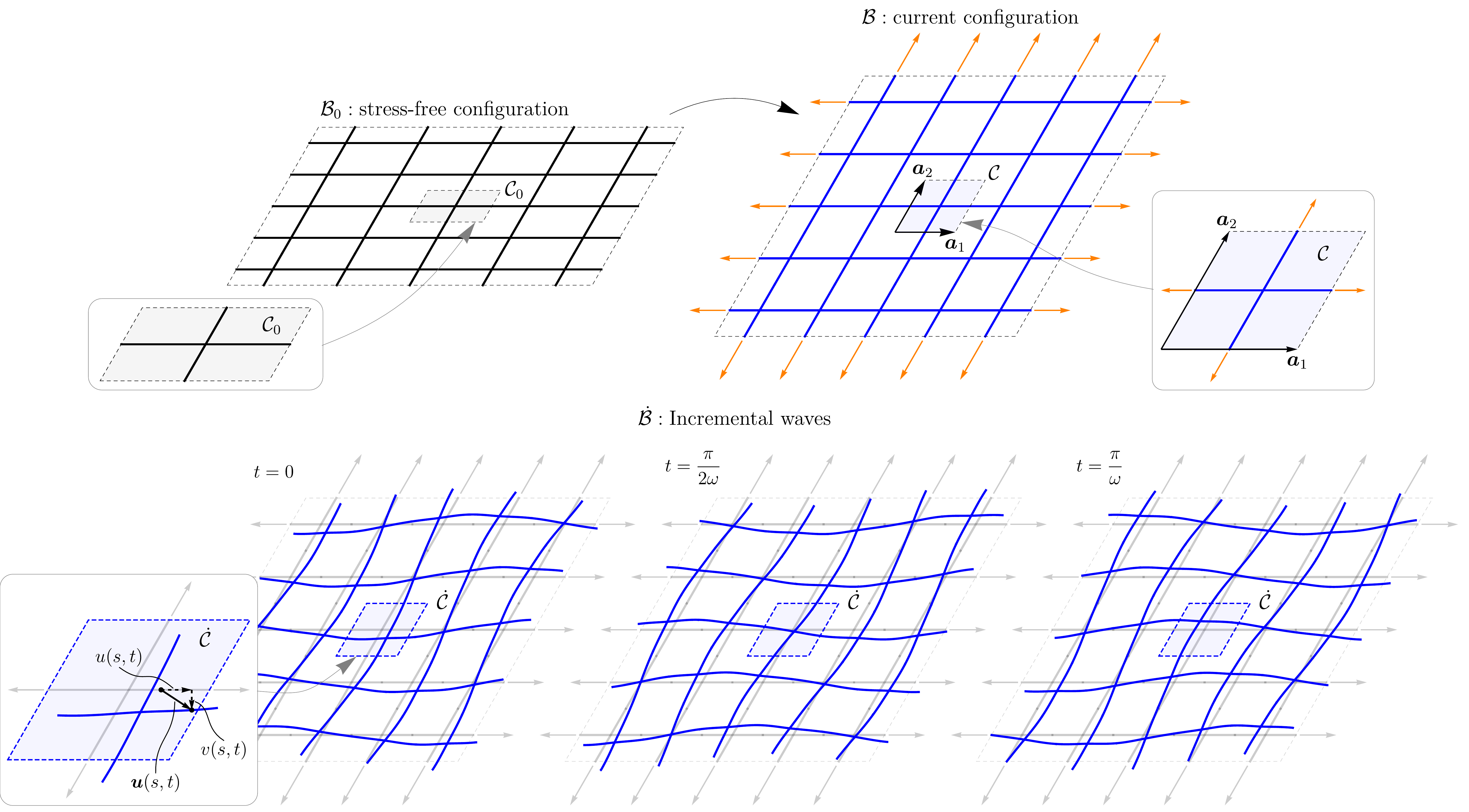}
    \caption{\label{fig:lattice_configurations}
        A periodic two-dimensional lattice of (axially and flexurally deformable) elastic rods is considered, preloaded from the stress-free configuration $\mB_0$ (upper part, on the left) by means of a pure axial loading state, transforming $\mB_0$ to the current, preloaded configuration $\mB$ (upper part, on the right).
        The latter configuration, used as reference in an updated Lagrangian description, can be represented as the tessellation of a single unit cell along the vectors of the direct basis $\{\ba_1,\ba_2\}$.
        The incremental dynamic response (lower part where an incremental deformation $\dot{\mB}$ is shown at three different instants of time) is defined on $\mB$ by the incremental displacement field of each rod $\bu(s,t)$, here decomposed in an axial and transverse component, $u(s,t)$ and $v(s,t)$.
    }
\end{figure}

An infinite two-dimensional lattice structure is considered, composed of nonlinear elastic rods which are axially preloaded (or prestretched) from an unloaded reference configuration $\mB_0$ to an axially preloaded configuration $\mB$, used as reference in an updated Lagrangian formulation of incremental dynamics (Fig.~\ref{fig:lattice_configurations}).
The configuration $\mB$ is assumed to be undistorted and described by the tessellation of a single unit cell along the vectors of the direct basis $\{\ba_1,\ba_2\}$, as shown in Fig.~\ref{fig:lattice_configurations}.
In the figure, in addition to the configurations $\mB_0$ and $\mB$, also incremental deformations $\dot{\mB}$ at three different instants of time are sketched, which highlight that the incremental analysis fully involves in-plane bending, stretching and shear.

By introducing a local coordinate $s_k$ for each rod of a given unit cell, the \textit{incremental} kinematics is described by means of the following fields
\begin{equation*}
    \bu_k(s_k,t)=\trans{\{ u_k(s_k,t), v_k(s_k,t) \}}, \qquad \forall k \in \{1,...,N_b\} \,,
\end{equation*}
where $N_b$ is the number of rods in the unit cell, and the two in-plane displacement components, respectively axial and transverse, are denoted by $u_k(s_k,t)$ and $v_k(s_k,t)$, while the rotation of the cross-section $\theta_k(s_k,t)$ is assumed to satisfy the unshearability condition $\theta_k(s_k,t)=v_k'(s_k,t)$\footnote{A dash will be used to denote differentiation with respect to the coordinate $s_k$.}.
Time-harmonic solutions are sought, so that, by introducing the circular frequency $\omega$, the dependence on time $t$ can be represented as
\begin{equation}
    \label{eq:time_harmonic}
    \bu_k(s_k,t) = \hat{\bu}_k(s_k)\, e^{-i\,\omega\,t} \qquad \forall k \in \{1,...,N_b\} \,,
\end{equation}
where $\hat{\bu}_k(s_k)=\trans{\{ \hat{u}_k(s_k), \hat{v}_k(s_k) \}}$ are functions of the coordinate $s_k$ only and $i=\sqrt{-1}$.

In the following the superscript symbol $\hat{~}$ will be omitted, but it will be tacitly assumed that all quantities depend on time as prescribed by equation \eqref{eq:time_harmonic}.

\subsection{Time-harmonic solution for a preloaded elastic rod}
\label{sec:time-harmonic_solution_beam}
The analytic representation for time-harmonic vibrations of a Rayleigh rod is briefly introduced.
In the framework of a linearized theory, the equations of motion governing the incremental time-harmonic dynamics of an axially pre-stretched Rayleigh rod are the following
\begin{subequations}
    \label{eq:governing_beam_EB}
    \begin{gather}
        \label{eq:governing_beam_EB_u}
        -\gamma(\lambda_0)\,\omega^2 u(s) - A(\lambda_0)\, u''(s) = 0 \,, \\
        \label{eq:governing_beam_EB_v}
        -\gamma(\lambda_0)\,\omega^2 v(s) + \gamma_r(\lambda_0)\,\omega^2 v''(s) + B(\lambda_0)\, v''''(s) - P(\lambda_0)\,v''(s) = 0 \,,
    \end{gather}
\end{subequations}
where $\gamma(\lambda_0)$ is the \textit{current} linear mass density, $\gamma_r(\lambda_0)$ is the \textit{current} rotational inertia, $\lambda_0$ is the axial pre-stretch and $P(\lambda_0)$ the corresponding axial preload (assumed positive in tension), while $A(\lambda_0)$ and $B(\lambda_0)$ are, respectively, the \textit{current} axial and bending stiffnesses.
The analytic derivation of equations~\eqref{eq:governing_beam_EB} is reported in Appendix~\ref{sec:linearized_elastica}.
Moreover, the derivation of the current stiffnesses $A(\lambda_0)$ and $B(\lambda_0)$ from strain-energy functions, as well as the their identification for rods made up of an incompressible nonlinear elastic material such as Mooney-Rivlin, can be found in Appendix~\ref{sec:incompressible_rod}.
In the following, the parameters $\gamma(\lambda_0)$, $\gamma_r(\lambda_0)$, $A(\lambda_0)$, and $B(\lambda_0)$ will simply be denoted as, $\gamma$, $\gamma_r$, $A$, and $B$, and treated as independent quantities for generality.

The substitution of Eq.~\eqref{eq:time_harmonic} into Eq.~\eqref{eq:governing_beam_EB} leads to a system of linear ODEs for the functions $u(s)$ and $v(s)$.
As the system is fully decoupled, the solution is easily obtained in the form
\begin{equation}
    \label{eq:u_v_sol}
    u(s) = \sum_{j=1}^2 C_{j}^u\, e^{i\,\beta_j^u\,s} \,, \qquad v(s) = \sum_{j=1}^4 C_{j}^v\, e^{i\,\beta_j^v\,s} \,,
\end{equation}
where $\{C_{1}^u,C_{2}^u,C_{1}^v,...,C_{4}^v\}$ are 6 arbitrary complex constants and the characteristic roots $\beta_j^{u}$ and $\beta_j^{v}$ are given by
\begin{equation*}
    \beta_{1,2}^u = \pm \frac{\tilde{\omega}}{l} \,, \qquad
    \beta_{1,2,3,4}^v= \pm \frac{1}{l\sqrt{2}}\sqrt{-p +r\,\tilde{\omega}^2 \pm \sqrt{p^2 + (4\Lambda^2 - 2\,p\,r)\,\tilde{\omega}^2 + r^2\,\tilde{\omega}^4}} \,,
\end{equation*}
with $l$ being the current length of the rod, $\tilde{\omega} = \omega\, l \sqrt{\gamma/A}$ the non-dimensional angular frequency, $p = P l^2/B$ the non-dimensional preload, $\Lambda=l/\sqrt{B/A}$ the slenderness of the rod, and $r = \gamma_r A/(\gamma B)$ is the dimensionless rotational inertia.

\subsection{Exact time-harmonic shape functions, mass and stiffness matrices}
\label{sec:shape_functions}
To facilitate the asymptotic expansion needed for implementing the homogenization scheme, it is instrumental to identify the 6 constants $\trans{\{C_{1}^u,C_{2}^u,C_{1}^v,...,C_{4}^v\}}$, with the \textit{degrees of freedom} at the rod's ends, represented through its nodal displacements.
This allows a dimensional reduction through a direct application of the compatibility conditions at the joints.

For any given rod of length $l$ the following notation for the nodal parameters is introduced
\begin{equation}
    \label{eq:nodaldisp}
    u(0) = u_1 \,, \quad v(0) = v_1 \,, \quad \theta(0) = \theta_1 \,, \quad u(l) = u_2 \,, \quad v(l) = v_2 \,, \quad \theta(l) = \theta_2 \,.
\end{equation}
Collecting the degrees of freedom at the two ends of the rod in the vector $\bq=\trans{\{u_1,v_1,\theta_1,u_2,v_2,\theta_2\}}$ yields
the solution of system \eqref{eq:nodaldisp} in the form
\begin{equation}
    \label{eq:time_harmonic_sf}
    \bu(s) = \bN(s;\omega, P)\, \bq \,.
\end{equation}
The 2-by-6 matrix $\bN(s;\omega, P)$ acts as a matrix of frequency-dependent and preload-dependent `shape functions' which is the \textit{exact} functional basis in which the time-harmonic response of the rod can be represented.
Equation~\eqref{eq:time_harmonic_sf} can also be considered as the definition of a `finite element' endowed with shape functions built from the exact solution.

The time-harmonic shape functions, Eq. (\ref{eq:time_harmonic_sf}), reduce to the quasi-static solution when $\omega\to0$, so that at vanishing preload and in the limit $\lim_{\omega\to0}\bN(s;\omega, 0)$, the usual shape functions for beam elements, employed for instance in \cite{phani_2006}, are recovered.
In the following $\bN(s;\omega, P)$ will be denoted as $\bN(s;\omega)$, to simplify notation.

By employing Eq.~\eqref{eq:time_harmonic_sf}, the exact mass and stiffness matrices for a rod subject to time-harmonic vibration can be computed.
For the $k$-th rod, the kinetic energy and the elastic strain energy at second-order are given by\footnote{A superimposed dot is used to denote time differentiation unless explicitly stated otherwise.}
\begin{subequations}
    \label{eq:kinetic_strain_energy}
    \begin{align}
         &
        \label{eq:kinetic_energy}
        \begin{aligned}
            \mT_k & = \frac{1}{2} \int_0^{l_k} \gamma_k \left(\dot{u}_k(s_k,t)^2+\dot{v}_k(s_k,t)^2\right) ds_k + \frac{1}{2} \int_0^{l_k} \gamma_{r,k} \dot{v}'_k(s_k,t)^2 ds_k \\
                  &
            =\frac{1}{2} \,\trans{\dot{\bq}_{k}(t)} \left(\int_0^{l_k}  \trans{\bN_{k}(s_k;\omega)}\bJ_{k}\, \bN_{k}(s_k;\omega) ds_k\right) \dot{\bq}_{k}(t)
            + \frac{1}{2} \,\trans{\dot{\bq}_{k}(t)} \left(\int_0^{l_k} \gamma_{r,k}\, \trans{\bb_{k}(s_k;\omega)} \bb_{k}(s_k;\omega) \,ds_k \right) \dot{\bq}_{k}(t) \,,
        \end{aligned} \\
         &
        \label{eq:strain_energy}
        \mE_k = \frac{1}{2} \int_0^{l_k} \left(A_k\,u'_k(s_k,t)^2 + B_k\,v''_k(s_k,t)^2\right) ds_k
        = \frac{1}{2} \,\trans{\bq_{k}(t)} \left(\int_0^{l_k} \trans{\bB_{k}(s_k;\omega)}\bE_k\,\bB_{k}(s_k;\omega) ds_k \right) \bq_{k}(t) \,,
    \end{align}
\end{subequations}
where $\bJ_{k}$ and $\bE_{k}$ are matrices collecting the inertia and stiffness terms, respectively, while $\bB_{k}(s_k;\omega)$ is the strain-displacement matrix, which are defined as
\begin{equation*}
    \bJ_{k}=\begin{bmatrix}
        \gamma_k & 0        \\
        0        & \gamma_k
    \end{bmatrix} ,
    \qquad
    \bE_{k}=\begin{bmatrix}
        A_k & 0   \\
        0   & B_k
    \end{bmatrix} ,
    \qquad
    \bB_{k}(s_k;\omega) = \begin{bmatrix}
        \deriv{}{s_k} & 0                 \\[2mm]
        0             & \deriv{^2}{s_k^2}
    \end{bmatrix} \bN_{k}(s_k;\omega) \,,
\end{equation*}
and $\bb_{k}(s_k;\omega)=\begin{bmatrix}0 & \deriv{}{s_k}\end{bmatrix}\bN_{k}(s_k;\omega)$ is a row vector containing the derivative of the shape functions corresponding to the transverse displacement $v$.
Note that the kinetic energy~\eqref{eq:kinetic_energy} accounts for translational as well as rotational inertia of the rod.
In addition, the contribution of the axial preload $P$ has to be included in the second-order potential energy as (details are provided in Appendix~\ref{sec:linearized_elastica})
\begin{equation}
    \label{eq:prestress_energy}
    \mV_k^g = \frac{1}{2} P_k \int_0^{l_k} v'_k(s_k,t)^2 \,ds_k
    = \frac{1}{2} \,\trans{\bq_{k}(t)} \left(P_k \int_0^{l_k} \trans{\bb_{k}(s_k;\omega)} \bb_{k}(s_k;\omega) \,ds_k \right) \bq_{k}(t) \,.
\end{equation}
By combining Eqs.~\eqref{eq:strain_energy} and~\eqref{eq:prestress_energy}, the second-order potential energy of the $k$-th rod is denoted as
\begin{equation}
    \label{eq:potential_energy}
    \mV_k = \mE_k + \mV_k^g \,.
\end{equation}

From Eqs.~\eqref{eq:kinetic_strain_energy},~\eqref{eq:prestress_energy} and~\eqref{eq:potential_energy} the frequency-dependent mass and stiffness matrices are naturally defined as
\begin{subequations}
    \label{eq:M_K_beam}
    \begin{align}
        \label{eq:M_beam}
        \bM_{k}(\omega) & = \int_0^{l_k} \trans{\bN_{k}(s_k;\omega)}\bJ_{k}\,\bN_{k}(s_k;\omega)\,ds_k + \int_0^{l_k} \gamma_{r,k}\, \trans{\bb_{k}(s_k;\omega)} \bb_{k}(s_k;\omega)\,ds_k \,, \\
        \label{eq:K_beam}
        \bK_{k}(\omega) & = \int_0^{l_k} \trans{\bB_{k}(s_k;\omega)}\bE_{k}\,\bB_{k}(s_k;\omega)\,ds_k +
        P_k \int_0^{l_k} \trans{\bb_{k}(s_k;\omega)} \bb_{k}(s_k;\omega) \,ds_k \,.
    \end{align}
\end{subequations}
The matrices for the quasi-static case can be obtained by evaluating the limit $\omega\to0$.
In particular, the quasi-static stiffness matrix $\lim_{\omega\to0}\bK_{k}(\omega)$ can be used to formulate the incremental equilibrium and analyze the bifurcation of preloaded lattices.

\subsection{Equations of motion for the unit cell}
\label{sec:equations_motion_cell}
As expressions~\eqref{eq:kinetic_strain_energy}--\eqref{eq:potential_energy} govern the \textit{incremental} dynamics of a single rod, the kinetic and potential energies of a single unit cell can be obtained through a summation of the contributions from each rod
\begin{equation*}
    \mT = \sum_{k=1}^{N_b} \mT_k \,, \qquad \mV = \sum_{k=1}^{N_b} \mV_k \,,
\end{equation*}
and the \textit{incremental} equations of motion for the unit cell can be derived from the Lagrangian
\begin{equation}
    \label{eq:lagrangian}
    \mL(\bq,\dot{\bq}) = \mT(\dot{\bq}) - \mV(\bq) \,,
\end{equation}
where $\bq$ is the vector collecting all the degrees of freedom of the unit cell.

The Lagrangian~\eqref{eq:lagrangian} leads to the following Euler-Lagrange equations
\begin{equation}
    \label{eq:equations_of_motion_cell}
    \bM(\omega)\,\ddot{\bq}(t) + \bK(\omega)\,\bq(t) = \bef(t) \,,
\end{equation}
where $\bM(\omega)$ and $\bK(\omega)$ are, respectively, the mass and stiffness matrices of the unit cell\footnote{The matrices $\bM(\omega)$ and $\bK(\omega)$ can be easily obtained by assembling the matrices expressed by Eqs.~\eqref{eq:M_K_beam} for all the rods in the unit cell.} and the force vector $\bef$ collects the \textit{incremental} forces acting on the boundary nodes of the unit cell.
Recall that due to the assumption of null body forces acting on the lattice, the nodes inside the unit cell are not externally loaded.

By recalling the time-harmonic assumption, Eq.~\eqref{eq:equations_of_motion_cell} is rewritten as
\begin{equation}
    \label{eq:system_isolated}
    \bA(\omega)\, \bq = \bef \,,
\end{equation}
with the definition $\bA(\omega)=-\omega^2\bM(\omega)+\bK(\omega)$.
Note that the dimension of the linear system~\eqref{eq:system_isolated} is $3N_j$, where $N_j$ is the number of nodes within the unit cell.

It is also important to recall that both matrices $\bM(\omega)$ and $\bK(\omega)$ depend also on the preloads $P_1,\dots,P_{N_b}$ of the rods composing the unit cell.
The state of preload will be conveniently denoted as a vector $\bP=\{P_1,\dots,P_{N_b}\}$.

\subsection{Bloch's theorem}
\label{sec:Bloch_theorem}
Wave propagation in an infinite periodic elastic medium can be effectively analyzed through the application of Bloch's theorem.
Essentially, the theorem states that the time-harmonic solutions of the equations of motion possess a modulation in space having the same periodicity of the medium, a condition expressed by the following requirement\footnote{Note that the representation~\eqref{eq:bloch_theorem} holds also when the displacement is replaced by fields defining the generalized forces internal to the rods.}
\begin{subequations}
    \label{eq:bloch_theorem}
    \begin{equation}
        \label{eq:bloch_theorem_1}
        \bu(\bx) = \bvarphi(\bx) \,e^{i\,\bk \scalp \bx} \,,
    \end{equation}
    where $\bk$ is the Bloch vector and the modulation $\bvarphi(\bx)$ is periodic with respect to the direct basis, so that it satisfies
    \begin{equation*}
        \bvarphi(\bx+n_j\ba_j)=\bvarphi(\bx) \qquad \forall \{n_1,n_2\}\in\Integers^2 \quad\forall\bx\in\Reals^2 \,.
    \end{equation*}
    Eq.~\eqref{eq:bloch_theorem_1} can be equivalently expressed as
    \begin{equation}
        \label{eq:bloch_theorem_2}
        \bu(\bx+n_j\ba_j) = \bu(\bx) \,e^{i\,\bk \scalp (n_j\ba_j)} \,.
    \end{equation}
\end{subequations}

Note that, in the case of a lattice made up of rods, the waveform $\bvarphi(\bx)$ (as well as the field $\bu(\bx,t)$) is described by the displacement of the rods forming the unit cell, and thus it is defined only for $\bx$ corresponding to the location of the structural elements.

The importance of the Bloch's theorem lies in the fact that it allows the response of an infinite and periodic structure to be described through the equations of motion for a unit cell, the latter complemented by suitable boundary conditions.
These so-called `Floquet-Bloch' conditions enforce the periodicity of the lattice, by relating nodal displacements and forces on the boundary of the unit cell, according to the property expressed by Eq.~\eqref{eq:bloch_theorem_2}.
The application of these conditions to periodic beam lattices is well-known~\cite{langley_1993, phani_2006} and is briefly summarized in the following.
\begin{figure}[htb!]
    \centering
    \begin{subfigure}{0.5\textwidth}
        \centering
        \caption{\label{fig:FB_joints_q}}
        \includegraphics[height=38mm]{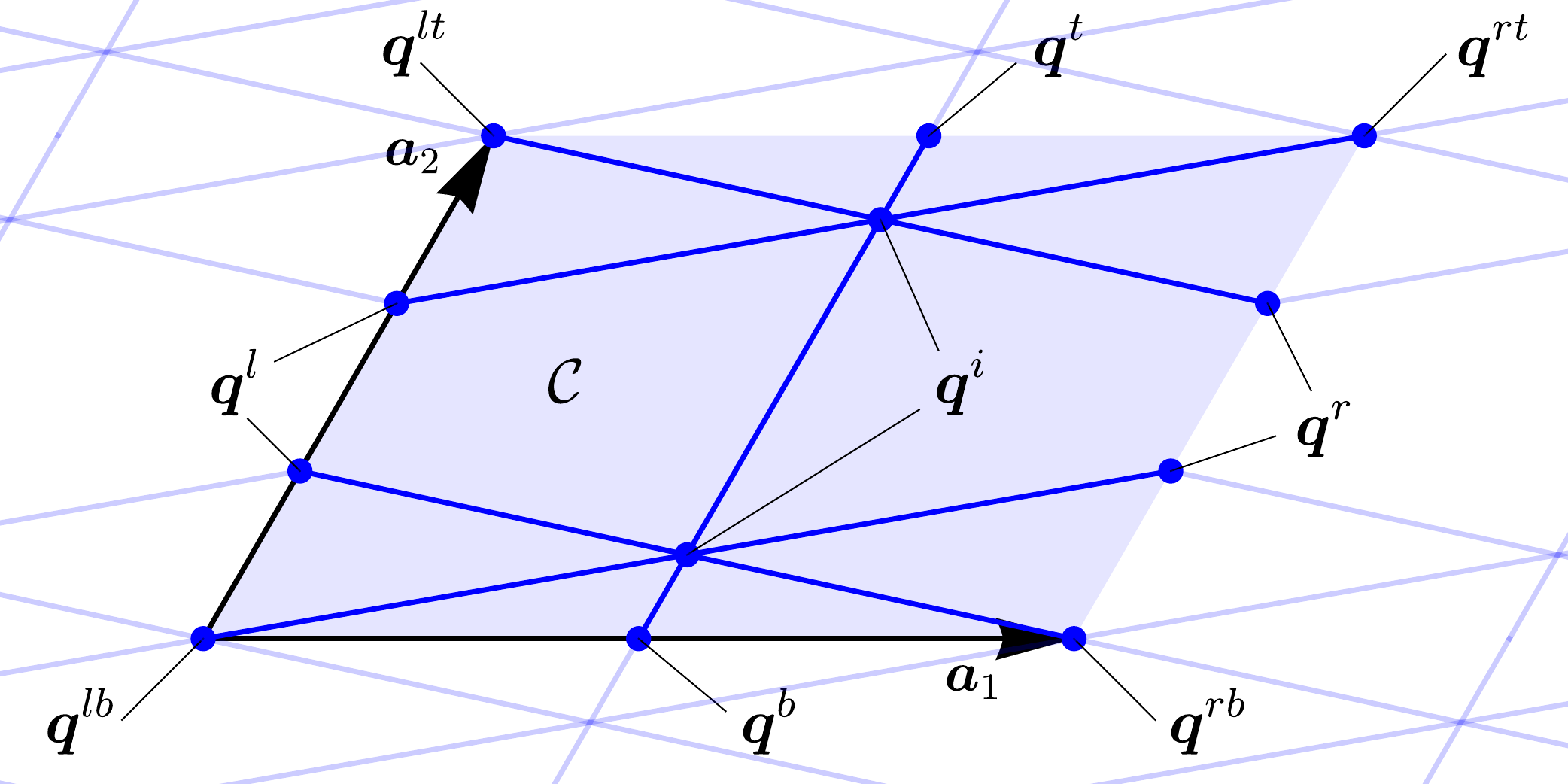}
    \end{subfigure}%
    \begin{subfigure}{0.4\textwidth}
        \centering
        \caption{\label{fig:FB_joints_f}}
        \includegraphics[height=38mm]{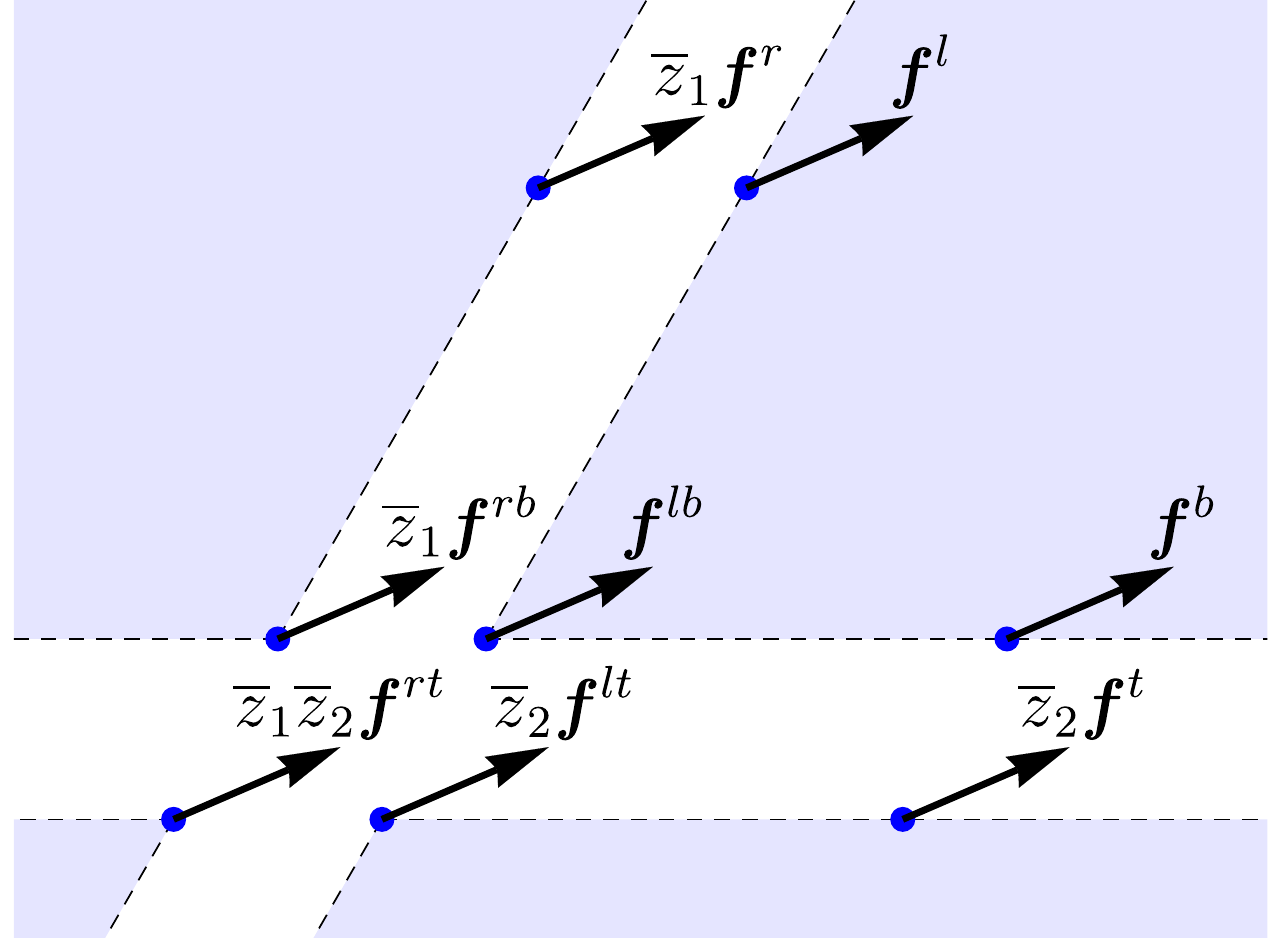}
    \end{subfigure}
    \caption{\label{fig:FB_joints}
        The imposition of the Floquet-Bloch conditions required by Eq.~\eqref{eq:bloch_theorem_2} is obtained using vector $\bq$, which collects the degrees of freedom of the unit cell.
        The vector $\bq$ is partitioned by distinguishing the sets of internal nodes $\bq^i$ from the nodes at the boundary, located at corners $\{\bq^{lb}, \bq^{lt}, \bq^{rb}, \bq^{rt}\}$ and on the edges $\{\bq^l, \bq^b, \bq^r, \bq^t\}$~(\subref{fig:FB_joints_q}).
        The corresponding force vector $\bef$ is partitioned in the same way~(\subref{fig:FB_joints_f}) and the forces acting on the boundary nodes have to satisfy equilibrium as well as the Floquet-Bloch conditions (the `shift factors' $\bar{z}_j$, $j=1,2$ are the complex conjugate of $z_j = e^{i\,\bk\scalp\ba_j}$ used in the Bloch representation, Eq.~\eqref{eq:bloch_theorem_2}).
    }
\end{figure}

Equation~\eqref{eq:bloch_theorem_2} requires
\begin{equation}
    \label{eq:FB_shifts}
    \bq_q = \bq_p \, e^{i\,\bk\scalp(\bx_q-\bx_p)} \,,
\end{equation}
for all pairs of nodes $\{p,q\}$ such that $\bx_q-\bx_p$ is an integer, linear combination of the lattice vectors $\{\ba_1,\ba_2\}$.
Therefore, the relations to be imposed on the boundary of the unit cell derive directly from Eq.~\eqref{eq:bloch_theorem_2}, evaluated at $\bx\in\partial\mC$ and for $n_j\in\{0,1\}$, hence obtaining
\begin{subequations}
    \label{eq:FB_conditions}
    \begin{equation}
        \label{eq:FB_conditions_long}
        \bq =
        \begin{Bmatrix}
            \bq^i    \\
            \bq^l    \\
            \bq^b    \\
            \bq^{lb} \\
            \bq^r    \\
            \bq^t    \\
            \bq^{rb} \\
            \bq^{lt} \\
            \bq^{rt}
        \end{Bmatrix}
        =
        \begin{bmatrix}
            \bI    & \bzero & \bzero & \bzero      \\
            \bzero & \bI    & \bzero & \bzero      \\
            \bzero & \bzero & \bI    & \bzero      \\
            \bzero & \bzero & \bzero & \bI         \\
            \bzero & z_1\bI & \bzero & \bzero      \\
            \bzero & \bzero & z_2\bI & \bzero      \\
            \bzero & \bzero & \bzero & z_1\bI      \\
            \bzero & \bzero & \bzero & z_2\bI      \\
            \bzero & \bzero & \bzero & z_1 z_2 \bI
        \end{bmatrix}
        \begin{Bmatrix}
            \bq^i \\
            \bq^l \\
            \bq^b \\
            \bq^{lb}
        \end{Bmatrix} ,
    \end{equation}
    succinctly written as
    \begin{equation}
        \label{eq:FB_conditions_short}
        \bq = \bZ(\bk)\, \bq^* \,,
    \end{equation}
\end{subequations}
where $\bZ(\bk)$ and $\bq^*$ are defined according to Eq.~\eqref{eq:FB_conditions_long}, and $z_j = e^{i\,\bk\scalp\ba_j}$, with $j=1,2$.
In Eq.~\eqref{eq:FB_conditions_long} the vector $\bq$ has been partitioned to denote the inner and boundary nodes according to the notation sketched in Fig.~\ref{fig:FB_joints}.
The same partitioning is also introduced for the force vector $\bef$.

A substitution of Eqs.~\eqref{eq:FB_conditions} into Eq.~\eqref{eq:system_isolated} provides
\begin{equation*}
    \bA(\omega)\bZ(\bk)\, \bq^* = \bef \,,
\end{equation*}
so that a left multiplication by $\conjtrans{\bZ(\bk)}$, where the superscript $\conjtrans{\,}$ denotes the complex conjugate transpose operation\footnote{
    The transpose of the conjugate of a matrix $\bM$ is defined~ as $\conjtrans{M_{ij}}=\conj{M}_{ji}$, where the bar denotes the complex conjugate.
}, leads to the reduced system
\begin{equation}
    \label{eq:system_f}
    \conjtrans{\bZ(\bk)}\bA(\omega)\bZ(\bk)\, \bq^* = \bef^* \,,
\end{equation}
where the following definition is introduced
\begin{equation*}
    \bef^* = \conjtrans{\bZ(\bk)} \bef =
    \begin{Bmatrix}
        \bef^i                     \\
        \bef^l + \conj{z}_1 \bef^r \\
        \bef^b + \conj{z}_2 \bef^t \\
        \bef^{lb} + \conj{z}_1 \bef^{rb} + \conj{z}_2 \bef^{lt} + \conj{z}_1 \conj{z}_2 \bef^{rt}
    \end{Bmatrix} .
\end{equation*}

Note that the dimension of system~\eqref{eq:system_f} is smaller than the dimension of system \eqref{eq:system_isolated}.
In fact, the imposition of the Floquet-Bloch conditions allows to express the equations of motion for the lattice only in terms of the reduced variables $\bq^*$ and $\bef^*$.

If external loads are not present in the infinite lattice (so that the boundary forces shown in Fig.~\ref{fig:FB_joints_f} are purely internal), Eq.~\eqref{eq:bloch_theorem_2} implies $\bef^*=\bzero$ and therefore the following homogeneous system of equations is obtained, governing Floquet-Bloch wave propagation within the lattice
\begin{equation}
    \label{eq:system}
    \bA^*(\omega,\bk) \, \bq^* = \bzero \,,
\end{equation}
where the matrix of the reduced system is
\begin{equation}
    \bA^*(\omega,\bk) = \conjtrans{\bZ(\bk)}\bA(\omega)\bZ(\bk) \,.
\end{equation}

Note that this matrix is non-symmetric, but for a conservative system is always Hermitian, so that $\bA^*(\omega,\bk)=\conjtrans{\bA^*(\omega,\bk)}$ and $\omega^2\in\Reals$.

\subsection{Generalized eigenvalue problem for the dynamics of an infinite lattice}
Eq.~\eqref{eq:system} defines a homogeneous linear system for the unknown vector $\bq^*$, in which the angular frequency $\omega$ and the wave vector $\bk$ are for the moment undetermined.

It is important to note that the matrix-valued complex function $\bA^*(\omega,\bk)$ depends on all mechanical and geometrical parameters of the unit cell.
Eq.~\eqref{eq:system} is homogeneous, so that all non-trivial solutions are obtained by imposing the condition
\begin{equation}
    \label{eq:dispersion}
    \det\,\bA^*(\omega,\bk) = 0 \,,
\end{equation}
that defines, although implicitly, the \textit{dispersion relation}, providing $\omega$ as a function of $\bk$ for an infinite and periodic grillage of preloaded elastic rods.

For every given wave vector $\bk$ and for each of the corresponding roots $\omega(\bk)$ obtained from Eq.~\eqref{eq:dispersion}, the non-trivial solutions of Eq.~\eqref{eq:system} provide the modes $\bq^*$ of the Floquet-Bloch waves propagating through the lattice at each frequency.
This means that the vector $\bq^*$ is an implicit function of $\omega$ and therefore of $\bk$, so that the dependence on $\bk$ can be made explicit and Eq.~\eqref{eq:system} is rewritten as
\begin{equation}
    \label{eq:system_dispersion}
    \bA^*(\omega(\bk),\bk)\,\bq^*(\omega(\bk),\bk)=\bzero \, ,
\end{equation}
an expression that will help clarifying the asymptotic expansion that will be performed in the next section.

\section{Dynamic homogenization of the lattice}
\label{sec:asymptotic_lattice_waves}
The scope of this section is the analysis of the low-frequency/long-wavelength asymptotics of the lattice dynamics, which will enable the identification of the effective continuum material.
Identification requires first appropriate reference to the equivalent continuum.

\subsection{Wave propagation in a prestressed elastic continuum}
\label{sec:wave_propagation_Cauchy}
Before introducing the homogenization technique, it is worth recalling the fundamental equations governing incremental wave propagation in a prestressed hyperelastic continuum.
An appropriate form of the equations for the prestressed continuum has to be selected, to result compatible with the formulation of the lattice dynamics introduced in Section~\ref{sec:system_governing_equations}.
Specifically, it is observed that the equations of motion for the lattice are
\begin{enumerate*}[label=(\roman*)]
    \item obtained in the context of a linearized theory, and
    \item referred to a preloaded reference configuration.
\end{enumerate*}
Therefore, the dynamics of the unknown `equivalent' continuum has to be formulated in the context of the incremental theory of nonlinear elasticity by means of a relative Lagrangian description~\cite{bigoni_2012}.
This is based on \textit{incremental constitutive laws}, relating the increment of the first Piola-Kirchhoff stress, $\dot{\bS}$, to the gradient of incremental displacement, $\bL=\grad{\bu}$, as
\begin{equation}
    \label{eq:incremental_constitutive_law}
    \dot{\bS} = \fC[\bL] \,,
\end{equation}
through the elasticity tensor $\fC$
\begin{equation}
    \label{eq:constitutive_operator}
    \fC = \fE + \bI \boxtimes \bT , \qquad \mbox{in components} \qquad \fC_{ijkl} = \fE_{ijkl} + \delta_{ik} T_{jl} \,,
\end{equation}
where $\bT$ is the Cauchy stress, defining here the \textit{prestress}, and $\fE$ is a fourth-order tensor endowed with the left and right minor symmetries and the major symmetry, while $\fC$ lacks the minor symmetries.

Eq.~\eqref{eq:constitutive_operator} implies that the number of unknown components of $\fE$ is at most 6 for a 2D material and 21 in the 3D case.
Moreover, the (symmetric) prestress $\bT$ introduces 3 further components in 2D and 6 in 3D, so that $\fC$ is characterized by 9 for a 2D material, and 27 in the 3D case, independent components.

In the absence of body forces, the incremental equations of motion for the continuum can be written in the usual form
\begin{equation}
    \label{eq:equations_of_motion_continuum}
    \Div{\dot{\bS}} = \rho\,\ddot{\bu} \,,
\end{equation}
where $\bu$ is the incremental displacement field and $\rho>0$ the mass density.

Assuming the usual plane wave representation for incremental displacement, $\bu(\bx,t) = \ba \exp [i(\bk \scalp \bx - \widehat{\omega}\,t)]$, Eq.~\eqref{eq:equations_of_motion_continuum} leads to the following eigenvalue problem
\begin{equation}
    \label{eq:eigenvalue_problem_cauchy}
    \left(k^2 \bA^{(\fC)}(\bn) -\rho \, \widehat{\omega}^2\bI \right) \ba = \bzero \,,
\end{equation}
where $k = \norm{\bk}$ is the wave number and $\bn$ is a unit vector defining the propagation direction, $\bk = k\,\bn$.
Eq.~\eqref{eq:eigenvalue_problem_cauchy} governs the wave propagation in a homogeneous elastic material whose acoustic tensor $\bA^{(\fC)}(\bn)$ is defined as
\begin{equation*}
    A^{(\fC)}_{pr}(\bn) = n_q\,\fC_{pqrs}\,n_s \,.
\end{equation*}
The eigenvectors $\ba$ represent the wave amplitudes, while the eigenvalues $\widehat{\omega}^2$ are the roots of the characteristic equation
\begin{equation}
    \label{eq:characteristic_equation_cauchy}
    \det\left(k^2 \bA^{(\fC)}(\bn)-\rho \, \widehat{\omega}^2\bI\right) = 0 \,.
\end{equation}
Note that $\widehat{\omega}(\bk)$ is introduced to be differentiated from $\omega(\bk)$, so that the former defines the angular frequency of a wave propagating through the equivalent homogeneous elastic continuum, while the latter the dispersion relation for the lattice.

\subsection{Asymptotic expansion of Floquet-Bloch waves}
\label{sec:wave_expansion}
A perturbation method is now developed for the equations governing wave propagation in a lattice made of prestressed elastic rods, through a generalization of the technique proposed by Born \cite{born_1955} for lattices involving only point-like mass interactions.
A rigorous link is established between the low-frequency solutions of Eq.~\eqref{eq:system_dispersion} and spectral characteristics of the equivalent continuum governed by Eq.~\eqref{eq:eigenvalue_problem_cauchy}.

Suppose for a moment that the grillage of rods would satisfy the following conditions:
\begin{itemize}
    \item a \textit{linear dispersion relation}, obtained by writing $\bk = k \,\bn$ (with $\bn$ being a unit vector),
          \begin{equation}
              \label{eq:omega_linear}
              \omega(k\,\bn) = c_{\bn}\,k \,,
          \end{equation}
          where $c_{\bn}$ is the wave speed and depends only on the direction $\bn$;
    \item a \textit{uniform spatial modulation}
          \begin{equation}
              \label{eq:modulation_constant}
              \bvarphi_{\bk}(\bx)=\ba(\bk) \, ,
          \end{equation}
\end{itemize}
in this case wave propagation in the grillage would be non-dispersive and with uniform amplitude, a situation which cannot occur for every $\bk$, but is instead typical of a homogeneous medium.
However, conditions~\eqref{eq:omega_linear}--\eqref{eq:modulation_constant} can be met in the limit $k\to0$, so that in this case an equivalent material can be defined and vector $\ba$ in Eq.~\eqref{eq:modulation_constant} becomes the eigenmode defined by Eq.~\eqref{eq:eigenvalue_problem_cauchy} and $\widehat{\omega} = c_{\bn}\,k$.
This is how the homogenization will be performed.

The linear relation~\eqref{eq:omega_linear} can be considered as the first-order term of the asymptotic expansion of the dispersion relation $\omega(k\,\bn)$ centered at $k=0$, where  $\omega=0$ and $\bk=\bzero$, and along the direction $\bn$ in the $\bk$-space.
This asymptotic expansion, truncated at the $N$-th term, is
\begin{equation}
    \label{eq:omega_expansion}
    \omega(k\,\bn) \sim \omega_{\bn}^{(1)}\,k + \omega_{\bn}^{(2)}\,k^2 +...+\omega_{\bn}^{(N)}\,k^N = \mS_{\bn}^N (\omega)(k) \,,
\end{equation}
where the $\mO(k^0)$ term vanishes in the above expansion because the point $\{\omega,\bk\}=\bzero$ satisfies the dispersion equation~\eqref{eq:dispersion}.
This follows from the fact that, setting $\{\omega,\bk\}=\bzero$, Eq.~\eqref{eq:system_dispersion} becomes
\begin{equation}
    \label{eq:A0}
    \bA^*(\bzero)\,\bq^*(\bzero) = \conjtrans{\bZ(\bzero)}\bK(0)\bZ(\bzero)\,\bq^*(\bzero) = \bzero \,,
\end{equation}
whose non-trivial solutions $\bq^*(\bzero)$ are represented by 2 rigid-body translations (rigid-body rotations are excluded because the matrix $\bZ(\bzero)$ prescribes \textit{equal} displacements on corresponding sides of the unit cell).
The two rigid-body translations, plus the rigid-body rotation, are contained in the nullspace of $\bK(0)$.

To be more precise, the dimension of $\ker\bA^*(\bzero)$ is \textit{at least} 2.
In fact, any other deformation mode besides the translations, possibly contained in $\ker\bA^*(\bzero)$, is a zero-energy mode (called also with the pictoresque name `floppy mode'~\cite{mao_2018,zhang_2018}).
These modes are excluded for the purpose of the asymptotic analysis, so that $\ker\bA^*(\bzero)$ contains \textit{only} two (in the present 2D formulation) rigid-body translations.
This restriction does not affect the generality of the formulation, as the analysis of floppy modes can always be recovered in the limit of vanishing stiffness of appropriate structural elements.
Note also that sometimes floppy modes can be eliminated or introduced simply playing with the prestress state (which may induce stiffening or softening~\cite{pellegrino_1986,pellegrino_1990}).

The rigid-body translations represent the limit of the eigenmodes $\bvarphi_{\bk}(\bx)$ for $\norm{\bk}\to0$.
Their derivation requires first the construction of an asymptotic expansion for $\bvarphi_\bk(\bx)$, in complete analogy to Eq.~\eqref{eq:omega_expansion},
\begin{equation}
    \label{eq:phi_expansion}
    \bvarphi_{k\,\bn}(\bx) \sim \bvarphi^{(0)}_{\bn}(\bx) + \bvarphi^{(1)}_{\bn}(\bx)\,k + \bvarphi^{(2)}_{\bn}(\bx)\,k^2 +...+ \bvarphi^{(N)}_{\bn}(\bx)\,k^N
    =\mS_{\bn}^N (\bvarphi)(k) \,,
\end{equation}
and then the computation of the limit $k\to0$.
As a result, the zeroth-order term of the waveform $\bvarphi^{(0)}_{\bn}$ is indeed \textit{uniform} in space (independent of $\bx$), and therefore the acoustic properties of the equivalent elastic continuum have to satisfy
\begin{subequations}
    \label{eq:omega_phi_match}
    \begin{align}
        \label{eq:omega_match}
        \widehat{\omega}(k\,\bn) & = c_{\bn}\,k = \omega_{\bn}^{(1)}\,k \,,                 \\
        \label{eq:phi_match}
        \ba(k\,\bn)              & = \bvarphi^{(0)}_{\bn} \qquad \forall \bn\in\Reals^2 \,,
    \end{align}
\end{subequations}
conditions which define an `acoustic equivalence' between the lattice and the continuum. Note that equation (\ref{eq:omega_match}) defines that
$\omega$ and $\widehat{\omega}$ coincide at first-order.

An effective method to obtain the series expansions~\eqref{eq:omega_expansion} and~\eqref{eq:phi_expansion} is outlined in the following.
As the waveform $\bvarphi_{\bk}(\bx)$ for a lattice made up of rods is governed by the vector of degrees of freedom $\bq^*$, solution of the eigenvalue problem~\eqref{eq:system_dispersion}, the expansion of $\bq^*(\omega(\bk),\bk)$ is performed along an arbitrary direction $\bn$ in the $\bk$-space
\begin{equation}
    \label{eq:q_expansion}
    \bq^*(\omega(k\,\bn),k\,\bn) \sim \bq^{*(0)}_{\bn} + \bq^{*(1)}_{\bn}\,k + \bq^{*(2)}_{\bn}\,k^2 + ... \,,
\end{equation}
so that the first term $\bq^{*(0)}_{\bn}$ can be used to identify the left-hand side of Eq.~\eqref{eq:phi_match}.
To this end, the matrix $\bA^*(\omega(\bk),\bk)$ is expanded as
\begin{equation}
    \label{eq:A_expansion}
    \bA^*(\omega(k\,\bn),k\,\bn) \sim \bA^{*(0)}_{\bn} + \bA^{*(1)}_{\bn}\,k + \bA^{*(2)}_{\bn}\,k^2 + ... \,,
\end{equation}
so that the eigenvalue problem~\eqref{eq:system_dispersion} is rewritten through a substitution of the series representations~\eqref{eq:q_expansion} and~\eqref{eq:A_expansion} as
\begin{equation}
    \label{eq:system_expansion}
    \left(\bA^{*(0)}_{\bn} + \bA^{*(1)}_{\bn}\,k + \bA^{*(2)}_{\bn}\,k^2 + ...\right)\left(\bq^{*(0)}_{\bn} + \bq^{*(1)}_{\bn}\,k + \bq^{*(2)}_{\bn}\,k^2 + ...\right) = \bzero \,.
\end{equation}

Since the dispersion relation $\omega(\bk)$ is formally inserted in the above expansions, Eq.~\eqref{eq:system_expansion} has to be satisfied for every value of $k$, which means that the left-hand side has to vanish at every order in $k$.
Thus the following sequence of linear systems is obtained
\begin{equation}
    \label{eq:system_sequence}
    \begin{aligned}
         & \mO(k^0): \quad \bA^{*(0)}_{\bn}\,\bq^{*(0)}_{\bn} = \bzero \,,                                                                           \\
         & \mO(k^1): \quad \bA^{*(0)}_{\bn}\,\bq^{*(1)}_{\bn} + \bA^{*(1)}_{\bn}\,\bq^{*(0)}_{\bn} = \bzero \,,                                      \\
         & \mO(k^2): \quad \bA^{*(0)}_{\bn}\,\bq^{*(2)}_{\bn} + \bA^{*(1)}_{\bn}\,\bq^{*(1)}_{\bn} + \bA^{*(2)}_{\bn}\,\bq^{*(0)}_{\bn} = \bzero \,, \\
         & \quad\vdots \qquad \qquad\qquad \qquad\qquad\qquad \vdots                                                                                 \\
         & \mO(k^j): \quad \bA^{*(0)}_{\bn}\,\bq^{*(j)}_{\bn} + \sum_{h=1}^{j} \bA^{*(h)}_{\bn}\,\bq^{*(j-h)}_{\bn} = \bzero \qquad \forall j>0  \,, \\
    \end{aligned}
\end{equation}
which has to be solved for the unknown vectors $\bq^{*(j)}_{\bn}$.
It is clear that the computation of these vectors starts from the solution of the zeroth-order equation and then, sequentially, the higher-order terms are to be obtained.
At the $j$-th order, the matrix of the linear system is $\bA^{*(0)}_{\bn}$ and multiplies the unknown vector $\bq^{*(j)}_{\bn}$, so that the constant term (not involving the unknown $\bq^{*(j)}_{\bn}$) contains all the previously determined vectors $\{\bq^{*(0)}_{\bn},...,\bq^{*(j-1)}_{\bn}\}$.
Moreover, it is important to observe that the terms $\bA^{*(j)}_{\bn}$ in the expansion~\eqref{eq:A_expansion} can be computed explicitly once the series $\mS_{\bn}^N (\omega)(k)$ has been determined.

It is recalled that, as shown by Eq.~\eqref{eq:A0}, the matrix of each linear system $\bA^{*(0)}_{\bn}$ is singular and it has a two-dimensional nullspace spanned by two linearly independent vectors, $\bt_1$ and $\bt_2$, which represent the two in-plane rigid-body translations\footnote{
    Since $\bt_1$ and $\bt_2$ describe two arbitrary rigid translations, $\bt_1$ and $\bt_2$ can be conveniently chosen as the rigid translations aligned parallel to $\be_1$ and $\be_2$, respectively.}.
Thus, every linear combination in the form
\begin{equation}
    \label{eq:q0}
    \bq^{*(0)}_{\bn} = \alpha_1\bt_1 + \alpha_2\bt_2 \qquad \forall \{\alpha_1,\alpha_2\}\in\Reals^2 \,,
\end{equation}
is a solution of the zeroth-order equation in~\eqref{eq:system_sequence}.
This implies that the matrix $\bA^{*(0)}_{\bn}$ is not invertible, so that the solvability of the $j$-th linear system depends on the form of its right-hand side, which has to satisfy the following condition, known as the Fredholm alternative theorem
\begin{equation}
    \label{eq:solvability}
    \by \scalp \sum_{h=1}^{j} \bA^{*(h)}_{\bn}\,\bq^{*(j-h)}_{\bn}  = 0 \qquad \forall \by\in\ker\bA^{*(0)}_{\bn} \qquad \forall j>0 \,,
\end{equation}
or, equivalently, using Eq.~\eqref{eq:q0}, the condition
\begin{equation*}
    \bt_1 \scalp  \sum_{h=1}^{j} \bA^{*(h)}_{\bn}\,\bq^{*(j-h)}_{\bn}   = 0 \,, \qquad \bt_2 \scalp  \sum_{h=1}^{j} \bA^{*(h)}_{\bn}\,\bq^{*(j-h)}_{\bn} = 0 \qquad \forall j>0 \,.
\end{equation*}

In principle, Eqs.~\eqref{eq:system_sequence} and~\eqref{eq:solvability} are sufficient to compute the series representations~\eqref{eq:q_expansion} and~\eqref{eq:omega_expansion}, thus making conditions~\eqref{eq:omega_phi_match} explicit.

\subsection{The acoustic tensor for a lattice of elastic rods}
\label{sec:acoustic_tensor_lattice}
The perturbation method outlined in Section~\ref{sec:wave_expansion} is general enough to provide, up to the desired order, the series representation of the acoustic properties of a preloaded lattice subject to incremental dynamics.

It will be proved in the following that it is always possible to employ the above-described perturbation technique to construct an eigenvalue problem governing the propagation of waves in a lattice (where elements are subject to both axial and flexural deformation) in the low-frequency and long-wavelength regime.
In particular, this eigenvalue problem will possess the following properties:
\begin{enumerate}[label=(\roman*)]
    \item the eigenvalues identify the first-order term $\omega_{\bn}^{(1)}$ of both acoustic branches of the dispersion relation;
    \item the eigenvectors govern the zeroth-order term of the Floquet-Bloch waveform for both the two acoustic waves, through coefficients $\{\alpha_1,\alpha_2\}$ in the linear combination~\eqref{eq:q0};
    \item the algebraic structure of the problem is \textit{exactly equivalent} to that governing wave propagation in a prestressed elastic material, Eq.~\eqref{eq:eigenvalue_problem_cauchy}.
\end{enumerate}

\textit{The construction of the above eigenvalue problem allows the rigorous definition of the `acoustic tensor for a lattice of elastic rods' and from the latter the identification of the elasticity tensor representing a material equivalent to the lattice}.
In fact, this eigenvalue problem defines eigenvalues and eigenvectors satisfying the conditions of acoustic equivalence, Eq.~\eqref{eq:omega_phi_match}.

In order to construct the eigenvalue problem, the solution of the sequence of the linear systems~\eqref{eq:system_sequence} is obtained up to the order $\mO(k^2)$.
The equations involve the following terms of the series~\eqref{eq:A_expansion}
\begin{equation}
    \label{eq:A0_A1_A2}
    \begin{aligned}
        \bA^{*(0)}       & = \conjtrans{{\bZ^{(0)}}}\bK^{(0)}\bZ^{(0)} \,,                                                                                         \\
        \bA^{*(1)}_{\bn} & = \conjtrans{{\bZ^{(0)}}}\bK^{(0)}\bZ^{(1)}_{\bn} + \conjtrans{{\bZ^{(1)}_{\bn}}}\bK^{(0)}\bZ^{(0)} \,,                                 \\
        \bA^{*(2)}_{\bn} & = \conjtrans{{\bZ^{(1)}_{\bn}}}\bK^{(0)}\bZ^{(1)}_{\bn} + \conjtrans{{\bZ^{(0)}}}\bK^{(0)}\bZ^{(2)}_{\bn} +                             \\
                         & \quad+\conjtrans{{\bZ^{(2)}_{\bn}}}\bK^{(0)}\bZ^{(0)} - \conjtrans{{\bZ^{(0)}}}\bM^{(0)}\bZ^{(0)} \left(\omega_{\bn}^{(1)}\right)^2 \,,
    \end{aligned}
\end{equation}
where a series expansion has been introduced for the matrices $\bK(\omega(k\,\bn))$, $\bM(\omega(k\,\bn))$ and $\bZ(k\,\bn)$ as $k\to0$.
It is important to note that:
\begin{enumerate}[label=(\roman*)]
    \item up to the order $\mO(k^2)$, only the zeroth-order terms of the matrices $\bK$ and $\bM$ (which correspond to the quasi-static limit, $\bK^{(0)} = \lim_{\omega\to0}\bK(\omega)$, $\bM^{(0)}=\lim_{\omega\to0}\bM(\omega)$) are present;
    \item the zeroth-order matrix $\bZ^{(0)}$, and consequently $\bA^{*(0)}$, is independent of the direction $\bn$ (owing to continuity of $\bZ(\bk)$);
          while $\bA^{*(1)}_{\bn}$ is linear in $\bn$ and $\bA^{*(2)}_{\bn}$ is quadratic in $\bn$;
    \item the linear term $\omega_{\bn}^{(1)}$ starts to appear at order $\mO(k^2)$.
\end{enumerate}

In the following, the first and second-order equations in the sequence of equations~\eqref{eq:system_sequence} are considered and their solvability conditions derived, Eq.~\eqref{eq:solvability}.
By means of Eq.~\eqref{eq:q0}, the first-order equation in the sequence~\eqref{eq:system_sequence} reads as
\begin{equation}
    \label{eq:system_first_order}
    \mO(k^1): \quad \bA^{*(0)}\,\bq^{*(1)}_{\bn} + \bA^{*(1)}_{\bn}(\alpha_1\bt_1+\alpha_2\bt_2) = \bzero \,,
\end{equation}
and its solvability condition requires
\begin{equation*}
    \bt_1 \scalp \bA^{*(1)}_{\bn}(\alpha_1\bt_1+\alpha_2\bt_2) = 0 \,, \qquad
    \bt_2 \scalp \bA^{*(1)}_{\bn}(\alpha_1\bt_1+\alpha_2\bt_2) = 0 \,,
\end{equation*}
two conditions which are always satisfied.
In fact, a use of Eq.~\eqref{eq:A0_A1_A2}$_2$ yields
\begin{equation*}
    \bt_j\scalp\bA^{*(1)}_{\bn}\bt_i = \bt_j \scalp \left(\conjtrans{{\bZ^{(0)}}}\bK^{(0)}\bZ^{(1)}_{\bn} + \conjtrans{{\bZ^{(1)}_{\bn}}}\bK^{(0)}\bZ^{(0)}\right)\bt_i \,,
\end{equation*}
a scalar product which vanishes because $\bZ^{(0)}\bt_i$ is a rigid-body translation, so that it cannot produce any stress, hence $\bK^{(0)}\bZ^{(0)}\bt_i=\bzero$.
Since Eq.~\eqref{eq:system_first_order} is always solvable, all its solutions can be expressed in the form
\begin{equation}
    \label{eq:q1}
    \bq^{*(1)}_{\bn} = \alpha_1\bt_1'(\bn) + \alpha_2\bt_2'(\bn) \qquad \forall \{\alpha_1,\alpha_2\}\in\Reals^2 \,,
\end{equation}
where $\bt_1'$ and $\bt_2'$ are the solutions of the following two linear systems
\begin{equation*}
    \bA^{*(0)}\,\bt_1'(\bn) + \bA^{*(1)}_{\bn}\bt_1 = \bzero \,, \qquad \bA^{*(0)}\,\bt_2'(\bn) + \bA^{*(1)}_{\bn}\bt_2 = \bzero \,.
\end{equation*}
Note that $\bt_1'$ and $\bt_2'$ are defined up to an arbitrary rigid-body translation.

By employing Eqs.~\eqref{eq:q0} and~\eqref{eq:q1}, the linear system of order $\mO(k^2)$ reads as
\begin{equation}
    \label{eq:system_second_order}
    \mO(k^2): \quad \bA^{*(0)}\,\bq^{*(2)}_{\bn} = -\alpha_1 \left(\bA^{*(1)}_{\bn}\bt_1'(\bn) + \bA^{*(2)}_{\bn}\bt_1\right)-\alpha_2 \left(\bA^{*(1)}_{\bn}\bt_2'(\bn) + \bA^{*(2)}_{\bn}\bt_2\right) \,,
\end{equation}
which (because $\bA^{*(0)}$ is singular) admits a solution if and only if the right-hand side is orthogonal to both $\bt_1$ and $\bt_2$, namely
\begin{equation*}
    \begin{aligned}
        \alpha_1 \left(\bA^{*(1)}_{\bn}\bt_1'(\bn) + \bA^{*(2)}_{\bn}\bt_1\right)\scalp\bt_1+\alpha_2 \left(\bA^{*(1)}_{\bn}\bt_2'(\bn) + \bA^{*(2)}_{\bn}\bt_2\right)\scalp\bt_1 = 0 \,, \\
        \alpha_1 \left(\bA^{*(1)}_{\bn}\bt_1'(\bn) + \bA^{*(2)}_{\bn}\bt_1\right)\scalp\bt_2+\alpha_2 \left(\bA^{*(1)}_{\bn}\bt_2'(\bn) + \bA^{*(2)}_{\bn}\bt_2\right)\scalp\bt_2 = 0 \,,
    \end{aligned}
\end{equation*}
that in matrix form can be written as\footnote{
    Note that vectors $\bt_j'(\bn)$ may contain an arbitrary rigid-body translation. This would apparently lead to a non-uniqueness in the form of Eq.~\eqref{eq:system_solvability_second_order}, because the terms $\bt_j\scalp\bA^{*(1)}_{\bn}\bt_i'(\bn)$ are present. This lack of uniqueness is only apparent, because $\bt_j\scalp\bA^{*(1)}_{\bn}\bt_i=0$.
}
\begin{equation}
    \label{eq:system_solvability_second_order}
    \begin{bmatrix}
        \left(\bA^{*(1)}_{\bn}\bt_1'(\bn) + \bA^{*(2)}_{\bn}\bt_1\right)\scalp\bt_1 & \left(\bA^{*(1)}_{\bn}\bt_2'(\bn) + \bA^{*(2)}_{\bn}\bt_2\right)\scalp\bt_1 \\
        \left(\bA^{*(1)}_{\bn}\bt_1'(\bn) + \bA^{*(2)}_{\bn}\bt_1\right)\scalp\bt_2 & \left(\bA^{*(1)}_{\bn}\bt_2'(\bn) + \bA^{*(2)}_{\bn}\bt_2\right)\scalp\bt_2
    \end{bmatrix}
    \begin{Bmatrix}
        \alpha_1 \\
        \alpha_2
    \end{Bmatrix}=
    \begin{Bmatrix}
        0 \\
        0
    \end{Bmatrix} .
\end{equation}

Up to order $\mO(k^1)$ the coefficients $\{\alpha_1,\alpha_2\}$ and the linear term $\omega_{\bn}^{(1)}$ remain \textit{completely arbitrary}, but now they have to satisfy system~\eqref{eq:system_solvability_second_order} in order to make Eq.~\eqref{eq:system_second_order} solvable.
In fact, the homogeneous system~\eqref{eq:system_solvability_second_order} represents an eigenvalue problem with eigenvectors $\{\alpha_1,\alpha_2\}$ and eigenvalues $\left(\omega_{\bn}^{(1)}\right)^2$.
To see this point more explicitly, expressions~\eqref{eq:A0_A1_A2} can be substituted into Eq.~\eqref{eq:system_solvability_second_order} to obtain\footnote{The resulting expression has been simplified using again the property $\bK^{(0)}\bZ^{(0)}\bt_i=\bzero$.}
\begin{equation}
    \label{eq:eigenvalue_problem_omega1}
    \begin{aligned}
        \overbrace{
            \begin{bmatrix}
                \bt_1\scalp\tilde{\bK}^{(1)}_{\bn}\bt_1 + \bt_1'(\bn)\scalp\tilde{\bK}^{(0)}\bt_1'(\bn) & \bt_2\scalp\tilde{\bK}^{(1)}_{\bn}\bt_1 + \bt_2'(\bn)\scalp\tilde{\bK}^{(0)}\bt_1'(\bn) \\
                \bt_1\scalp\tilde{\bK}^{(1)}_{\bn}\bt_2 + \bt_1'(\bn)\scalp\tilde{\bK}^{(0)}\bt_2'(\bn) & \bt_2\scalp\tilde{\bK}^{(1)}_{\bn}\bt_2 + \bt_2'(\bn)\scalp\tilde{\bK}^{(0)}\bt_2'(\bn)
            \end{bmatrix}}^{\bXi}
        \begin{Bmatrix}
            \alpha_1 \\
            \alpha_2
        \end{Bmatrix} + \\[1mm]
        -\left(\omega_{\bn}^{(1)}\right)^2
        \underbrace{
            \begin{bmatrix}
                \bt_1\scalp\tilde{\bM}^{(0)}\bt_1 & \bt_2\scalp\tilde{\bM}^{(0)}\bt_1 \\
                \bt_1\scalp\tilde{\bM}^{(0)}\bt_2 & \bt_2\scalp\tilde{\bM}^{(0)}\bt_2
            \end{bmatrix}}_{\bGamma}
        \begin{Bmatrix}
            \alpha_1 \\
            \alpha_2
        \end{Bmatrix} =
        \begin{Bmatrix}
            0 \\
            0
        \end{Bmatrix} ,
    \end{aligned}
\end{equation}
with the following definitions
\begin{equation*}
    \tilde{\bK}^{(0)} = \conjtrans{{\bZ^{(0)}}}\bK^{(0)}\bZ^{(0)} \,, \quad
    \tilde{\bK}^{(1)}_{\bn} = \conjtrans{{\bZ^{(1)}_{\bn}}}\bK^{(0)}\bZ^{(1)}_{\bn} \,, \quad
    \tilde{\bM}^{(0)} = \conjtrans{{\bZ^{(0)}}}\bM^{(0)}\bZ^{(0)} \,.
\end{equation*}

Eq.~\eqref{eq:eigenvalue_problem_omega1} is an eigenvalue problem, and the following properties can be deduced:
\begin{enumerate}[label=(\roman*)]
    \item As the matrices $\tilde{\bK}^{(0)}$, $\tilde{\bK}^{(1)}_{\bn}$ and $\tilde{\bM}^{(0)}$ are real and symmetric, also the matrices $\bXi$ and $\bGamma$ are real and symmetric, hence the eigenvalues $\left(\omega_{\bn}^{(1)}\right)^2$ are real;
    \item Since $\bZ^{(0)}$, $\bt_1$ and $\bt_2$ are independent of $\bn$ and $\bZ^{(1)}_{\bn}$, $\bt_1'(\bn)$ and $\bt_2'(\bn)$ are all linear in $\bn$, each component of the 2-by-2 matrix $\bXi$ is a quadratic form in $\bn$;
    \item The components of the 2-by-2 matrix $\bGamma$ admit the following simplifications
          \begin{equation*}
              \bt_2\scalp\tilde{\bM}^{(0)}\bt_1 = \bt_1\scalp\tilde{\bM}^{(0)} \bt_2 = 0 \,,
              \quad \bt_1\scalp\tilde{\bM}^{(0)}\bt_1 = \bt_2\scalp\tilde{\bM}^{(0)}\bt_2= \rho\, \lvert\mC\lvert \,,
          \end{equation*}
          where $\rho$ is the average mass density
          \begin{equation}
              \rho = \frac{1}{\lvert\mC\lvert}\sum_{k=1}^{N_b} \gamma_k\,l_k \,,
          \end{equation}
          and $\lvert\mC\lvert$ the area of the unit cell.
          Note that the matrix $\bGamma$ contains only terms in the form $\bt_i\scalp\tilde{\bM}^{(0)}\bt_j$, where $\bt_i$ are rigid translations, and therefore \textit{the rotational inertia of the rods plays no role} (recall the definition~\eqref{eq:M_beam}).
\end{enumerate}

Finally the eigenvalue problem~\eqref{eq:eigenvalue_problem_omega1} can be written in the standard form
\begin{equation}
    \label{eq:eigenvalue_problem_omega1_compact}
    \left[\bA^{\text{(L)}}(\bn) - \rho \left(\omega_{\bn}^{(1)}\right)^2 \bI \right] \balpha = \bzero \,,
\end{equation}
where $\balpha = \trans{\{\alpha_1,\alpha_2\}}$ and tensor $\bA^{\text{(L)}}(\bn)$ reads as
\begin{equation}
    \label{eq:acoustic_tensor_lattice}
    \bA^{\text{(L)}}(\bn) = \frac{1}{\lvert\mC\lvert}
    \begin{bmatrix}
        \bt_1\scalp\tilde{\bK}^{(1)}_{\bn}\bt_1 + \bt_1'(\bn)\scalp\tilde{\bK}^{(0)}\bt_1'(\bn) & \bt_2\scalp\tilde{\bK}^{(1)}_{\bn}\bt_1 + \bt_2'(\bn)\scalp\tilde{\bK}^{(0)}\bt_1'(\bn) \\
        \bt_1\scalp\tilde{\bK}^{(1)}_{\bn}\bt_2 + \bt_1'(\bn)\scalp\tilde{\bK}^{(0)}\bt_2'(\bn) & \bt_2\scalp\tilde{\bK}^{(1)}_{\bn}\bt_2 + \bt_2'(\bn)\scalp\tilde{\bK}^{(0)}\bt_2'(\bn)
    \end{bmatrix} .
\end{equation}

It is important to note at this stage, that the eigenvalue problem~\eqref{eq:eigenvalue_problem_omega1_compact} has exactly the same structure of Eq.~\eqref{eq:eigenvalue_problem_cauchy}.
Furthermore, tensor $\bA^{\text{(L)}}(\bn)$, the `acoustic tensor of the lattice', uniquely defines the eigenvalues and eigenvectors appearing on the right-hand side of the equivalence conditions~\eqref{eq:omega_phi_match}, so that $\widehat{\omega} = c_{\bn}\,k = \omega_{\bn}^{(1)} k$ and $\ba = \bvarphi_{\bn}^{(0)} = \balpha$.
\textit{This implies that the acoustic equivalence holds if and only if the `acoustic tensor of the lattice' coincides with the acoustic tensor of the continuum material.}
Therefore, the effective elastic continuum has to satisfy the acoustic equivalence condition (valid for every unit vector $\bn$)
\begin{equation}
    \label{eq:equivalence_acoustic_tensors}
    \bA^{(\fC)}(\bn) = \bA^{\text{(L)}}(\bn) \,.
\end{equation}

It is important to note that the equivalence condition has been obtained without introducing restrictive assumptions on the lattice structure, so that the homogenization method is completely general and \textit{includes a generic state of axial preload acting on the lattice}.
Moreover, the presented technique can easily be extended to three-dimensional lattices.

\subsection{Identification of the continuum equivalent to a preloaded lattice}
\label{sec:identification_constitutive_tensor}
The perturbation method developed in Section~\ref{sec:asymptotic_lattice_waves} leads to the determination of the acoustic tensor of an effective prestressed elastic continuum, equivalent to the low-frequency response of a preloaded grillage of rods.
As the method is entirely based on the dynamics of the periodic medium, the acoustic tensor is obtained directly, without any prior computation of the effective constitutive tensor, which is instead traditional in standard energy-based homogenization techniques~\cite{willis_2002,hutchinson_2006,elsayed_2010,bacca_2013,abdoul-anziz_2018}.

In this section the steps for retrieving the incremental (or `tangent') constitutive tensor $\fC$ are outlined from the acoustic tensor given by Eq.~\eqref{eq:equivalence_acoustic_tensors}.

As the condition~\eqref{eq:equivalence_acoustic_tensors} has to hold for an arbitrary direction of propagation, it can equivalently be expressed by applying the Hessian with respect to $\bn$ on both sides of Eq.~\eqref{eq:equivalence_acoustic_tensors} to obtain
\begin{equation}
    \label{eq:hessian_acoustic_tensor}
    \fC_{ikjl} + \fC_{iljk} = \frac{\partial^2 A^{\text{(L)}}_{ij}(\bn)}{\partial n_k \partial n_l} \,,
\end{equation}
where the right-hand side can be regarded as a data defined by the lattice structure, namely, the Hessian of tensor~\eqref{eq:acoustic_tensor_lattice}.
By considering the symmetry with respect to the $\{k,l\}$ indices, Eq.~\eqref{eq:hessian_acoustic_tensor} provides a linear system of 54 equations in a three-dimensional setting or 12 equations in a two-dimensional setting, while the rank of the system is found to be 26 or 8, respectively.
By recalling that the unknown tensor $\fC$ has the form~\eqref{eq:constitutive_operator}, it is clear that, if the system is solvable, all but one of the unknown components of $\fC$ can be determined as these are 27 for a three-dimensional lattice and 9 for a two-dimensional.

In order to solve for the identification, results obtained by Max Born~\cite{born_1955} can now be generalized to prove that
\begin{enumerate*}[label=(\roman*)]
    \item the system is solvable when the equations of motion of the lattice satisfy the rotational invariance and
    \item the solution is unique, except for the spherical part of the prestress (i.e. $\tr{\bT}/3$ in 3D and $\tr{\bT}/2$ in 2D) which remains undetermined for the system~\eqref{eq:hessian_acoustic_tensor}.
    However, the homogenized prestress tensor $\bT$, and hence its spherical part, can be directly obtained by averaging the tractions along the boundary of the unit cell, as will be done for the lattice considered in Section~\ref{sec:grid}.
\end{enumerate*}

\section{Stability of prestressed lattices of elastic rods, strong ellipticity, and ellipticity of the effective continuum}
\label{sec:ellipticity_stability}
The homogenization technique developed in the previous section allows the determination of a prestress-sensitive elastic solid which captures the effective behavior of the preloaded lattice.
The step is crucial for the investigation of material instabilities and bifurcations so that the possibility of designing a lattice representative of a material with special characteristics can be analyzed.
The present section outlines the theoretical framework for the stability analysis that will be applied to both the grid and its continuous approximation.

\paragraph{Lattice bifurcations} are governed by the value of the preload state $\bP=\{P_1,\dots,P_{N_b}\}$ and they can exhibit deformation modes with different wavelengths.
When the wavelength becomes infinite, a `global' or `macro' bifurcation occurs, otherwise the bifurcation is called `microscopic'.
The systematic investigation of bifurcations occurring in the lattice can be conducted by analyzing the incremental equilibrium of the lattice~\cite{hutchinson_2006}.
To this end, the formulation of the lattice dynamics outlined in Section~\ref{sec:system_governing_equations} can be directly specialized for the stability analysis by considering the quasi-static limit.

In the limit of vanishing frequency, Eq.~\eqref{eq:system} yields
\begin{equation}
    \label{eq:equilibrium_local_buckling}
    \conjtrans{\bZ(\bk)} \bK^{(0)}(\bP)\, \bZ(\bk) \, \bq^* = \bzero \,,
\end{equation}
where the dependence of the static stiffness $\bK^{(0)}(\bP)$ on the prestress state $\bP$ has been made explicit.

For a given $\bk$, the associated preload state $\bP$ leading to a bifurcation can be obtained by searching for non-trivial solutions of the incremental equilibrium~\eqref{eq:equilibrium_local_buckling}.
Hence, by introducing the notation $\bK^{*(0)}(\bP,\bk)=\conjtrans{\bZ(\bk)} \bK^{(0)}(\bP)\, \bZ(\bk)$, a bifurcation becomes possible when
\begin{equation}
    \label{eq:det_local_buckling}
    \det \bK^{*(0)}(\bP,\bk) = 0 \,.
\end{equation}
Note that, as the matrix $\bK^{*(0)}(\bP,\bk)$ is Hermitian, the determinant~\eqref{eq:det_local_buckling} is always real.
Moreover, the periodicity of $\bZ(\bk)$ implies that this determinant is periodic in the $\bk$-space with period $[0,2\pi]{\times}[0,2\pi]$ in the basis $\{\bb_1,\bb_2\}$, reciprocal to $\{\ba_1,\ba_2\}$, so that  $\bb_i\scalp\ba_j=\delta_{ij}$.

In order to construct the \textit{stability domain} of a lattice, the \textit{critical} (in other words, first) bifurcation needs to be selected by solving Eq.~\eqref{eq:det_local_buckling} for the smallest preload spanning over all possible wavelengths.
Specifically, by introducing the unit vector $\hat{\bP}$, which singles out a direction in the preload space, the prestress state is defined as $\gamma \hat{\bP}$ for a radial loading.
Therefore, the critical bifurcation corresponds to the value $\gamma_{\text{B}}$ defined as
\begin{equation}
    \label{eq:local_buckling_multiplier_simpler}
    \gamma_{\text{B}} = \inf_{\gamma \geq 0}
    \left\{ \gamma \, \Big\lvert \, \det\bK^{*(0)}(\gamma\hat{\bP},\, \eta_1\bb_1+\eta_2\bb_2) = 0\,,\, 0 < \eta_1 < 2\pi\,,\, 0 < \eta_2 < 2\pi \right\} ,
\end{equation}
where the periodicity of $\bK^{*(0)}(\bP,\bk)$ is used to conveniently restrict to one period the search for the infimum over the $\bk$-space.
It is worth noting that for a vanishing wave vector, Eq.~\eqref{eq:det_local_buckling} is always satisfied regardless of the preload state, because the nullspace of $\bK^{*(0)}(\bP,\bzero)$ always contains rigid-body translations.
These trivial solutions clearly need not be considered.

\paragraph{Strong ellipticity} (SE) enforces uniqueness of the incremental problem of a homogeneous and homogeneously deformed material subject to prescribed incremental displacement on the whole boundary \cite{hill_1962} and corresponds to the positive definiteness of the acoustic tensor (associated to the incremental constitutive tensor $\fC$) defined with reference to every unit vectors $\bn$ and $\bg$ as
\begin{equation}
    \bA^{(\fC)}(\bn)\,\bg = \fC[\bg\otimes\bn]\,\bn \,.
\end{equation}
When the prestress state is null and except in the case of an extreme material, where the stiffness of the rods becomes vanishing small~\cite{gourgiotis_2016}, the homogenized material response is strongly elliptic,
\begin{equation}
    \label{eq:strong_ellipticity}
    \bg \scalp \bA^{(\fC)}(\bn)\, \bg > 0 \quad \forall \bg\neq\bzero \quad \forall \bn \neq\bzero \,.
\end{equation}

\paragraph{Failure of ellipticity,} which characterizes the onset of a localization of deformation in the equivalent continuum, corresponds to a macro (or global) instability, where the bifurcation is characterized by an infinitely long wavelength (when compared to the period of the lattice structure).
The homogenized material is elliptic (E) as long as the the acoustic tensor $\bA^{(\fC)}(\bn)$ is non-singular for every pair of unit vectors $\bn$ and $\bg$, namely,
\begin{equation}
    \label{eq:ellipticity}
    \bA^{(\fC)}(\bn)\,\bg \neq \bzero \,.
\end{equation}
When the acoustic tensor becomes singular, a localization of deformation may occur corresponding to a dyad $\bg \otimes \bn$.
The localization is called `shear band' in the special case $\bg\scalp\bn=0$, or `compaction band' or `splitting mode' when $\bg\scalp\bn=\pm 1$.

The prestress-dependent stiffness implies that the homogenized acoustic tensor~\eqref{eq:acoustic_tensor_lattice} is in turn a function of the axial preloads $\bP$ in the rods, so that the notation $\bA^{(\fC)}(\bP, \bn)$ is introduced.
Therefore, using again the previously defined unit vector $\hat{\bP}$ and with reference to an infinite material (or to a material with prescribed displacement on the whole boundary) bifurcations are excluded as long as the response remains strongly elliptic.
Failure of this condition determines a simultaneous failure of ellipticity, which corresponds to the value $\gamma_{\text{E}}$ defined as
\begin{equation}
    \label{eq:failure_E_multiplier}
    \gamma_{\text{E}} = \min_{\gamma\geq0}\left\{ \gamma \, \Big\lvert \, \min_{\bn,\norm{\bn}=1} \left[\det\bA^{(\fC)}(\gamma\hat{\bP},\bn)\right]=0\right\}.
\end{equation}

\paragraph{Relation between bifurcations in the lattice and in the effective continuum} is that failure of ellipticity of the latter corresponds to long-wavelength bifurcations of the former, $\norm{\bk}\to0$, while \textit{all} bifurcations are scanned through equation~\eqref{eq:local_buckling_multiplier_simpler}, a circumstance which implies $\gamma_{\text{B}}\leq\gamma_{\text{E}}$.
Moreover, whenever $\gamma_{\text{B}}<\gamma_{\text{E}}$ \textit{the bifurcation occurs at microscopic level and is not detectable in the homogenized material, which can still be strongly elliptic}~\cite{geymonat_1993,triantafyllidis_1993,lopez-pamies_2006}.

\section{Derivation of the incremental constitutive tensor, failure of ellipticity and micro-bifurcation for a preloaded elastic grid}
\label{sec:grid}
In order to demonstrate the effectiveness of the homogenization method developed in Section~\ref{sec:asymptotic_lattice_waves}, bifurcation and loss of ellipticity are investigated in a preloaded two-dimensional grid lattice of elastic rods.
The grillage will be directly analyzed with the Floquet-Bloch technique reviewed in Section \ref{sec:system_governing_equations} and results will be compared to those obtained on the equivalent elastic material, Section \ref{sec:asymptotic_lattice_waves}.
\begin{figure}[htb!]
    \centering
    \begin{subfigure}{0.49\textwidth}
        \centering
        \caption{\label{fig:geometry_grid}}
        \includegraphics[height=52mm]{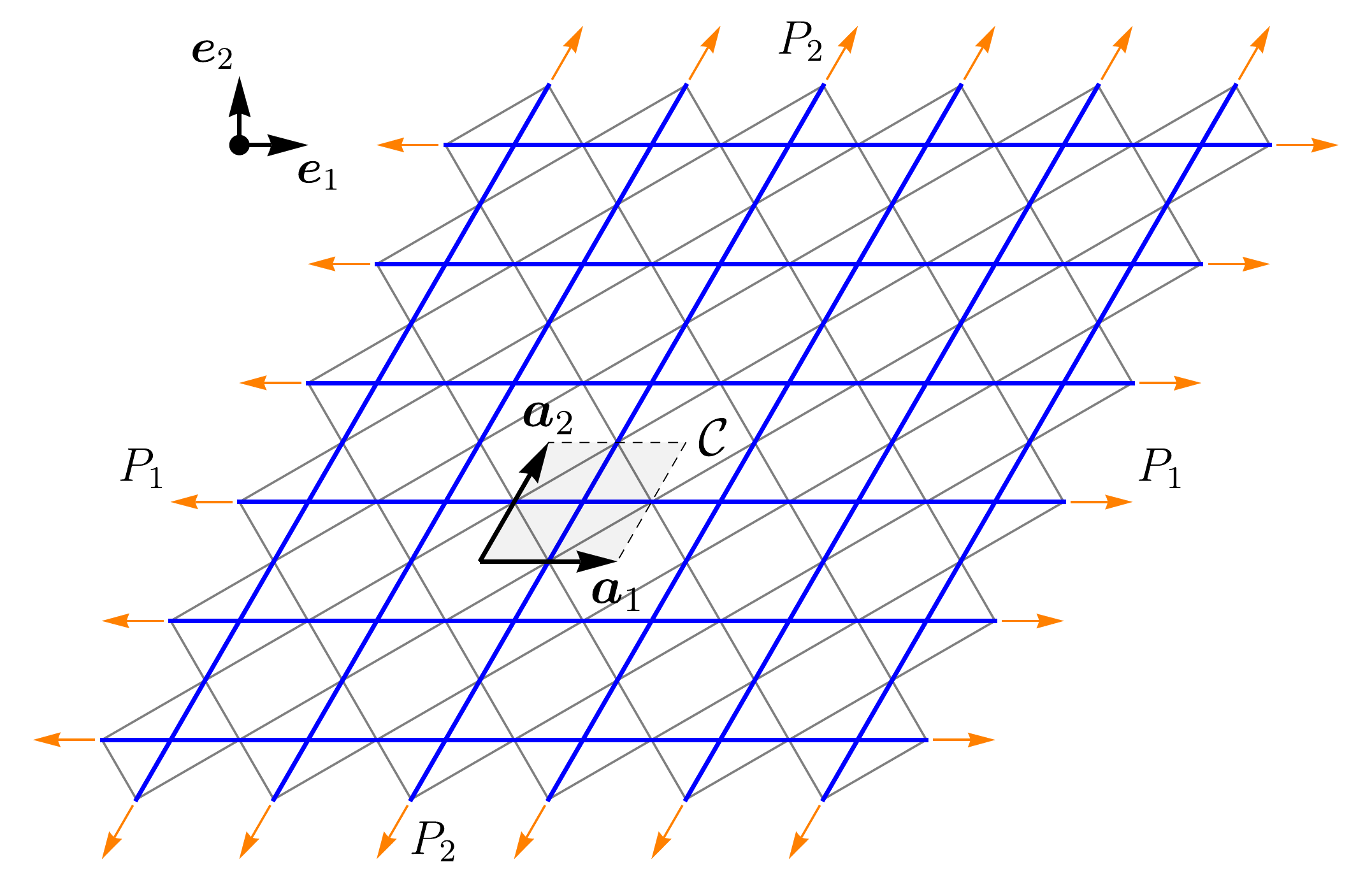}
    \end{subfigure}\hspace{3mm}%
    \begin{subfigure}{0.45\textwidth}
        \centering
        \caption{\label{fig:geometry_grid_unit_cell}}
        \includegraphics[height=52mm]{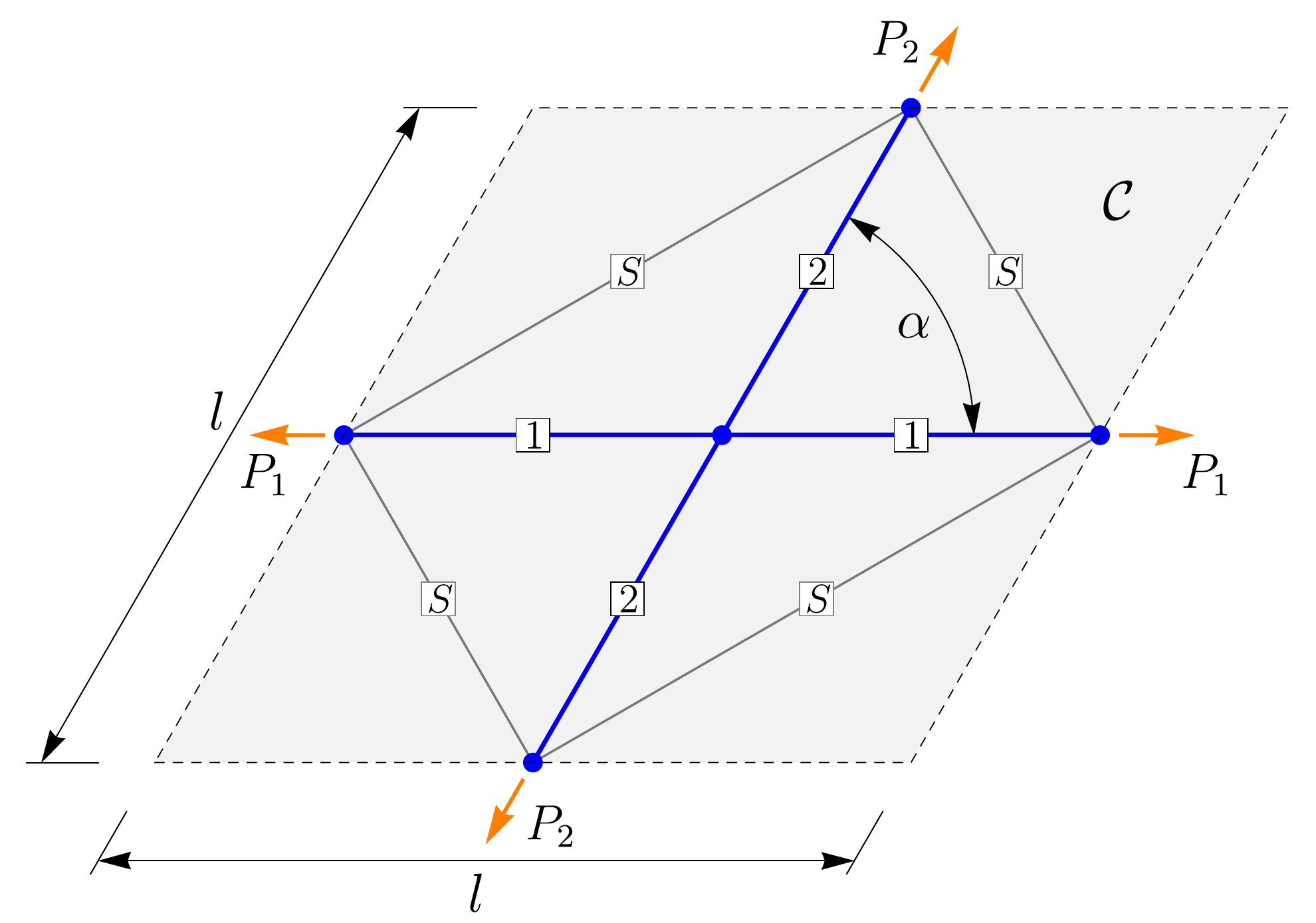}
    \end{subfigure}%
    \caption{\label{fig:geometry_grid_and_unit_cell}
        Current configuration of a rhombic lattice of preloaded elastic rods~(\subref{fig:geometry_grid}), with the associated unit cell $\mC$~(\subref{fig:geometry_grid_unit_cell}).
        The direct basis of the lattice is denoted by the pair of vectors $\{\ba_1,\ba_2\}$ (\subref{fig:geometry_grid}).
        Labels 1, 2, and $S$ denote the horizontal rods, the inclined rods, and the diagonal springs, respectively~(\subref{fig:geometry_grid_unit_cell}).
        The spring stiffness, the axial and flexural rigidity of the rods, the preloads $P_1$ and $P_2$, as well as the grid angle $\alpha$ can all be varied to investigate different incremental responses.
    }
\end{figure}

The geometry of the current, prestressed configuration of the lattice, selected to apply the previously developed formalism, is sketched in Fig.~\ref{fig:geometry_grid_and_unit_cell}.
This is composed of a rhombic grid (of side $l$) of elastic rods, inclined at an angle $\alpha$, and characterized by the following non-dimensional parameters $A_2=A_1=A$, $\Lambda_1 = l \sqrt{A/B_1}$, $\Lambda_2 = l \sqrt{A/B_2}$, where the subscript $1$ and $2$ are relative to the horizontal and inclined rods, as depicted in Fig.~\ref{fig:geometry_grid_unit_cell}.
For simplicity, the linear mass density is assumed to be the same for both rods $\gamma_1=\gamma_2=\gamma$, while the rotational inertia has been shown in Section~\ref{sec:acoustic_tensor_lattice} to be negligible in the homogenization scheme.
The direct basis of the periodic structure is denoted by the pair of vectors $\{\ba_1,\ba_2\}$ whose representation with respect to the basis $\{\be_1,\be_2\}$ (see Fig.~\ref{fig:geometry_grid}) is
\begin{equation*}
    \ba_1 = l\,\be_1 \,, \qquad \ba_2 = l\,(\be_1\cos\alpha+\be_2\sin\alpha) \,,
\end{equation*}
while the reciprocal basis $\{\bb_1,\bb_2\}$ is defined as $\ba_i\scalp\bb_j=\delta_{ij}$, so that
\begin{equation*}
    \bb_1 = (\be_1 - \be_2\cot\alpha)/l \,, \qquad \bb_2 = \be_2\csc\alpha/l \,.
\end{equation*}
Therefore, the wave vector can be written as
\begin{equation}
    \label{eq:reciprocal_basis}
    \bk= \eta_1 \bb_1 + \eta_2 \bb_2 \,,
\end{equation}
where $\eta_{1}$ and $\eta_{2}$ are dimensionless components.

The `skewed' grid resulting from the above description is also stiffened by a diagonal bracing realized with linear springs\footnote{
    These springs can be thought as added after the lattice has been deformed or as deformed together with the lattice. In the former case further assumptions need not be introduced, while in the latter, the effects of the preload on the springs has to be neglected in the interest of simplicity. The diagonal springs are used in this example to show that microscopic instabilities may occur before macroscopic.}
connecting the midpoints of the horizontal and inclined rods, as sketched in Fig.~\ref{fig:geometry_grid_unit_cell}.
The stiffness of the springs is assumed constant $k_{\text{S}}=\kappa A/l$, with $\kappa$ being a dimensionless measure of stiffness.

In the configuration shown in Fig.~\ref{fig:geometry_grid_and_unit_cell}, the lattice is subject to a preload state defined by the axial forces $P_1$ and $P_2$, made dimensionless respectively as $p_1 = {P_1 l^2}/{B_1}$ and $p_2 = {P_2 l^2}/{B_2}$, so that a deformed configuration of a lattice is defined by the parameter set $\{p_1,p_2,\Lambda_1,\Lambda_2,\kappa,\alpha\}$.
Note also that the considered lattice structure includes, as a special case, the rectangular grid analyzed in~\cite{triantafyllidis_1993}.

Failure of ellipticity, macroscopic and microscopic bifurcations, and stability of the preloaded lattice are investigated in the following subsections.
\begin{enumerate}[label=(\roman*)]
    \item The acoustic tensor $\bA^{(\fC)}(\bn)$ of the homogenized continuum is derived analytically as an explicit function of the parameter set $\{p_1,p_2,\Lambda_1,\Lambda_2,\kappa,\alpha\}$ (the corresponding constitutive tensor $\fC$ is also determined $\fC$);
    \item Loss of ellipticity is analyzed for cubic, orthotropic, and fully anisotropic lattices by identifying the prestress states leading to a vanishing eigenvalue of the acoustic tensor, and computing the corresponding eigenvector defining the localization mode;
    \item The stability domains in the $\{p_1,p_2\}$-space and its dependence on lattice parameters $\{\Lambda_1,\Lambda_2,\kappa,\alpha\}$ is determined;
    \item The bifurcation modes are evaluated for the most relevant configurations of the grid, so to clarify the difference between macro and micro bifurcation.
\end{enumerate}

The analysis of the lattice response near the identified stability thresholds is addressed in Section~\ref{sec:dynamic_forced_response} and~\ref{sec:static_forced_response}.

\subsection{Acoustic tensor, eigenvalues, eigenvectors, and ellipticity domain}
\label{sec:acoustic_tensor_grid}
With reference to the orthonormal basis defined by the two unit vectors $\{\be_1,\be_2\}$, the acoustic tensor $\bA^{(\fC)}(\bn)$ for the continuum equivalent, in a homogenized sense, to the lattice shown in Fig.~\ref{fig:geometry_grid_and_unit_cell} is represented as
\begin{equation}
    \label{eq:acoustic_tensor_grid}
    \bA^{(\fC)}(\bn) = A^{(\fC)}_{11}(\bn)\,\be_1\otimes\be_1 + A^{(\fC)}_{12}(\bn)\,\be_1\otimes\be_2 + A^{(\fC)}_{21}(\bn)\,\be_2\otimes\be_1 + A^{(\fC)}_{22}(\bn)\,\be_2\otimes\be_2 \,,
\end{equation}
where the components, computed via Eq.~\eqref{eq:acoustic_tensor_lattice}, are expressed as follows
\begin{equation*}
    \begin{aligned}
        A^{(\fC)}_{11}(\bn)                       & = \left(h_{1111}\,n_1^2 + h_{1112}\,n_1 n_2 + h_{1122}\,n_2^2\right) A/l \,, \\
        A^{(\fC)}_{12}(\bn) = A^{(\fC)}_{21}(\bn) & = \left(h_{1211}\,n_1^2 + h_{1212}\,n_1 n_2 + h_{1222}\,n_2^2\right) A/l \,, \\
        A^{(\fC)}_{22}(\bn)                       & = \left(h_{2211}\,n_1^2 + h_{2212}\,n_1 n_2 + h_{2222}\,n_2^2\right) A/l \,,
    \end{aligned}
\end{equation*}
with the coefficients $h_{ijkl}$ being function of the parameter set $\{p_1,p_2,\Lambda_1,\Lambda_2,\kappa,\alpha\}$.
The contribution of the rods' grid and the springs are denoted as $h_{ijkl}^{\text{G}}$ and $h_{ijkl}^{\text{S}}$, respectively, so that
\begin{equation}
    \label{eq:acoustic_tensor_grid_contributions}
    h_{ijkl}(p_1,p_2,\Lambda_1,\Lambda_2,\kappa,\alpha) = h_{ijkl}^{\text{G}} (p_1,p_2,\Lambda_1,\Lambda_2,\alpha) + h_{ijkl}^{\text{S}}(\kappa,\alpha) \,.
\end{equation}
The nonlinear dependence on the prestress $\bp = \{p_1, p_2\}$ causes the full expression for the functions $h_{ijkl}^{\text{G}}$ to be quite lengthy and therefore the complete result is omitted (but all components of the constitutive tensor are resported in Appendix~\ref{sec:homogenized_constitutive_tensor_grid}), while the first-order expansion with respect to $\bp$ is (components that have to be equal by symmetry are not reported)
\begin{dgroup*}[style={\footnotesize},breakdepth={20}]
    \begin{dmath*}
        h_{1111}^{\text{G}} \sim \frac{12 \sin\alpha \cos^2\alpha}{\Lambda_1^2+\Lambda_2^2}+\csc\alpha+\cos^3\alpha \cot\alpha
        +p_1 \frac{\Lambda_1^2 \sin\alpha \cos^2\alpha}{5 \left(\Lambda_1^2+\Lambda_2^2\right)^2}
        +p_2 \frac{\sin\alpha \cos^2\alpha \left(5 \Lambda_1^4+10 \Lambda_1^2 \Lambda_2^2+6 \Lambda_2^4\right)}{5 \Lambda_2^2 \left(\Lambda_1^2+\Lambda_2^2\right)^2}
        ,
    \end{dmath*}
    \begin{dmath*}
        h_{1112}^{\text{G}} \sim \frac{2 \cos\alpha \left(\cos^2\alpha \left(\Lambda_1^2+\Lambda_2^2-12\right)+12\right)}{\Lambda_1^2+\Lambda_2^2}
        +p_1 \frac{2 \Lambda_1^2 \sin^2\alpha \cos\alpha}{5 \left(\Lambda_1^2+\Lambda_2^2\right)^2}
        +p_2 \frac{2 \sin^2\alpha \cos\alpha \left(5 \Lambda_1^4+10 \Lambda_1^2 \Lambda_2^2+6 \Lambda_2^4\right)}{5 \Lambda_2^2 \left(\Lambda_1^2+\Lambda_2^2\right)^2}
        ,
    \end{dmath*}
    \begin{dmath*}
        h_{1122}^{\text{G}} \sim \frac{12 \sin^3\alpha}{\Lambda_1^2+\Lambda_2^2}+\sin\alpha \cos^2\alpha
        +p_1 \frac{\Lambda_1^2 \sin^3\alpha}{5 \left(\Lambda_1^2+\Lambda_2^2\right)^2}
        +p_2 \frac{\sin^3\alpha \left(5 \Lambda_1^4+10 \Lambda_1^2 \Lambda_2^2+6 \Lambda_2^4\right)}{5 \Lambda_2^2 \left(\Lambda_1^2+\Lambda_2^2\right)^2}
        ,
    \end{dmath*}
    \begin{dmath*}
        h_{1211}^{\text{G}} \sim \frac{\cos\alpha \left(\cos^2\alpha \left(\Lambda_1^2+\Lambda_2^2-12\right)+12\right)}{\Lambda_1^2+\Lambda_2^2}
        +p_1 \frac{\Lambda_1^2 \sin^2\alpha \cos\alpha}{5 \left(\Lambda_1^2+\Lambda_2^2\right)^2}
        +p_2 \frac{\Lambda_2^4 \cos\alpha-\cos^3\alpha \left(5 \Lambda_1^4+10 \Lambda_1^2 \Lambda_2^2+6 \Lambda_2^4\right)}{5 \Lambda_2^2 \left(\Lambda_1^2+\Lambda_2^2\right)^2}
        ,
    \end{dmath*}
    \begin{dmath*}
        h_{1212}^{\text{G}} \sim \frac{\sin\alpha \left(\cos (2 \alpha ) \left(\Lambda_1^2+\Lambda_2^2-12\right)+\Lambda_1^2+\Lambda_2^2\right)}{\Lambda_1^2+\Lambda_2^2}
        -p_1 \frac{\Lambda_1^2 \sin\alpha \cos (2 \alpha )}{5 \left(\Lambda_1^2+\Lambda_2^2\right)^2}
        +p_2 \frac{\sin\alpha \left(\Lambda_2^4-2 \cos^2\alpha \left(5 \Lambda_1^4+10 \Lambda_1^2 \Lambda_2^2+6 \Lambda_2^4\right)\right)}{5 \Lambda_2^2 \left(\Lambda_1^2+\Lambda_2^2\right)^2}
        ,
    \end{dmath*}
    \begin{dmath*}
        h_{1222}^{\text{G}} \sim \frac{\sin^2\alpha \cos\alpha \left(\Lambda_1^2+\Lambda_2^2-12\right)}{\Lambda_1^2+\Lambda_2^2}
        -p_1 \frac{\Lambda_1^2 \sin^2\alpha \cos\alpha}{5 \left(\Lambda_1^2+\Lambda_2^2\right)^2}
        +p_2 -\frac{\sin^2\alpha \cos\alpha \left(5 \Lambda_1^4+10 \Lambda_1^2 \Lambda_2^2+6 \Lambda_2^4\right)}{5 \Lambda_2^2 \left(\Lambda_1^2+\Lambda_2^2\right)^2}
        ,
    \end{dmath*}
    \begin{dmath*}
        h_{2211}^{\text{G}} \sim \frac{12 \sin^3\alpha}{\Lambda_1^2+\Lambda_2^2}+\sin\alpha \cos^2\alpha
        +p_1 \left( \frac{\Lambda_1^2 \sin^3\alpha}{5 \left(\Lambda_1^2+\Lambda_2^2\right)^2}+\frac{\csc\alpha}{\Lambda_1^2} \right)
        +p_2 \frac{4 \cos (2 \alpha ) \left(5 \Lambda_1^4+10 \Lambda_1^2 \Lambda_2^2+4 \Lambda_2^4\right)+(\cos (4 \alpha )+3) \left(5 \Lambda_1^4+10 \Lambda_1^2 \Lambda_2^2+6 \Lambda_2^4\right)}{40 \Lambda_2^2 \left(\Lambda_1^2+\Lambda_2^2\right)^2 \sin\alpha}
        ,
    \end{dmath*}
    \begin{dmath*}
        h_{2212}^{\text{G}} \sim \frac{\sin\alpha \sin (2 \alpha ) \left(\Lambda_1^2+\Lambda_2^2-12\right)}{\Lambda_1^2+\Lambda_2^2}
        -p_1 \frac{2 \Lambda_1^2 \sin^2\alpha \cos\alpha}{5 \left(\Lambda_1^2+\Lambda_2^2\right)^2}
        +p_2 \frac{\cos\alpha \left(\cos (2 \alpha ) \left(5 \Lambda_1^4+10 \Lambda_1^2 \Lambda_2^2+6 \Lambda_2^4\right)+5 \Lambda_1^4+10 \Lambda_1^2 \Lambda_2^2+4 \Lambda_2^4\right)}{5 \Lambda_2^2 \left(\Lambda_1^2+\Lambda_2^2\right)^2}
        ,
    \end{dmath*}
    \begin{dmath*}
        h_{2222}^{\text{G}} \sim \frac{12 \sin\alpha \cos^2\alpha}{\Lambda_1^2+\Lambda_2^2}+\sin^3\alpha
        +p_1 \frac{\Lambda_1^2 \sin\alpha \cos^2\alpha}{5 \left(\Lambda_1^2+\Lambda_2^2\right)^2}
        +p_2 \frac{\sin\alpha \cos^2\alpha \left(5 \Lambda_1^4+10 \Lambda_1^2 \Lambda_2^2+6 \Lambda_2^4\right)}{5 \Lambda_2^2 \left(\Lambda_1^2+\Lambda_2^2\right)^2}
        .
    \end{dmath*}
\end{dgroup*}
The components $h_{ijkl}^{\text{S}}$, ruling the effect of diagonal springs, can be written as
\begin{align*}
    h^{\text{S}}_{1111} & = \kappa \frac{5+3 \cos (2 \alpha)}{4\sin\alpha}  \,,                              &
    h^{\text{S}}_{1112} & = 2 \kappa  \cos \alpha  \,,                                                       &
    h^{\text{S}}_{1211} & = \kappa  \cos \alpha \,,                                                            \\
    h^{\text{S}}_{1212} & = \kappa  \sin \alpha \,,                                                          &
    h^{\text{S}}_{1122} & = h^{\text{S}}_{2211} = h^{\text{S}}_{2222} = \frac{1}{2} \kappa  \sin \alpha  \,, &
    h^{\text{S}}_{1222} & = h^{\text{S}}_{2212} = 0 \,.
\end{align*}

For the special case of a square grid $\alpha=\pi/2$ and in the absence of prestress ($\bp=\bzero$), the acoustic tensor can be further simplified to
\begin{align*}
    \lim_{\bp\to\bzero} \left.\bA^{(\fC)}(\bn)\right\rvert_{\alpha=\pi/2} & =
    \frac{A}{l}\left(n_1^2+\frac{12 n_2^2}{\Lambda_1^2+\Lambda_2^2}\right)\,\be_1\otimes\be_1 +                                                                                         \\
                                                                          & + \frac{A}{l}\frac{12 n_1 n_2}{\Lambda_1^2+\Lambda_2^2}\,\left(\be_1\otimes\be_2 + \be_2\otimes\be_1\right)
    + \frac{A}{l}\left(\frac{12 n_1^2}{\Lambda_1^2+\Lambda_2^2}+n_2^2\right)\,\be_2\otimes\be_2 \,.
\end{align*}

As shown in Section~\ref{sec:acoustic_tensor_lattice}, the acoustic tensor resulting from homogenization is symmetric, which implies that its eigenvalues are always real.
This means that, letting $\{c_1^2, c_2^2\}$ be the eigenvalues of a symmetric $\bA^{(\fC)}(\bn)$, (SE) is equivalent to the strict positiveness of the eigenvalues, $c_1^2>0,\, c_2^2>0$, Eq.~\eqref{eq:strong_ellipticity}, while (E) is equivalent to the condition of non-vanishing eigenvalues, $c_1^2\neq0,\, c_2^2\neq0$ (for all unit vectors $\bn$), Eq.~\eqref{eq:ellipticity}.
It can be directly verified that in the absence of preload, $\bp\to\bzero$, the considered grid has $c_1^2>0$ and $c_2^2>0$, so that (SE) holds.

The objective is now to characterize \textit{failure of (E)} by studying the eigenvalues of the acoustic tensor~\eqref{eq:acoustic_tensor_grid} as functions of the preload state applied to the grillage.
To this end, solutions are sought for the following loss of ellipticity condition
\begin{equation}
    \label{eq:ellipticity_loss}
    c_1^2(\bn,\bp)\, c_2^2(\bn,\bp) = 0 \,,
\end{equation}
where $\bn$ is the usual unit vector defining the direction of propagation and $\bp=\{p_1,p_2\}$ is a vector simply collecting the preload parameters.
In Eq.~\eqref{eq:ellipticity_loss} the dependence on the geometric parameters $\{\alpha,\Lambda_1,\Lambda_2\}$ is omitted for brevity and moreover, without loss of generality, it is assumed that $c_1^2\leq c_2^2$.
For every solution $\{\bn_{\text{E}},\bp_{\text{E}}\}$ of Eq.~\eqref{eq:ellipticity_loss}, the eigenvector $\bg_{\text{E}}=\bg(\bn_{\text{E}})$ associated to the vanishing eigenvalue can be computed.
Vectors $\bn_{\text{E}}$ and $\bg_{\text{E}}$ will be respectively referred as the \textit{direction} (more precisely, the normal to) and deformation \textit{mode} of the strain localization band.
\begin{table}[htb!]
    \centering
    \begin{tabular}{lllll}
        \toprule
        \textit{Geometry}      & \textit{Slenderness}         & \textit{Symmetry} & $\bp_{\text{E}}$    & $\theta_{\text{cr}}$     \\ \midrule
        Square $\alpha=\pi/2$  & $\Lambda_1=\Lambda_2=10$     & Cubic             & $\{-5.434,-5.434\}$ & $0^\circ,90^\circ$       \\
                               & $\Lambda_1=7,\,\Lambda_2=15$ & Orthotropic       & $\{-2.071,-2.071\}$ & $0^\circ$                \\
        Rhombus $\alpha=\pi/3$ & $\Lambda_1=\Lambda_2=10$     & Orthotropic       & $\{-5.345,-5.345\}$ & $88.2^\circ,151.8^\circ$ \\
                               & $\Lambda_1=7,\,\Lambda_2=15$ & Anisotropic       & $\{-2.043,-2.043\}$ & $151.4^\circ$            \\
        \bottomrule
    \end{tabular}
    \caption{\label{tab:cases_analyzed}
        Loss of ellipticity for different geometric configurations of the preloaded grid-like lattice (Fig.~\ref{fig:geometry_grid_and_unit_cell}) in the absence of diagonal springs ($\kappa=0$).
        The symmetry class is referred to the unloaded configuration.
        The preload $\bp_{\text{E}}$ and the localization direction $\bn_{\text{E}}=\cos(\theta_{\text{cr}})\be_1+\sin(\theta_{\text{cr}})\be_2$ are obtained by solving the loss of ellipticity condition~\eqref{eq:ellipticity_loss} (assuming a radial path $\bp=\{p_1,p_1\}$).
    }
\end{table}
\begin{figure}[htb!]
    \centering
    \begin{subfigure}{0.46\textwidth}
        \centering
        \caption{\label{fig:eigenvalue_square_10_10}$p_1=p_2=-5.434,\,\Lambda_1=\Lambda_2=10,\,\alpha=\pi/2$}
        \includegraphics[width=0.98\linewidth]{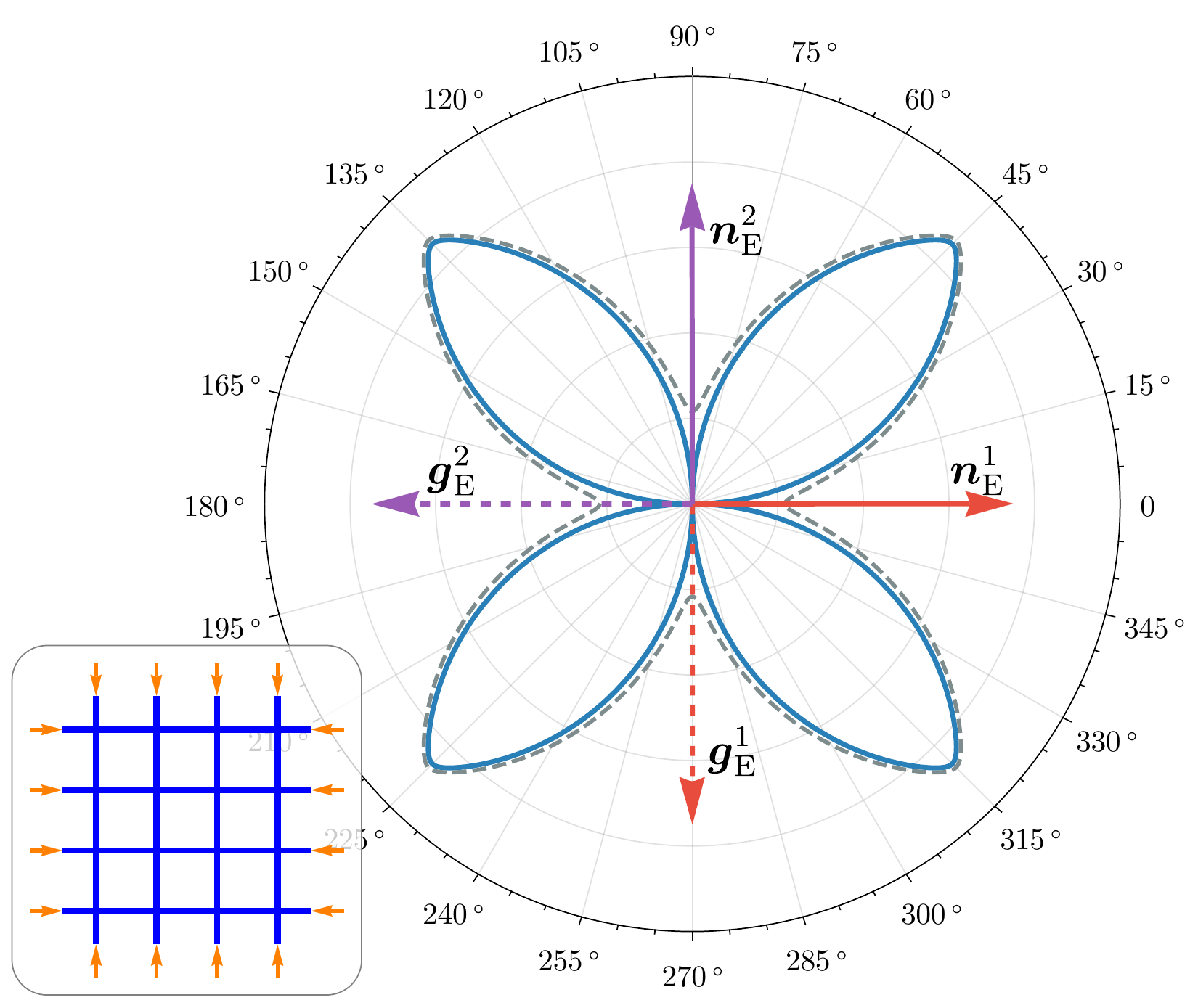}
    \end{subfigure}%
    \begin{subfigure}{0.46\textwidth}
        \centering
        \caption{\label{fig:eigenvalue_square_7_15}$p_1=p_2=-2.071,\,\Lambda_1=7,\, \Lambda_2=15,\,\alpha=\pi/2$}
        \includegraphics[width=0.98\linewidth]{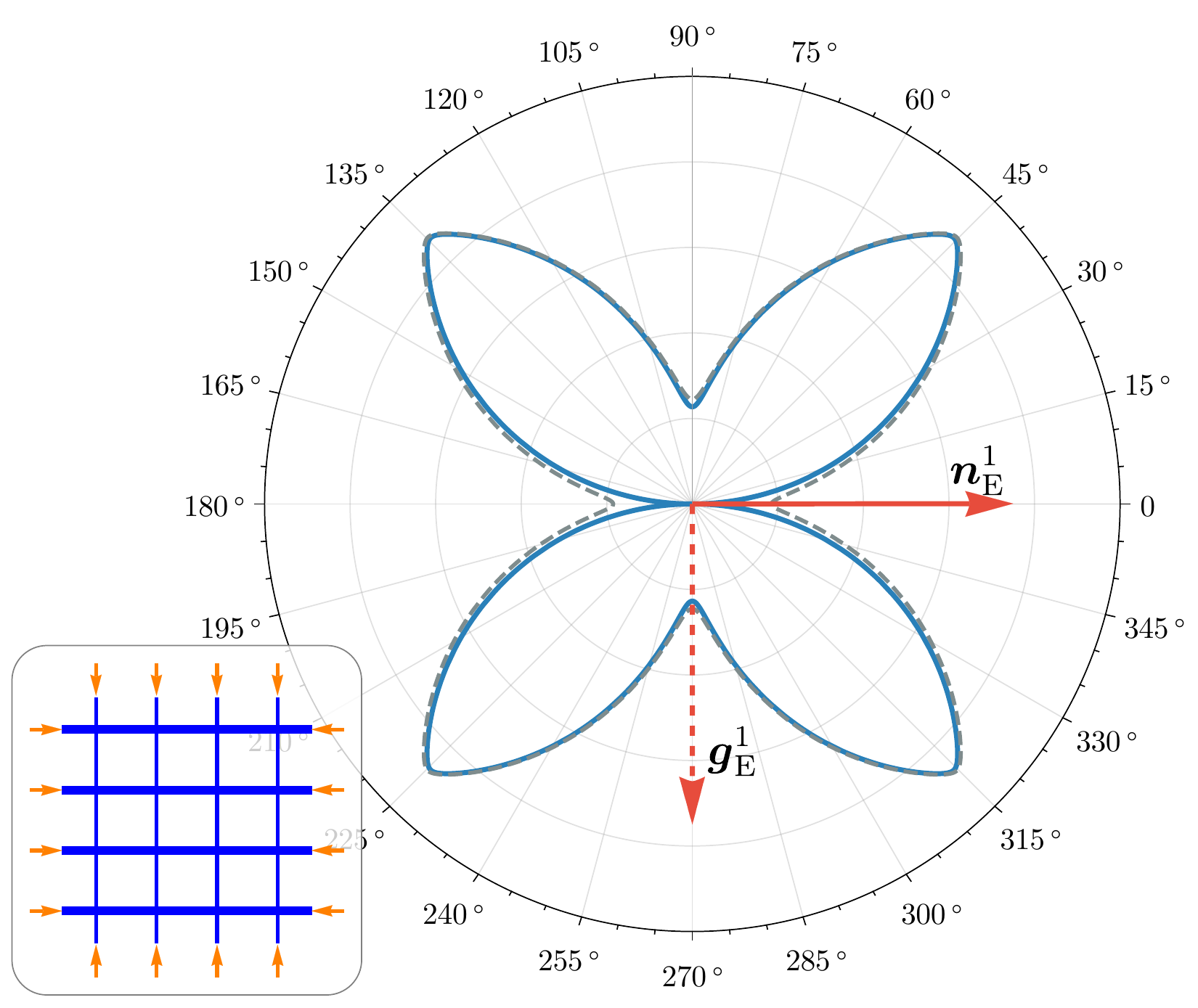}
    \end{subfigure}\\ \vspace{4mm}
    \begin{subfigure}{0.46\textwidth}
        \centering
        \caption{\label{fig:eigenvalue_rhombus_10_10}$p_1=p_2=-5.345,\,\Lambda_1=\Lambda_2=10,\,\alpha=\pi/3$}
        \includegraphics[width=0.98\linewidth]{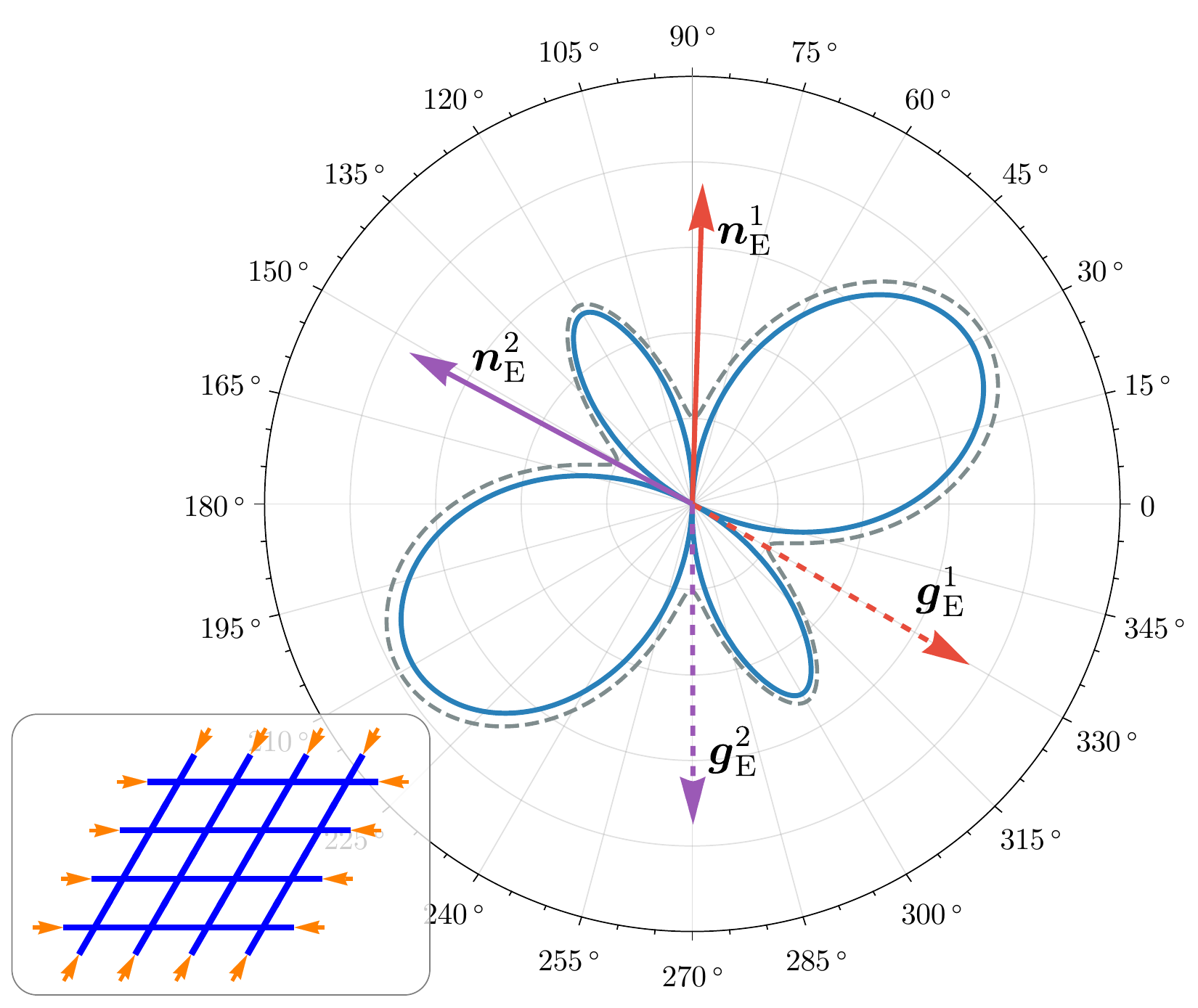}
    \end{subfigure}%
    \begin{subfigure}{0.46\textwidth}
        \centering
        \caption{\label{fig:eigenvalue_rhombus_7_15}$p_1=p_2=-2.043,\,\Lambda_1=7,\, \Lambda_2=15,\,\alpha=\pi/3$}
        \includegraphics[width=0.98\linewidth]{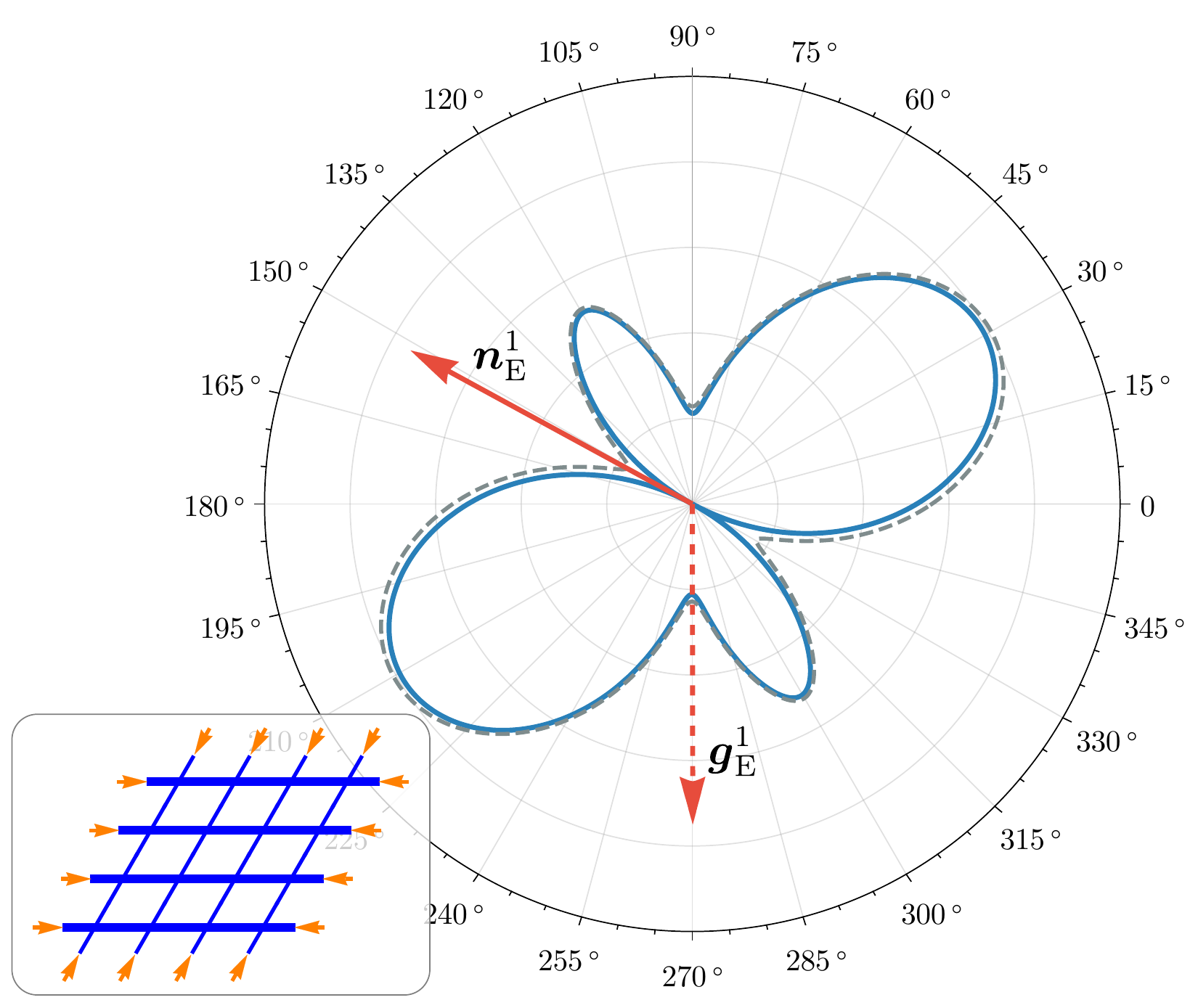}
    \end{subfigure}%
    \caption{\label{fig:eigenvalue_square_rhombus}
    Polar plots of the square root of the lowest eigenvalue of the acoustic tensor~\eqref{eq:acoustic_tensor_grid}, as a function of $\bn$, for a prestress state at $50\%$ (dashed gray line) and $100\%$ (continuous blue line) of the limit value for ellipticity loss ($\bp_{\text{E}}$ reported in Table~\ref{tab:cases_analyzed}).
    A square lattice and a rhombic lattice are considered with $\Lambda_1=\Lambda_2=10$ and $\Lambda_1=7$, $\Lambda_2=15$.
    In the orthotropic cases, both for squared and rhombic lattices, the loss of ellipticity is characterized by the simultaneous vanishing of the eigenvalue of the acoustic tensor along two directions $\bn^1_{\text{E}}$ and $\bn^2_{\text{E}}$ (the associated wave amplitudes are reported as $\bg^1_{\text{E}}$ and $\bg^2_{\text{E}}$).
    The fully anisotropic cases, displays the vanishing of the eigenvalue along only one direction $\bn^1_{\text{E}}$.
    Note that for the rhombic lattice, the relative orientation of $\bn_{\text{E}}$ with respect to $\bg_{\text{E}}$ shows that the mode of localization is neither a pure shear nor a pure expansion wave, but a mixing of the two.
    Conversely, the square lattice always reaches loss of ellipticity through the formation of bands of pure shear strain.
    }
\end{figure}

It follows from the symmetry of $\bA^{(\fC)}(\bn)$ that, starting from the unloaded state $\bp=\bzero$ with rods of finite stiffness and continuously varying the prestress, the material remains both (SE) and (E) until both conditions simultaneously fail.
Therefore, solutions of Eq.~\eqref{eq:ellipticity_loss} are sought as pairs $\{\bn_{\text{E}},\bp_{\text{E}}\}$ such that $\bp_{\text{E}}$ represents the terminal point of a path starting at $\bp=\bzero$ and entirely contained in the (SE) domain;
in other words, $\bp_{\text{E}}$ is on both boundaries of (SE) and (E).
The set of these points $\bp_{\text{E}}$ is, with a little abuse\footnote{
    In fact, the elliptic boundary as referred to in this article is the part of this boundary which is coincident with the boundary of strong ellipticity.
},
referred to as the \textit{elliptic boundary}.

In order to explore loss of ellipticity for lattice configurations characterized by different symmetry classes, a square and a rhombic grid are considered, respectively with $\alpha=\pi/2$ and $\alpha=\pi/3$.
For both examples, the slenderness $\Lambda_1=\Lambda_2=10$, and $\Lambda_1=7$, $\Lambda_2=15$ are selected.

In Table~\ref{tab:cases_analyzed}, for each geometry considered, the first solution to Eq.~\eqref{eq:ellipticity_loss} for equal prestress components $p_1=p_2$ is reported, together with the associated directions of localization, denoted as $\bn_{\text{E}}=\cos(\theta_{\text{cr}})\be_1+\sin(\theta_{\text{cr}})\be_2$.
Note that the symmetry class is referred here to the unloaded configuration, so that the symmetry of incremental response may change as an effect of loading.
With the assumed values for grid angle $\alpha$ and slenderness, the cubic, orthotropic, and fully anisotropic cases (10 components of tensor $\fC$, which correspond to 6 independent parameters of $\fE$ plus three components of the prestress $\bT$ in the case of planar elasticity) can be investigated.

In order to better visualize the direction $\bn_{\text{E}}$ and the associated mode $\bg_{\text{E}}$,
a polar plot of the square root of the lowest eigenvalue $c_1(\bn,\bp)$ is reported
in Fig.~\ref{fig:eigenvalue_square_rhombus},
for the cases listed in Table~\ref{tab:cases_analyzed}, at two levels of preload, namely $0.5\,\bp_{\text{E}}$ (dashed gray line) and $\bp_{\text{E}}$ (continuous blue line).
In Fig.~\ref{fig:eigenvalue_square_10_10} the square lattice with $\Lambda_1=\Lambda_2=10$ is subject to an isotropic prestress in the two directions, $p_1=p_2$, and therefore the cubic symmetry is maintained in the prestressed state.
Owing to this symmetry, ellipticity is lost along two orthogonal directions $\bn^1_{\text{E}}$ and $\bn^2_{\text{E}}$.
Moreover, the associated wave amplitudes $\bg^1_{\text{E}}$ and $\bg^2_{\text{E}}$ are perpendicular to the vectors $\bn^1_{\text{E}}$ and $\bn^2_{\text{E}}$ respectively, hence indicating that the modes of localization are pure shear waves, the so-called \textit{shear bands}.

For the orthotropic square lattice ($\Lambda_1=7$, $\Lambda_2=15$), the polar plot is given in Fig.~\ref{fig:eigenvalue_square_7_15}.
In this case, owing to the orthotropy, waves propagating along the horizontal and vertical direction possess different velocities and therefore ellipticity is lost when the smallest of these velocities vanishes, leading to a single shear band (in this case with a normal $\bn^1_{\text{E}}$ aligned parallel to the horizontal direction).

Quite remarkably, \textit{the shear wave responsible for the ellipticity loss is the one propagating along the direction of the `stiffer' elastic link} (having the lowest slenderness), \textit{while intuitively a `shear mechanism' would be expected in the direction of the `soft' elastic link}.
This effect will be confirmed and explained further with the computation of the forced response in Section~\ref{sec:dynamic_forced_response}.
Moreover, it is worth noting that the shear band directions for the square lattice, both cubic and orthotropic, are aligned parallel to the directions of the rods forming the lattice.

For the rhombic lattice with $\Lambda_1=\Lambda_2=10$ and isotropic preload, shown in Fig.~\ref{fig:eigenvalue_rhombus_10_10}, the mechanical behavior is orthotropic and therefore two directions of localization are obtained.
The associated wave amplitudes $\bg^1_{\text{E}}$ and $\bg^2_{\text{E}}$ both have respectively a component orthogonal and parallel to the vectors $\bn^1_{\text{E}}$ and $\bn^2_{\text{E}}$, so that a `mixture' of shear and compression waves is involved.

The fully anisotropic version for the rhombic lattice (Fig.~\ref{fig:eigenvalue_rhombus_7_15}) can be obtained by changing the slenderness values ($\Lambda_1=7$, $\Lambda_2=15$), so that one of the two localizations is suppressed, while the other is preserved.
It is also worth noting that, in contrast to the square case, the directions of localization for the rhombic lattice are not perfectly aligned parallel to the rods' normal, instead, they result slightly inclined, as will be confirmed by the computation of the forced response reported in Section~\ref{sec:dynamic_forced_response}.

\subsection{Constitutive tensor and prestress for the effective continuum}
\label{sec:grid_C_T}
The prestress tensor $\bT$, equivalent in the continuum to the preload forces $\bP$ in the elastic lattice, can be obtained by computing the average normal and tangential tractions along the faces with unit normal $\be_1$ and $\be_2$.
With reference to Fig.~\ref{fig:geometry_grid_unit_cell} the following expression is obtained
\begin{equation}
    \label{eq:grid_T_guessed}
    \bT = \left( \frac{P_1}{l \sin\alpha} + \frac{P_2 \cos^2\alpha}{l \sin \alpha} \right) \be_1\otimes\be_1 + \frac{P_2 \cos\alpha}{l} \left(\be_1\otimes\be_2 + \be_2\otimes\be_1\right) + \frac{P_2 \sin\alpha}{l} \be_2\otimes\be_2 \,.
\end{equation}

As explained in Section~\ref{sec:identification_constitutive_tensor}, once the spherical part of the prestress is known, the effective incremental constitutive tensor $\fC$ can be computed from the homogenized acoustic tensor $\bA^{(\fC)}$ by solving the linear system~\eqref{eq:hessian_acoustic_tensor}.
For completeness the full expression of $\fC$ for the preloaded grid as function of the parameter set $\{p_1,p_2,\Lambda_1,\Lambda_2,\kappa,\alpha\}$ is reported in Appendix~\ref{sec:homogenized_constitutive_tensor_grid}.

\subsection{Loss of ellipticity vs micro-bifurcation}
\label{sec:grid_ellipticity_local_buckling}
With reference to the lattice sketched in Fig.~\ref{fig:geometry_grid_unit_cell},
the value of the prestress state, which is critical for bifurcation of the grid is determined by employing conditions~\eqref{eq:failure_E_multiplier} and~\eqref{eq:local_buckling_multiplier_simpler}, and computing numerically the prestress multipliers $\gamma_{\text{E}}$ and $\gamma_{\text{B}}$.
Results are presented as \textit{uniqueness or stability domains} in the non-dimensional prestress space $\{p_1,p_2\}$ by fixing the set of geometrical and mechanical parameters $\{\alpha,\Lambda_1,\Lambda_2,\kappa\}$.
\begin{figure}[!htb]
    \centering
    \begin{subfigure}[t]{0.32\textwidth}
        \centering
        \caption{\label{fig:macro_bifurcation_stab}Uniqueness domain $(\kappa=0)$}
        \includegraphics[width=\linewidth]{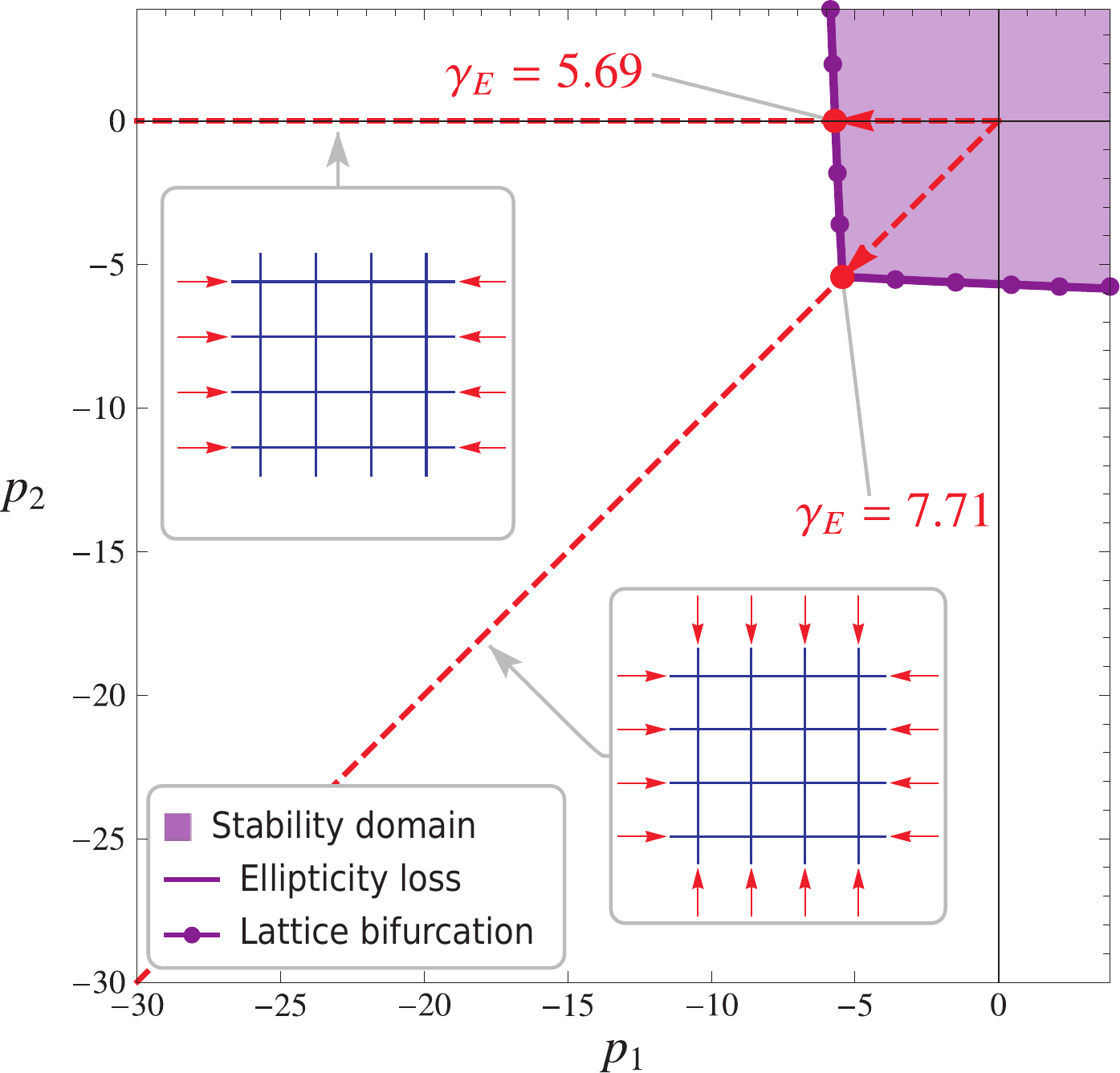}
    \end{subfigure}
    \hspace*{1mm}
    \begin{subfigure}[t]{0.31\textwidth}
        \centering
        \caption{\label{fig:macro_bifurcation_surf1}Equibiaxial compression}
        \includegraphics[width=0.95\linewidth]{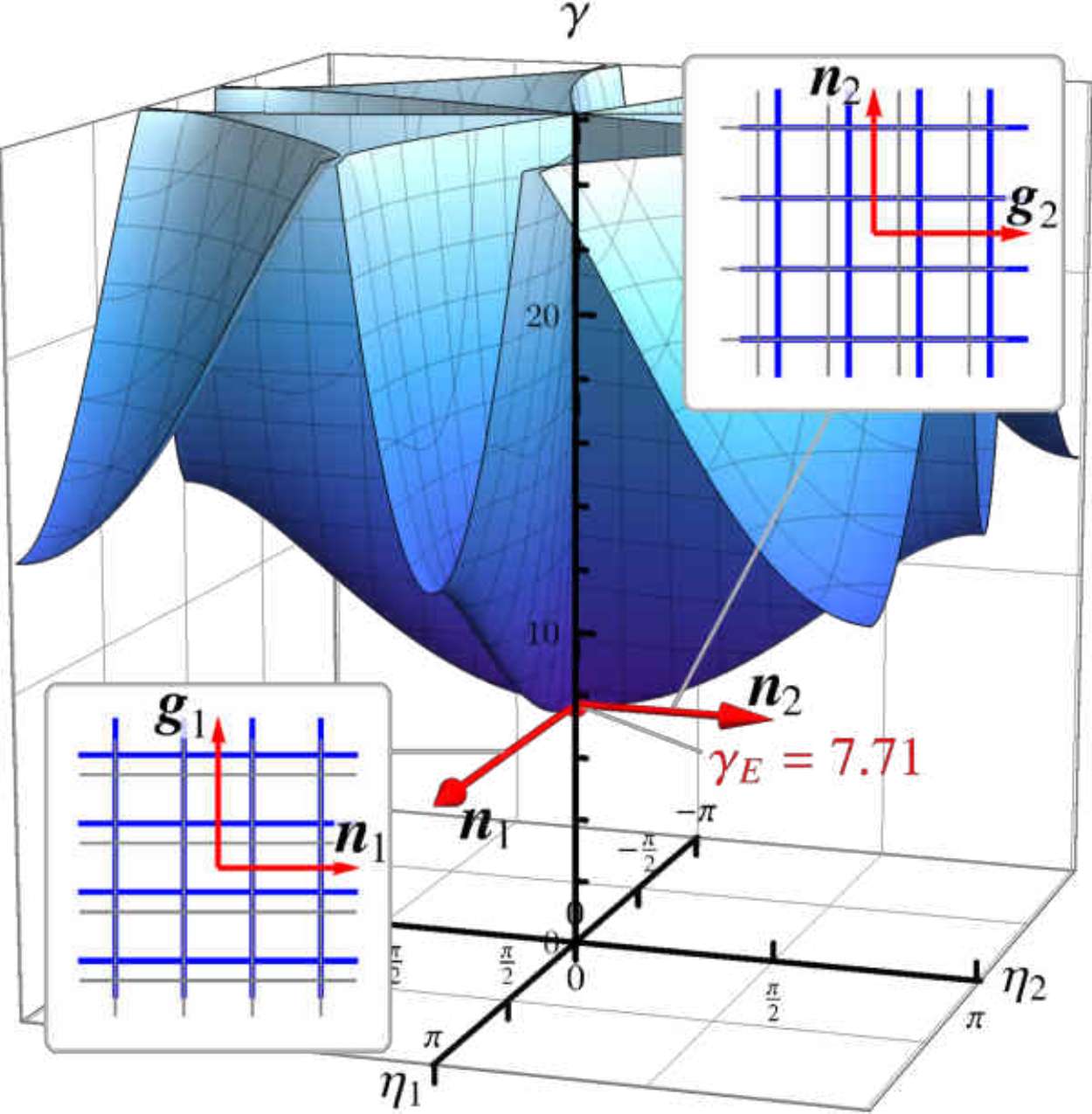}
    \end{subfigure}
    \begin{subfigure}[t]{0.31\textwidth}
        \centering
        \caption{\label{fig:macro_bifurcation_surf2}Uniaxial compression}
        \includegraphics[width=0.95\linewidth]{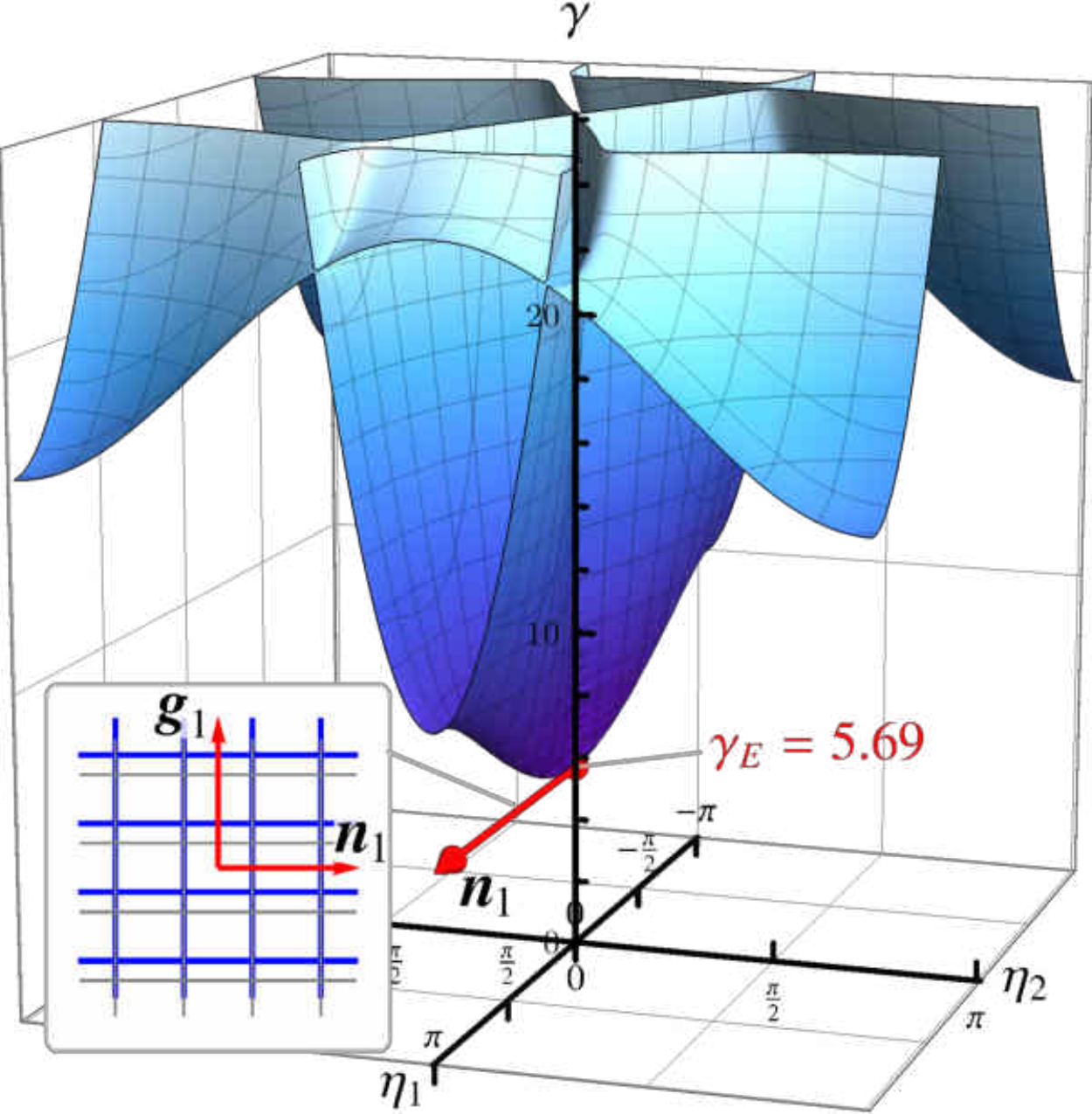}
    \end{subfigure}\\ \vspace{4mm}
    \begin{subfigure}[t]{0.32\textwidth}
        \centering
        \caption{\label{fig:macro_micro_bifurcation_stab}Uniqueness domain $(\kappa=0.2)$}
        \includegraphics[width=\linewidth]{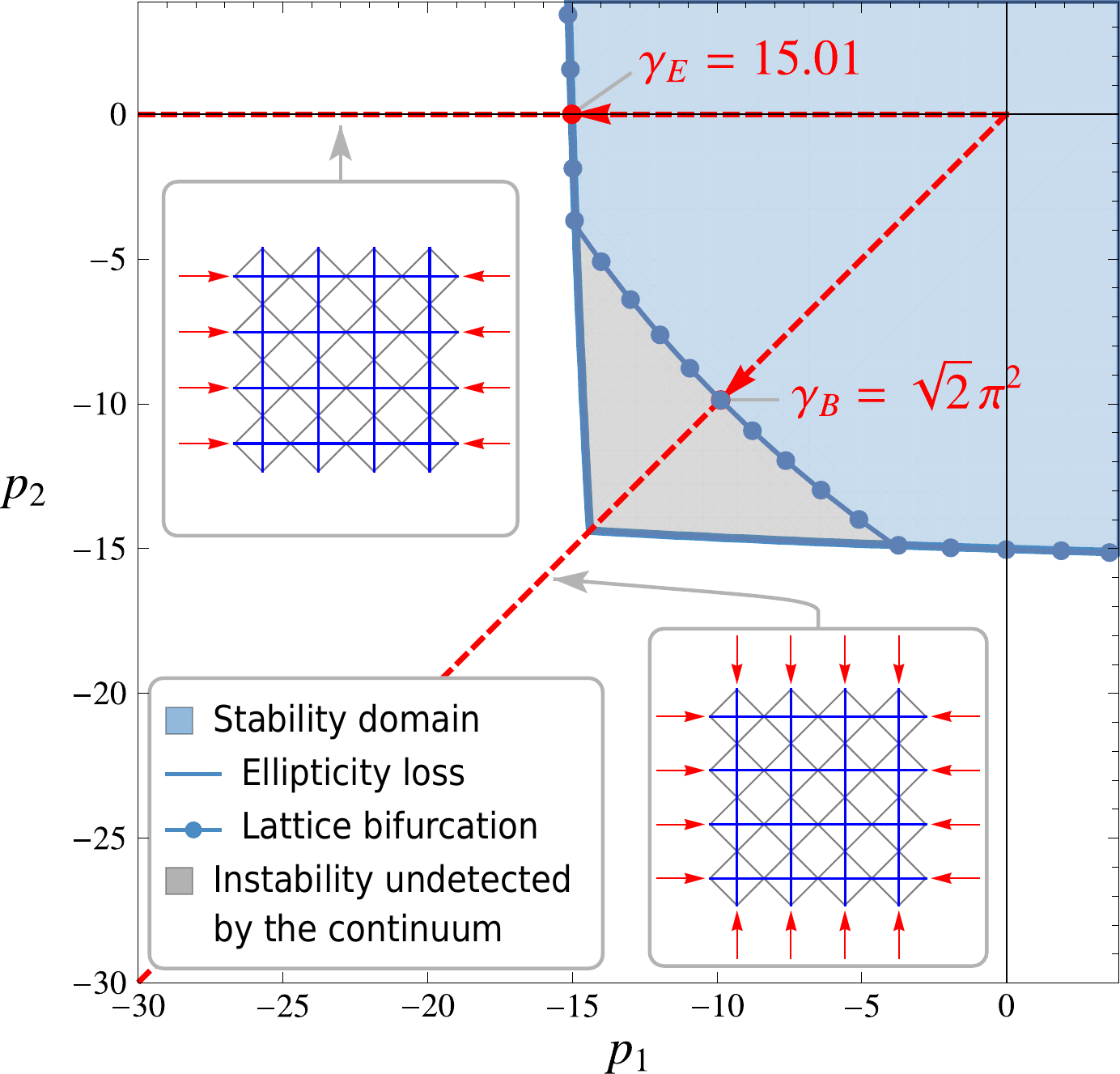}
    \end{subfigure}
    \hspace*{1mm}
    \begin{subfigure}[t]{0.31\textwidth}
        \centering
        \caption{\label{fig:macro_micro_bifurcation_surf1}Equibiaxial compression}
        \includegraphics[width=0.95\linewidth]{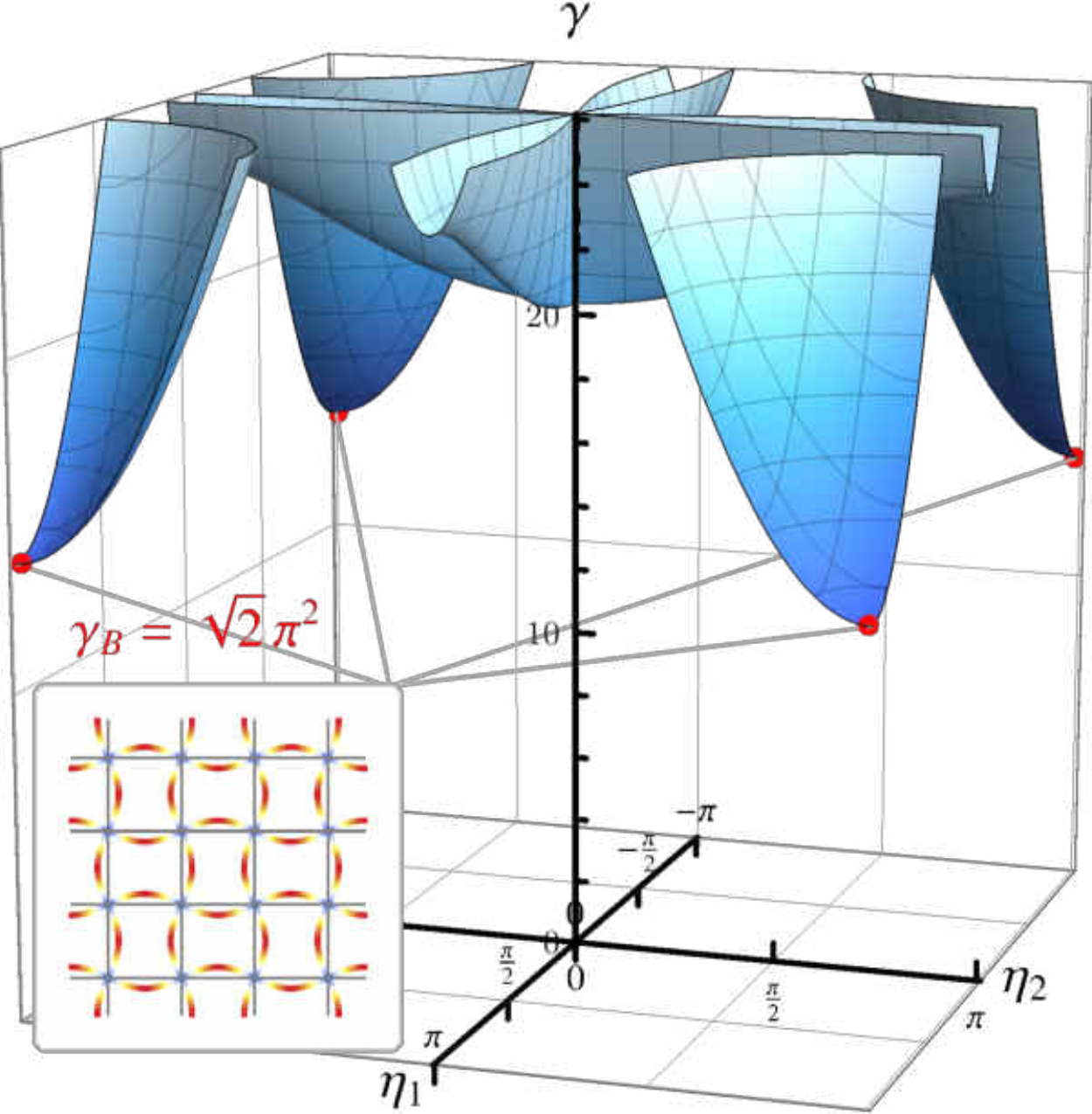}
    \end{subfigure}
    \begin{subfigure}[t]{0.31\textwidth}
        \centering
        \caption{\label{fig:macro_micro_bifurcation_surf2}Uniaxial compression}
        \includegraphics[width=0.95\linewidth]{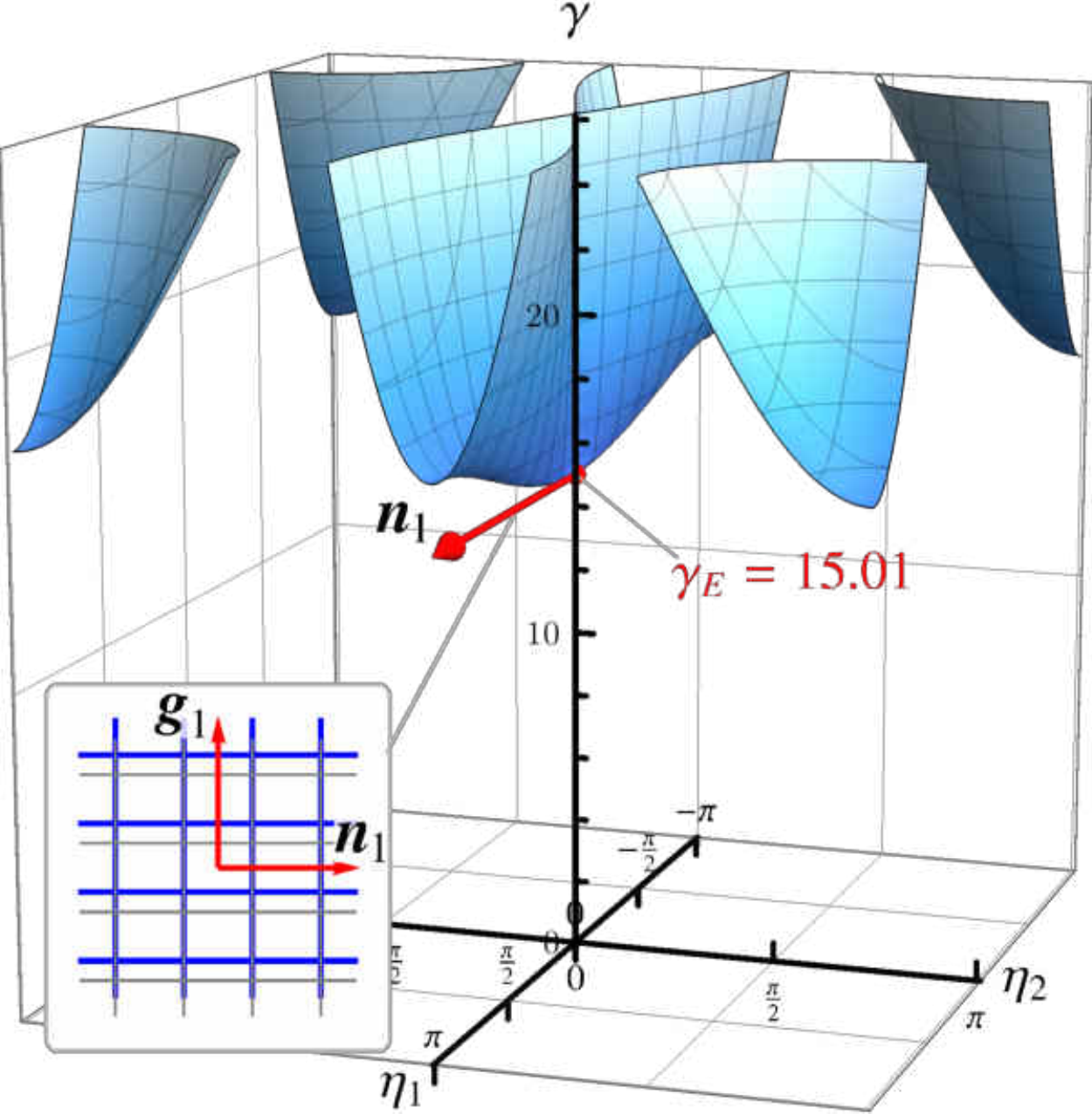}
    \end{subfigure}
    \caption{\label{fig:macro_micro_bifurcation}
        (\subref{fig:macro_bifurcation_stab}) and (\subref{fig:macro_micro_bifurcation_stab}):
        Uniqueness/stability domains in the loading space $\{p_1,p_2\}$ for a square grid (with equal slenderness of the rods $\Lambda_1 = \Lambda_2 = 10$), when diagonal springs are absent (upper part) and present (with spring stiffness $\kappa=0.2$,  lower part). The continuous line denotes failure of ellipticity in the continuum, while the dotted line represents critical bifurcation in the grid.
        (\subref{fig:macro_bifurcation_stab}): there is only one uniqueness domain (purple) for both the grid and its continuum counterpart.
        (\subref{fig:macro_micro_bifurcation_stab}): the uniqueness domain for the grid (blue) is smaller than the strong ellipticity domain (blue, plus gray), so that micro-bifurcation may occur before macro-bifurcation.
        (\subref{fig:macro_bifurcation_surf1},~\subref{fig:macro_bifurcation_surf2}) and ~(\subref{fig:macro_micro_bifurcation_surf1},~\subref{fig:macro_micro_bifurcation_surf2}) bifurcation surfaces in the space $\{\eta_1,\eta_2,\gamma\}$, referred to the specific radial loading paths shown as red dashed lines in (\subref{fig:macro_bifurcation_stab}) and (\subref{fig:macro_micro_bifurcation_stab}).
        The insets show the critical modes.
    }
\end{figure}
The boundary of the stability domain identifies the `critical', namely, the first bifurcation of the incremental equilibrium of the lattice.
Therefore, this domain is the intersection of the domains corresponding to the exclusion of micro and macro bifurcations, so that one or the other instability may, depending on the parameters and on the prestress, be the first encountered in the loading path.

The dependence on the parameters $\{\alpha,\,\Lambda_1,\,\Lambda_2,\,\kappa\}$ has been analyzed by considering two grid configurations that will be referred to as the \textit{orthotropic grid}, with equal slenderness $\Lambda_1=\Lambda_2=10$, and the \textit{anisotropic grid}, characterized by different slenderness values, $\Lambda_1=7$ and $\Lambda_2=15$.
For each lattice, the influence of the rods' inclination is explored by setting $\alpha=\pi/2, \pi/3, \pi/4, \pi/6$, while the stiffness of the springs is investigated in the range $\kappa \in [0,1]$. In this way, the influence of the diagonal bracing on the critical bifurcation mode is analyzed.

Macroscopic (infinite wavelength) and microscopic (finite wavelength) bifurcations are investigated in Fig.~\ref{fig:macro_micro_bifurcation} for the orthotropic grid with $\alpha=\pi/2$.
Here, critical bifurcation loads $p_1$ and $p_2$ are reported for the cases in which diagonal springs are absent ($\kappa=0$, Fig.~\ref{fig:macro_bifurcation_stab},~\subref{fig:macro_bifurcation_surf1},~\subref{fig:macro_bifurcation_surf2}) and for a spring stiffness $\kappa=0.2$ (Fig.~\ref{fig:macro_micro_bifurcation_stab},~\subref{fig:macro_micro_bifurcation_surf1},~\subref{fig:macro_micro_bifurcation_surf2}).

There are two uniqueness (or stability) domains, regions in the prestress state where bifurcation is excluded, one for the grid and one for its equivalent continuum.
For the continuum, uniqueness is represented by strong ellipticity and failure of this coincides with loss of ellipticity.
The latter, in turn, always corresponds to a bifurcation in the grid with a mode of infinite wavelength.
When this mode is critical, a macro-instability occurs in the lattice, so that continuum and grid display the same behavior.
If, however, a micro-instability is critical for the grid, this always occurs when the continuum is still strongly elliptic.
For this reason, in all figures, the domain of strong ellipticity is reported (whose boundary represents the condition for which the infimum of Eq.~\eqref{eq:local_buckling_multiplier_simpler} is attained at $\bk = \bzero$), together with the continuous-dotted contour representing bifurcation in the grid at either $\bk=\bzero$ or $\bk\neq\bzero$.

The uniqueness domains (Fig.~\ref{fig:macro_bifurcation_stab} and~\ref{fig:macro_micro_bifurcation_stab}) have been computed by solving equation~\eqref{eq:local_buckling_multiplier_simpler} for radial loading paths in the non-dimensional load space $\{p_1,p_2\}$.
The location of the infimum can be visualized, by fixing the loading direction as $\bp=\gamma\,\hat{\bp}$, and then by numerically computing the bifurcation surface defined as $\det\bK^*(\gamma\hat{\bp},\,\eta_1\bb_1+\eta_2\bb_2) = 0$ in the space $\{\eta_1,\eta_2,\gamma\}$.
Two radial paths are considered in Fig.~\ref{fig:macro_bifurcation_stab} and~\ref{fig:macro_micro_bifurcation_stab}, namely, equibiaxial $\hat{\bp}=\{-1/\sqrt{2},-1/\sqrt{2}\}$ and uniaxial $\hat{\bp}=\{-1,0\}$ compression (red dashed lines), and the corresponding bifurcation surfaces are reported in Fig.~\ref{fig:macro_bifurcation_surf1},~\subref{fig:macro_bifurcation_surf2} and Fig.~\ref{fig:macro_micro_bifurcation_surf1},~\subref{fig:macro_micro_bifurcation_surf2}, respectively.

In the absence of diagonal springs, Fig.~\ref{fig:macro_bifurcation_stab} reports the strong ellipticity domain in the solid equivalent to the lattice, showing that (for every loading direction $\hat{\bp}$) a macro-bifurcation, in other words an ellipticity loss (referred to the dyad $\bn\otimes\bg$), is always reached before micro-bifurcation.

For the two radial loading paths shown in Fig.~\ref{fig:macro_bifurcation_stab}, the bifurcation surfaces Figs.~\ref{fig:macro_bifurcation_surf1},\subref{fig:macro_bifurcation_surf2}, show that the minimum values of the load multiplier $\gamma$ are attained at $\{\eta_1,\eta_2\} = \{0,0\}$, which corresponds to a macro-bifurcation for the lattice (associated to an infinite wavelength mode), so that the critical prestress multipliers $\gamma_{\text{E}}=7.71$ and $\gamma_{\text{E}}=5.69$ lie on the border of ellipticity loss.
The two bifurcations correspond respectively to two orthogonal modes and a single mode.

The presence of diagonal springs complicates the situation as reported in Fig.~\ref{fig:macro_micro_bifurcation_stab}.
In this case the uniqueness/stability domains show that micro-bifurcations may sometimes occur within the region of strong ellipticity, which is for instance the case of equibiaxial compression (radial path inclined at $45^\circ$) and not the case of uniaxial compression (horizontal radial path).
In fact, for equibiaxial compression a critical micro-bifurcation occurs, so that Fig.~\ref{fig:macro_micro_bifurcation_surf1} shows that the minimum value of the load multiplier, $\gamma_{\text{B}}=\sqrt{2} \pi^2$, is attained at four points, $\{\eta_1,\eta_2\} = \{\pm \pi,\pm \pi\}$, all associated to a bifurcation mode with a finite wavelength, as shown in the inset.
For uniaxial compression, Fig.~\ref{fig:macro_micro_bifurcation_surf2}, a macro-bifurcation of the grid occurs at $\gamma_{\text{E}} = 15.01$ and the tangent to the bifurcation surface at the origin singles out the infinite-wavelength bifurcation mode (shown in the inset and appearing as a rigid translation).
\begin{figure}[htb!]
    \centering
    \begin{subfigure}{0.49\textwidth}
        \centering
        \caption{\label{fig:ellipticity_domains_10_10_pi2}$\alpha=\pi/2$}
        \includegraphics[width=0.95\linewidth]{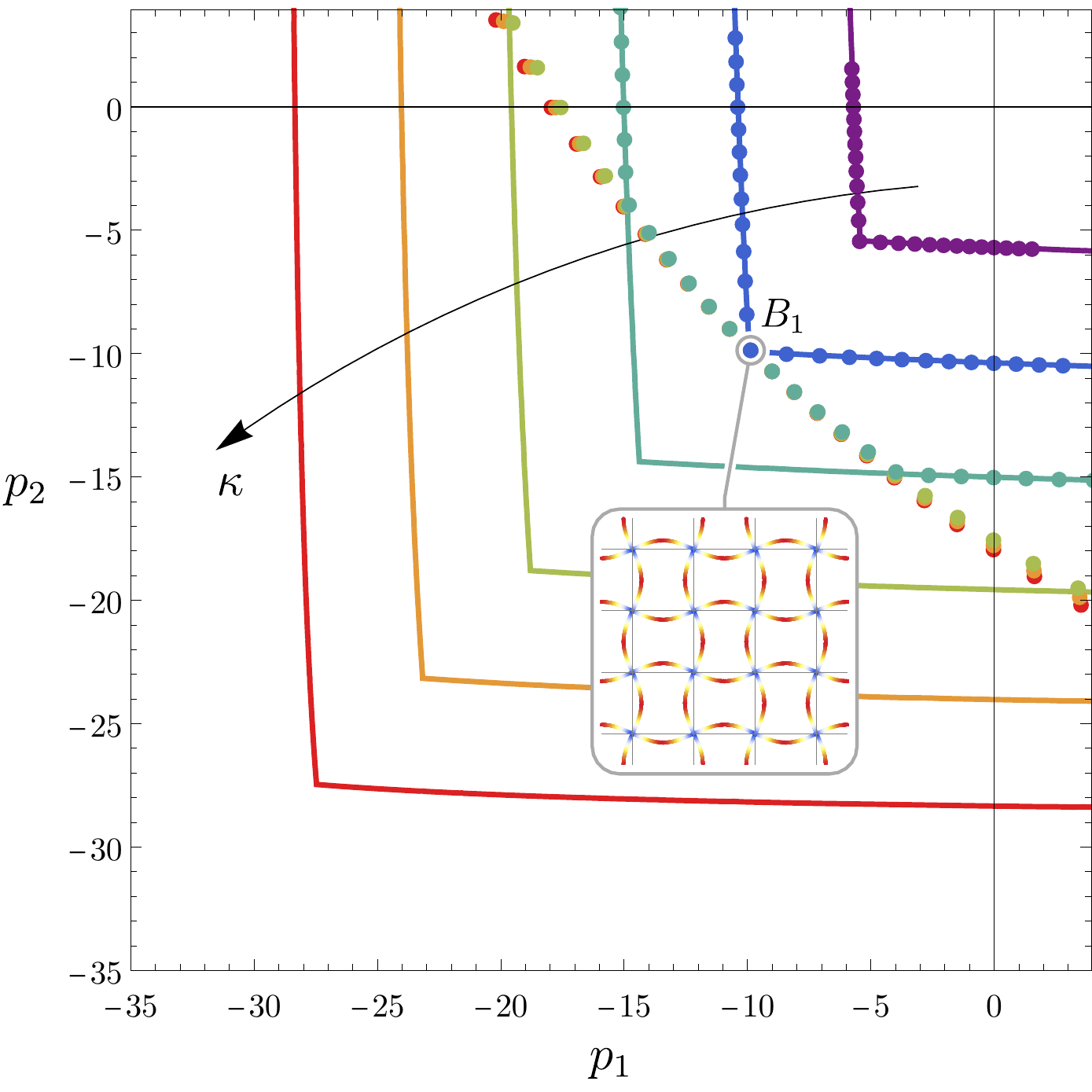}
    \end{subfigure}
    \begin{subfigure}{0.49\textwidth}
        \centering
        \caption{\label{fig:ellipticity_domains_10_10_pi3}$\alpha=\pi/3$}
        \includegraphics[width=0.95\linewidth]{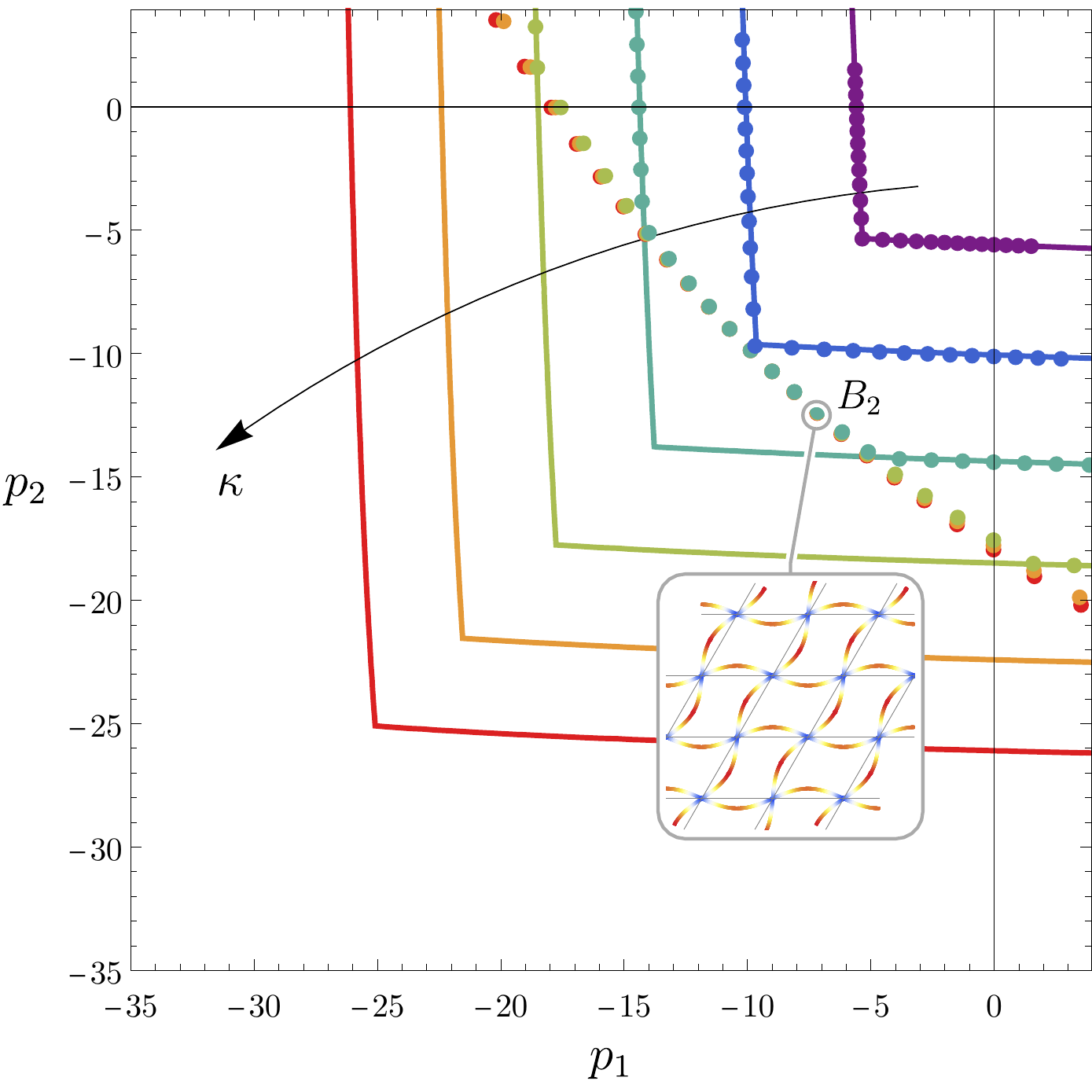}
    \end{subfigure}\\
    \begin{subfigure}{0.49\textwidth}
        \centering
        \caption{\label{fig:ellipticity_domains_10_10_pi4}$\alpha=\pi/4$}
        \includegraphics[width=0.95\linewidth]{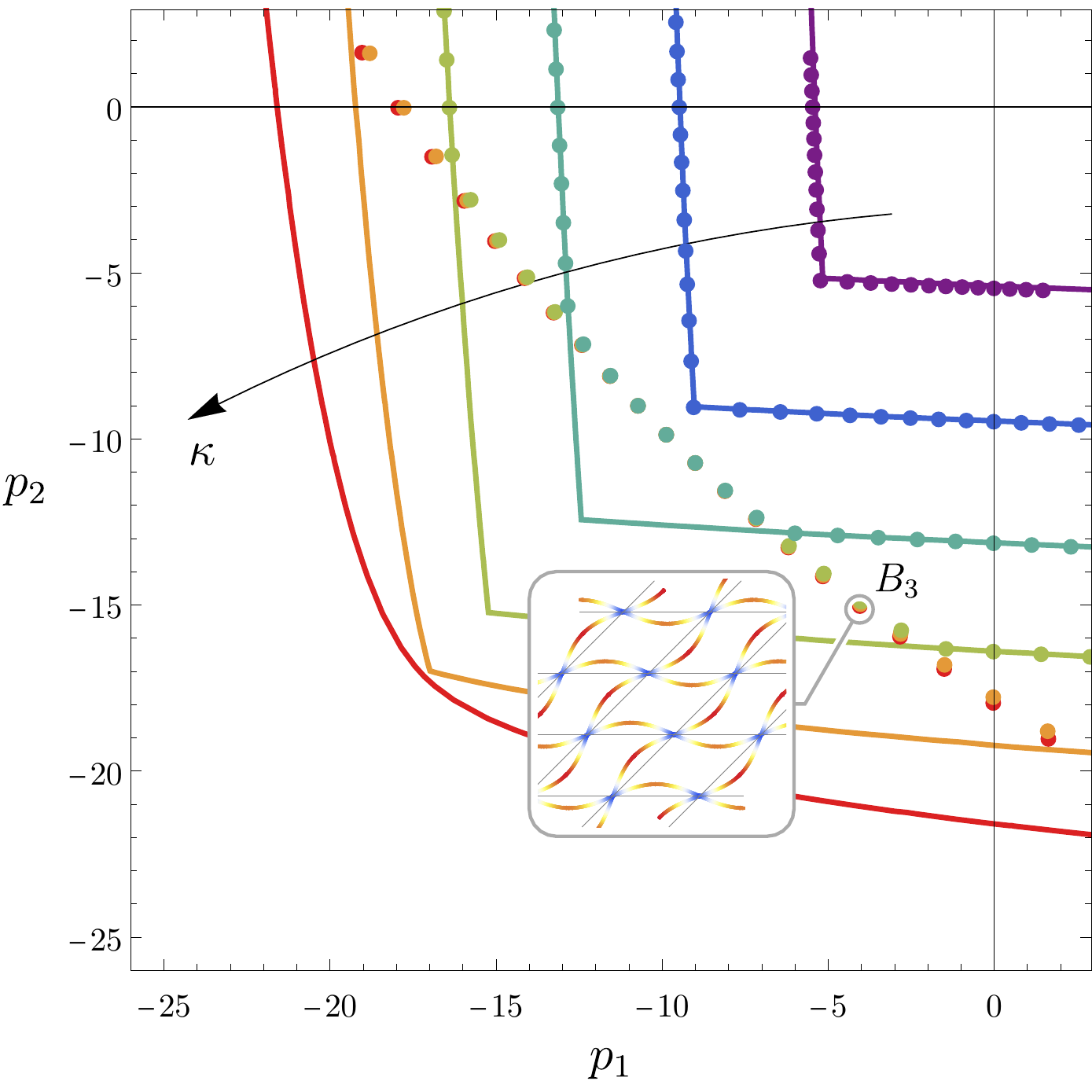}
    \end{subfigure}
    \begin{subfigure}{0.49\textwidth}
        \centering
        \caption{\label{fig:ellipticity_domains_10_10_pi6}$\alpha=\pi/6$}
        \includegraphics[width=0.95\linewidth]{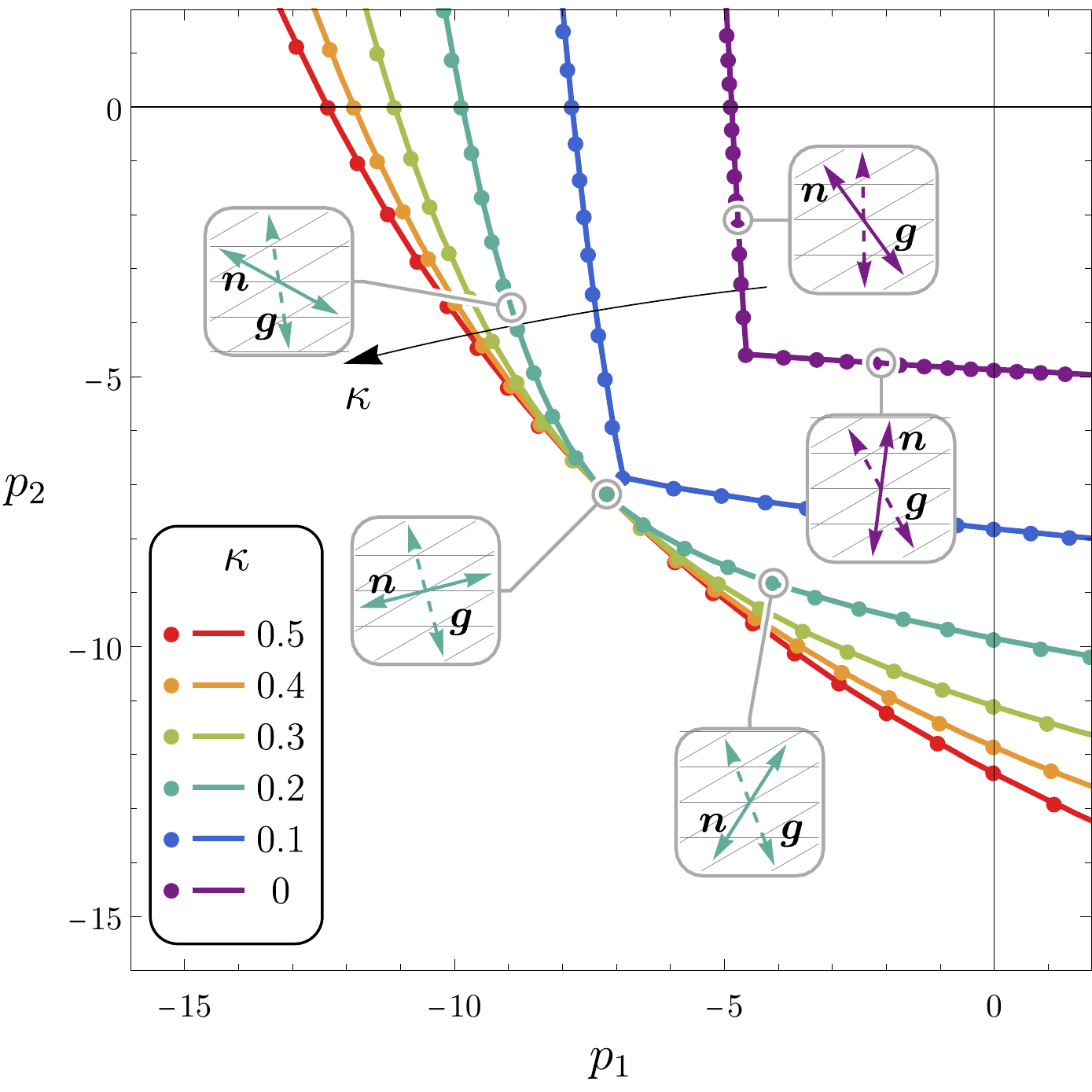}
    \end{subfigure}%
    \caption{\label{fig:ellipticity_domains_10_10}
        Strong ellipticity domains (corresponding to macro-bifurcations, continuous lines) and lines of occurrence of micro-bifurcation in the lattice (circular markers) for an orthotropic lattice of prestressed elastic rods with $\Lambda_1=\Lambda_2=10$, at different rod angles $\alpha$ and stiffness $\kappa$ of the diagonal springs.
        Points $B_1,B_2,B_3$ on the stability boundaries have been selected for the computation of the associated critical bifurcation mode (shown in the insets and details reported in
        Table~\ref{tab:buckling_modes_points}).
        At small grid angles, for instance $\alpha=\pi/6$, failure of ellipticity coincides with micro-bifurcation in the lattice, so that the bifurcation mode is always characterized by an infinite wavelength.
        For these grid configurations, the direction of ellipticity loss exhibits a \textit{`super-sensitivity'} with respect to the load directionality, shown in the insets of part~(\subref{fig:ellipticity_domains_10_10_pi6}), where the critical dyads $\bn \otimes \bg$ for failure of ellipticity are reported.
    }
\end{figure}

\begin{figure}[htb!]
    \centering
    \begin{subfigure}{0.49\textwidth}
        \centering
        \caption{\label{fig:ellipticity_domains_7_15_pi2}$\alpha=\pi/2$}
        \includegraphics[width=0.95\linewidth]{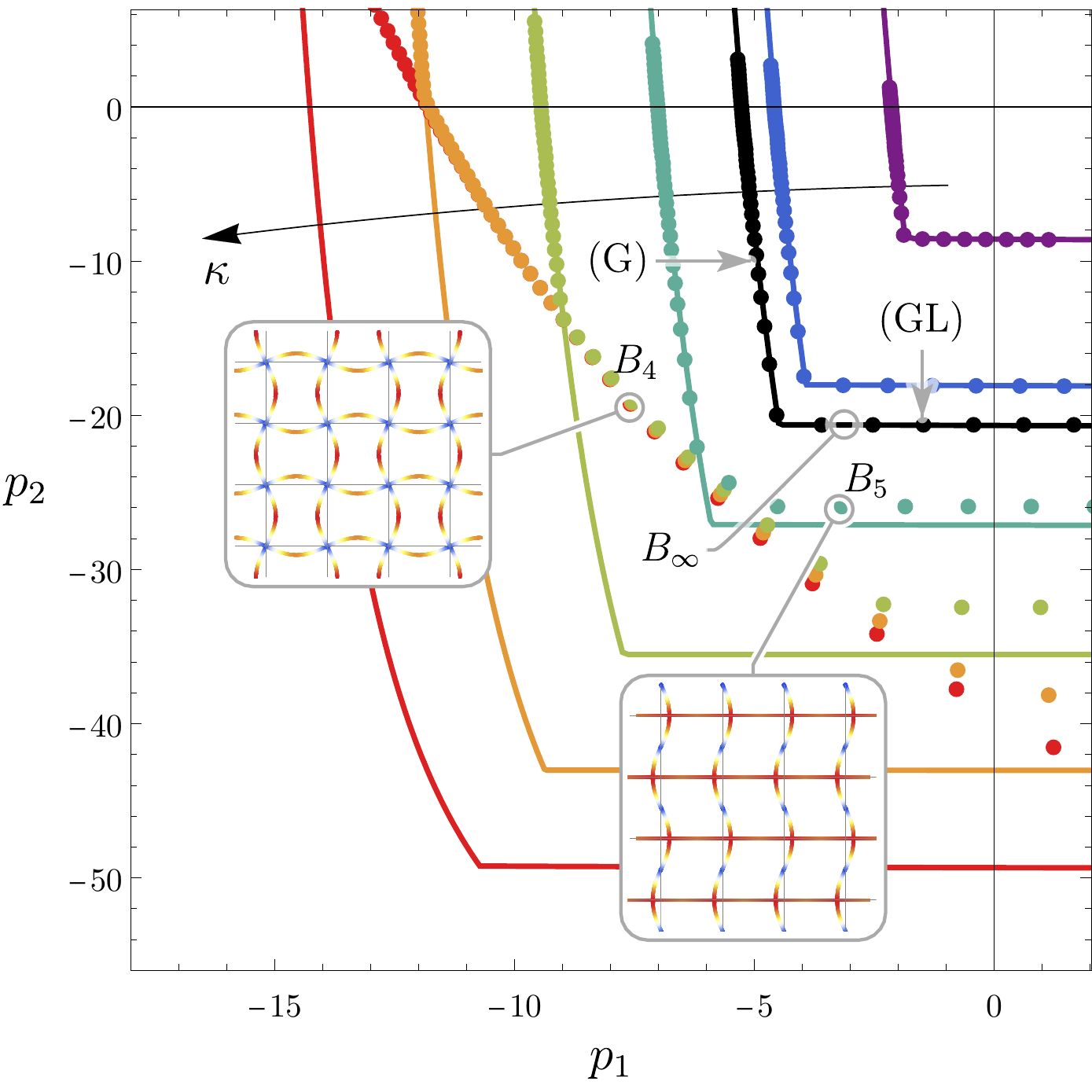}
    \end{subfigure}
    \begin{subfigure}{0.49\textwidth}
        \centering
        \caption{\label{fig:ellipticity_domains_7_15_pi3}$\alpha=\pi/3$}
        \includegraphics[width=0.95\linewidth]{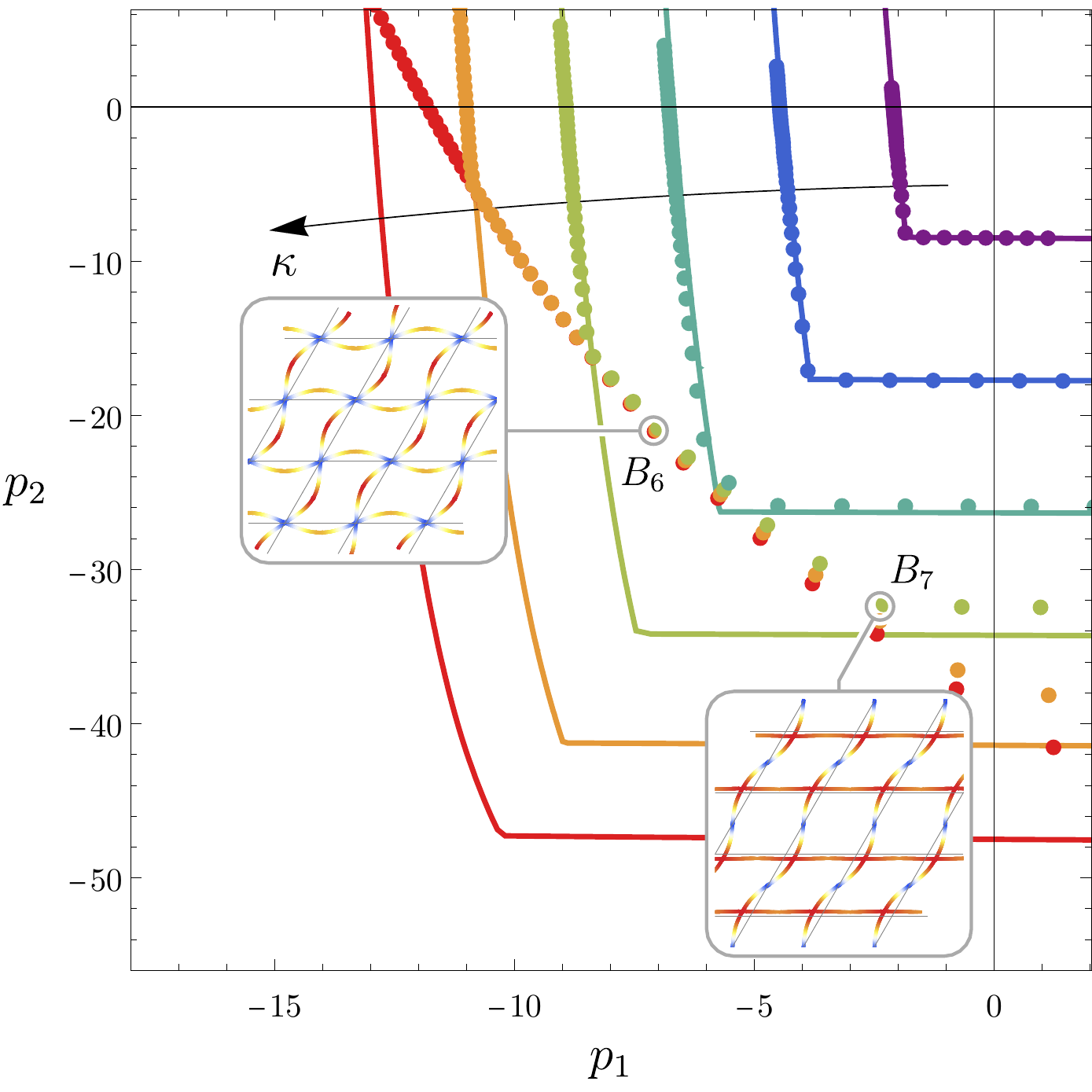}
    \end{subfigure}\\
    \begin{subfigure}{0.49\textwidth}
        \centering
        \caption{\label{fig:ellipticity_domains_7_15_pi4}$\alpha=\pi/4$}
        \includegraphics[width=0.95\linewidth]{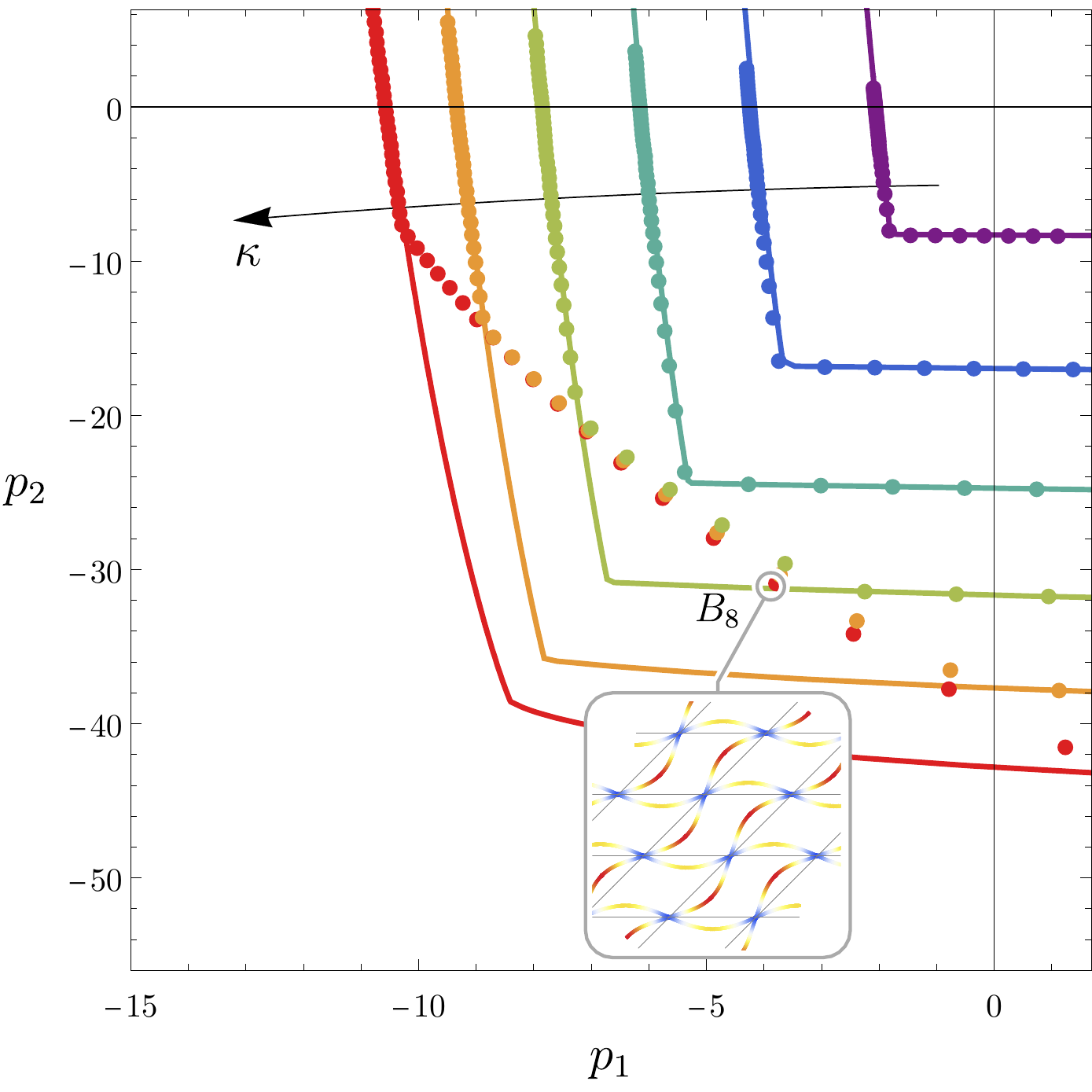}
    \end{subfigure}
    \begin{subfigure}{0.49\textwidth}
        \centering
        \caption{\label{fig:ellipticity_domains_7_15_pi6}$\alpha=\pi/6$}
        \includegraphics[width=0.95\linewidth]{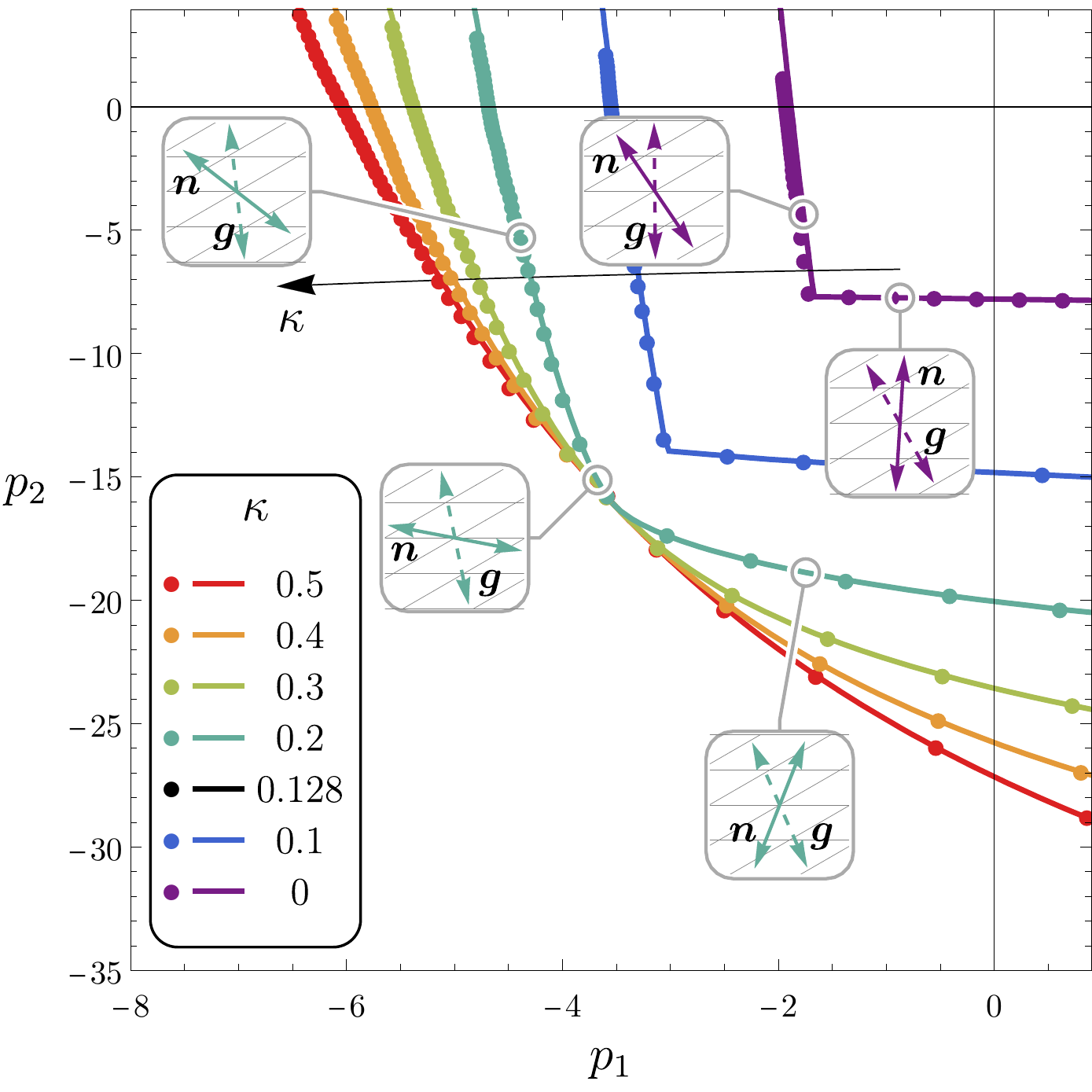}
    \end{subfigure}%
    \caption{\label{fig:ellipticity_domains_7_15}
        As for Fig. \ref{fig:ellipticity_domains_10_10}, except that $\Lambda_1=7,\,\Lambda_2=15$ and that the points on the stability boundary for which the critical bifurcation modes have been computed are labeled $B_4,B_5,B_6,B_7,B_8$.
        Note also that, the typical microscopic bifurcation modes of the anisotropic grid exhibit widely different deformations, dictated by the prestress direction (see insets in parts~\subref{fig:ellipticity_domains_7_15_pi2}, \subref{fig:ellipticity_domains_7_15_pi3}, \subref{fig:ellipticity_domains_7_15_pi4} corresponding to points labeled $B_4,B_5,B_6,B_7,B_8$ and compare for instance mode $B_4$ to $B_5$ or $B_6$ to $B_7$).
        In part~(\subref{fig:ellipticity_domains_7_15_pi2}) the black stability boundary denoted with the label (GL) identifies a set a special conditions where both global and local instabilities occur simultaneously at the same prestress level.
        This phenomenon is investigated in detail in Fig.~\ref{fig:infinite_buckling_modes}.
    }
\end{figure}

Further results on uniqueness domains for the orthotropic and the anisotropic grid are reported in Figs.~\ref{fig:ellipticity_domains_10_10} and~\ref{fig:ellipticity_domains_7_15}, respectively.
The strong ellipticity boundary (corresponding to macro-bifurcation) in the equivalent solid is denoted with a continuous line, while the circular markers identify the line for critical micro-bifurcation in the grid.
Moreover, critical bifurcation modes have been reported in insets of Figs.~\ref{fig:ellipticity_domains_10_10} and~\ref{fig:ellipticity_domains_7_15}, which refer to some specific points on the stability boundary (labelled as $B_1,B_2,B_3$ in the former figure and $B_4,B_5,B_6,B_7,B_8$ in the latter).
The critical loads and the critical wave vectors for each bifurcation mode are reported in Table~\ref{tab:buckling_modes_points}.
\begin{table}[htb!]
    \centering
    \begin{tabular}{c|ccccccc}
        \toprule
        Label      & $\Lambda_1$ & $\Lambda_2$ & $\alpha$ & $\kappa$ & $p_1$    & $p_2$    & $\bk_{\text{cr}}$            \\
        \midrule
        $B_1$      & 10          & 10          & $\pi/2$  & $0.2$    & $-\pi^2$ & $-\pi^2$ & $\pi\bb_1+\pi\bb_2$          \\
        $B_2$      & 10          & 10          & $\pi/3$  & $0.3$    & $-7.16$  & $-12.40$ & $\pi\bb_1+\pi\bb_2$          \\
        $B_3$      & 10          & 10          & $\pi/4$  & $0.7$    & $-4.05$  & $-15.13$ & $\pi\bb_1+\pi\bb_2$          \\
        $B_4$      & 7           & 15          & $\pi/2$  & $0.4$    & $-7.72$  & $-18.64$ & $\pi\bb_1+\pi\bb_2$          \\
        $B_5$      & 7           & 15          & $\pi/2$  & $0.2$    & $-3.41$  & $-25.91$ & $\pi\bb_2$                   \\
        $B_6$      & 7           & 15          & $\pi/3$  & $0.3$    & $-6.98$  & $-20.93$ & $\pi\bb_1+\pi\bb_2$          \\
        $B_7$      & 7           & 15          & $\pi/3$  & $0.3$    & $-2.12$  & $-32.40$ & $\pi\bb_2$                   \\
        $B_8$      & 7           & 15          & $\pi/4$  & $0.5$    & $-4.00$  & $-30.37$ & $\pi\bb_1+\pi\bb_2$          \\
        $B_\infty$ & 7           & 15          & $\pi/2$  & $0.128$  & $-3.44$  & $-20.62$ & $\eta_2\bb_2\,\forall\eta_2$ \\
        \bottomrule
    \end{tabular}
    \caption{
        Critical bifurcation modes $\bk_{\text{cr}}$ for several configurations of the orthotropic ($\Lambda_1=\Lambda_2=10$) and anisotropic ($\Lambda_1=7,\,\Lambda_2=15$) lattice.
        Plots of the corresponding deformation fields are reported as insets in Figs.~\ref{fig:ellipticity_domains_10_10} and~\ref{fig:ellipticity_domains_7_15}.
    }
    \label{tab:buckling_modes_points}
\end{table}

From Figs.~\ref{fig:ellipticity_domains_10_10} and~\ref{fig:ellipticity_domains_7_15} the following features can be highlighted.
\begin{itemize}
    \item The stable region is unbounded for tensile (positive) preload and bounded when both the preloads are compressive (negative);
          this is an expected feature, as the contribution of a tensile preload to the potential energy, Eq.~\eqref{eq:prestress_energy}, is positive definite;
    \item The elliptic boundary appears to be smooth everywhere except at \textit{a corner point};
    \item The anisotropy induced by different values of slenderness causes the corner to move;
          the elliptic region is reduced in size along the direction of the smallest slenderness;
    \item For the orthotropic grid the strong ellipticity boundary is symmetric with respect to the bisector defined by the condition $p_1=p_2$, which is the principal direction of orthotropy for the grid when $\Lambda_1=\Lambda_2$ (a symmetry which is broken for the anisotropic grid);
    \item For every value of the grid angle $\alpha$, the effect of the diagonal springs essentially consists in an enlargement of the strong ellipticity region (see the arrow in Fig. \ref{fig:ellipticity_domains_7_15} denoting increasing values of stiffness $\kappa$);
    \item The stiffening induced by increasing the spring stiffness $\kappa$ is much more effective for nearly orthogonal grids ($\alpha\approx\pi/2$) than for small values of inclination $\alpha$ (compare Fig.~\ref{fig:ellipticity_domains_10_10_pi2} to Fig.~\ref{fig:ellipticity_domains_10_10_pi6} and Fig.~\ref{fig:ellipticity_domains_7_15_pi2} to Fig.~\ref{fig:ellipticity_domains_7_15_pi6});
    \item For every value of the spring stiffness $\kappa$, the deviation from orthogonality of the grid always reduces the size of the strong ellipticity region, so that the largest strong ellipticity region is attained for $\alpha=\pi/2$.
\end{itemize}

The stability boundaries (circular markers in Fig.~\ref{fig:ellipticity_domains_10_10} and~\ref{fig:ellipticity_domains_7_15}), evidence the following characteristics.
\begin{itemize}
    \item At small values of spring stiffness $\kappa$, the first bifurcation is always global, so that \textit{the strong ellipticity and the stability boundaries coincide independently of the prestress direction}; a feature visible for $\kappa=0$ and 0.1 (purple and blue lines in Fig.~\ref{fig:ellipticity_domains_10_10_pi2} and Fig.~\ref{fig:ellipticity_domains_7_15_pi2});
    \item An increase in the spring stiffness $\kappa$ leads to a first bifurcation of local type (the critical mode is characterized by a finite wavelength), \textit{so that the stability region lies inside the elliptic boundary};
    \item Fig.~\ref{fig:ellipticity_domains_10_10_pi6} and Fig.~\ref{fig:ellipticity_domains_7_15_pi6} show that, at sufficiently small values of grid angle (for instance at $\alpha=\pi/6$), \textit{failure of strong ellipticity dictates the first bifurcation independently of the stiffness of the diagonal springs} (see circular markers of the stability boundary overlapping with the elliptic boundary);
    \item The typical \textit{microscopic bifurcation modes of the orthotropic grillage are characterized by a deformation involving bending with the nodes only
              subject to rotations} (see insets in Fig.~\ref{fig:ellipticity_domains_10_10_pi2}, \subref{fig:ellipticity_domains_10_10_pi3}, \subref{fig:ellipticity_domains_10_10_pi4} corresponding to the points labeled as $B_1,B_2,B_3$);
    \item Typical \textit{microscopic bifurcation modes of the anisotropic grid exhibit widely different incremental deformations. These are dictated by the prestress direction} (see insets in Fig.~\ref{fig:ellipticity_domains_7_15_pi2}, \subref{fig:ellipticity_domains_7_15_pi3}, \subref{fig:ellipticity_domains_7_15_pi4} corresponding to points labeled $B_4,B_5,B_6,B_7,B_8$ and compare for instance mode $B_4$ to $B_5$ or $B_6$ to $B_7$).
    \item A bifurcation always occurs for every lattice geometry at an equibiaxial load $\{p_1,p_2\}=\{-\pi^2,-\pi^2\}$ (point $B_1$ in Fig.~\ref{fig:ellipticity_domains_10_10_pi2}) regardless of the values of $\Lambda_1$, $\Lambda_2$, $\kappa$, and $\alpha$.
          This bifurcation can be explained by the fact that the normalized load $P l^2/B=-\pi^2$ corresponds to the buckling load of a simply supported Euler-Bernoulli beam, and thus, when all the rods of an arbitrary grid are prestressed at this level, a purely flexural buckling mode becomes available (shown in the inset of Fig.~\ref{fig:ellipticity_domains_10_10_pi2}).
\end{itemize}

Despite the complex influence of the geometrical and mechanical parameters on the stability of the prestressed grillage, two important `transitions' characterize the nature of the first bifurcation, namely:
\begin{enumerate}[label=(\roman*)]
    \item a \textit{macro-to-micro} transition of the critical bifurcation mode occurs at increasing stiffness of the diagonal springs $\kappa$;
    \item a \textit{micro-to-macro} transition of the critical bifurcation mode occurs at decreasing rods' inclination $\alpha$.
\end{enumerate}
The above transitions will be exploited in Section~\ref{sec:dynamic_forced_response} and~\ref{sec:static_forced_response} to investigate the incremental response induced by perturbations applied to a lattice preloaded close to a bifurcation (both global and local bifurcations will be considered).

\subsection{A single localization band with a highly tunable inclination}
\label{sec:tunable_localization}
A remarkable characteristic is associated to the micro-to-macro bifurcation transition obtained at decreasing angle $\alpha$, namely, a \textit{super-sensitivity of the localization band normal, represented by the unit vector $\bn$, with respect to the state of preload, while the localization mode $\bg$ results only weakly affected}.

For instance, at $\alpha=\pi/6$ and sufficiently high spring stiffness $\kappa$, the insets in Figs.~\ref{fig:ellipticity_domains_10_10_pi6} and \ref{fig:ellipticity_domains_7_15_pi6} show that the relative inclinations between the localization band normal $\bn$ and the localization mode $\bg$ strongly vary as a function of the preload state in the lattice.

When the spring stiffness vanishes, $\kappa=0$, the localization band is inclined near the angles $0$ and $\pi/6$, which represent the grid inclination,
so that failure of ellipticity occurs in a direction $\bn$ that is almost orthogonal to the rods.
On the contrary, at $\kappa=0.2$ a single localization band occurs, whose inclination strongly depends on the load directionality and is essentially unrelated to the underlying grid pattern (shown in the insets for $\kappa=0.2$, Figs.~\ref{fig:ellipticity_domains_10_10_pi6} and~\ref{fig:ellipticity_domains_7_15_pi6}).
\textit{The super-sensitivity of the localization direction provides an enhanced tunability of the macroscopic localization pattern by means of a simple modification of the preload applied to the lattice}.

It is worth noting that the localization direction can also be designed by constructing a lattice with a suitable value of rods' angle $\alpha$, but with this approach the localization direction would not be easily reconfigurable, as the structure geometry would be defined in advance.

\subsection{Infinite set of bifurcation wavelengths in a lattice: perfect equivalence with the continuum}
\label{sec:infinite_buckling_modes}
Loss of ellipticity in a solid involves simultaneously infinite modes of every wavelength, while the corresponding condition in the lattice \textit{usually} involves only one mode of infinite wavelength.
\begin{figure}[htb!]
    \centering
    \centering
    \begin{subfigure}{0.495\textwidth}
        \centering
        \caption{\label{fig:contour_3d_infinite_buckling_modes}}
        \includegraphics[width=0.91\linewidth]{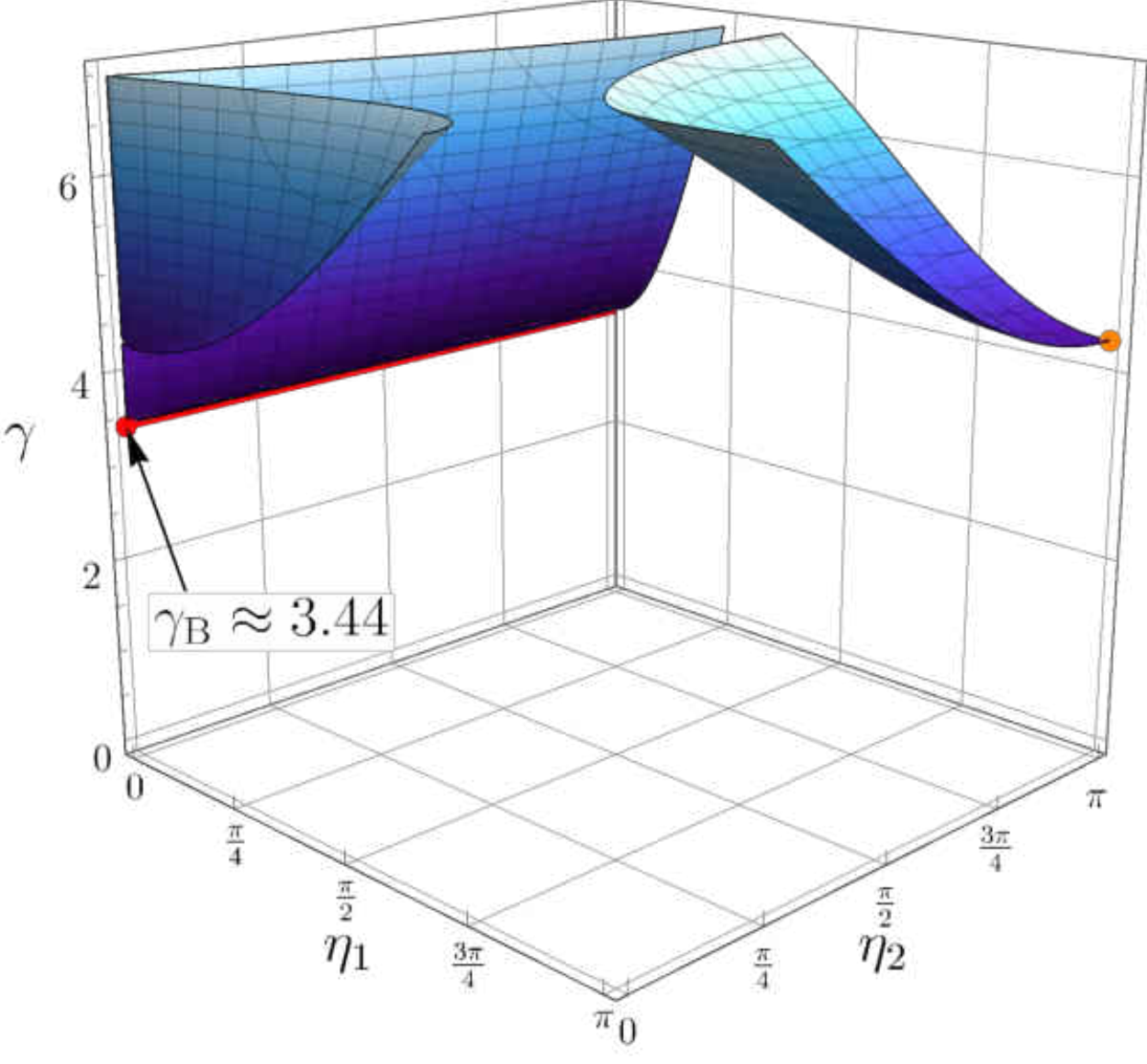}
    \end{subfigure}
    \begin{subfigure}{0.495\textwidth}
        \centering
        \caption{\label{fig:contour_2d_infinite_buckling_modes}}
        \includegraphics[width=\linewidth]{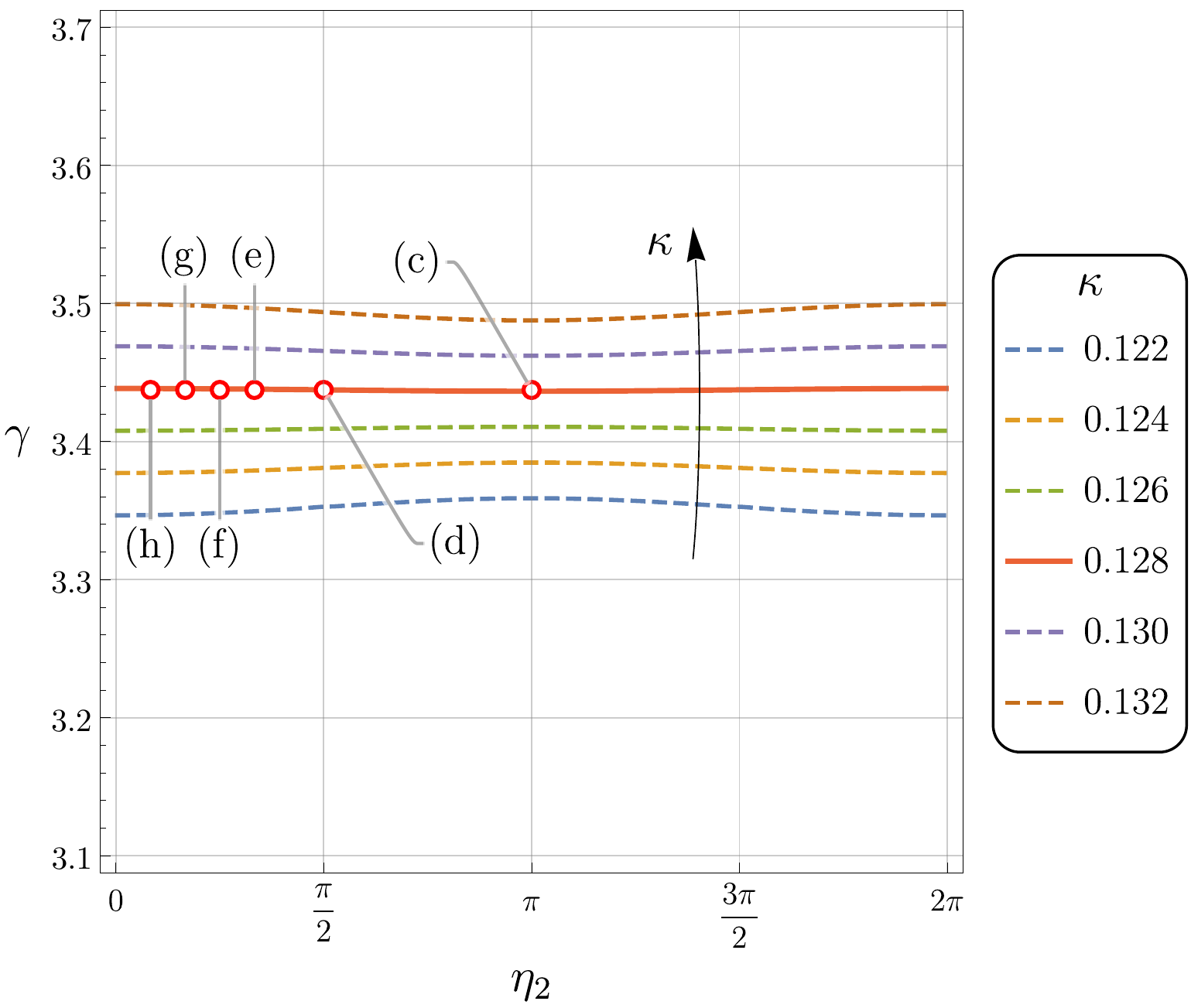}
    \end{subfigure}\\ \vspace{4mm}
    \begin{subfigure}{0.32\textwidth}
        \centering
        \caption{\label{fig:buckling_mode_7_15_k2_pi1}$\bk_{\text{cr}}=\pi\,\bb_2$}
        \includegraphics[width=0.98\linewidth]{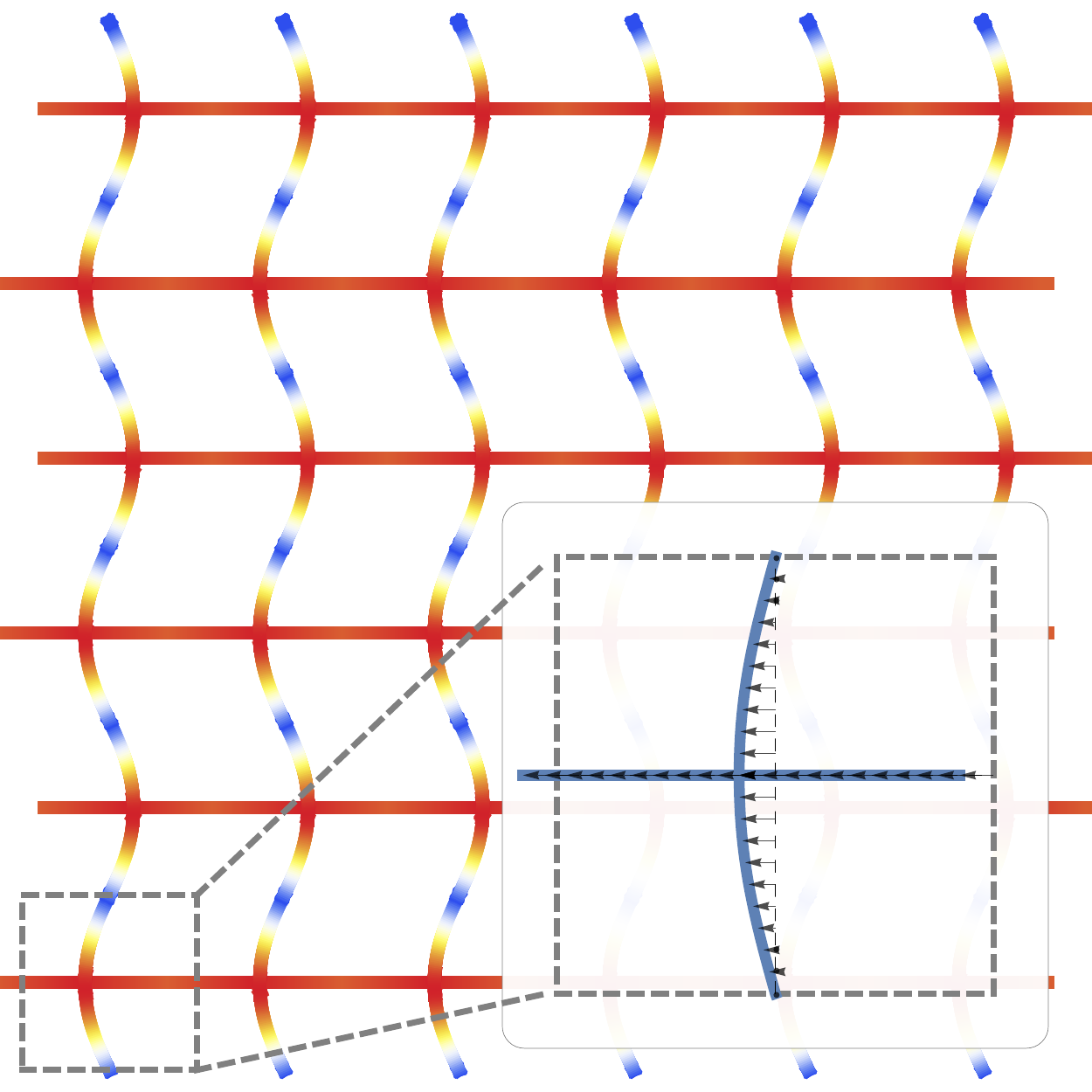}
    \end{subfigure}
    \begin{subfigure}{0.32\textwidth}
        \centering
        \caption{\label{fig:buckling_mode_7_15_k2_pi2}$\bk_{\text{cr}}=\pi/2\,\bb_2$}
        \includegraphics[width=0.98\linewidth]{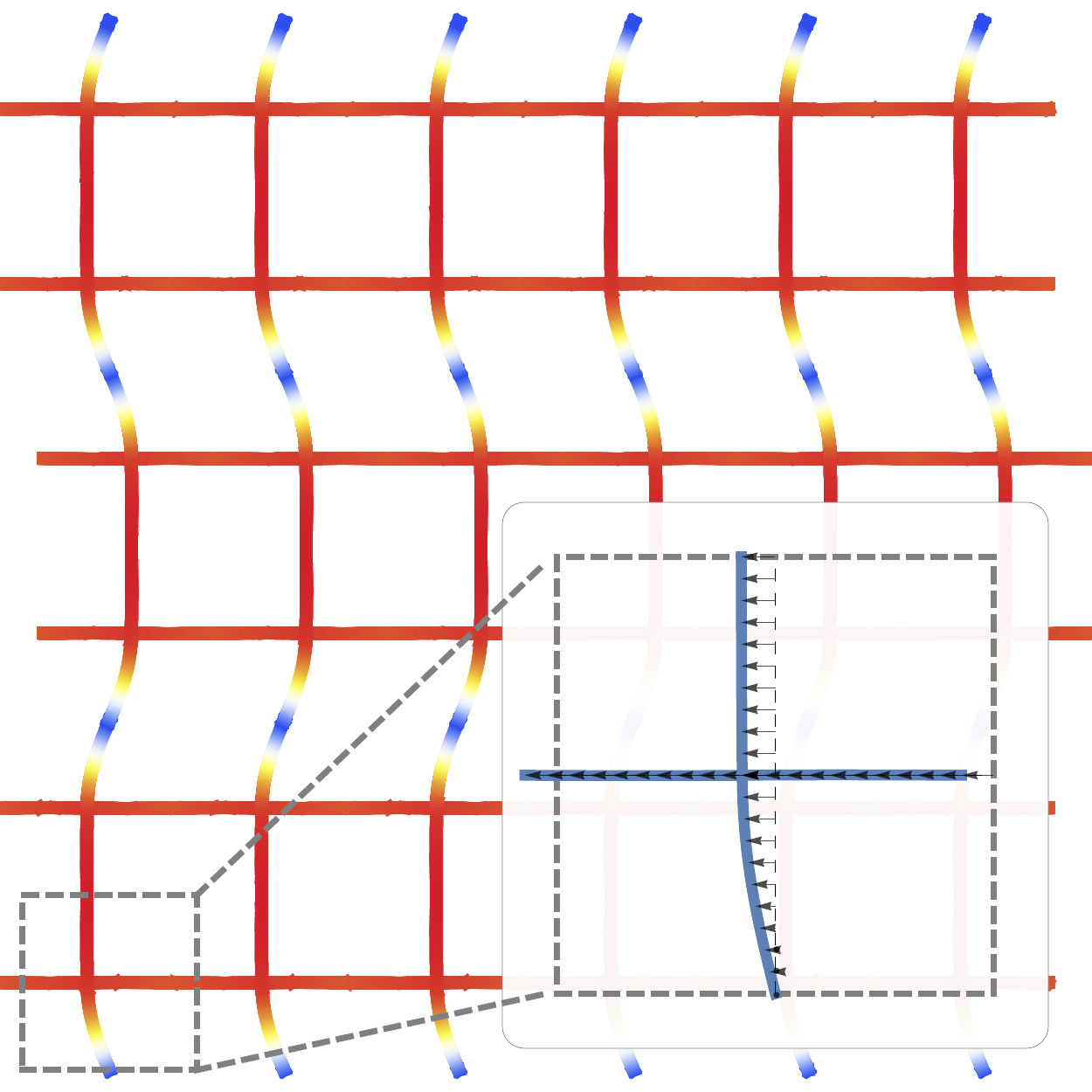}
    \end{subfigure}
    \begin{subfigure}{0.32\textwidth}
        \centering
        \caption{\label{fig:buckling_mode_7_15_k2_pi3}$\bk_{\text{cr}}=\pi/3\,\bb_2$}
        \includegraphics[width=0.98\linewidth]{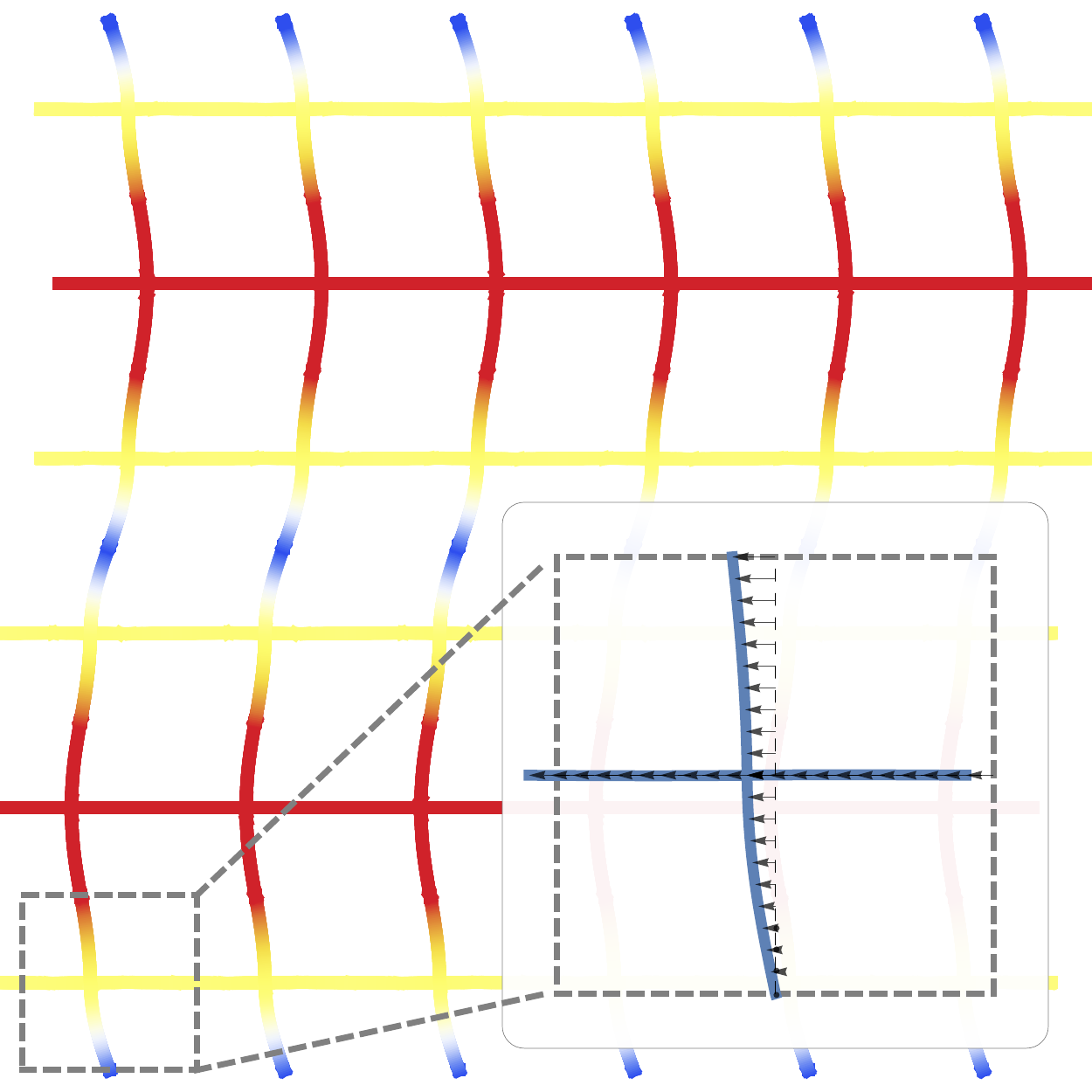}
    \end{subfigure}\\ \vspace{4mm}
    \begin{subfigure}{0.32\textwidth}
        \centering
        \caption{\label{fig:buckling_mode_7_15_k2_pi4}$\bk_{\text{cr}}=\pi/4\,\bb_2$}
        \includegraphics[width=0.98\linewidth]{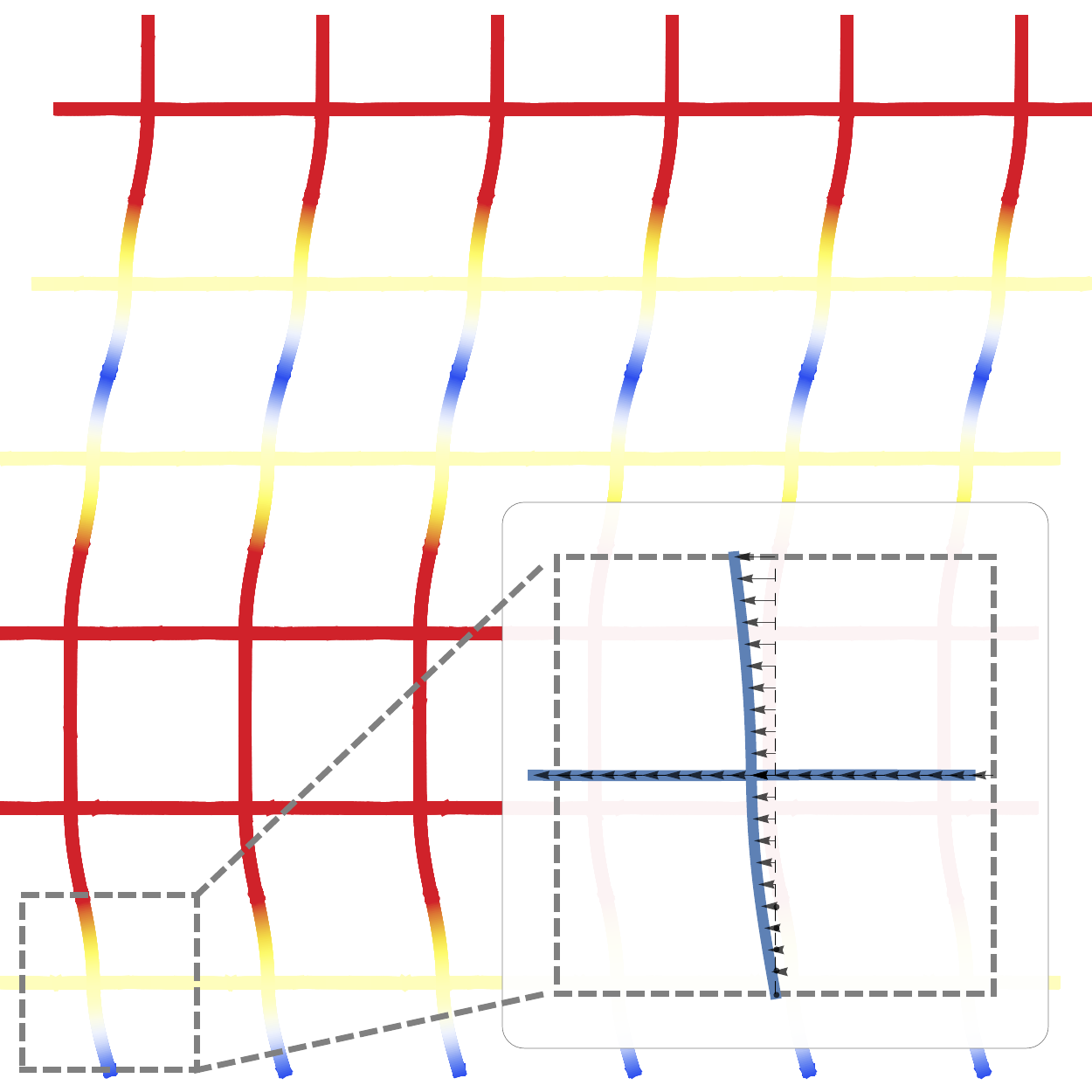}
    \end{subfigure}
    \begin{subfigure}{0.32\textwidth}
        \centering
        \caption{\label{fig:buckling_mode_7_15_k2_pi6}$\bk_{\text{cr}}=\pi/6\,\bb_2$}
        \includegraphics[width=0.98\linewidth]{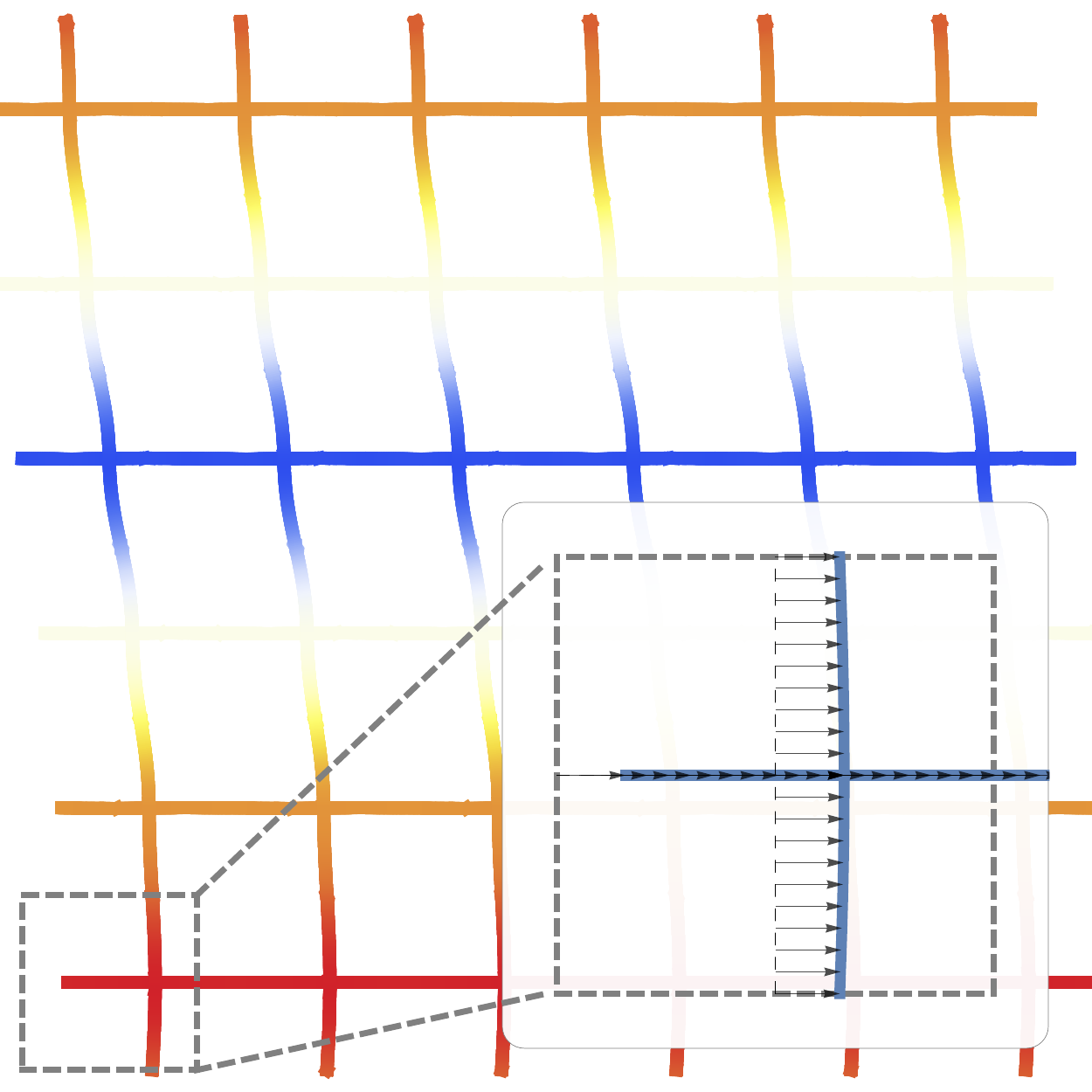}
    \end{subfigure}
    \begin{subfigure}{0.32\textwidth}
        \centering
        \caption{\label{fig:buckling_mode_7_15_k2_pi12}$\bk_{\text{cr}}=\pi/12\,\bb_2$}
        \includegraphics[width=0.97\linewidth]{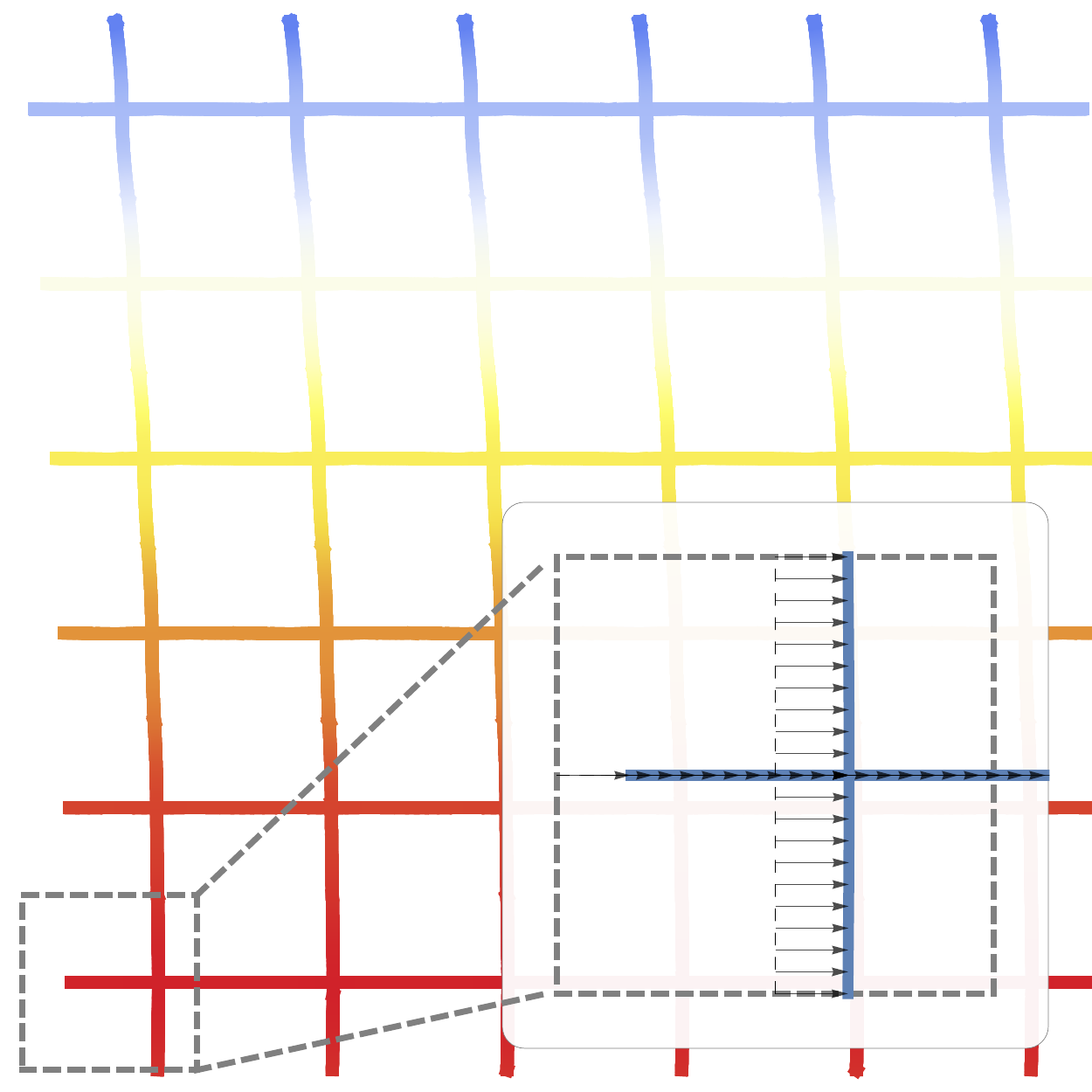}
    \end{subfigure}
    \caption{\label{fig:infinite_buckling_modes}
        Conditions showing a perfect equivalence between the lattice and the corresponding continuum, so that when the latter looses ellipticity, the former exhibits bifurcation occurring with infinite modes covering all wavelengths of Bloch type (these are greater than $2l$ and correspond to $\bk_{\text{cr}} \leq \pi\,\bb_2$), a situation which is revealed by the flat line (highlighted in red) in the bifurcation diagram (\subref{fig:contour_3d_infinite_buckling_modes}).
        The perfect equivalence is obtained through accurate tuning of the stiffness of the diagonal springs ($\kappa \approx 0.128$ for a square grid with $\Lambda_1=7$ and $\Lambda_2=15$ and a loading $\{p_1,p_2\} \approx 3.44\{-1,-6\}$).
        Part~(\subref{fig:contour_2d_infinite_buckling_modes}) represents a section of the bifurcation surface at $\eta_1=0$ detailing the flat minimum of the curve occurring at $\kappa \approx 0.128$.
        Parts (\subref{fig:buckling_mode_7_15_k2_pi1})--(\subref{fig:buckling_mode_7_15_k2_pi12}) present selected bifurcation modes documenting a transition at increasing wavelength of the bifurcation modes from a local~(\subref{fig:buckling_mode_7_15_k2_pi1}) to a shear-band~(\subref{fig:buckling_mode_7_15_k2_pi12}) bifurcation.
    }
\end{figure}
In this sense, the equivalent continuum displays a response differing from the lattice, a circumstance which may be expected as a consequence of the homogenization procedure.

Surprisingly, it is shown in the following that special conditions can be found in which the lattice bifurcates similarly to the equivalent continuum, by displaying infinite modes, covering every wavelength.
In this case a \textit{perfect equivalence between the bifurcation in the grillage and failure of ellipticity in the effective continuum occurs}.

For a square grid (with $\alpha=\pi/2$, $\Lambda_1=7$, and $\Lambda_2=15$), the perfect equivalence can be reached at a fixed value of load by varying the stiffness of the diagonal springs $\kappa$, thus obtaining $\kappa\approx0.128$.
This value was calculated by numerically solving equation~\eqref{eq:local_buckling_multiplier_simpler} between $\kappa=0.1$ and $\kappa=0.2$, because these two values pinpoint the threshold of separation between macro and micro bifurcation.
Moreover, to obtain this special feature, the loading path must be appropriately selected, as shown in Fig.~\ref{fig:ellipticity_domains_7_15_pi2}.
Indeed, along the curved boundary denoted as $\text{(G)}$ in Fig.~\ref{fig:ellipticity_domains_7_15_pi2} the bifurcation mode is unique and involves only the infinite wavelength (macro bifurcation), while on the boundary denoted as $\text{(GL)}$ \textit{an infinite number of bifurcation modes of arbitrary wavelength is present for every critical loading state}, as detailed for the point $B_\infty$ in Fig.~\ref{fig:infinite_buckling_modes}.

With reference to the reciprocal basis~\eqref{eq:reciprocal_basis}, a three-dimensional plot of the bifurcation surface in the space $\{\eta_1,\eta_2,\gamma\}$ ($\gamma$ is the loading multiplier) is reported in Fig.~\ref{fig:contour_3d_infinite_buckling_modes}.
Points on the surface satisfy the vanishing of the determinant in Eq.~\eqref{eq:local_buckling_multiplier_simpler} and are calculated only for the loading path $\{p_1,p_2\}=\gamma\{-1,-6\}$.
The lowest, i.e. critical, bifurcation occurs at $\gamma=\gamma_{\text{B}}$ (a value represented in the figure as a red segment).
A section of this surface at $\eta_1=0$ is reported in Fig.~\ref{fig:contour_2d_infinite_buckling_modes} to show the dependence of the critical multiplier on the stiffness $\kappa$.
In particular, for $\kappa<0.128$ the critical wave vector is $\bk_{\text{cr}}=0$ (macro instability), while for $\kappa>0.128$ the critical wave vector is $\bk_{\text{cr}}=\pi\,\bb_2$ (micro instability).
Finally and most importantly, for $\kappa=0.128$ every wave vector of the form $\bk_{\text{cr}}=\eta_2\,\bb_2$ (with arbitrary $\eta_2$)\footnote{
    Note that, due to the periodic structure of the lattice, the shortest wavelength of Bloch type is equal to $2\, l$, corresponding to $\bk_{\text{cr}} = \pi\,\bb_2$.
}
identifies a different bifurcation mode occurring at the same load multiplier $\gamma_{\text{B}}\approx3.44$.
Within this infinite set of bifurcation modes, a few bifurcation modes are reported (the labelled points on the red contour of Fig.~\ref{fig:contour_2d_infinite_buckling_modes}), to show the transition of the bifurcation mode from a local bifurcation (Fig.~\ref{fig:buckling_mode_7_15_k2_pi1}) to a global shear-band type instability (Fig.~\ref{fig:buckling_mode_7_15_k2_pi12}).

\subsection{Macro and micro bifurcations as degeneracies of the dispersion relation}
\label{sec:macro_micro_bifurcations_dynamics}
The effect of the diagonal reinforcement (springs labeled with $S$ in Fig.~\ref{fig:geometry_grid_and_unit_cell}) on the bifurcation of the lattice has been systematically investigated in Section~\ref{sec:grid_ellipticity_local_buckling}, where it has been shown to play a fundamental role in determining the wavelength critical for bifurcation.
\begin{figure}[htb!]
    \centering
    \begin{subfigure}{0.245\textwidth}
        \centering
        \includegraphics[width=0.95\linewidth]{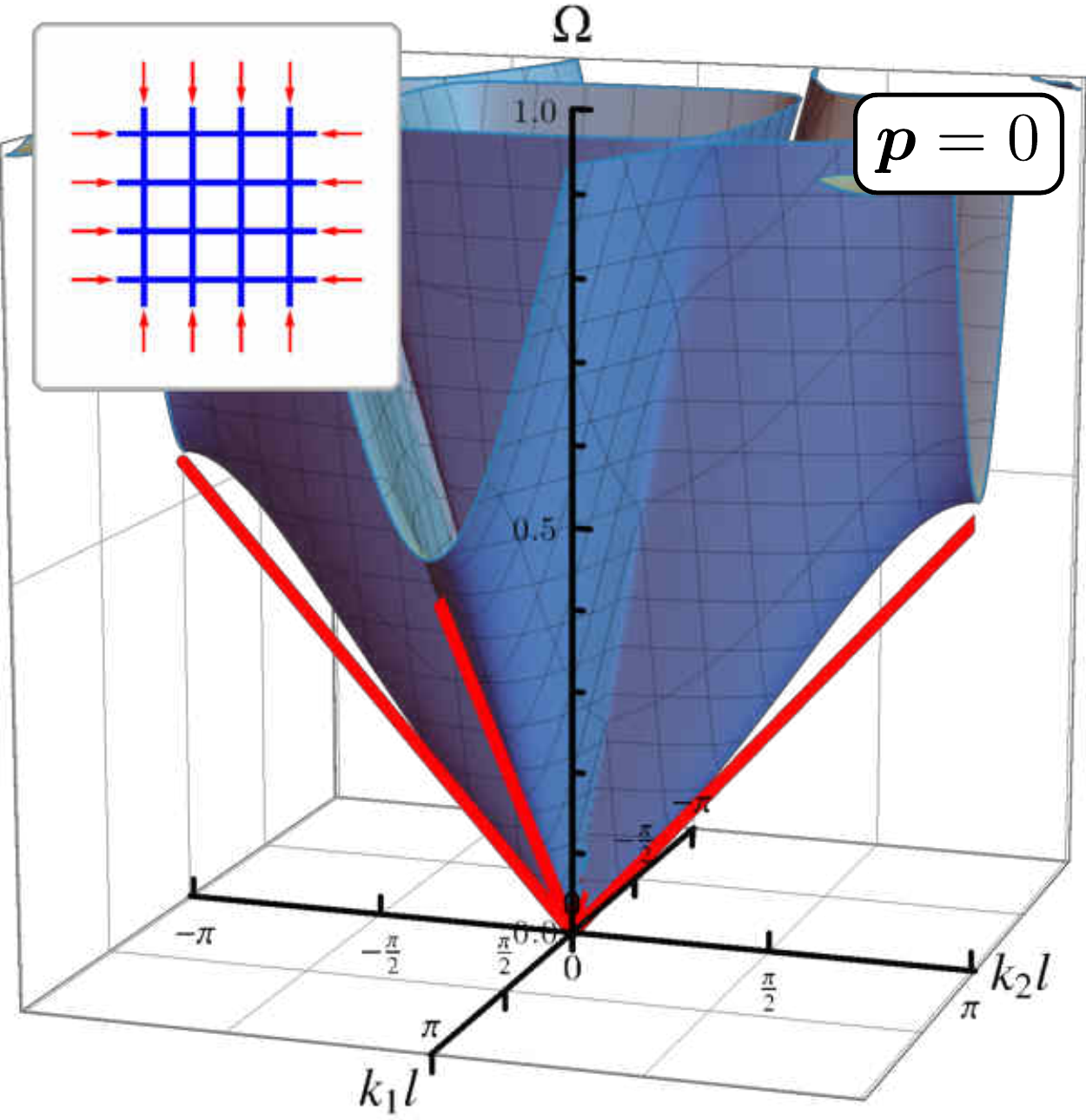}
    \end{subfigure}%
    \begin{subfigure}{0.245\textwidth}
        \centering
        \includegraphics[width=0.95\linewidth]{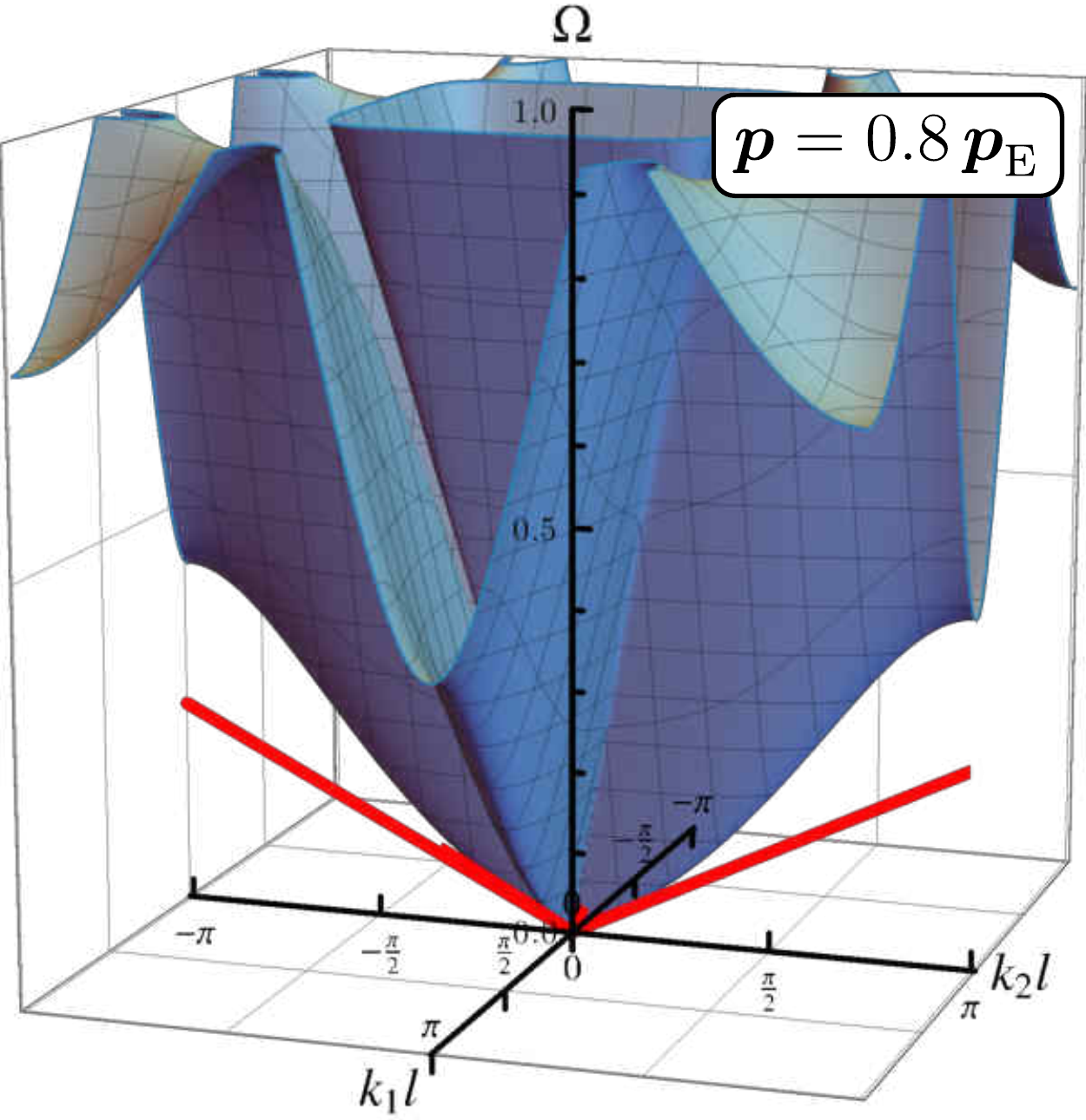}
    \end{subfigure}%
    \begin{subfigure}{0.245\textwidth}
        \centering
        \includegraphics[width=0.95\linewidth]{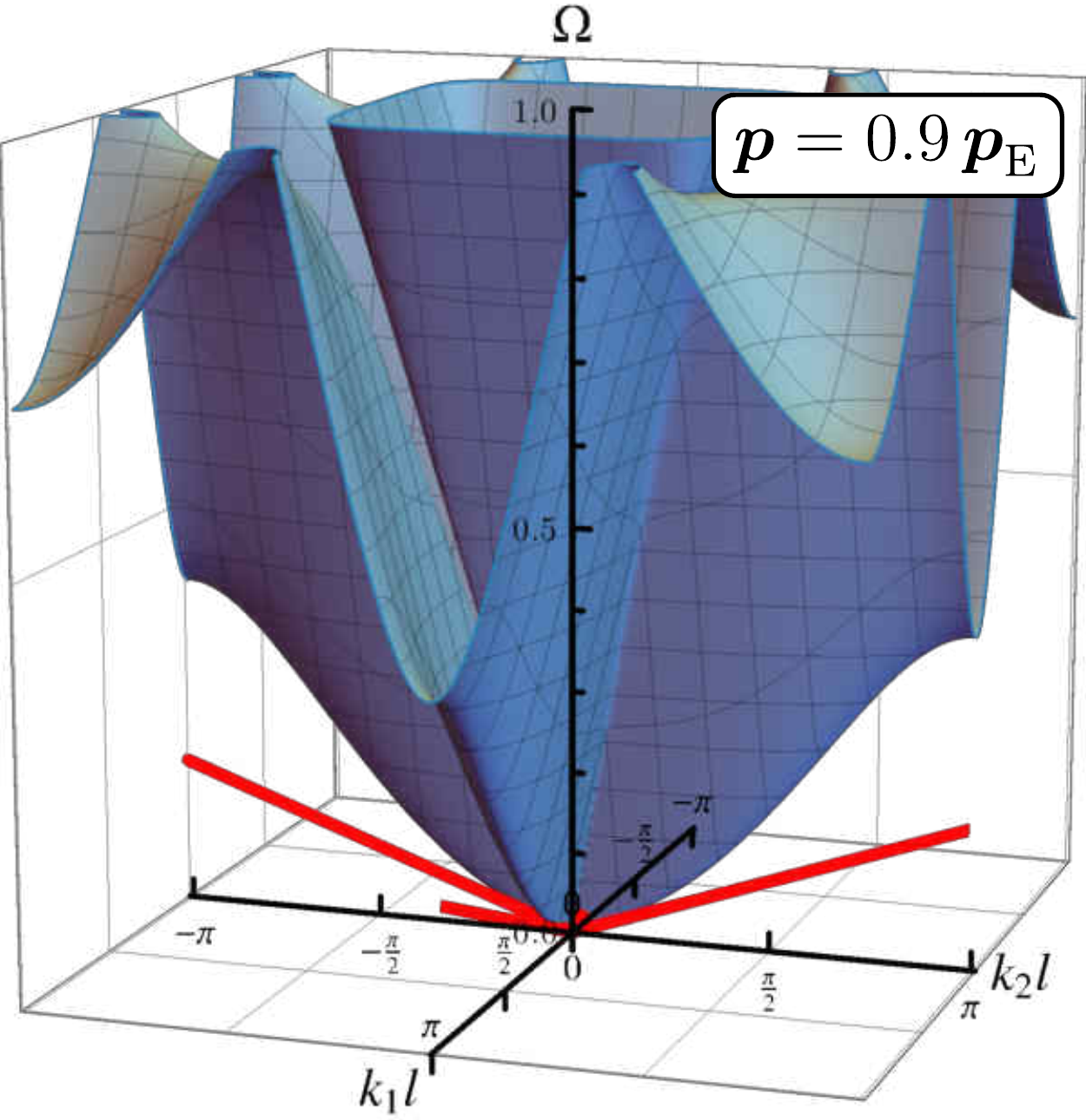}
    \end{subfigure}%
    \begin{subfigure}{0.245\textwidth}
        \centering
        \includegraphics[width=0.95\linewidth]{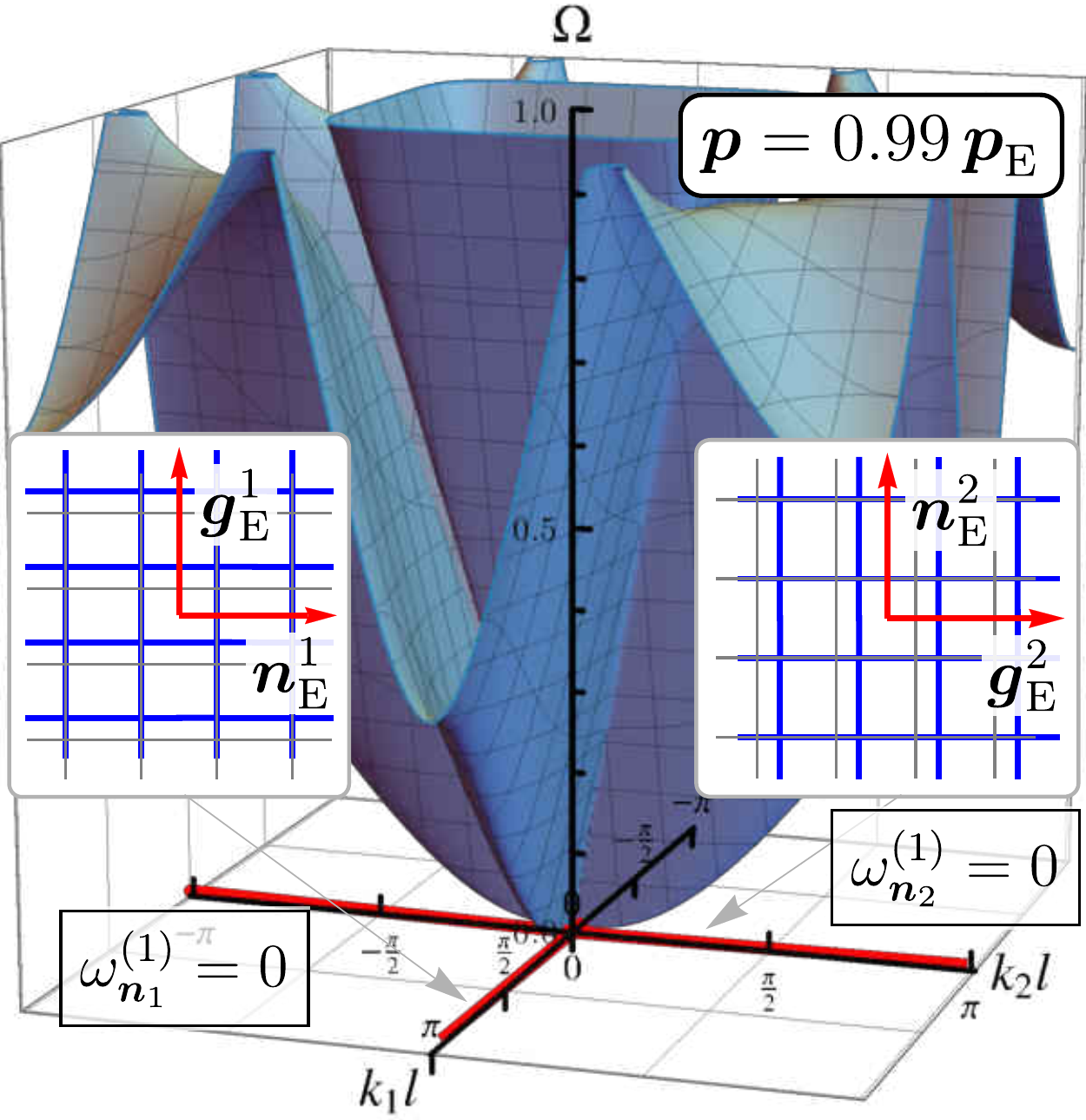}
    \end{subfigure}\\ \vspace{2mm}
    \begin{subfigure}{0.245\textwidth}
        \centering
        \includegraphics[width=0.95\linewidth]{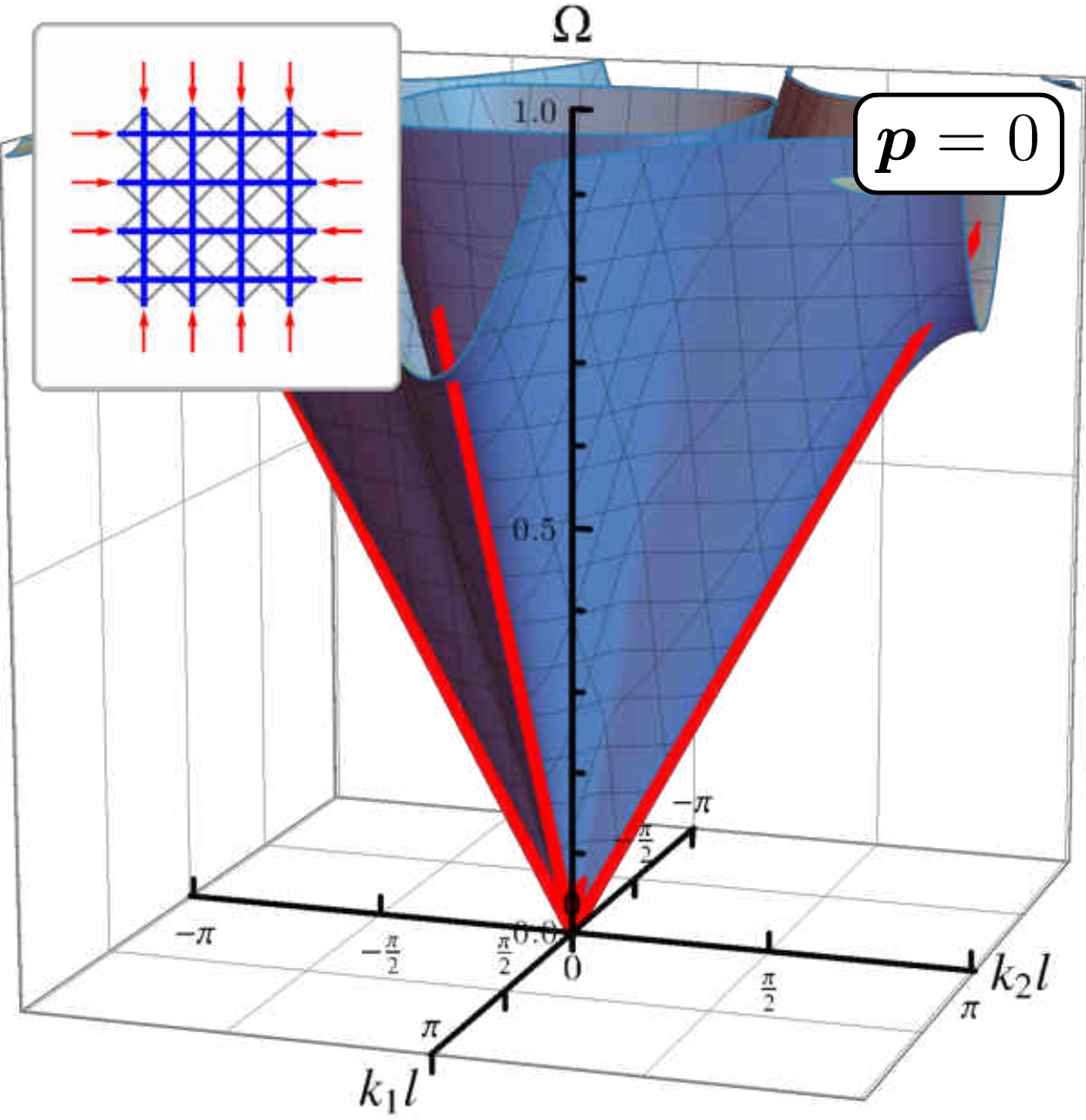}
    \end{subfigure}%
    \begin{subfigure}{0.245\textwidth}
        \centering
        \includegraphics[width=0.95\linewidth]{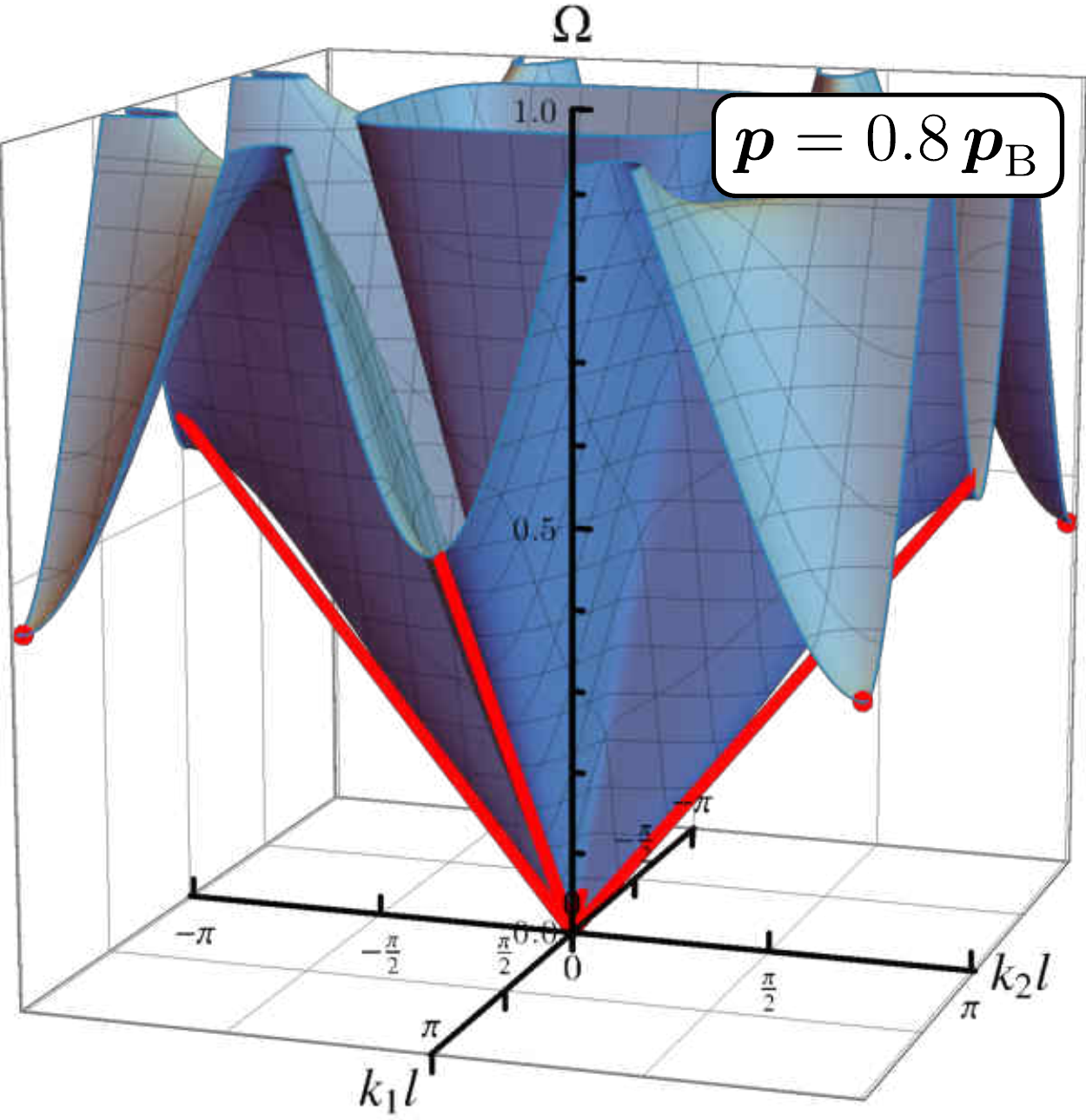}
    \end{subfigure}%
    \begin{subfigure}{0.245\textwidth}
        \centering
        \includegraphics[width=0.95\linewidth]{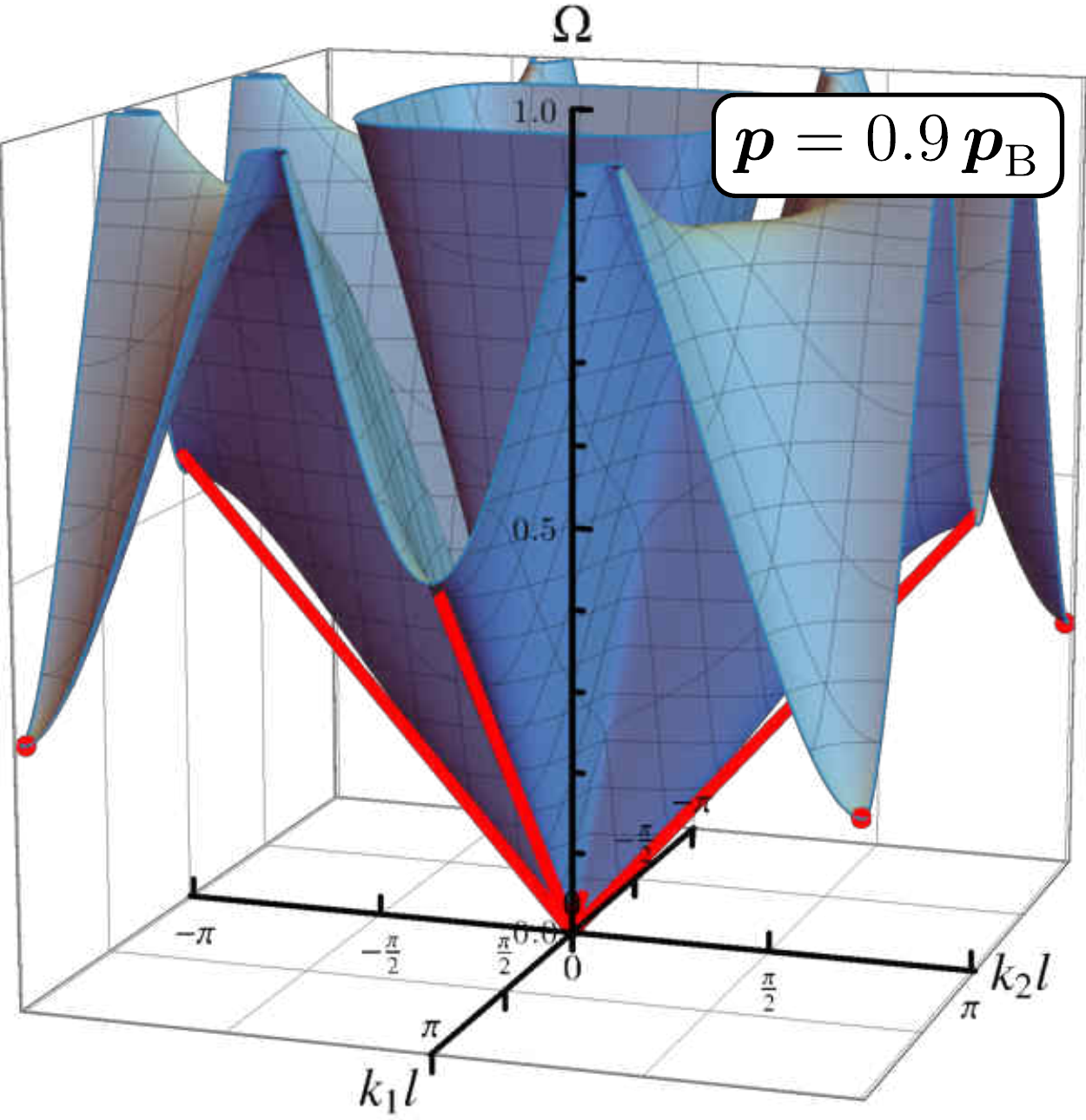}
    \end{subfigure}%
    \begin{subfigure}{0.245\textwidth}
        \centering
        \includegraphics[width=0.95\linewidth]{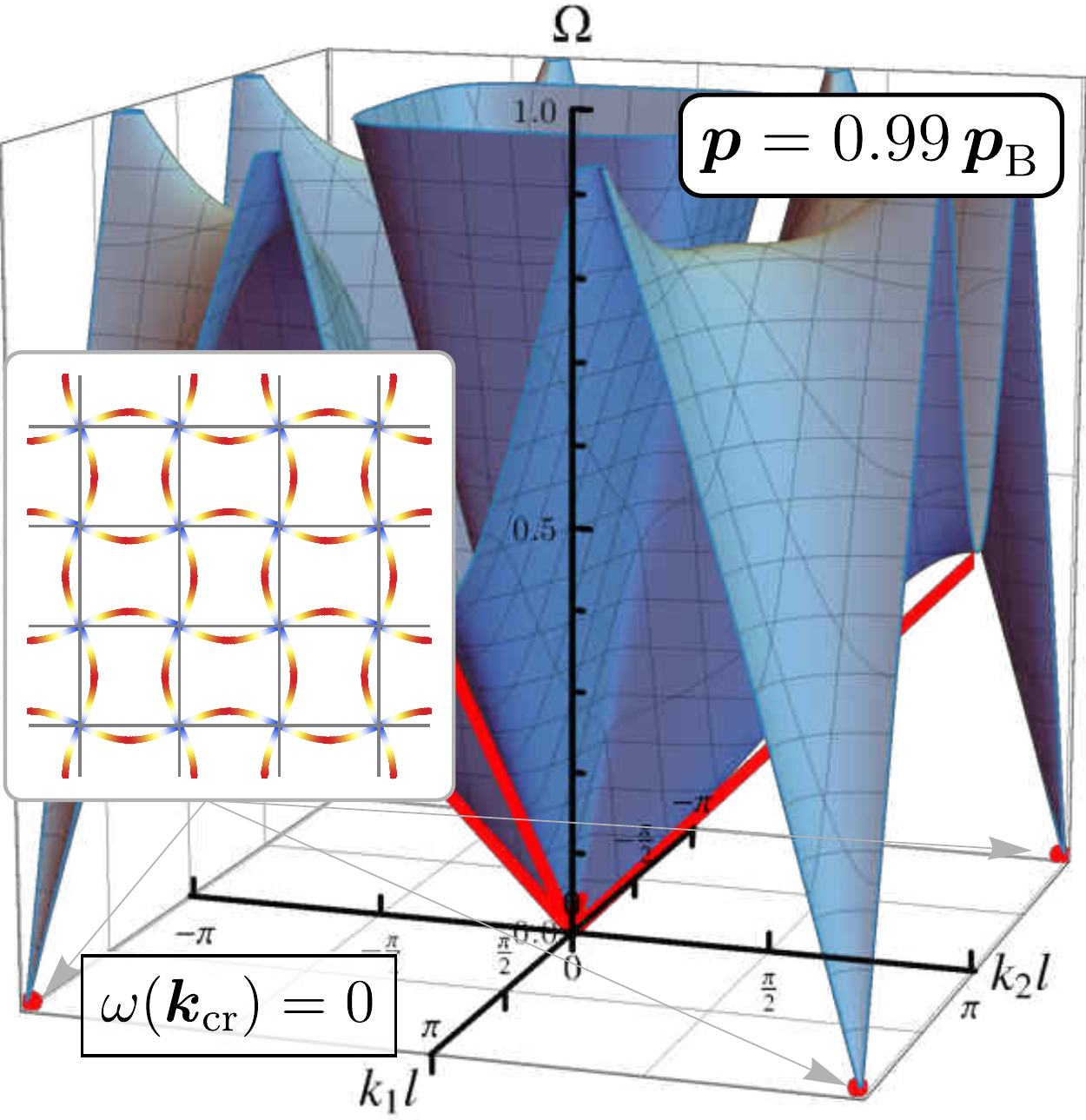}
    \end{subfigure}
    \caption{\label{fig:macro_micro_bifurcation_dynamics}
        The dispersion surfaces, computed for states of preload of increasing magnitude (from left to right), demonstrate the difference between macroscopic (upper part) and microscopic (lower part) bifurcations occurring in a square grid of elastic rods.
        The stiffness of the diagonal springs can be tuned to cause a switching of the critical bifurcation mode from macroscopic (low spring stiffness) to microscopic (high spring stiffness).
        The four surfaces reported in the upper part refer to equibiaxial compression of a square grid with $\Lambda_1=\Lambda_2=10$ and not reinforced with springs.
        In this case an infinite-wavelength bifurcation occurs at vanishing slope of the acoustic branches at the origin (when the preload reaches a critical value $\bp_{\text{E}}=\{-5.434,-5.434\}$).
        The four surfaces reported in the lower part refer to the same grid but reinforced with springs ($\kappa=0.4$), which induces a `stiffening' of the acoustic branches at the origin. This triggers a critical bifurcation at a preload $\bp_{\text{B}}=\{-\pi^2,-\pi^2\}$, when the lowering of the dispersion surface causes the generation of a zero-frequency wave with non-null wave vector (corresponding to a finite wavelength buckling).
    }
\end{figure}
Specifically, it has been demonstrated that an increase in the spring stiffness induces a transition of the critical bifurcation from macroscopic to microscopic, and in particular the bifurcation is characterized by an infinite wavelength when $\kappa=0$.

On the other hand, the time-harmonic formulation of Section~\ref{sec:system_governing_equations} and~\ref{sec:asymptotic_lattice_waves} can be leveraged to provide a dynamic interpretation to lattice instabilities different from the one obtained via the quasi-static approach of Section~\ref{sec:ellipticity_stability}.
The difference between macroscopic and microscopic bifurcation will be specially focused, as the latter is lost in the homogenization approach.
The homogenization scheme introduced in Section~\ref{sec:asymptotic_lattice_waves} proves that the long-wavelength asymptotics for waves propagating in the lattice is governed by the acoustic tensor of the effective medium, Eq.~\eqref{eq:acoustic_tensor_lattice}. Therefore, it becomes now clear that a macro-bifurcation in the lattice has to be equivalent to failure of ellipticity in its equivalent continuum.
Hence, a macro-bifurcation occurs when the velocity of the acoustic long waves of the lattice vanishes along some directions.
Moreover, a clear interpretation of short-wavelength bifurcations (micro-bifurcations) is also provided by the analysis of the dispersion relation of the lattice~\eqref{eq:dispersion}, interpreted now as a function of the preload state. The latter can be used to identify the condition of buckling in the lattice as the `propagation' of a Bloch wave at vanishing frequency.
In fact, regardless of the critical wavelength, macro and micro bifurcations can be visualized by plotting the evolution of the dispersion surfaces along a loading path up to loss of stability.

The essential difference between the two kinds of bifurcation is exemplified in Fig.~\ref{fig:macro_micro_bifurcation_dynamics} for two square grids ($\alpha=\pi/2$ and $\Lambda_1=\Lambda_2=10$), one without diagonal springs (upper row in the figure) and the other with $\kappa=0.4$ (lower row in the figure), subject to equibiaxial compression ($p_1=p_2$) of increasing magnitude (from left to right in the figure).
The dispersion surfaces (plotted in the non-dimensional space $\{k_1 l,k_2 l,\Omega\}$ with $\Omega=\omega\,l\sqrt{\gamma/A}$ and $\gamma_r=0$) show that \textit{the macro-bifurcation in the grid without springs occurs with the progressive lowering, and eventually vanishing, of the slope of the acoustic branches at the origin, while the dispersion surface attains non-null frequency for every other wave vector}.
On the contrary, the micro-bifurcation occurring in the grid reinforced with springs is characterized by a non-vanishing slope of the acoustic branches at the origin.
Moreover, because of the preload-induced lowering of the dispersion surface, a zero-frequency wave is generated with a non-null wave vector, which corresponds to a finite wavelength bifurcation.
These dispersion surfaces can be considered the dynamic counterpart of the bifurcation surfaces presented in Section~\ref{sec:grid_ellipticity_local_buckling}.

\section{Time-harmonic forced response near the elliptic boundary}
\label{sec:dynamic_forced_response}
The analysis of the homogenized continuum, equivalent to a preloaded grid of elastic rods (presented in Section~\ref{sec:grid}) predicts that the incremental response can display strain localizations due to prestress-induced loss of ellipticity.
However, while the relation between localization and failure of ellipticity is well-known in a continuum, it is not equally clear why a `global bifurcation' in the grillage should correspond to a localization of motion.
The scope of this section is a definitive clarification of this important point through a perturbative approach, in which a perturbing agent in terms of a pulsating concentrated force is applied both to the lattice and its continuum approximation and the results in terms of incremental displacement maps compared.

It should also be remarked that the perturbative approach employed is designed to capture the \textit{onset} of strain localization, while its \textit{development} during a deformation path, after its appearance, will later be influenced by the post-critical behavior (e.g. see~\cite{kyriakides_1995}).
Nonetheless, loss of ellipticity indicates a long-wavelength bifurcation eigenmode of the lattice and will be shown to correspond to a localization of incremental strain when a perturbing agent, as a concentrated force, is superimposed on a prestressed state \textit{close} to the macro-bifurcation threshold.
\begin{figure}[htb!]
    \centering
    \begin{subfigure}{0.265\textwidth}
        \centering
        \includegraphics[height=40.5mm]{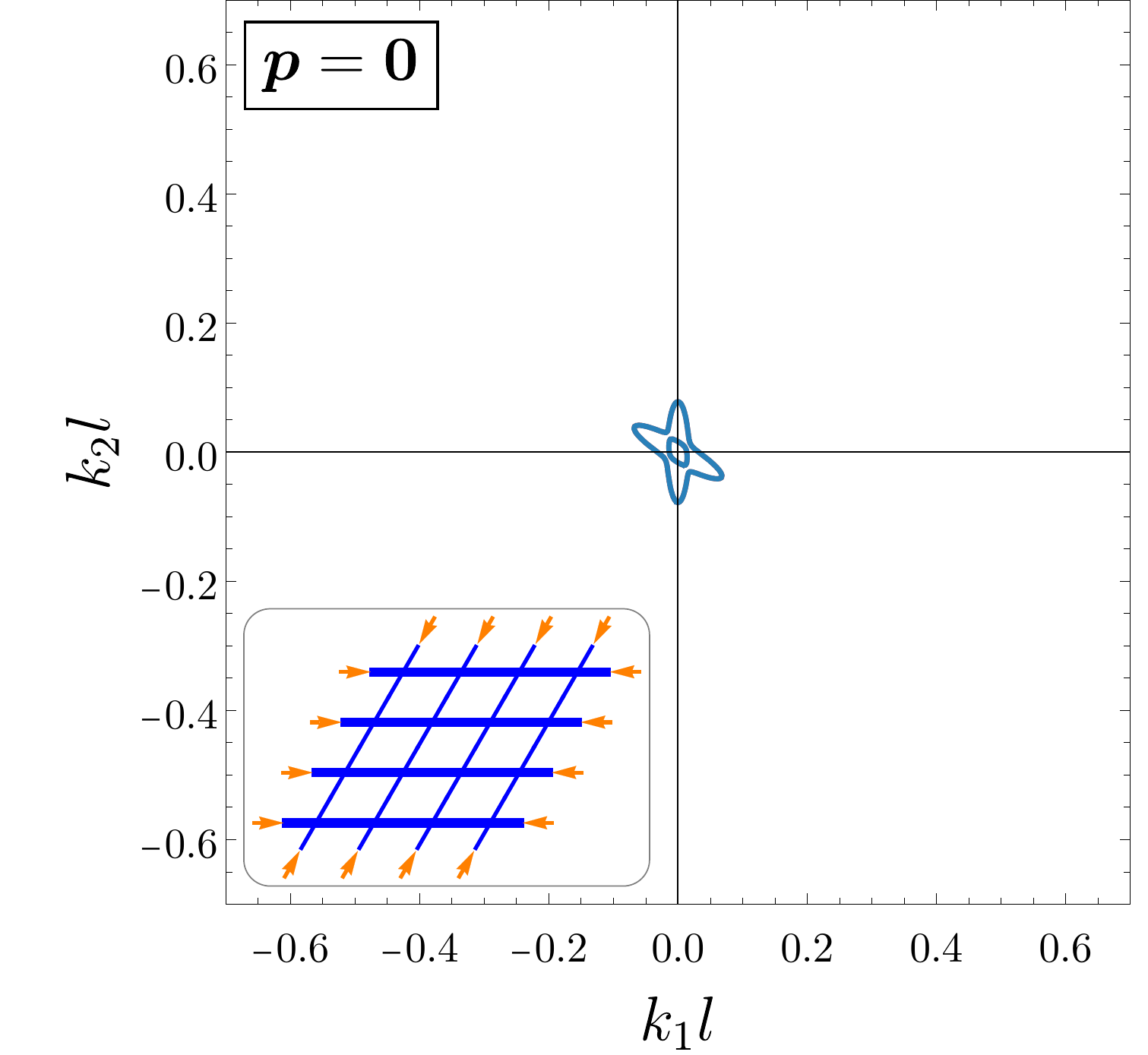}
    \end{subfigure}%
    \begin{subfigure}{0.245\textwidth}
        \centering
        \includegraphics[height=40.5mm]{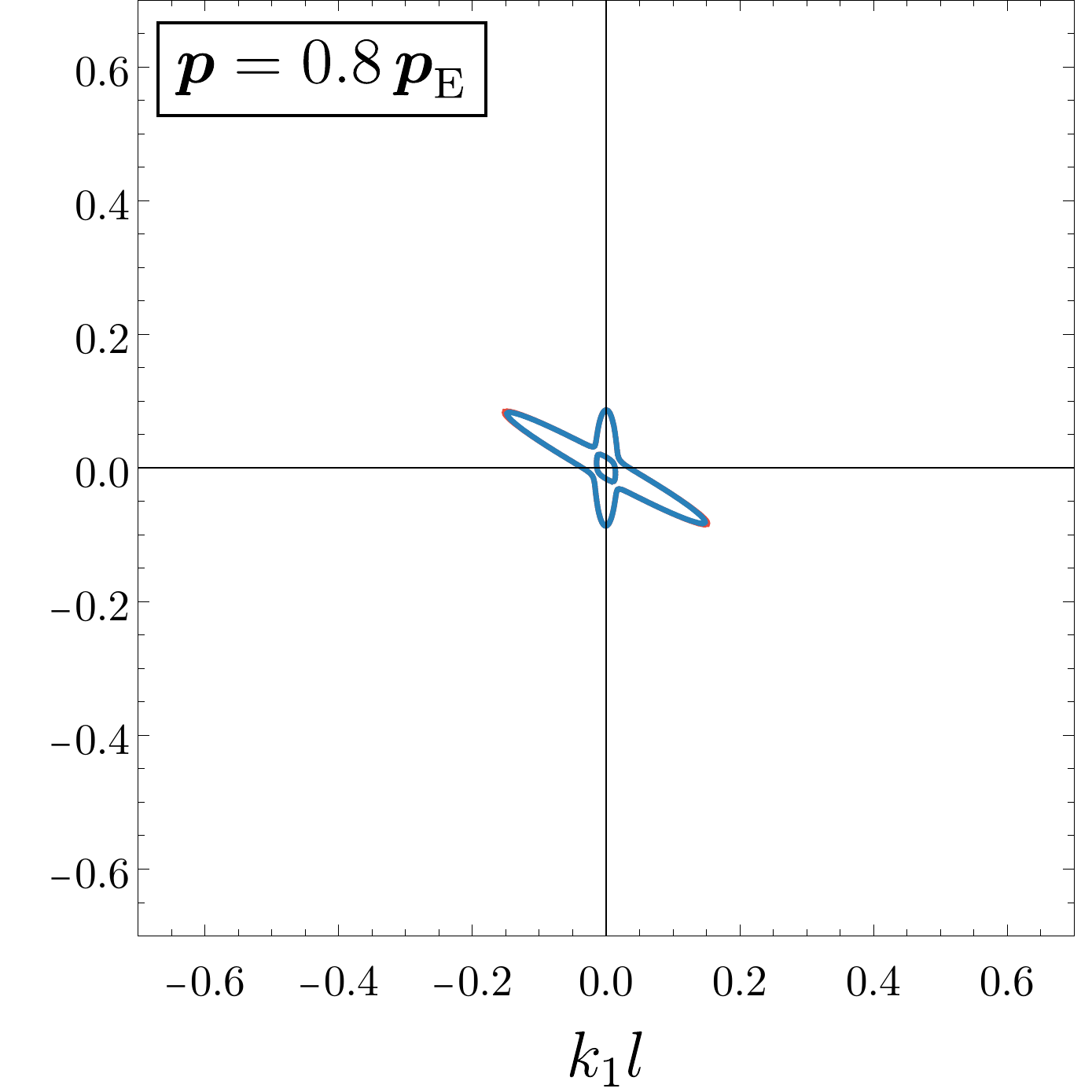}
    \end{subfigure}%
    \begin{subfigure}{0.245\textwidth}
        \centering
        \includegraphics[height=40.5mm]{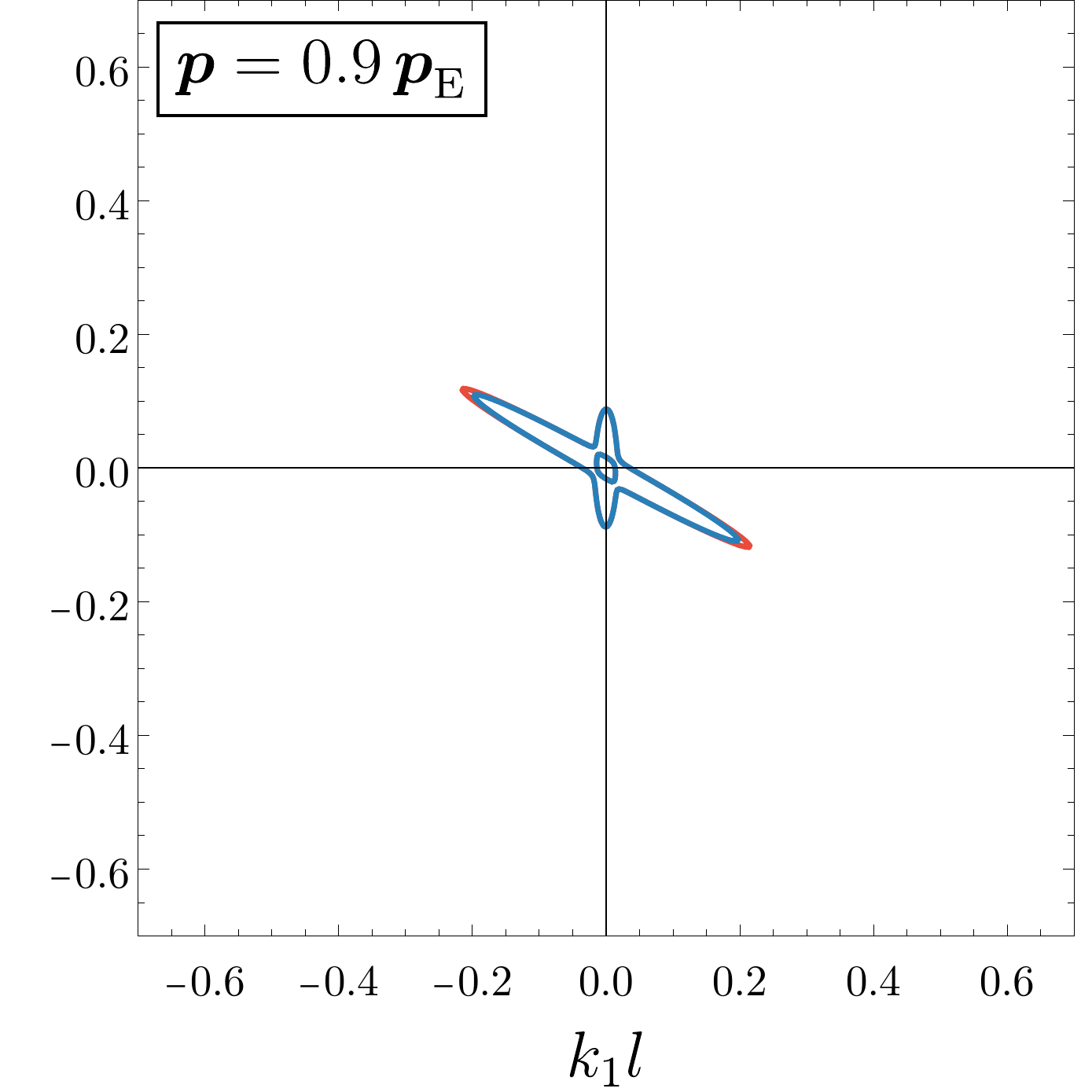}
    \end{subfigure}%
    \begin{subfigure}{0.245\textwidth}
        \centering
        \includegraphics[height=40.5mm]{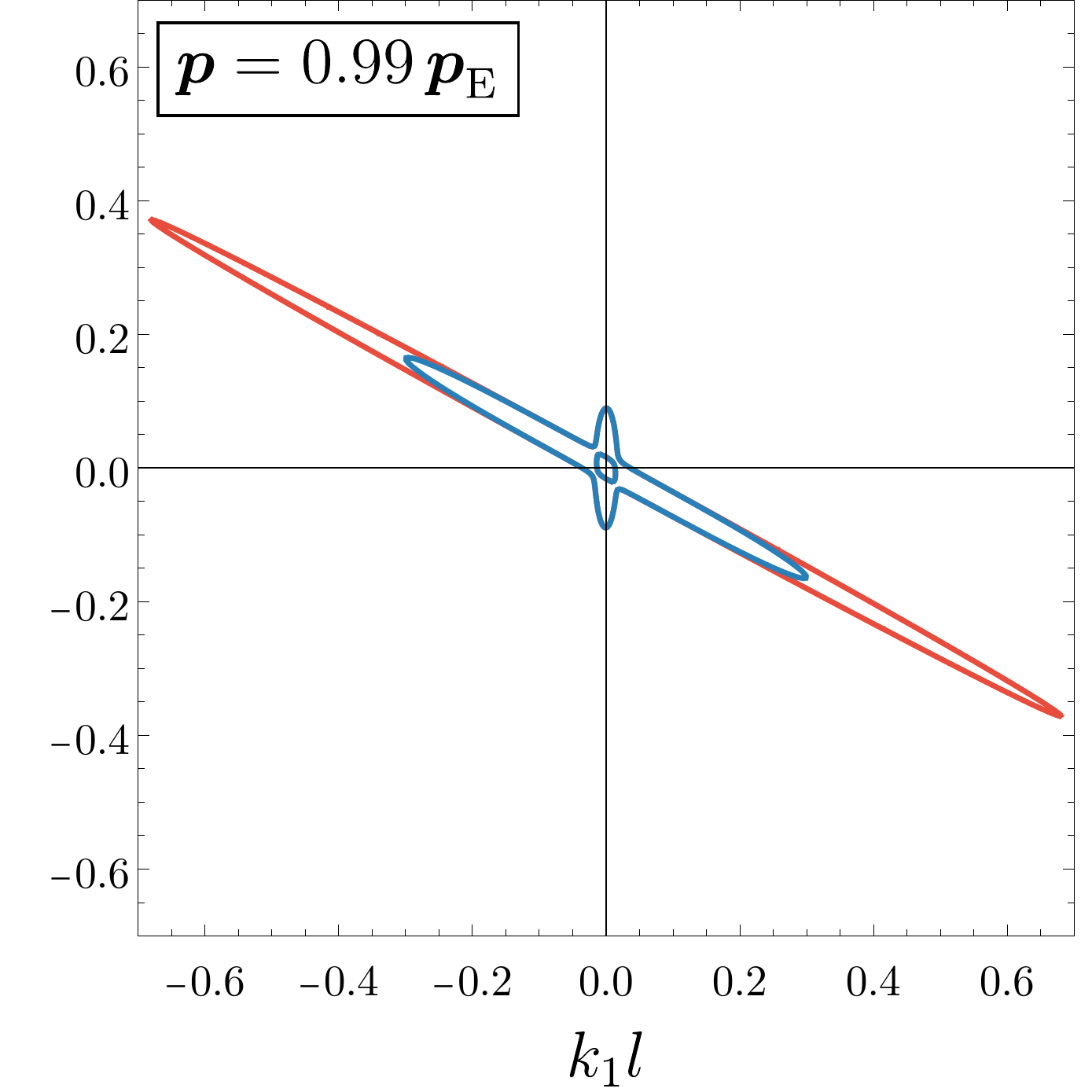}
    \end{subfigure}\\ \vspace{2mm}
    \begin{subfigure}{0.265\textwidth}
        \centering
        \includegraphics[height=38mm]{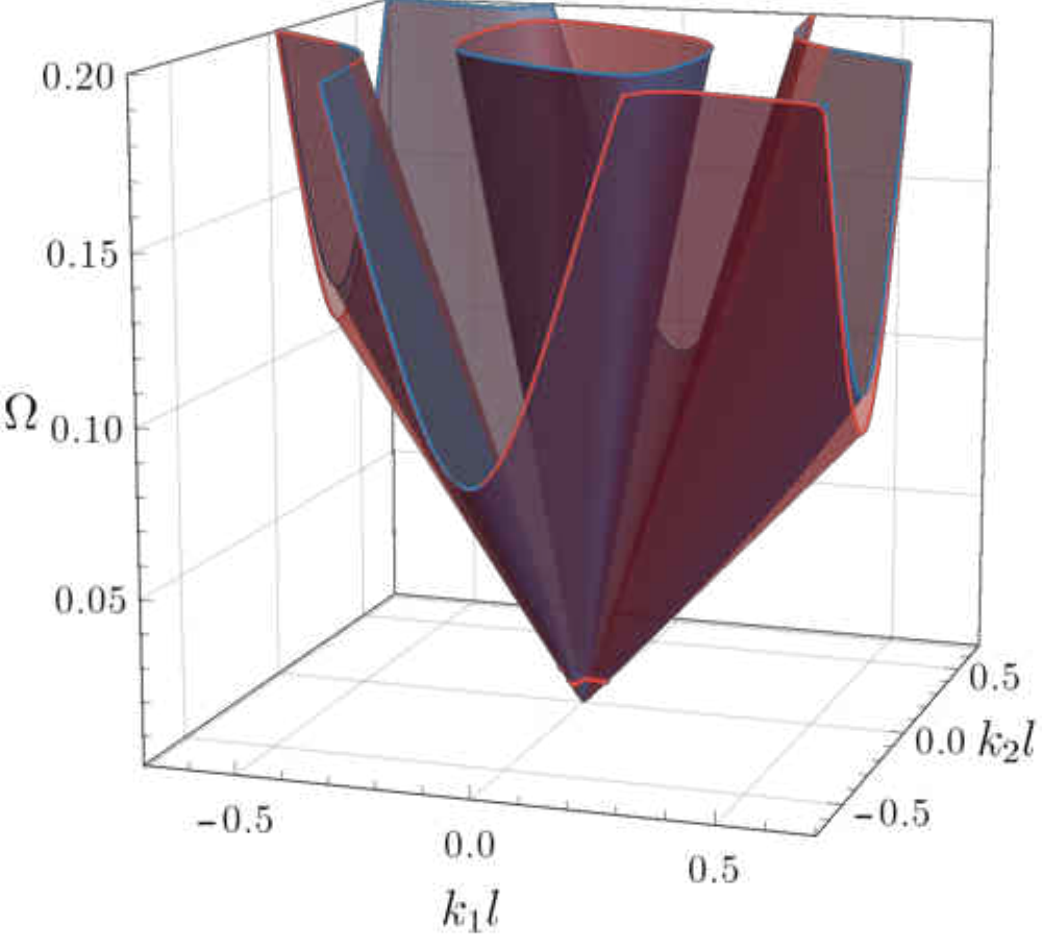}
    \end{subfigure}%
    \begin{subfigure}{0.245\textwidth}
        \centering
        \includegraphics[height=38mm]{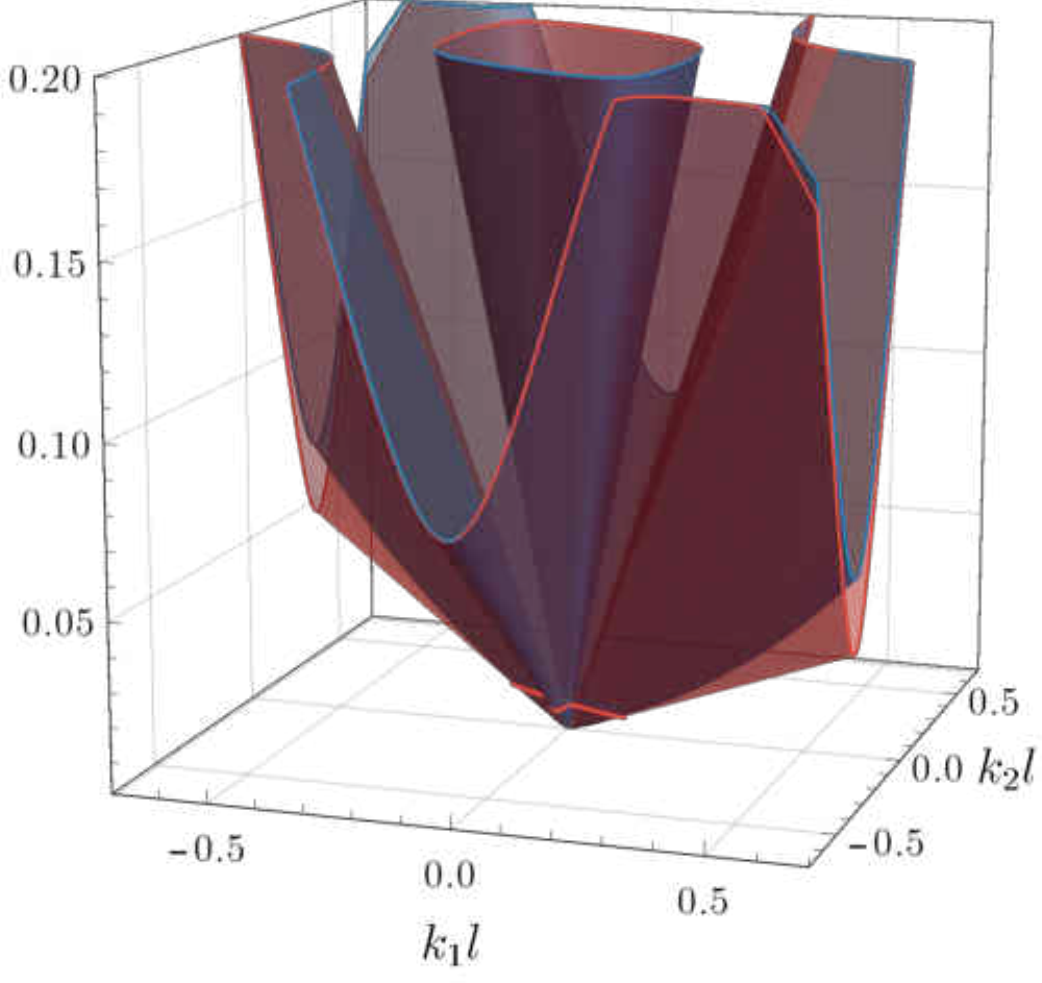}
    \end{subfigure}%
    \begin{subfigure}{0.245\textwidth}
        \centering
        \includegraphics[height=38mm]{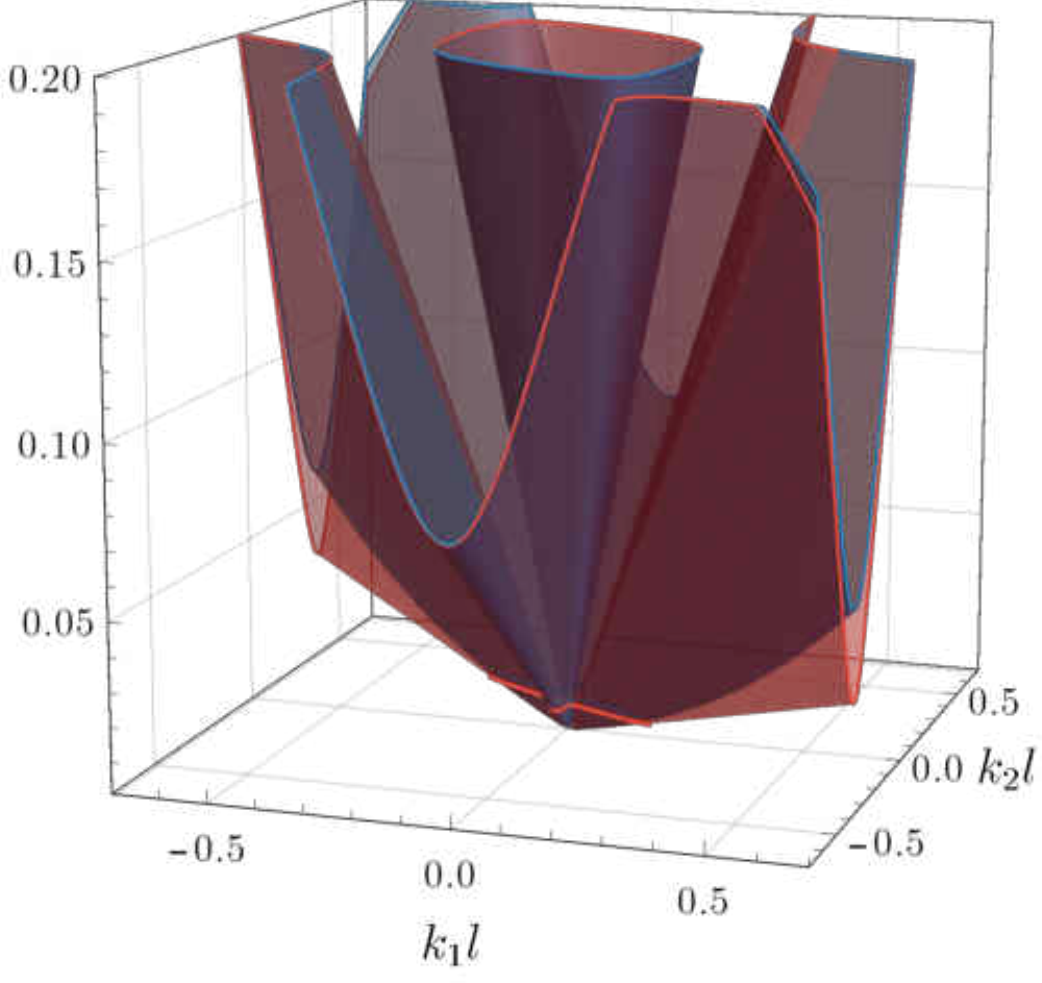}
    \end{subfigure}%
    \begin{subfigure}{0.245\textwidth}
        \centering
        \includegraphics[height=38mm]{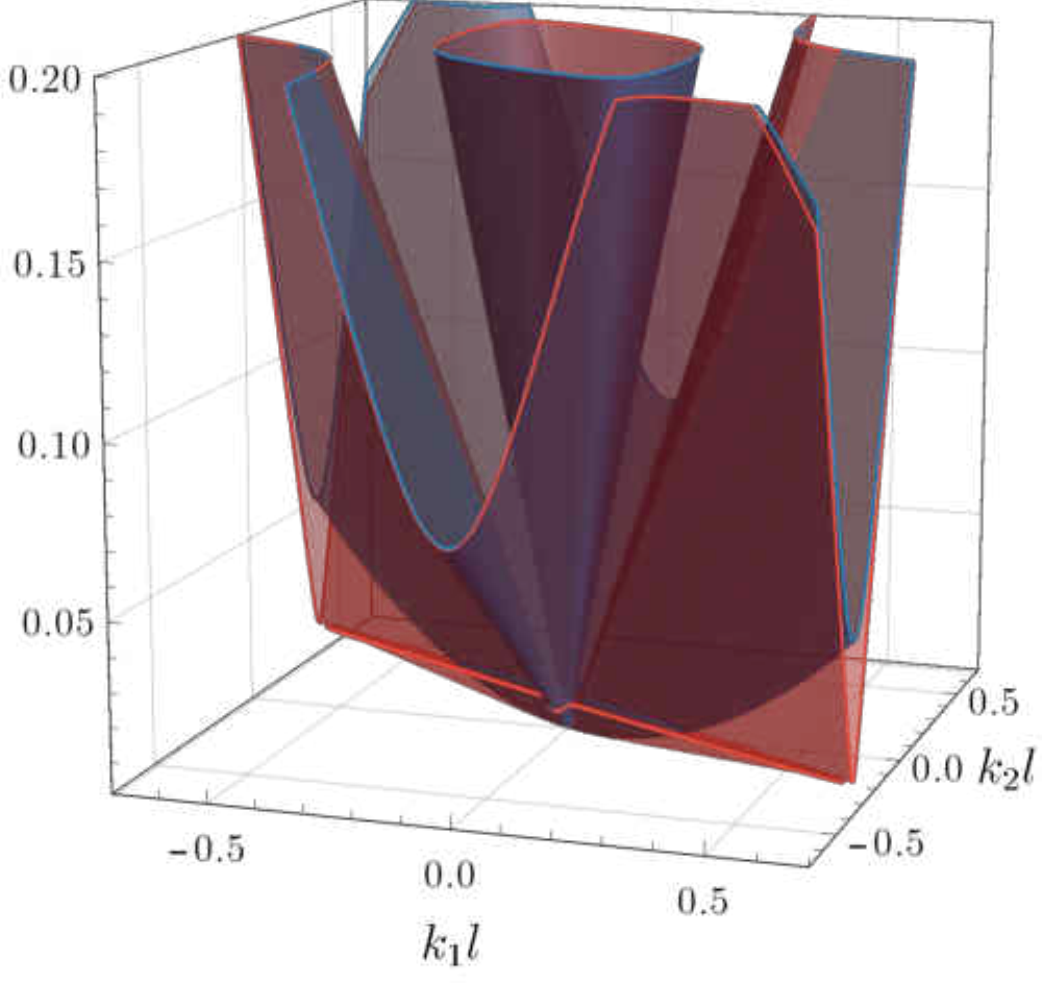}
    \end{subfigure}%
    \caption{\label{fig:contours_rhombus_7_15}
        Slowness contours and dispersion surfaces for a rhombic anisotropic lattice with $\Lambda_1=7,\,\Lambda_2=15$ (in blue) and for the effective elastic continuum (in red) at frequency $\Omega=0.01$.
        The evolution of the contours and dispersion surfaces induced by a compressive preload of equal components $p_1=p_2$ along two inclined directions (the considered loading path is depicted in Fig.~\ref{fig:ellipticity_domains_7_15}) demonstrates that the nonlinear dispersion of the lattice is negligible, except when the material is very close to the elliptic boundary, namely, for a prestress above $0.9\,\bp_{\text{E}}$.
        The comparison between the behaviors of the lattice and its equivalent continuum shows a great agreement.
    }
\end{figure}
This will provide validation to the homogenization scheme, showing that localization occurs both in the lattice and in the equivalent solid when a macroscopic bifurcation occurs.
To this purpose, the actual low-frequency forced response of the lattice (simulated numerically with a finite element technique) and the time-harmonic Green's function (which can be found in~\cite{piccolroaz_2006}) for the equivalent continuum are compared at increasing levels of preload, so that the elliptic boundary is approached.

The comparison is performed for the four geometric configurations reported in Table~\ref{tab:cases_analyzed} and for four prestress levels, namely $\{0,0.8,0.9,0.99\}\,\bp_{\text{E}}$ (with $\bp_{\text{E}}$ being the prestress state leading to ellipticity loss).
The lattice response is numerically analyzed using the COMSOL Multiphysics\textsuperscript{\circledR} finite element program in the frequency response mode.
A square finite-size computational window with a width of 350 unit cells (of dimension $350\,l$, with $l$ denoting the cell edge) is considered, with a perfectly matched layer (PML) along the boundaries, so that here waves are not reflected, rather absorbed, and the response of an infinite body is simulated.
The governing equation for the prestressed Euler-Bernoulli rod, Eq.~\eqref{eq:governing_beam_EB_v}, used in the finite element scheme has been implemented by modifying the bending moment contribution with an additional geometric term given by the load multiplied by the transverse displacement of the rod.
With regard to the computational mesh, the rod's length $l$ is discretized in 10 finite elements with cubic shape functions.
The selected mesh has been defined by testing three different mesh refinements, namely 5, 10, and 20 elements for $l$.
Eventually, the mesh with 10 elements has been selected, as 20 provided no significant improvement, but a substantial computational burden.

A pulsating concentrated force, applied in-plane, is considered acting at the center of the computational domain.
For a given load, the complex displacement field $\bu(\bx)$, with horizontal and vertical components $u_1(x_1,x_2)$ and $u_2(x_1,x_2)$, is computed and the results are plotted in terms of the modulus of the displacement associated to its real part only, $\delta_R(x_1,x_2) = \sqrt{(\Re u_1)^2+(\Re u_2)^2}$ (the plots of the imaginary part of the displacement is omitted for brevity).

In all the following analyses the frequency of the pulsating force is set to be $\Omega=\omega\,l\sqrt{\gamma/A}=0.01$, a low value providing a reasonable match, in term of acoustic properties, between the effective continuum and the lattice.
In fact, the mismatch between the two is different from zero for any non-vanishing frequency, although becomes zero in the limit $\Omega\to0$.
When the elliptic boundary is approached, this mismatch is expected to become wider for those waves which propagate parallel to the direction of ellipticity loss.
This is easily explained by the fact that, as the \textit{linear} term in the dispersion relation tends to vanish (in a direction $\bn_{\text{E}}$), the \textit{nonlinear} dispersion of the lattice becomes non-negligible at any \textit{non-vanishing} frequency.

By considering for instance the rhombic anisotropic grid ($\Lambda_1=7,\,\Lambda_2=15$), the deviation between the responses of the lattice (reported in blue in Fig.~\ref{fig:contours_rhombus_7_15}) and its equivalent continuum (reported in red in Fig.~\ref{fig:contours_rhombus_7_15}) can be visualized in terms of slowness contours computed at the frequency $\Omega=0.01$.
By comparing the contours for the four preload states, it can be appreciated that these are superimposed up $0.9\,\bp_{\text{E}}$, so that the nonlinear dispersion of the lattice becomes non-negligible only when the material is very close to the elliptic boundary, namely, at a preload $0.99\,\bp_{\text{E}}$, and only for waves close to the direction of ellipticity loss.
It is also worth noting that when $\bp\approx\bp_{\text{E}}$, the slowness contour pertinent to the lattice (reported in blue) is always contained inside the contour relative to the continuum (reported in red), so that the nonlinear dispersion implies that waves speeds are slightly \textit{higher} for the lattice than for the effective elastic medium.

\subsection{Square lattice}
\label{sec:square_lattice}
Cubic and orthotropic square grids are considered, subject to a pulsating diagonal force (inclined at $45^\circ$ with respect to the rods' axes), with the purpose of revealing the emergence of strain localizations.
\begin{figure}[htb!]
    \centering
    \begin{subfigure}{0.24\textwidth}
        \centering
        \phantomsubcaption{\label{fig:square_10_10_force_d_0}}
        \includegraphics[width=0.98\linewidth]{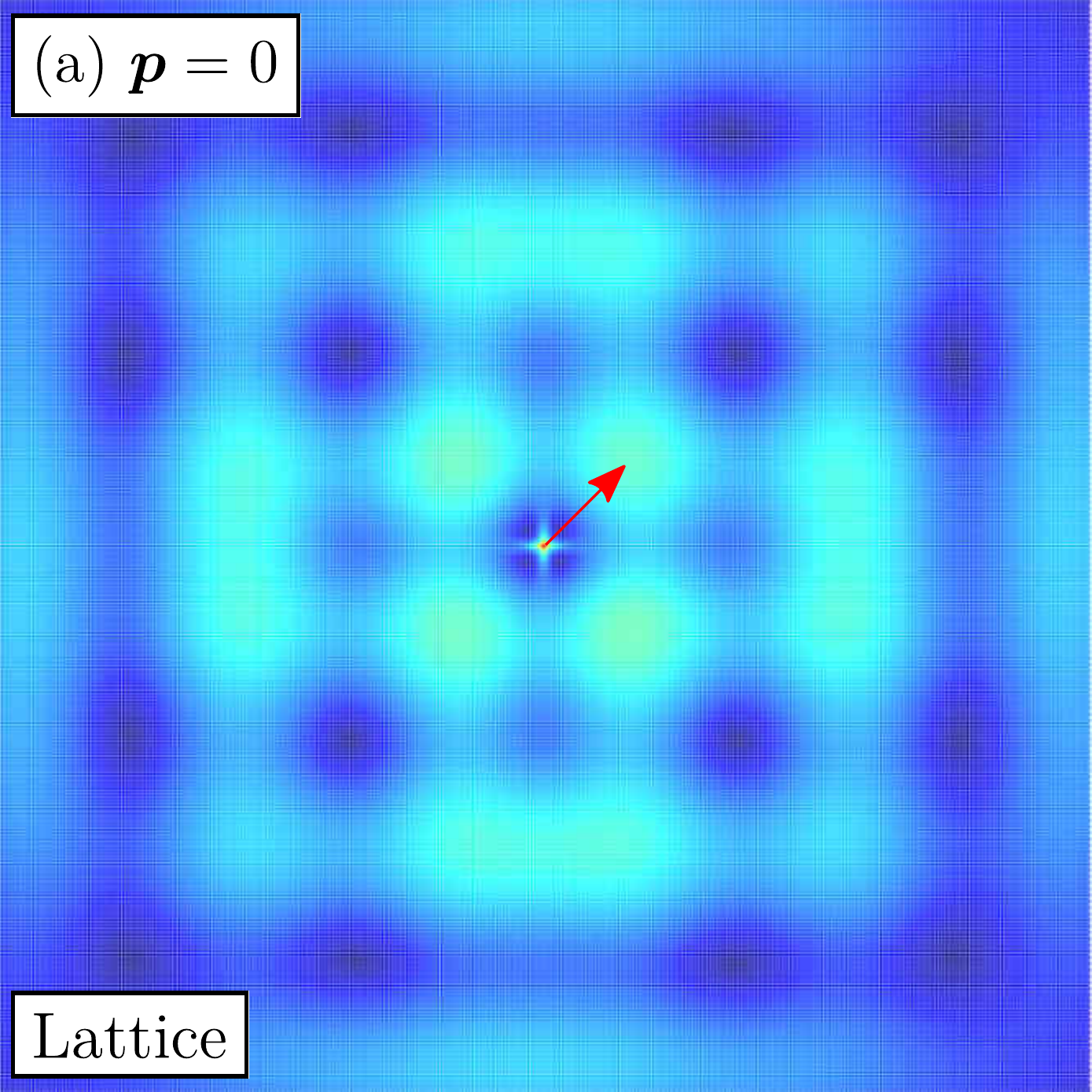}
    \end{subfigure}
    \begin{subfigure}{0.24\textwidth}
        \centering
        \phantomsubcaption{\label{fig:square_10_10_force_d_80}}
        \includegraphics[width=0.98\linewidth]{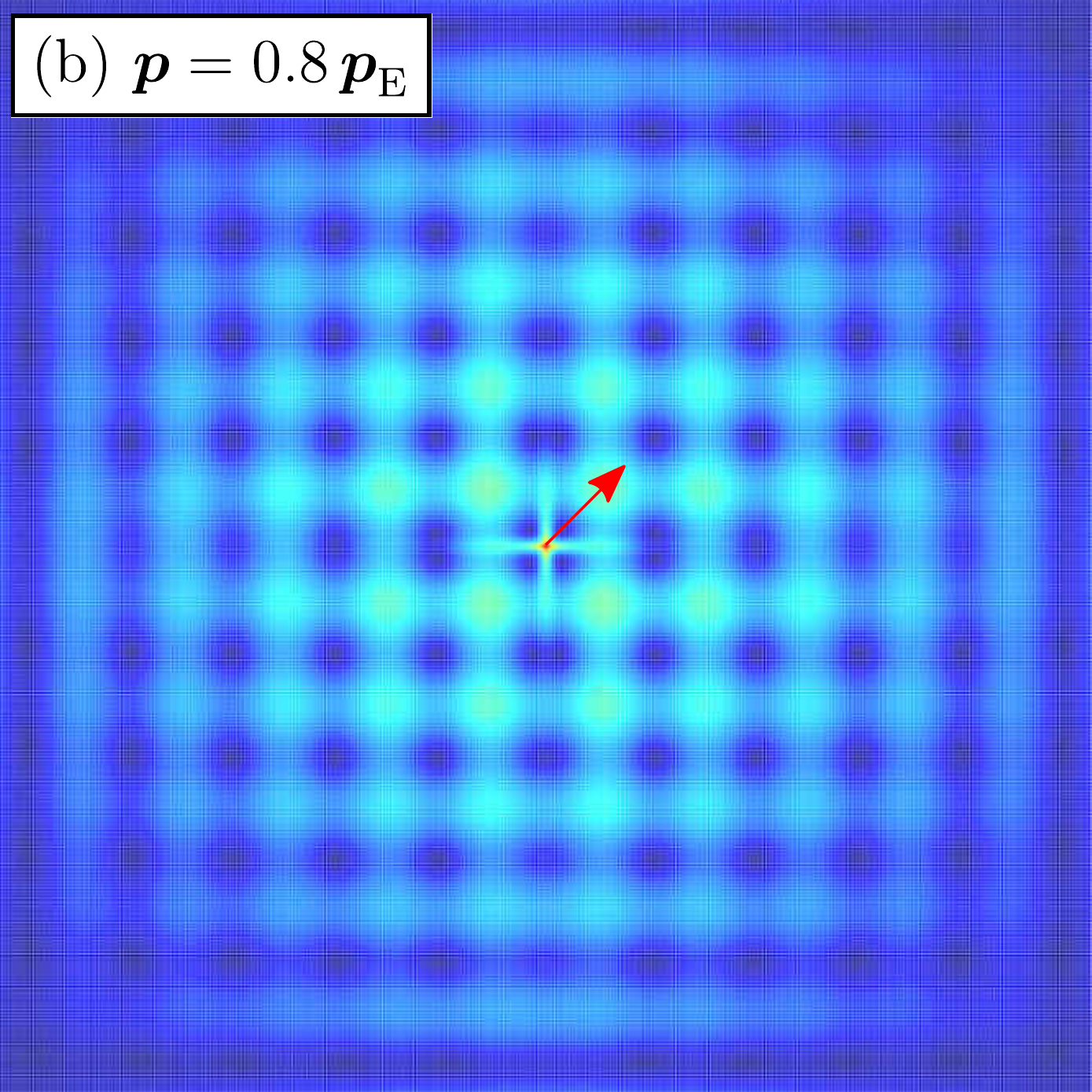}
    \end{subfigure}
    \begin{subfigure}{0.24\textwidth}
        \centering
        \phantomsubcaption{\label{fig:square_10_10_force_d_90}}
        \includegraphics[width=0.98\linewidth]{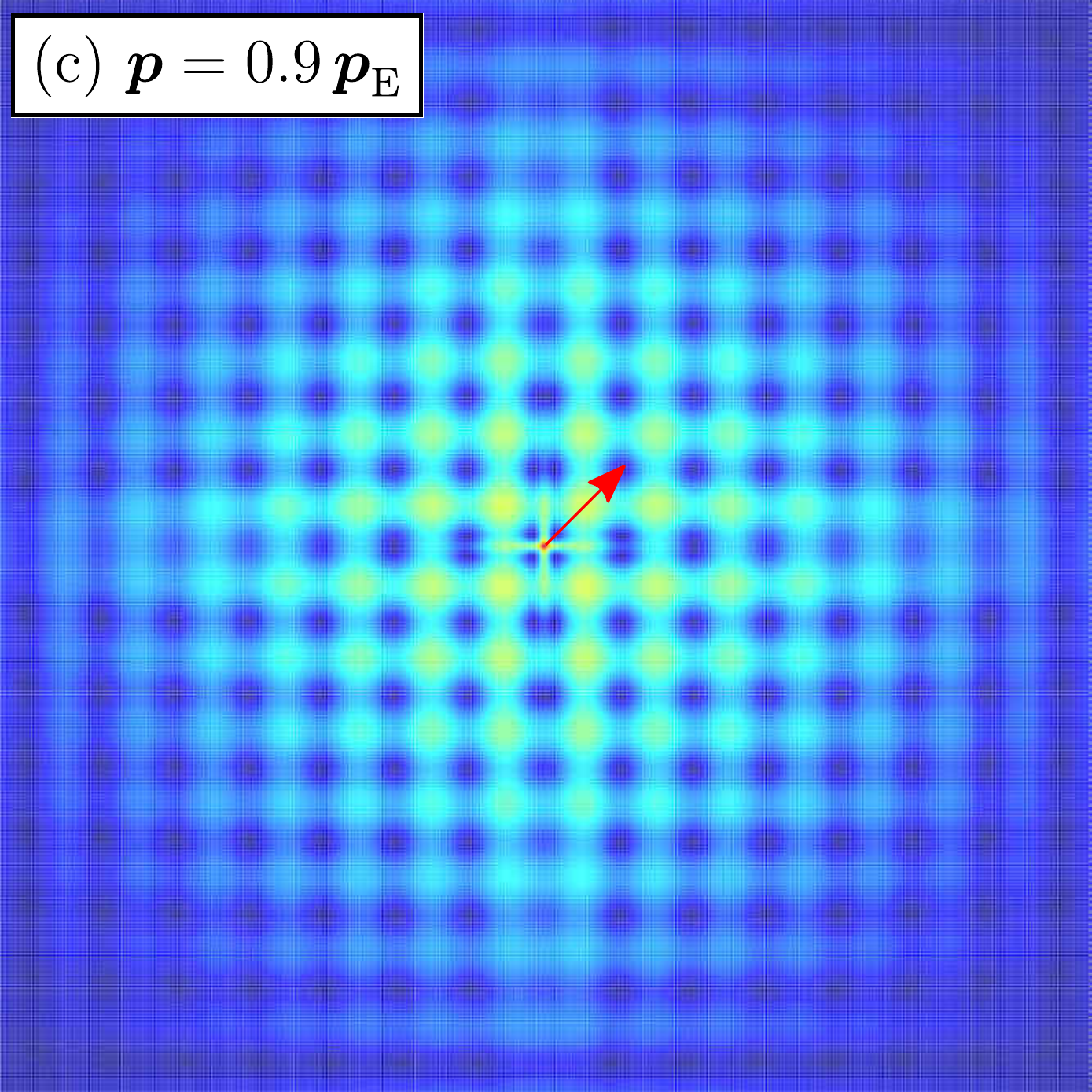}
    \end{subfigure}
    \begin{subfigure}{0.24\textwidth}
        \centering
        \phantomsubcaption{\label{fig:square_10_10_force_d_99}}
        \includegraphics[width=0.98\linewidth]{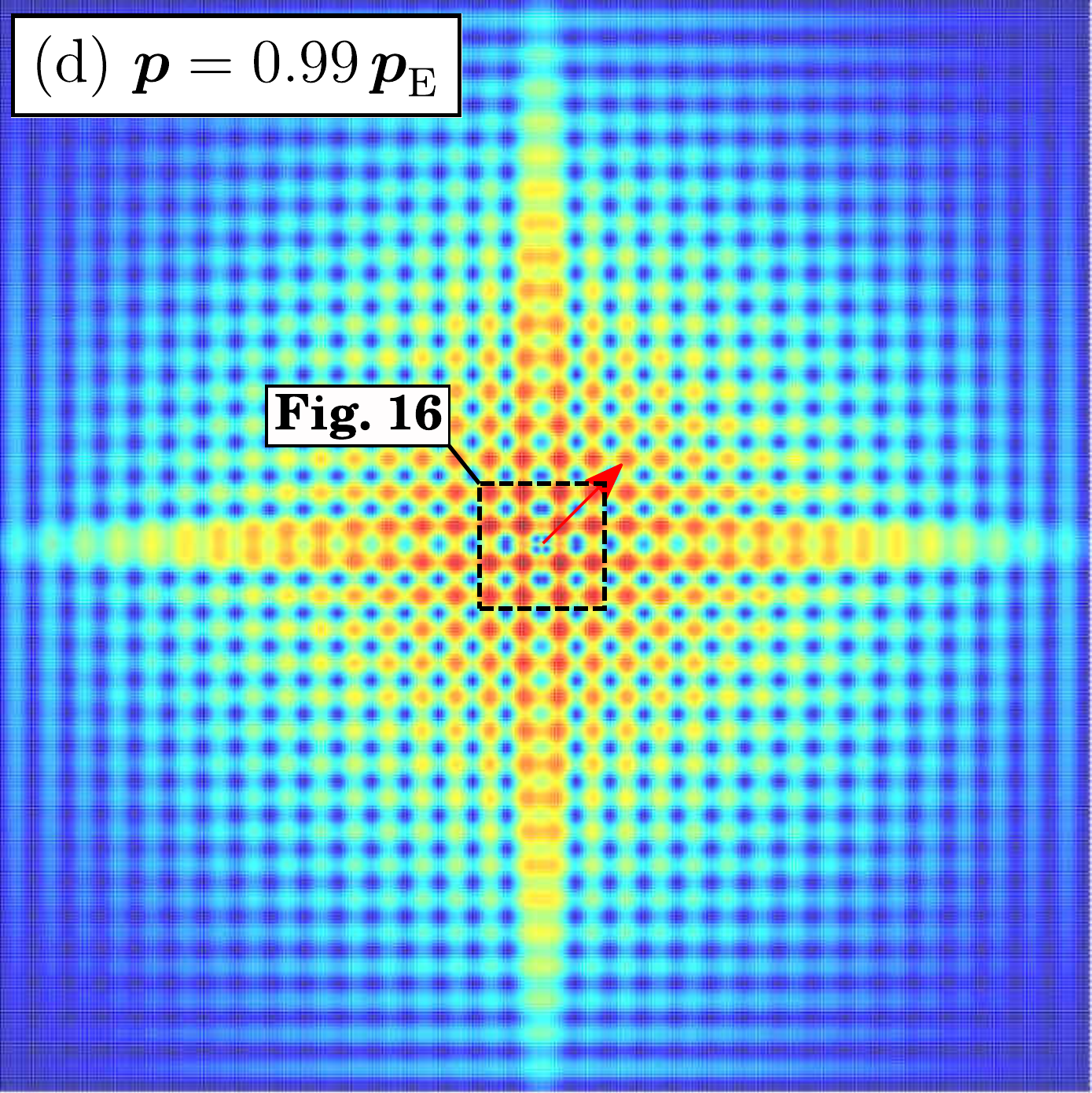}
    \end{subfigure}\\
    \vspace{0.01\linewidth}
    \centering
    \begin{subfigure}{0.24\textwidth}
        \centering
        \phantomsubcaption{\label{fig:square_10_10_force_d_0_gf}}
        \includegraphics[width=0.98\linewidth]{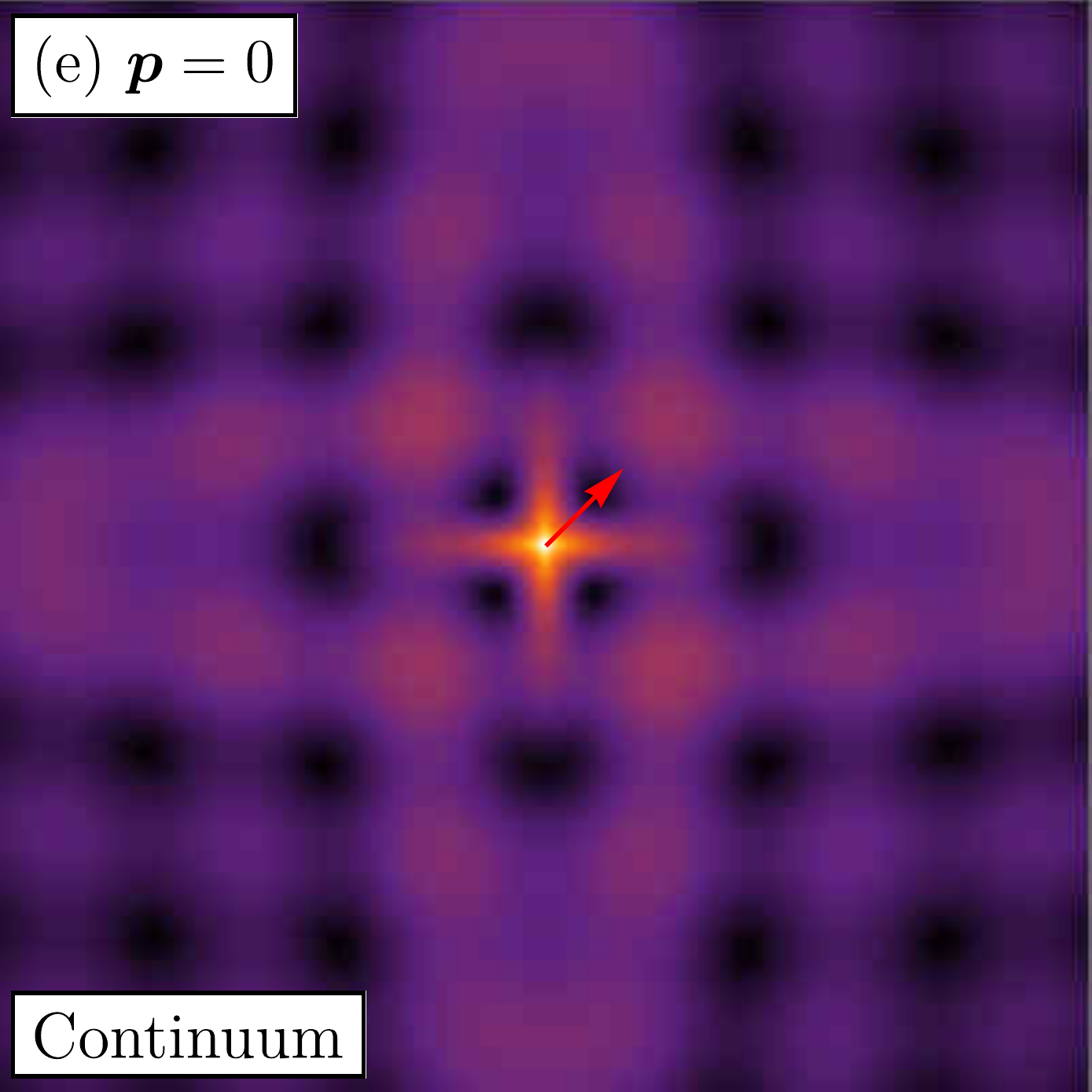}
    \end{subfigure}
    \begin{subfigure}{0.24\textwidth}
        \centering
        \phantomsubcaption{\label{fig:square_10_10_force_d_80_gf}}
        \includegraphics[width=0.98\linewidth]{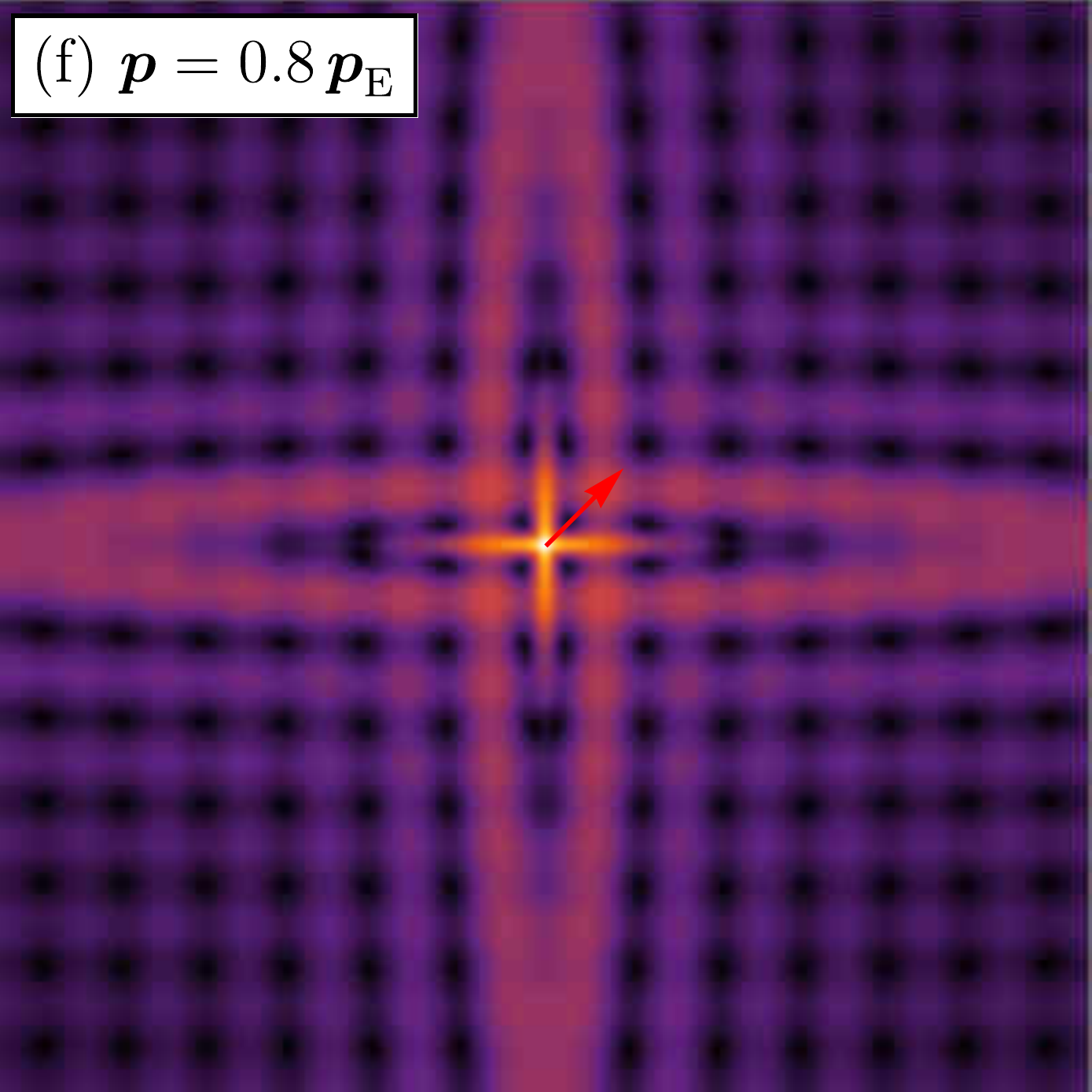}
    \end{subfigure}
    \begin{subfigure}{0.24\textwidth}
        \centering
        \phantomsubcaption{\label{fig:square_10_10_force_d_90_gf}}
        \includegraphics[width=0.98\linewidth]{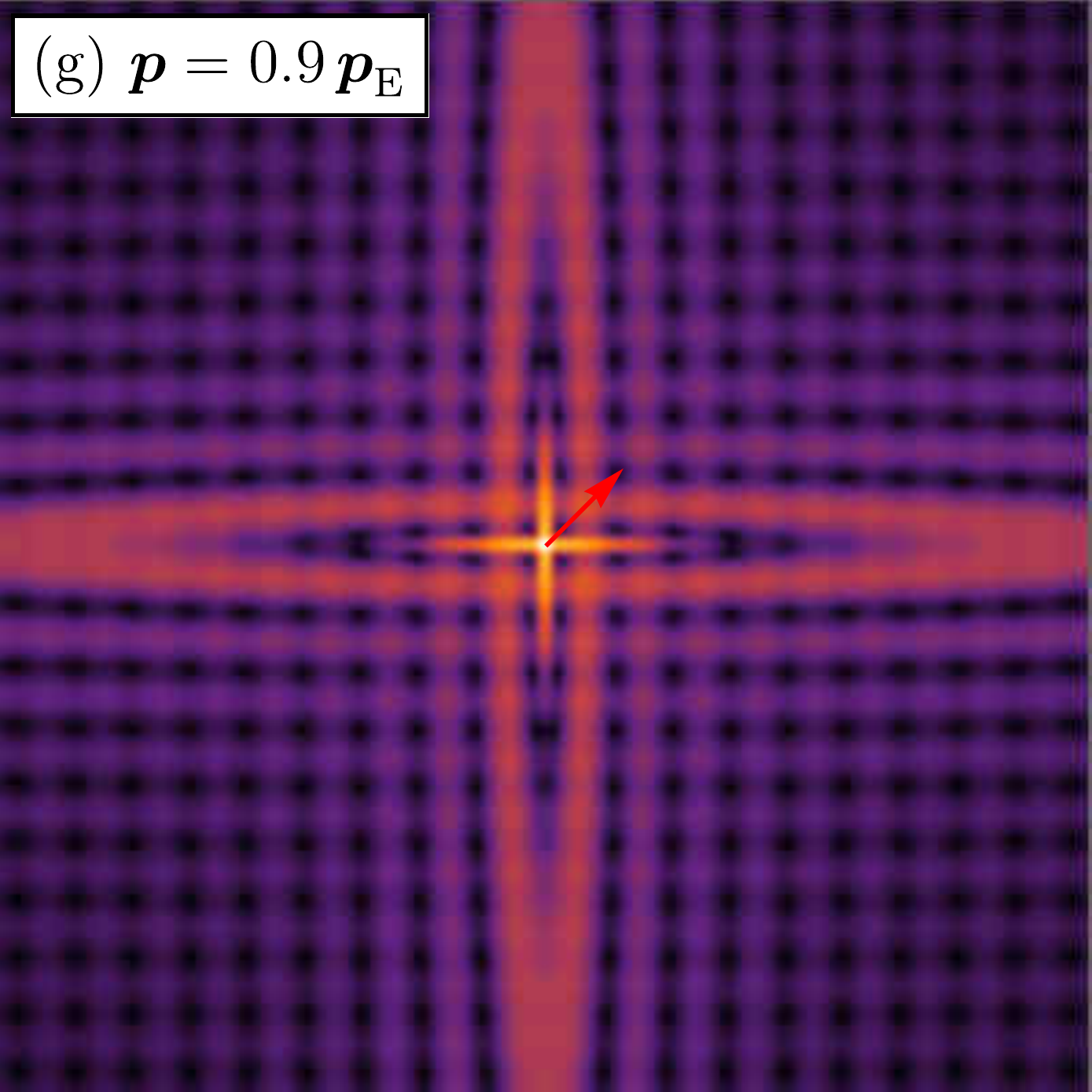}
    \end{subfigure}
    \begin{subfigure}{0.24\textwidth}
        \centering
        \phantomsubcaption{\label{fig:square_10_10_force_d_99_gf}}
        \includegraphics[width=0.98\linewidth]{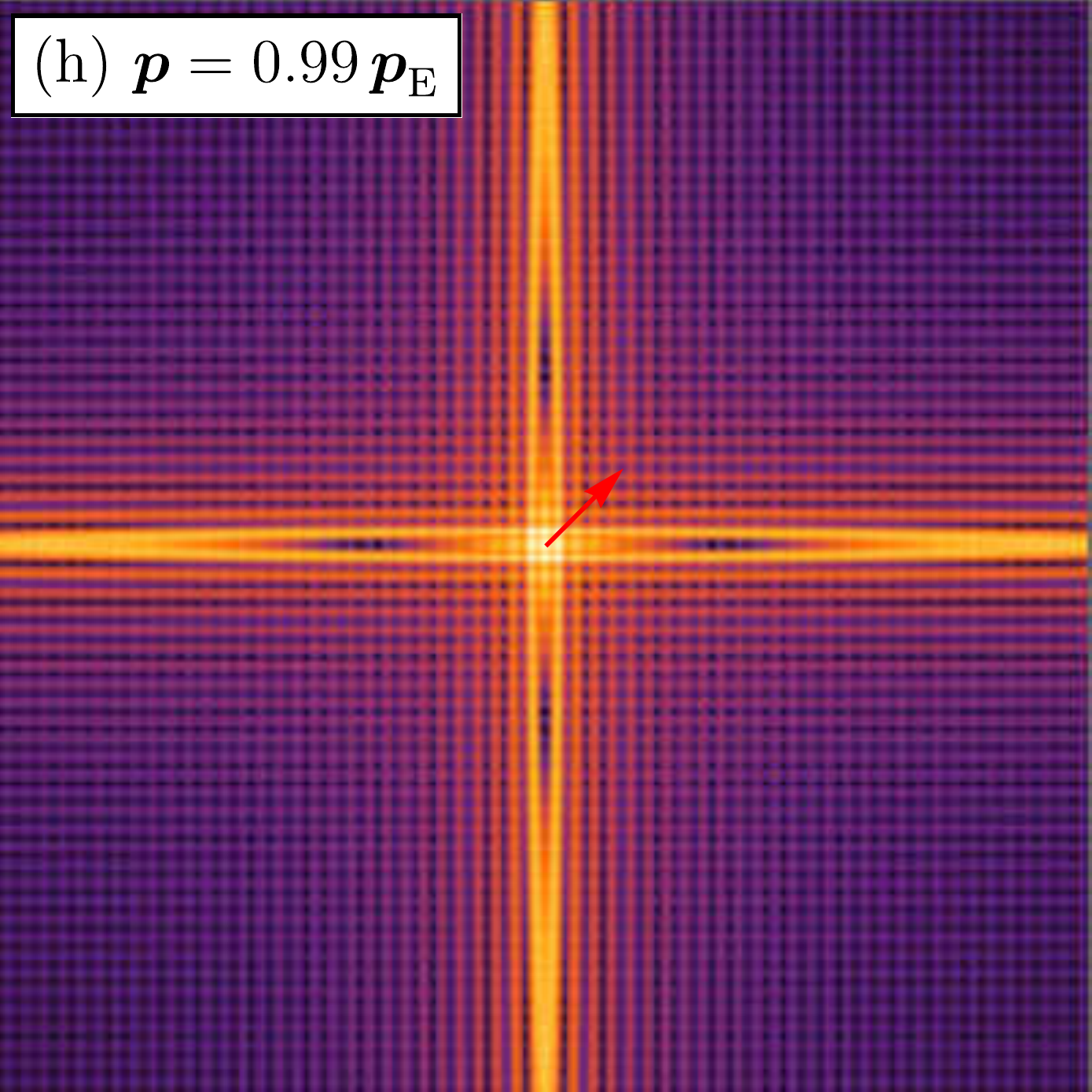}
    \end{subfigure}
    \caption{\label{fig:square_10_10_force}
        The displacement field generated by a pulsating diagonal force (denoted with a red arrow and applied to the square lattice with \textit{cubic} symmetry, $\Lambda_1=\Lambda_2=10$) is simulated via f.e.m., see (\subref{fig:square_10_10_force_d_0})--(\subref{fig:square_10_10_force_d_99}), and compared to the response of the homogenized continuum, see (\subref{fig:square_10_10_force_d_0_gf})--(\subref{fig:square_10_10_force_d_99_gf}), at different levels of prestress $\bp$.
        Note the emergence of two orthogonal shear bands, aligned parallel to the directions predicted for failure of ellipticity, see Fig.~\ref{fig:eigenvalue_square_10_10}.
    }
\end{figure}
\begin{figure}[htb!]
    \centering
    \begin{subfigure}{0.24\textwidth}
        \centering
        \caption*{$t=0$}
        \includegraphics[width=0.98\linewidth]{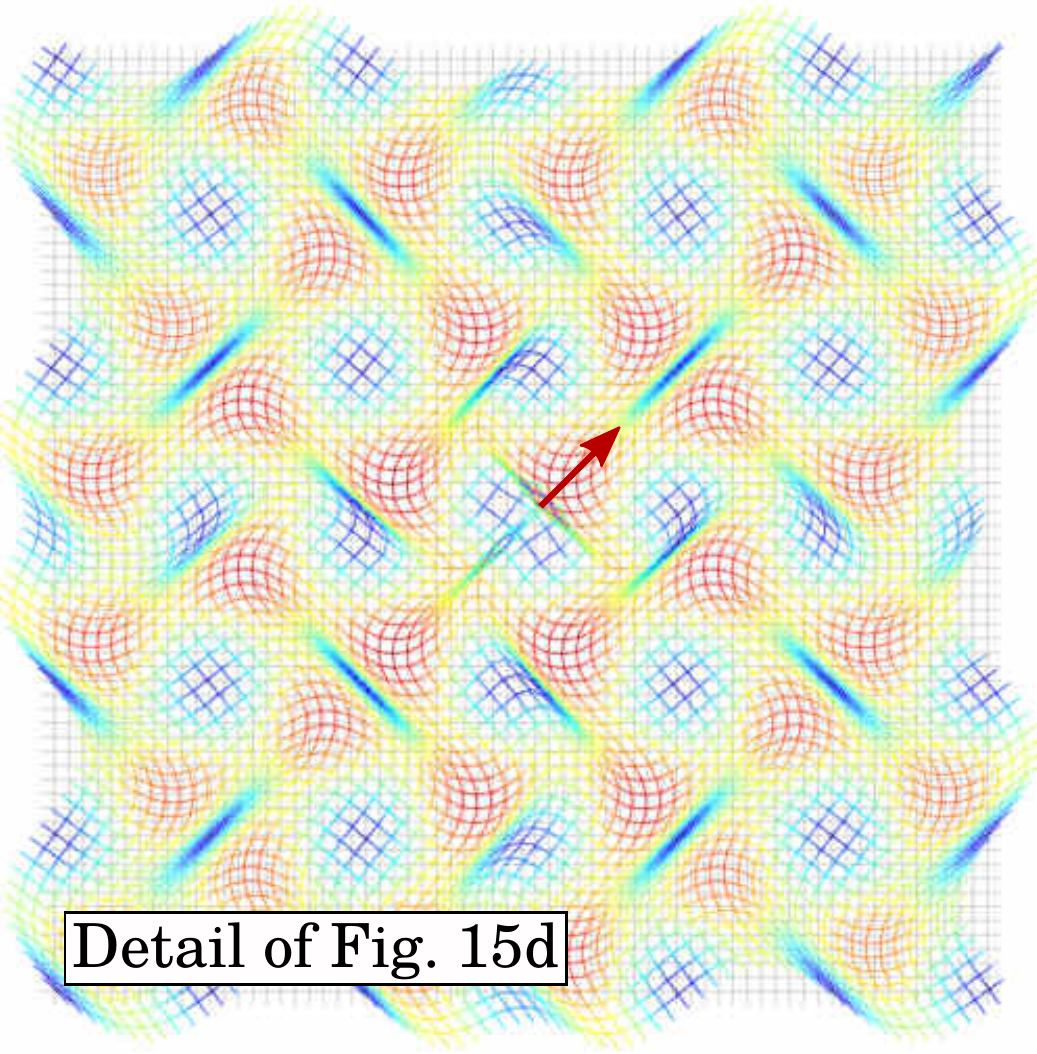}
    \end{subfigure}
    \begin{subfigure}{0.24\textwidth}
        \centering
        \caption*{$t=\frac{\pi}{2\omega}$}
        \includegraphics[width=0.98\linewidth]{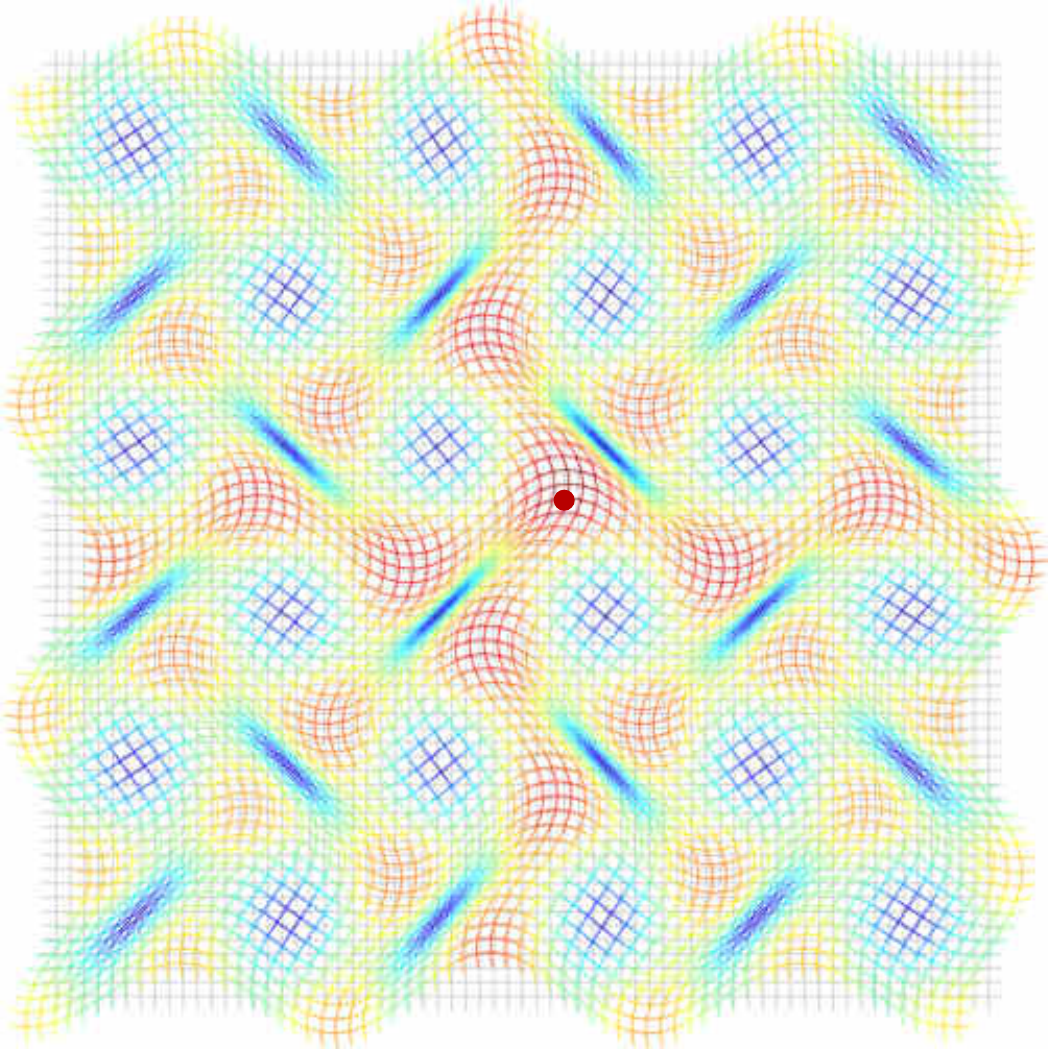}
    \end{subfigure}
    \begin{subfigure}{0.24\textwidth}
        \centering
        \caption*{$t=\frac{\pi}{\omega}$}
        \includegraphics[width=0.98\linewidth]{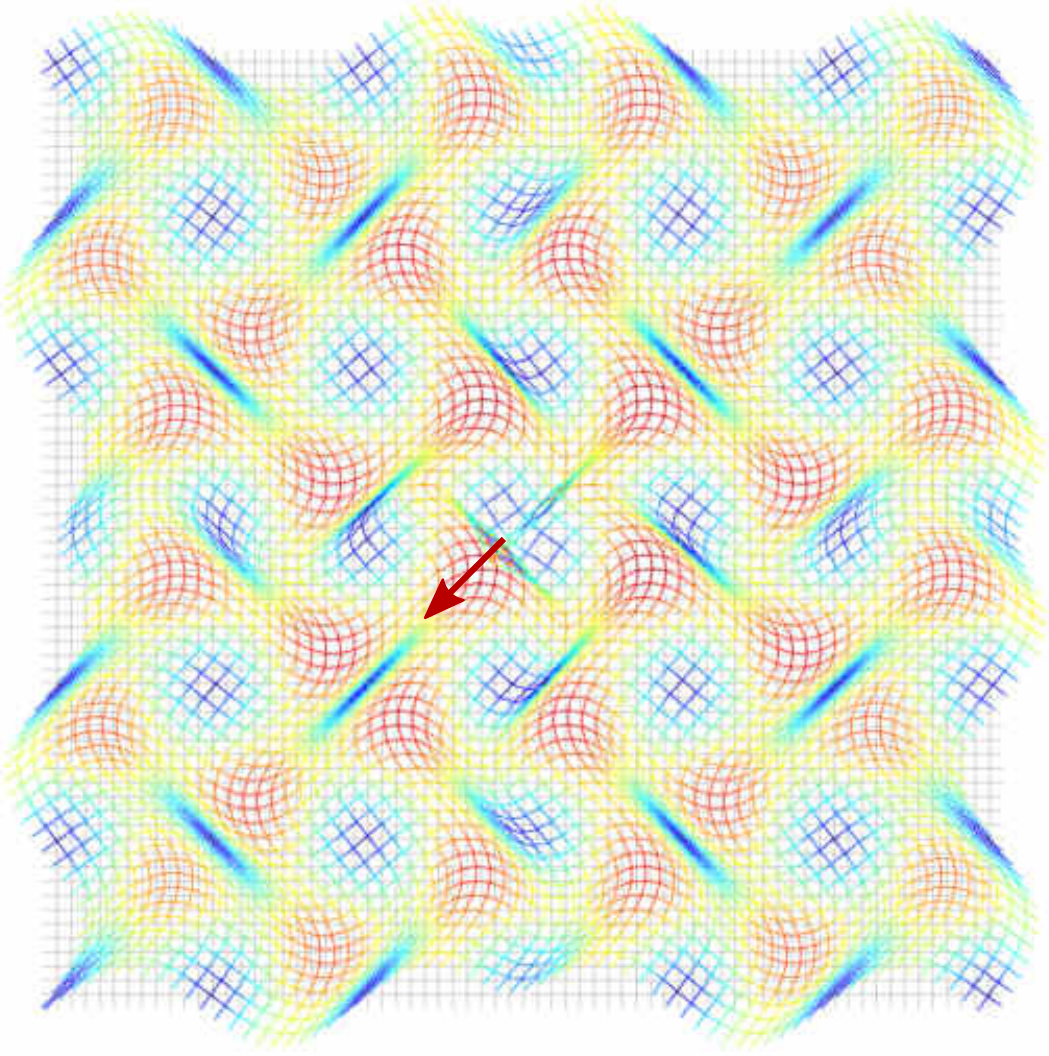}
    \end{subfigure}
    \begin{subfigure}{0.24\textwidth}
        \centering
        \caption*{$t=\frac{3\pi}{2\omega}$}
        \includegraphics[width=0.98\linewidth]{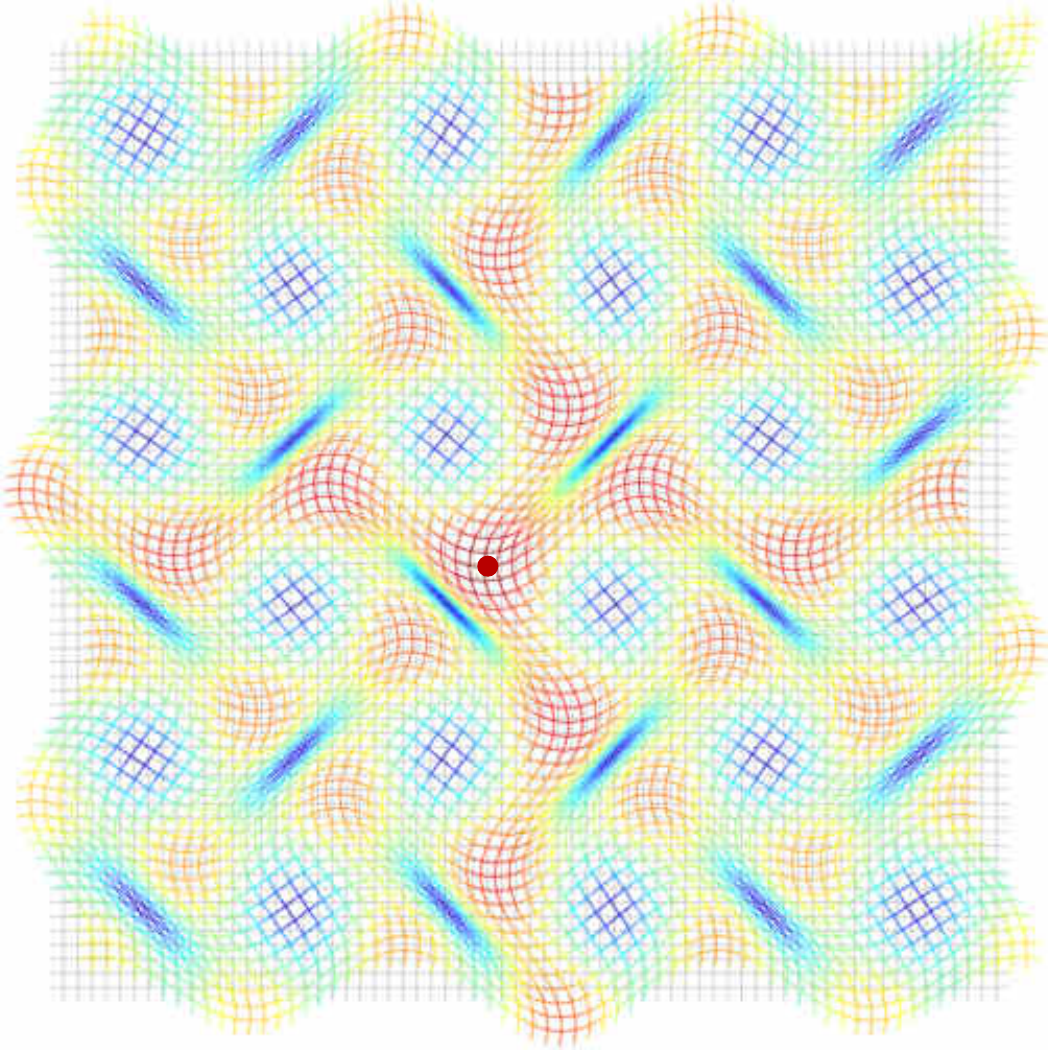}
    \end{subfigure}
    \caption{\label{fig:square_10_10_deformed}
        Deformed configurations of a square lattice with \textit{cubic} symmetry ($\Lambda_1=\Lambda_2=10$) near the point of application of a pulsating diagonal force (denoted with a red arrow) at a level of prestress close to the elliptic boundary ($\bp=0.99\,\bp_{\text{E}}$).
        The figure reports the zoomed view of the region shown in Fig.~\ref{fig:square_10_10_force_d_99} and depicts four snapshots of the forced response of the lattice at different instants of time.
        The pattern shows a motion resulting from the superposition of two shear localizations induced by the dynamic loading.
    }
\end{figure}

The case of cubic symmetry ($\Lambda_1=\Lambda_2=10$) is analyzed in Fig.~\ref{fig:square_10_10_force}, where the displacement field, numerically computed for the square grid (subject to a pulsating concentrated force, upper row), is compared to the response of the homogenized continuum (subject to the same concentrated force, solved via Green's function, lower row), for four values of prestress (increasing from left to right, $p_1=p_2=\{0,-3.347,-4.891,-5.380\}$).
As the elliptic boundary is approached, the emergence of two strain localizations becomes evident and confirms the predicted vanishing of an eigenvalue (wave speed) reported in Fig.~\ref{fig:eigenvalue_square_10_10}.
Snapshots of the displacement map at different instants of time, during the dynamic response of the grid and near the point of application of the pulsating force, reveal the actual localization mode activated by the applied diagonal force.
In fact, deformed configurations calculated in the grid (the zone is indicated in Fig.~\ref{fig:square_10_10_force_d_99}) through a finite element simulation and plotted in Fig.~\ref{fig:square_10_10_deformed} display a characteristic motion resulting from the superposition of two shear localizations emanating from the loading point.
\begin{figure}[htb!]
    \centering
    \begin{subfigure}{0.24\textwidth}
        \centering
        \phantomsubcaption{\label{fig:square_7_15_force_d_0}}
        \includegraphics[width=0.98\linewidth]{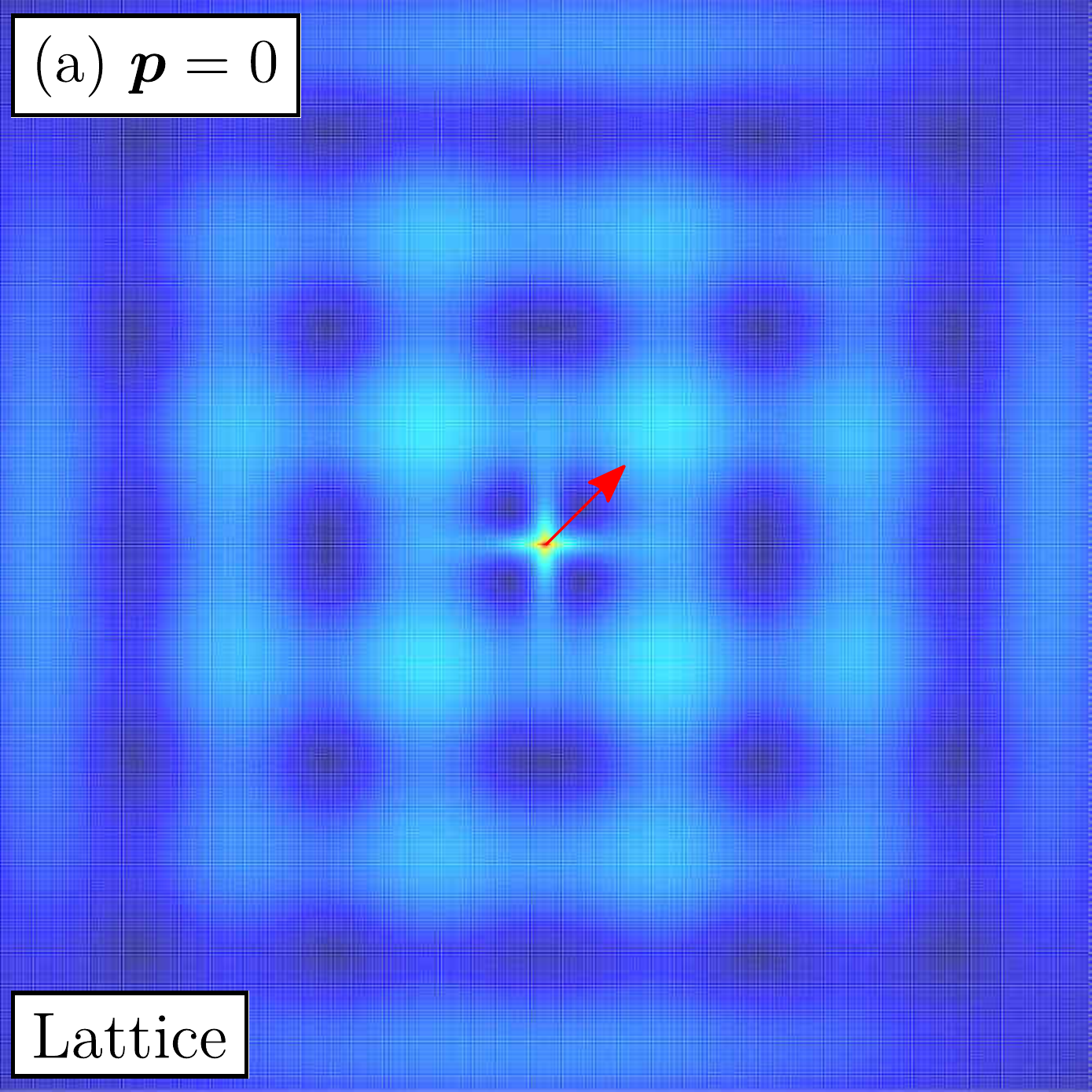}
    \end{subfigure}
    \begin{subfigure}{0.24\textwidth}
        \centering
        \phantomsubcaption{\label{fig:square_7_15_force_d_80}}
        \includegraphics[width=0.98\linewidth]{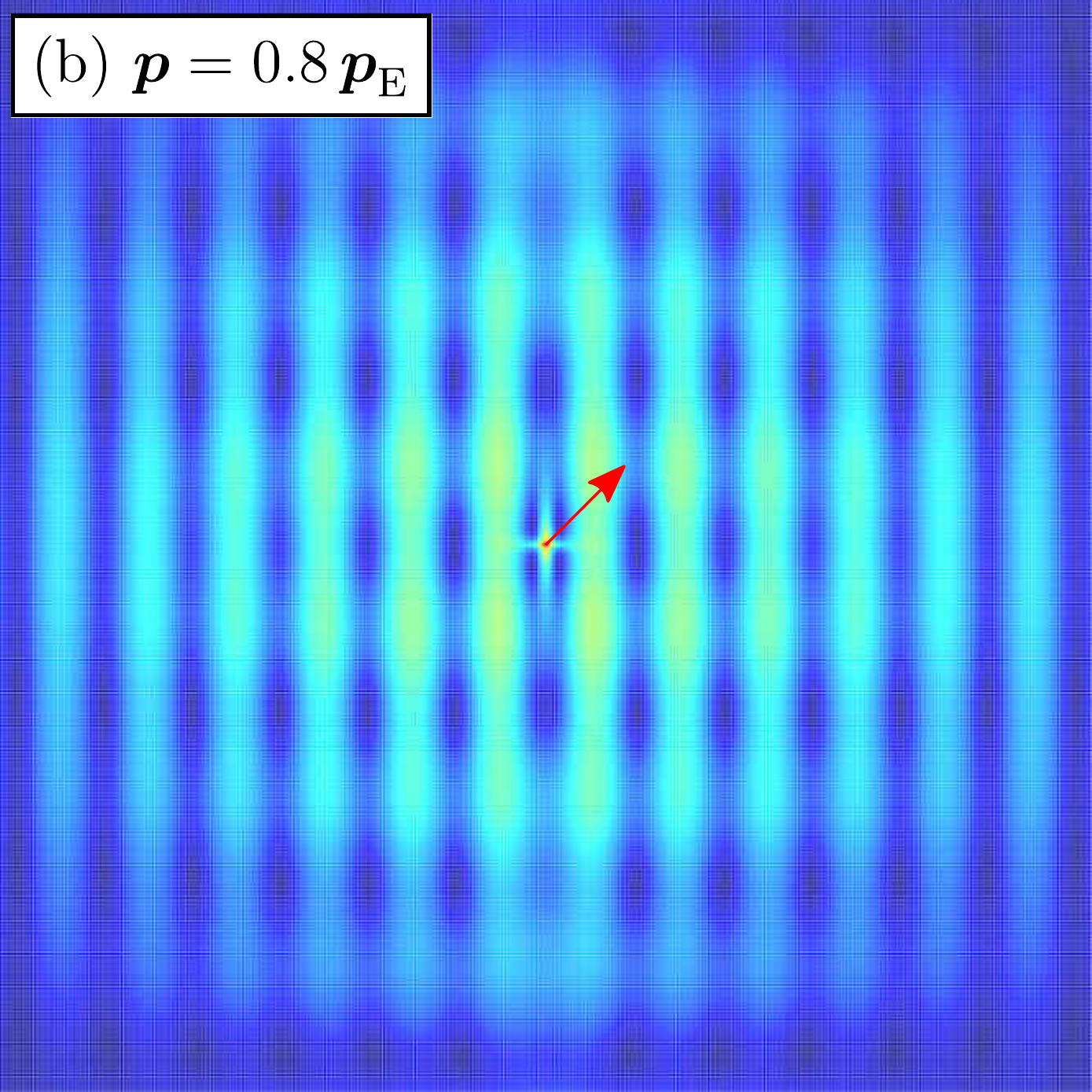}
    \end{subfigure}
    \begin{subfigure}{0.24\textwidth}
        \centering
        \phantomsubcaption{\label{fig:square_7_15_force_d_90}}
        \includegraphics[width=0.98\linewidth]{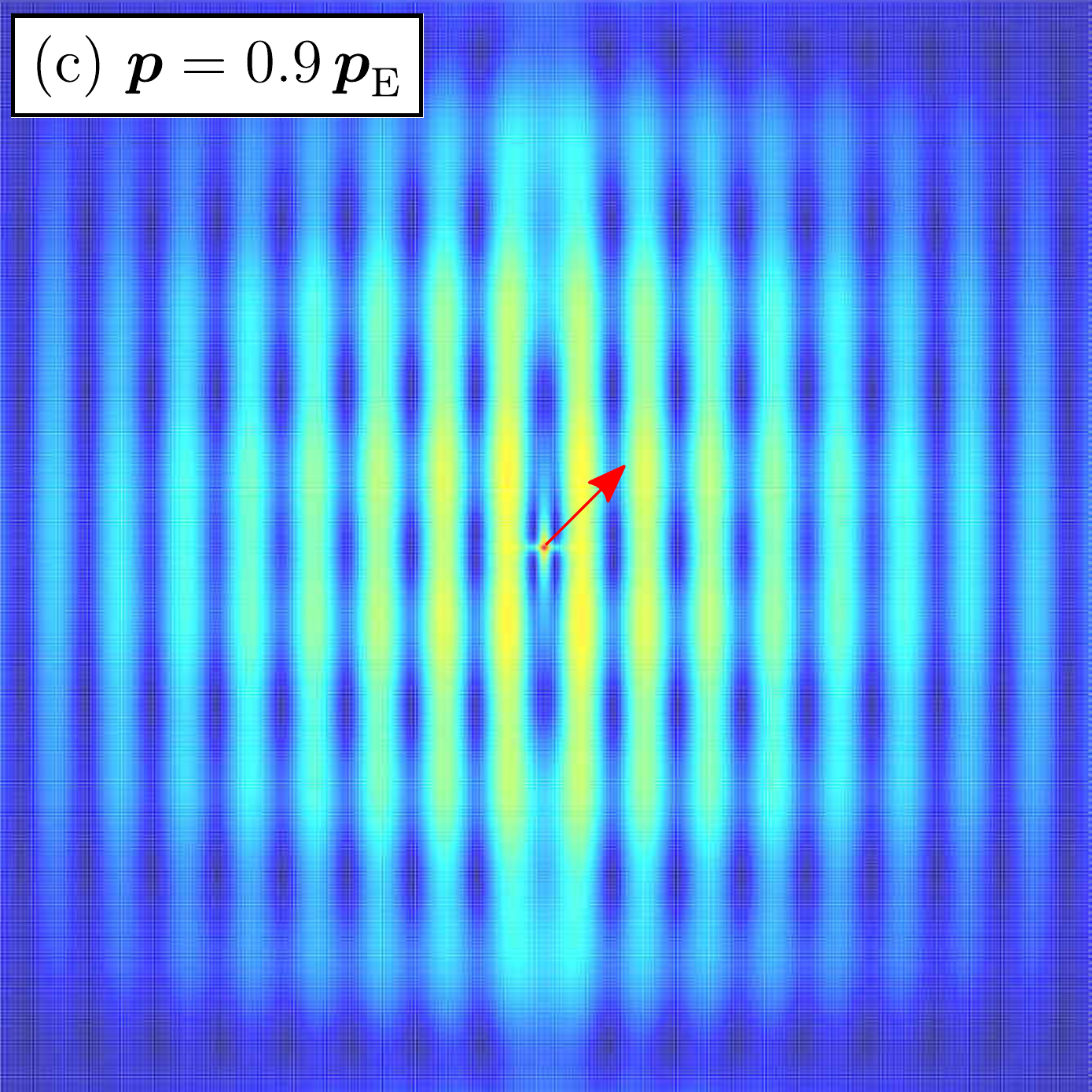}
    \end{subfigure}
    \begin{subfigure}{0.24\textwidth}
        \centering
        \phantomsubcaption{\label{fig:square_7_15_force_d_99}}
        \includegraphics[width=0.98\linewidth]{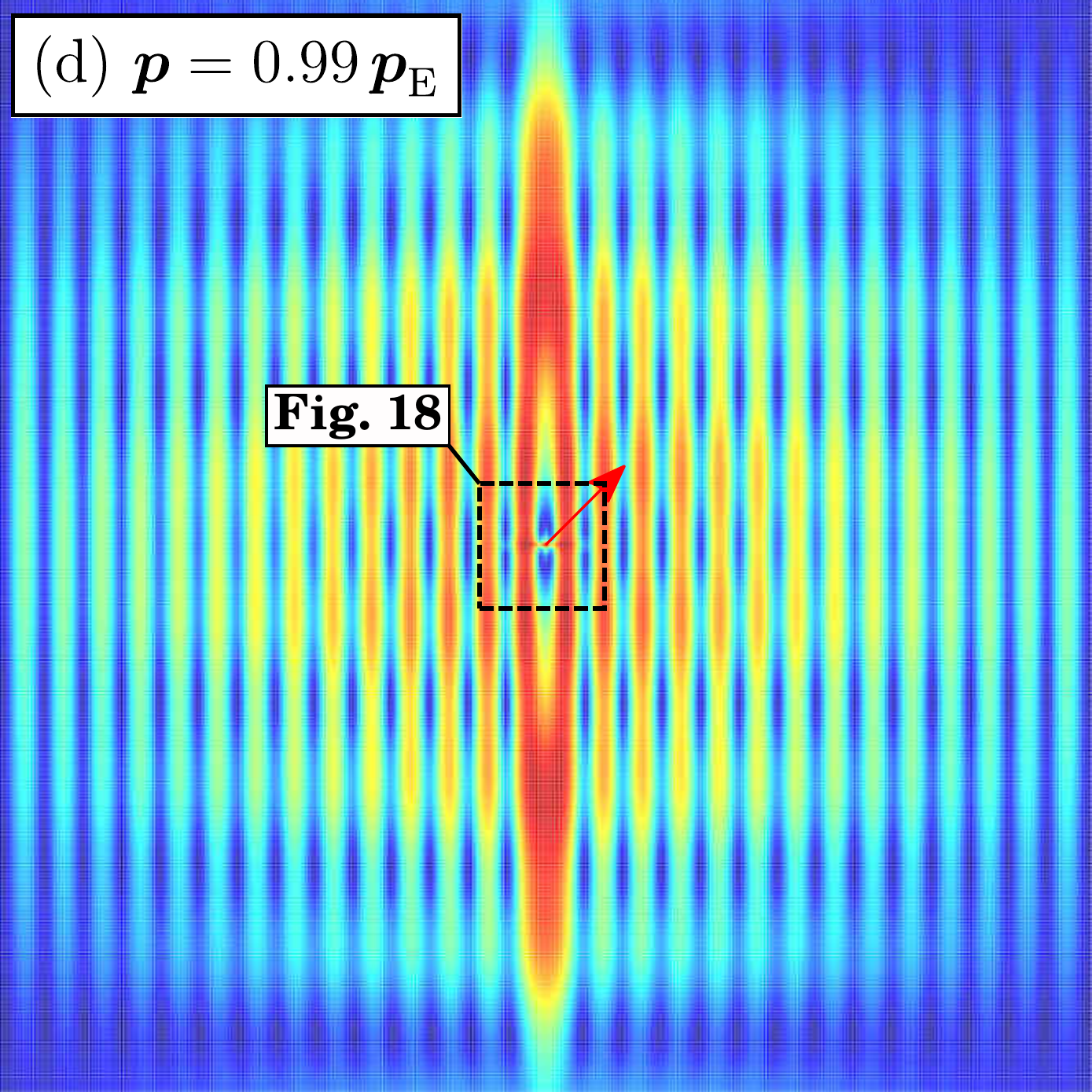}
    \end{subfigure}\\
    \vspace{0.01\linewidth}
    \begin{subfigure}{0.24\textwidth}
        \centering
        \phantomsubcaption{\label{fig:square_7_15_force_d_0_gf}}
        \includegraphics[width=0.98\linewidth]{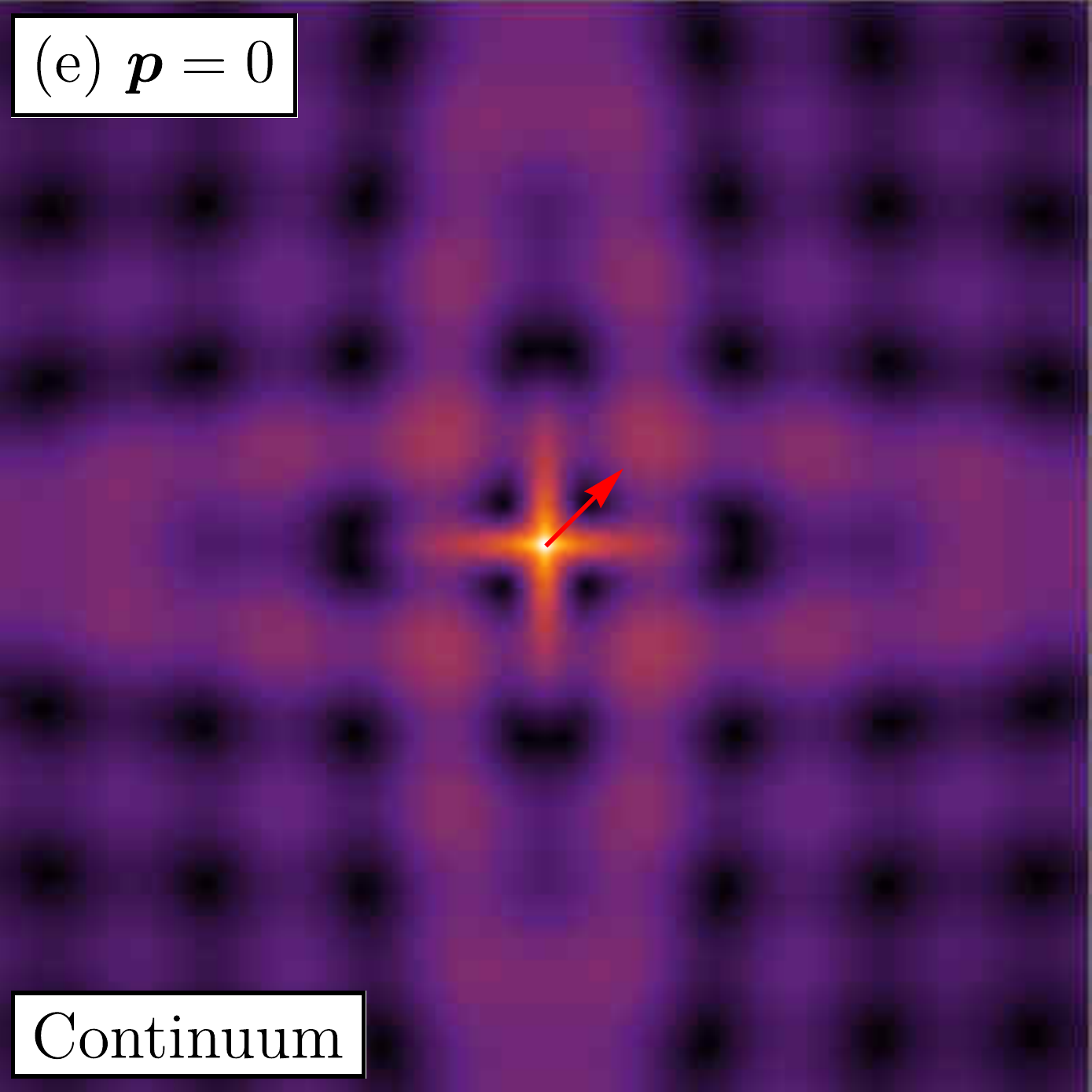}
    \end{subfigure}
    \begin{subfigure}{0.24\textwidth}
        \centering
        \phantomsubcaption{\label{fig:square_7_15_force_d_80_gf}}
        \includegraphics[width=0.98\linewidth]{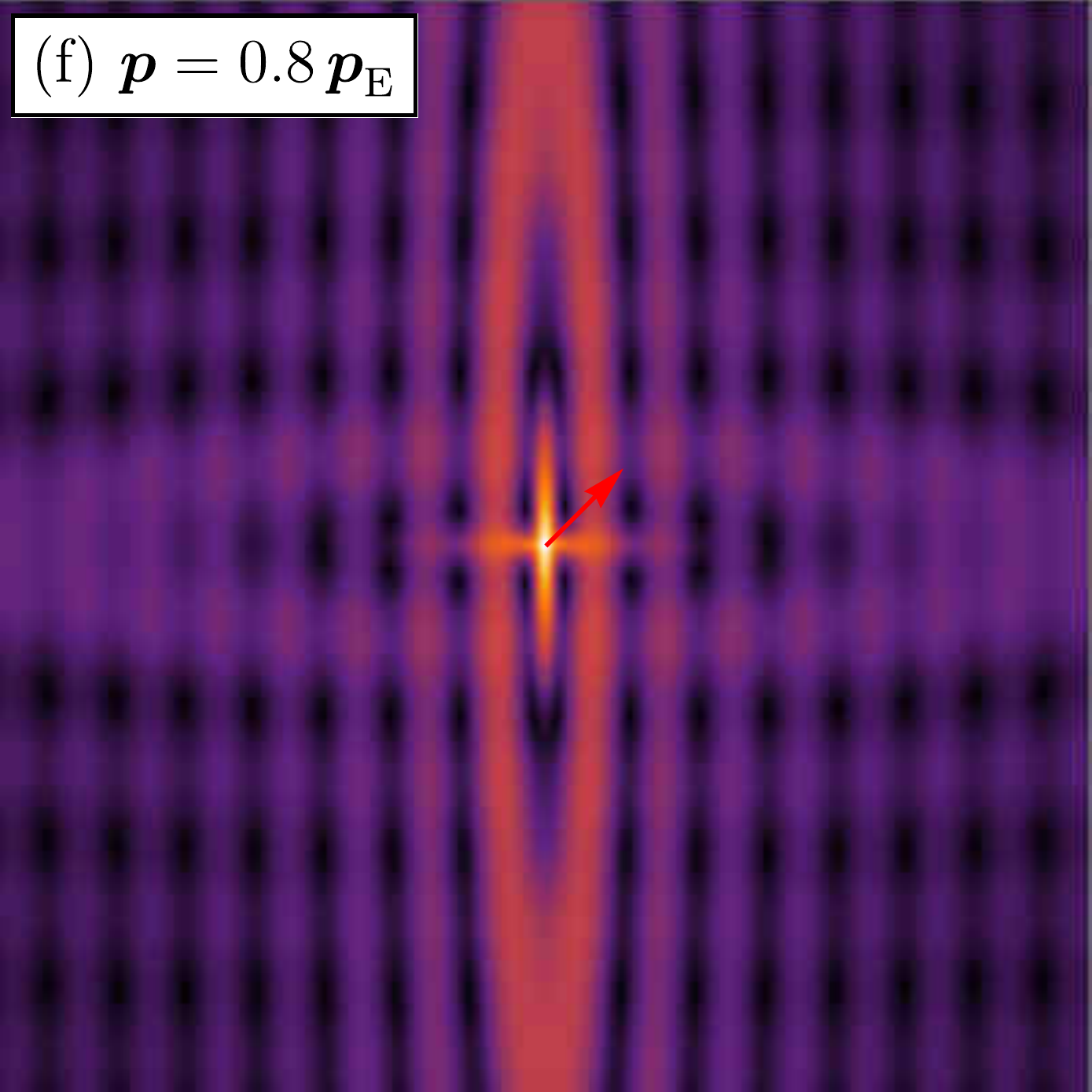}
    \end{subfigure}
    \begin{subfigure}{0.24\textwidth}
        \centering
        \phantomsubcaption{\label{fig:square_7_15_force_d_90_gf}}
        \includegraphics[width=0.98\linewidth]{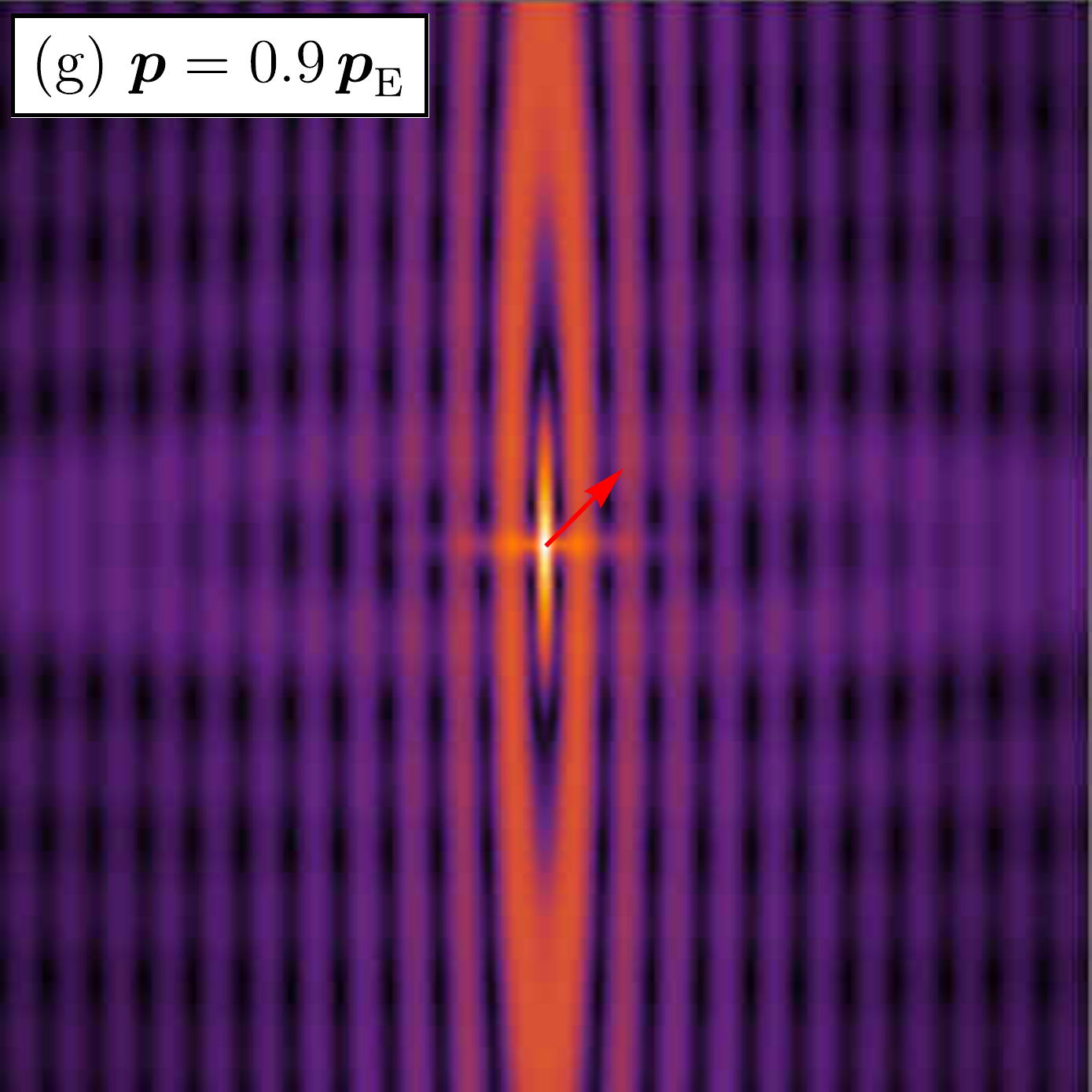}
    \end{subfigure}
    \begin{subfigure}{0.24\textwidth}
        \centering
        \phantomsubcaption{\label{fig:square_7_15_force_d_99_gf}}
        \includegraphics[width=0.98\linewidth]{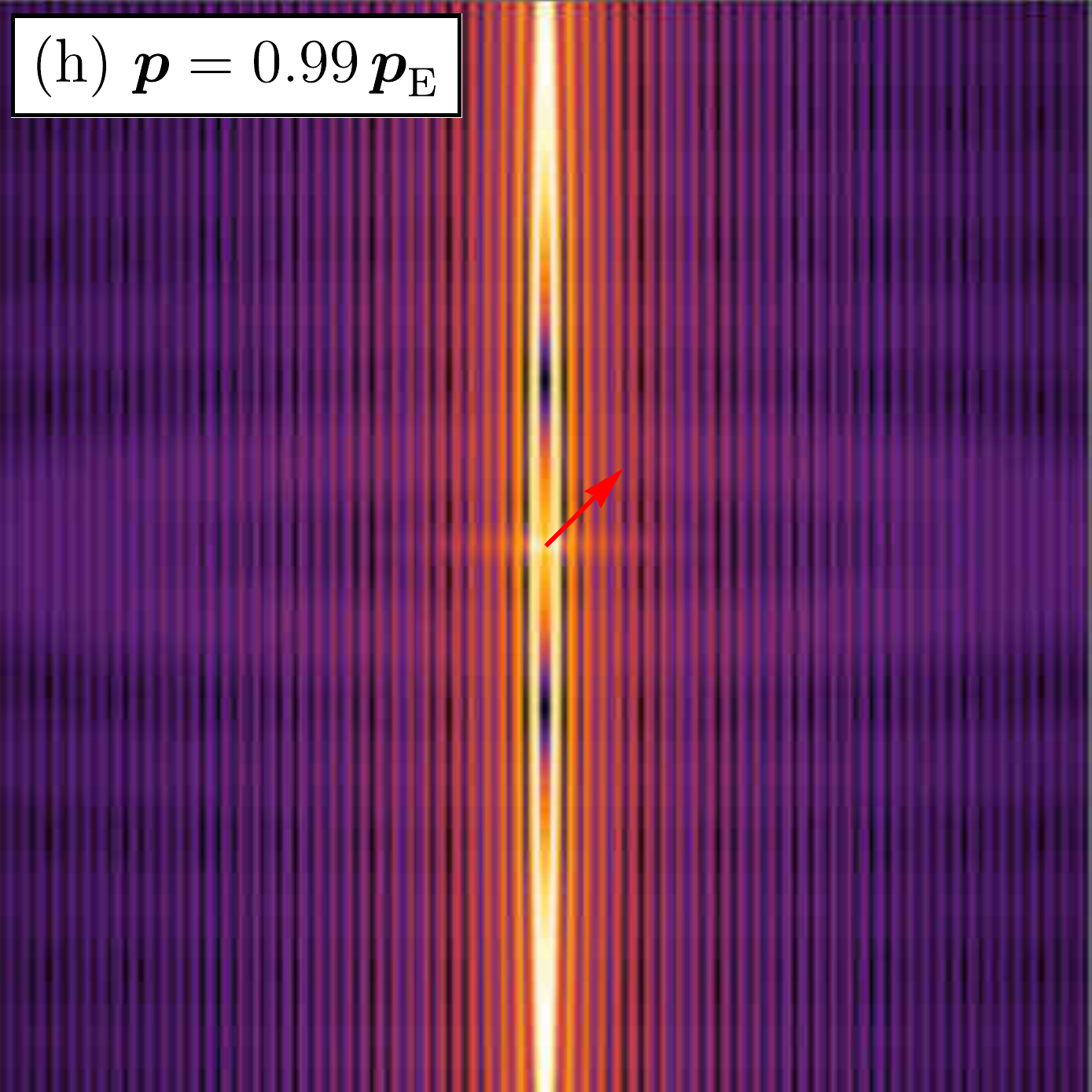}
    \end{subfigure}
    \caption{\label{fig:square_7_15_force}
        As for Fig.~\ref{fig:square_10_10_force}, but for an \textit{orthotropic} square lattice ($\Lambda_1=7\,, \Lambda_2=15$).
        Note the emergence of a vertical shear band, as predicted in Fig.~\ref{fig:eigenvalue_square_7_15}.
    }
\end{figure}

\begin{figure}[htb!]
    \centering
    \begin{subfigure}{0.24\textwidth}
        \centering
        \caption*{$t=0$}
        \includegraphics[width=0.98\linewidth]{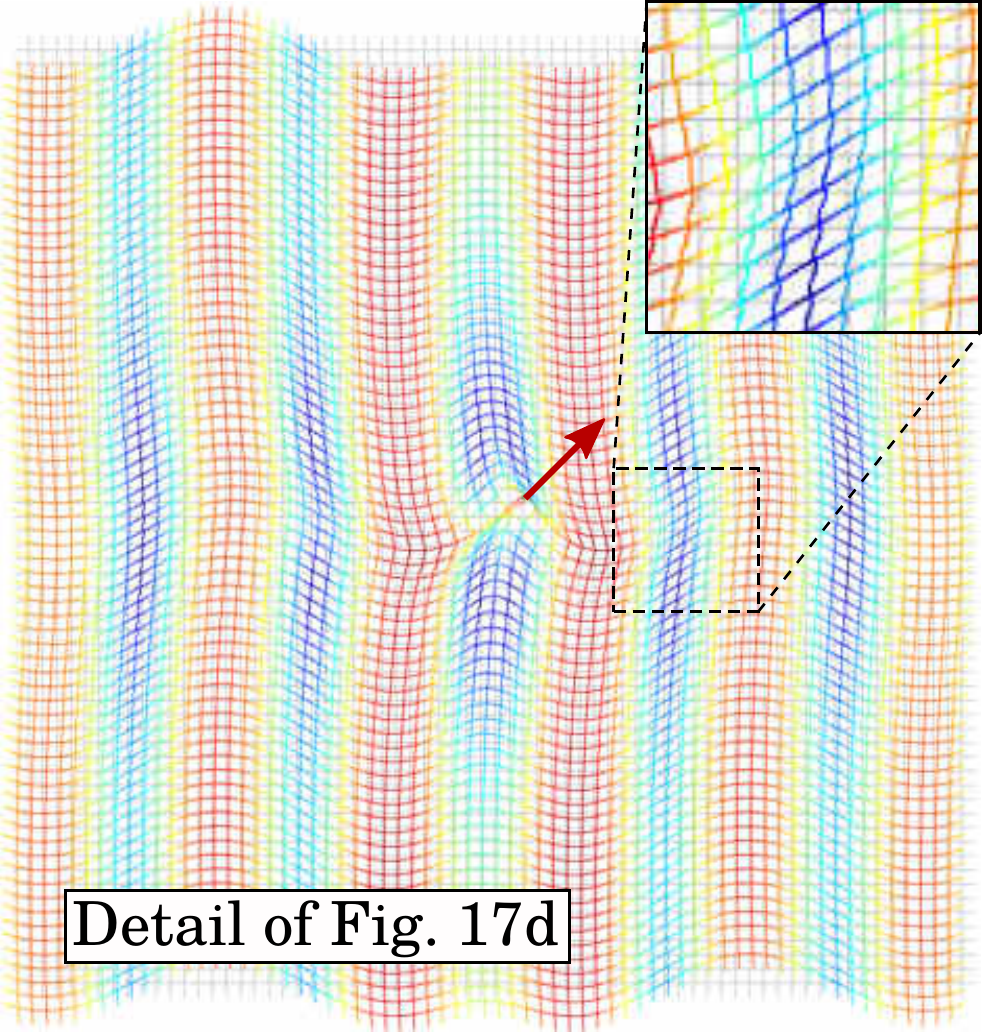}
    \end{subfigure}
    \begin{subfigure}{0.24\textwidth}
        \centering
        \caption*{$t=\frac{\pi}{2\omega}$}
        \includegraphics[width=0.98\linewidth]{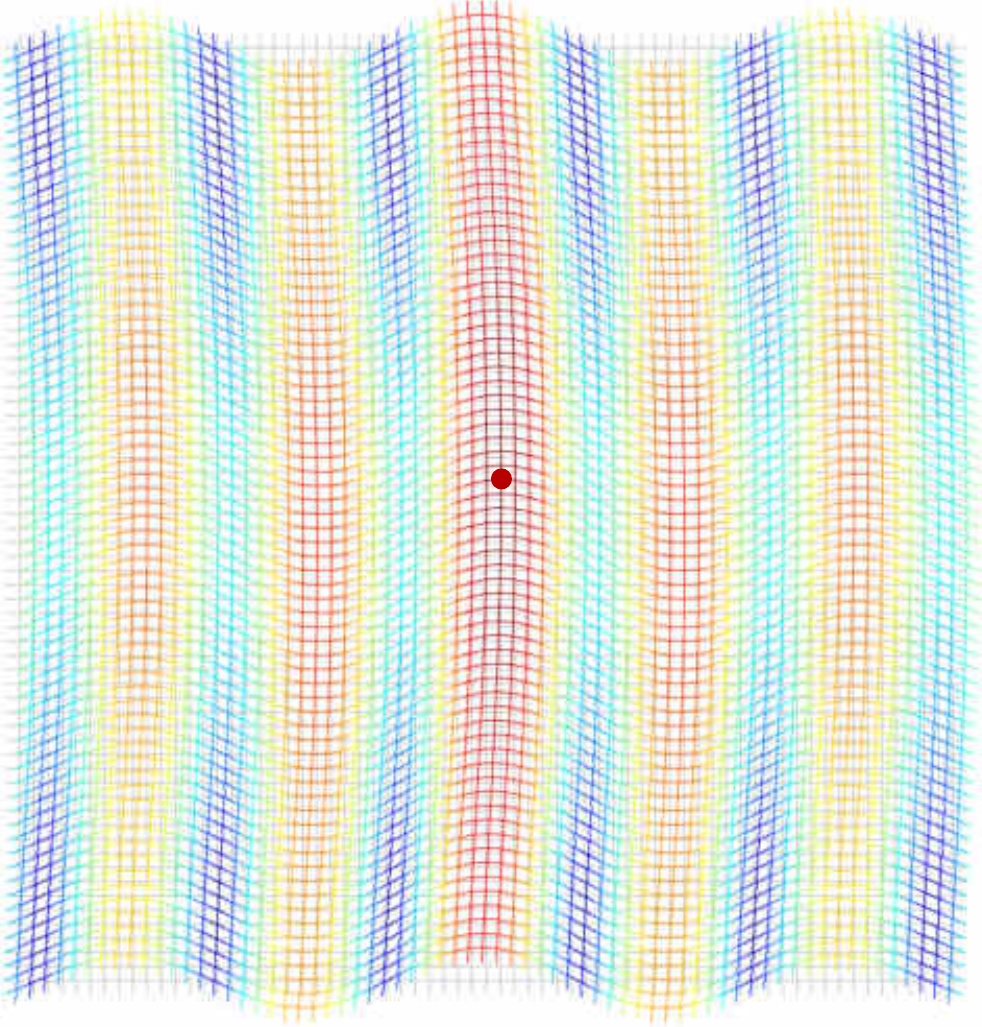}
    \end{subfigure}
    \begin{subfigure}{0.24\textwidth}
        \centering
        \caption*{$t=\frac{\pi}{\omega}$}
        \includegraphics[width=0.98\linewidth]{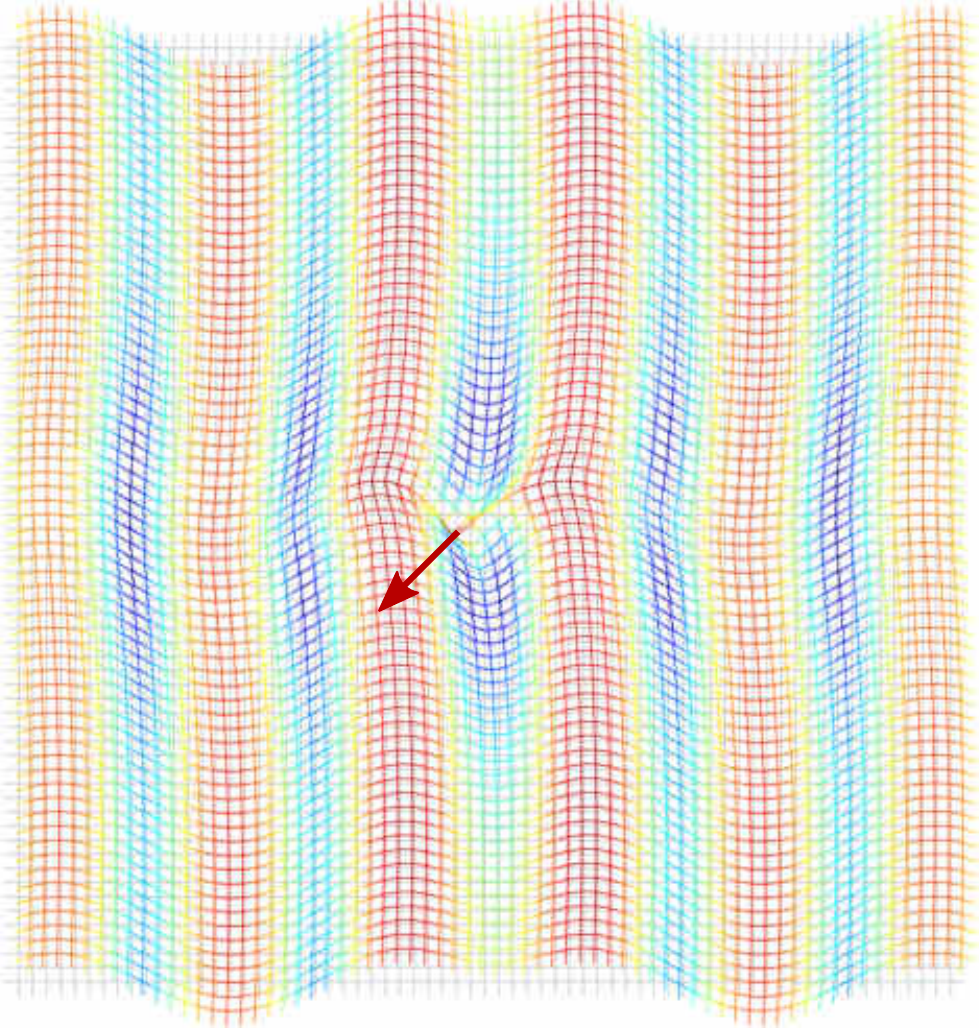}
    \end{subfigure}
    \begin{subfigure}{0.24\textwidth}
        \centering
        \caption*{$t=\frac{3\pi}{2\omega}$}
        \includegraphics[width=0.98\linewidth]{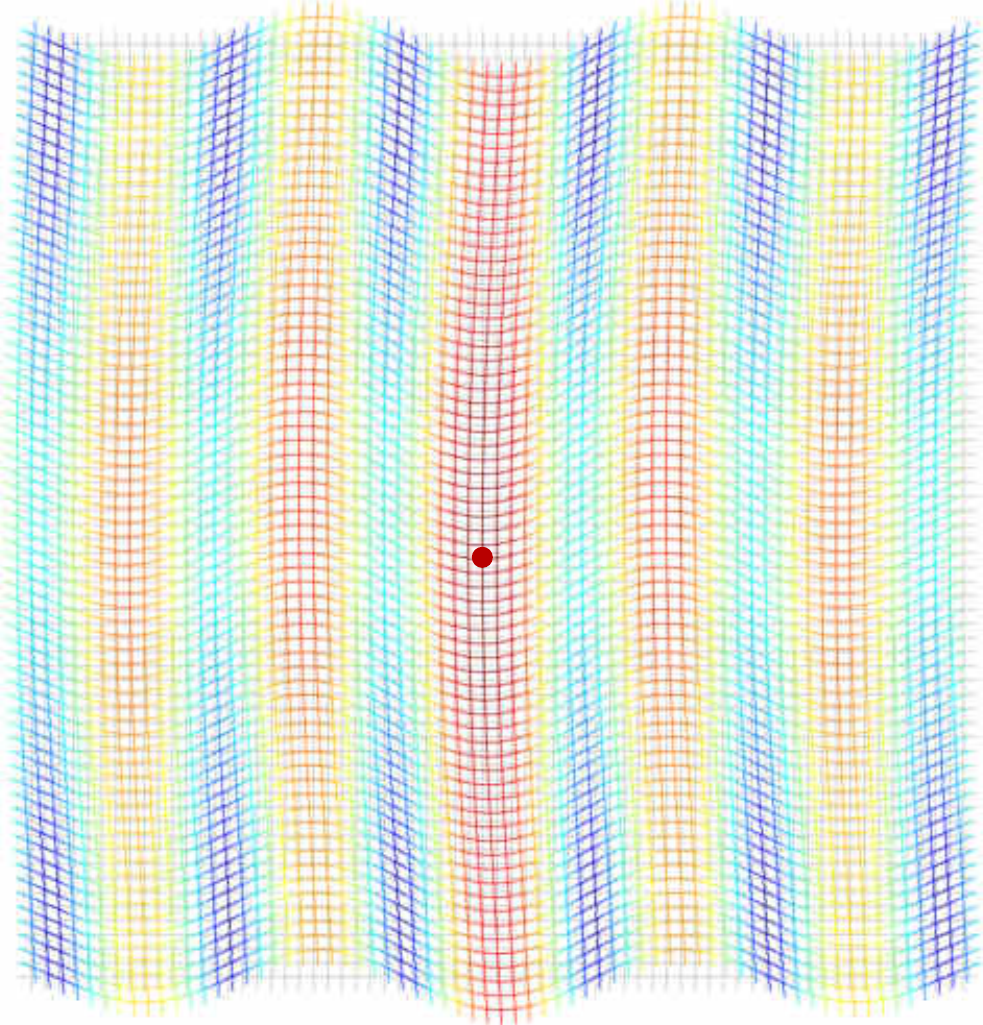}
    \end{subfigure}
    \caption{\label{fig:square_7_15_deformed}
        As for Fig~\ref{fig:square_10_10_deformed}, but for an \textit{orthotropic} square lattice ($\Lambda_1=7,\,\Lambda_2=15$).
        The zone reported in the figure is shown in Fig.~\ref{fig:square_7_15_force_d_99}.
        The pattern shows the characteristic motion of a vertical shear band, induced by the pulsating load.
        The localization emerges from a prevalent bending deformation of the `soft' vertical links and an approximately rigid rotation of the `stiff' horizontal rods allowed by a significant rotation of the grid nodes (see inset of zoomed region at $t=0$).
    }
\end{figure}

The square grid displays a strain localization in the form of a \textit{single} shear band when the slenderness values of the two orthogonal elastic links are set to be different, thus breaking the cubic symmetry, but preserving orthotropy.
The response of the grid with $\Lambda_1=7$ and $\Lambda_2=15$ is reported in Fig.~\ref{fig:square_7_15_force} for four preload states corresponding to $p_1=p_2=\{0,-1.657,-1.864,-2.050\}$.
As already revealed by Fig.~\ref{fig:eigenvalue_square_7_15}, a single vertical shear band emerges, thus confirming the counter-intuitive result obtained in the previous section, namely that \textit{the shear wave responsible for the ellipticity loss is the one propagating along the direction of the `stiffest' elastic link} (which possesses the lowest slenderness).
The mechanism underlying this effect is displayed by analyzing the actual deformed configuration of the grid reported in Fig.~\ref{fig:square_7_15_deformed} (plotted at different instants of time and obtained via f.e.m. simulations).
The figure, which refers to the zone indicated in Fig.~\ref{fig:square_7_15_force_d_99}, reveals that the vertical strain localization emerges from a prevalent bending deformation of the `soft' vertical links accompanied by an approximately rigid rotation of the `stiff' horizontal rods, which is allowed by a large rotation of the nodes (see also inset of zoomed region at $t=0$).
\begin{figure}[htb!]
    \centering
    \begin{subfigure}{0.4\textwidth}
        \centering
        \caption{\label{fig:square_10_10_force_d_FFT}Cubic square lattice ($\Lambda_1=\Lambda_2=10$)}
        \includegraphics[width=0.98\linewidth]{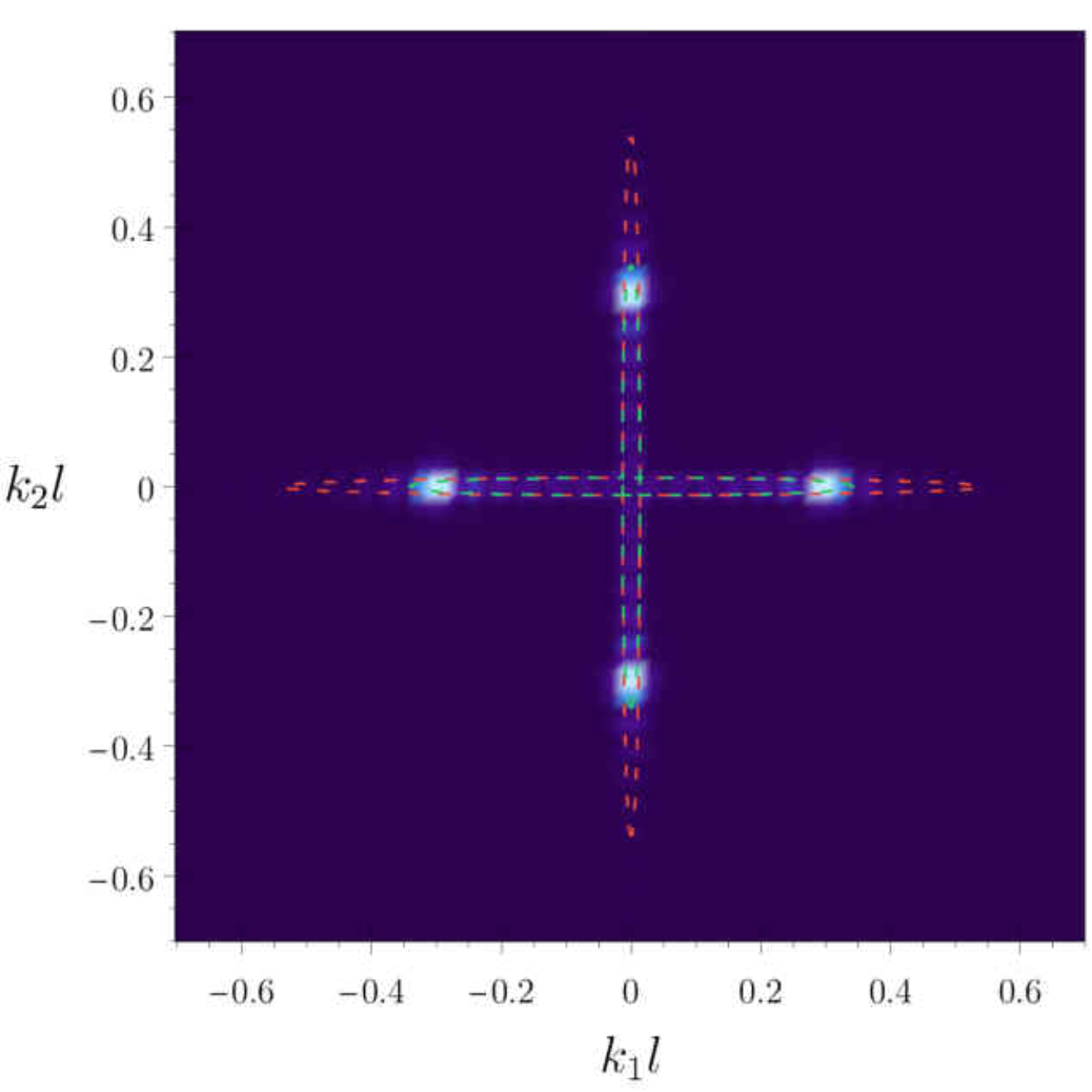}
    \end{subfigure} \hspace{2mm}
    \begin{subfigure}{0.4\textwidth}
        \centering
        \caption{\label{fig:square_7_15_force_d_FFT}Orthotropic square lattice ($\Lambda_1=7,\, \Lambda_2=15$)}
        \includegraphics[width=0.98\linewidth]{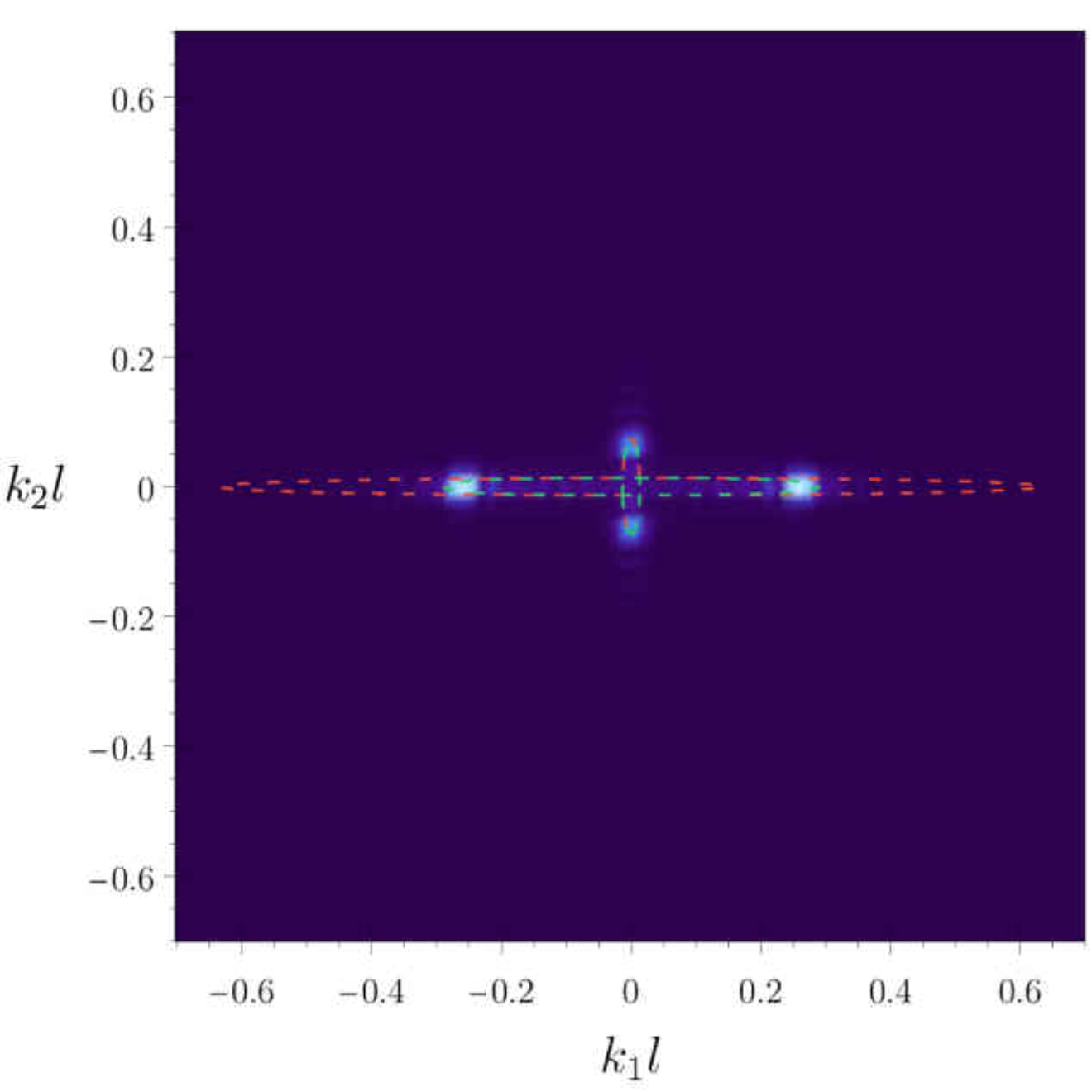}
    \end{subfigure}
    \caption{\label{fig:square_force_d_FFT}
        Fourier transform of the complex displacement fields for the cubic~(\subref{fig:square_10_10_force_d_FFT}) and orthotropic~(\subref{fig:square_7_15_force_d_FFT}) square lattice subject to a pulsating diagonal force, close to the elliptic boundary (fields reported in Fig.~\ref{fig:square_10_10_force_d_99} and~\ref{fig:square_7_15_force_d_99}, respectively).
        The slowness contours of the lattice (dashed green) and the effective continuum (dashed red) are superimposed to highlight the Bloch spectrum of waves excited by the forcing source.
        Note that the strong focus of the spectrum indicates that few plane waves, namely those `slow' waves that are close to cause the ellipticity loss, prevail in the response.
        This is expected for the orthotropic grid~(\subref{fig:square_7_15_force_d_FFT}) near the elliptic boundary, where the waves propagating vertically remain almost inactive, when compared to those propagating horizontally.
    }
\end{figure}

The comparison between the responses of the grid and of the homogenized continuum, presented in Figs.~\ref{fig:square_10_10_force} and~\ref{fig:square_7_15_force}, shows an almost perfect agreement from low to high prestress levels, up to values close to the elliptic boundary.
The agreement can be further tested by considering the lattice's complex displacement field (reported in the last column of Figs.~\ref{fig:square_10_10_force} and~\ref{fig:square_7_15_force}, prestressed at $\bp=0.99\,\bp_{\text{E}}$), computing its Fourier transform and superimposing this to the corresponding slowness contour (associated to the chosen frequency $\Omega = 0.01$).
This is reported in Fig.~\ref{fig:square_force_d_FFT}, where the Fourier transform shows that, for both considered square grids, the Bloch spectrum of waves excited by the diagonal load matches the slowness contour of the lattice (reported in green) and is also highly focused around the directions of ellipticity loss where it is at the maximum distance from the contour of the continuum (reported in red).
It is worth noting that the strong focus of the spectrum confirms the fact that few plane waves, namely those `slow' waves that are close to cause the ellipticity loss, prevail in the response, as it is expected for a material near the elliptic boundary.

\subsection{Rhombic lattice}
\label{sec:rhombic_lattice}
In the previous section, the square lattice was shown to display only localizations in the form of `pure' shear bands, i.e. in which shear strain prevails, perfectly aligned parallel to the elastic ligaments.
However, on the basis of the analysis performed in Section~\ref{sec:grid}, the formation of localizations is expected along different directions and with different deformation modes, when a rhombic grid is considered, $\alpha\neq\pi/2$.
\begin{figure}[htb!]
    \centering
    \begin{subfigure}{0.24\textwidth}
        \centering
        \phantomsubcaption{\label{fig:rhombus_10_10_force_h_0}}
        \includegraphics[width=0.98\linewidth]{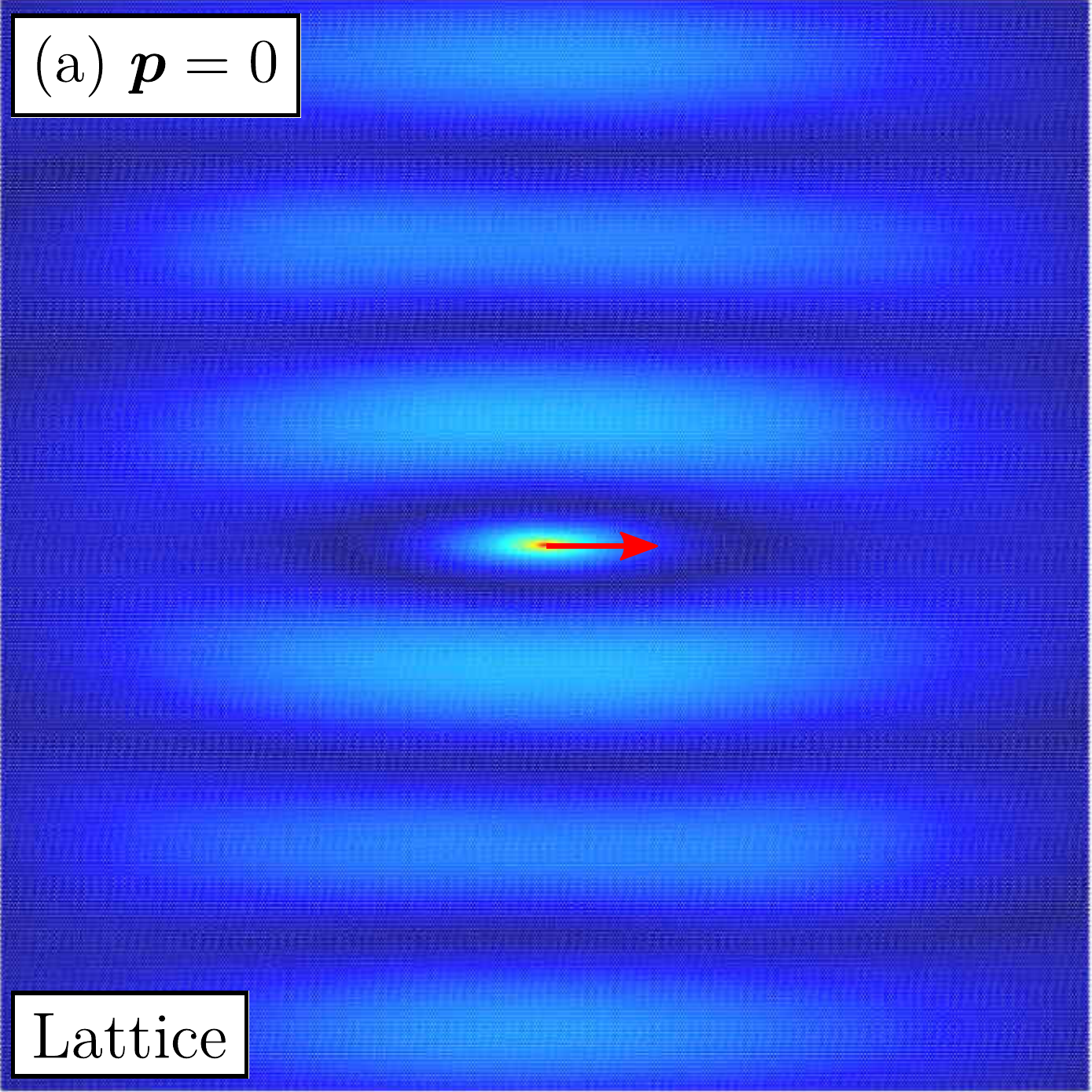}
    \end{subfigure}
    \begin{subfigure}{0.24\textwidth}
        \centering
        \phantomsubcaption{\label{fig:rhombus_10_10_force_h_80}}
        \includegraphics[width=0.98\linewidth]{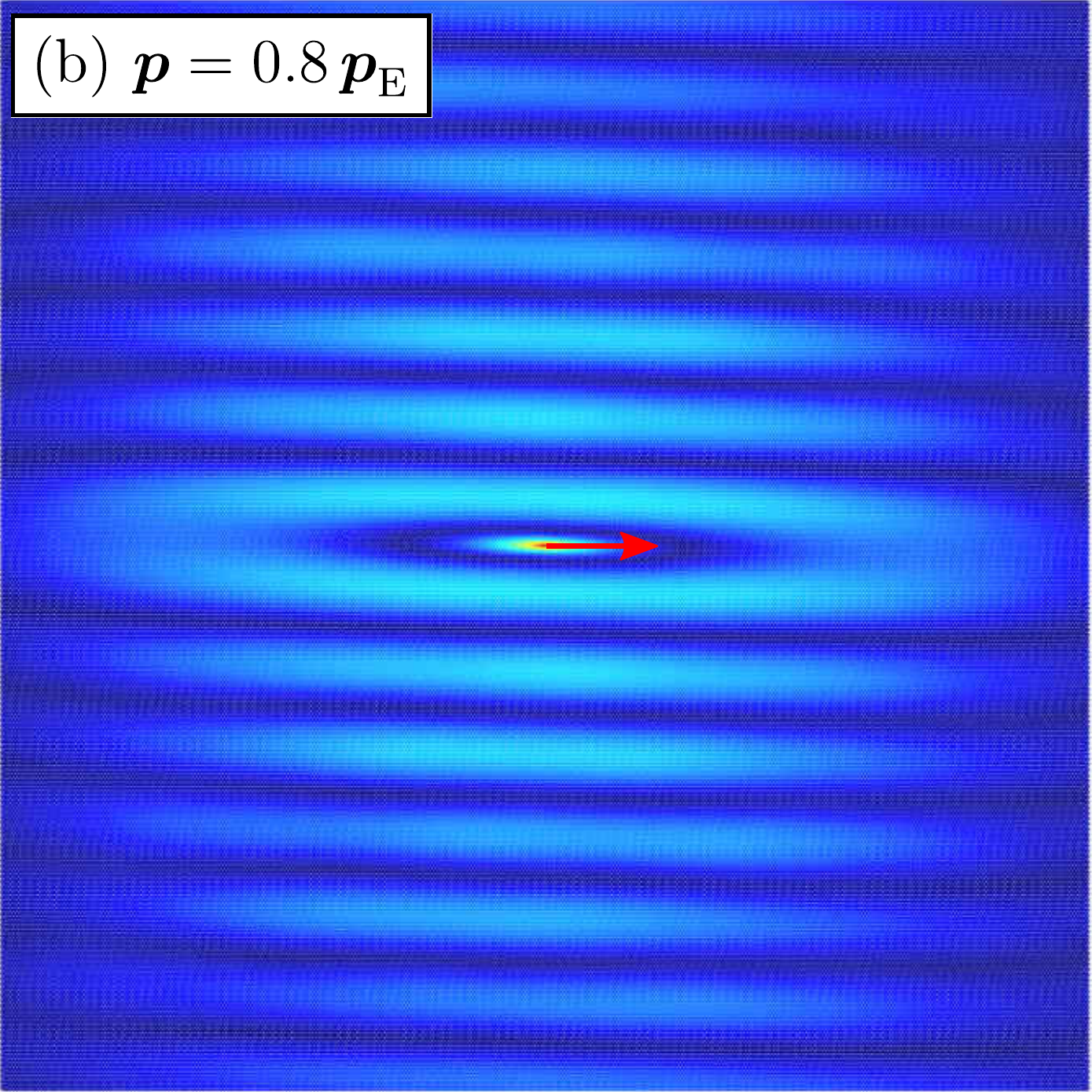}
    \end{subfigure}
    \begin{subfigure}{0.24\textwidth}
        \centering
        \phantomsubcaption{\label{fig:rhombus_10_10_force_h_90}}
        \includegraphics[width=0.98\linewidth]{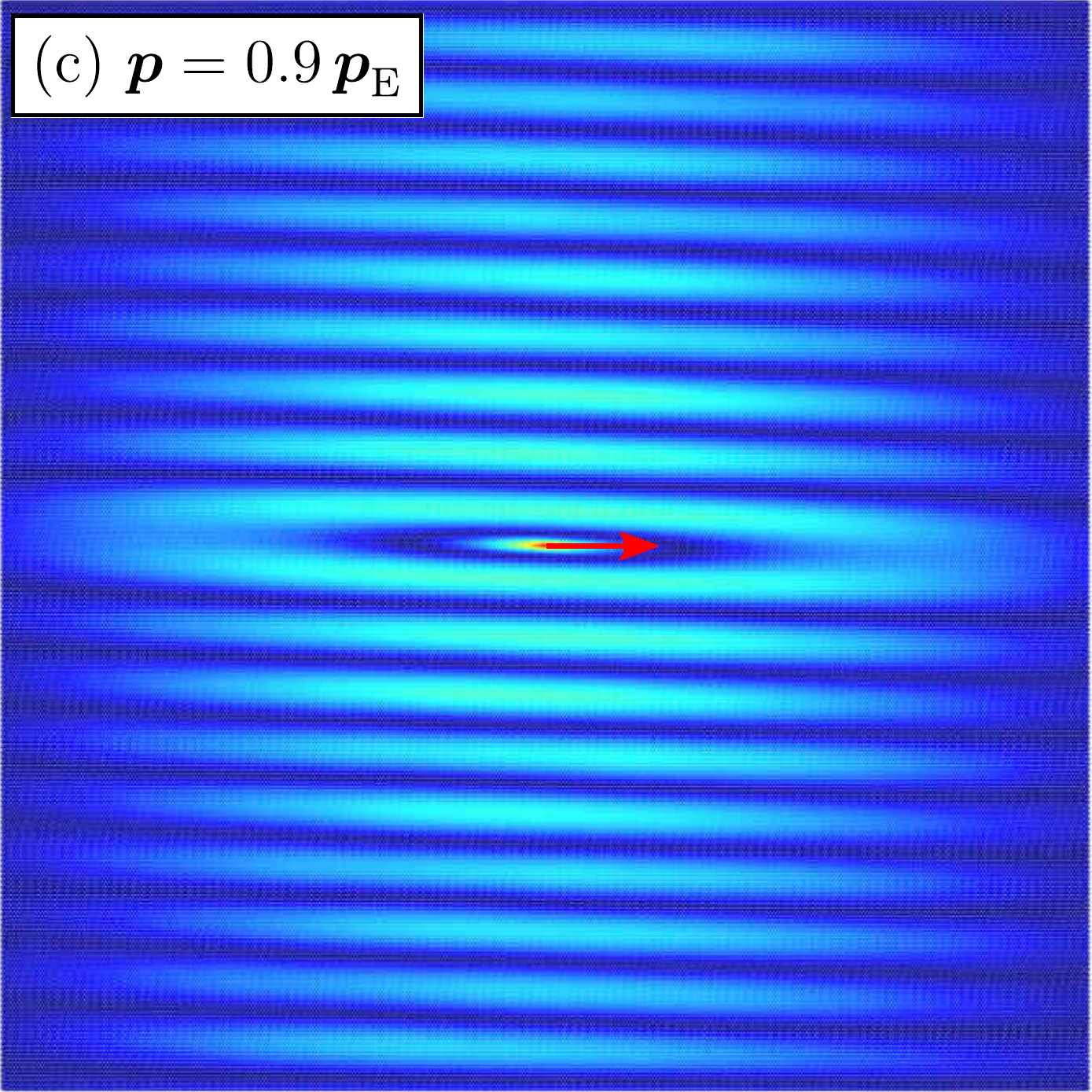}
    \end{subfigure}
    \begin{subfigure}{0.24\textwidth}
        \centering
        \phantomsubcaption{\label{fig:rhombus_10_10_force_h_99}}
        \includegraphics[width=0.98\linewidth]{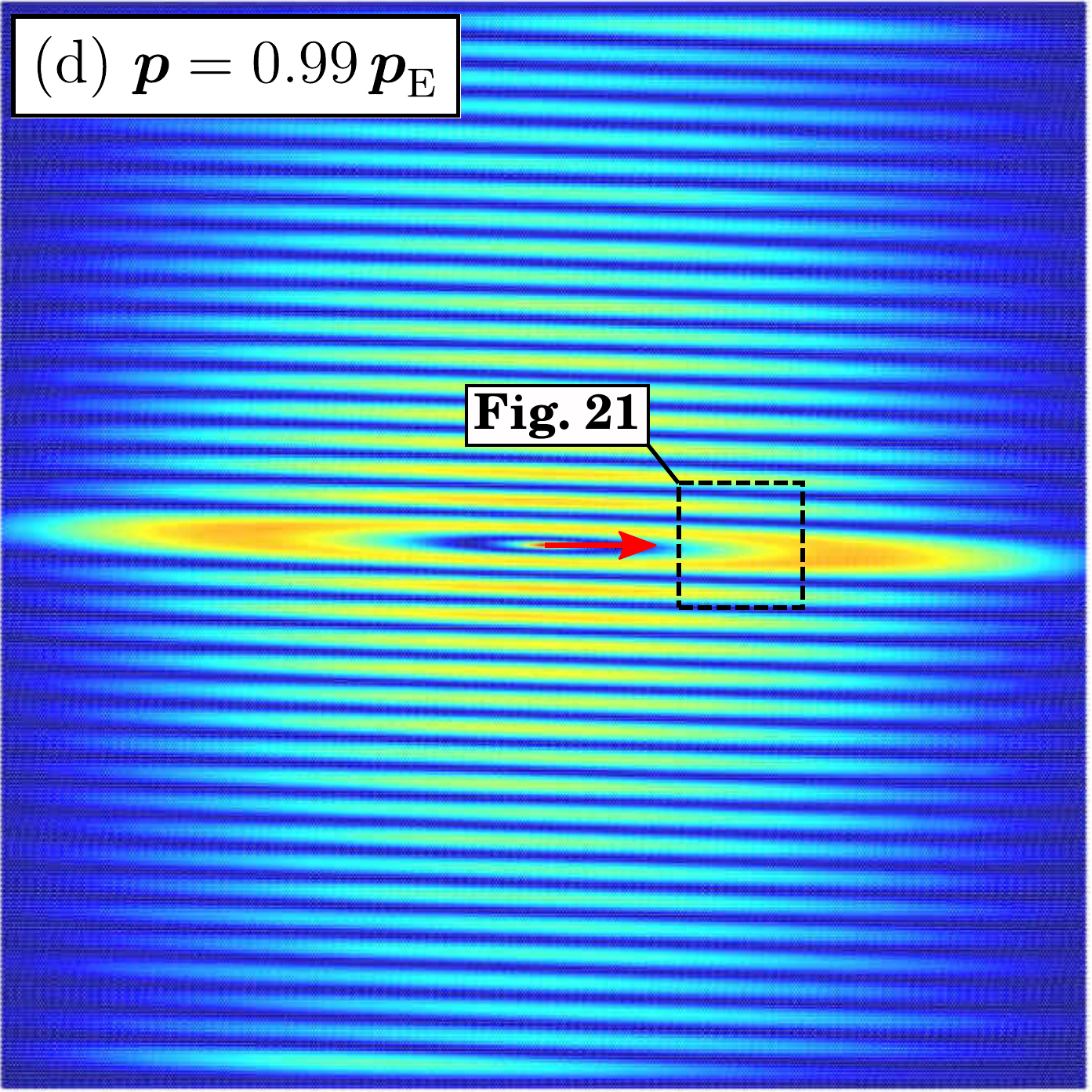}
    \end{subfigure}\\
    \vspace{0.01\linewidth}
    \begin{subfigure}{0.24\textwidth}
        \centering
        \phantomsubcaption{\label{fig:rhombus_10_10_force_h_0_gf}}
        \includegraphics[width=0.98\linewidth]{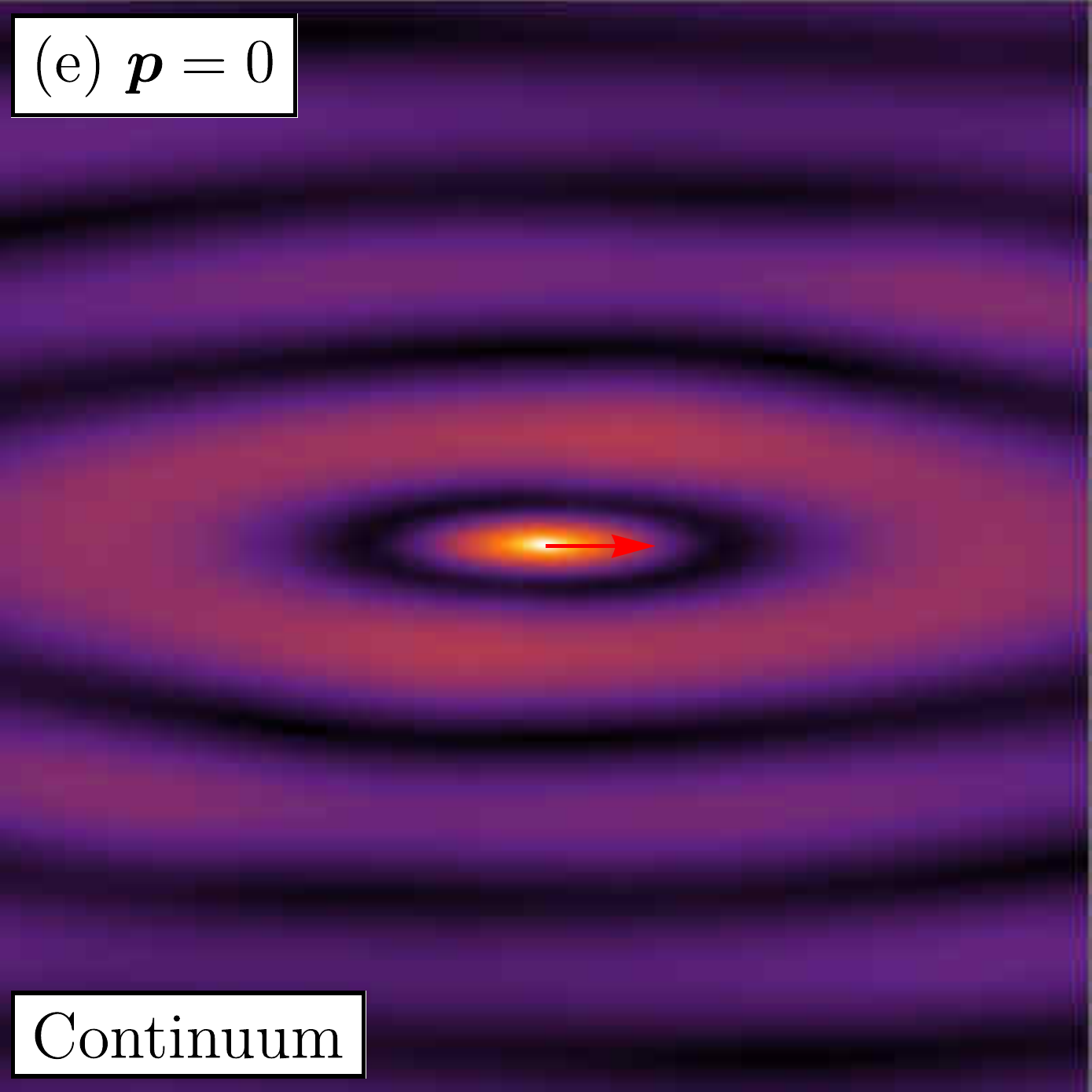}
    \end{subfigure}
    \begin{subfigure}{0.24\textwidth}
        \centering
        \phantomsubcaption{\label{fig:rhombus_10_10_force_h_80_gf}}
        \includegraphics[width=0.98\linewidth]{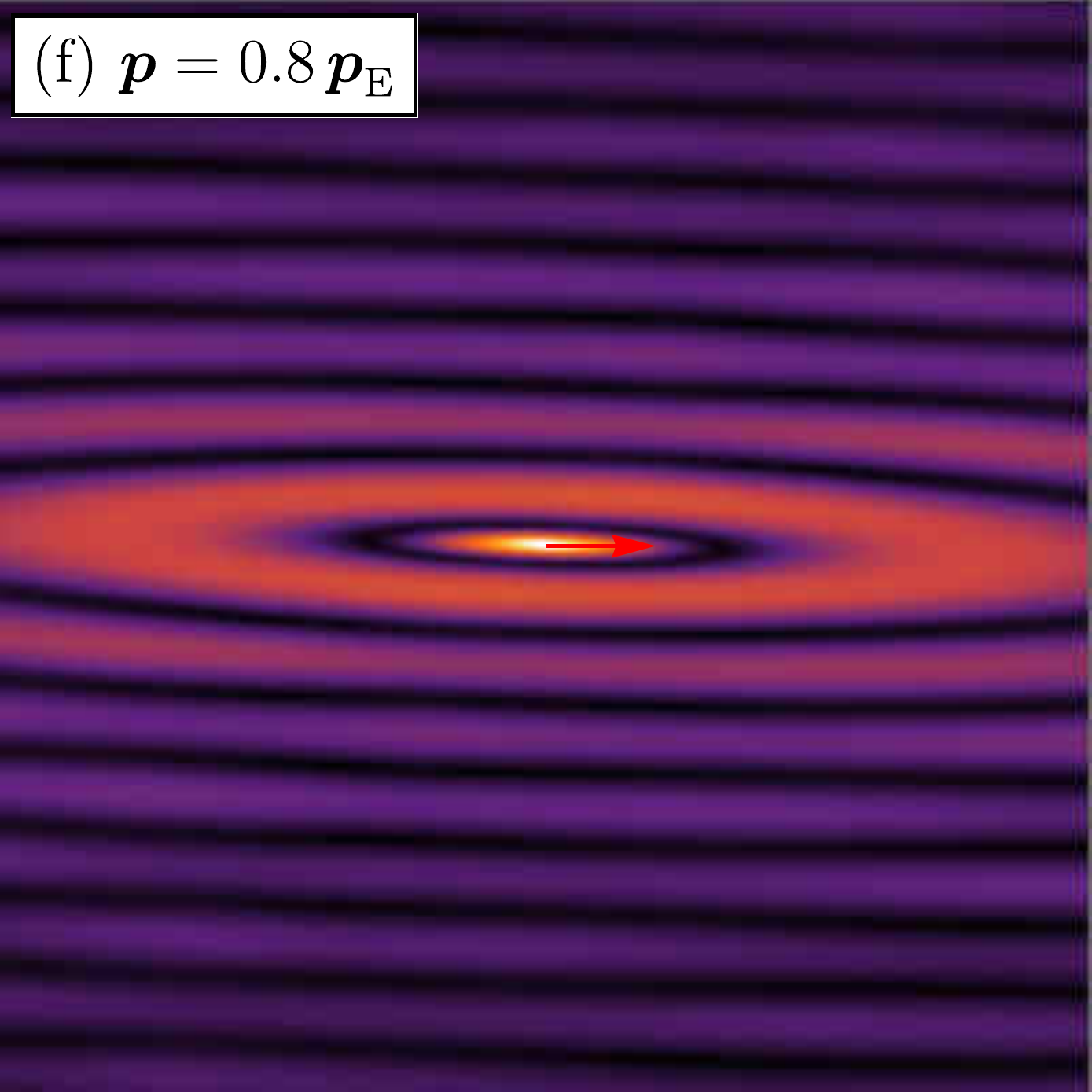}
    \end{subfigure}
    \begin{subfigure}{0.24\textwidth}
        \centering
        \phantomsubcaption{\label{fig:rhombus_10_10_force_h_90_gf}}
        \includegraphics[width=0.98\linewidth]{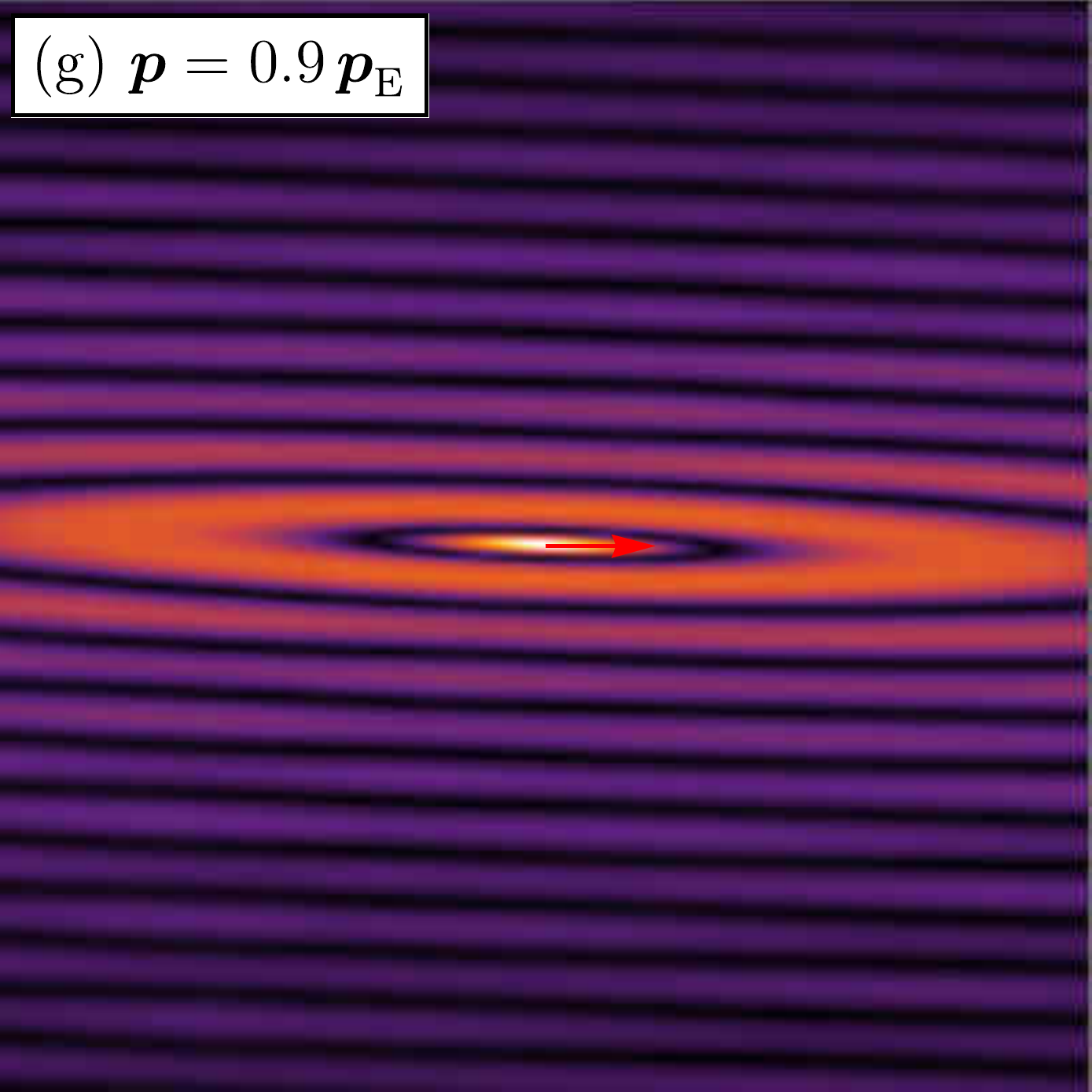}
    \end{subfigure}
    \begin{subfigure}{0.24\textwidth}
        \centering
        \phantomsubcaption{\label{fig:rhombus_10_10_force_h_99_gf}}
        \includegraphics[width=0.98\linewidth]{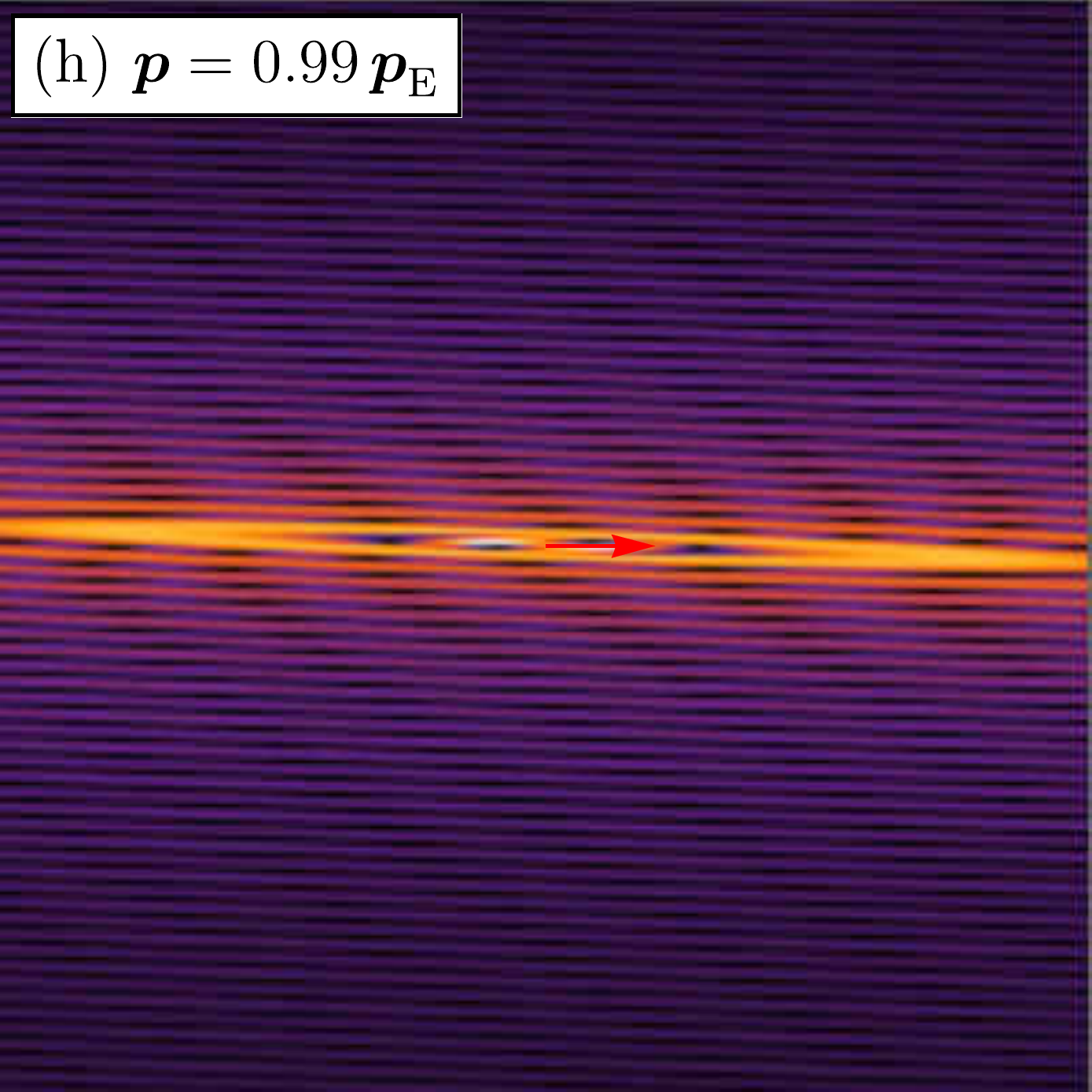}
    \end{subfigure}\\
    \vspace{0.015\linewidth}
    \begin{subfigure}{0.24\textwidth}
        \centering
        \phantomsubcaption{\label{fig:rhombus_10_10_force_v_0}}
        \includegraphics[width=0.98\linewidth]{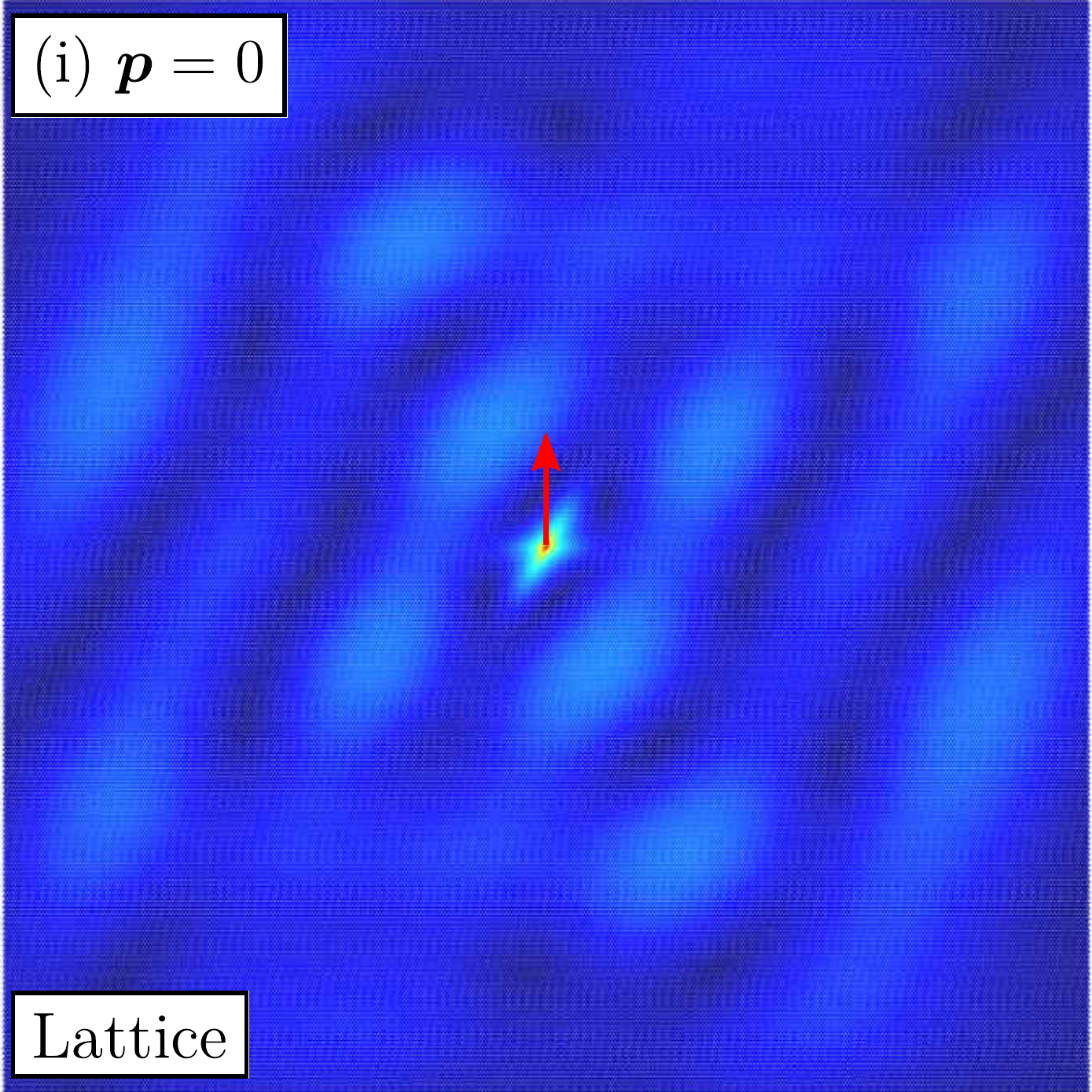}
    \end{subfigure}
    \begin{subfigure}{0.24\textwidth}
        \centering
        \phantomsubcaption{\label{fig:rhombus_10_10_force_v_80}}
        \includegraphics[width=0.98\linewidth]{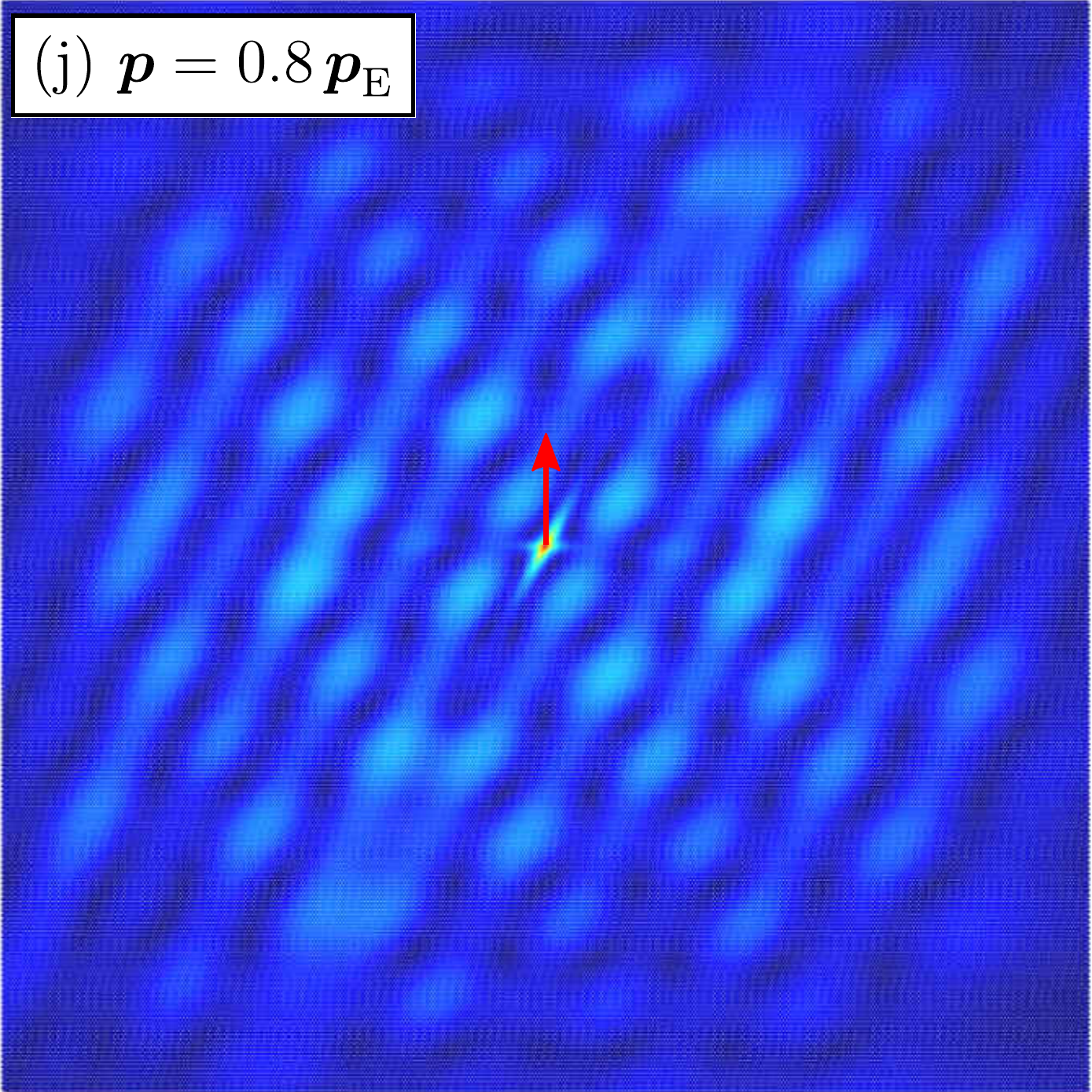}
    \end{subfigure}
    \begin{subfigure}{0.24\textwidth}
        \centering
        \phantomsubcaption{\label{fig:rhombus_10_10_force_v_90}}
        \includegraphics[width=0.98\linewidth]{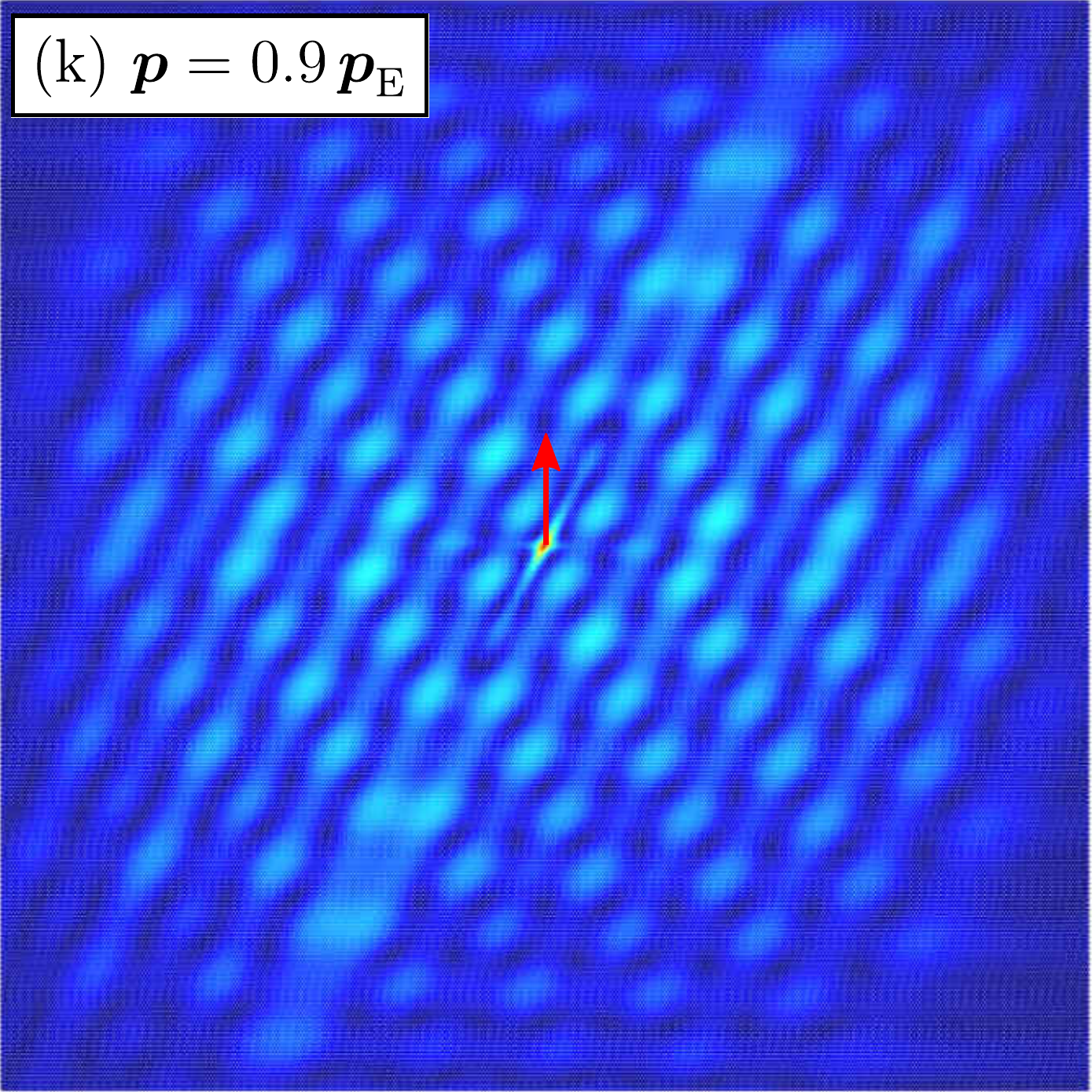}
    \end{subfigure}
    \begin{subfigure}{0.24\textwidth}
        \centering
        \phantomsubcaption{\label{fig:rhombus_10_10_force_v_99}}
        \includegraphics[width=0.98\linewidth]{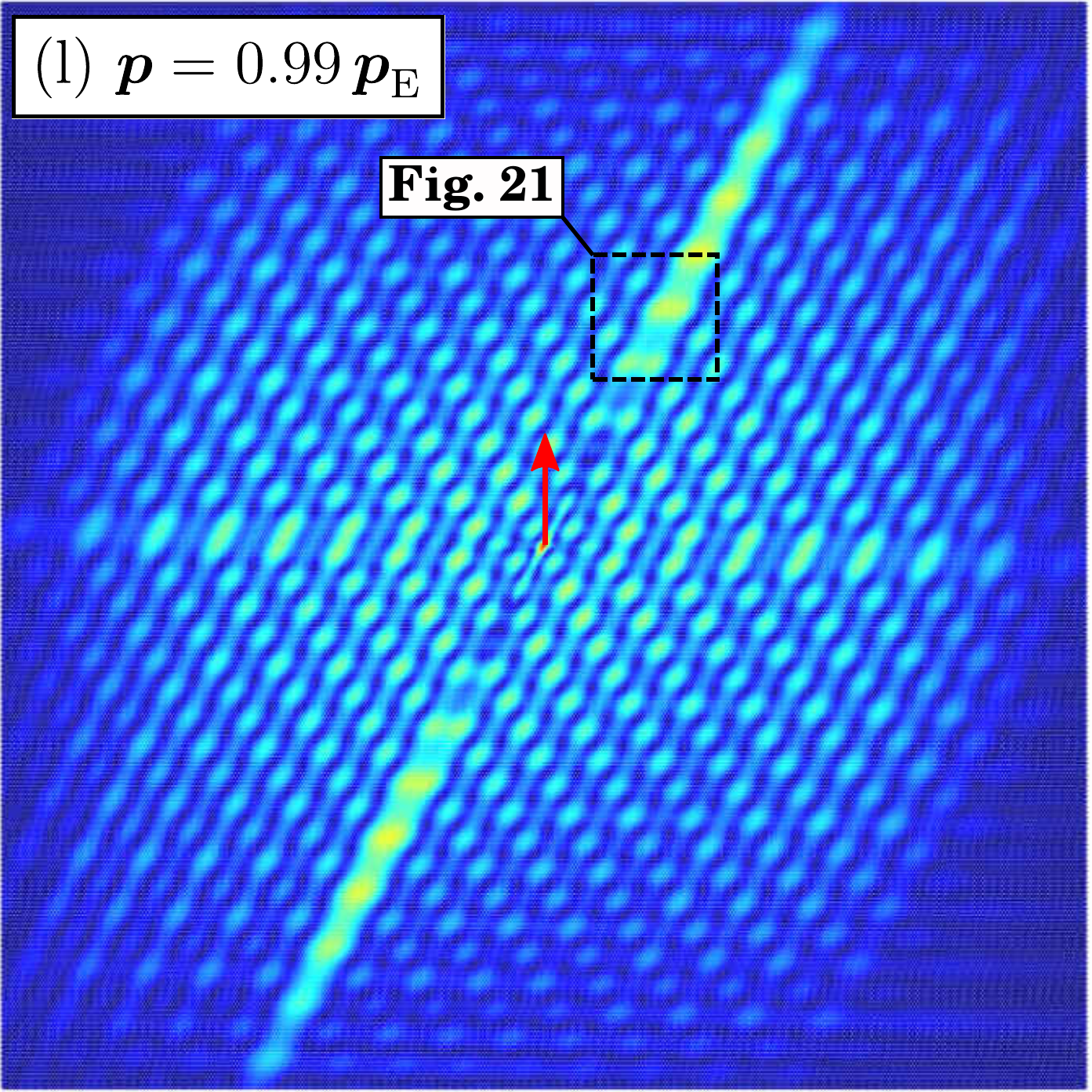}
    \end{subfigure}\\
    \vspace{0.01\linewidth}
    \begin{subfigure}{0.24\textwidth}
        \centering
        \phantomsubcaption{\label{fig:rhombus_10_10_force_v_0_gf}}
        \includegraphics[width=0.98\linewidth]{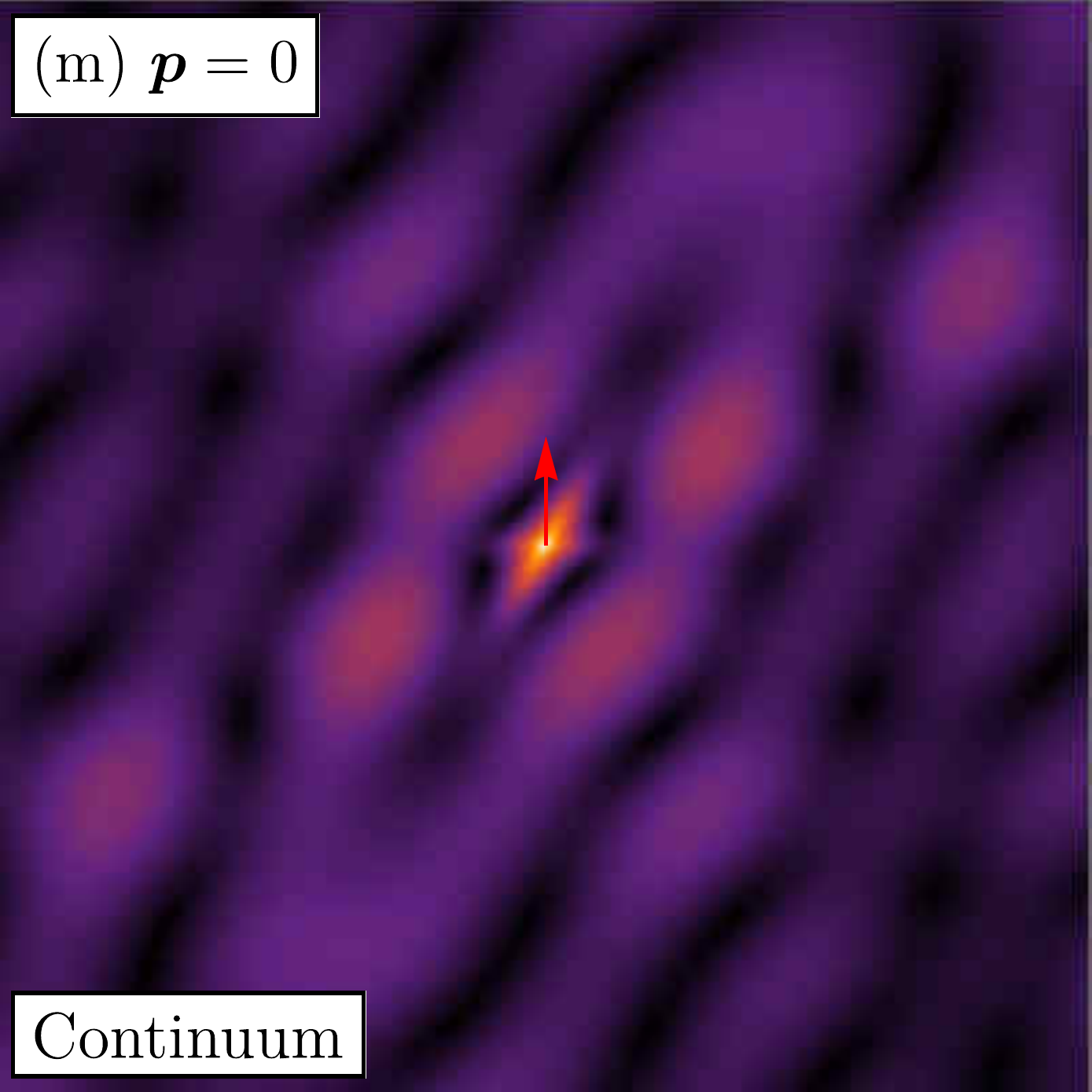}
    \end{subfigure}
    \begin{subfigure}{0.24\textwidth}
        \centering
        \phantomsubcaption{\label{fig:rhombus_10_10_force_v_80_gf}}
        \includegraphics[width=0.98\linewidth]{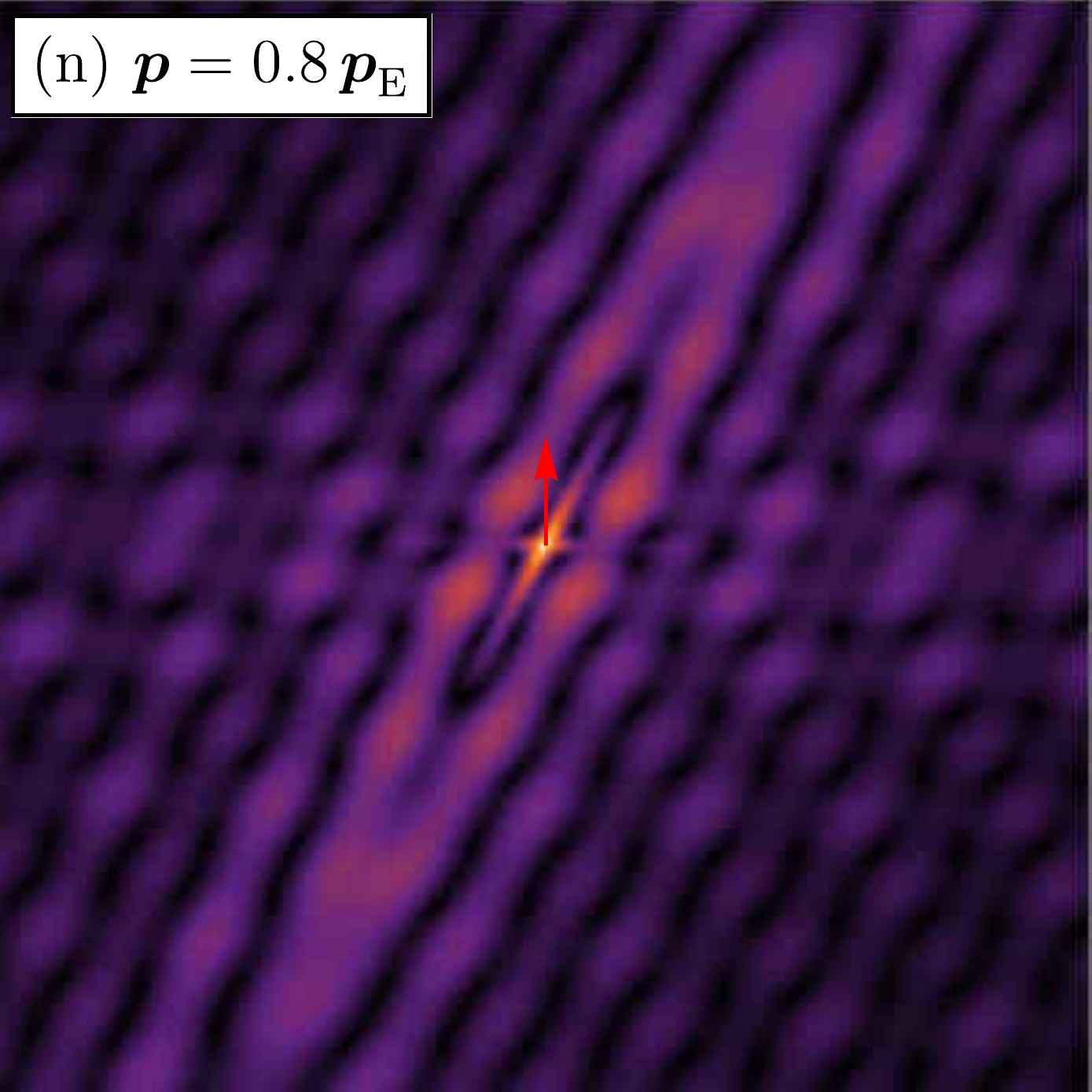}
    \end{subfigure}
    \begin{subfigure}{0.24\textwidth}
        \centering
        \phantomsubcaption{\label{fig:rhombus_10_10_force_v_90_gf}}
        \includegraphics[width=0.98\linewidth]{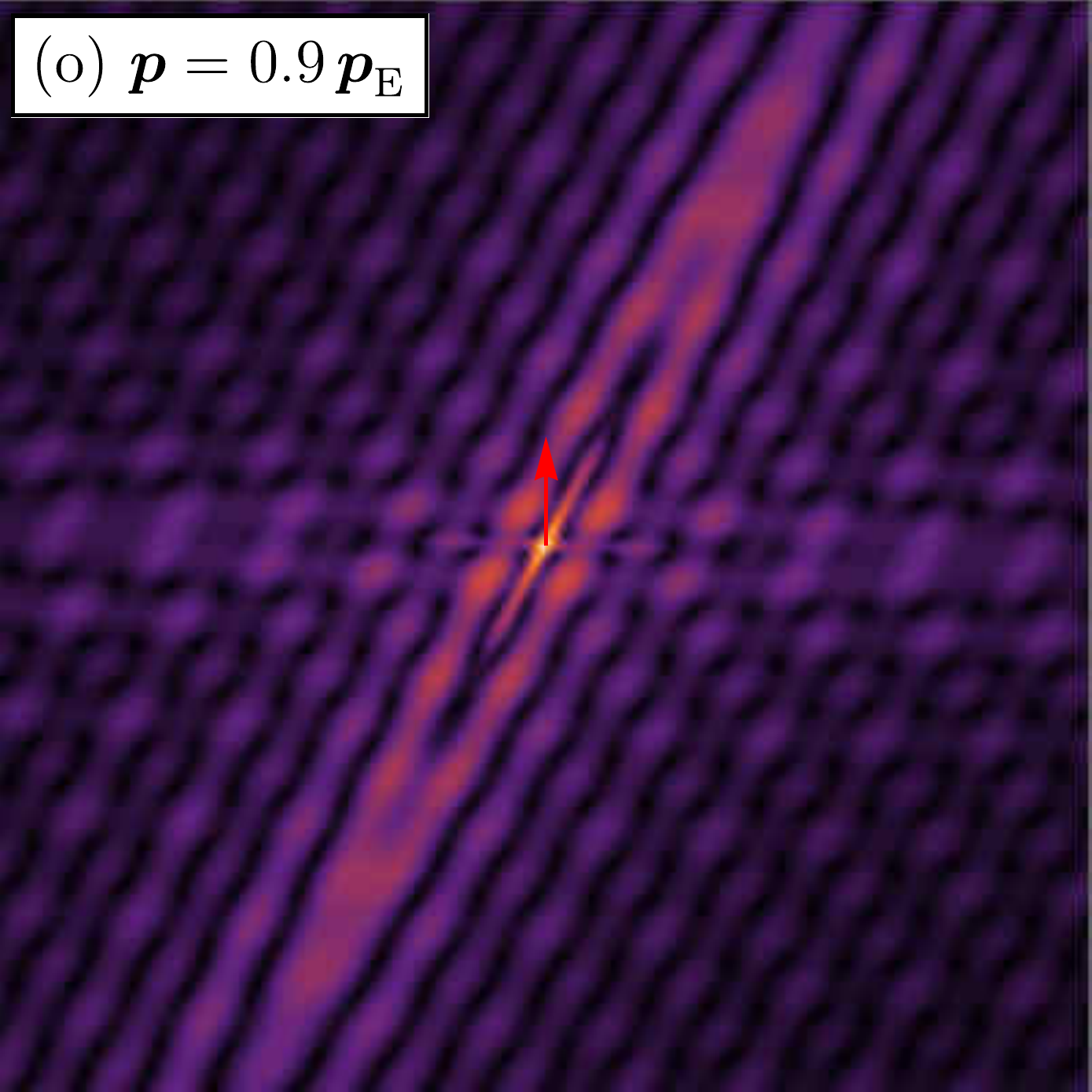}
    \end{subfigure}
    \begin{subfigure}{0.24\textwidth}
        \centering
        \phantomsubcaption{\label{fig:rhombus_10_10_force_v_99_gf}}
        \includegraphics[width=0.98\linewidth]{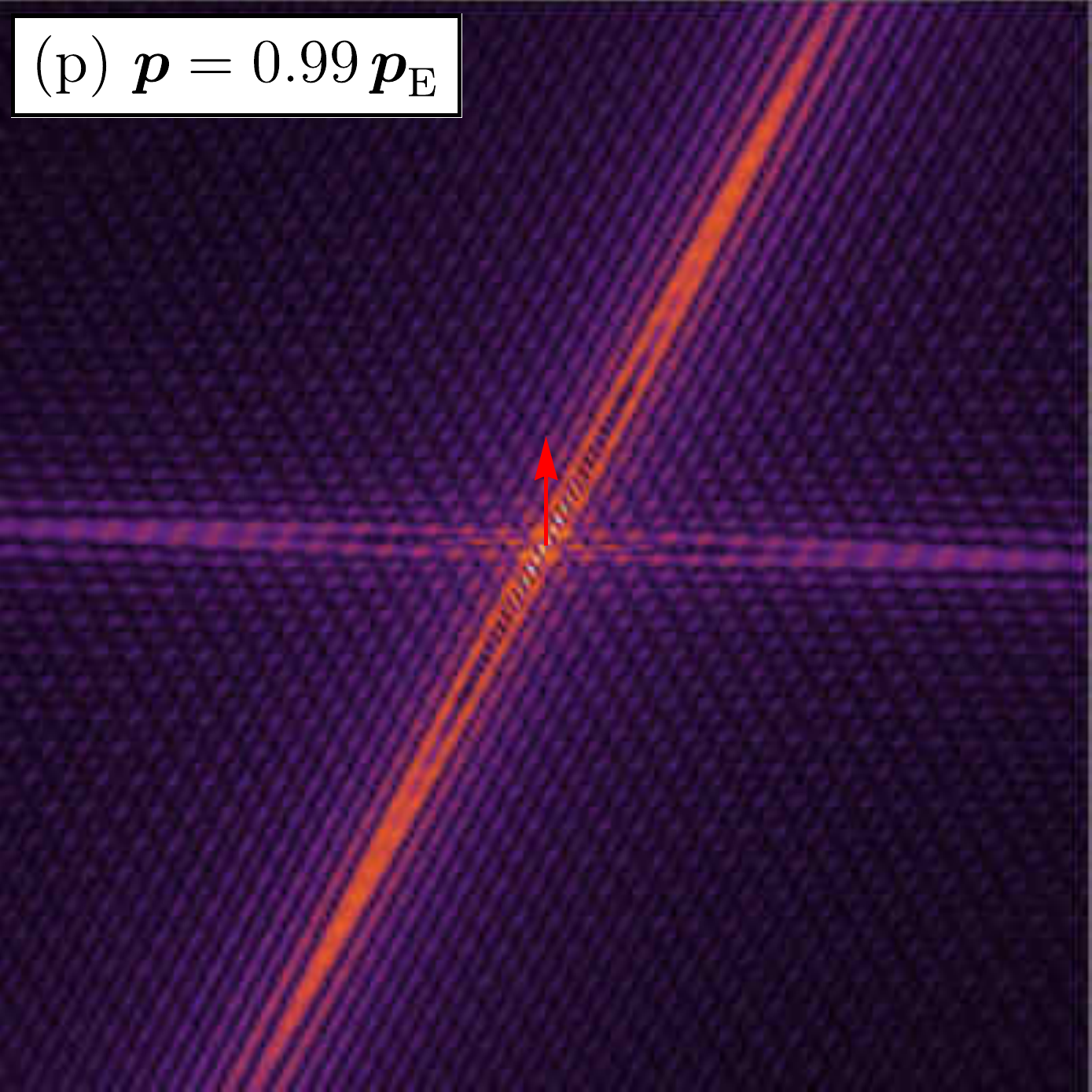}
    \end{subfigure}
    \caption{\label{fig:rhombus_10_10_force}
        The displacement field generated by a pulsating \textit{horizontal} force (denoted with a red arrow and applied to the \textit{orthotropic} rhombic lattice, $\Lambda_1=\Lambda_2=10$) is simulated via f.e.m., see (\subref{fig:rhombus_10_10_force_h_0})--(\subref{fig:rhombus_10_10_force_h_99}), and compared to the response of the homogenized continuum, see (\subref{fig:rhombus_10_10_force_h_0_gf})--(\subref{fig:rhombus_10_10_force_h_99_gf}), at different levels of prestress $\bp$.
        The same comparison is reported in (\subref{fig:rhombus_10_10_force_v_0})--(\subref{fig:rhombus_10_10_force_v_99}) and (\subref{fig:rhombus_10_10_force_v_0_gf})--(\subref{fig:rhombus_10_10_force_v_99_gf}) for a pulsating \textit{vertical} force.
        Even though ellipticity is lost along the two directions predicted in Fig.~\ref{fig:eigenvalue_rhombus_10_10}, the activation of strain localization depends on the orientation of the load, so that two bands are activated by the vertical force, while only one is generated by the horizontal load.
        Furthermore, note that the directions of the localization bands (with angles of the normal equal to $\theta_{\text{cr}}=88.2^\circ,151.8^\circ$) is slightly misaligned with respect to the rod's inclination.
    }
\end{figure}

In order to investigate the response of the orthotropic ($\Lambda_1=\Lambda_2=10$) and anisotropic ($\Lambda_1=7,\,\Lambda_2=15$) rhombic lattices ($\alpha=\pi/3$), both horizontal and vertical concentrated forces will be considered, so to observe a dependence of the number of strain localizations on the loading orientation.
Furthermore, in contrast to what happens in the case of the square grid, the directions of localization are expected to occur with a slight misalignment with respect to the directions of the rods, as predicted in Figs.~\ref{fig:eigenvalue_rhombus_10_10} and~\ref{fig:eigenvalue_rhombus_7_15}.

In Fig.~\ref{fig:rhombus_10_10_force} the displacement field computed via f.e.m. for the orthotropic rhombic lattice (horizontally and vertically loaded with a pulsating force and reported on first and third row from the top of the figure) is compared to the response of the homogenized continuum (reported in the second and fourth row) at four values of preload (increasing from left to right) $p_1=p_2=\{0,-4.276,-4.811,-5.292\}$.

A comparison between Figs.~\ref{fig:rhombus_10_10_force_h_0}--\ref{fig:rhombus_10_10_force_h_99} and Figs.~\ref{fig:rhombus_10_10_force_h_0_gf}--\ref{fig:rhombus_10_10_force_h_99_gf} and a comparison between Figs.~\ref{fig:rhombus_10_10_force_v_0}--\ref{fig:rhombus_10_10_force_v_99} and Figs.~\ref{fig:rhombus_10_10_force_v_0_gf}--\ref{fig:rhombus_10_10_force_v_99_gf}, shows an excellent agreement between the lattice response and its homogenized continuum counterpart, for each state of lattice's preload.
With reference to a prestress state $\bp=0.99\,\bp_{\text{E}}$ (last column on the right of the figure), while two localization bands are activated by the vertical force, only one is generated by the horizontal force.
\begin{figure}[htb!]
    \centering
    \begin{subfigure}{0.24\textwidth}
        \centering
        \caption*{$t=0$}
        \includegraphics[width=0.98\linewidth]{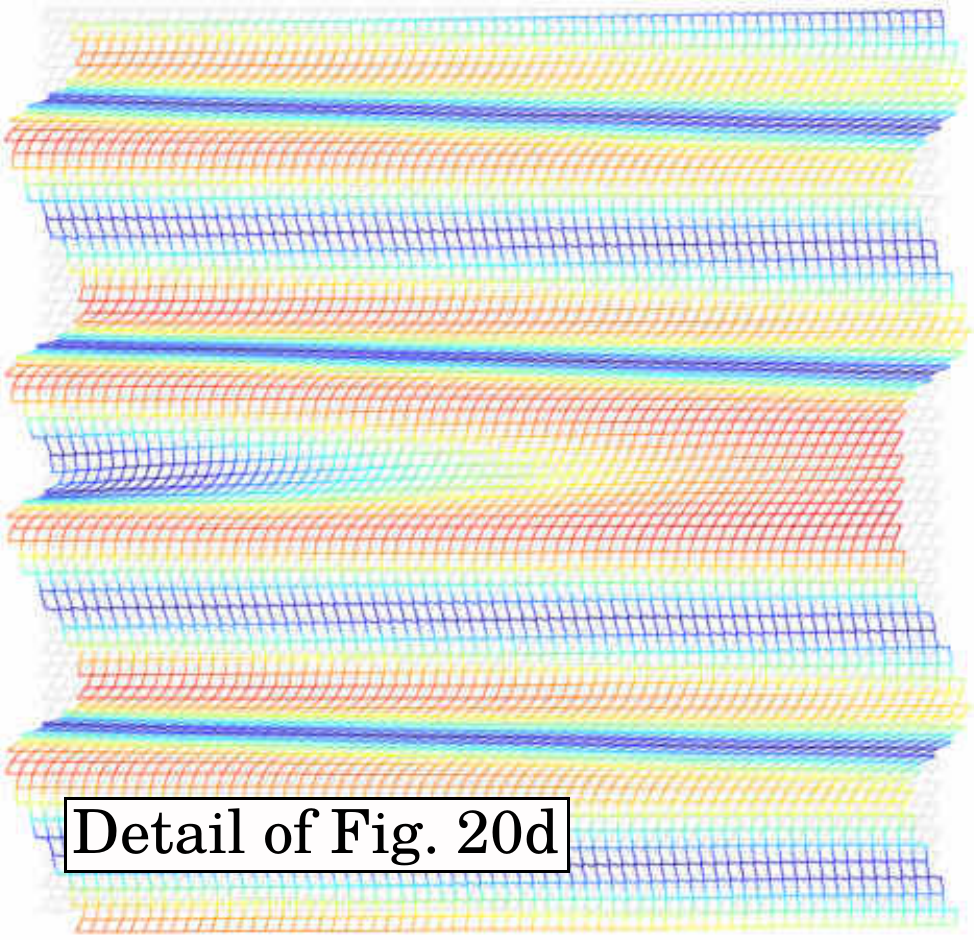}
    \end{subfigure}
    \begin{subfigure}{0.24\textwidth}
        \centering
        \caption*{$t=\frac{\pi}{2\omega}$}
        \includegraphics[width=0.98\linewidth]{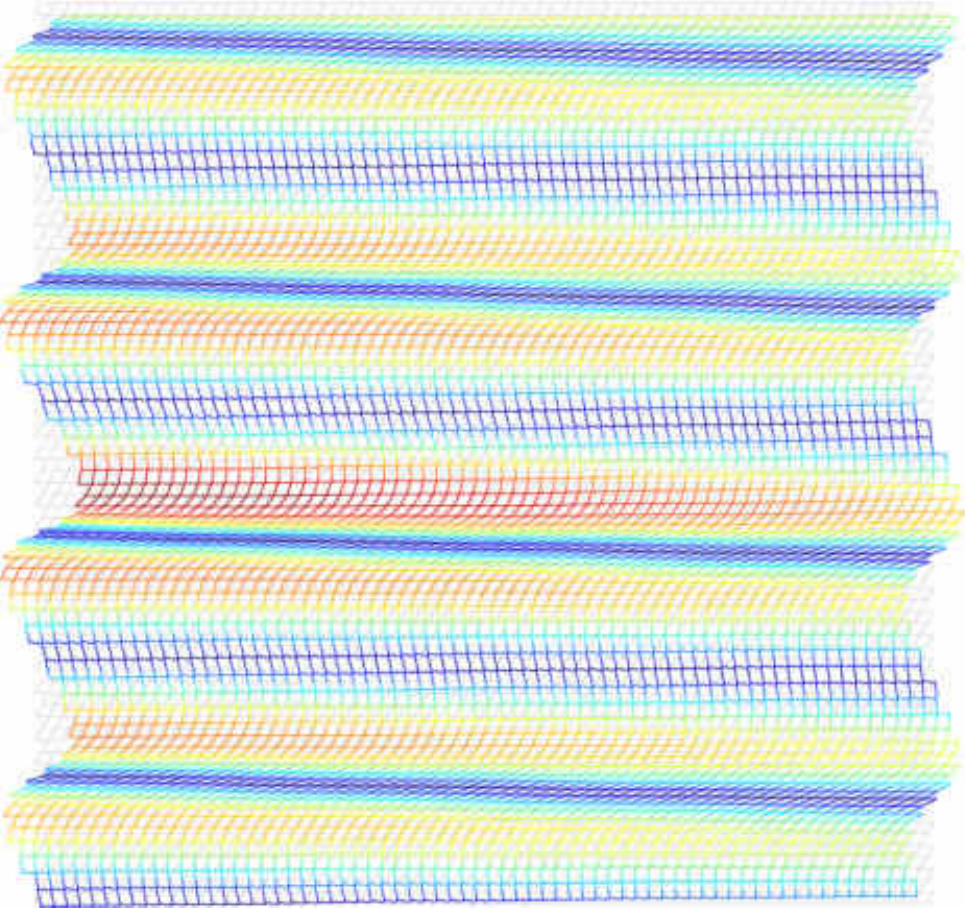}
    \end{subfigure}
    \begin{subfigure}{0.24\textwidth}
        \centering
        \caption*{$t=\frac{\pi}{\omega}$}
        \includegraphics[width=0.98\linewidth]{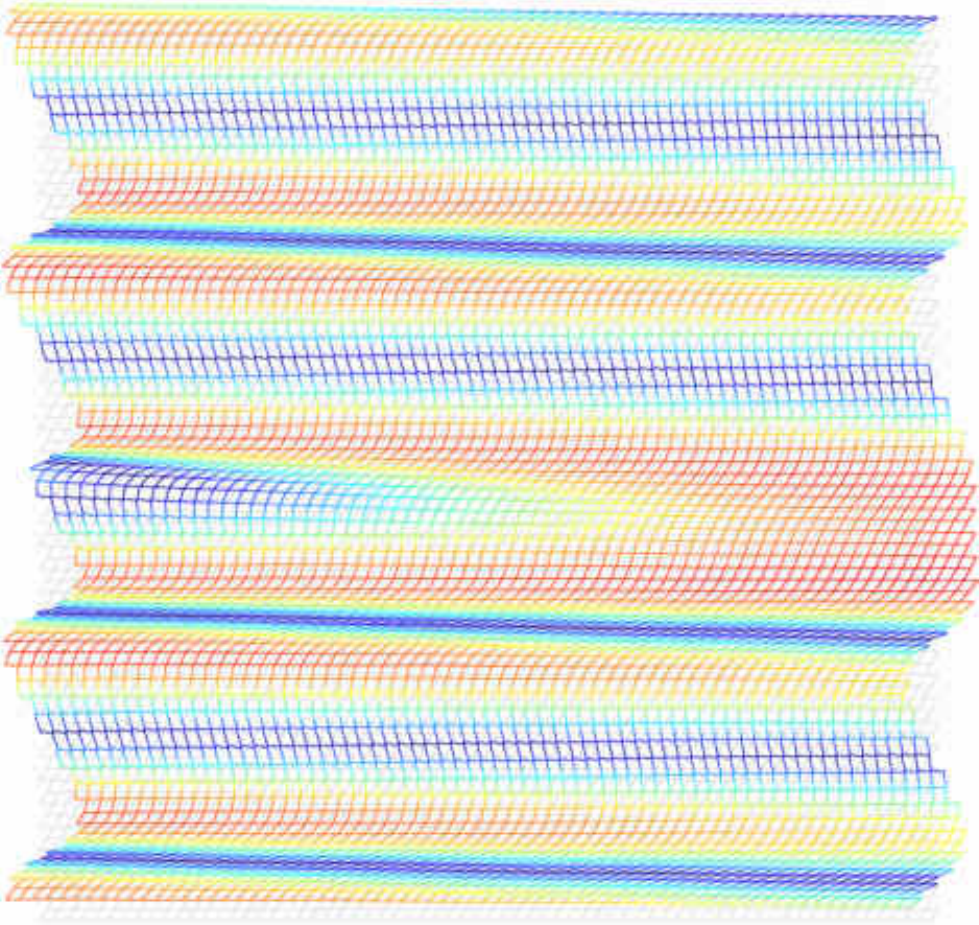}
    \end{subfigure}
    \begin{subfigure}{0.24\textwidth}
        \centering
        \caption*{$t=\frac{3\pi}{2\omega}$}
        \includegraphics[width=0.98\linewidth]{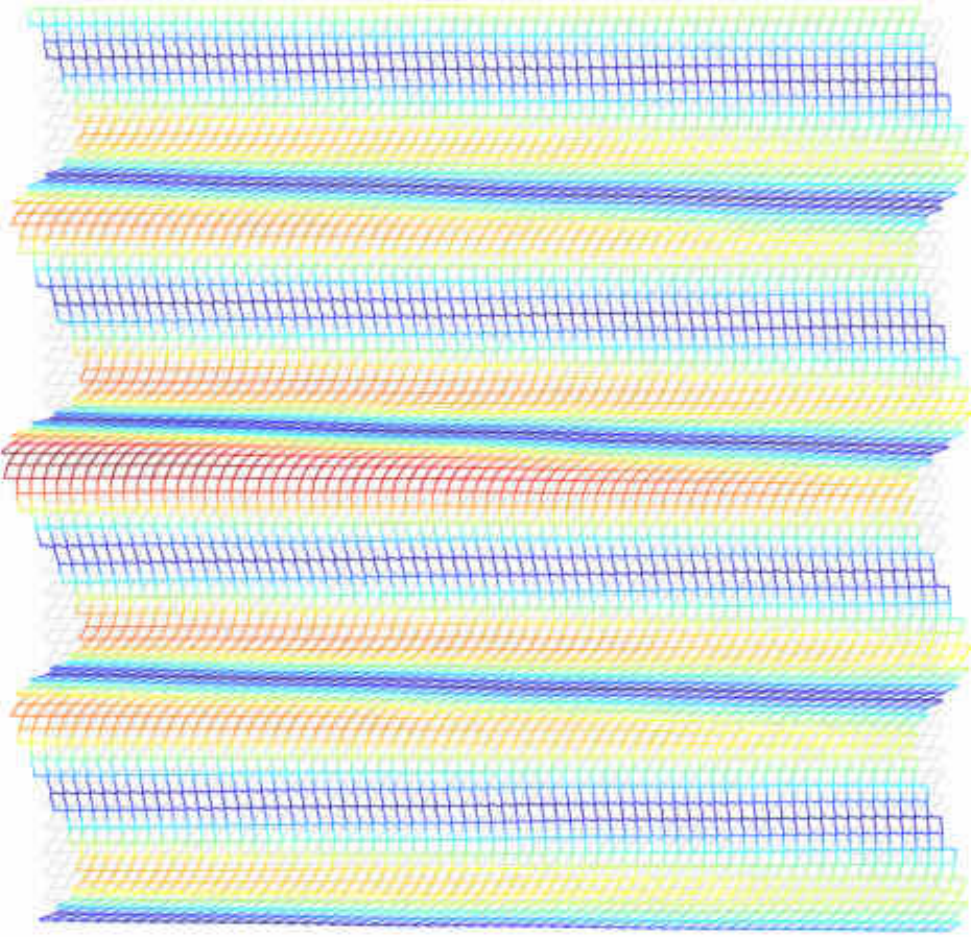}
    \end{subfigure}\\
    \vspace{1mm}
    \begin{subfigure}{0.24\textwidth}
        \centering
        \includegraphics[width=0.98\linewidth]{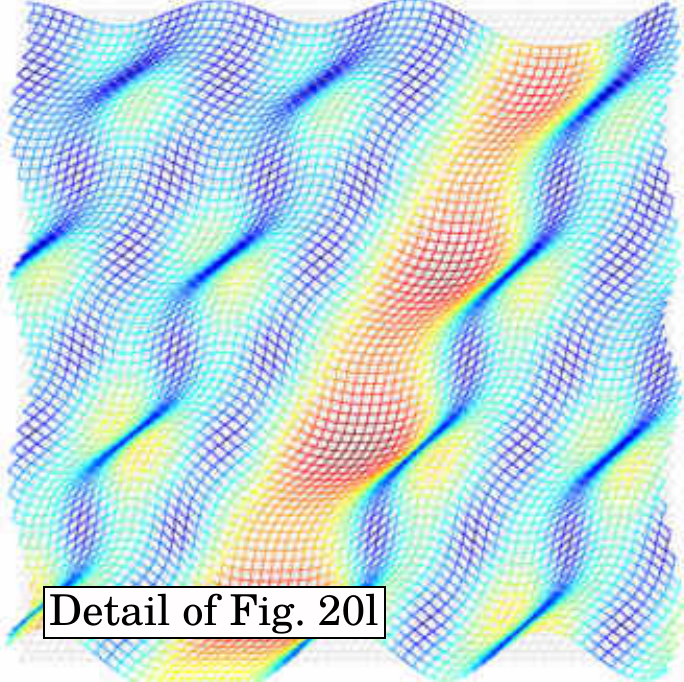}
    \end{subfigure}
    \begin{subfigure}{0.24\textwidth}
        \centering
        \includegraphics[width=0.98\linewidth]{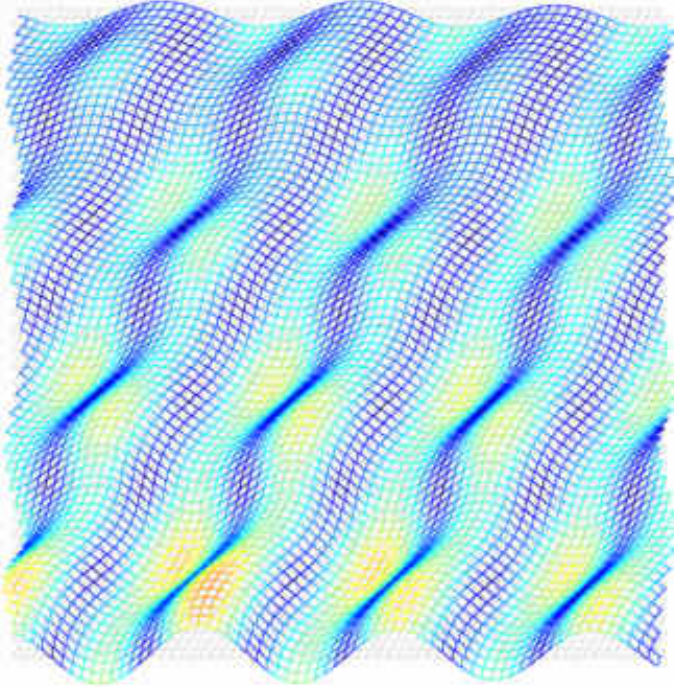}
    \end{subfigure}
    \begin{subfigure}{0.24\textwidth}
        \centering
        \includegraphics[width=0.98\linewidth]{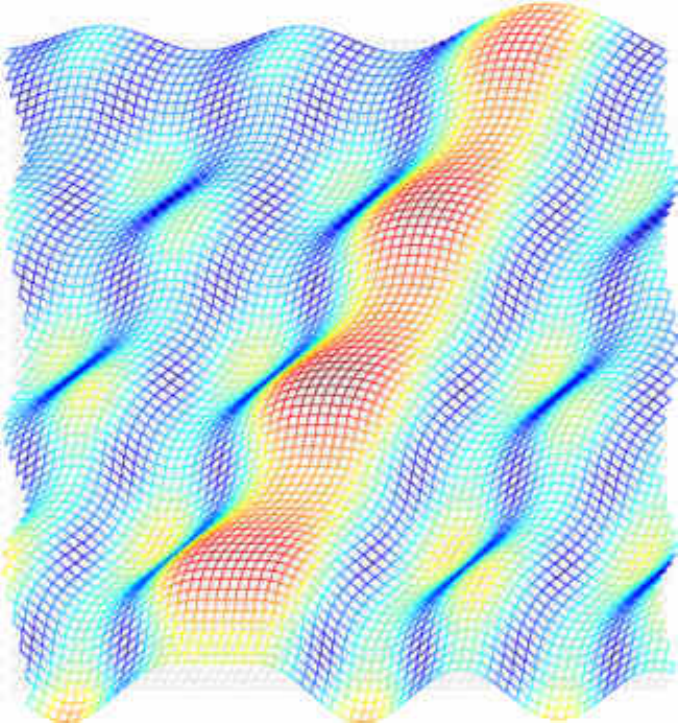}
    \end{subfigure}
    \begin{subfigure}{0.24\textwidth}
        \centering
        \includegraphics[width=0.98\linewidth]{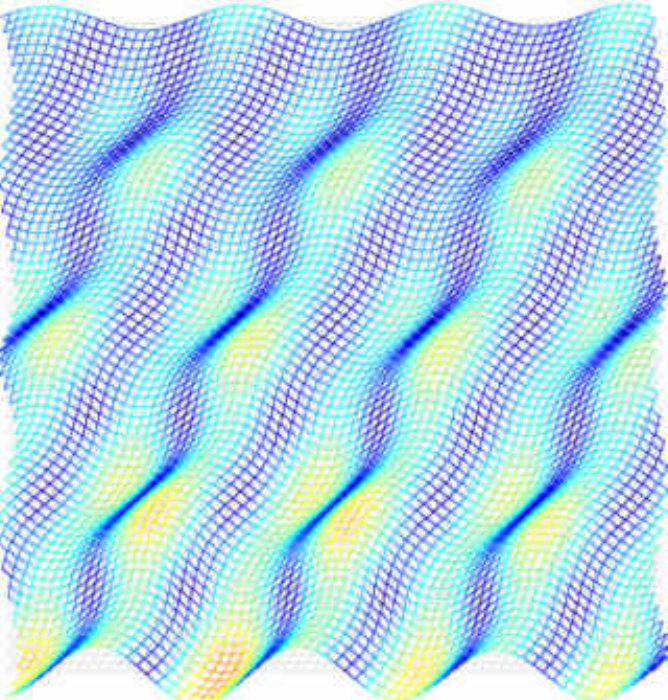}
    \end{subfigure}
    \caption{\label{fig:rhombus_10_10_deformed}
        Deformed configurations of the \textit{orthotropic} rhombic lattice ($\Lambda_1=\Lambda_2=10$) near the point of application of a pulsating concentrated force (the zones are indicated in Fig. \ref{fig:rhombus_10_10_force} \subref{fig:rhombus_10_10_force_h_99} and \subref{fig:rhombus_10_10_force_v_99}), at a level of prestress close to the elliptic boundary ($\bp=0.99\,\bp_{\text{E}}$).
        The pattern on the first (the second) row shows the motion of the localization induced by the pulsating horizontal (vertical) load.
        Note that the effect of the load is the generation of bands almost parallel to each other and possessing almost constant amplitude, except for a modulation of the inclined localization (shown in the lower row).
        In both cases the deformation mode is mostly of shear-type even though an `expansion' component is also present, as indicated by the vectors $\bg_{\text{E}}$, Fig.~\ref{fig:eigenvalue_rhombus_10_10}.
    }
\end{figure}
Note also that a slight misalignment between the localization direction and the rod angle remains hardly visible \textit{until the material is close to elliptic boundary} (compare for example the case of vanishing prestress, Fig.~\ref{fig:rhombus_10_10_force_h_0}, to the case $\bp=0.99\,\bp_{\text{E}}$, Fig.~\ref{fig:rhombus_10_10_force_h_99}).
\begin{figure}[htb!]
    \centering
    \begin{subfigure}{0.4\textwidth}
        \centering
        \caption{\label{fig:rhombus_10_10_force_h_FFT}Horizontal loading}
        \includegraphics[width=0.98\linewidth]{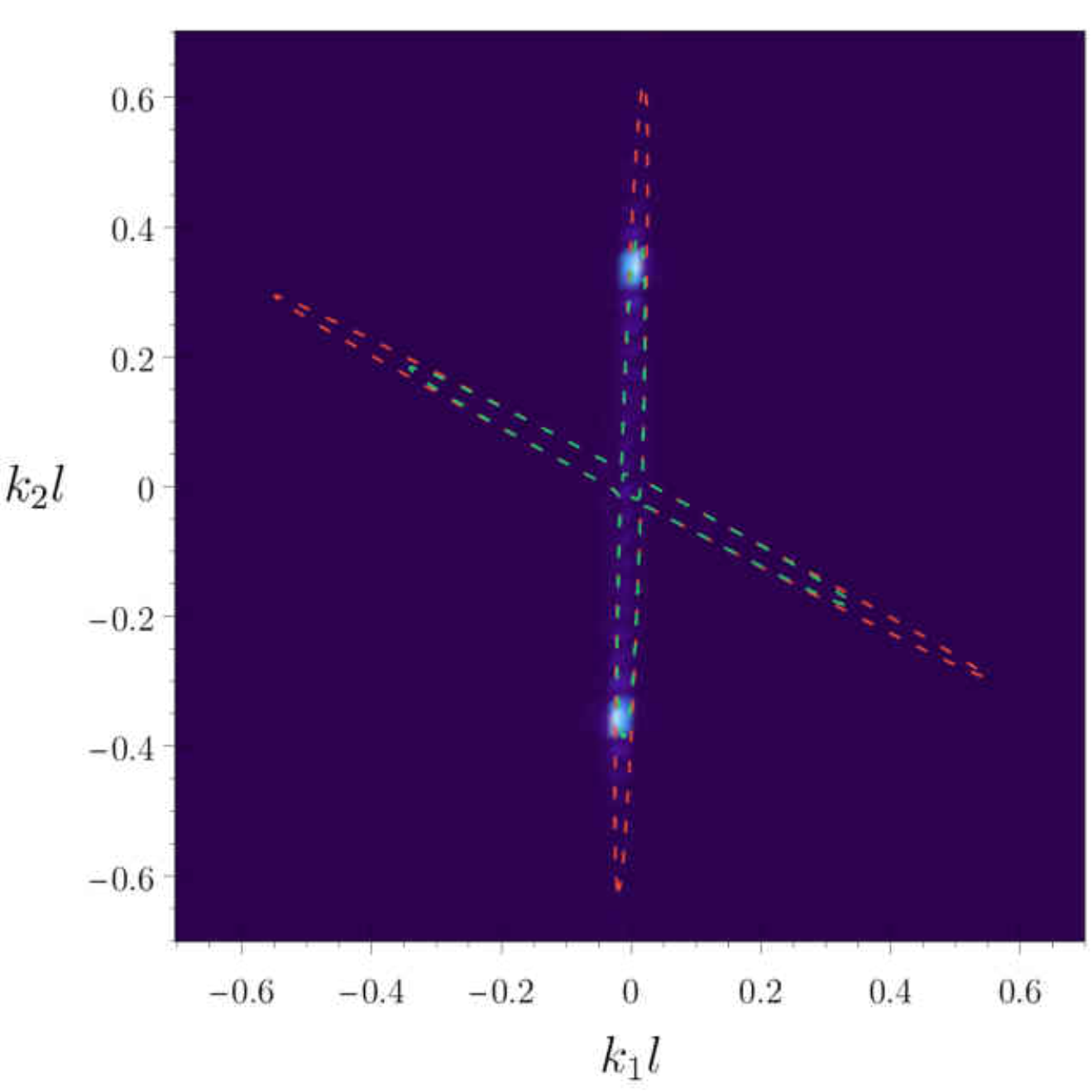}
    \end{subfigure} \hspace{2mm}
    \begin{subfigure}{0.4\textwidth}
        \centering
        \caption{\label{fig:rhombus_10_10_force_v_FFT}Vertical loading}
        \includegraphics[width=0.98\linewidth]{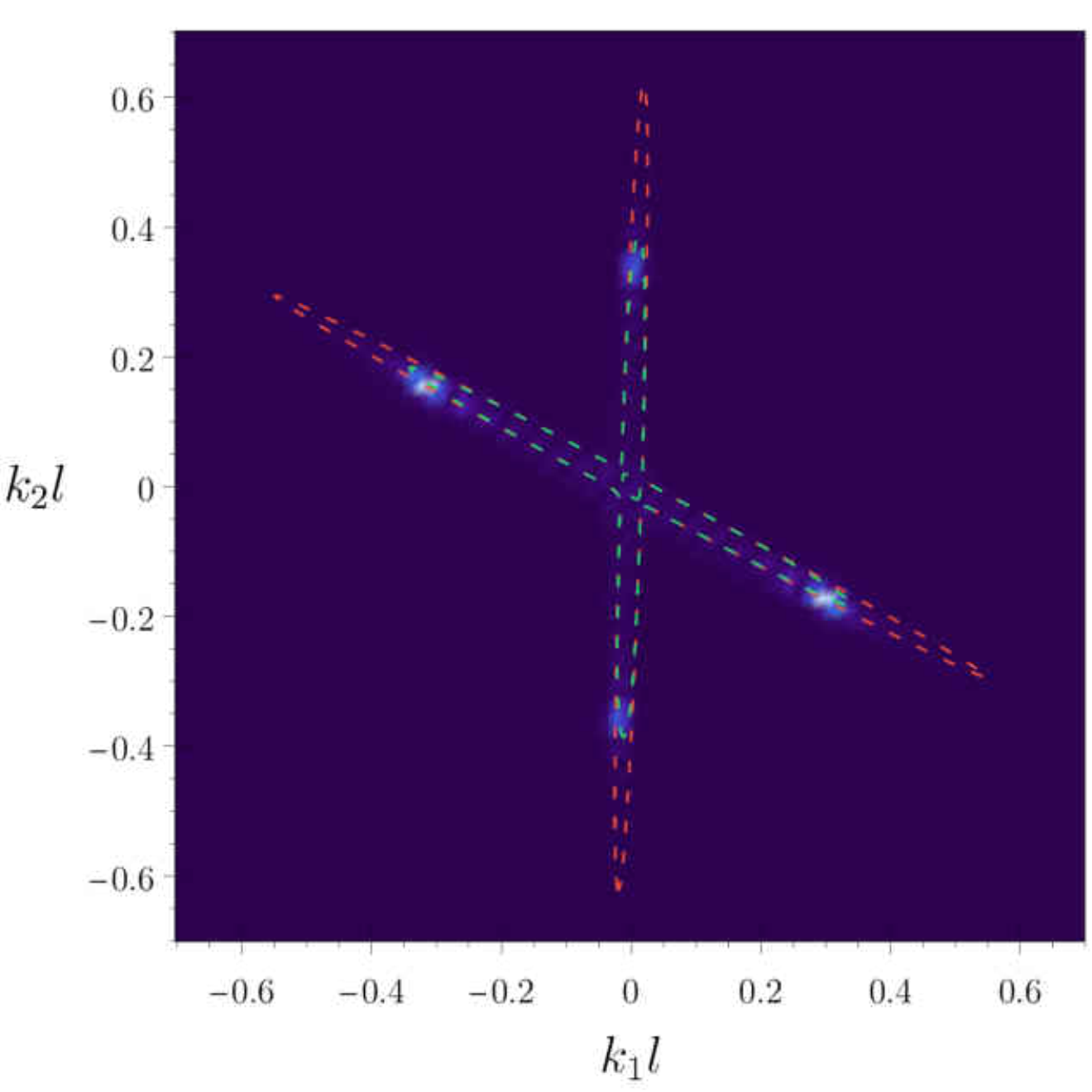}
    \end{subfigure}
    \caption{\label{fig:rhombus_10_10_force_FFT}
        Fourier transform of the complex displacement fields of the \textit{orthotropic} rhombic lattice (reported in Fig. \ref{fig:rhombus_10_10_force}  \subref{fig:rhombus_10_10_force_h_99} and \subref{fig:rhombus_10_10_force_v_99}, with $\Lambda_1=\Lambda_2=10$) subject to an horizontal~(\subref{fig:rhombus_10_10_force_h_FFT}) and vertical~(\subref{fig:rhombus_10_10_force_v_FFT}) pulsating force applied at a prestress close to the elliptic boundary, $\bp=0.99\,\bp_{\text{E}}$.
        The slowness contours of the lattice (dashed green) and of the effective continuum (dashed red) are superimposed to highlight the Bloch spectrum of waves excited by the forcing source.
        The sharp peaks in the contours are aligned to the directions of ellipticity loss as predicted in Fig.~\ref{fig:eigenvalue_rhombus_10_10}.
        In~(\subref{fig:rhombus_10_10_force_h_FFT}) the horizontal concentrated force only activates waves propagating at $\theta_{\text{cr}}=88.2^\circ$, while in~(\subref{fig:rhombus_10_10_force_v_FFT}) the vertical concentrated force generates four peaks along the directions $\theta_{\text{cr}}=88.2^\circ, 151.8^\circ$.
        This is in excellent agreement with the responses reported in Figs.~\ref{fig:rhombus_10_10_force_h_99} and~\ref{fig:rhombus_10_10_force_v_99}.
    }
\end{figure}
The localization modes are analyzed in Fig.~\ref{fig:rhombus_10_10_deformed} by inspecting the lattice deformation computed via f.e.m. at different temporal instants through snapshots taken in regions near the loading point (Figs.~\ref{fig:rhombus_10_10_force} (\subref{fig:rhombus_10_10_force_h_99}) and (\subref{fig:rhombus_10_10_force_v_99}) show the zones considered).

A comparison between the localization band induced by the horizontal load (upper row of Fig.~\ref{fig:rhombus_10_10_deformed}) and that generated by the vertical one (second row in Fig.~\ref{fig:rhombus_10_10_deformed}) shows that the (almost) horizontal band is characterized by an almost perfectly straight wavefront, while the inclined band displays a \textit{periodic modulation along the front}.
This modulation is due to the superposition of the two localization patterns that are activated by the vertical force, where the inclined band prevails over the almost horizontal one, as can be seen in Figs.~\ref{fig:rhombus_10_10_force_v_99} and~\ref{fig:rhombus_10_10_force_v_99_gf}.
\begin{figure}[htb!]
    \centering
    \begin{subfigure}{0.24\textwidth}
        \centering
        \phantomsubcaption{\label{fig:rhombus_7_15_force_h_0}}
        \includegraphics[width=0.98\linewidth]{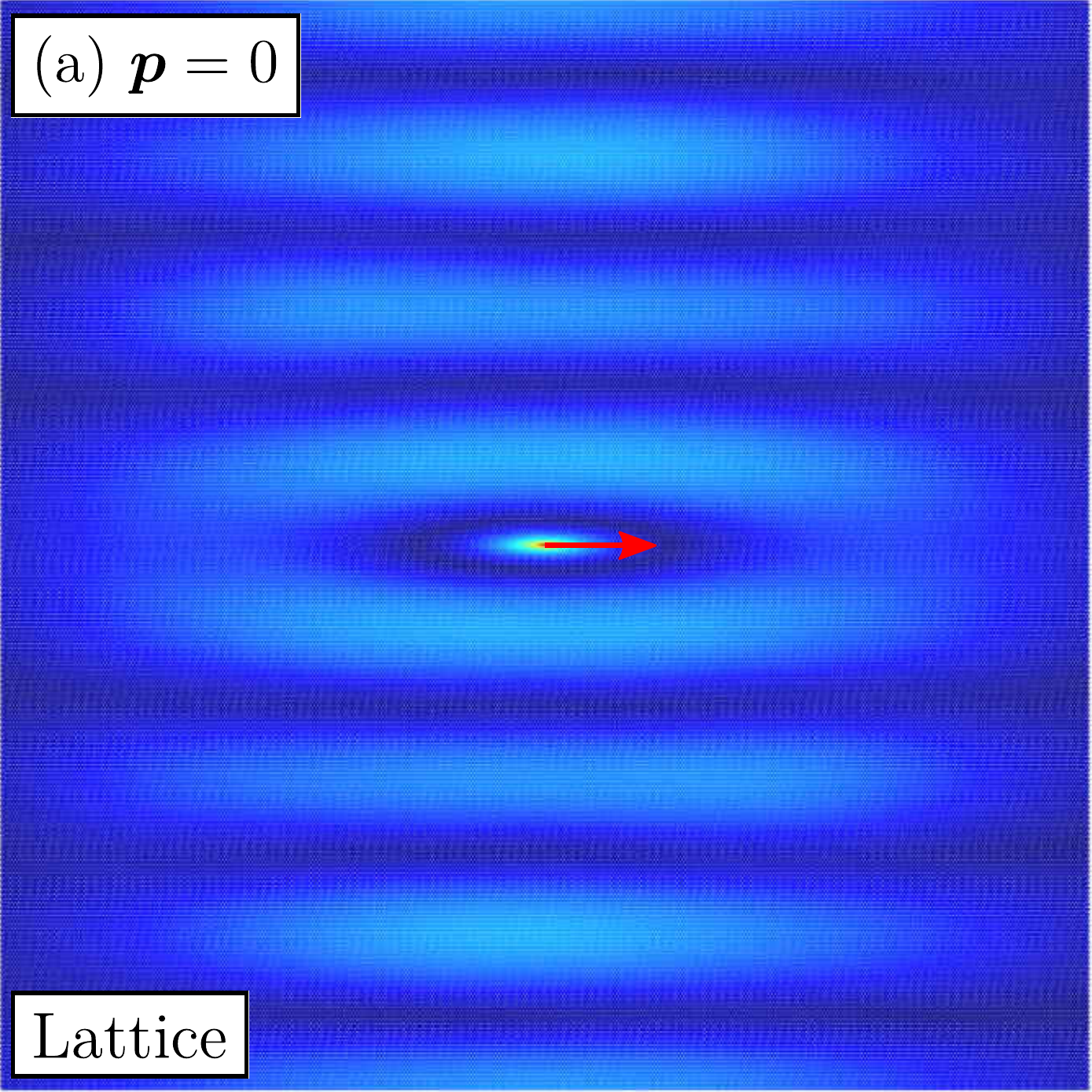}
    \end{subfigure}
    \begin{subfigure}{0.24\textwidth}
        \centering
        \phantomsubcaption{\label{fig:rhombus_7_15_force_h_80}}
        \includegraphics[width=0.98\linewidth]{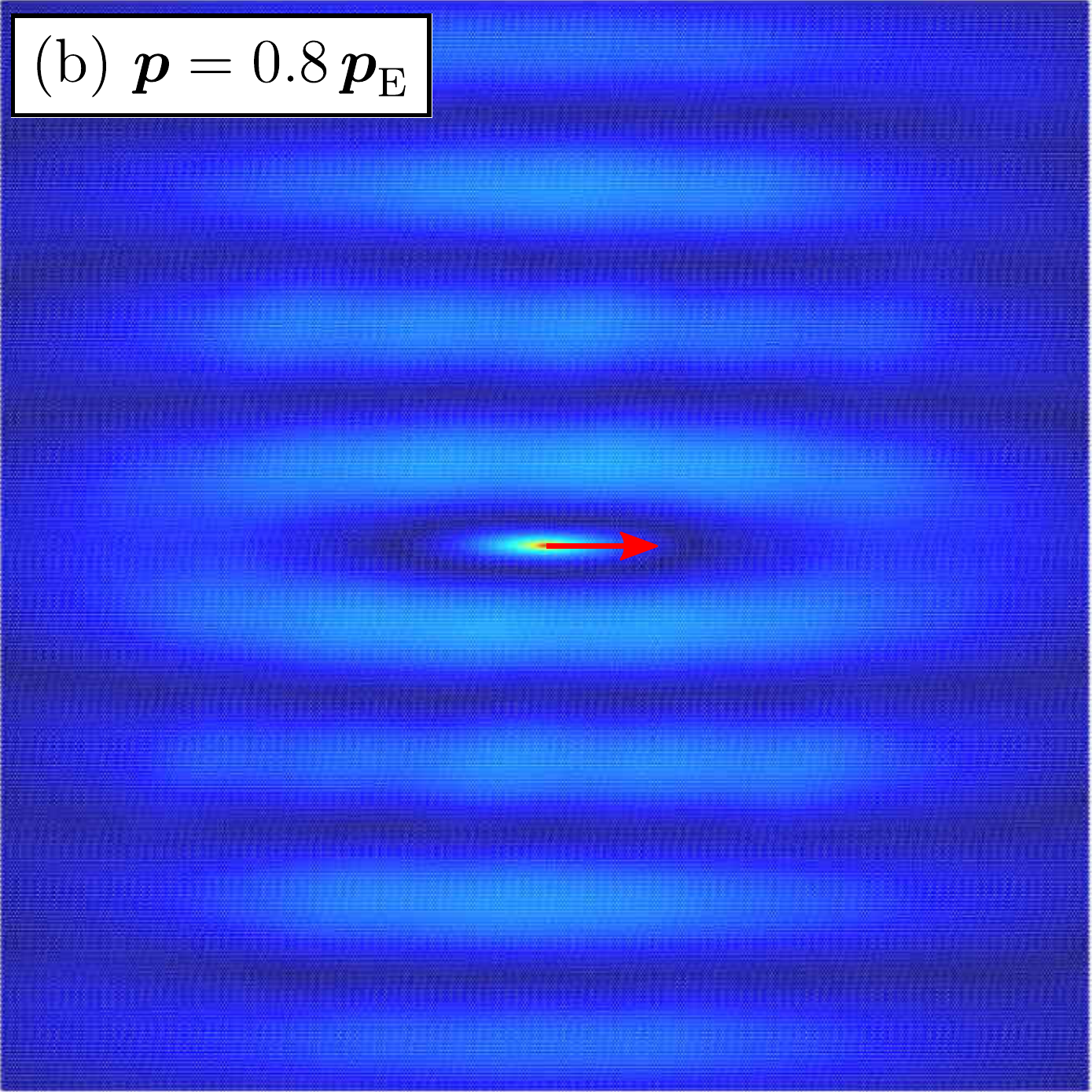}
    \end{subfigure}
    \begin{subfigure}{0.24\textwidth}
        \centering
        \phantomsubcaption{\label{fig:rhombus_7_15_force_h_90}}
        \includegraphics[width=0.98\linewidth]{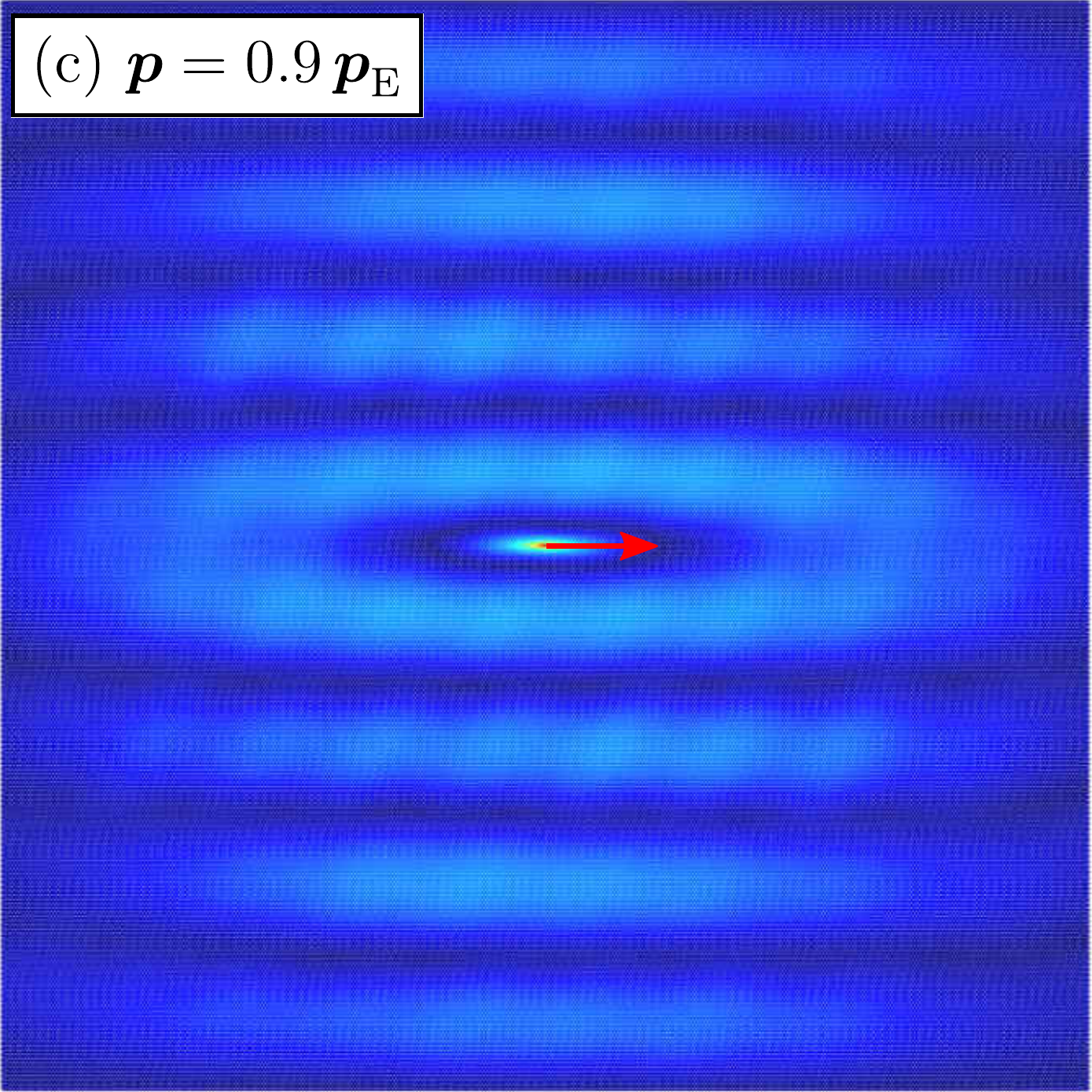}
    \end{subfigure}
    \begin{subfigure}{0.24\textwidth}
        \centering
        \phantomsubcaption{\label{fig:rhombus_7_15_force_h_99}}
        \includegraphics[width=0.98\linewidth]{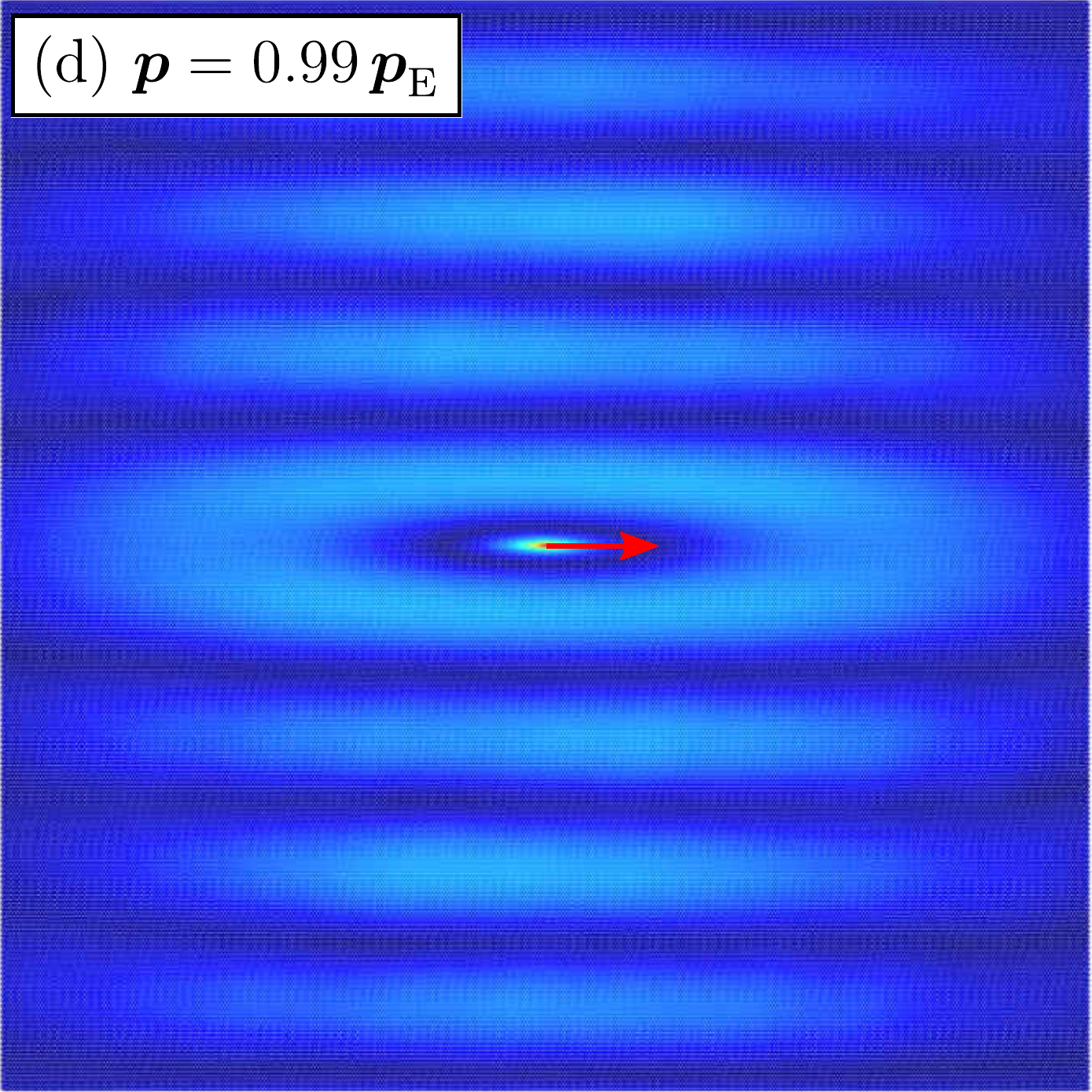}
    \end{subfigure}\\
    \vspace{0.01\linewidth}
    \begin{subfigure}{0.24\textwidth}
        \centering
        \phantomsubcaption{\label{fig:rhombus_7_15_force_h_0_gf}}
        \includegraphics[width=0.98\linewidth]{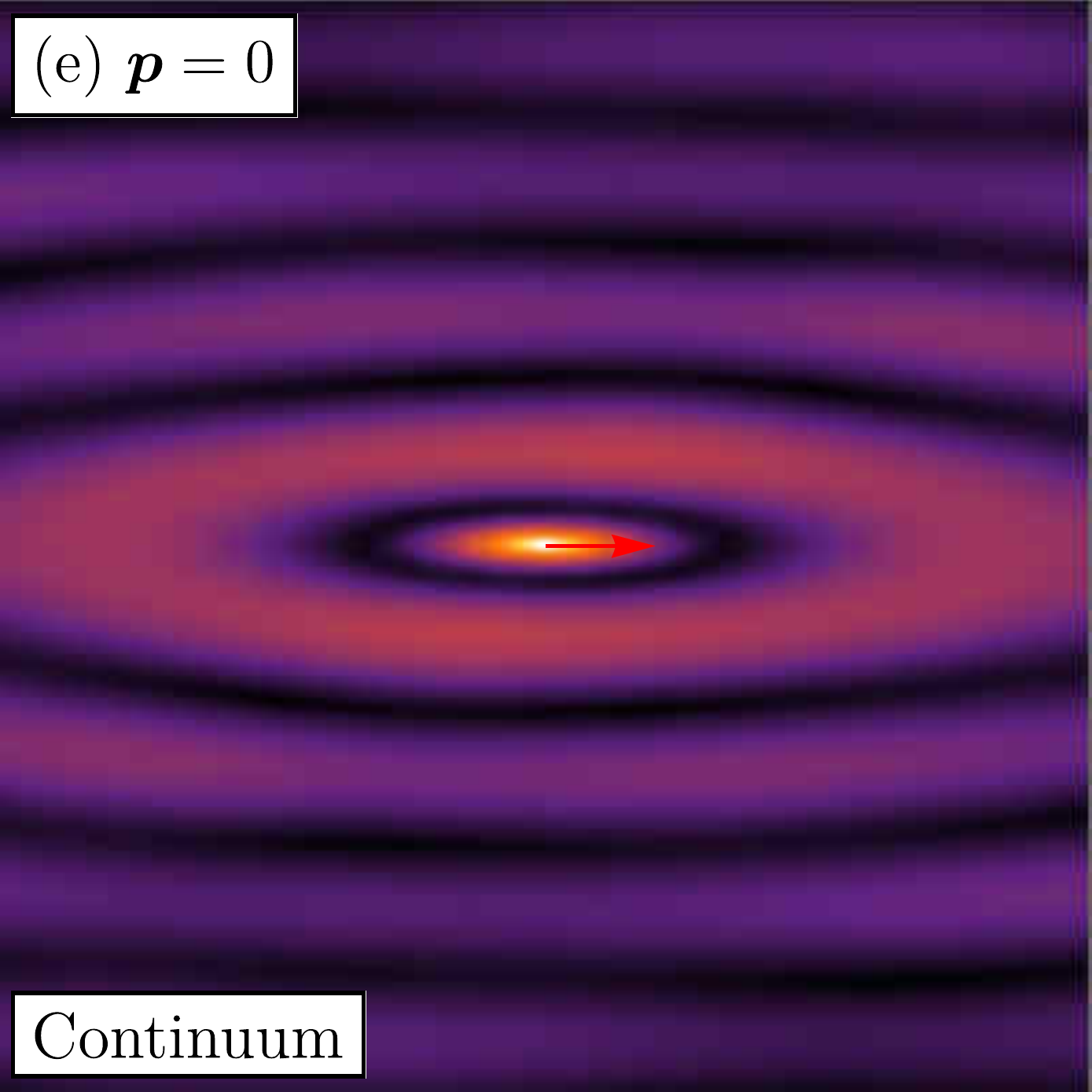}
    \end{subfigure}
    \begin{subfigure}{0.24\textwidth}
        \centering
        \phantomsubcaption{\label{fig:rhombus_7_15_force_h_80_gf}}
        \includegraphics[width=0.98\linewidth]{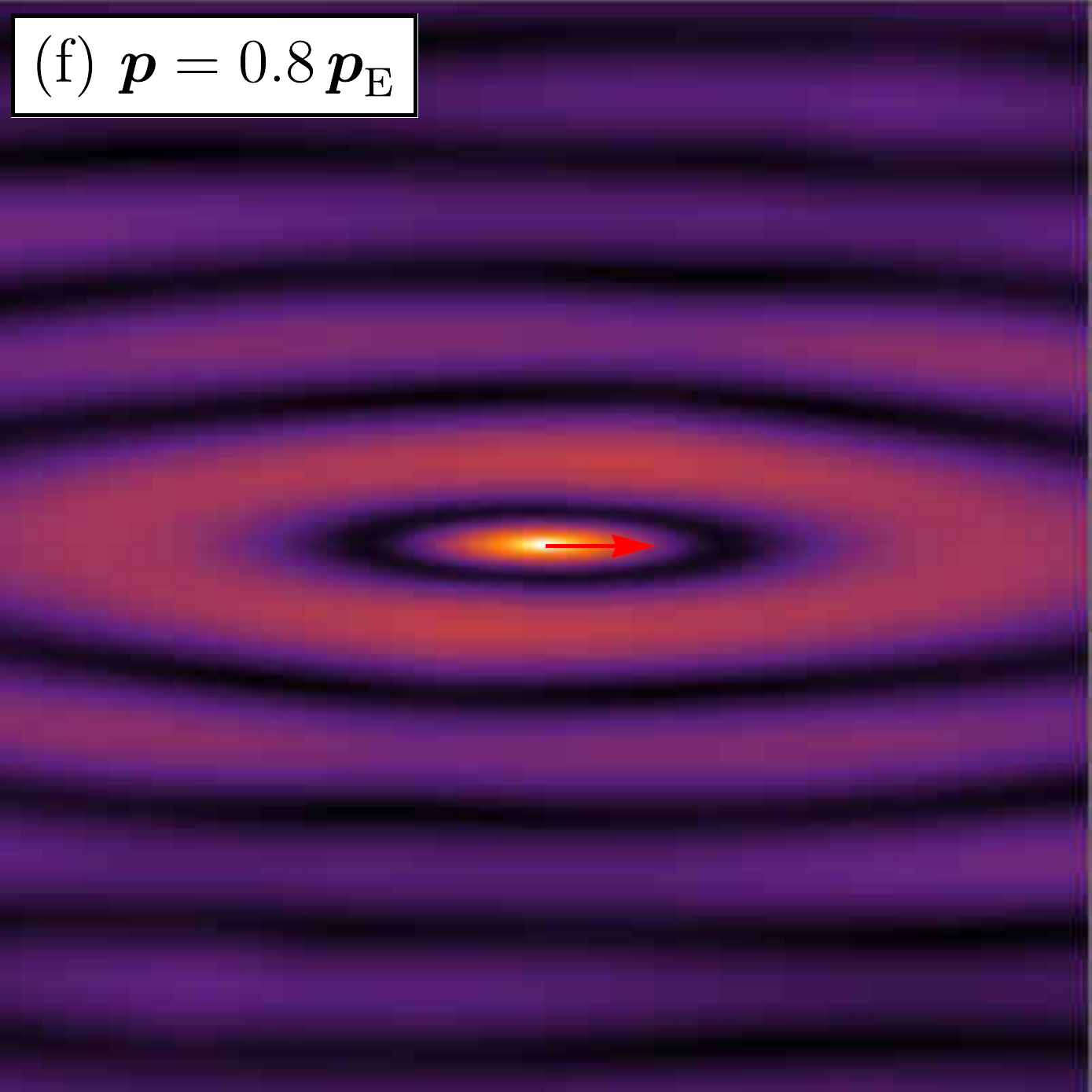}
    \end{subfigure}
    \begin{subfigure}{0.24\textwidth}
        \centering
        \phantomsubcaption{\label{fig:rhombus_7_15_force_h_90_gf}}
        \includegraphics[width=0.98\linewidth]{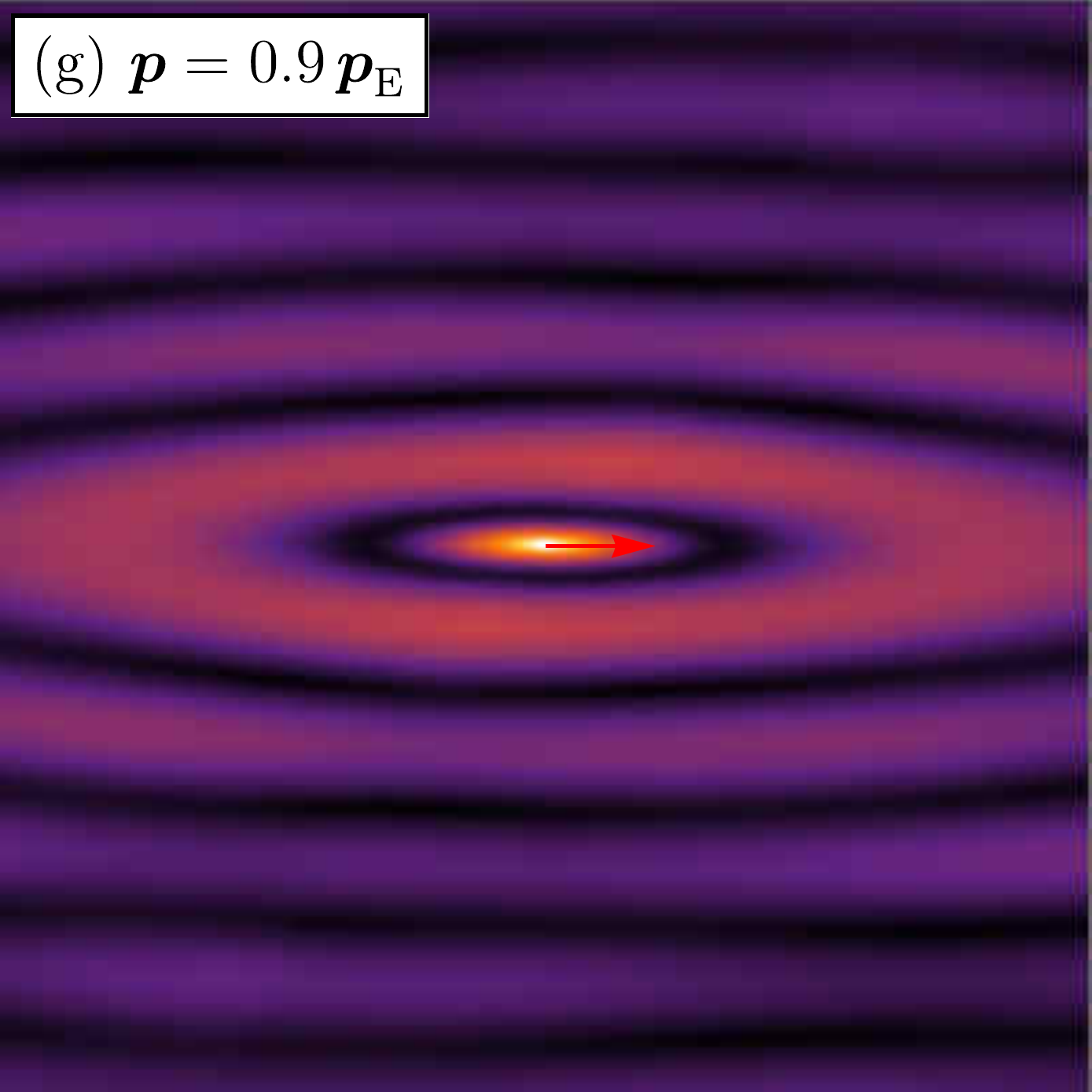}
    \end{subfigure}
    \vspace{0.015\linewidth}
    \begin{subfigure}{0.24\textwidth}
        \centering
        \phantomsubcaption{\label{fig:rhombus_7_15_force_h_99_gf}}
        \includegraphics[width=0.98\linewidth]{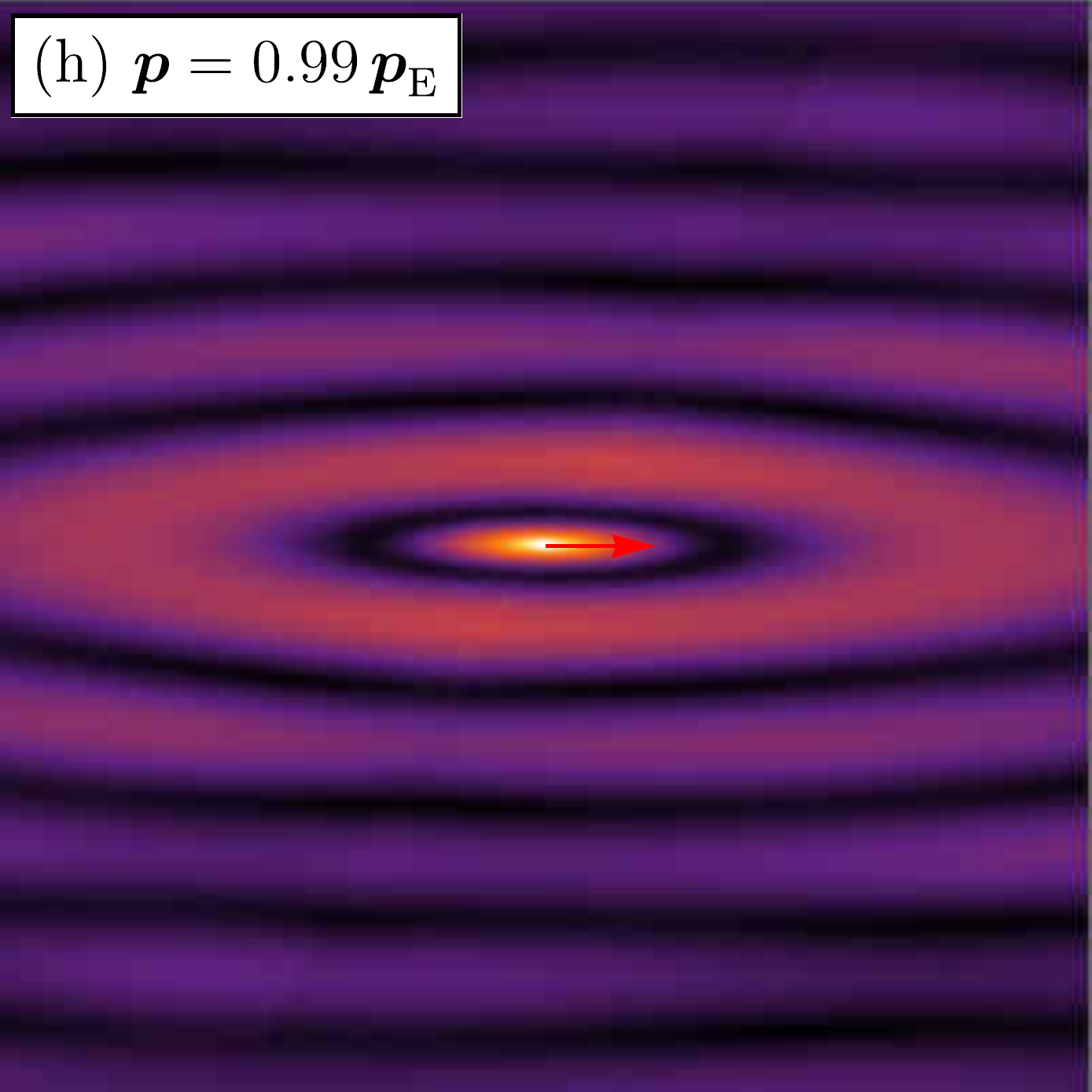}
    \end{subfigure}\\
    \begin{subfigure}{0.24\textwidth}
        \centering
        \phantomsubcaption{\label{fig:rhombus_7_15_force_v_0}}
        \includegraphics[width=0.98\linewidth]{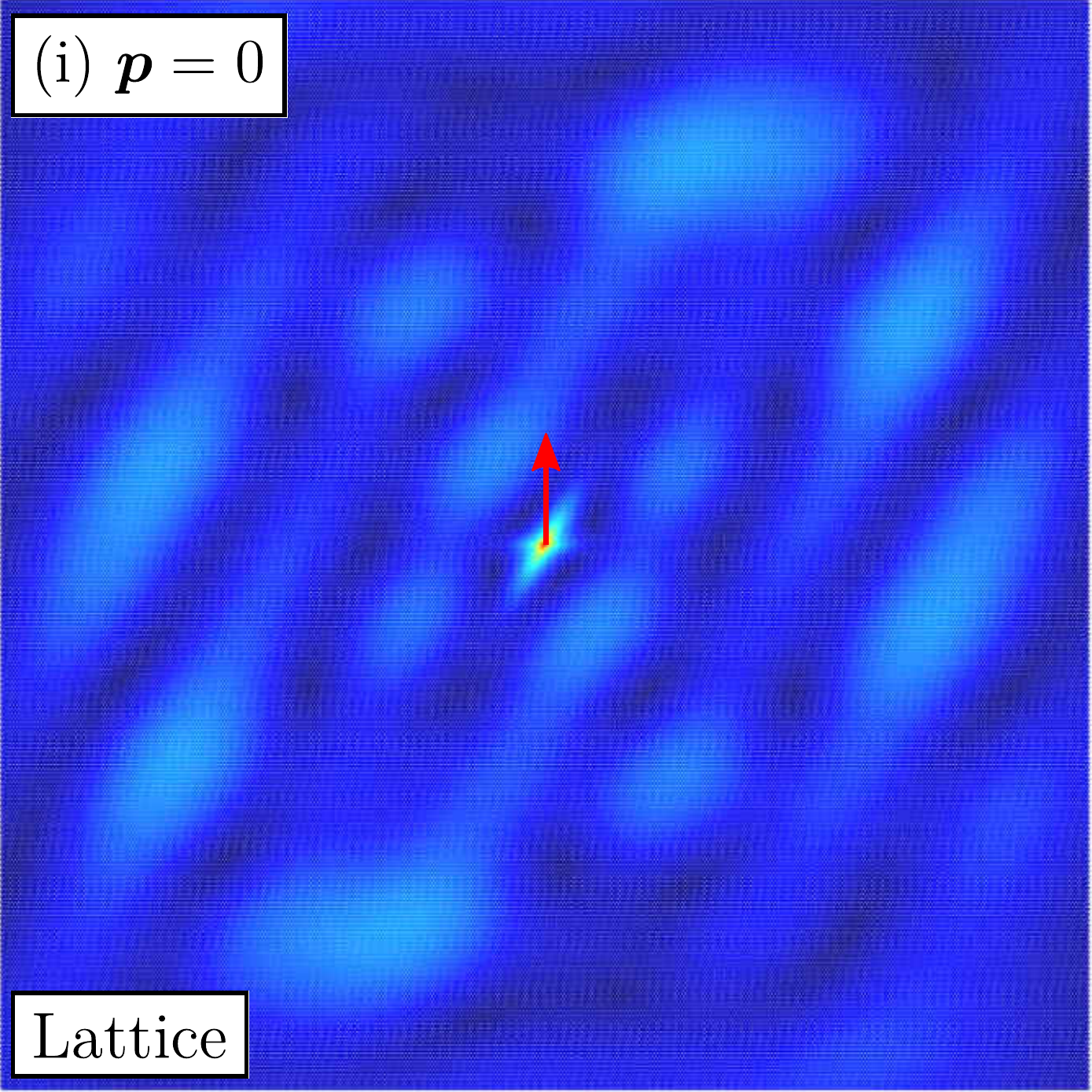}
    \end{subfigure}
    \begin{subfigure}{0.24\textwidth}
        \centering
        \phantomsubcaption{\label{fig:rhombus_7_15_force_v_80}}
        \includegraphics[width=0.98\linewidth]{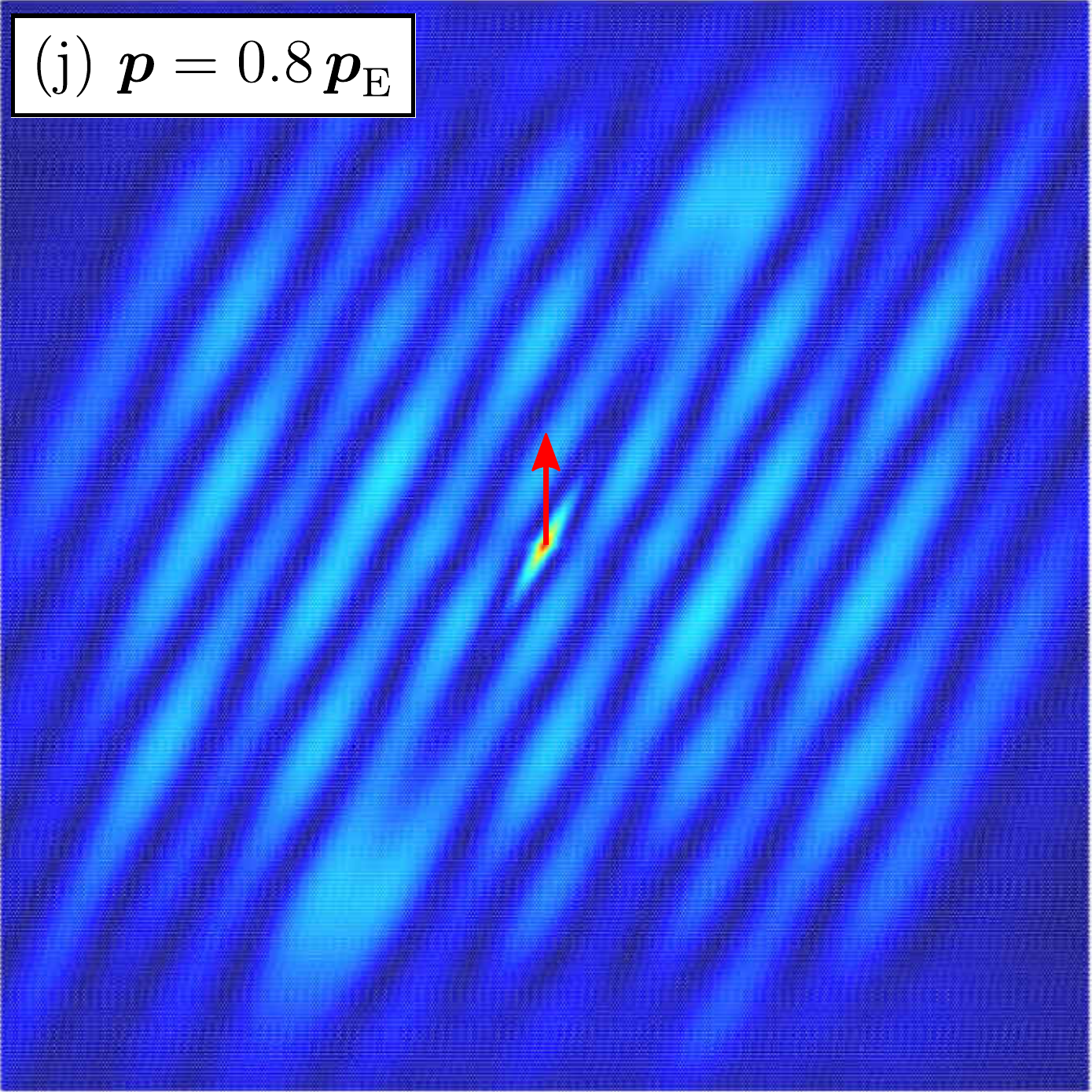}
    \end{subfigure}
    \begin{subfigure}{0.24\textwidth}
        \centering
        \phantomsubcaption{\label{fig:rhombus_7_15_force_v_90}}
        \includegraphics[width=0.98\linewidth]{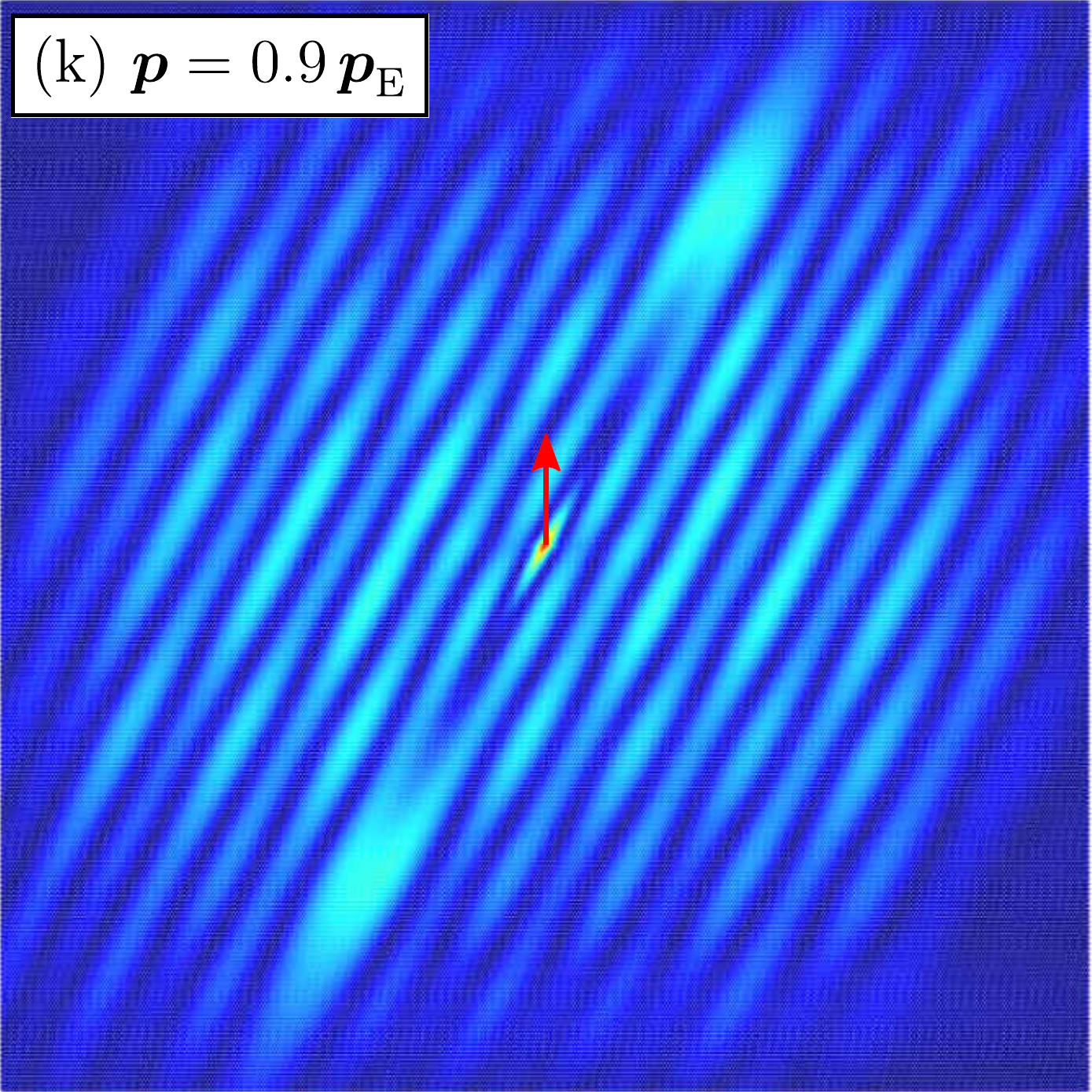}
    \end{subfigure}
    \begin{subfigure}{0.24\textwidth}
        \centering
        \phantomsubcaption{\label{fig:rhombus_7_15_force_v_99}}
        \includegraphics[width=0.98\linewidth]{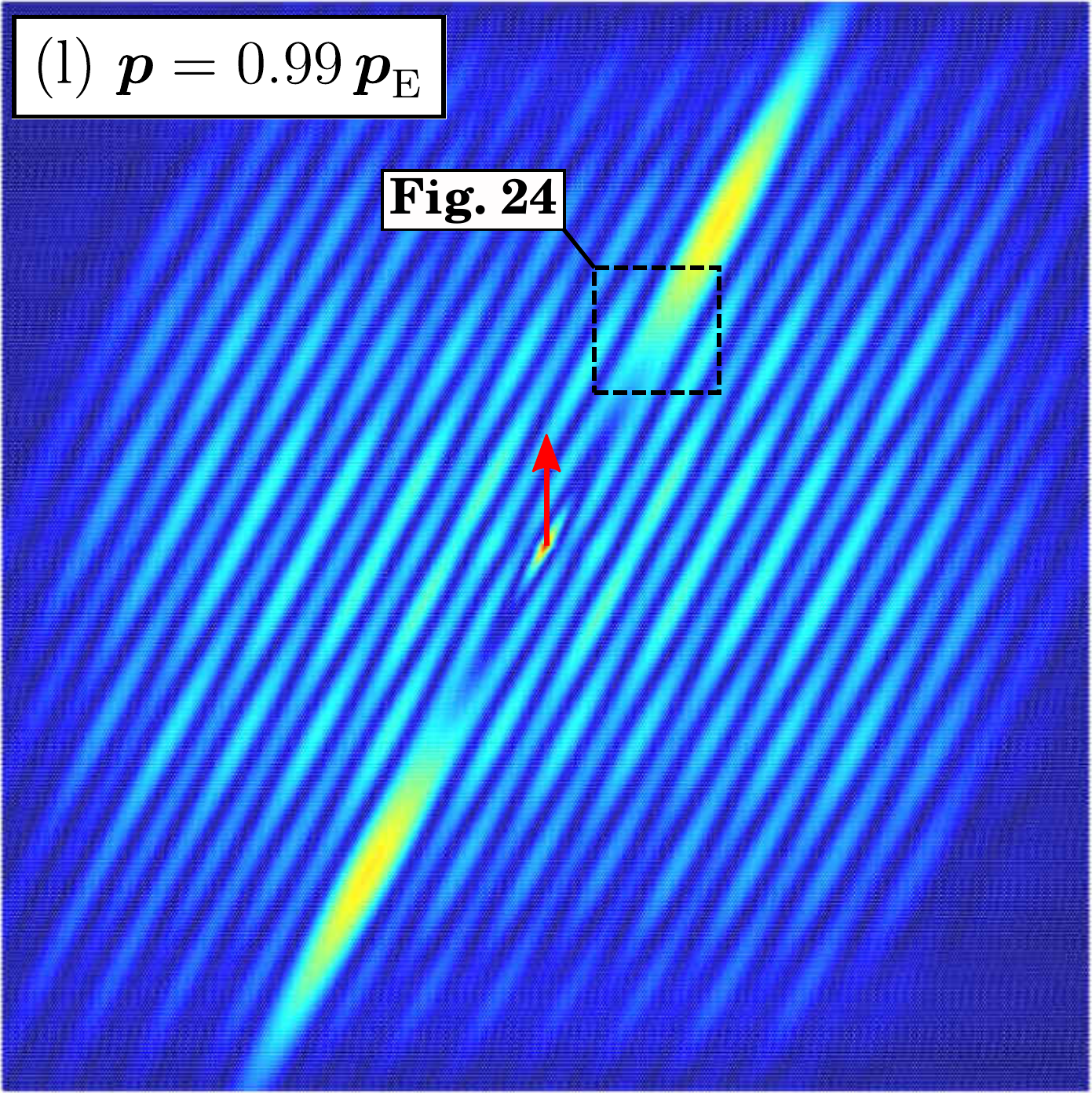}
    \end{subfigure}\\
    \vspace{0.01\linewidth}
    \begin{subfigure}{0.24\textwidth}
        \centering
        \phantomsubcaption{\label{fig:rhombus_7_15_force_v_0_gf}}
        \includegraphics[width=0.98\linewidth]{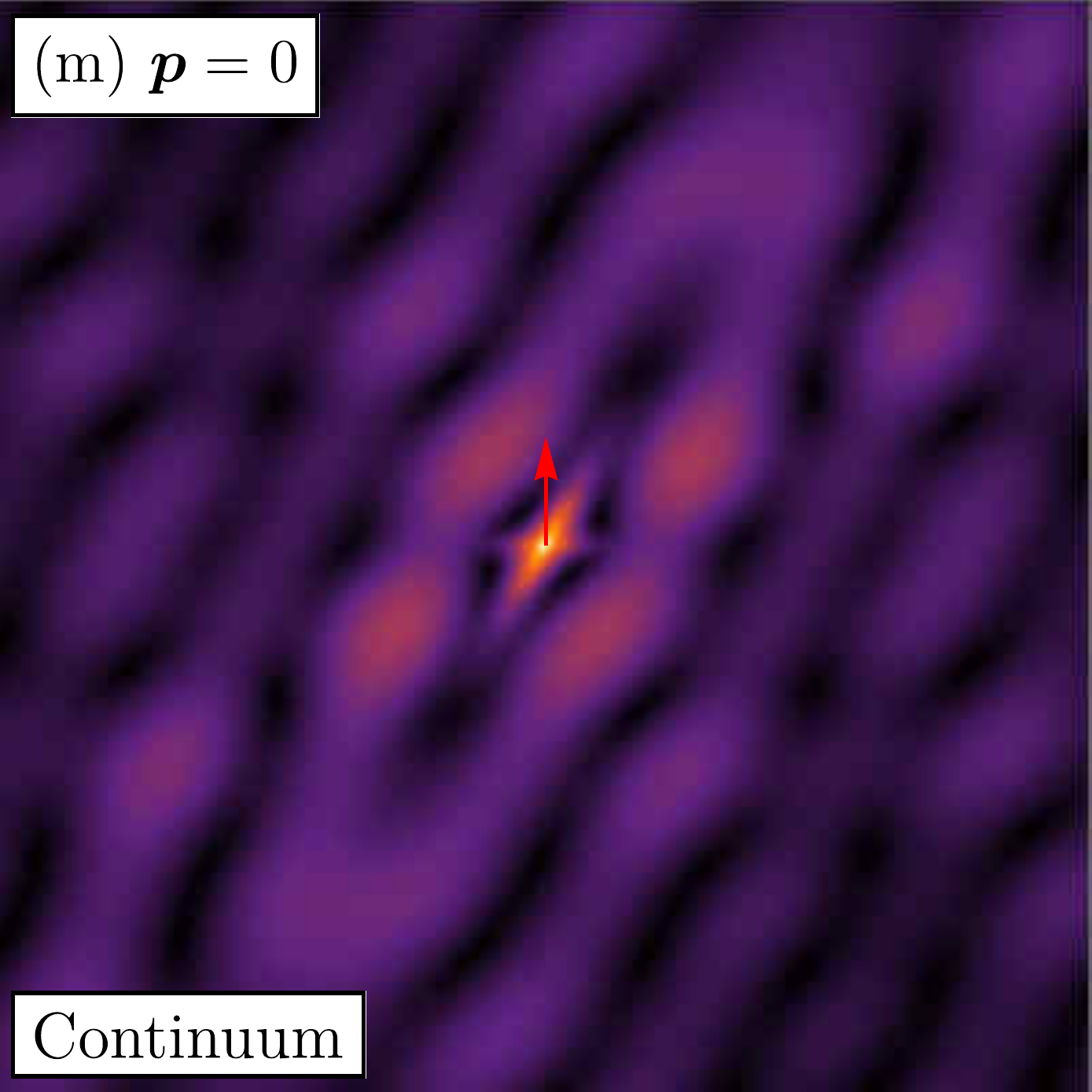}
    \end{subfigure}
    \begin{subfigure}{0.24\textwidth}
        \centering
        \phantomsubcaption{\label{fig:rhombus_7_15_force_v_80_gf}}
        \includegraphics[width=0.98\linewidth]{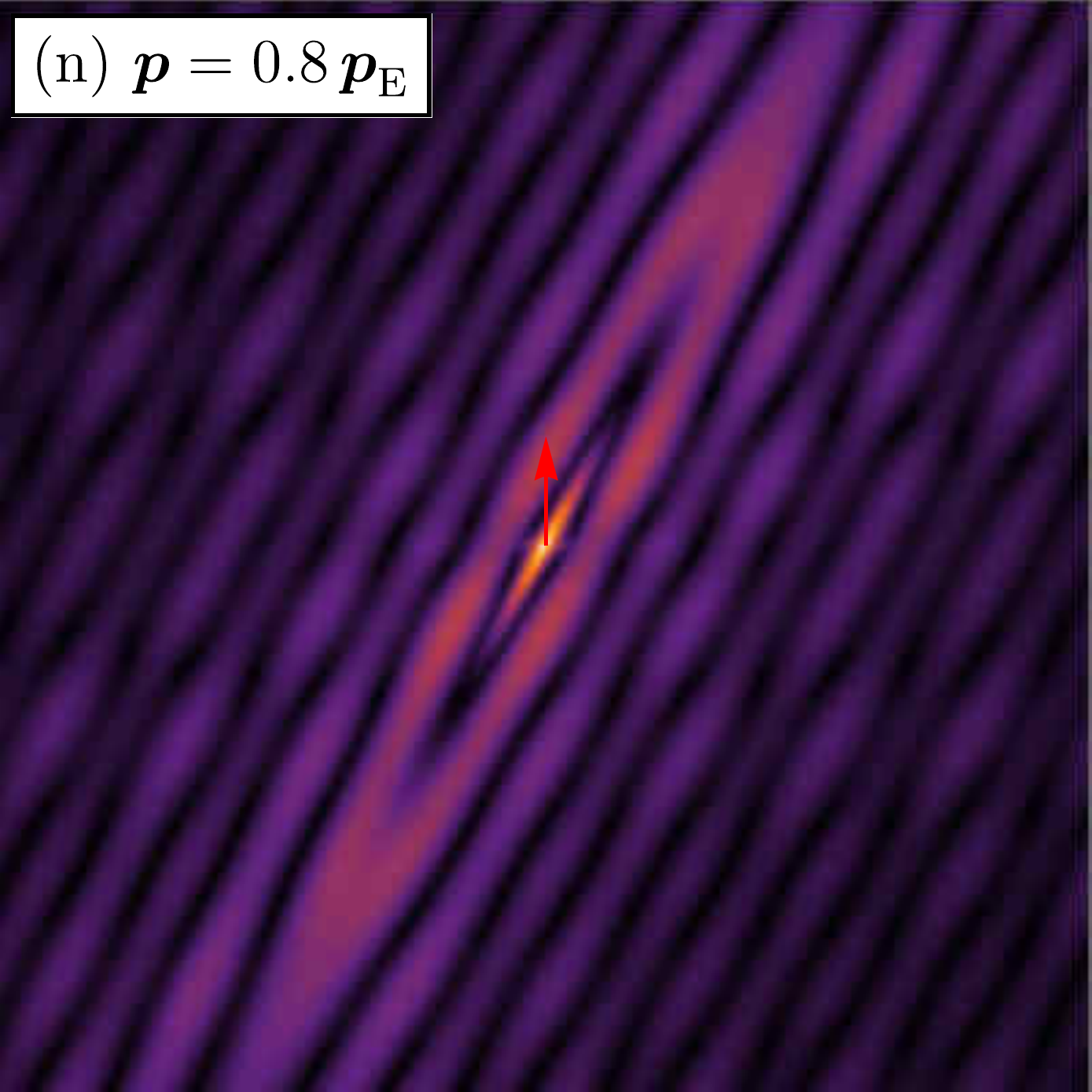}
    \end{subfigure}
    \begin{subfigure}{0.24\textwidth}
        \centering
        \phantomsubcaption{\label{fig:rhombus_7_15_force_v_90_gf}}
        \includegraphics[width=0.98\linewidth]{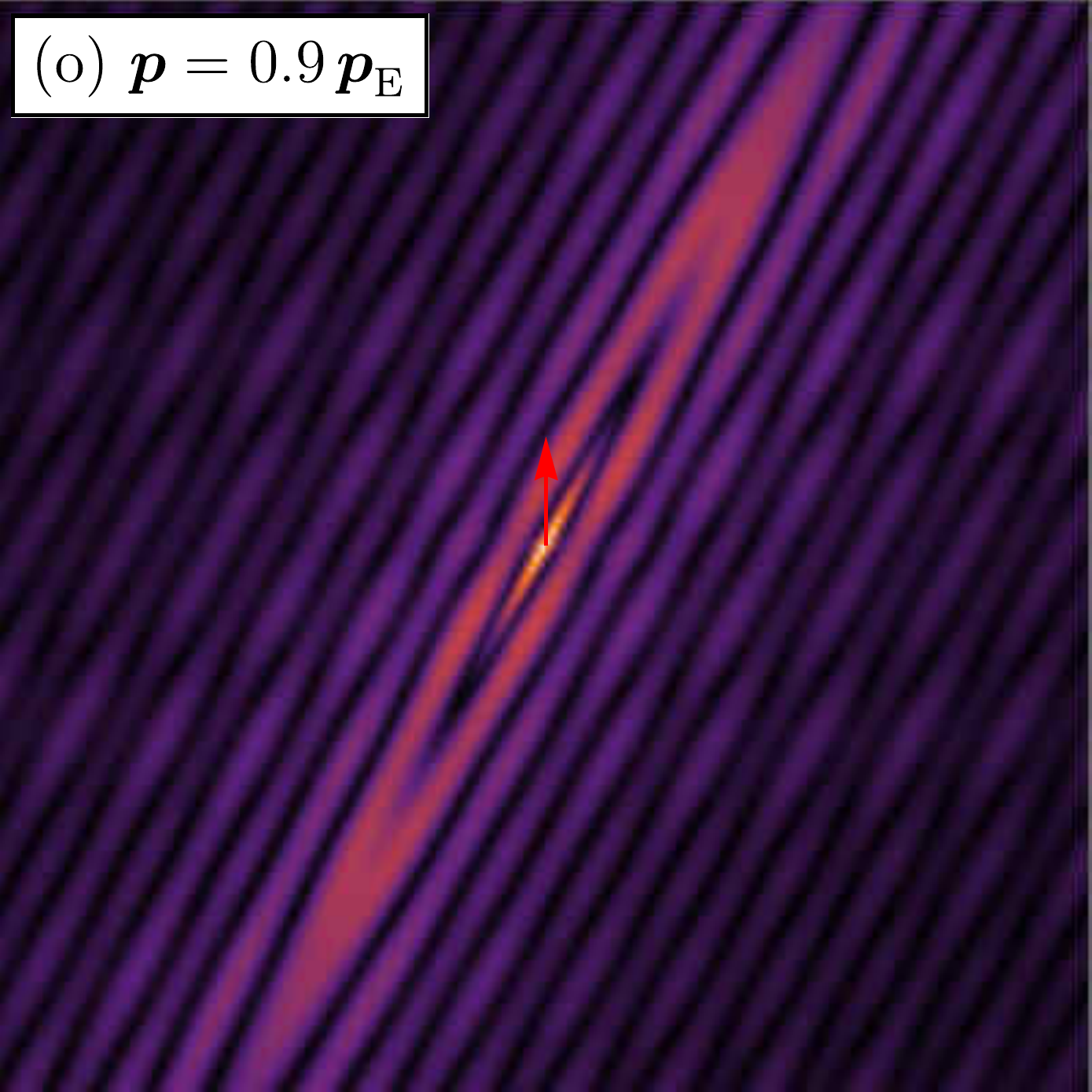}
    \end{subfigure}
    \begin{subfigure}{0.24\textwidth}
        \centering
        \phantomsubcaption{\label{fig:rhombus_7_15_force_v_99_gf}}
        \includegraphics[width=0.98\linewidth]{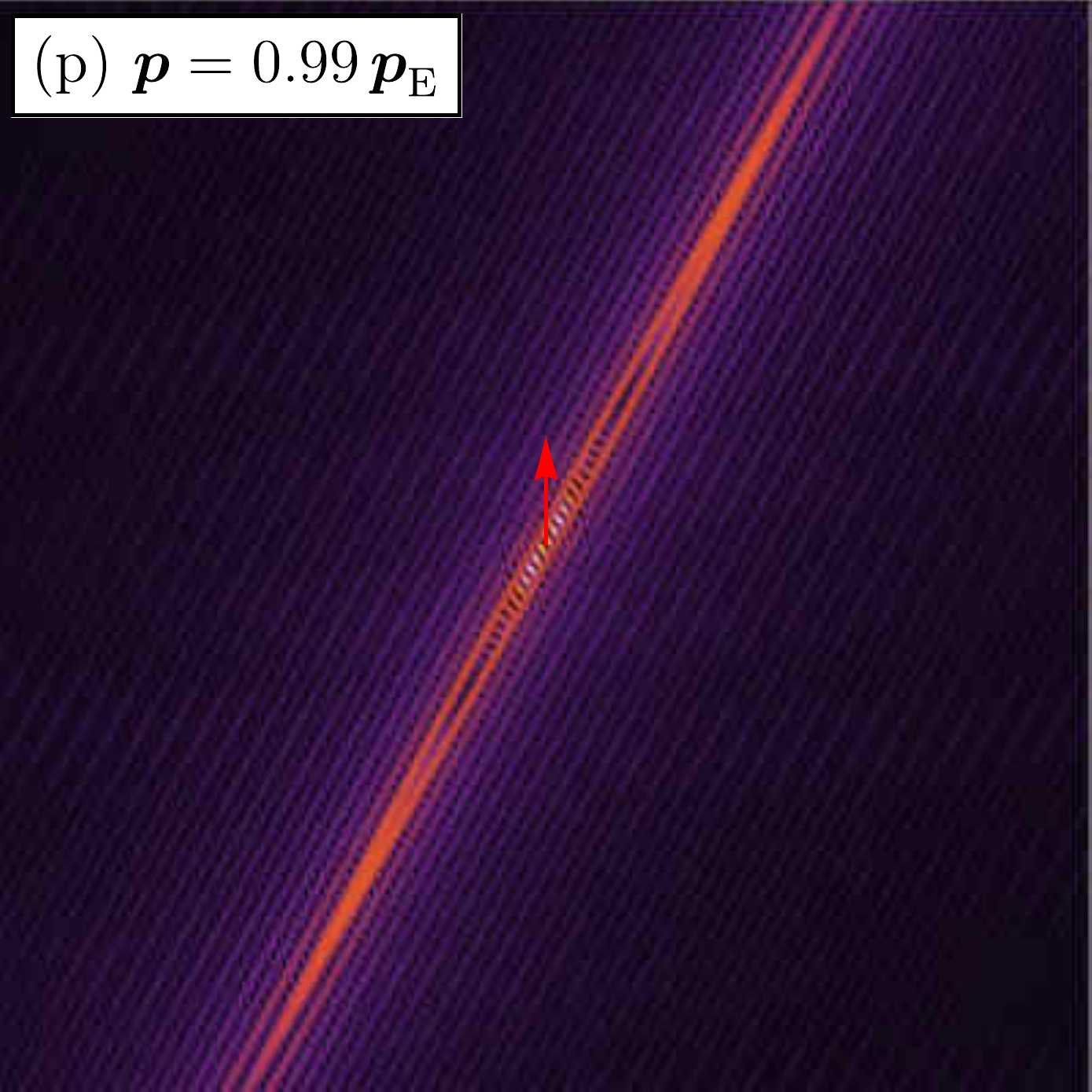}
    \end{subfigure}
    \caption{\label{fig:rhombus_7_15_force}
        As for Fig. \ref{fig:rhombus_10_10_force}, except that an \textit{anisotropic} rhombic lattice ($\Lambda_1=7,\,\Lambda_2=15$) is considered.
        Note that, even though ellipticity is lost along the direction predicted in Fig.~\ref{fig:eigenvalue_rhombus_7_15} (with normal angle $\theta_{\text{cr}}=151.4^\circ$), the activation of the strain localization depends on the orientation of the applied pulsating force, so that the band is activated by the vertical force while localization is inhibited when the grid is loaded horizontally.
    }
\end{figure}

The relative contribution of the two localizations can be further investigated through a Fourier transform of the lattice response, to be compared with the Bloch spectrum generated by the forcing source.
Fig.~\ref{fig:rhombus_10_10_force_FFT} shows the Fourier transform of the field generated in the rhombic grid when the material is close to the elliptic boundary ($\bp=0.99\,\bp_{\text{E}}$).
Fig.~\ref{fig:rhombus_10_10_force_h_FFT} and~\ref{fig:rhombus_10_10_force_v_FFT} correspond, respectively, to the Fourier transform of Fig.~\ref{fig:rhombus_10_10_force_h_99} and Fig.~\ref{fig:rhombus_10_10_force_v_99}.
The two sharp peaks of Fig.~\ref{fig:rhombus_10_10_force_h_FFT} clearly show that the source is emanating pure plane waves propagating almost vertically ($\theta_{\text{cr}}=88.2^\circ$) (the slight tilt exactly matches the sub-horizontal wavefronts of the response).
Instead, the four peaks of Fig.~\ref{fig:rhombus_10_10_force_v_FFT} demonstrate that two families of plane waves are activated: the prevailing ones propagate along the inclined direction ($\theta_{\text{cr}}=151.8^\circ$) while the vertically-propagating waves result dimmer (in agreement with the response of Fig.~\ref{fig:rhombus_10_10_force_v_99}).

Results pertaining to an \textit{anisotropic} rhombic lattice ($\Lambda_1=7,\,\Lambda_2=15$, while the other parameters are equal to those used to generate Fig.~\ref{fig:rhombus_10_10_force}, relative to an orthotropic rhombic lattice) are reported in Fig.~\ref{fig:rhombus_7_15_force} for four values of prestress, $p_1=p_2=\{0,-1.634,-1.839,-2.023\}$.
\begin{figure}[htb!]
    \centering
    \begin{subfigure}{0.24\textwidth}
        \centering
        \caption*{$t=0$}
        \includegraphics[width=0.98\linewidth]{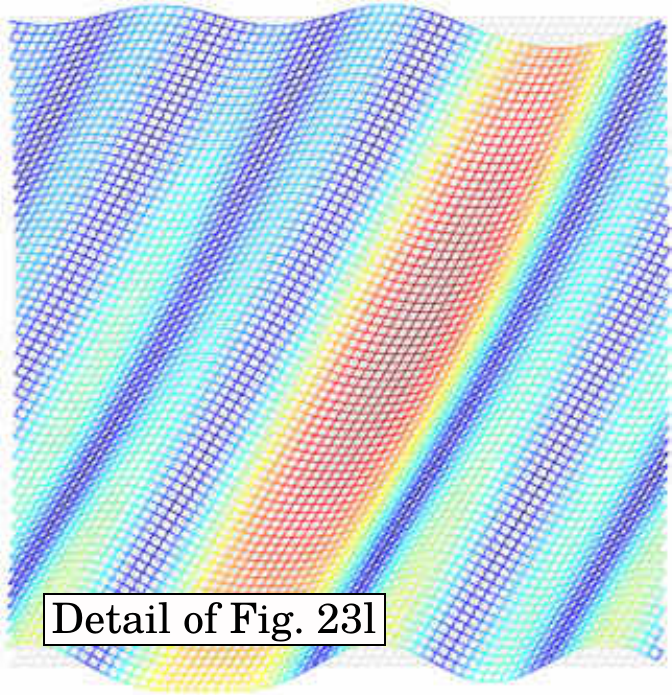}
    \end{subfigure}
    \begin{subfigure}{0.24\textwidth}
        \centering
        \caption*{$t=\frac{\pi}{2\omega}$}
        \includegraphics[width=0.98\linewidth]{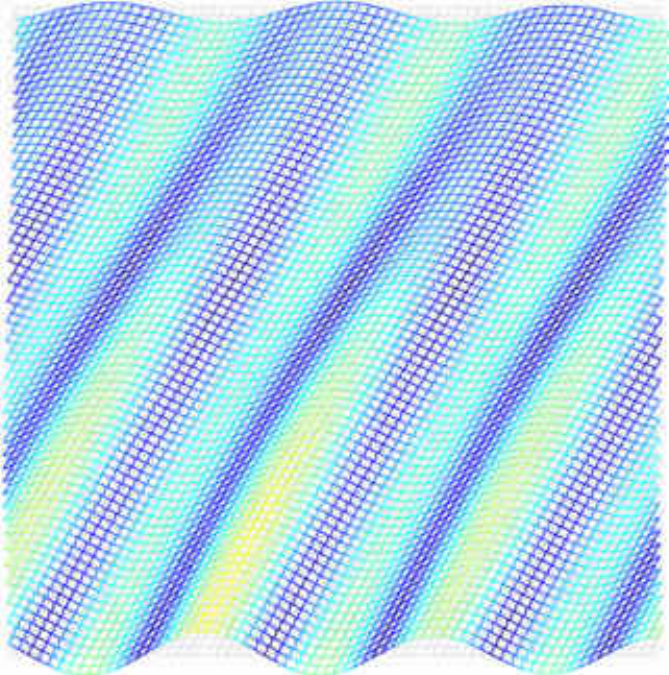}
    \end{subfigure}
    \begin{subfigure}{0.24\textwidth}
        \centering
        \caption*{$t=\frac{\pi}{\omega}$}
        \includegraphics[width=0.98\linewidth]{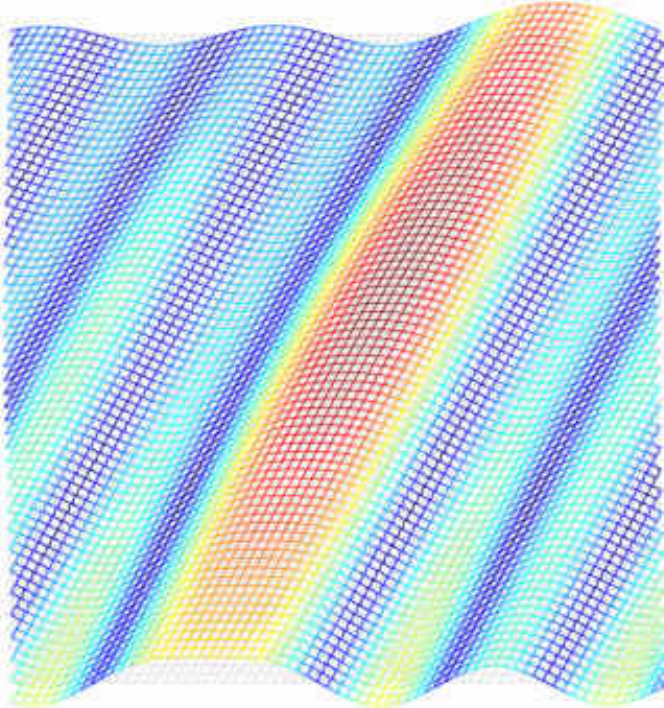}
    \end{subfigure}
    \begin{subfigure}{0.24\textwidth}
        \centering
        \caption*{$t=\frac{3\pi}{2\omega}$}
        \includegraphics[width=0.98\linewidth]{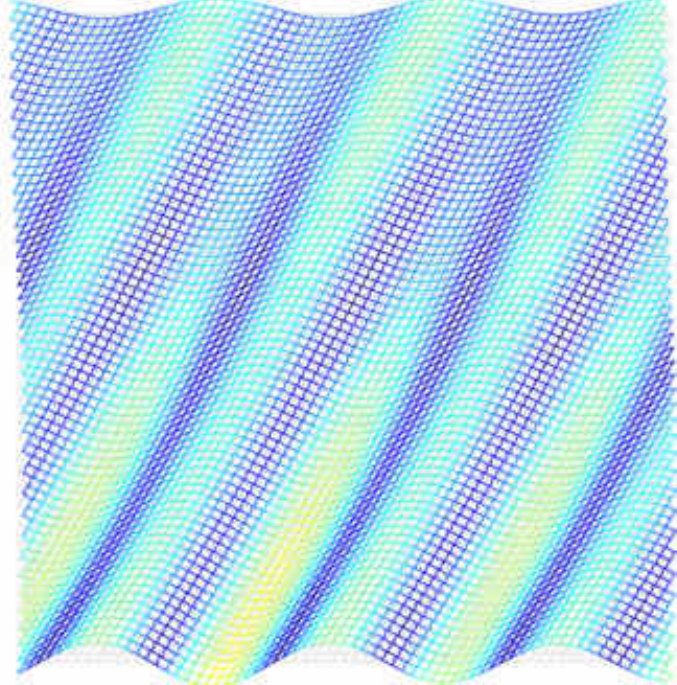}
    \end{subfigure}
    \caption{\label{fig:rhombus_7_15_deformed}
        As for Fig.~\ref{fig:rhombus_10_10_deformed}, except that an \textit{anisotropic} rhombic lattice ($\Lambda_1=7,\,\Lambda_2=15$) is considered.
        The zone reported in the figure is shown in Fig.~\ref{fig:rhombus_7_15_force_v_99}.
        The pattern shows the motion of the localization induced by the pulsating vertical load, which generates waves emanating outwards from the localization band, parallel to each other, and possessing an almost constant amplitude.
        The deformation mode is mostly of shear-type even though an `expansion' component is also present, as indicated by the vector $\bg_{\text{E}}$ reported in Fig.~\ref{fig:eigenvalue_rhombus_7_15}.
    }
\end{figure}

For a completely anisotropic material only a single localization is expected to occur and in the case of the anisotropic grid considered the localization direction has been predicted in Fig.~\ref{fig:eigenvalue_rhombus_7_15} to occur at an inclination angle $\theta_{\text{cr}}=151.4^\circ$ of the band normal.
However, similarly to the case of the orthotropic grid, the activation of the localization depends on the orientation of the perturbing force.
This can be observed by comparing the lattice response generated by a horizontal and a vertical pulsating force, both reported in Fig.~\ref{fig:rhombus_7_15_force} and showing that strain localization is absent when a horizontal force is applied, regardless of the prestress level (see Figs.~\ref{fig:rhombus_7_15_force_h_0}--\ref{fig:rhombus_7_15_force_h_99_gf}). On the other hand, the vertical concentrated force triggers an inclined localization when the material is brought close to ellipticity loss (see Figs.~\ref{fig:rhombus_7_15_force_v_0}--\ref{fig:rhombus_7_15_force_v_99_gf}).

\begin{figure}[htb!]
    \centering
    \begin{subfigure}{0.4\textwidth}
        \centering
        \caption{\label{fig:rhombus_7_15_force_h_FFT}Horizontal loading}
        \includegraphics[width=0.98\linewidth]{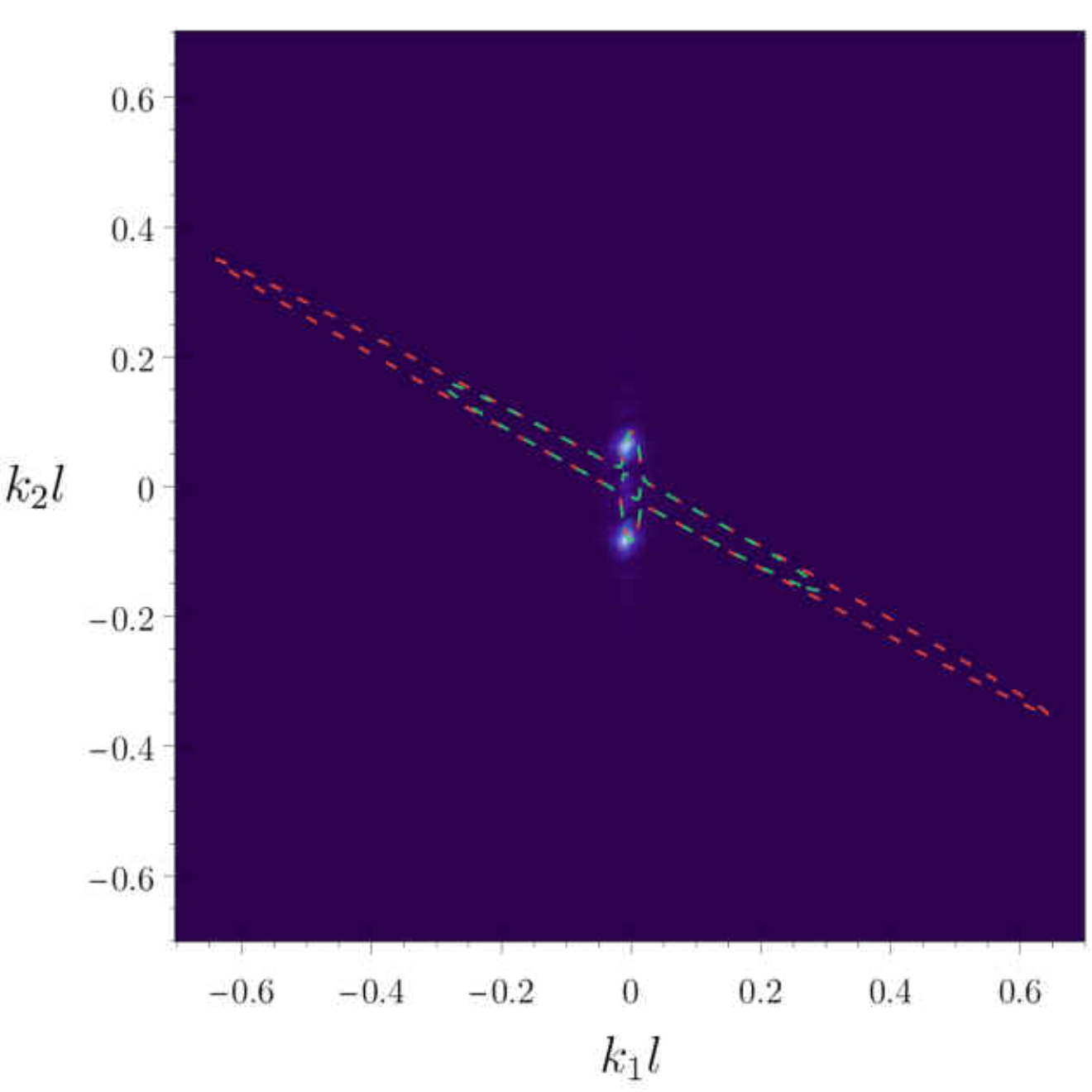}
    \end{subfigure} \hspace{2mm}
    \begin{subfigure}{0.4\textwidth}
        \centering
        \caption{\label{fig:rhombus_7_15_force_v_FFT}Vertical loading}
        \includegraphics[width=0.98\linewidth]{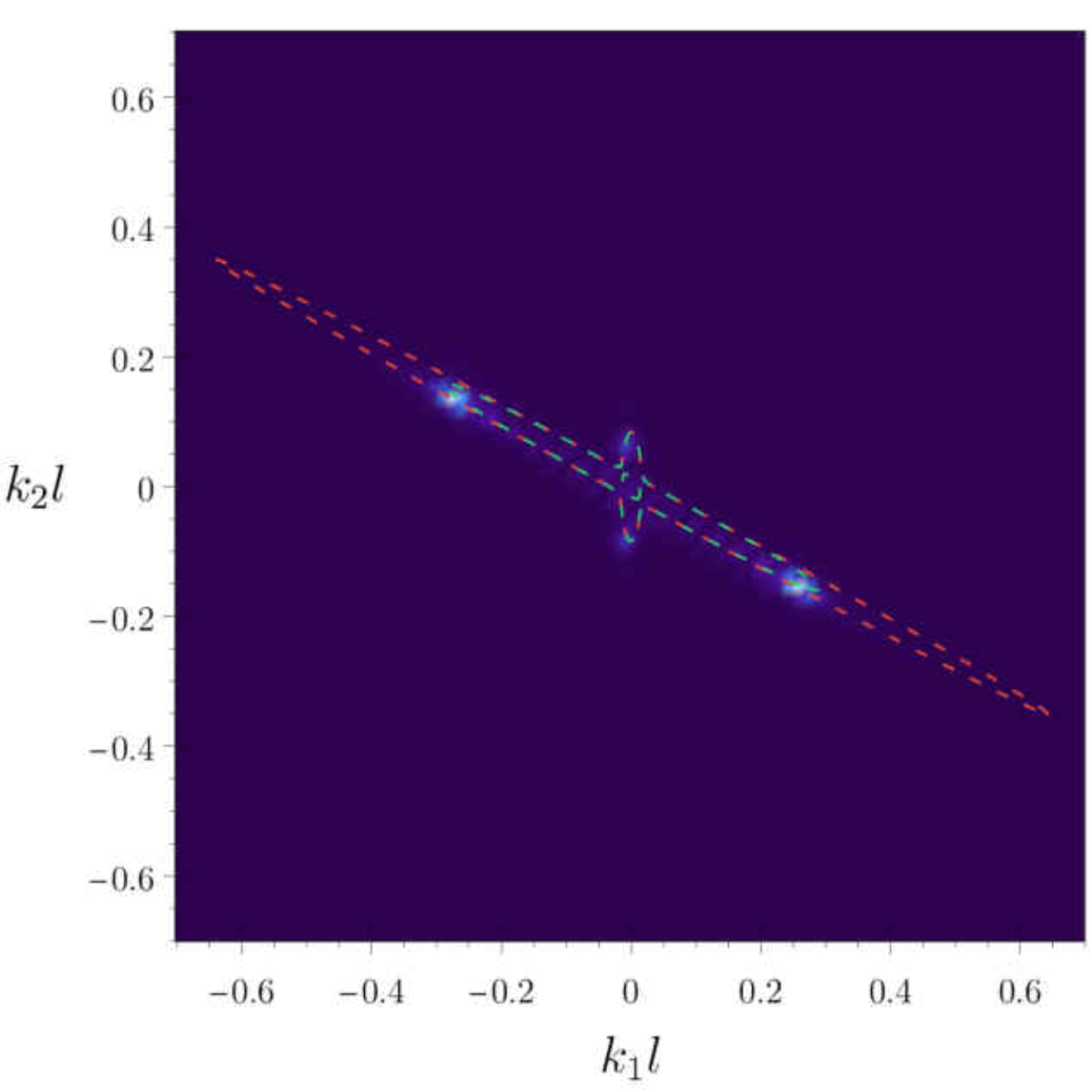}
    \end{subfigure}
    \caption{\label{fig:rhombus_7_15_force_FFT}
        As for Fig. \ref{fig:rhombus_10_10_force_FFT}, except that an \textit{anisotropic} rhombic lattice ($\Lambda_1=7,\,\Lambda_2=15$) is considered and the Fourier transform refers to the displacement fields reported in Figs. \ref{fig:rhombus_7_15_force_h_99} and \ref{fig:rhombus_7_15_force_v_99}.
        Localization is absent when the pulsating force is horizontal,~(\subref{fig:rhombus_7_15_force_h_FFT}), as the waves corresponding to the elongated tips of the contours remain inactive.
        The vertical force triggers two peaks along the direction $\theta_{\text{cr}}=151.4^\circ$, which is in fact the direction normal to the localization band,~(\subref{fig:rhombus_10_10_force_v_FFT}).
        The Fourier transform is in perfect agreement with the responses shown in Figs.~\ref{fig:rhombus_7_15_force_h_99} and ~\ref{fig:rhombus_7_15_force_v_99}.
    }
\end{figure}

Results reported in Figs.~\ref{fig:rhombus_7_15_deformed} and~\ref{fig:rhombus_7_15_force_FFT}, referred to the \textit{anisotropic} rhombic lattice, have been obtained with the same setting of Figs.~\ref{fig:rhombus_10_10_deformed} and~\ref{fig:rhombus_10_10_force_FFT}, referring to the orthotropic case.

Fig.~\ref{fig:rhombus_7_15_deformed} shows that, as only one localization band is present, the deformation pattern is characterized by the generation of essentially straight wavefronts propagating outwards from the localization band.
The generation of these parallel waves is perfectly captured by the sharp peaks in the Fourier transform of the lattice response, reported in Fig.~\ref{fig:rhombus_7_15_force_FFT}.
Fig.~\ref{fig:rhombus_7_15_force_h_FFT} and~\ref{fig:rhombus_7_15_force_v_FFT} correspond, respectively, to the Fourier transform of Fig.~\ref{fig:rhombus_7_15_force_h_99} and Fig.~\ref{fig:rhombus_7_15_force_v_99}.
The two light spots in Fig.~\ref{fig:rhombus_7_15_force_v_FFT}, superimposed to two tips of the contour aligned parallel to the direction $\theta_{\text{cr}}=151.4^\circ$, denote the peaks of the Fourier transform.
These clearly shows that the response induced by the vertical force involves pure plane waves propagating with fronts inclined at $\theta_{\text{cr}}=151.4^\circ-90^\circ=61.4^\circ$ with respect to the horizontal axis.

It is also important to note that waves do not propagate vertically when the load is vertical while these become the only propagation mode when the load is horizontal (see the peaks on the short tips of the contour in Fig.~\ref{fig:rhombus_7_15_force_h_FFT}).
This is in agreement with the fact that localization is not generated by the horizontal force and the `slow waves' leading the homogenized continuum to failure of ellipticity remain inactive.

\section{Incremental static response: macro and micro localization}
\label{sec:static_forced_response}
The correlation between the \textit{static} incremental response of the lattice and of the equivalent solid is of great interest and is now investigated close to the conditions of instability.
Following the perturbative approach, the response of the lattice to an applied static concentrated load, in the form of a force dipole, is numerically evaluated via finite elements and compared to the response of the equivalent solid, also subject to the same force dipole.
The latter is constructed by means of the Green's function associated to the operator governing the incremental equilibrium, $\diver\fC[\grad(\bullet)]$, \cite{bigoni_2012}, and resulting in
\begin{equation}
    \label{eq:Green_function}
    \mG(\hat{\bx}) = -\frac{1}{4\pi^2} \oint_{\abs{\bn}=1} \left(\bA^{(\fC)}(\bn)\right)^{-1} \log\abs{\hat{\bx}\scalp\bn} \,,
\end{equation}
where the position vector $\bx$ has been made dimensionless through division by the rod's length $l$, so that $\hat{\bx}=\bx/l$.
Note that $\mG=\trans{\mG}$ due to the symmetry of the acoustic tensor.

The numerical results are obtained via a static analysis adopting the same computational setup described in Section~\ref{sec:dynamic_forced_response}.
As the simulations are meant to be compared to the infinite-body Green's function, the size of the domain has been calibrated in order to minimize boundary disturbances with clamped conditions at the four edges of the square domain.

The investigation presented below reveals that:
\begin{enumerate}[label=(\roman*)]
    \item The localization of deformation connected to macro bifurcation in the lattice and to failure of ellipticity in the equivalent solid are strictly similar;
    \item The lattice response close to a micro-bifurcation evidences a \textit{`microscopic' type of localization}, which remains completely undetected in the homogenized material.
\end{enumerate}
These two different mechanical behaviors are analyzed by exploiting the macro-to-micro transition of the first bifurcation mode, which is controlled by the increase in the stiffness of the diagonal springs of the lattice considered in Section~\ref{sec:grid}.
Hence, in Section~\ref{sec:macroscopic_localization} the lattice is considered in the absence of diagonal springs ($\kappa=0$), while in Section~\ref{sec:microscopic_localization} the lattice is reinforced with springs of stiffness $\kappa=0.4$.

\subsection{Macroscopic localizations on the verge of ellipticity loss}
\label{sec:macroscopic_localization}
The lattice configurations selected for the following analysis are reported in Table~\ref{tab:cases_analyzed}, together with the values of the preload $\bp_{\text{E}}$ for loss of ellipticity in the effective continuum.
\begin{figure}[htb!]
    \centering
    \begin{subfigure}{0.24\textwidth}
        \centering
        \phantomcaption{\label{fig:square_10_10_dipole_0}}
        \includegraphics[width=0.98\linewidth]{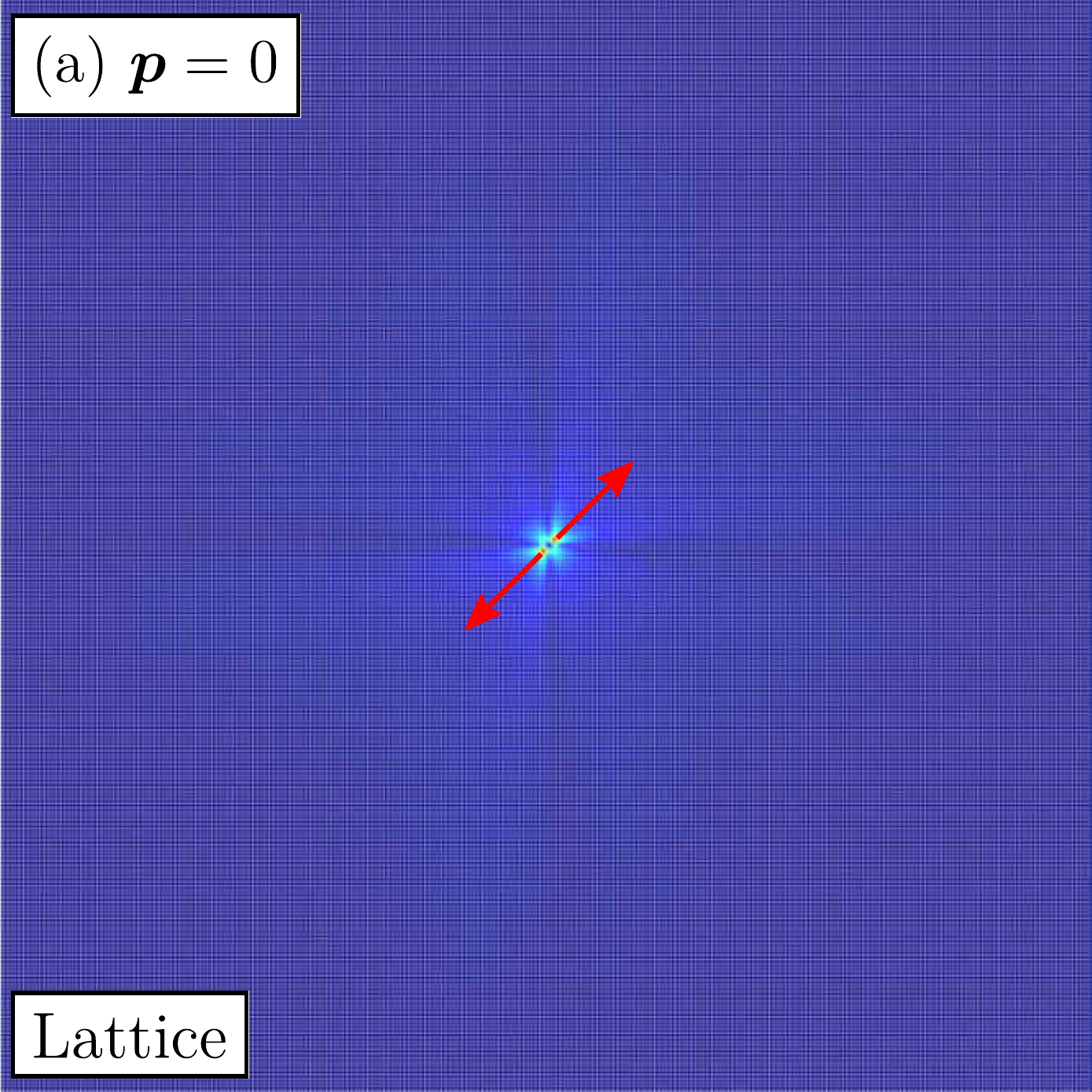}
    \end{subfigure}
    \begin{subfigure}{0.24\textwidth}
        \centering
        \phantomcaption{\label{fig:square_10_10_dipole_80}}
        \includegraphics[width=0.98\linewidth]{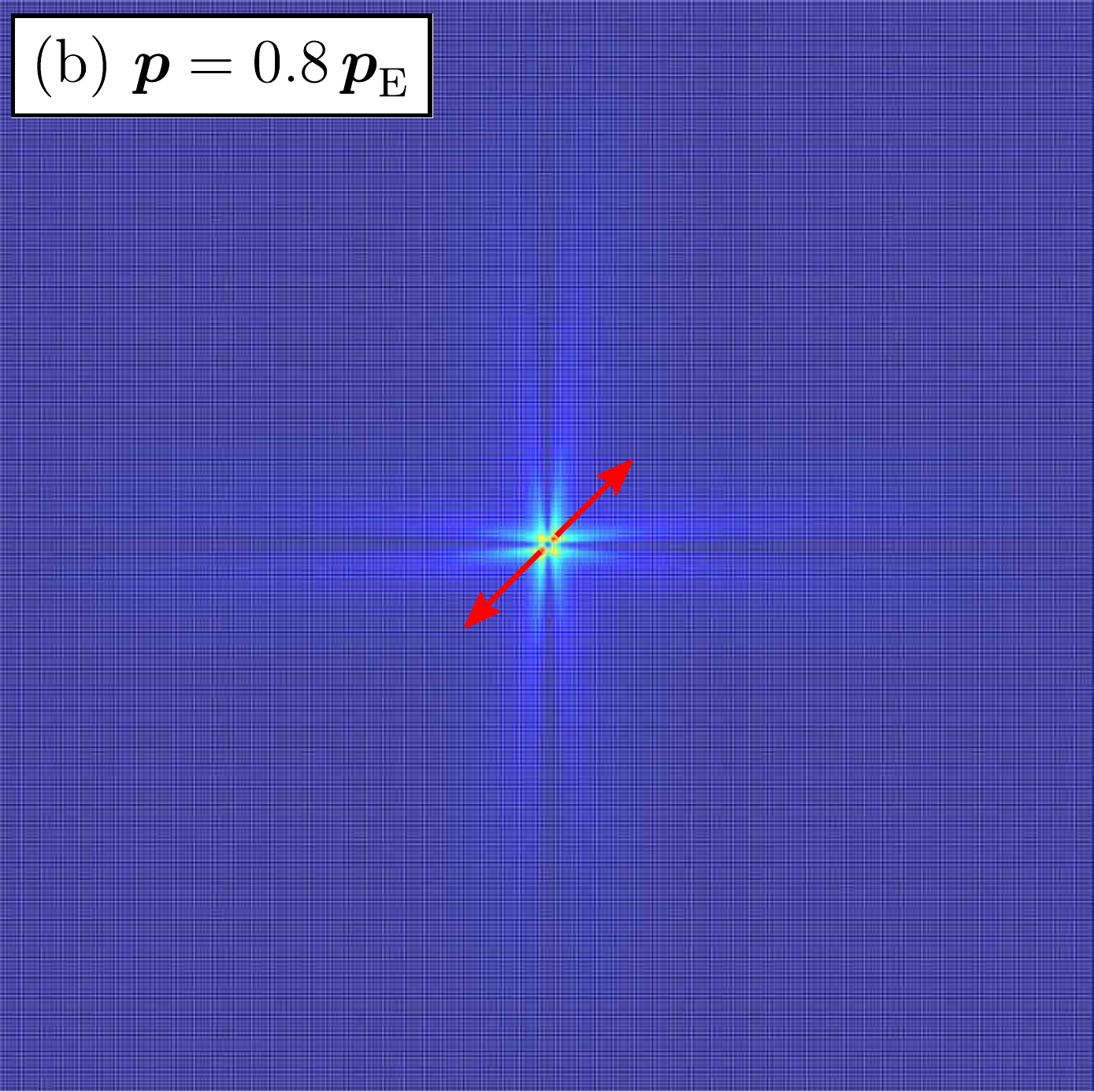}
    \end{subfigure}
    \begin{subfigure}{0.24\textwidth}
        \centering
        \phantomcaption{\label{fig:square_10_10_dipole_90}}
        \includegraphics[width=0.98\linewidth]{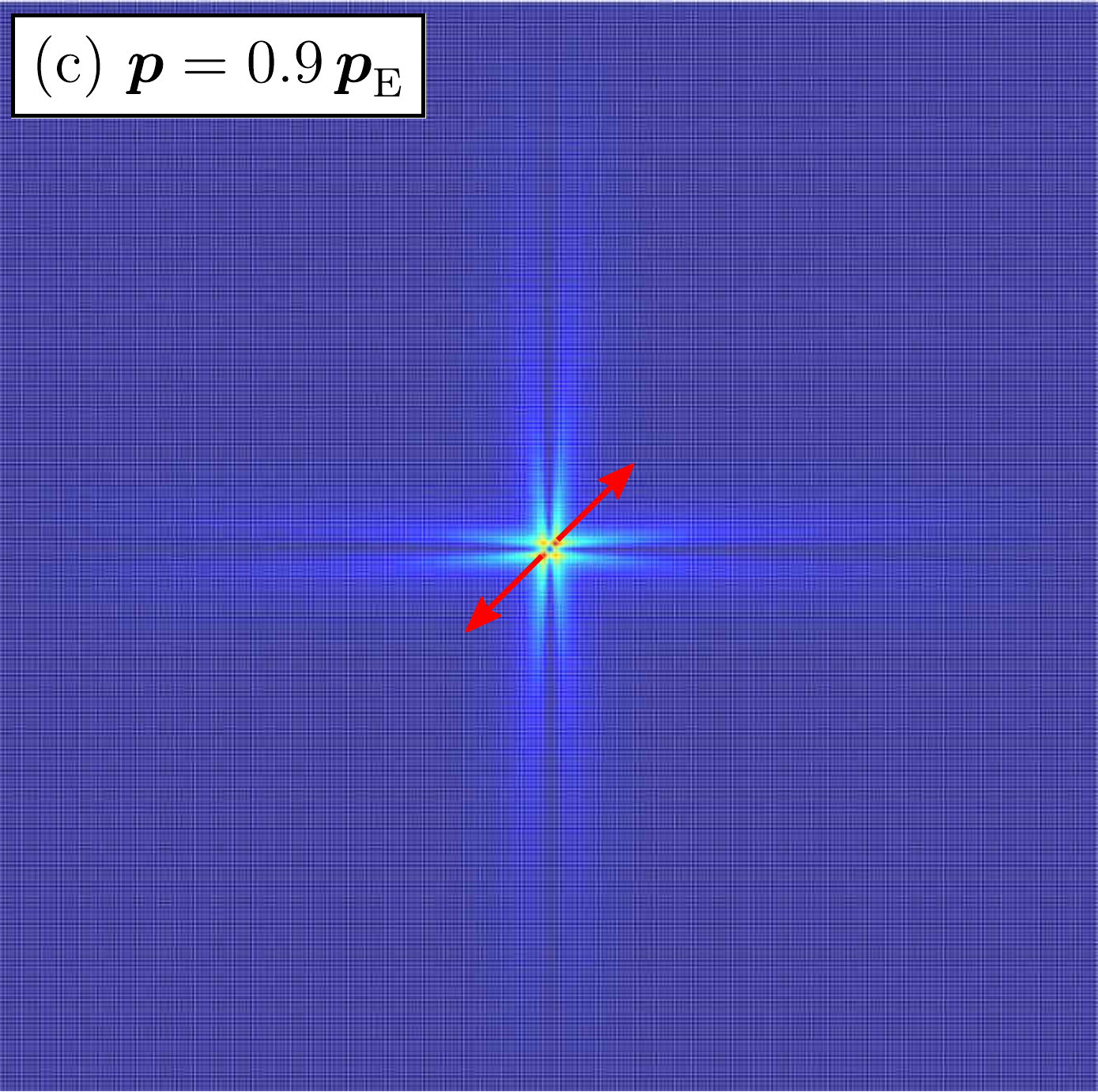}
    \end{subfigure}
    \begin{subfigure}{0.24\textwidth}
        \centering
        \phantomcaption{\label{fig:square_10_10_dipole_99}}
        \includegraphics[width=0.98\linewidth]{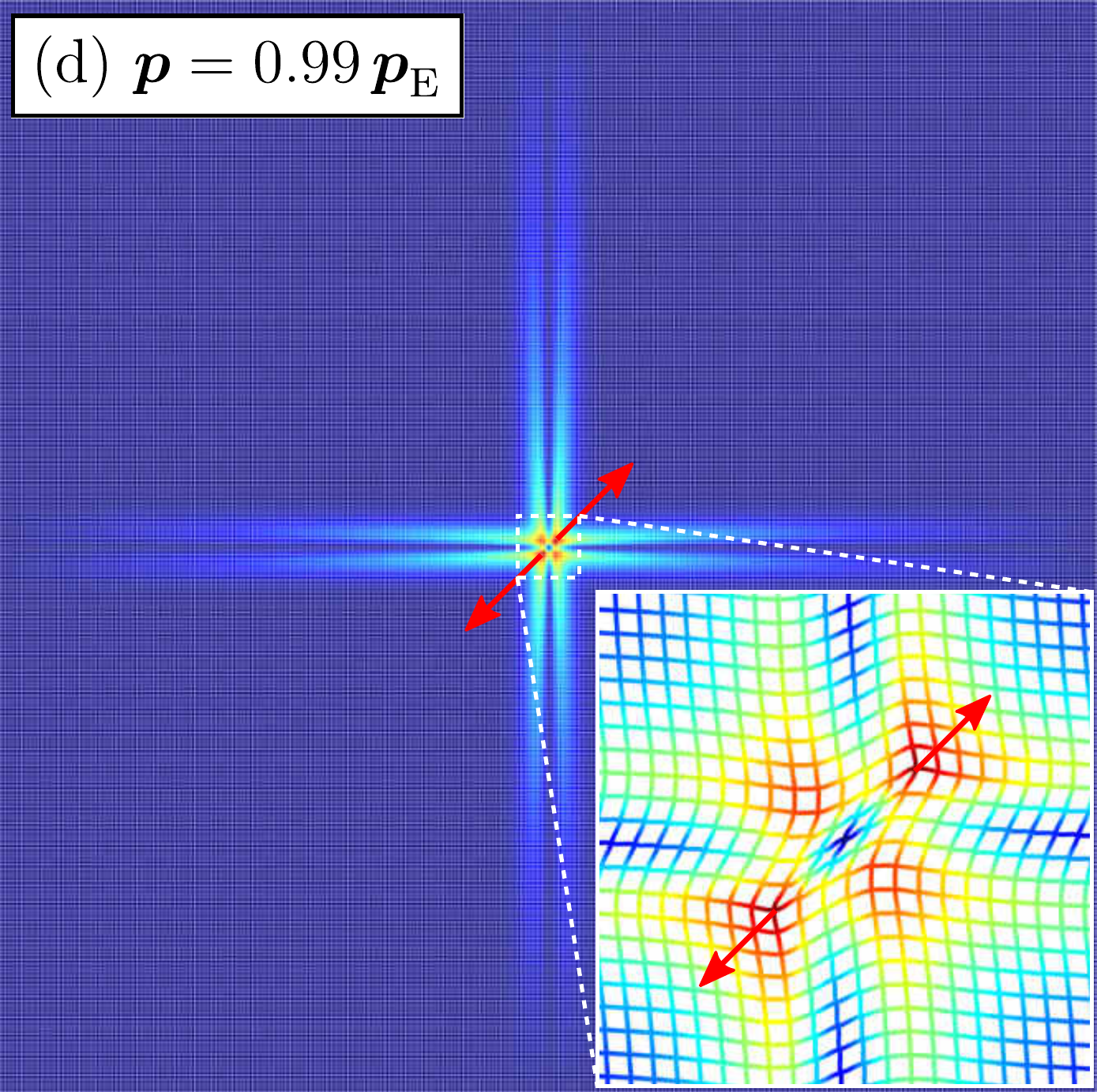}
    \end{subfigure}\\
    \vspace{0.01\linewidth}
    \begin{subfigure}{0.24\textwidth}
        \centering
        \phantomcaption{\label{fig:square_10_10_dipole_0_gf}}
        \includegraphics[width=0.98\linewidth]{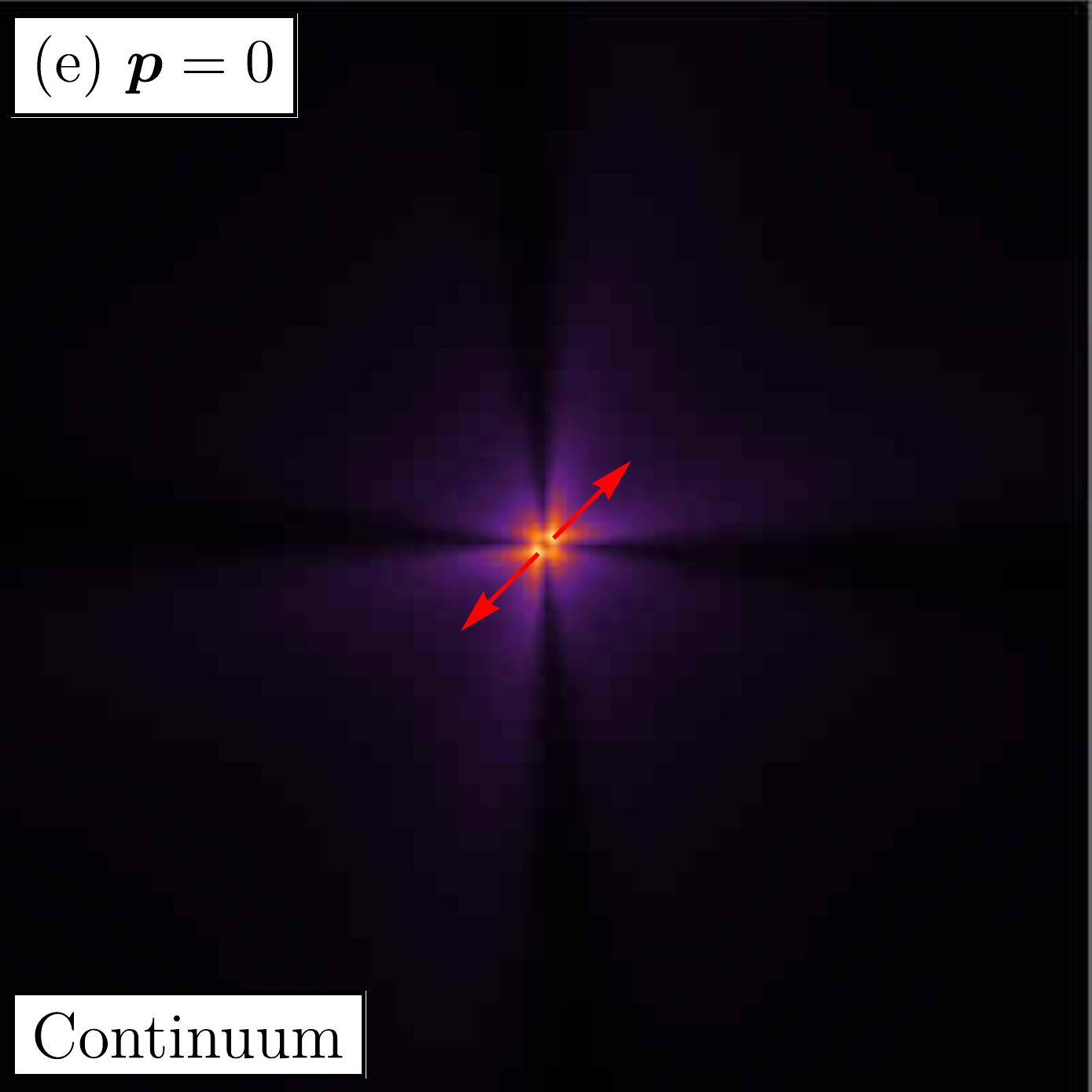}
    \end{subfigure}
    \begin{subfigure}{0.24\textwidth}
        \centering
        \phantomcaption{\label{fig:square_10_10_dipole_80_gf}}
        \includegraphics[width=0.98\linewidth]{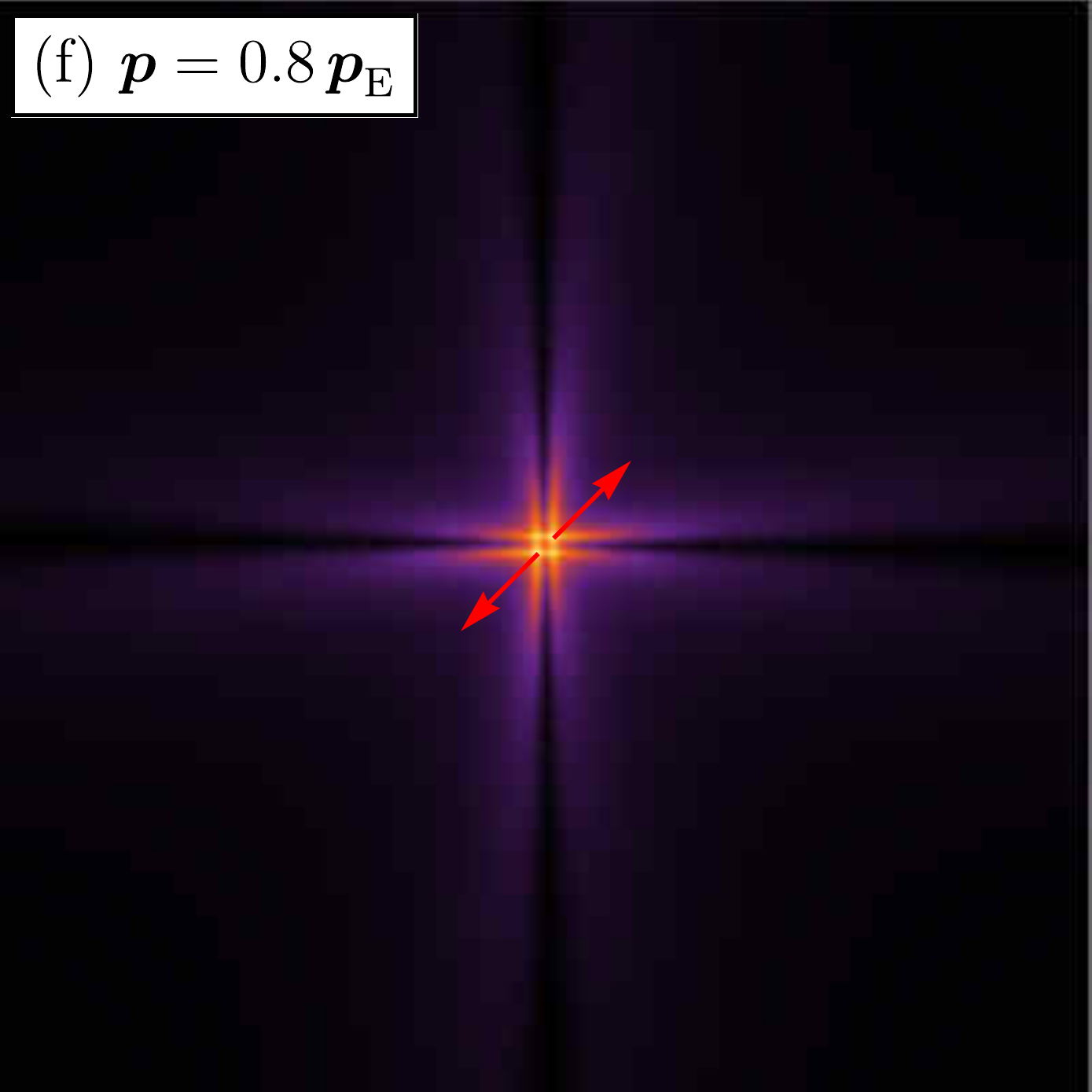}
    \end{subfigure}
    \begin{subfigure}{0.24\textwidth}
        \centering
        \phantomcaption{\label{fig:square_10_10_dipole_90_gf}}
        \includegraphics[width=0.98\linewidth]{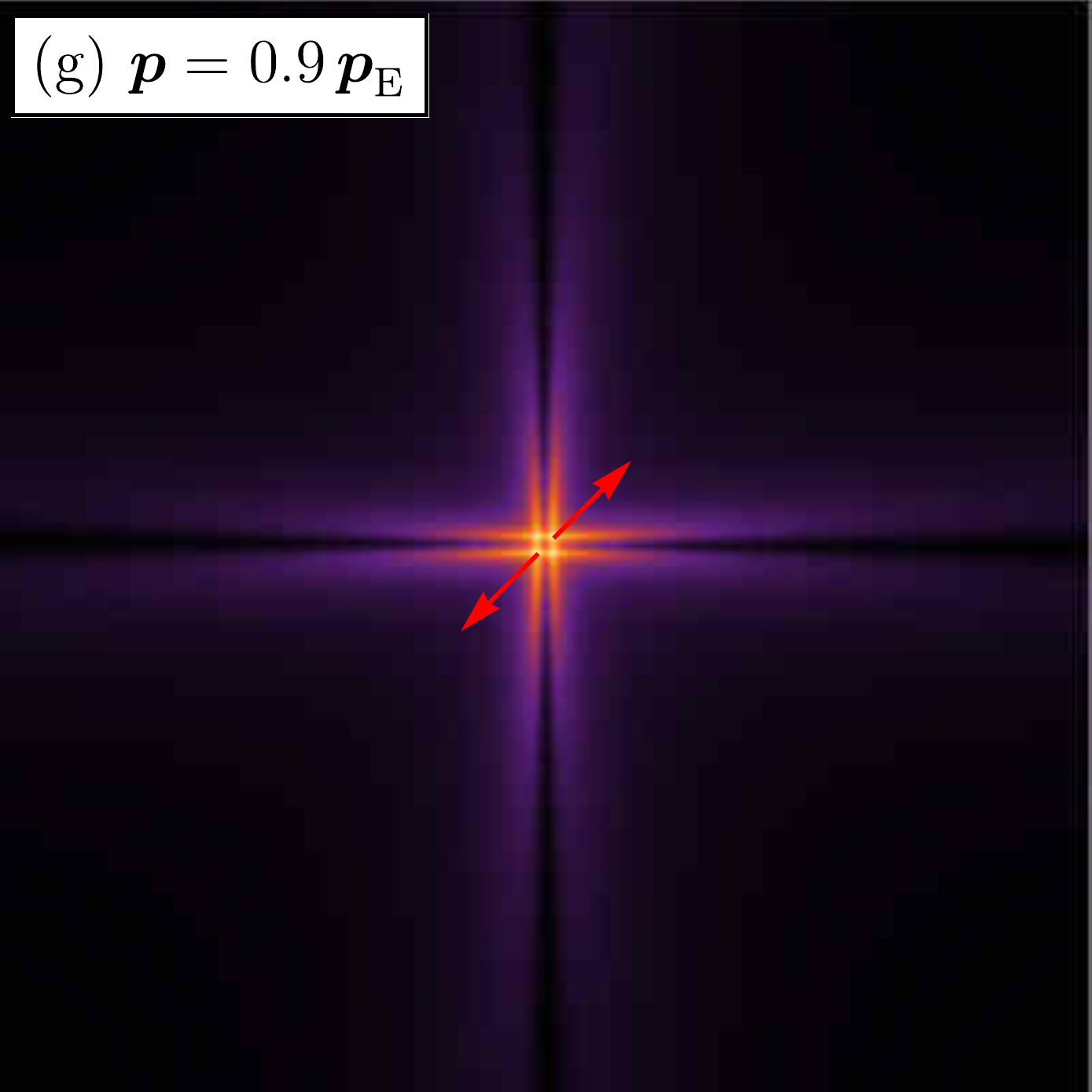}
    \end{subfigure}
    \begin{subfigure}{0.24\textwidth}
        \centering
        \phantomcaption{\label{fig:square_10_10_dipole_99_gf}}
        \includegraphics[width=0.98\linewidth]{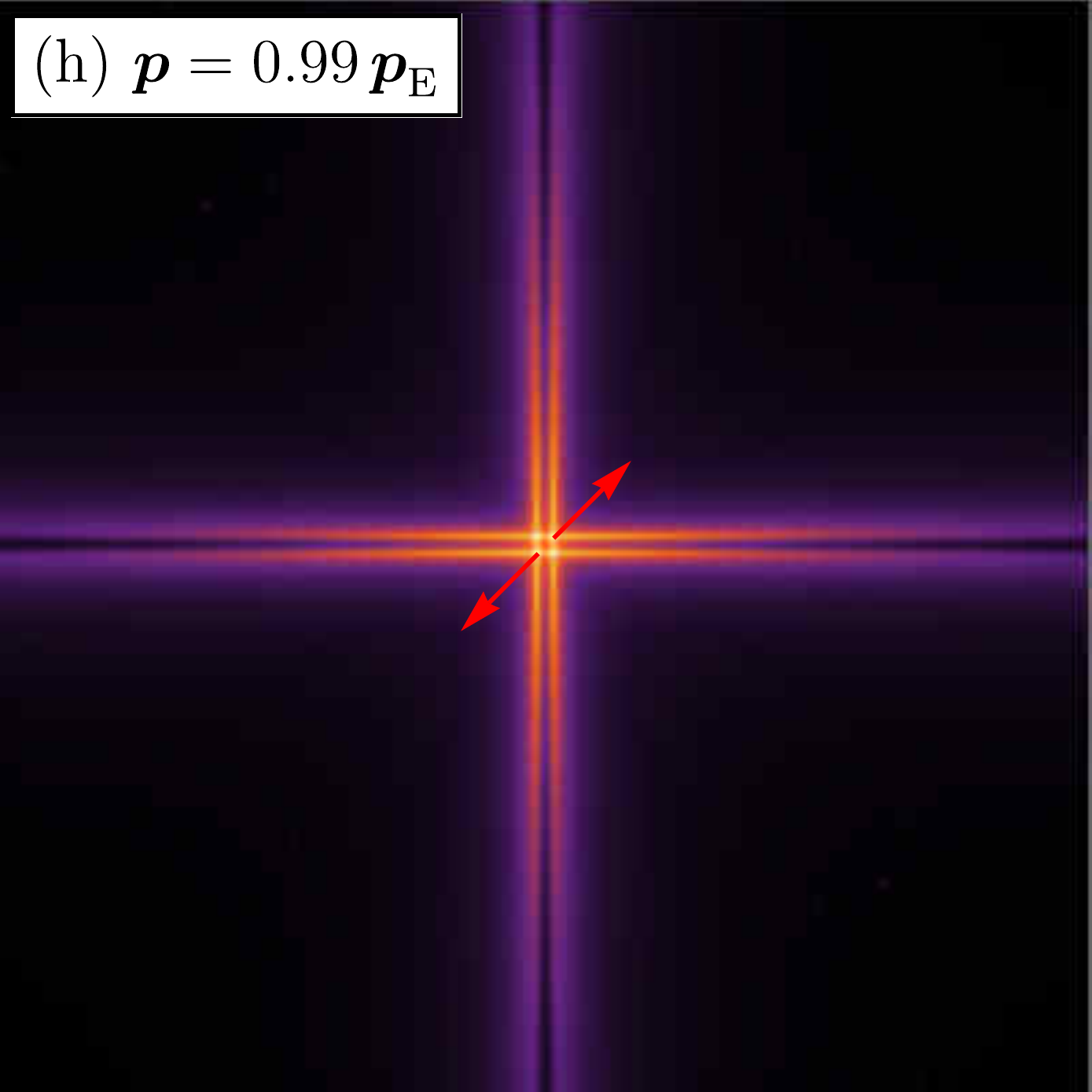}
    \end{subfigure}
    \caption{\label{fig:square_10_10_dipole}
        Progressive emergence, at increasing load, of two orthogonal shear bands in the displacement field generated by a diagonal force dipole. The dipole is applied to a square lattice (with cubic symmetry, $\Lambda_1=\Lambda_2=10$, upper part,  \subref{fig:square_10_10_dipole_0}--\subref{fig:square_10_10_dipole_99}, simulated via f.e.m.) and compared to the response of its equivalent continuum (lower part  \subref{fig:square_10_10_dipole_0_gf}--\subref{fig:square_10_10_dipole_99_gf}).
        From left to right the load increases towards failure of strong ellipticity $\bp_{\text{E}}$.
        Shear bands are aligned parallel to the directions predicted at failure of ellipticity ($\theta_{\text{cr}}=0^\circ,90^\circ$).
    }
\end{figure}
\begin{figure}[htb!]
    \centering
    \begin{subfigure}{0.24\textwidth}
        \centering
        \phantomcaption{\label{fig:square_7_15_dipole_0}}
        \includegraphics[width=0.98\linewidth]{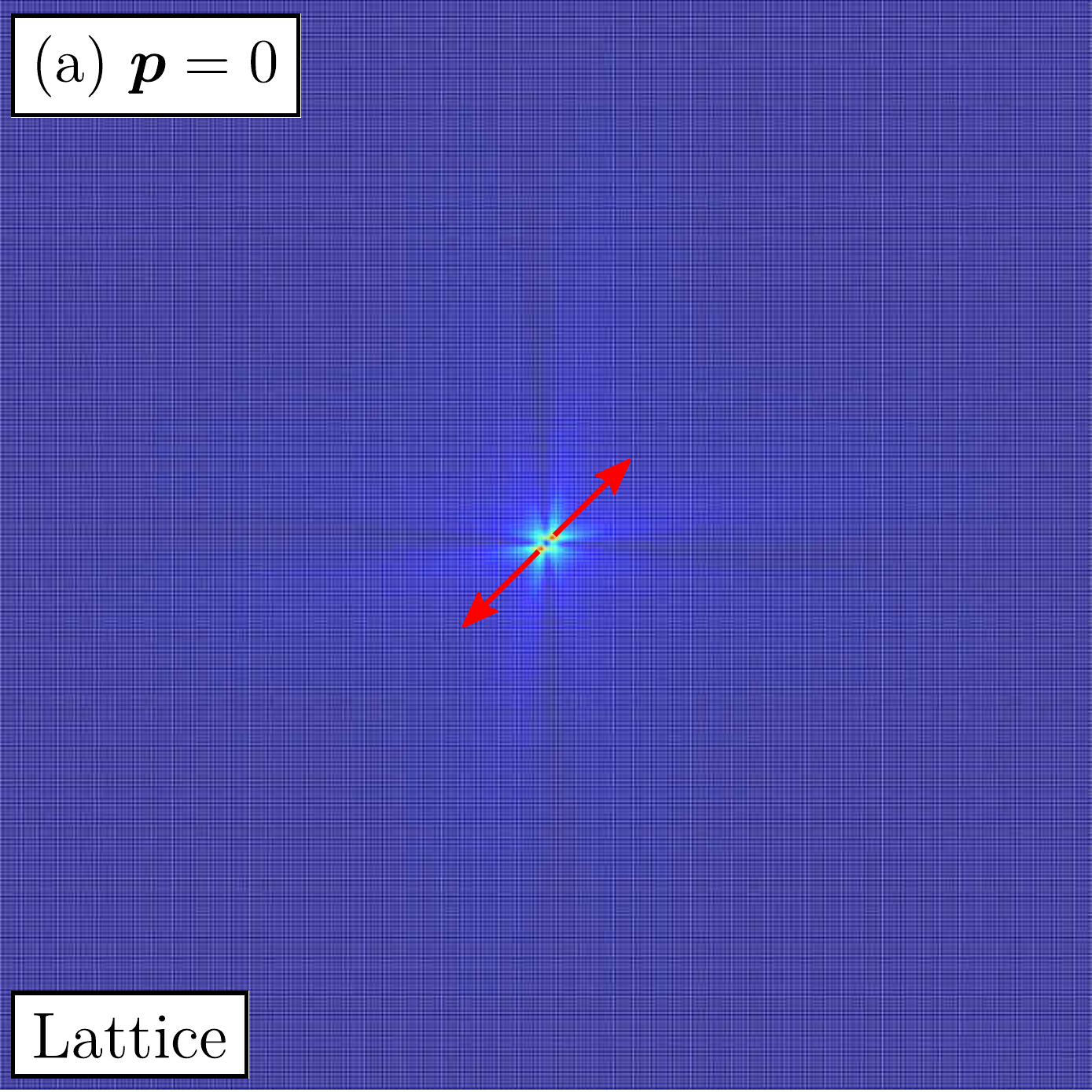}
    \end{subfigure}
    \begin{subfigure}{0.24\textwidth}
        \centering
        \phantomcaption{\label{fig:square_7_15_dipole_80}}
        \includegraphics[width=0.98\linewidth]{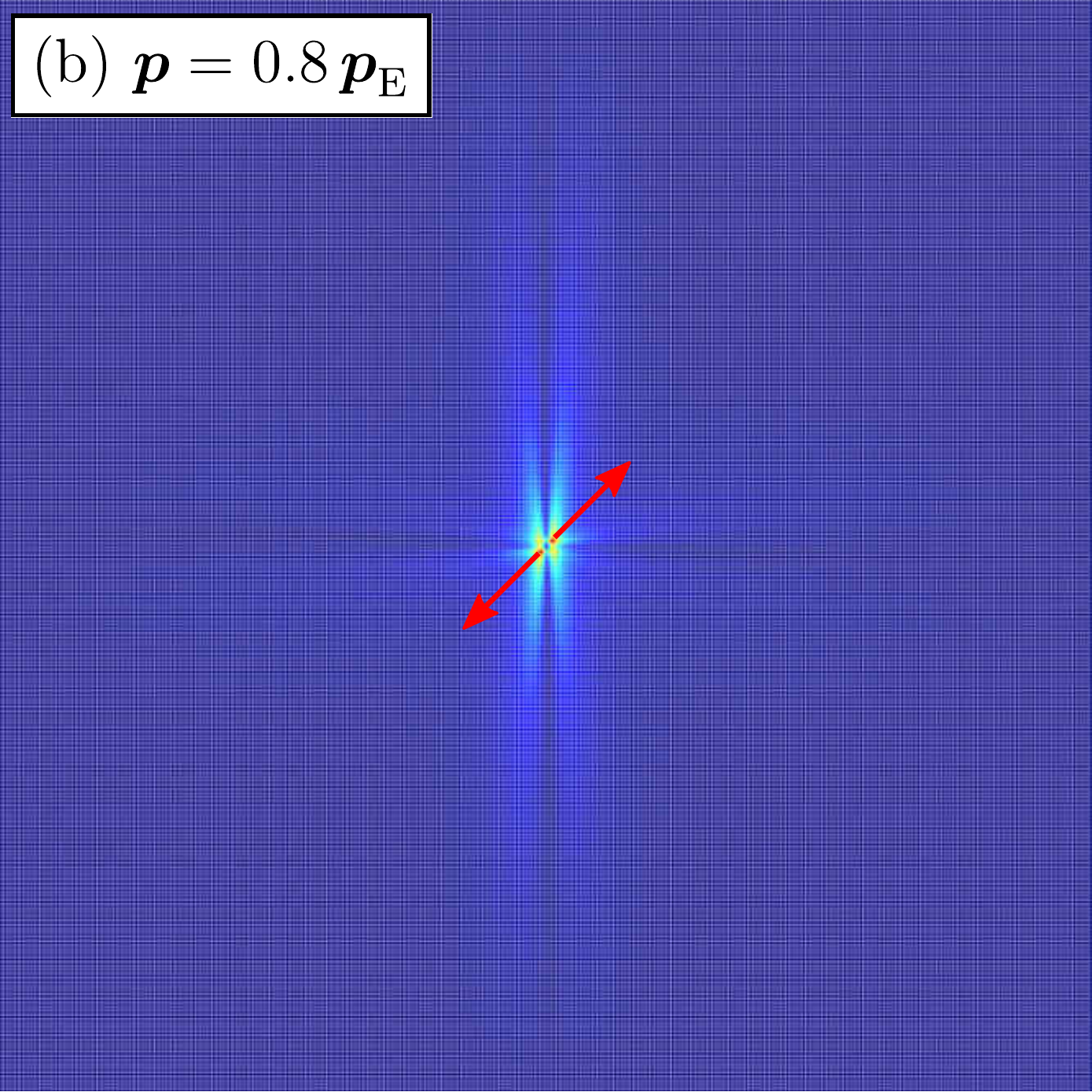}
    \end{subfigure}
    \begin{subfigure}{0.24\textwidth}
        \centering
        \phantomcaption{\label{fig:square_7_15_dipole_90}}
        \includegraphics[width=0.98\linewidth]{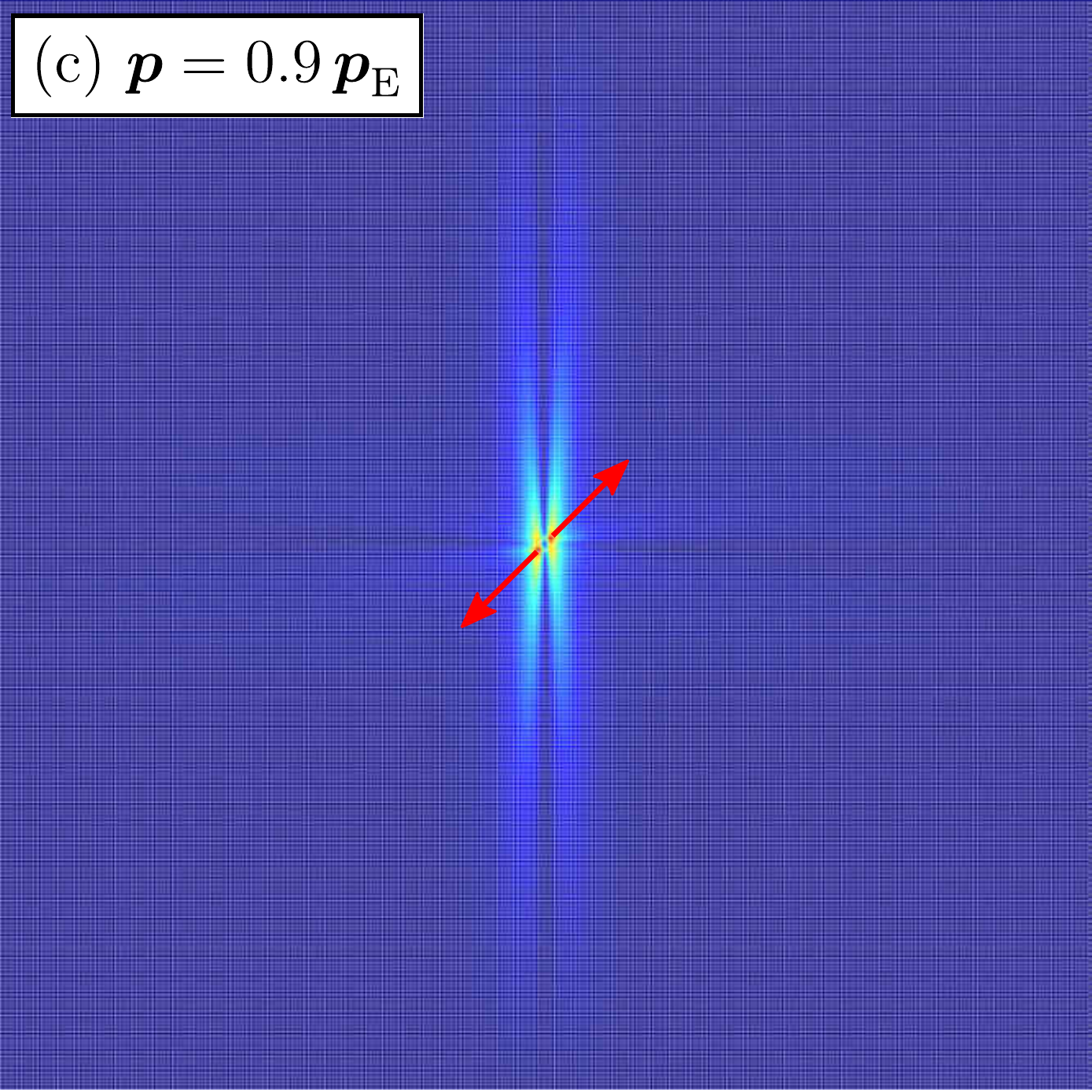}
    \end{subfigure}
    \begin{subfigure}{0.24\textwidth}
        \centering
        \phantomcaption{\label{fig:square_7_15_dipole_99}}
        \includegraphics[width=0.98\linewidth]{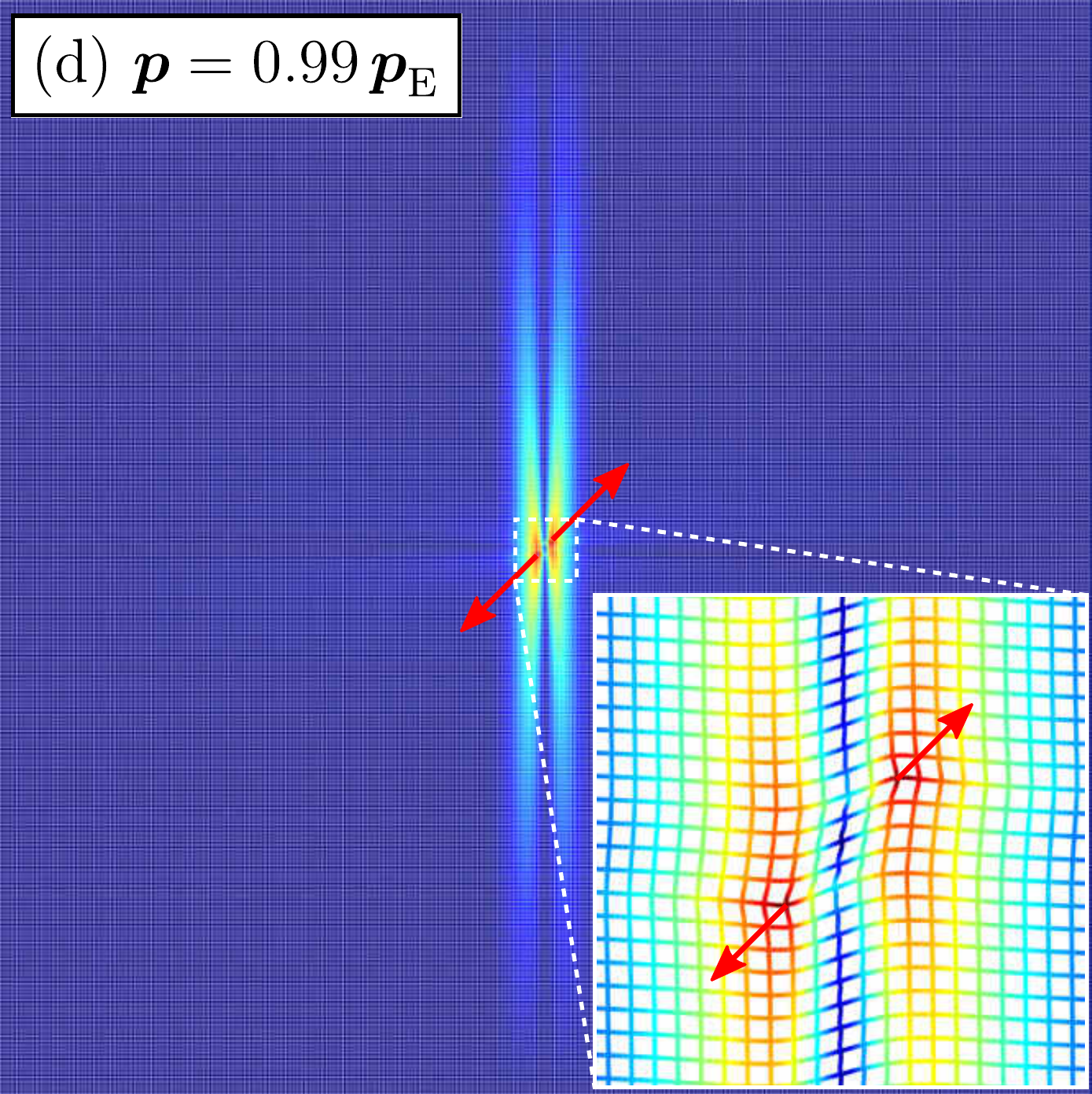}
    \end{subfigure}\\
    \vspace{0.01\linewidth}
    \begin{subfigure}{0.24\textwidth}
        \centering
        \phantomcaption{\label{fig:square_7_15_dipole_0_gf}}
        \includegraphics[width=0.98\linewidth]{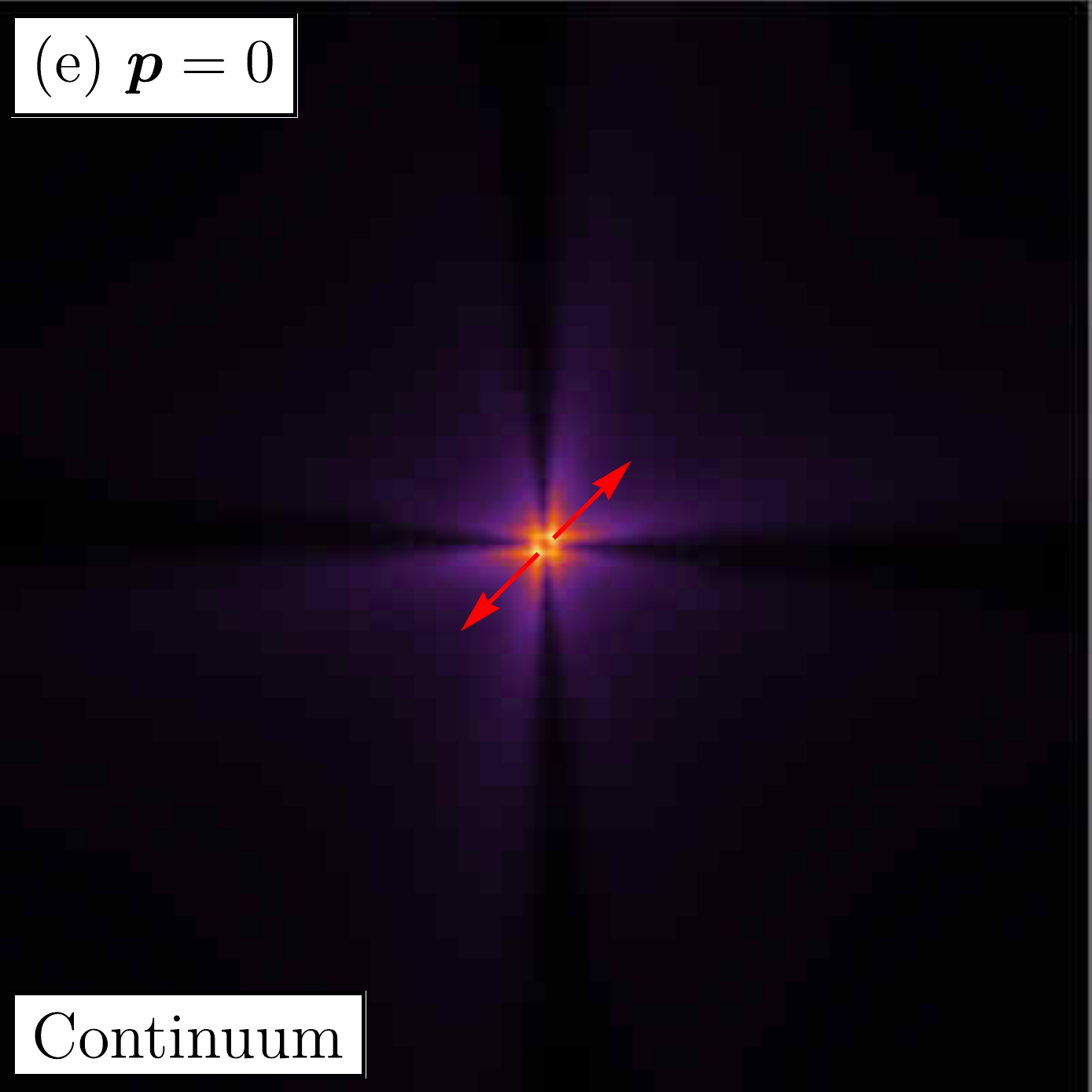}
    \end{subfigure}
    \begin{subfigure}{0.24\textwidth}
        \centering
        \phantomcaption{\label{fig:square_7_15_dipole_80_gf}}
        \includegraphics[width=0.98\linewidth]{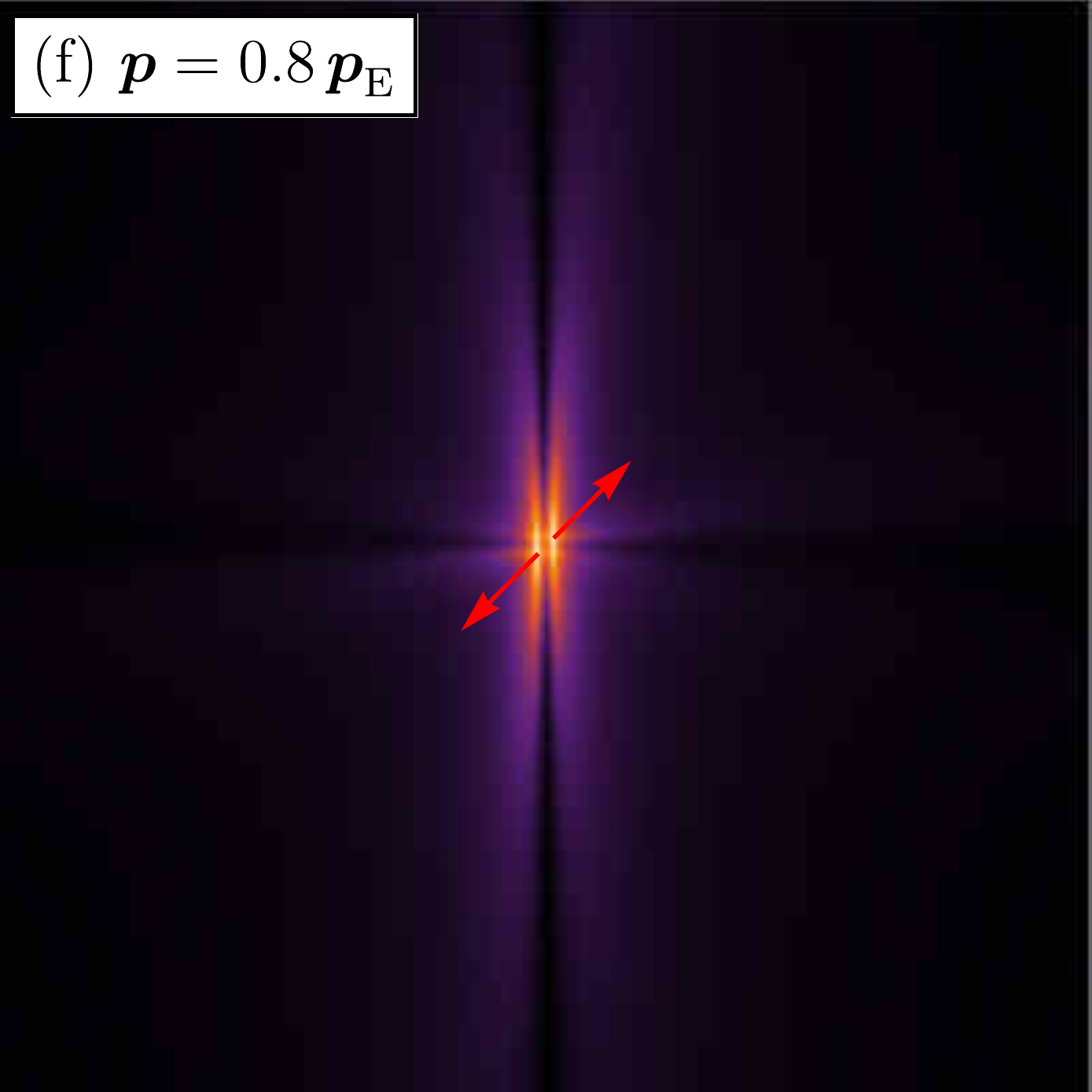}
    \end{subfigure}
    \begin{subfigure}{0.24\textwidth}
        \centering
        \phantomcaption{\label{fig:square_7_15_dipole_90_gf}}
        \includegraphics[width=0.98\linewidth]{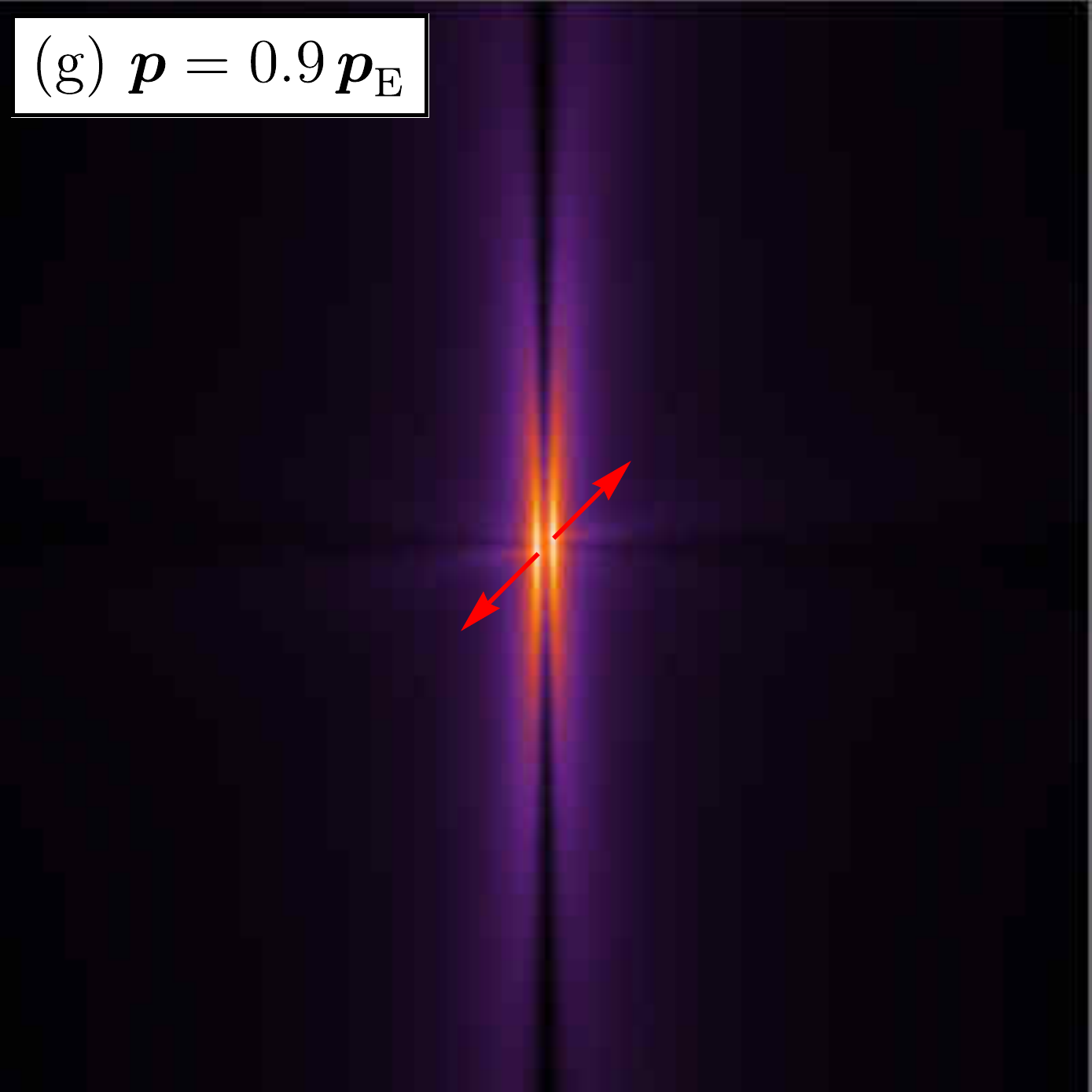}
    \end{subfigure}
    \begin{subfigure}{0.24\textwidth}
        \centering
        \phantomcaption{\label{fig:square_7_15_dipole_99_gf}}
        \includegraphics[width=0.98\linewidth]{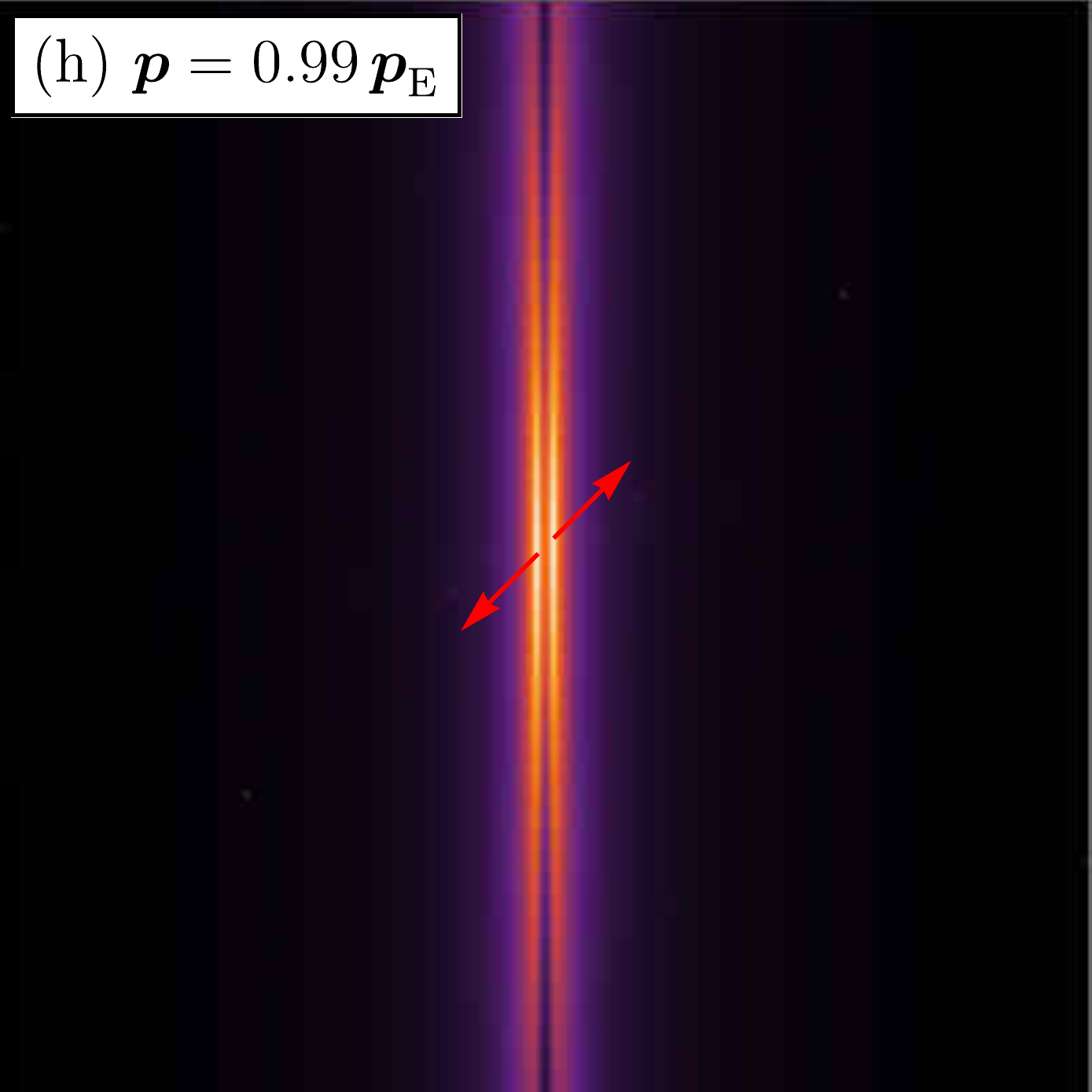}
    \end{subfigure}
    \caption{\label{fig:square_7_15_dipole}
        As for Fig. \ref{fig:square_10_10_dipole}, but for an orthotropic square lattice ($\Lambda_1=7,\,\Lambda_2=15$), where a single and vertical, $\theta_{\text{cr}}=0^\circ$, shear band forms.
    }
\end{figure}

\begin{figure}[htb!]
    \centering
    \begin{subfigure}{0.24\textwidth}
        \centering
        \phantomcaption{\label{fig:rhombus_10_10_dipole_0}}
        \includegraphics[width=0.98\linewidth]{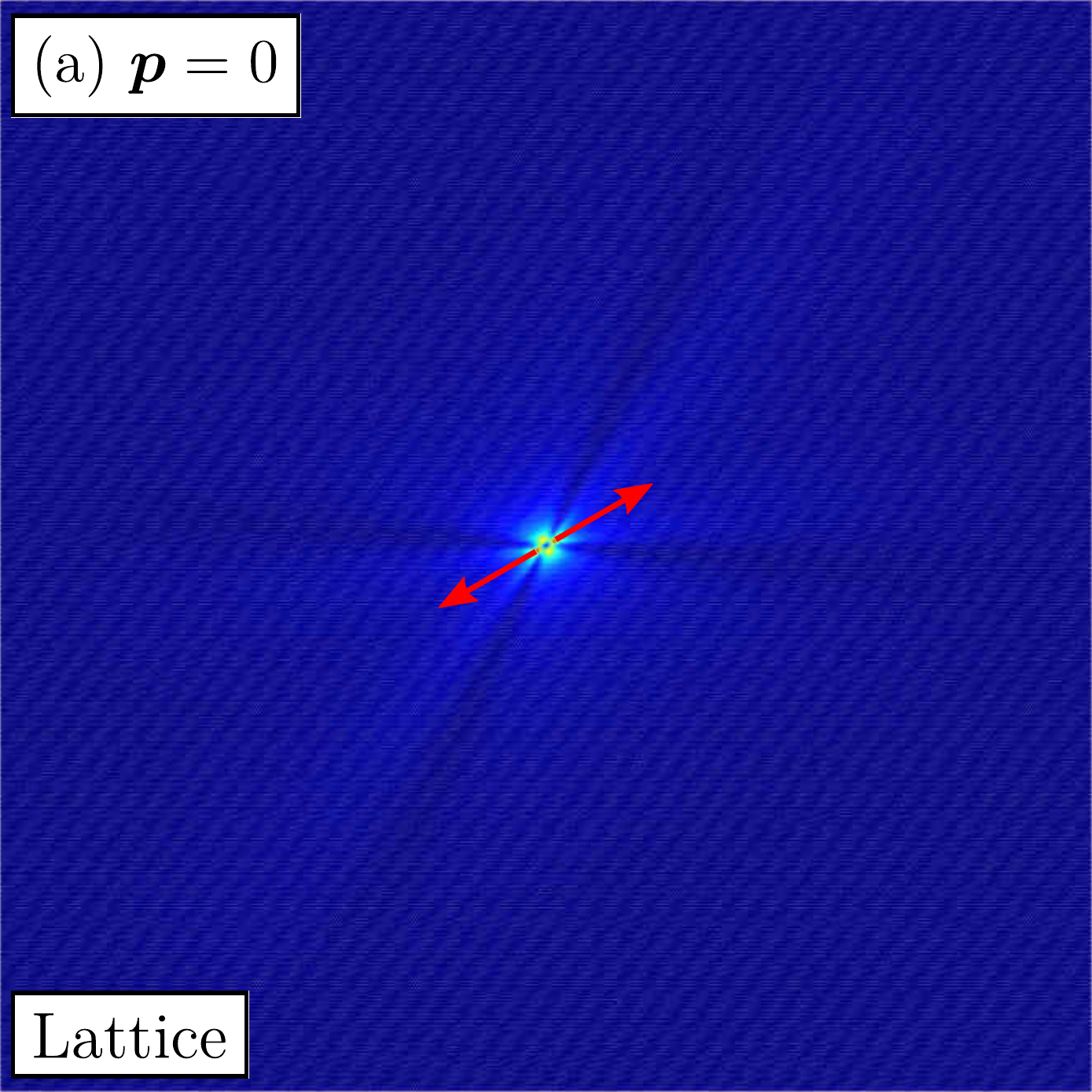}
    \end{subfigure}
    \begin{subfigure}{0.24\textwidth}
        \centering
        \phantomcaption{\label{fig:rhombus_10_10_dipole_80}}
        \includegraphics[width=0.98\linewidth]{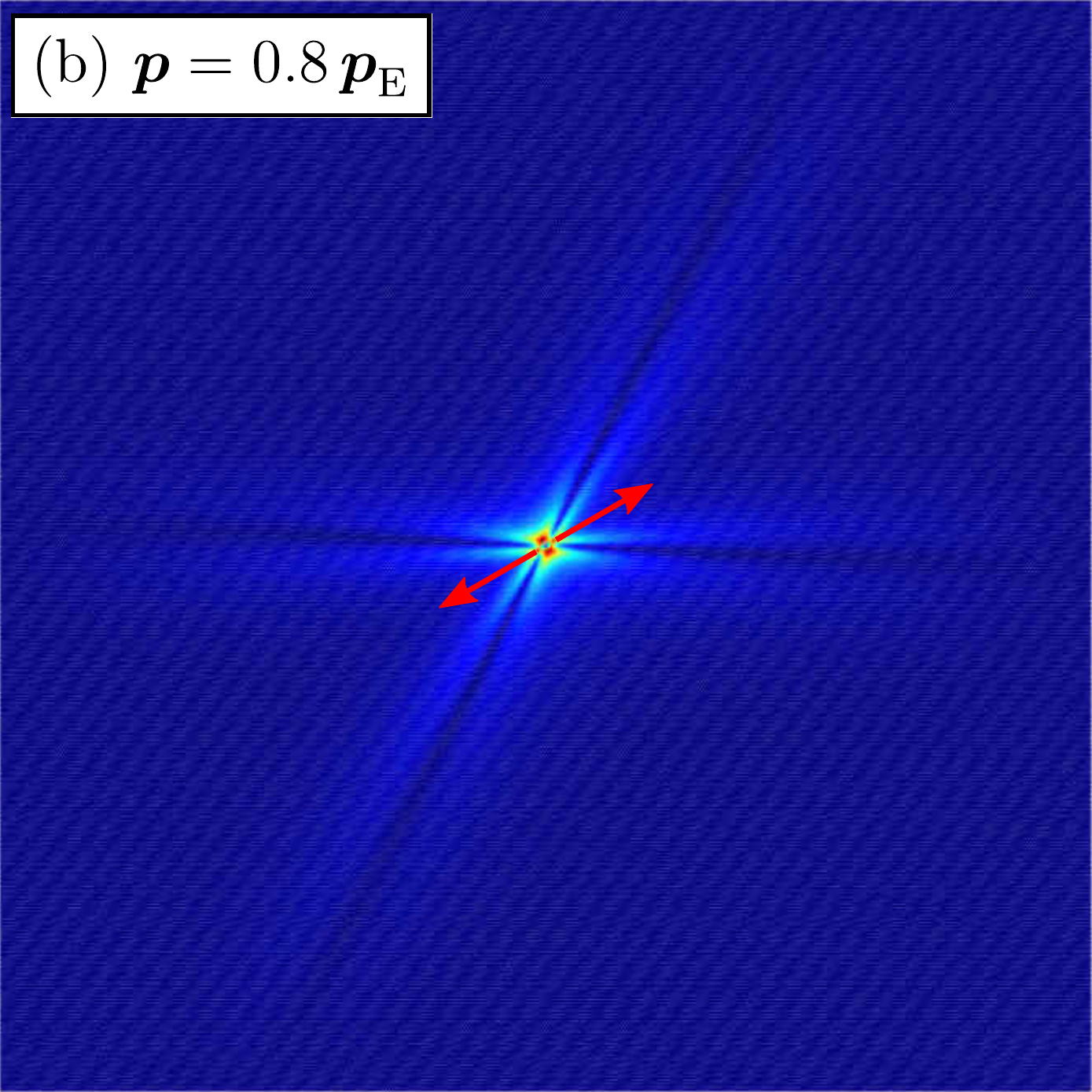}
    \end{subfigure}
    \begin{subfigure}{0.24\textwidth}
        \centering
        \phantomcaption{\label{fig:rhombus_10_10_dipole_90}}
        \includegraphics[width=0.98\linewidth]{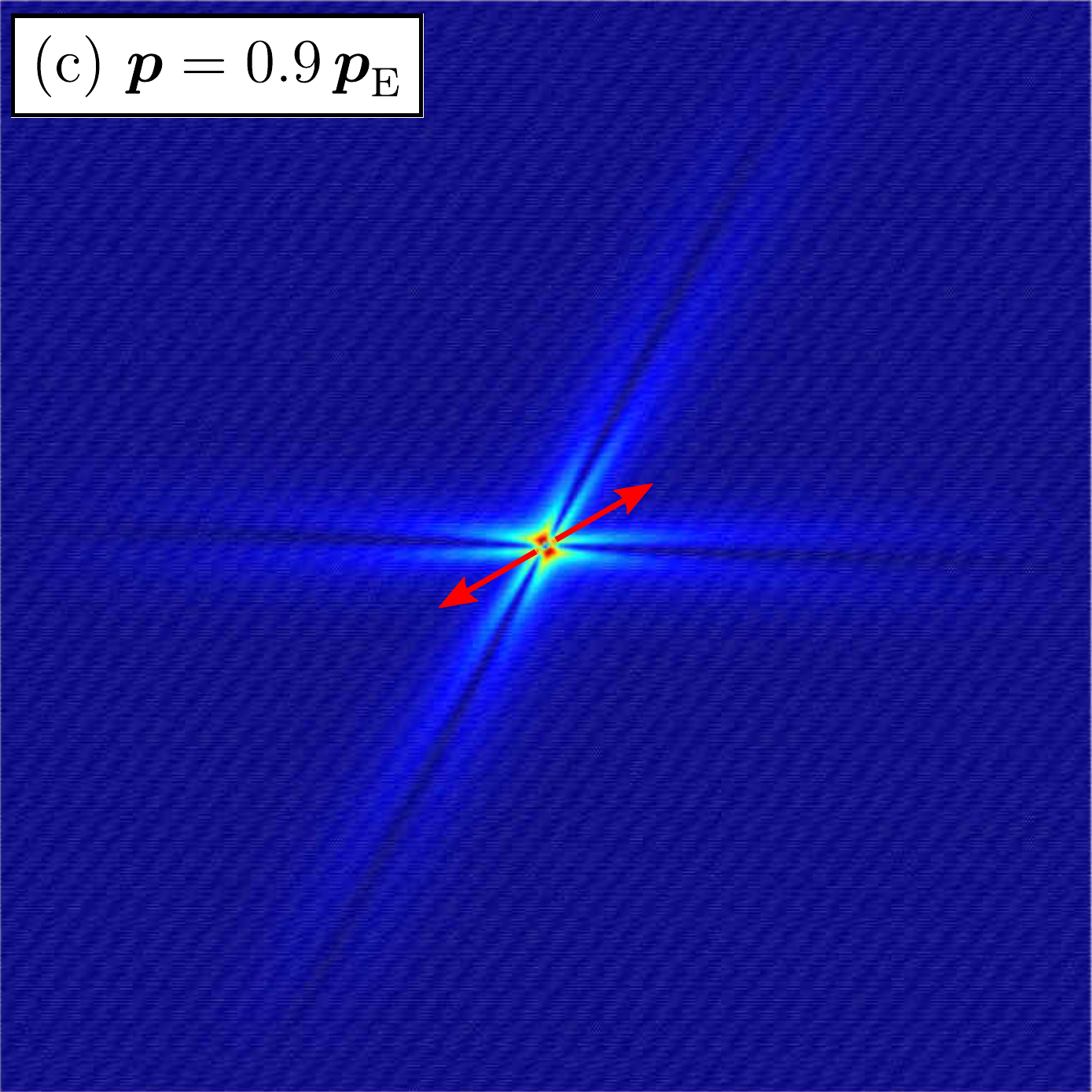}
    \end{subfigure}
    \begin{subfigure}{0.24\textwidth}
        \centering
        \phantomcaption{\label{fig:rhombus_10_10_dipole_99}}
        \includegraphics[width=0.98\linewidth]{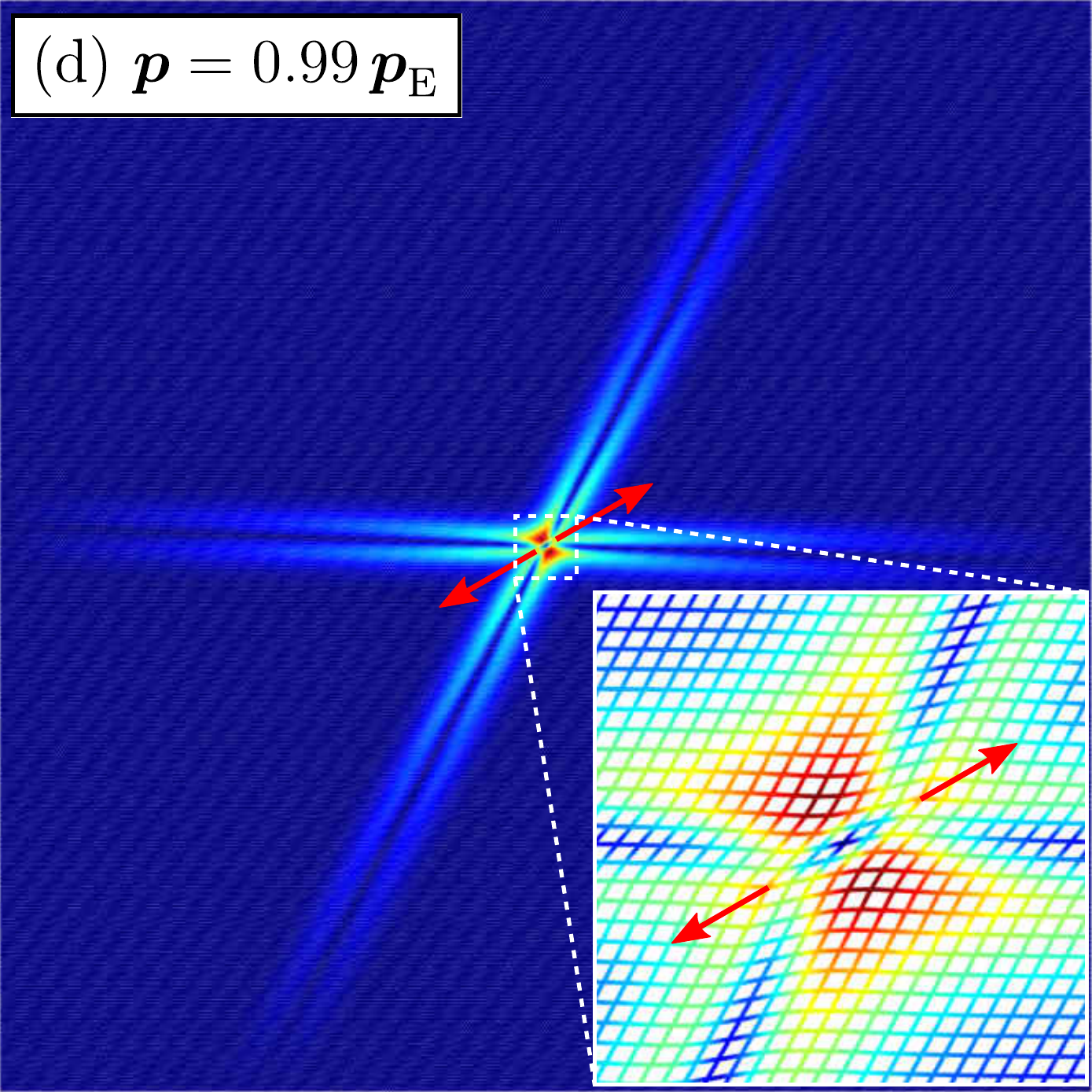}
    \end{subfigure}\\
    \vspace{0.01\linewidth}
    \begin{subfigure}{0.24\textwidth}
        \centering
        \phantomcaption{\label{fig:rhombus_10_10_dipole_0_gf}}
        \includegraphics[width=0.98\linewidth]{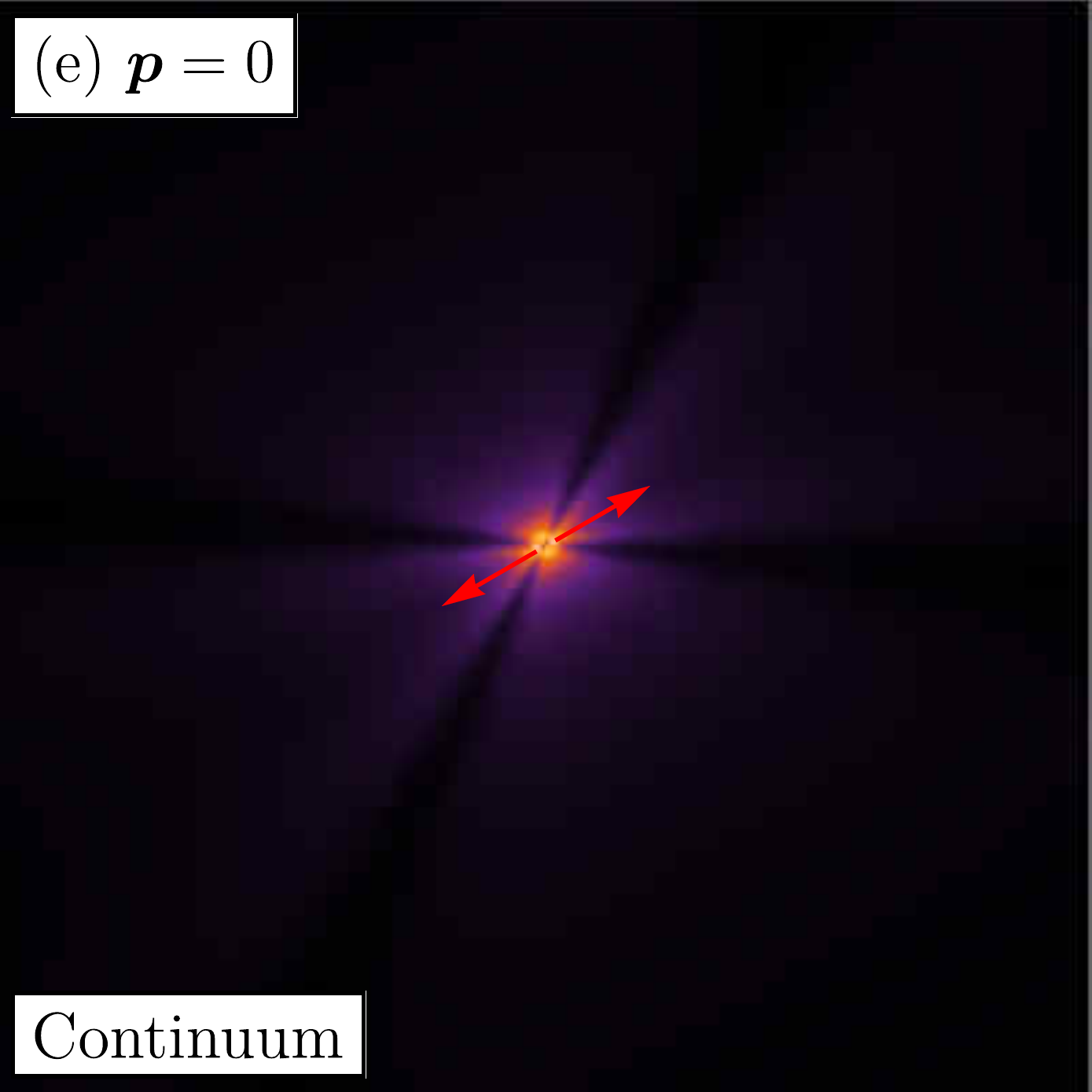}
    \end{subfigure}
    \begin{subfigure}{0.24\textwidth}
        \centering
        \phantomcaption{\label{fig:rhombus_10_10_dipole_80_gf}}
        \includegraphics[width=0.98\linewidth]{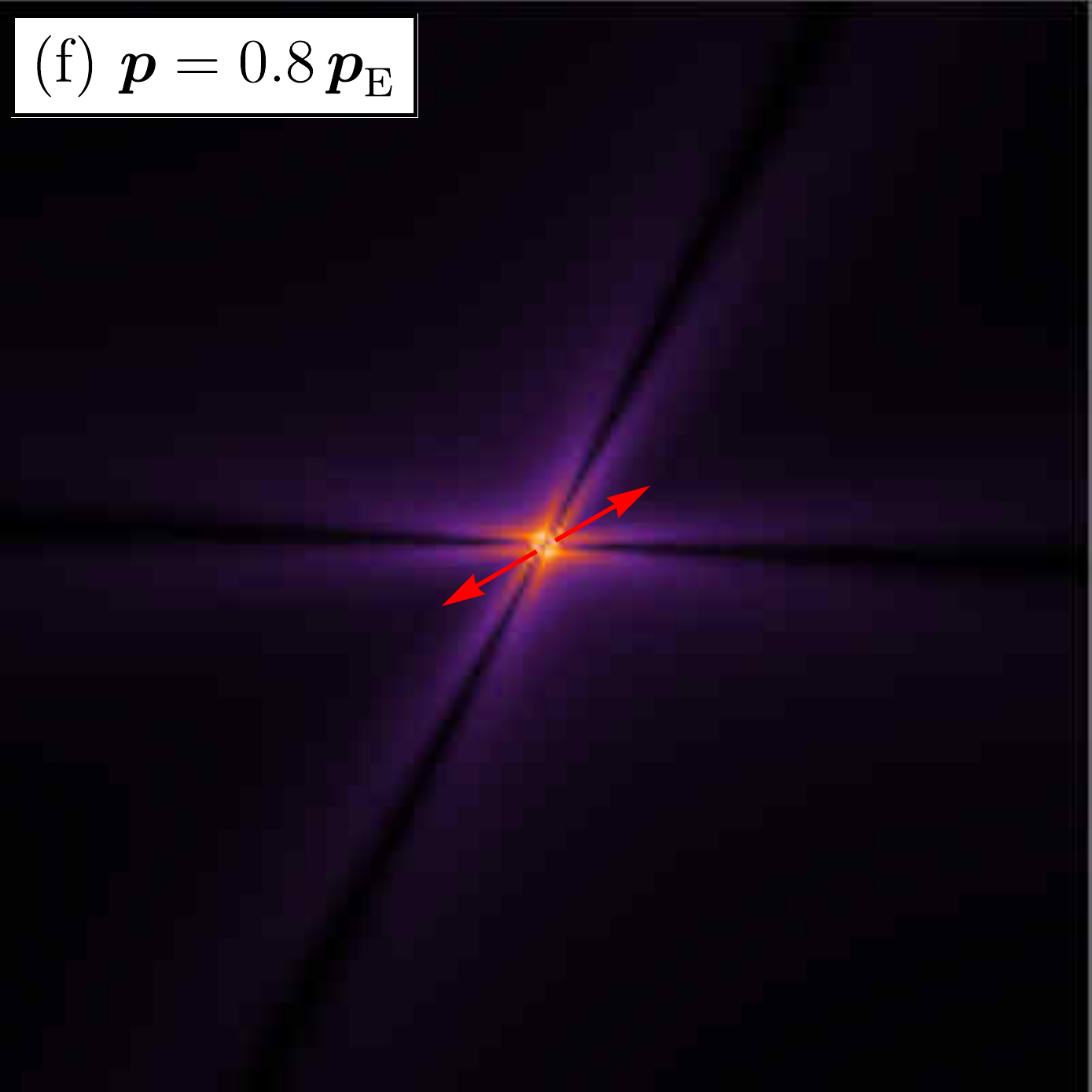}
    \end{subfigure}
    \begin{subfigure}{0.24\textwidth}
        \centering
        \phantomcaption{\label{fig:rhombus_10_10_dipole_90_gf}}
        \includegraphics[width=0.98\linewidth]{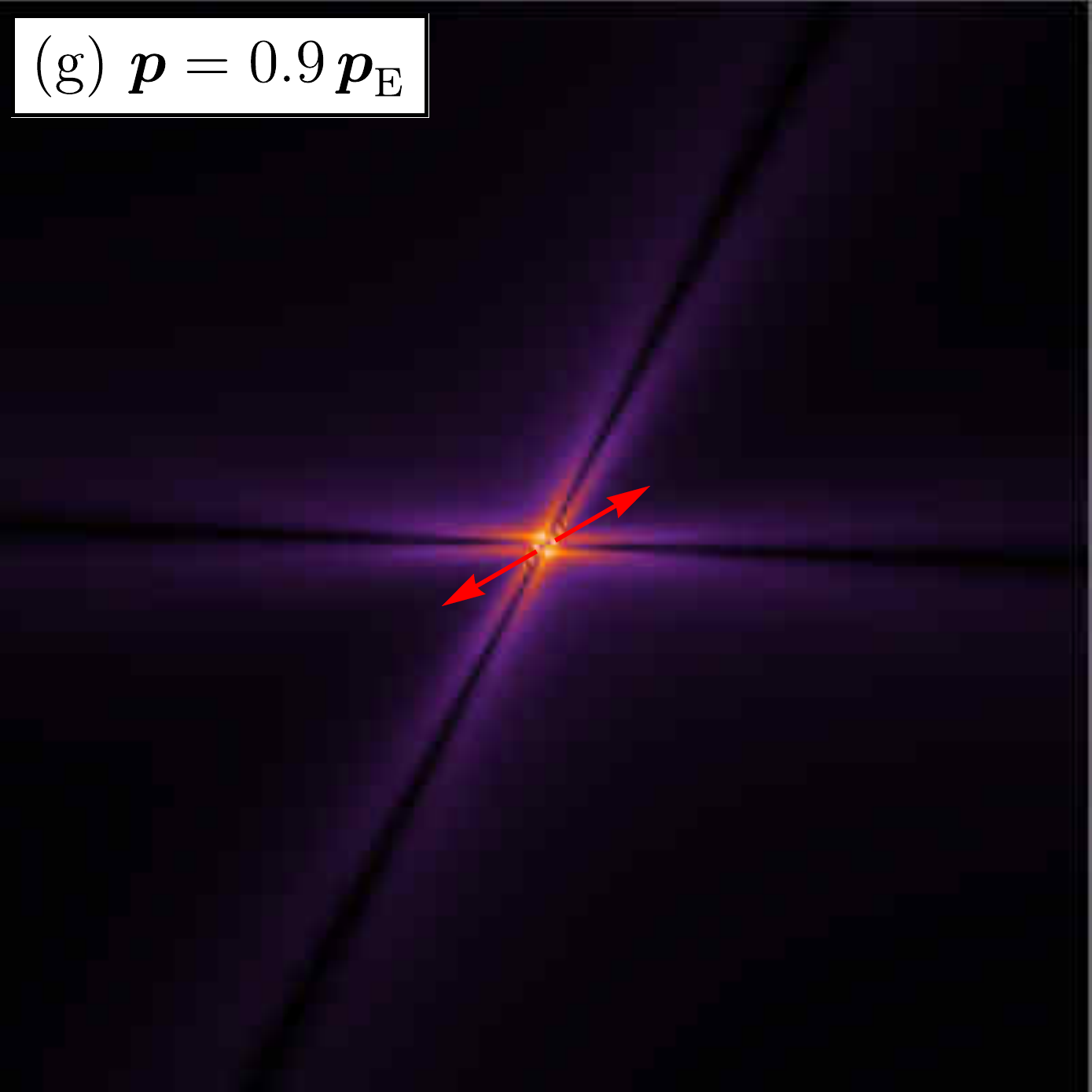}
    \end{subfigure}
    \begin{subfigure}{0.24\textwidth}
        \centering
        \phantomcaption{\label{fig:rhombus_10_10_dipole_99_gf}}
        \includegraphics[width=0.98\linewidth]{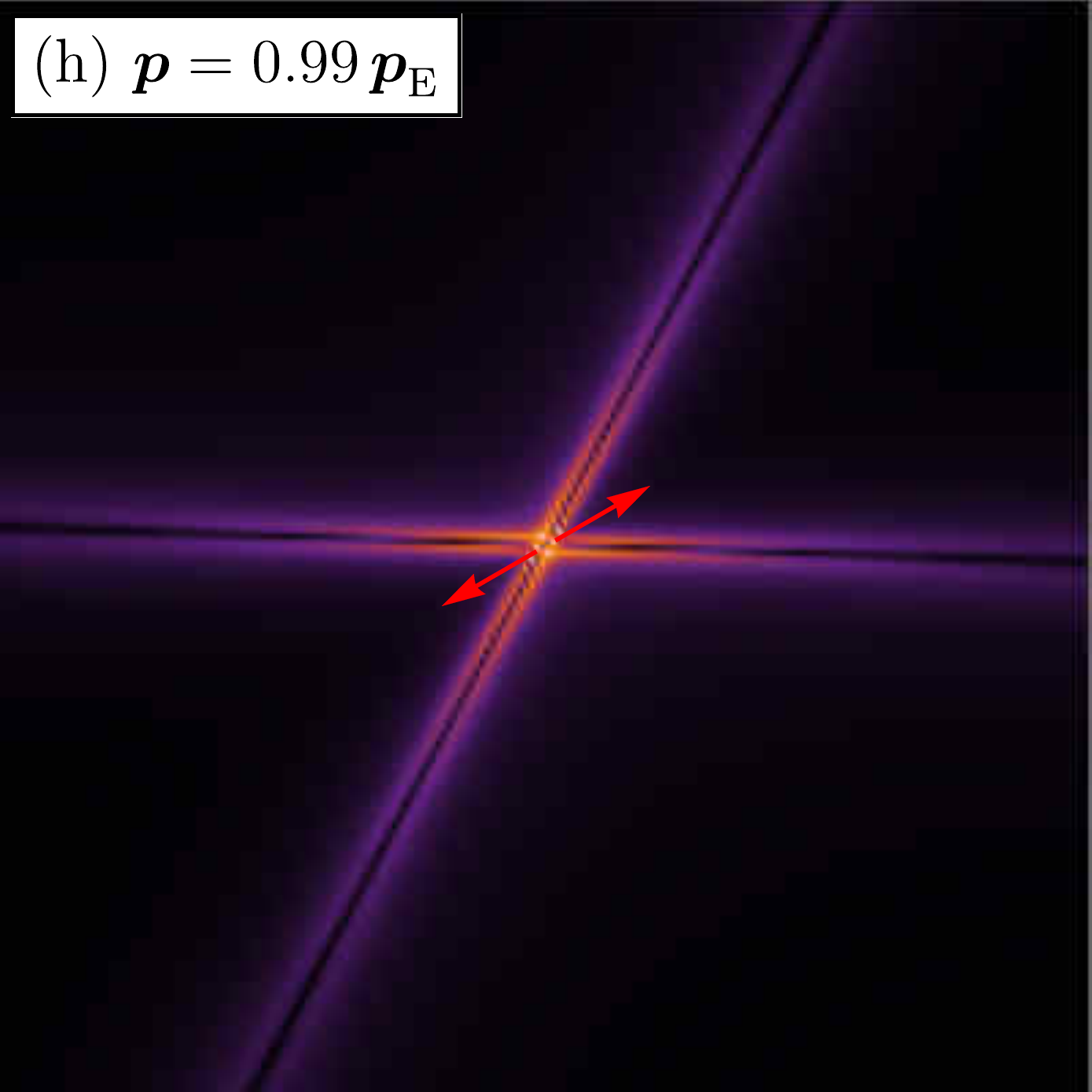}
    \end{subfigure}
    \caption{\label{fig:rhombus_10_10_dipole}
        As for Fig. \ref{fig:square_10_10_dipole}, but for an orthotropic rhombic lattice ($\Lambda_1=\Lambda_2=10$), where the localization bands are inclined ($\theta_{\text{cr}}=88.2^\circ, 151.8^\circ$).
        In contrast to the square grid, the inclination of the localization bands \textit{do not} coincide with the direction of the rods.
    }
\end{figure}

\begin{figure}[htb!]
    \centering
    \begin{subfigure}{0.24\textwidth}
        \centering
        \phantomcaption{\label{fig:rhombus_7_15_dipole_0}}
        \includegraphics[width=0.98\linewidth]{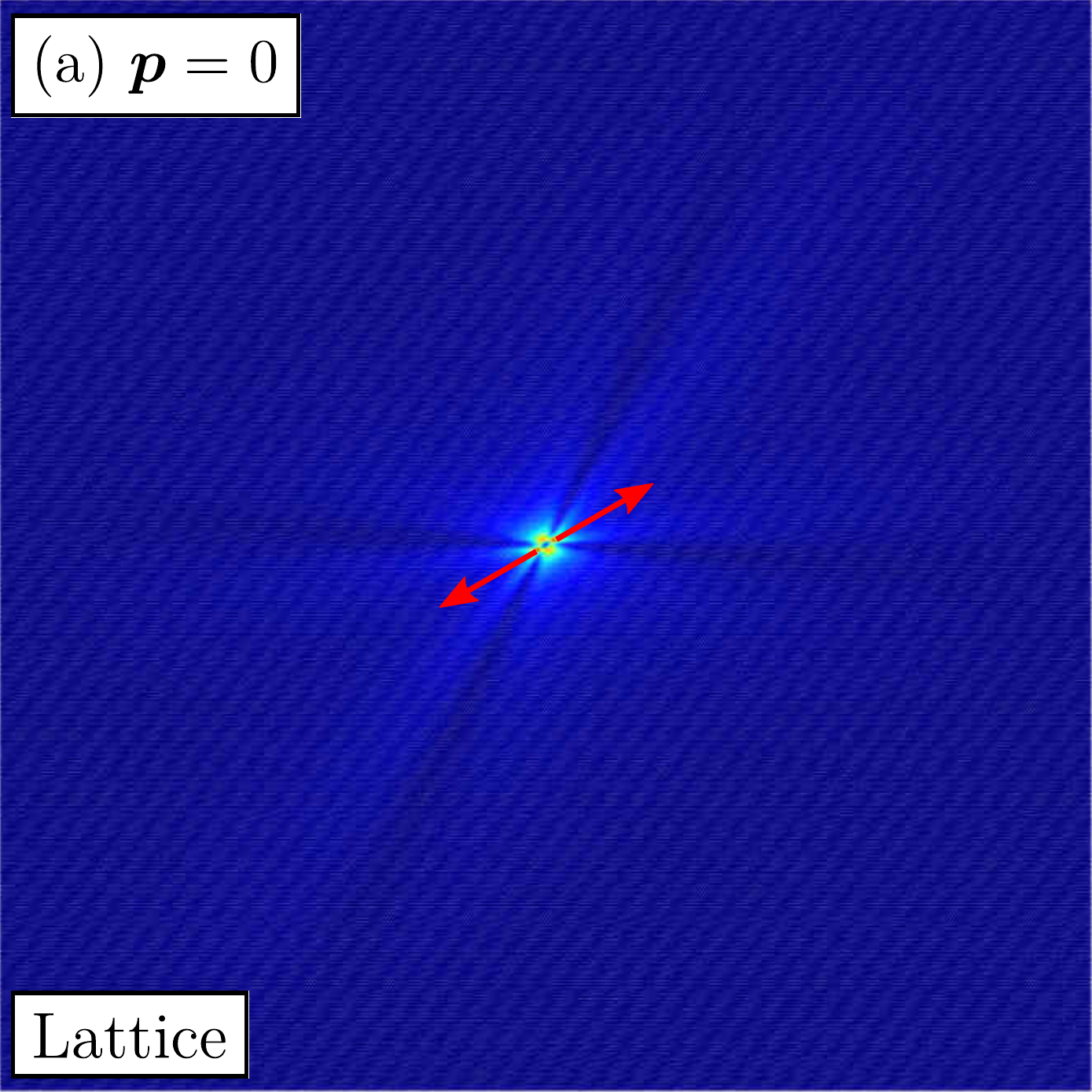}
    \end{subfigure}
    \begin{subfigure}{0.24\textwidth}
        \centering
        \phantomcaption{\label{fig:rhombus_7_15_dipole_80}}
        \includegraphics[width=0.98\linewidth]{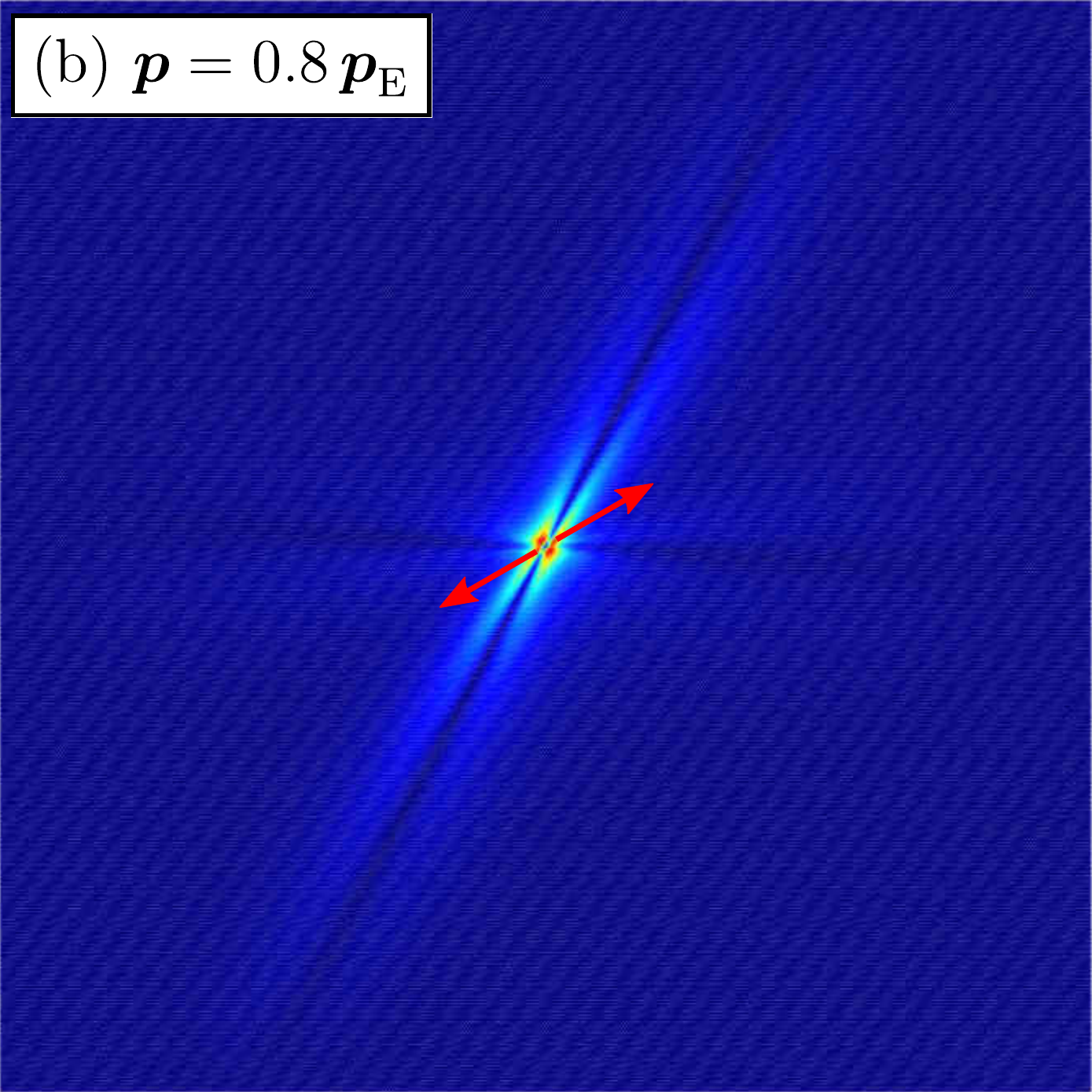}
    \end{subfigure}
    \begin{subfigure}{0.24\textwidth}
        \centering
        \phantomcaption{\label{fig:rhombus_7_15_dipole_90}}
        \includegraphics[width=0.98\linewidth]{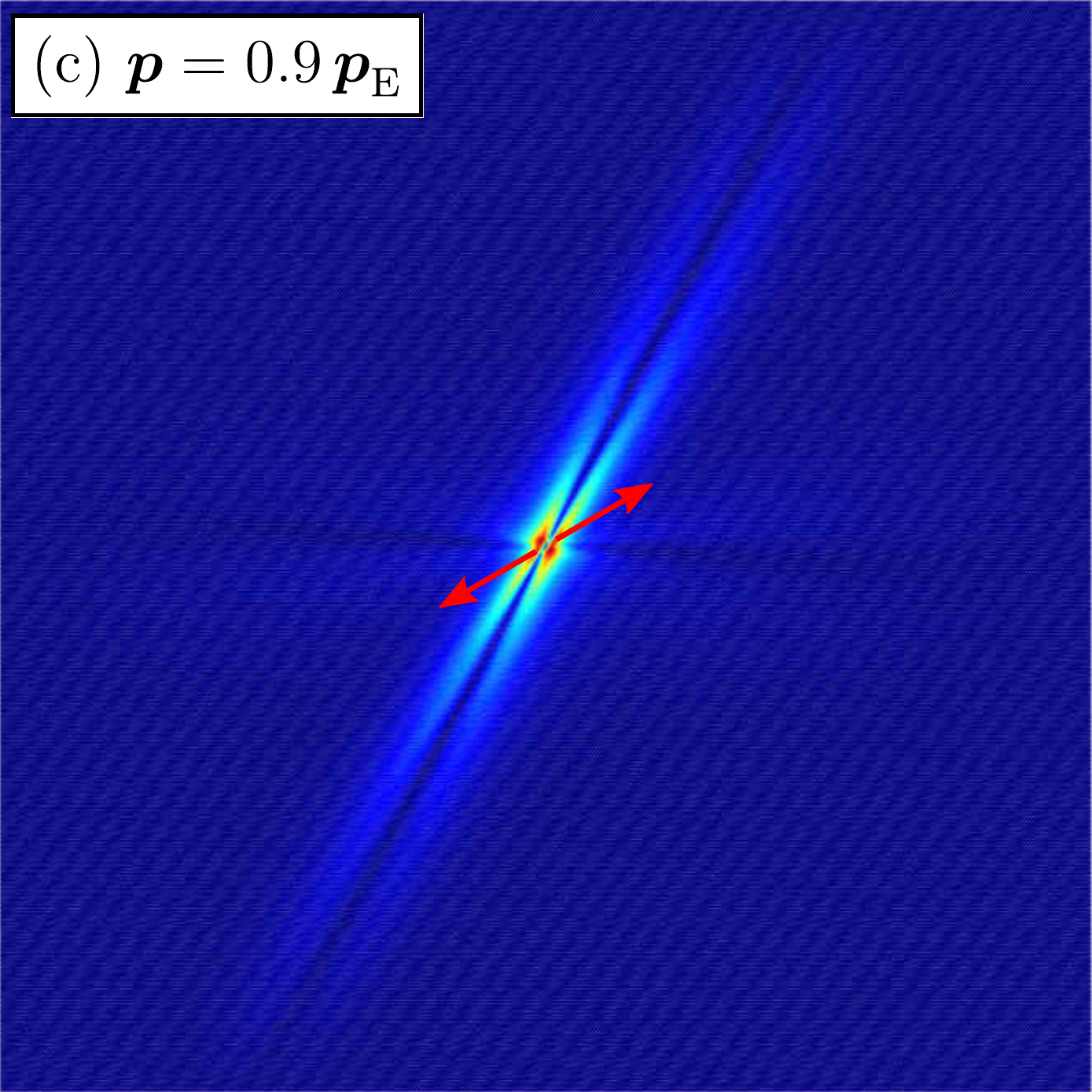}
    \end{subfigure}
    \begin{subfigure}{0.24\textwidth}
        \centering
        \phantomcaption{\label{fig:rhombus_7_15_dipole_99}}
        \includegraphics[width=0.98\linewidth]{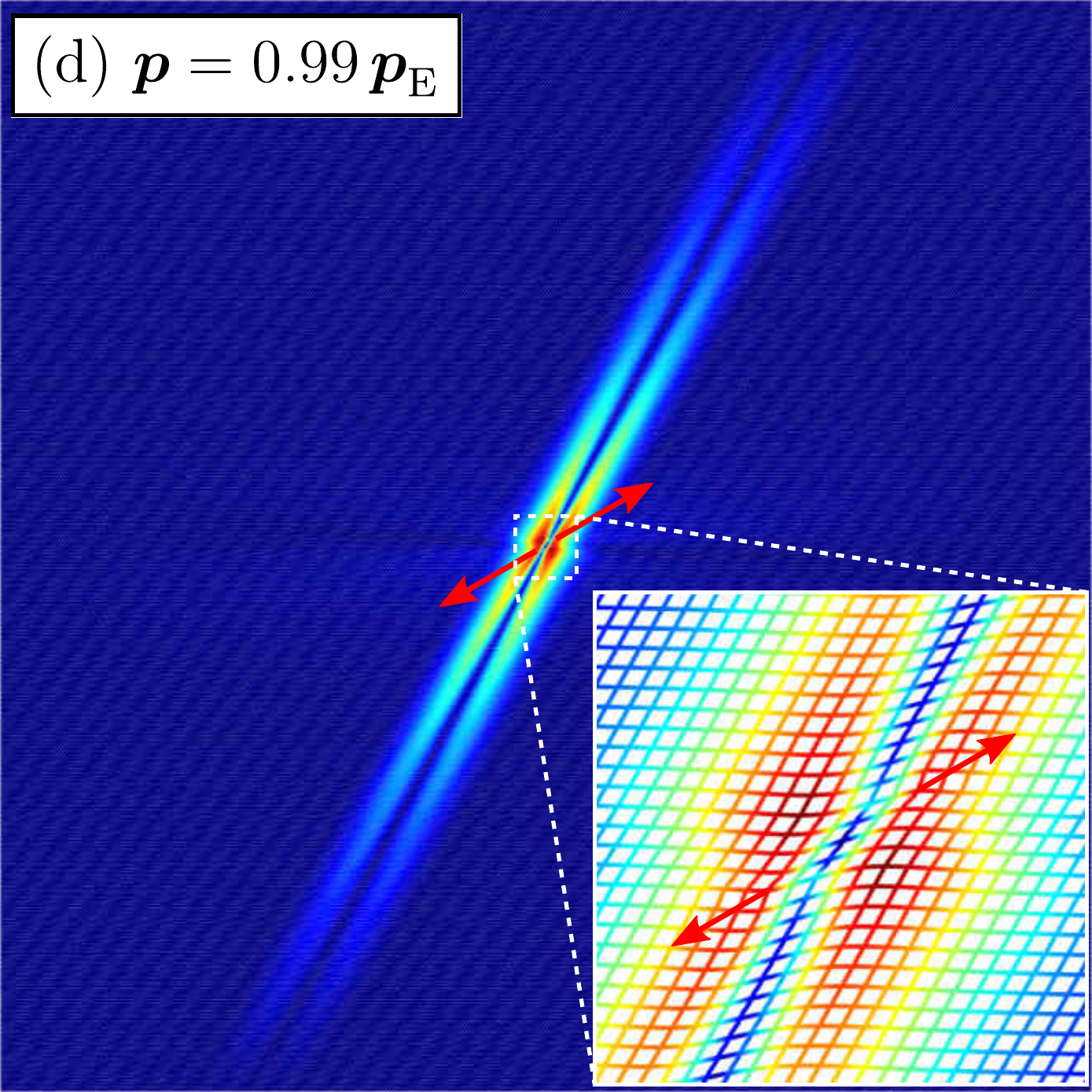}
    \end{subfigure}\\
    \vspace{0.01\linewidth}
    \begin{subfigure}{0.24\textwidth}
        \centering
        \phantomcaption{\label{fig:rhombus_7_15_dipole_0_gf}}
        \includegraphics[width=0.98\linewidth]{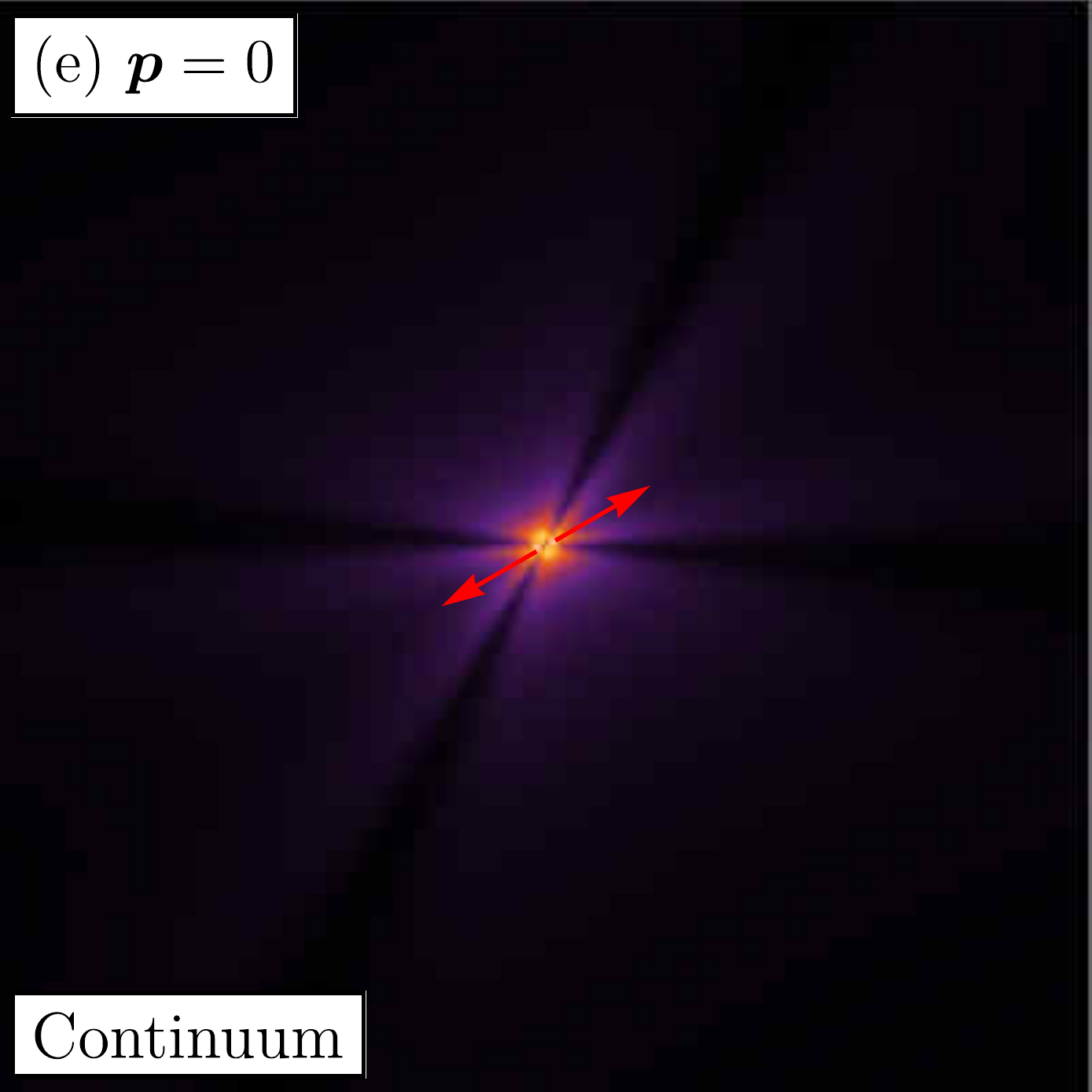}
    \end{subfigure}
    \begin{subfigure}{0.24\textwidth}
        \centering
        \phantomcaption{\label{fig:rhombus_7_15_dipole_80_gf}}
        \includegraphics[width=0.98\linewidth]{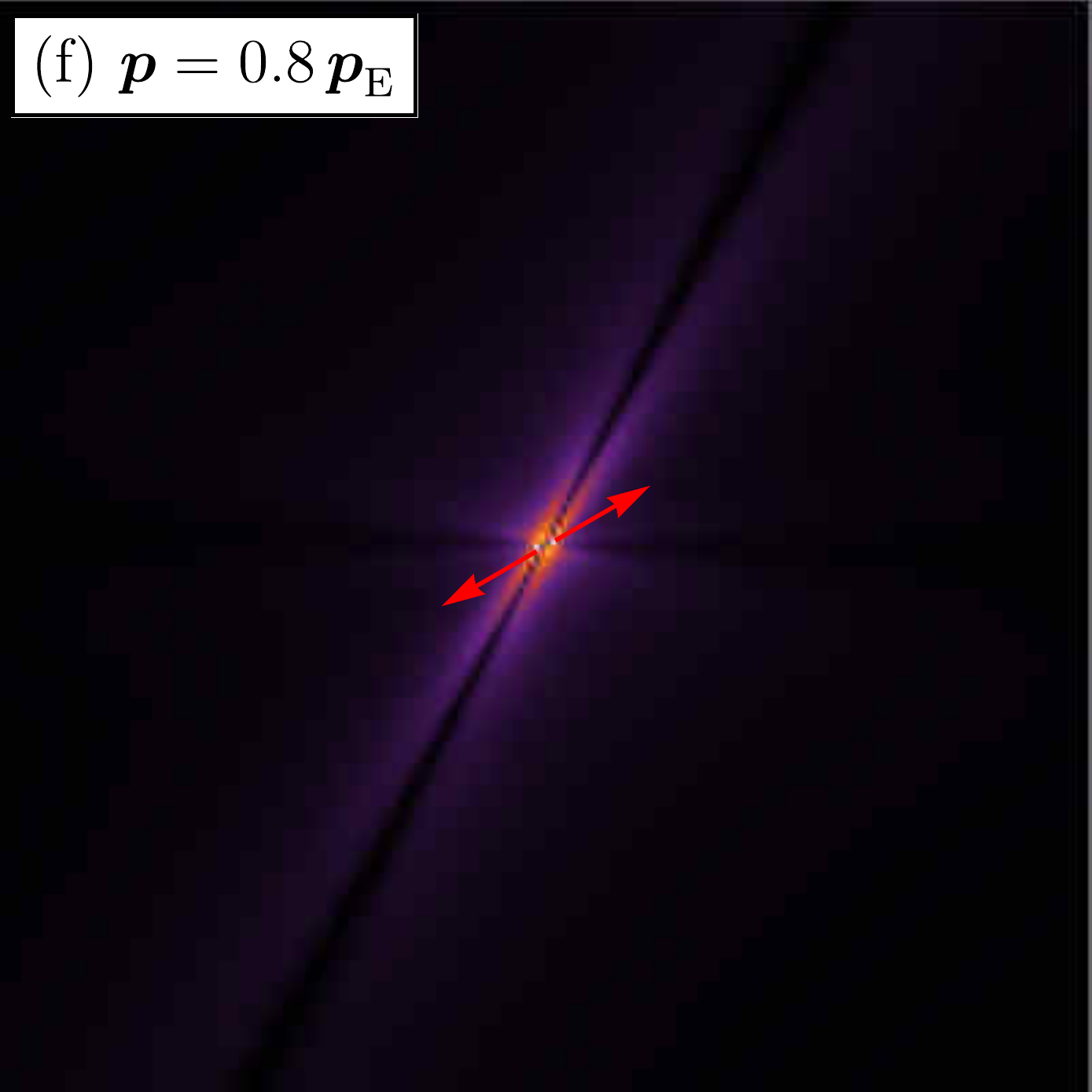}
    \end{subfigure}
    \begin{subfigure}{0.24\textwidth}
        \centering
        \phantomcaption{\label{fig:rhombus_7_15_dipole_90_gf}}
        \includegraphics[width=0.98\linewidth]{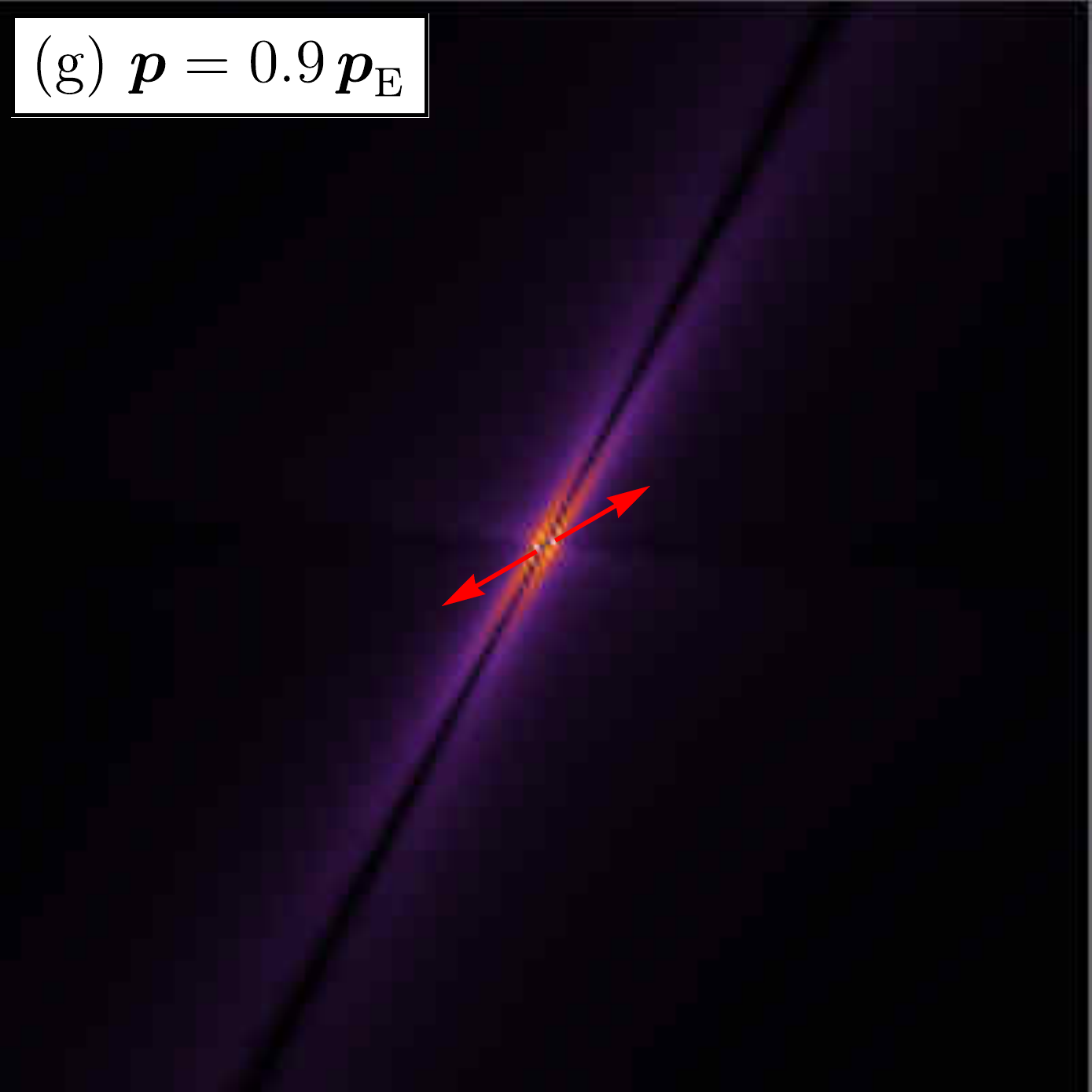}
    \end{subfigure}
    \begin{subfigure}{0.24\textwidth}
        \centering
        \phantomcaption{\label{fig:rhombus_7_15_dipole_99_gf}}
        \includegraphics[width=0.98\linewidth]{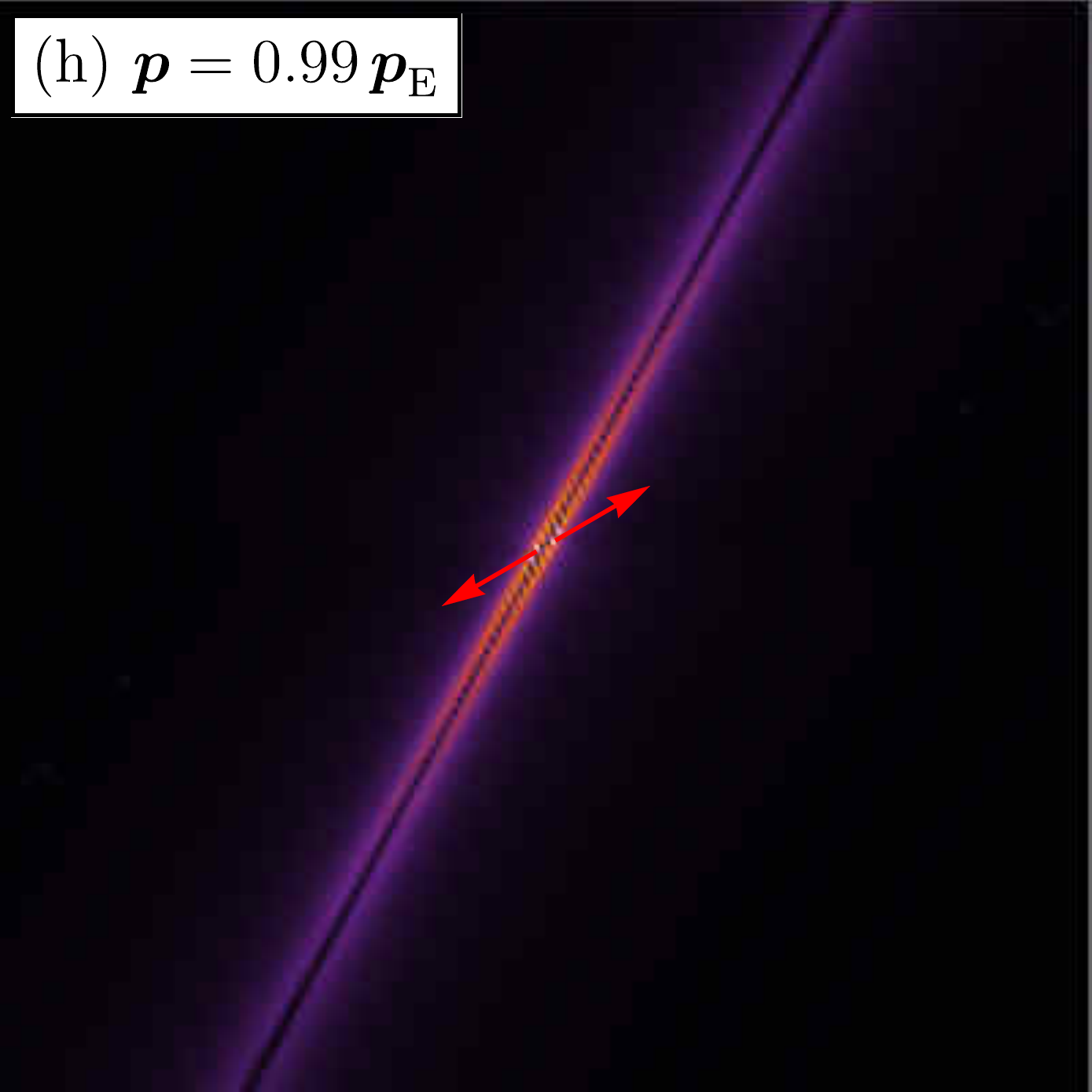}
    \end{subfigure}
    \caption{\label{fig:rhombus_7_15_dipole}
        As for Fig. \ref{fig:square_10_10_dipole}, but for an anisotropic rhombic lattice ($\Lambda_1=7, ~\Lambda_2=15$), where a single inclined localization band forms  ($\theta_{\text{cr}}=151.4^\circ$).
    }
\end{figure}

In the static regime, a comparison is presented between the response of the grillage loaded with a concentrated force dipole and a dipole Green's function for the effective solid.
As for the dynamic analysis of Section~\ref{sec:dynamic_forced_response}, the comparison is presented in terms of maps of incremental displacements (contour plots in Figs.~\ref{fig:square_10_10_dipole}--\ref{fig:rhombus_7_15_dipole}), where the color scale in the grid has been conveniently normalized according to the maximum value of the computed displacements.
In the upper part of the figures, results pertaining to the grid are presented, while, in the lower part, results refer to the equivalent continuum, obtained via homogenization.
The figures from left to right correspond to the application of increasing preloads, which approach the strong ellipticity boundary in the equivalent solid.
Insets placed in parts (d) of each figure ($\bp=0.99\bp_{\text{E}}$) illustrate a magnification of the lattice deformation in the neighborhood of the loading zone.
These details highlight the microscopic deformation patterns associated to the extreme mechanical response of the grid when loaded closely to the elliptic boundary.

In the situations analyzed in Figs. \ref{fig:square_10_10_dipole}--\ref{fig:rhombus_7_15_dipole}, the equivalent solid is found to be fully representative of the lattice structure. Therefore, approaching failure of ellipticity, the perturbative approach reveals, both in the continuum and in the real lattice, the formation of single or double bands in which incremental deformation localizes.
The bands can be horizontal, vertical or inclined.
The correspondence between the behavior of the grid and of its equivalent continuum is found to be excellent, so that the maps reported in the upper part of the figures are practically identical to the corresponding maps in the lower part of the figures.

\subsection{Micro-bifurcation in the lattice and effects on the equivalent solid}
\label{sec:microscopic_localization}
Micro-bifurcations occurring when the equivalent solid is still in the strong ellipticity range are investigated in this section, with reference to an equibiaxially compressed square grillage with cubic symmetry $\Lambda_1=\Lambda_2=10$ and diagonal springs of stiffness $\kappa=0.4$.
With the assumed spring stiffness, a \textit{microscopic} bifurcation is critical, as it occurs when the equivalent solid is still strongly elliptic.
\begin{figure}[htb!]
    \centering
    \begin{subfigure}{0.32\textwidth}
        \centering
        \phantomcaption{\label{fig:square_10_10_k_04_quadrupole}}
        \includegraphics[width=0.98\linewidth]{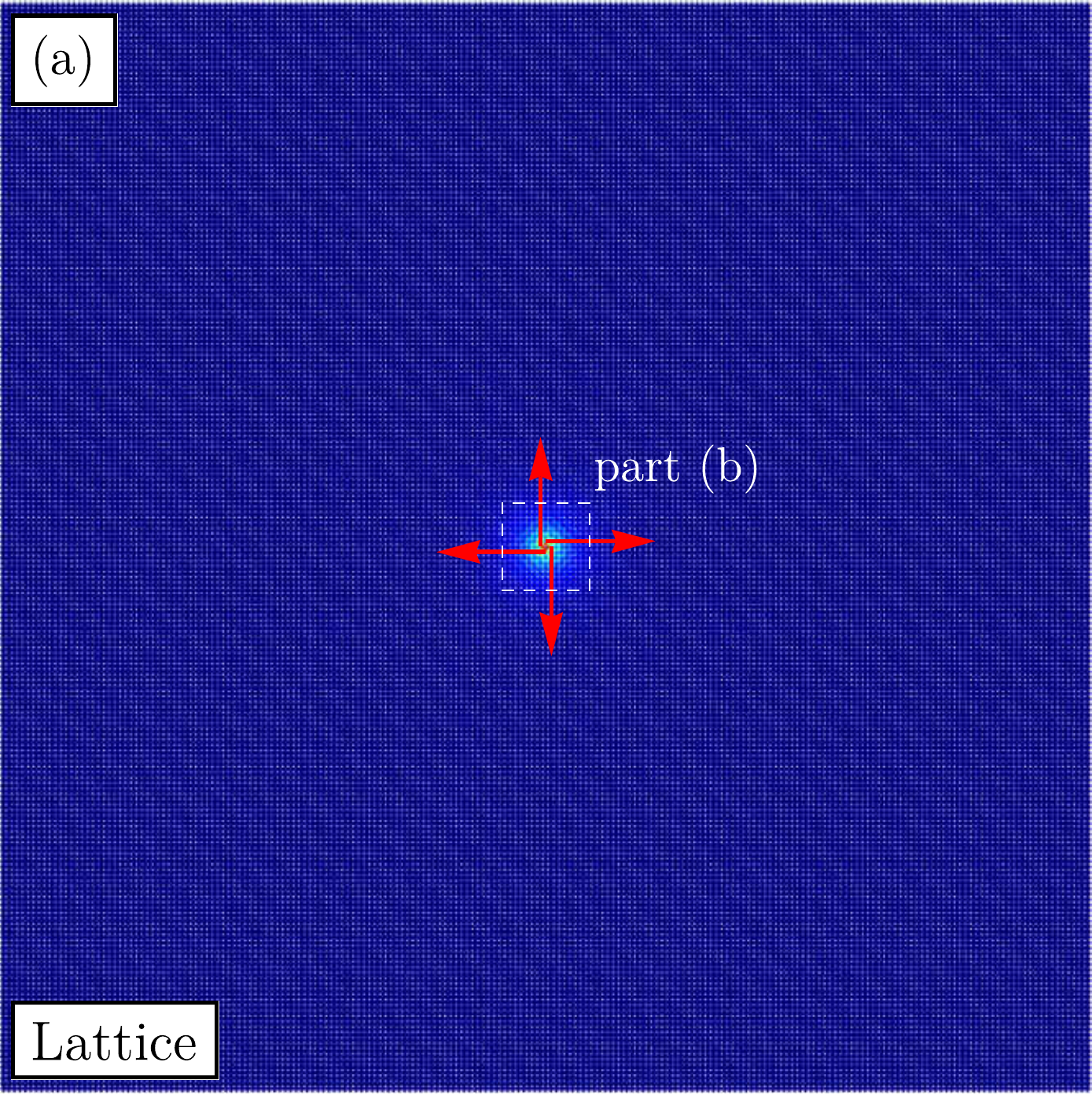}
    \end{subfigure}
    \begin{subfigure}{0.32\textwidth}
        \centering
        \phantomcaption{\label{fig:square_10_10_k_04_quadrupole_zoom}}
        \includegraphics[width=0.98\linewidth]{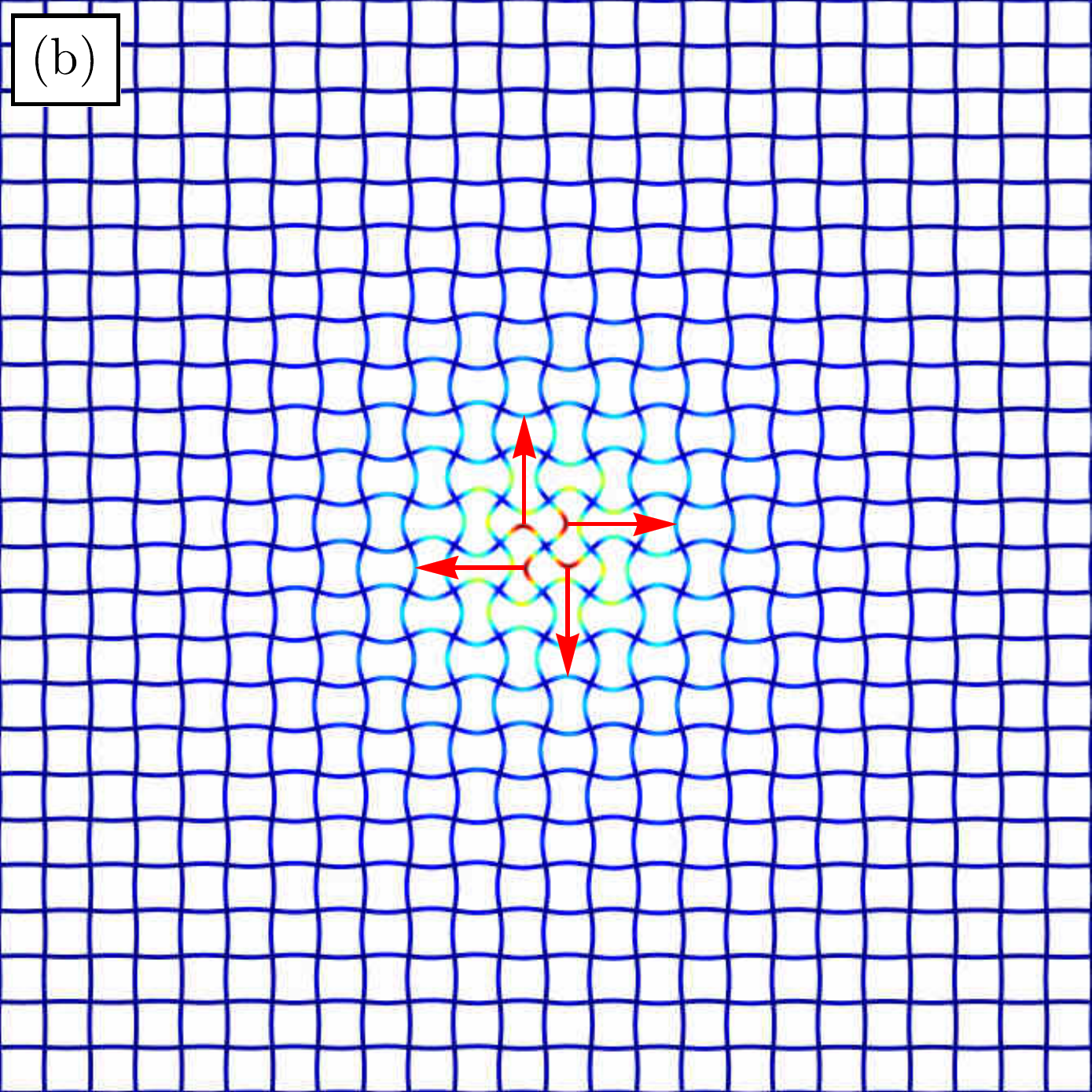}
    \end{subfigure}\\
    \vspace{0.01\linewidth}
    \begin{subfigure}{0.32\textwidth}
        \centering
        \phantomcaption{\label{fig:square_10_10_k_04_quadrupole_gf}}
        \includegraphics[width=0.98\linewidth]{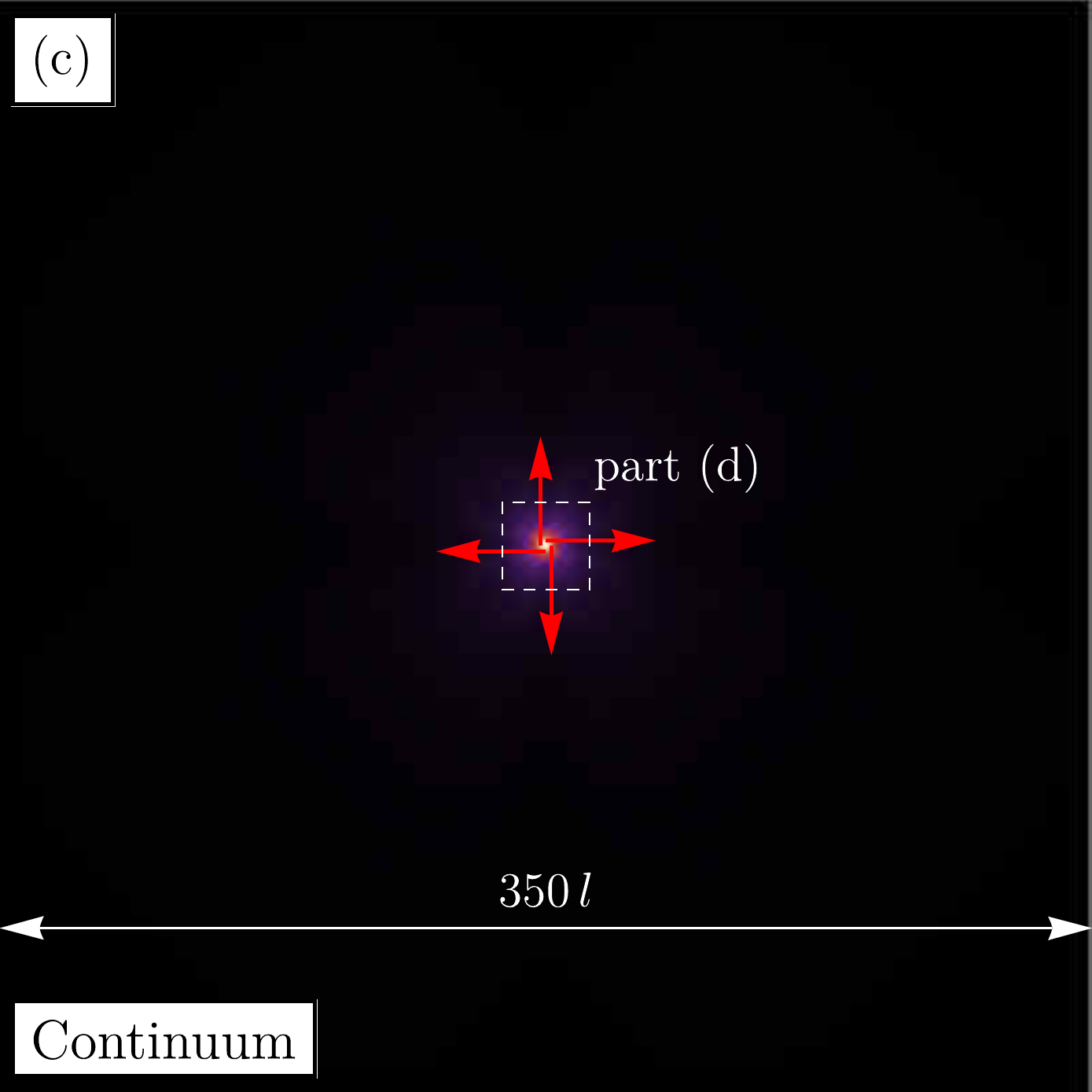}
    \end{subfigure}
    \begin{subfigure}{0.32\textwidth}
        \centering
        \phantomcaption{\label{fig:square_10_10_k_04_quadrupole_zoom_gf}}
        \includegraphics[width=0.98\linewidth]{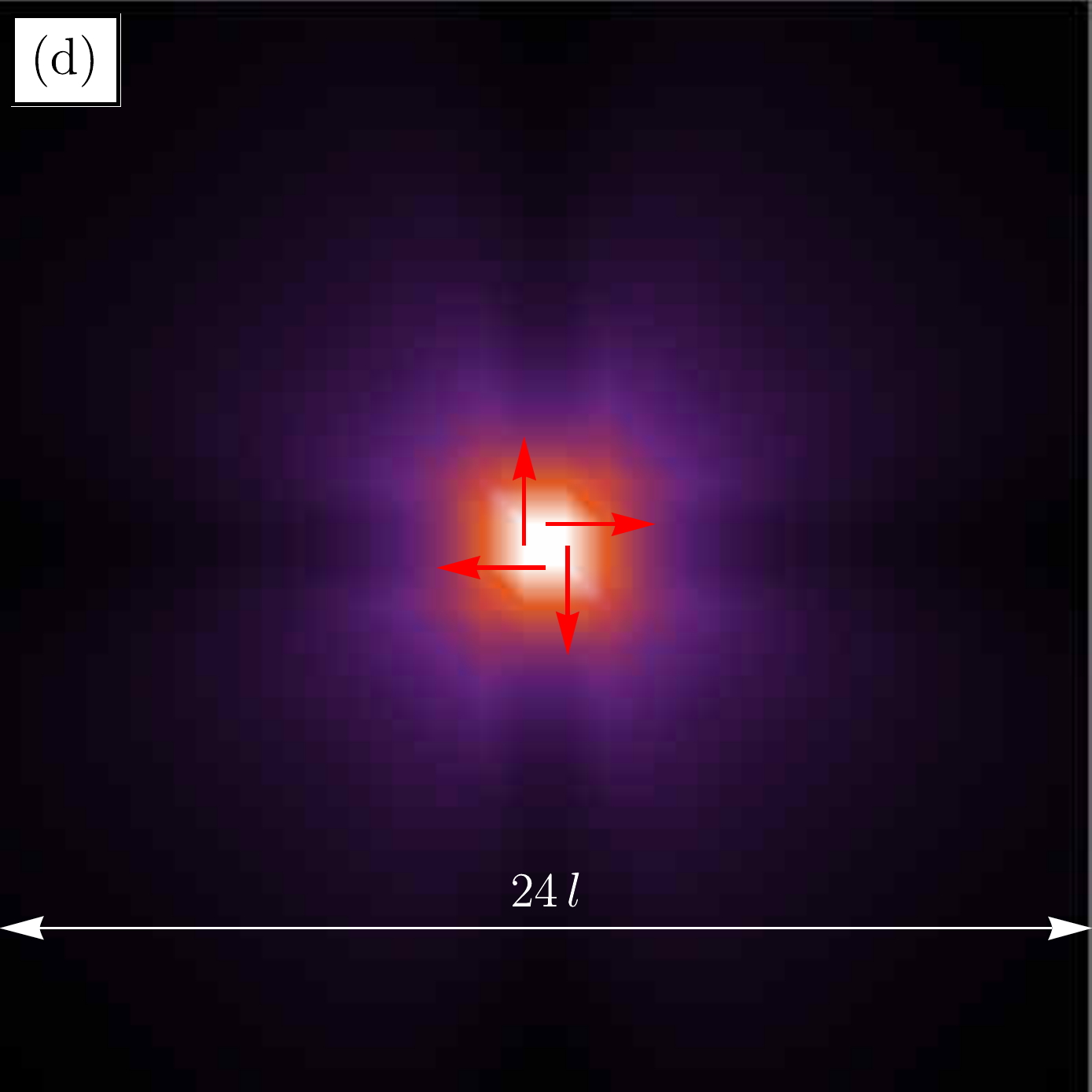}
    \end{subfigure}
    \caption{\label{fig:square_10_10_k_04_local_buckling}
        Microscopic localization of the bifurcation mode in the incremental displacement of a square lattice (cubic symmetry, $\Lambda_1=\Lambda_2=10$, upper part) and in its equivalent continuum (lower part). The prestress is an equibiaxial compression corresponding to bifurcation, $\bp_{\text{B}}=\{-\pi^2,-\pi^2\}$. A `quadrupole' of forces is applied at the midpoints of the rods.
        The quadrupole activates a highly localized `rotational' bifurcation mode (labeled as in $B_1$ in Fig.~\ref{fig:ellipticity_domains_10_10_pi2} and Table~\ref{tab:buckling_modes_points}), which leaves the lattice and the equivalent solid almost undeformed at a global level, but the lattice evidences a predominant \textit{inter-node deformation} at the scale of the unit cell.
    }
\end{figure}

The incremental displacement maps in the lattice \textit{at} the critical load for micro-bifurcation and in its equivalent continuum (still strongly elliptic) are shown in Fig.~\ref{fig:square_10_10_k_04_local_buckling}.
The incremental displacement is generated by the application of a force quadrupole.
The upper parts (lower parts) of the figure refer to the grid (to the continuum) and the parts on the right are a magnification of the zone near the force quadrupole.

The figure shows that the incremental response of the prestressed lattice is highly localized, so that only a strong magnification reveals buckling of the elastic rods.
Even if the equivalent continuum is not at bifurcation, its distribution of displacements somehow resembles that in the lattice, so that the homogenization may still be representative of the response of the discrete structure, even though the \textit{inter-node deformation} cannot be captured.
\begin{figure}[htb!]
    \centering
    \begin{subfigure}{55.5mm}
        \centering
        \phantomcaption{\label{fig:square_10_10_k_04_p_100_dipole}}
        \includegraphics[width=\linewidth]{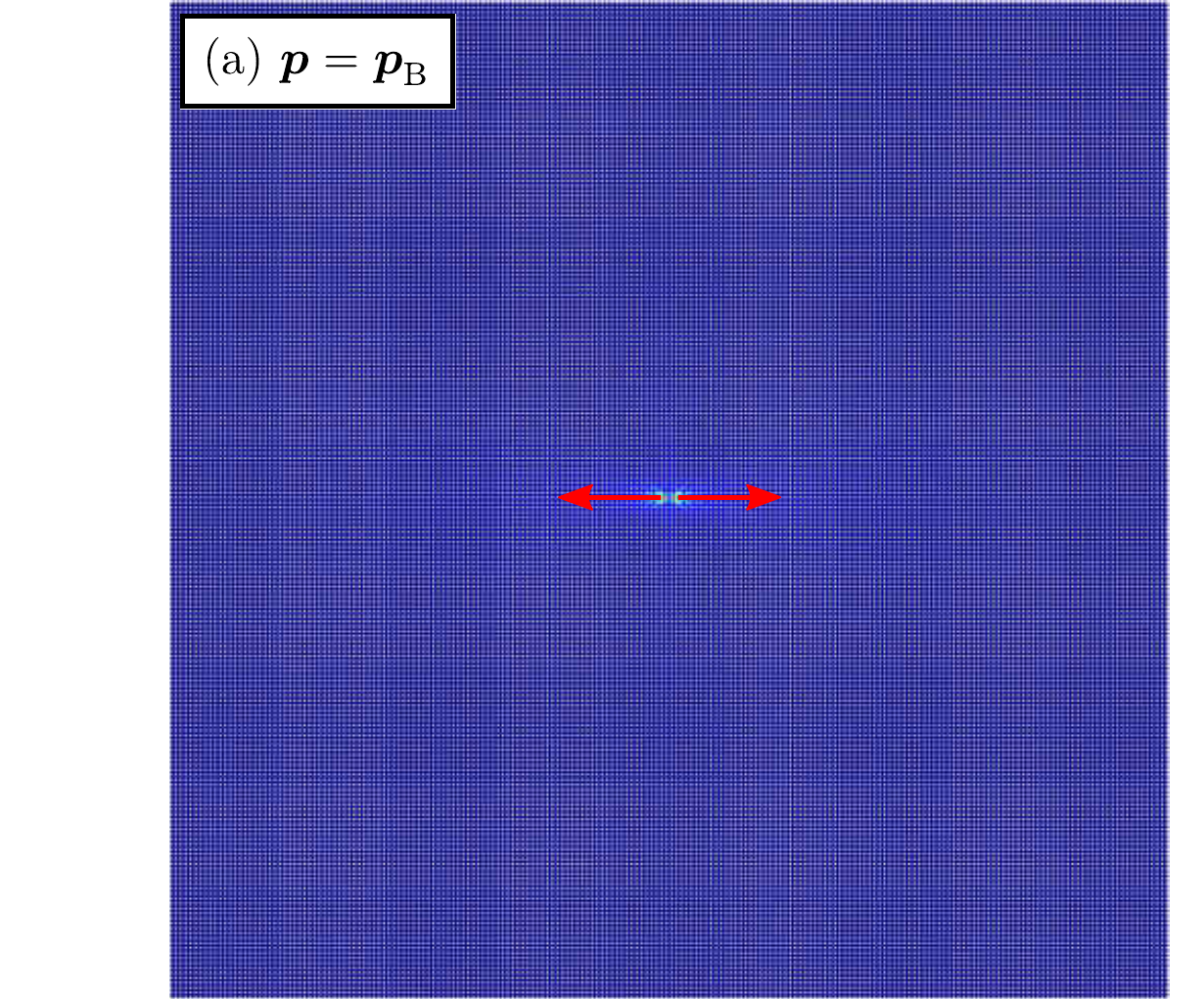}
    \end{subfigure}
    \begin{subfigure}{48.3mm}
        \centering
        \phantomcaption{\label{fig:square_10_10_k_04_p_105_dipole}}
        \includegraphics[width=\linewidth]{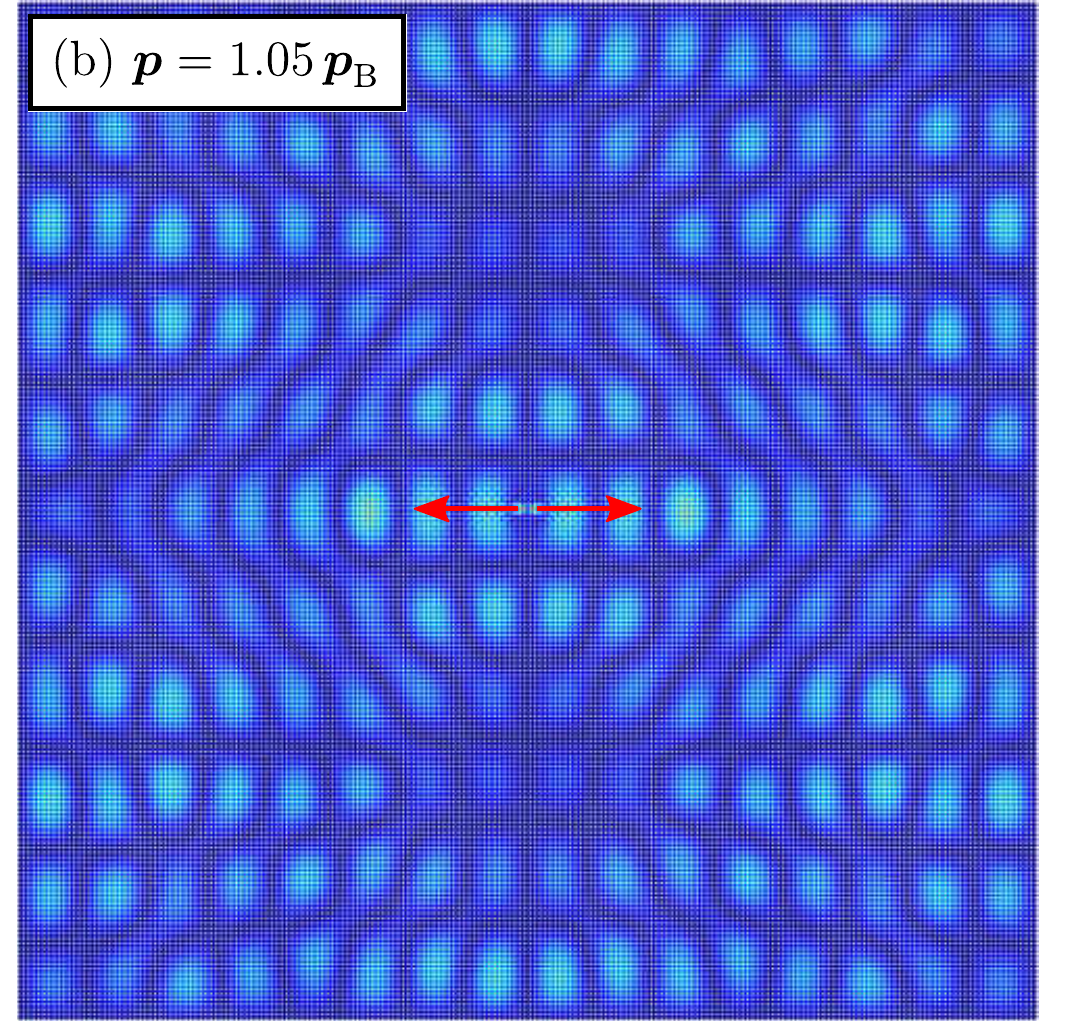}
    \end{subfigure}
    \begin{subfigure}{48.3mm}
        \centering
        \phantomcaption{\label{fig:square_10_10_k_04_p_110_dipole}}
        \includegraphics[width=\linewidth]{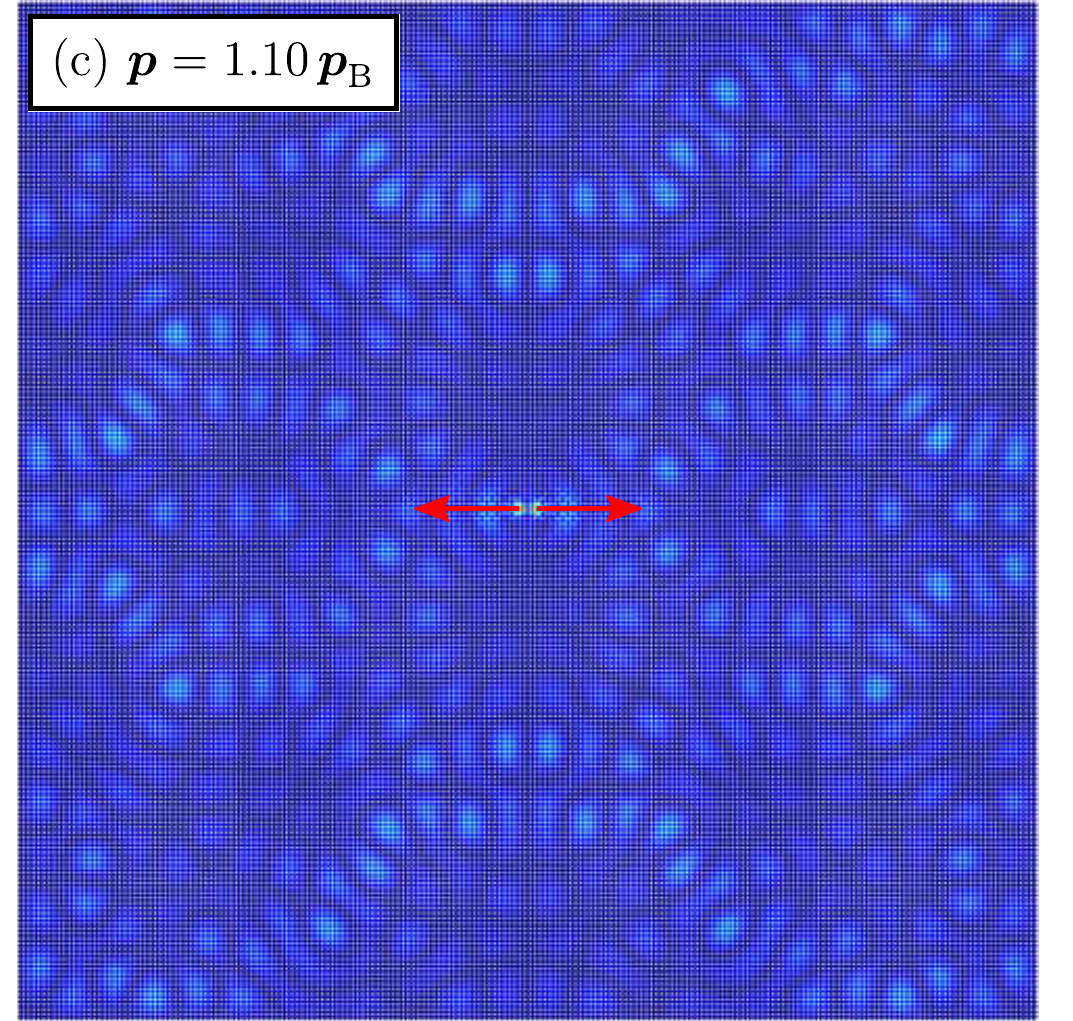}
    \end{subfigure}\\
    \vspace{2mm}
    \centering
    \begin{subfigure}{55.5mm}
        \centering
        \phantomcaption{\label{fig:square_10_10_k_04_p_100_dipole_FFT}}
        \includegraphics[width=\linewidth]{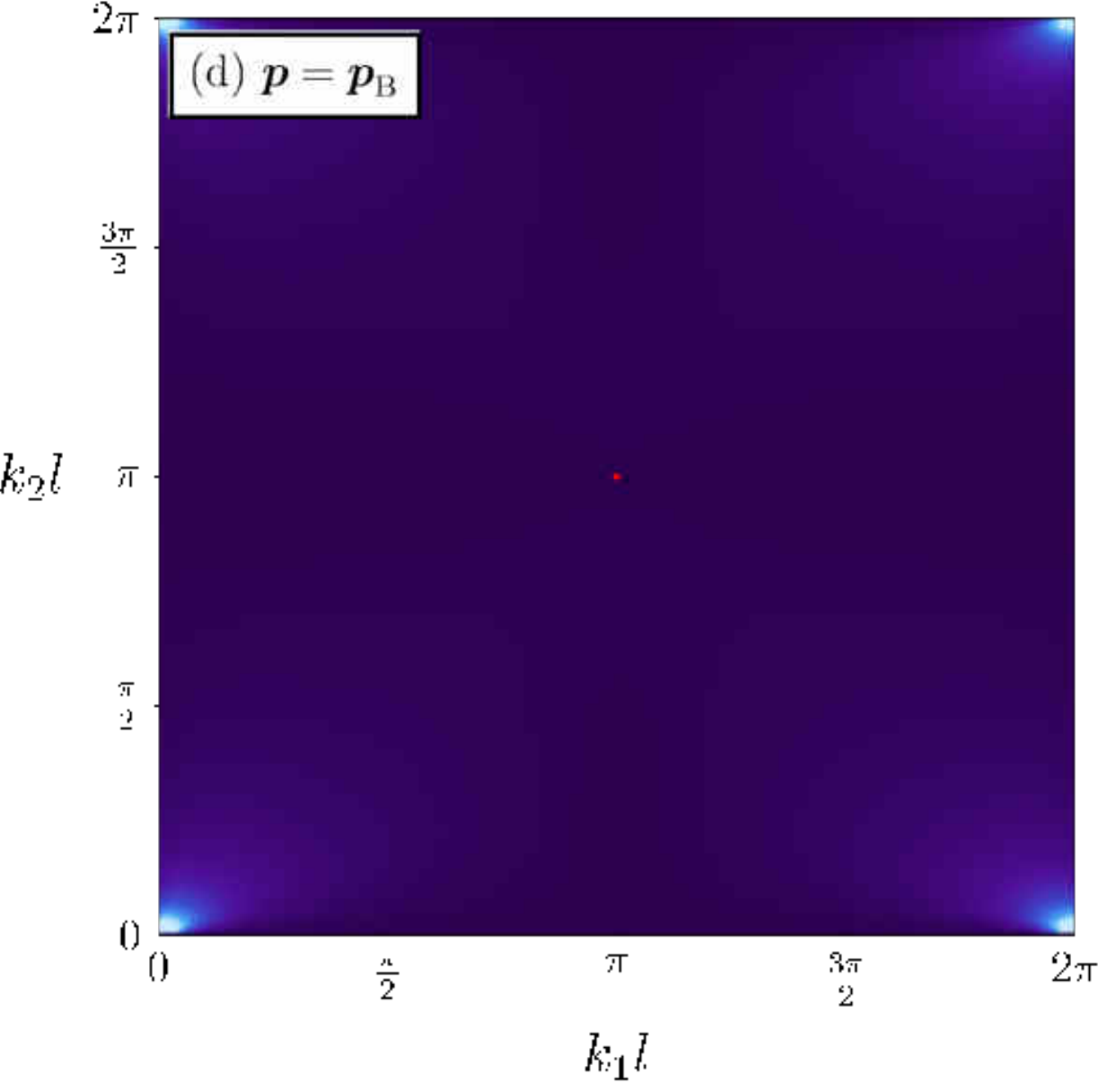}
    \end{subfigure}
    \begin{subfigure}{48.3mm}
        \centering
        \phantomcaption{\label{fig:square_10_10_k_04_p_105_dipole_FFT}}
        \includegraphics[width=\linewidth]{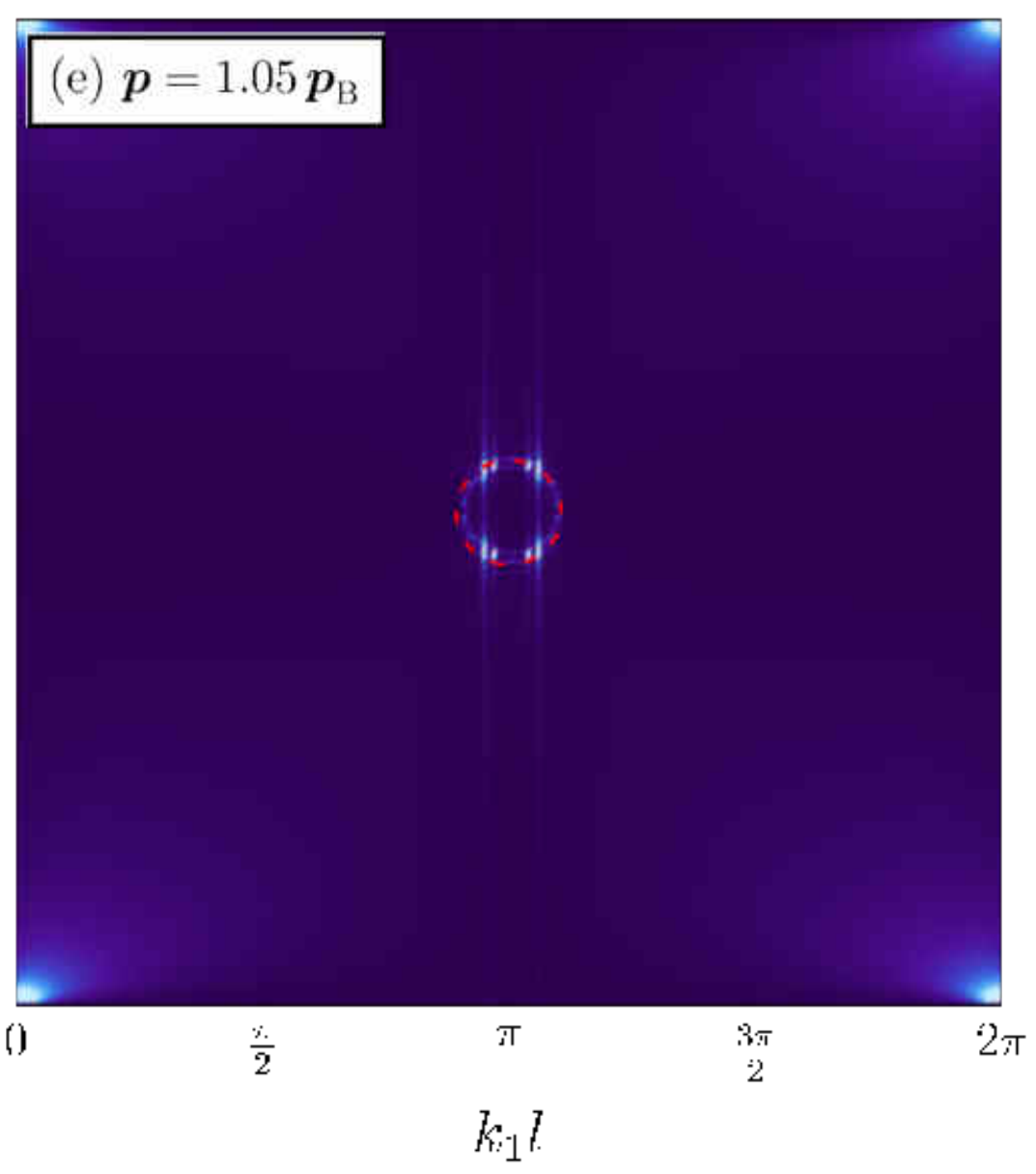}
    \end{subfigure}
    \begin{subfigure}{48.3mm}
        \centering
        \phantomcaption{\label{fig:square_10_10_k_04_p_110_dipole_FFT}}
        \includegraphics[width=\linewidth]{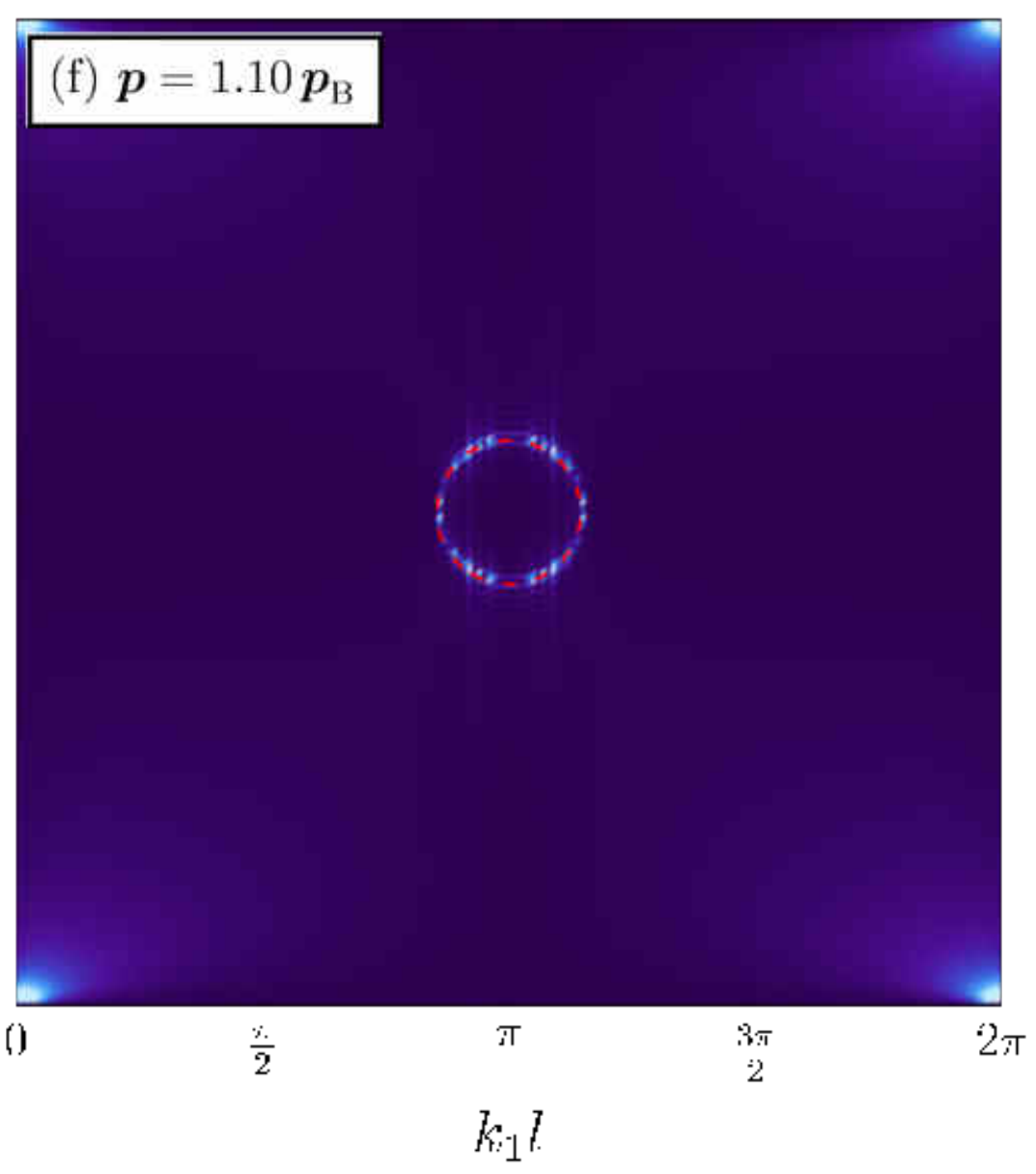}
    \end{subfigure}
    \caption{\label{fig:square_10_10_k_04_dipole_local_buckling}
        F.e.m. results for the displacement map (\subref{fig:square_10_10_k_04_p_100_dipole})--(\subref{fig:square_10_10_k_04_p_110_dipole}) and the corresponding Fourier transform (\subref{fig:square_10_10_k_04_p_100_dipole_FFT})--(\subref{fig:square_10_10_k_04_p_110_dipole_FFT}) showing the response of a lattice at a preload corresponding to microscopic instability, $\bp=\bp_{\text{B}}$ and beyond, $\bp=1.05\bp_{\text{B}}, ~ 1.10 \bp_{\text{B}}$.
        The slowness contour at null frequency, evaluated through the bifurcation condition is superimposed in red.
        While at the critical load the perturbation is so localized that is almost invisible, at higher loads an `explosive' instability involving the whole lattice and extending up to the boundary of the domain is clearly observed.
    }
\end{figure}

The situation depicted in Fig.~\ref{fig:square_10_10_k_04_local_buckling} completely changes when the lattice is loaded with forces beyond the critical value for micro bifurcation in the lattice, as shown in Fig.~\ref{fig:square_10_10_k_04_dipole_local_buckling}.
This figure refers only to the grid, loaded now with a horizontal force dipole, applied at three different biaxial compression preloadings.
In particular, a preload coincides with the critical load $\bp=\bp_{\text{B}}$ for micro-buckling and the other two, $\bp=1.05\,\bp_{\text{B}}$ and $\bp=1.10\,\bp_{\text{B}}$, are beyond.

This figure reports displacement maps (upper part) and the corresponding Fourier transform (obtained via FFT of nodal displacements, lower part), with superimposed slowness contours at null frequency.
The slowness contour (highlighted in red in the figure) was obtained from the bifurcation condition, Eq.~\eqref{eq:det_local_buckling}.
The fact that the slowness contour is superimposed to the peaks of the transform (reported white in the figure), is a validation of the good correspondence between calculations performed via Floquet-Bloch and finite element simulations.

The following conclusion can be drawn from Fig.~\ref{fig:square_10_10_k_04_dipole_local_buckling}.
\textit{
    While at micro-bifurcation an incremental perturbation remains confined and highly localized in the equivalent continuum, an `explosive instability' is found for the grillage.
    This instability does not decay and extends to the whole domain occupied by the structure.
    This is a special behavior which remains unobserved in the equivalent continuum (still strongly elliptic) and cannot be revealed through homogenization.
}

\section{Conclusions}
\label{sec:conclusions}
An analytic formulation has been developed for the time-harmonic dynamics of a grillage of elastic rods (equipped with distributed mass density and rotational inertia), subject to axial forces of an arbitrary amount and incrementally loaded in the plane.
Increments are unprescribed, so that incremental axial and shear forces and bending moments are involved.
The formulation leads, through an asymptotic expansion of Floquet-Bloch waves, to a low-frequency approximation for an equivalent prestressed elastic material.

The developed technique has been employed to systematically analyze arbitrary lattice geometries and preloaded configurations, therefore predicting both local and global material instabilities, in other words, micro-buckling and strain localization.
Loss of ellipticity has been analyzed for a skewed grid, to
\begin{enumerate*}[label=(\roman*)]
    \item explore cubic, orthotropic and fully anisotropic homogenized material responses,
    \item compute the elliptic domain for the homogenized continuum as a function of lattice parameters,
    \item analyze the structure of the acoustic branches close to ellipticity loss, and
    \item investigate forced vibrations (both in physical and Fourier spaces) revealing low-frequency wave localizations.
\end{enumerate*}

Loss of ellipticity has been analyzed both in quasi-static and dynamic conditions, the former situation obtained from the latter in the limit of vanishing frequency.
In particular, strain localization has been found to evidence the following features.

Quasi-static loading:
\begin{itemize}
    \item For all the analyzed grids, (i) the elliptic region is unbounded for tensile preload and bounded for compressive;
          (ii) deviations of the grid angle from orthogonality reduce the size of the elliptic region;
          (iii) the elliptic boundary is smooth everywhere except at a corner point.
    \item Several geometries of shear bands and localization bands have been detected.
          (i) A single shear band may form parallel or inclined to the grid, or
          (ii) two shear bands may occur, sometimes where the ellipticity domain forms a corner.
    \item Playing with the stiffness of the elements forming the grillage, it is possible to determine the occurrence of the first bifurcation, which may be macroscopic or microscopic.
          In the former case, it is detected in the homogenized continuum, in the latter it is not.
    \item For special grid geometries, a super-sensitivity of the localization band inclination has been found with respect to the state of preload.
          Super-sensitivity provides an enhanced tunability to be used in the design of materials to exhibit given localization patterns.
    \item Special conditions can be found in which the lattice bifurcates similarly to the equivalent continuum, namely, simultaneously displaying infinite modes, covering every wavelength.
          In this case, a perfect equivalence is obtained between the bifurcation in the grid and failure of ellipticity in the effective continuum.
\end{itemize}

Time-harmonic dynamics:
\begin{itemize}
    \item The rotational inertia of the elastic rods does not contribute to the definition of the prestressed elastic solid equivalent to the grillage.
    \item The homogenization scheme based on time-harmonic dynamics proves that the long-wavelength asymptotics for waves propagating in the lattice is governed by the acoustic tensor of the effective medium.
          This aspect definitely clarifies that a macro-bifurcation in the lattice has to be equivalent to failure of ellipticity in its equivalent continuum.
    \item Short-wavelength (or micro-) bifurcations are provided by the analysis of the dispersion relation of the lattice, interpreted now as a function of the axial preload state in the rods.
          Buckling in the lattice becomes the `propagation' of a Bloch wave at vanishing frequency.
          Macro-bifurcation in the grid occurs with the progressive lowering, and eventually vanishing, of the slope of the acoustic branches at the origin, while the dispersion surface attains non-null frequency for every other wave vector.
          On the contrary, micro-bifurcation is characterized by a non-vanishing slope of the acoustic branches at the origin, but the preload-induced lowering of the dispersion surface causes the generation of a zero-frequency wave with non-null wave vector (corresponding to a finite wavelength buckling).
    \item For square grid geometries, the shear wave responsible for the ellipticity loss is the one propagating along the direction of the stiffer elastic link (with the lowest slenderness).
          This conclusion is counterintuitive, as a shear mechanism would be expected to be generated in the direction of the soft elastic links.
    \item Shear bands have been investigated through a Fourier transform of the lattice response, evidencing the Bloch spectrum generated by the forcing source, to be compared with the corresponding slowness contour generated by the eigenmode analysis.
          In conditions close to the ellipticity loss, sharp peaks in the Fourier transform demonstrate that the pulsating force is emanating pure plane waves, which is the `signature' of strain localization in a dynamic context.
    \item The asymptotic homogenization scheme is performed near the vanishing frequency.
          It is therefore believed to be more closely representative of the lattice when the frequency of the pulsating force is sufficiently low.
          When the elliptic boundary is approached, the mismatch in the acoustic properties, between the lattice and its effective continuum approximation, has been found to become wider for those waves which propagate parallel to the direction of ellipticity loss.
\end{itemize}

Depending on the lattice geometry and preload state, a micro-bifurcation may occur in the grillage while the equivalent continuum is still strongly elliptic.
This bifurcation passes undetected via homogenization, but may become dominant.
In fact, when the preload in the grid is higher than that critical for micro-bifurcation, an `explosive' instability may occur.
Such instability may start at a point as the effect of a perturbation and evidence an unbounded growth.
This circumstance has been vividly demonstrated through the perturbative approach to material instability.

It can be concluded, in closure, that homogenization of the incremental response of a grillage of elastic rods, axially preloaded to an arbitrary amount, provides an excellent tool for the design of cellular elastic materials of tunable properties and capable of extremely localized deformations occurring within their elastic range.

\section*{Acknowledgements}
Financial support is acknowledged from:
the ERC Advanced Grant `Instabilities and nonlocal multiscale modeling of materials' ERC-2013-ADG-340561-INSTABILITIES (G.B. and L.C.),
PRIN 2015 2015LYYXA8-006 and ARS01-01384-PROSCAN (D.B. and A.P.).
The authors also acknowledge support from the Italian Ministry of Education, University and Research (MIUR) in the frame of the `Departments of Excellence' grant L. 232/2016.

\printbibliography

\appendix
\numberwithin{equation}{section}            %
\numberwithin{figure}{section}              %

\section{Linearized equations of motion for an axially pre-stretched elastica}
\label{sec:linearized_elastica}
A model for an axially stretchable Rayleigh elastic rod can be obtained through a linearization (around a stretched equilibrium configuration) of the equations governing the dynamics of large deflections and flexure of the elastica endowed with rotational inertia.
A local axial coordinate $x_0$ is introduced to single out points of the straight, undeformed, and stress-free configuration of the rod.
This configuration is assumed as reference, so that the potential and kinetic energies are defined as
\begin{subequations}
    \label{eq:potential_kinetic_energy_elastica}
    \begin{align}
        \mV & = \int_0^{l_0} \left( \psi_\lambda(\lambda) + \psi_\chi(\chi) - P\,u'(x_0,t) \right) dx_0 \,,                                                     \\
        \mT & = \frac{1}{2} \int_0^{l_0} \left(\gamma_0 \left(\dot{u}(x_0,t)^2 + \dot{v}(x_0,t)^2\right) + \gamma_{r,0}\,\dot{\theta}(x_0,t)^2 \right) dx_0 \,,
    \end{align}
\end{subequations}
where $l_0$, $\gamma_0$, and $\gamma_{r,0}$ are the initial length, linear mass density, and rotational inertia, while $\psi_\lambda$ and $\psi_\chi$ are strain-energy functions for, respectively, axial and flexural deformations.
The axial stretch $\lambda$ and the curvature $\chi$ are defined by the kinematics of an extensible, but unshearable, elastica as
\begin{subequations}
    \label{eq:strain_measures_elastica}
    \begin{align}
        \label{eq:axial_strain}
        \lambda & = (1+u'(x_0,t))\cos{\theta(x_0,t)} + v'(x_0,t)\sin{\theta(x_0,t)} \,,                                 \\
        \label{eq:curvature}
        \chi    & = \theta'(x_0,t) = \deriv{}{x_0} \left(\arctan\left(\frac{v'(x_0,t)}{1 + u'(x_0,t)}\right)\right) \,,
    \end{align}
\end{subequations}
where in~\eqref{eq:curvature} the unshearability constraint $\theta = \arctan\left(\frac{v'}{1+u'}\right)$ has been explicitly introduced and the symbol $'$ indicates differentiation with respect to the first argument of the function, in this case $x_0$.

A second-order expansion of the functionals~\eqref{eq:potential_kinetic_energy_elastica}, with respect to the independent displacement fields $\{u, v\}$, around the deformed configuration $\{u_0,v_0\} =\{(\lambda_0-1) x_0, 0\}$, yields the linearized response of the rod. The linearization is around a \textit{straight, but axially stretched, configuration}.
A substitution of Eq.~\eqref{eq:strain_measures_elastica} into Eq.~\eqref{eq:potential_kinetic_energy_elastica} and neglection of an arbitrary constant, leads to the following expansion
\begin{subequations}
    \label{eq:potential_kinetic_energy_elastica_second_order}
    \begin{align}
        \label{eq:potential_energy_elastica_second_order}
         & \begin{aligned}
            \mV(u_0+\delta u, v_0 + \delta v) \sim
             & \int_0^{l_0} \left(\psi_\lambda'(\lambda_0)-P\right)\delta u'(x_0,t) dx_0 +                  \\
             & +\frac{1}{2} \int_0^{l_0} \psi_\lambda''(\lambda_0)\delta u'(x_0,t)^2 dx_0 +                 \\
             & +\frac{1}{2} \int_0^{l_0}\left( \frac{\psi_\lambda'(\lambda_0)}{\lambda_0}\delta v'(x_0,t)^2
            + \frac{\psi_\chi''(0)}{\lambda_0^2}\delta v''(x_0,t)^2 \right) dx_0 \,,
        \end{aligned} \\
         & \begin{aligned}
            \mT(u_0+\delta u, v_0 + \delta v) \sim
            \frac{1}{2} \int_0^{l_0} \left( \gamma_0 \left(\delta\dot{u}(x_0,t)^2 + \delta\dot{v}(x_0,t)^2\right) + \frac{\gamma_{r,0}}{\lambda_0^2}\delta\dot{v}'(x_0,t)^2 \right)dx_0 \,,
        \end{aligned}
    \end{align}
\end{subequations}
where the `residual' bending moment in the unloaded configuration is assumed to be zero, $\psi_\chi'(0) = 0$, and \lq$\delta$' denotes a small variation.

The vanishing of the first-order term in Eq.~\eqref{eq:potential_energy_elastica_second_order}, occurring when the configuration $\{u_0,v_0\} =\{(\lambda_0 -1) x_0, 0\}$ satisfies equilibrium, implies that the prestretch $\lambda_0$ is the solution of the condition $\psi_\lambda'(\lambda_0)-P=0$.
This indicates that the axial load $P$ is equal to the axial pre-load.
Moreover, the second-order term in Eq.~\eqref{eq:potential_energy_elastica_second_order} involves the strain energy functions only in terms of second derivatives, $\psi_\lambda''(\lambda_0)$ and $\psi_\chi''(0)$, evaluated on the straight stretched configuration.

It is now instrumental to update the reference configuration from the stress-free configuration to the stretched configuration, so that the second-order functional~\eqref{eq:potential_kinetic_energy_elastica_second_order} can be adopted to govern the incremental response of the rod.
To this purpose, the variable of integration is changed from $x_0$ to the current stretched coordinate $s = \lambda_0 x_0$, so that the fields $\{u, v\}$ become functions of $s$.
The second-order terms in Eqs.~\eqref{eq:potential_kinetic_energy_elastica_second_order} can now be written as
\begin{subequations}
    \label{eq:potential_kinetic_energy_elastica_second_order_updated}
    \begin{align}
         & \begin{aligned}
            \mV(u_0+\delta u, v_0 + \delta v) \sim
             & \frac{1}{2} \int_0^{l} \psi_\lambda''(\lambda_0)\lambda_0\, \delta u'(s,t)^2 ds + \\
             & +\frac{1}{2} \int_0^{l}\left( P\delta v'(s,t)^2 +
            \psi_\chi''(0)\lambda_0\, \delta v''(s,t)^2 \right) ds \,,
        \end{aligned} \\
         & \begin{aligned}
            \mT(u_0+\delta u, v_0 + \delta v) \sim
            \frac{1}{2} \int_0^{l} \left( \frac{\gamma_0}{\lambda_0} \left(\delta\dot{u}(s,t)^2 + \delta\dot{v}(s,t)^2\right) + \frac{\gamma_{r,0}}{\lambda_0}\delta\dot{v}'(s,t)^2 \right) ds\,,
        \end{aligned}
    \end{align}
\end{subequations}
where $l=\lambda_0 l_0$ denotes the current length of the rod and the symbol $'$ now indicates differentiation with respect to $s$\footnote{
    Note that, with a little abuse of notation, the symbols for the functions $\{u,v\}$ have been maintained even though the independent variable has changed from $x_0$ to $s$.
}.
The variations $\delta u'(s)$ and $\delta v''(s)$ are, respectively, the incremental axial strain and curvature, so that the corresponding coefficients are the \textit{current values of axial and bending stiffness}, so that they can be concisely denoted as $\psi_\lambda''(\lambda_0)\lambda_0 = A(\lambda_0)$ and $\psi_\chi''(0)\lambda_0 = B(\lambda_0)$, both functions of the current axial stretch $\lambda_0$.

The second-order functionals, Eqs. (\ref{eq:potential_kinetic_energy_elastica_second_order_updated}),  describe the incremental response of the axially pre-stretched and pre-loaded rod.
Therefore, the equations of motion governing the incremental displacements can be derived via the following functionals
\begin{subequations}
    \label{eq:potential_kinetic_energy_elastica_incremental}
    \begin{align}
        \label{eq:potential_energy_elastica_incremental}
         & \mV(u, v) =
        \frac{1}{2} \int_0^{l} A(\lambda_0)\, u'(s,t)^2 ds
        +\frac{1}{2} \int_0^{l}\left( P v'(s,t)^2 +
        B(\lambda_0)\, v''(s,t)^2 \right) ds \,, \\
        \label{eq:kinetic_energy_elastica_incremental}
         & \mT(u, v) =
        \frac{1}{2} \int_0^{l} \left( \frac{\gamma_0}{\lambda_0} \left(\dot{u}(s,t)^2 + \dot{v}(s,t)^2\right) + \frac{\gamma_{r,0}}{\lambda_0}\dot{v}'(s,t)^2 \right) ds \,,
    \end{align}
\end{subequations}
where the fields $\{u(s,t),v(s,t)\}$ are current incremental fields.
Note that the initial linear mass density is divided by the prestretch, indicating that the \textit{current} density governs the incremental inertia of the rod. In fact, mass conservation requires $\gamma_0/\lambda_0=\gamma(\lambda_0)$, where $\gamma$ is the current linear mass density of the stretched rod.
Similarly, the \textit{current} rotational inertia is denoted as $\gamma_{r,0}/\lambda_0=\gamma_r(\lambda_0)$.

The governing equations~\eqref{eq:governing_beam_EB} are directly obtained through the application of Hamilton's principle to the Lagrangian $\mL=\mT-\mV$ constructed using the second-order functionals~\eqref{eq:potential_kinetic_energy_elastica_incremental}.

\subsection{Example of a rod made up of an incompressible hyperelastic material}
\label{sec:incompressible_rod}
The incremental potential, Eq.~\eqref{eq:potential_energy_elastica_incremental}, has been derived with reference to the elastica defined by two arbitrary strain-energy functions defining the current stiffnesses $A(\lambda_0)$ and $B(\lambda_0)$.
It is now shown that these two parameters can be evaluated for an incompressible non-linear elastic material, selected to model the rods.
To this purpose the analysis is developed in the static regime.

An initially isotropic, rectangular block of incompressible elastic material is considered, deformed under plane strain and subject to a uniaxial state of stress in-plane, $T_1\neq0, T_2=0$.
Its incremental constitutive response can be described through \cite{bigoni_2012}
\begin{equation*}
    \dot{S}_{11} = (2\mu_* - T_1) \deriv{u_1}{x_1} + \dot{p} \,, \qquad
    \dot{S}_{22} = 2\mu_* \deriv{u_2}{x_2} + \dot{p} \,,
\end{equation*}
where $\dot{S}_{ij}$ is the increment of the first Piola-Kirchhoff stress, $u_i$ the incremental displacement, $\mu_*$ the incremental modulus (corresponding to shearing inclined at $45^\circ$ with respect to the stress axes), and $\dot{p}$ the incremental Lagrange multiplier associated to the incompressibility constraint.
Assuming that plane stress prevails incrementally, $\dot{S}_{22} = 0$, and using the incompressibility constraint, $\dot{p}$ can be eliminated to yield
\begin{equation}
    \label{eq:S_11}
    \dot{S}_{11} = (4\mu_* - T_1) \deriv{u_1}{x_1} \,.
\end{equation}

The incremental equilibrium equation along the $x_1$ direction
\begin{equation*}
    \deriv{\dot{S}_{11}}{x_1} + \deriv{\dot{S}_{12}}{x_2} = 0 \,,
\end{equation*}
can be integrated over the current thickness $h$ of the block, so that a subsequent substitution of Eq.~\eqref{eq:S_11} leads to
\begin{equation}
    \label{eq:axial_incremental_equilibrium}
    \int_{-h/2}^{h/2} \deriv{\dot{S}_{11}}{x_1} \,dx_2 = (4\mu_* - T_1) \int_{-h/2}^{h/2} \nderiv{u_1}{x_1}{2}\,dx_2 = 0 \,,
\end{equation}
where the assumption of vanishing traction at $x_2=\pm h/2$ has been used.

The incremental flexural equilibrium can also be retrieved.
To this purpose, for a perturbation from the current uniaxial stress state, Biot~\cite{biot_1965} has shown that the incremental equilibrium requires
\begin{equation}
    \label{eq:flexural_incremental_equilibrium}
    \nderiv{}{x_1}{2} \int_{-h/2}^{h/2} x_2 \dot{S}_{11}\,dx_2 + T_1 \nderiv{}{x_1}{2} \int_{-h/2}^{h/2} u_2\,dx_2 = 0,
\end{equation}
where the first integral can be recognized to be the incremental bending moment.

By adopting the incremental kinematics of an Euler-Bernoulli beam (satisfying the unshearability condition)
\begin{equation}
    \label{eq:kinematics_EB}
    u_1(x_1,x_2) = u(x_1) - x_2 \deriv{v(x_1)}{x_1} \,, \qquad
    u_2(x_1,x_2) = v(x_1) \,,
\end{equation}
and using Eq.~\eqref{eq:S_11}, the axial and flexural equilibrium equations~\eqref{eq:axial_incremental_equilibrium} and~\eqref{eq:flexural_incremental_equilibrium} become
\begin{subequations}
    \label{eq:incremental_equilibrium}
    \begin{align}
        \label{eq:incremental_equilibrium_u} & (4\mu_* - T_1) h \nderiv{u(x_1)}{x_1}{2} = 0 \,,                                             \\
        \label{eq:incremental_equilibrium_v} & (4\mu_* - T_1)\frac{h^3}{12} \nderiv{v(x_1)}{x_1}{4} - T_1 h \nderiv{v(x_1)}{x_1}{2} = 0 \,.
    \end{align}
\end{subequations}
By noting that $T_1 h$ is the resultant axial load, so that $T_1 h=P$, a direct comparison between equations~\eqref{eq:incremental_equilibrium} and~\eqref{eq:governing_beam_EB} provides the identification of the current stiffnesses $A(\lambda_0)$ and $B(\lambda_0)$ as
\begin{equation}
    \label{eq:current_stiffnesses}
    A(\lambda_0) = (4\mu_*(\lambda_0) - T_1(\lambda_0)) h(\lambda_0) \,, \qquad B(\lambda_0) = (4\mu_*(\lambda_0) - T_1(\lambda_0)) h(\lambda_0)^3/12 \,,
\end{equation}
where the explicit dependence on the current pre-stretch $\lambda_0$ has been highlighted.
For instance, for a Mooney-Rivlin material $\mu_*=\mu_0(\lambda_0^2+\lambda_0^{-2})/2$ and $T_1=\mu_0(\lambda_0^2-\lambda_0^{-2})$. Therefore, expressions~\eqref{eq:current_stiffnesses} become
\begin{equation*}
    A(\lambda_0) = \mu_0(\lambda_0 + 3\lambda_0^{-3})h_0 \,, \qquad B(\lambda_0) = \mu_0(\lambda_0^{-1} + 3\lambda_0^{-5})h_0^3/12 \,,
\end{equation*}
where $h_0=h/\lambda_0$ is the initial thickness of the block and $\mu_0$ the initial shear modulus of the material.

\section{Full expression for the effective constitutive tensor}
\label{sec:homogenized_constitutive_tensor_grid}
The complete analytic expression for the effective constitutive tensor of the lattice analyzed in Section~\ref{sec:grid} is here reported.
The resulting tensor is made dimensionless as follows
\begin{subequations}
    \begin{equation}
        \label{eq:grid_C}
        \fC =  \frac{A}{l} \, \Bar{\fC}(\underbrace{p_1,\,p_2}_{\text{prestress}},\underbrace{\Lambda_1,\,\Lambda_2,\,\kappa,\,\alpha}_{\text{microstructure}}) \,,
    \end{equation}
    where the non-dimensional tensor-valued function $\Bar{\fC}(p_1,p_2,\Lambda_1,\Lambda_2,\kappa,\alpha)$ can be decomposed as
    \begin{equation}
        \label{eq:grid_C_contributions}
        \Bar{\fC}(p_1,p_2,\Lambda_1,\Lambda_2,\kappa,\alpha) = \Bar{\fC}^{\text{G}}(p_1,p_2,\Lambda_1,\Lambda_2,\alpha) + \Bar{\fC}^{\text{S}}(\kappa,\alpha) \,,
    \end{equation}
    with $\Bar{\fC}^{\text{G}}$ and $\Bar{\fC}^{\text{S}}$ being, respectively, the contribution of the rod's grid and the diagonal springs.
\end{subequations}
The components of $\Bar{\fC}^{\text{G}}$ in the basis $\{\be_1,\be_2\}$ read (components equal by symmetry are omitted)
\begin{dgroup*}[style={\footnotesize},breakdepth={20}]
    \begin{dmath*}
        \Bar{\fC}^{\text{G}}_{1111} = \frac{1}{2 d \sin\alpha}
        \left(\sinh \left(\frac{\sqrt{p_2}}{2}\right) \left(\sqrt{p_1} p_2 \phi \cosh \left(\frac{\sqrt{p_1}}{2}\right) \left( \cos (4 \alpha) \left(\Lambda_1^2-p_2 \phi\right)+4 \Lambda_1^2 \cos (2 \alpha)+11\Lambda_1^2+p_2 \phi\right)
            -2 \sinh \left(\frac{\sqrt{p_1}}{2}\right) \left( \cos (4 \alpha) \left(\Lambda_1^2 \left(p_1+p_2 \phi\right)-p_2^2 \phi^2\right)
                +\Lambda_1^2 \left(p_1+p_2\phi\right) \left(4 \cos (2 \alpha) +11\right)
                +p_2^2 \phi^2\right)\right) + p_1 \sqrt{p_2} \sinh \left(\frac{\sqrt{p_1}}{2}\right) \cosh \left(\frac{\sqrt{p_2}}{2}\right) \left(\cos (4 \alpha) \left(\Lambda_1^2-p_2 \phi\right)+4 \Lambda_1^2 \cos (2 \alpha)+ 11\Lambda_1^2+p_2 \phi\right)\right)
        ,
    \end{dmath*}
    \begin{dmath*}
        \Bar{\fC}^{\text{G}}_{1122} = \frac{4 \sin\alpha \cos^2\alpha}{d}
        \left(\sinh \left(\frac{\sqrt{p_2}}{2}\right) \left(\sqrt{p_1} p_2   \phi \cosh \left(\frac{\sqrt{p_1}}{2}\right) \left(\Lambda_1^2-p_2 \phi\right)-2 \sinh \left(\frac{\sqrt{p_1}}{2}\right) \left(\Lambda_1^2 \left(p_1+p_2   \phi\right)-p_2^2 \phi^2\right)\right)+p_1 \sqrt{p_2} \sinh \left(\frac{\sqrt{p_1}}{2}\right) \cosh \left(\frac{\sqrt{p_2}}{2}\right) \left(\Lambda_1^2-p_2 \phi\right)\right)
        ,
    \end{dmath*}
    \begin{dmath*}
        \Bar{\fC}^{\text{G}}_{1112} =
        \frac{-2 \cos\alpha}{d}
        \left( \sinh \left(\frac{\sqrt{p_2}}{2}\right) \left(2 \sinh \left(\frac{\sqrt{p_1}}{2}\right) \left(\cos (2 \alpha) \left(\Lambda_1^2 \left(p_1+p_2 \phi\right)-p_2^2 \phi^2\right)+\Lambda_1^2 \left(p_1+p_2 \phi\right)+p_2^2 \phi^2\right)-\sqrt{p_1} p_2 \phi \cosh \left(\frac{\sqrt{p_1}}{2}\right) \left(\cos (2 \alpha) \left(\Lambda_1^2-p_2 \phi\right)+\Lambda_1^2+p_2 \phi\right)\right)+p_1 \sqrt{p_2} \sinh \left(\frac{\sqrt{p_1}}{2}\right) \cosh \left(\frac{\sqrt{p_2}}{2}\right) \left(\cos (2 \alpha) \left(p_2 \phi-\Lambda_1^2\right)-\Lambda_1^2-p_2 \phi\right)\right)
        ,
    \end{dmath*}
    \begin{dmath*}
        \Bar{\fC}^{\text{G}}_{1121} = \frac{4 \cos\alpha}{d}
        \left(  \sinh \left(\frac{\sqrt{p_2}}{2}\right) \left(2 \sinh \left(\frac{\sqrt{p_1}}{2}\right) \left(p_1 p_2 \phi-\cos^2\alpha \left(\Lambda_1^2 \left(p_1+p_2   \phi\right)-p_2^2   \phi^2\right)\right)+\sqrt{p_1} p_2   \phi \cos^2\alpha \cosh \left(\frac{\sqrt{p_1}}{2}\right) \left(\Lambda_1^2-p_2 \phi\right)\right)+p_1 \sqrt{p_2}   \cos^2\alpha \sinh \left(\frac{\sqrt{p_1}}{2}\right) \cosh \left(\frac{\sqrt{p_2}}{2}\right) \left(\Lambda_1^2-p_2 \phi\right)\right)
        ,
    \end{dmath*}
    \begin{dmath*}
        \Bar{\fC}^{\text{G}}_{2222} =
        \frac{2 \sin\alpha }{d} \left(  \sinh \left(\frac{\sqrt{p_2}}{2}\right) \left(\sqrt{p_1} p_2  \phi \cosh \left(\frac{\sqrt{p_1}}{2}\right) \left(\cos (2 \alpha) \left(p_2 \phi-\Lambda_1^2\right)+\Lambda_1^2+p_2 \phi\right)-2 \sinh \left(\frac{\sqrt{p_1}}{2}\right) \left(-\cos (2 \alpha) \left(\Lambda_1^2 \left(p_1+p_2   \phi\right)-p_2^2   \phi^2\right)+\Lambda_1^2 \left(p_1+p_2   \phi\right)+p_2^2   \phi^2\right)\right)+p_1 \sqrt{p_2}   \sinh \left(\frac{\sqrt{p_1}}{2}\right) \cosh \left(\frac{\sqrt{p_2}}{2}\right) \left(\cos (2 \alpha) \left(p_2 \phi-\Lambda_1^2\right)+\Lambda_1^2+p_2 \phi\right)\right)
        ,
    \end{dmath*}
    \begin{dmath*}
        \Bar{\fC}^{\text{G}}_{2212} =
        \frac{4   \sin^2\alpha \cos\alpha }{d}
        \left(\sinh \left(\frac{\sqrt{p_2}}{2}\right) \left(\sqrt{p_1} p_2   \phi \cosh \left(\frac{\sqrt{p_1}}{2}\right) \left(\Lambda_1^2-p_2 \phi\right)-2 \sinh \left(\frac{\sqrt{p_1}}{2}\right) \left(\Lambda_1^2 \left(p_1+p_2   \phi\right)-p_2^2   \phi^2\right)\right)+p_1 \sqrt{p_2} \sinh \left(\frac{\sqrt{p_1}}{2}\right) \cosh \left(\frac{\sqrt{p_2}}{2}\right) \left(\Lambda_1^2-p_2 \phi\right)\right)
        ,
    \end{dmath*}
    \begin{dmath*}
        \Bar{\fC}^{\text{G}}_{2221} =
        \frac{2   \cos\alpha }{d}
        \left(\sinh \left(\frac{\sqrt{p_2}}{2}\right) \left(\sqrt{p_1} p_2   \phi \cosh \left(\frac{\sqrt{p_1}}{2}\right) \left(\cos (2 \alpha) \left(p_2 \phi-\Lambda_1^2\right)+\Lambda_1^2+p_2 \phi\right)-2 \sinh \left(\frac{\sqrt{p_1}}{2}\right) \left(-\cos (2 \alpha) \left(\Lambda_1^2 \left(p_1+p_2   \phi\right)-p_2^2   \phi^2\right)+\Lambda_1^2 \left(p_1+p_2   \phi\right)+p_2 \phi \left(2 p_1+p_2   \phi\right)\right)\right)+p_1 \sqrt{p_2} \sinh \left(\frac{\sqrt{p_1}}{2}\right) \cosh \left(\frac{\sqrt{p_2}}{2}\right) \left(\cos (2 \alpha) \left(p_2 \phi-\Lambda_1^2\right)+\Lambda_1^2+p_2 \phi\right)\right)
        ,
    \end{dmath*}
    \begin{dmath*}
        \Bar{\fC}^{\text{G}}_{1212} = \frac{-2  \sin\alpha }{d}
        \left(\sinh \left(\frac{\sqrt{p_2}}{2}\right) \left(2 \sinh \left(\frac{\sqrt{p_1}}{2}\right) \left(\cos (2 \alpha) \left(\Lambda_1^2 \left(p_1+p_2   \phi\right)-p_2^2   \phi^2\right)+\Lambda_1^2 \left(p_1+p_2   \phi\right)+p_2^2   \phi^2\right)-\sqrt{p_1} p_2   \phi \cosh \left(\frac{\sqrt{p_1}}{2}\right) \left(\cos (2 \alpha) \left(\Lambda_1^2-p_2 \phi\right)+\Lambda_1^2+p_2 \phi\right)\right)+p_1 \sqrt{p_2} \sinh \left(\frac{\sqrt{p_1}}{2}\right) \cosh \left(\frac{\sqrt{p_2}}{2}\right) \left(\cos (2 \alpha) \left(p_2 \phi-\Lambda_1^2\right)-\Lambda_1^2-p_2 \phi\right)\right)
        ,
    \end{dmath*}
    \begin{dmath*}
        \Bar{\fC}^{\text{G}}_{1221} = \frac{-4   \sin\alpha }{d}
        \left(\sinh \left(\frac{\sqrt{p_2}}{2}\right) \left(2 \sinh \left(\frac{\sqrt{p_1}}{2}\right) \left(\cos^2\alpha \left(\Lambda_1^2 \left(p_1+p_2 \phi\right)-p_2^2 \phi^2\right)-p_1 p_2 \phi\right)+\sqrt{p_1} p_2 \phi \cos^2\alpha \cosh \left(\frac{\sqrt{p_1}}{2}\right) \left(p_2 \phi-\Lambda_1^2\right)\right)+p_1 \sqrt{p_2} \cos^2\alpha \sinh \left(\frac{\sqrt{p_1}}{2}\right) \cosh \left(\frac{\sqrt{p_2}}{2}\right) \left(p_2 \phi-\Lambda_1^2\right)\right)
        ,
    \end{dmath*}
    \begin{dmath*}
        \Bar{\fC}^{\text{G}}_{2121} = \frac{p_1 \sqrt{p_2} }{d}
        \sinh \left(\frac{\sqrt{p_1}}{2}\right) \cosh \left(\frac{\sqrt{p_2}}{2}\right) \left( \sin\alpha \left(\Lambda_1^2-5 p_2 \phi\right)+ \sin (3 \alpha) \left(\Lambda_1^2-p_2 \phi\right)+4 \csc (\alpha) \left(p_1+p_2   \phi\right)\right)-2 \sin\alpha \sinh \left(\frac{\sqrt{p_2}}{2}\right) \left(2 \sinh \left(\frac{\sqrt{p_1}}{2}\right) \left( \cos (2 \alpha) \left(\Lambda_1^2 \left(p_1+p_2 \phi\right)-p_2^2 \phi^2\right)+2 \csc^2(\alpha) \left(p_1+p_2   \phi\right)^2+\Lambda_1^2 \left(p_1+p_2 \phi\right)-p_2 \phi \left(4 p_1+3 p_2   \phi\right)\right)-\sqrt{p_1} p_2 \phi \cosh \left(\frac{\sqrt{p_1}}{2}\right) \left(\cos (2 \alpha) \left(\Lambda_1^2-p_2 \phi\right)+2 \csc ^2(\alpha) \left(p_1+p_2 \phi\right)+ \left(\Lambda_1^2-3 p_2 \phi\right)\right)\right)
        ,
    \end{dmath*}
\end{dgroup*}
where
\begin{dmath*}[style={\footnotesize}]
    d =
    e^{-\frac{1}{2} \left(\sqrt{p_1}+\sqrt{p_2}\right)}
    \Lambda_1^2 \left(\left(e^{\sqrt{p_1}} \left(\sqrt{p_1}-2\right)+\sqrt{p_1}+2\right) \left(e^{\sqrt{p_2}}-1\right) p_2   \phi-2 \left(e^{\sqrt{p_1}}-1\right) p_1 \left(e^{\sqrt{p_2}}-1\right)+\left(e^{\sqrt{p_1}}-1\right) p_1 \left(e^{\sqrt{p_2}}+1\right) \sqrt{p_2}\right)
    .
\end{dmath*}
The constitutive tensor $\Bar{\fC}^{\text{S}}$ ruling the effect of diagonal springs can be written as
\begin{align*}
    \Bar{\fC}^{\text{S}}_{1111} & = \kappa \frac{5+3 \cos (2 \alpha)}{4\sin\alpha}  \,,                                                                                                                                                                      \\
    \Bar{\fC}^{\text{S}}_{1112} & = \Bar{\fC}^{\text{S}}_{1121} = \Bar{\fC}^{\text{S}}_{1211} = \Bar{\fC}^{\text{S}}_{1121} = \Bar{\fC}^{\text{S}}_{2111} = \kappa  \cos \alpha  \,,                                                                         \\
    \Bar{\fC}^{\text{S}}_{1122} & = \Bar{\fC}^{\text{S}}_{2211} = \Bar{\fC}^{\text{S}}_{1212} = \Bar{\fC}^{\text{S}}_{1221} = \Bar{\fC}^{\text{S}}_{2112} = \Bar{\fC}^{\text{S}}_{2121} = \Bar{\fC}^{\text{S}}_{2222} = \frac{1}{2} \kappa  \sin \alpha  \,, \\
    \Bar{\fC}^{\text{S}}_{1222} & = \Bar{\fC}^{\text{S}}_{2122} = \Bar{\fC}^{\text{S}}_{2212} = \Bar{\fC}^{\text{S}}_{2221} = 0 \,.
\end{align*}

\end{document}